\documentclass[sn-mathphys,Numbered]{sn-jnl}
\usepackage{graphicx}%
\usepackage{array}
\usepackage{multirow}%
\usepackage{mathtools}
\usepackage{amsmath,amssymb,amsfonts}%
\usepackage{amsthm}%
\usepackage{mathrsfs}%
\usepackage{xcolor}%
\usepackage{textcomp}%
\usepackage{manyfoot}%
\usepackage{algpseudocode}%
\usepackage{listings}%
\usepackage{float}
\usepackage{booktabs} 
\usepackage{adjustbox}
\usepackage[linesnumbered, ruled, vlined]{algorithm2e}
\usepackage{caption}
\usepackage{subcaption}
\usepackage{tcolorbox}
\usepackage{multicol}
\usepackage{tikz}
\usepackage[misc]{ifsym}
\usepackage{bbding}
\usepackage{thmtools,thm-restate}
\usepackage{varioref}
\usepackage{verbatim}
\usepackage{enumerate}
\usepackage{enumitem}
\usepackage{soul}
\usepackage{longtable}
\usepackage{url}
\usepackage{hyperref}
\hypersetup{
    colorlinks,
    citecolor=blue,
    filecolor=black,
    linkcolor=red,
    urlcolor=green
}

\theoremstyle{thmstyleone}%
\newtheorem{theorem}{Theorem}

\theoremstyle{thmstyletwo}%
\theoremstyle{thmstylethree}%
\newtheorem{definition}{Definition}%
\begin{document}
\title{A Regret-Aware Framework for Effective Social Media Advertising}
\author{\fnm{Poonam} \sur{Sharma}}\email{poonam.sharma@iitjammu.ac.in}
\author{\fnm{Dildar} \sur{Ali}}\email{2021rcs2009@iitjammu.ac.in}
\author*{\fnm{Suman} \sur{Banerjee}}\email{suman.banerjee@iitjammu.ac.in}
\affil{\orgdiv{Department of Computer Science and Engineering}, \orgname{Indian Institute of Technology Jammu}, \orgaddress{\postcode{181221}, \state{Jammu and Kashmir}, \country{India}}}
\maketitle
\begin{abstract}
\emph{Social Media Advertisement} has emerged as an effective approach for promoting the brands of a commercial house. Hence, many of them have started using this medium to maximize the influence among the users and create a customer base. In recent times, several companies have emerged as Influence Provider who provides views of advertisement content depending on the budget provided by the commercial house. In this process, the influence provider tries to exploit the information diffusion phenomenon of a social network, and a limited number of highly influential users are chosen and activated initially. Due to diffusion phenomenon, the hope is that the advertisement content will reach a large number of people. 
\par Now, consider that a group of advertisers is approaching an influence provider with their respective budget and influence demand. Now, for any advertiser, if the influence provider provides more or less influence, it will be a loss for the influence provider. It is an important problem from the influence provider's point of view, as it is important to allocate the seed nodes to the advertisers so that the loss is minimized. In this paper, we study this problem, which we formally referred to as \textbf{R}egret \textbf{M}inimization in \textbf{S}ocial \textbf{M}edia \textbf{A}dvertisement Problem. We propose a `noble regret model' that captures the aggregated loss encountered by the influence provider while allocating the seed nodes. We have shown that this problem is a computationally hard problem to solve. We have proposed three efficient heuristic solutions to solve our problem. All proposed solution approaches have been analyzed to understand their time and space requirements. They have been implemented with real-world social network datasets, and several experiments have been conducted. From the experiments, we have observed that the proposed solution approaches lead to the allocation of seed nodes among the advertisers, which results in much less regrets compared to many baseline methods. 
\end{abstract}

\section{Introduction} \label{Sec:Intro}
A social network is defined as an interconnected structure among a group of human beings created for social interactions \cite{wasserman1994social}. Due to the advancement of wireless internet technologies and hand-holding mobile devices, the use of social media has become an integral part of human life. In recent times, several social network datasets have been publicly available. As per the reported recent statistics, every year, the number of active social media users is increasing \footnote{\url{https://backlinko.com/social-media-users}}. Many problems have been studied in this domain, such as community detection \cite{chunaev2020community,papadopoulos2012community,javed2018community}, information diffusion \cite{kimura2006tractable,bakshy2012role}, influence maximization \cite{li2018influence,azaouzi2021new,banerjee2020survey}, opinion dynamics \cite{das2014modeling,de2016learning}, and many more. Commercial houses have exploited social media to promote their brands and make advertisements. In this case, the key phenomenon of social networks that have been exploited is the \emph{diffusion of information}  \cite{bakshy2012role}. This phenomenon says that every user of an online social network tends to share information with their neighbor. This way, the information, innovation, ideas, etc., spreads through the network and goes viral. Commercial houses have exploited this phenomenon to promote their brands and create their customer base \cite{richardson2002mining,domingos2001mining}. The key question that arises in this context is how we can select a limited number of highly influential users from the network such that the influence in the network gets maximized. This question remains an active area of research in the domain of Social Network Analysis. An enormous amount of research on this problem leads to a significant amount of literature \cite{li2018influence}, \cite{banerjee2020survey}, \cite{liang2023targeted}, \cite{azaouzi2021new}, \cite{chen2016robust}, \cite{chen2011influence}, \cite{peng2021dynamic}, \cite{cautis2019adaptive}.

\par In recent times, several media houses have emerged that conduct social media marketing. Technically, they are called \emph{Influence Provider}. In this approach, a set of commercial houses approaches an influence provider with their respective influence demand and budget. Here, the payment rule is as follows. If the influence provider provides the required or more influence, the advertiser will give the mentioned budget; otherwise, a partial payment (on a pro-rata basis) will be made to the influence provider. Now, if the influence provider provides more or less influence than the influence demand, it is a loss for the influence provider. This happens due to the following reasons. Consider the influence provider provides less influence to an advertiser, and as per the payment rule, the influence provider will receive a partial payment, which is a loss.

\begin{figure}[h!]

    \centering

    \includegraphics[width=\textwidth]{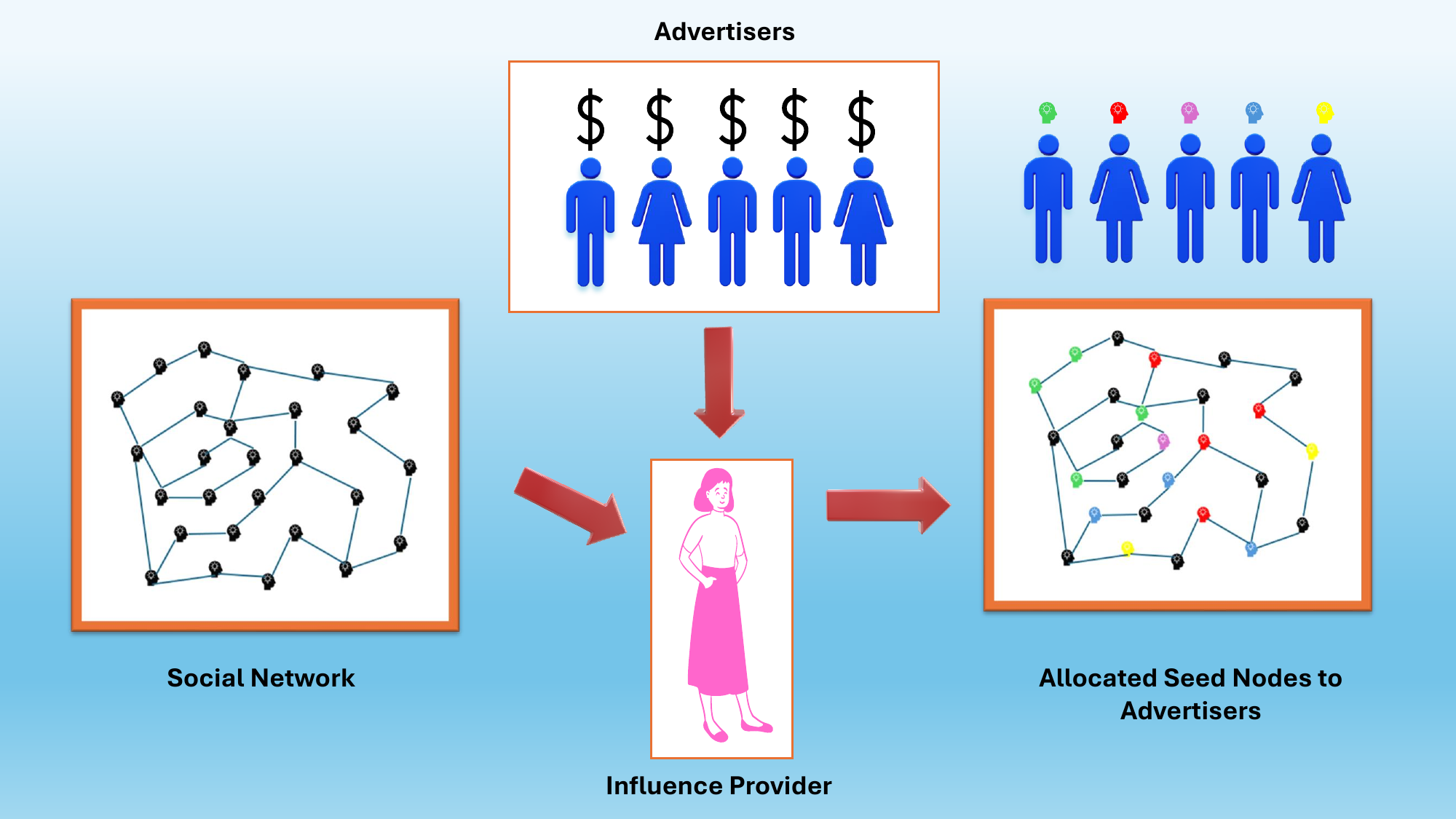} 

    \caption{A schematic diagram showing the allocation of seed nodes to advertisers in a social network.}

    \label{fig:schematic_diagram}

\end{figure}

On the other hand, if the influence provider provides more influence, then there is no additional payment for the extra influence. However, this extra influence can be provided to advertisers whose demand for influence has yet to be satisfied. Now, the task of the influence provider is to return respective seed sets to the advertisers so that this leads to the least regret. This is the key problem that we have studied in this paper. This process has been shown diagrammatically in Figure \ref{fig:schematic_diagram}.

\par `Regret' in the context of billboard advertisement has been studied, and some studies are available. To the best of our knowledge, Zhang et al. \cite{zhang2021minimizing} was the first to study the Regret Minimization Problem. They posed this problem as a discrete optimization problem and showed that this problem is \textsf{NP-hard} and had to approximate beyond any constant factor. They proposed two local search-based heuristic solutions for this problem. Later, Ali et al. \cite{ali2023efficient,ali2024toward} proposed several heuristic solutions for this problem, and from the experimental evaluations, it can be observed that the proposed solutions by Ali et al.\cite{ali2023efficient} lead to much less regret compared to the methods proposed by Zhang et al \cite{zhang2021minimizing}. However, to the best of my knowledge, regret in the context of social media advertisements has yet to be defined. In this paper, we bridge this gap by defining the notion of regret in the context of social media advertisement and propose several seed selection strategies to minimize the regret. Mainly, we make the following contributions:

\begin{itemize}
\item We have formally defined `Regret' in the context of social media advertisements.
\item We introduce and study the noble problem of \textbf{R}egret \textbf{M}inimization in \textbf{S}ocial \textbf{M}edia \textbf{A}dvertisement, and formally, we call it RMSMA Problem.
\item We formulate this problem as a discrete optimization problem and show that the problem is NP-hard and hard to approximate beyond any constant factor.
\item We propose two solution approaches for this problem with their detailed analysis and complexity calculation.
\item The proposed solution approaches have been experimentally evaluated to show their effectiveness and efficiency.
\end{itemize}

\par   The rest of the paper is organized as follows. Section \ref{Sec:Related_work} describes relevant studies from the literature. Section \ref{Sec:BPD} describes the relevant concepts and defines our problem formally. Section \ref{Sec:PS} describes the proposed solution approaches. The experimental evaluation has been described in Section \ref{Sec:Experimental_Details}. Finally, Section \ref{Sec:CFD} concludes our study and reports a few future research directions.

\section{Related Work}\label{Sec:Related_work}
In this section, we describe some relevant studies from the literature. Our study comes under the two broad domains, namely, influence and its analysis in social network and regret analysis in advertisement. In the following two subsections, we describe the relevant studies on both of this topics.
\subsection{Influence and its Analysis in Social Networks}
In past two decades, there has been significant amount of study on the information diffusion in social networks and its applications. Several diffusion models have been introduced and among many the most well studied problem in the context of information diffusion is the Problem of Influence Maximization. Given a social network, a diffusion model, and a positive integer $k$, this problem aims at identifying $k$ many highly influential users in the network such that their initial activation leads to the maximum influence in the network. Due to the direct application of this problem in the domain of viral marketing, this problem has been studied in different variants and a plethora of solution methodologies have been proposed in the literature.
\par To the best of our knowledge, Domingos and Reichardson were the first to introduce the problem of influence maximization in the context of viral marketing. Though, Kempe at al. \cite{DBLP:conf/kdd/KempeKT03,DBLP:conf/icalp/KempeKT05,DBLP:journals/toc/KempeKT15} were the first to study this problem as a discrete optimization problem. They showed that if the underlying diffusion model is linear threshold or independent cascade, the influence function becomes non-negative, monotone, and submodular. Also they showed that the iterative greedy algorithm which works based on marginal gain computation, leads to $(1-\frac{1}{e})$-factor approximation guarantee. Subsequently there are several studies and the proposed solution methodologies have been classified into many categories such as approximation algorithms, heuristic solution approaches, community-based solution approaches, reverse reachable set computation-based approaches, and many more. The solution methodologies that belongs to the approximation algorithm category, provides a seed set of a given cardinality whose expected influence is a constant factor approximation with respect to an optimal seed set of same cardinality. The first approximation algorithm was proposed by Kempe at al. \cite{DBLP:conf/kdd/KempeKT03}. Subsequently, Leskovic et al. \cite{DBLP:conf/kdd/LeskovecKGFVG07} proposed an approximation algorithm having approximation ratio same as one proposed by Kempe at al. \cite{DBLP:conf/kdd/KempeKT03}. However, the number of function evaluation performed by their method is much less. Hence, the method proposed by Leskovic et al. \cite{DBLP:conf/kdd/LeskovecKGFVG07}. Subsequently, there are many studies in this directions.
\par Subsequently, a different class of approximation algorithm has been proposed whose initial study was done by Brogs et al. \cite{DBLP:conf/soda/BorgsBCL14} which works based on the computation of reverse reachable sets. The basic idea of this approach is as follows. For every node, a set is constructed which contains the set of other nodes which can influence this node. This set is called as the reverse reachable set or RR set. For all the nodes of the network, once the reverse reachable sets are computed, a small subset of nodes are find out using some coverage algorithm such that the intersection of these nodes with RR sets of the nodes becomes non empty. Following the study by Brogs et al. \cite{DBLP:conf/soda/BorgsBCL14}, several approaches have been proposed in this direction such as TIM \cite{DBLP:conf/sigmod/TangXS14}, TIM+ \cite{DBLP:conf/sigmod/TangXS14}, IMM \cite{DBLP:conf/sigmod/TangSX15}, and many more. The other category of solution approaches can process a million scale social network in couple of seconds.  However, solution methodologies of this kind does not give any guarantee on the quality of solution. Solution methodology of this kind includes IRIE \cite{jung2012irie}, Staticgreedy \cite{cheng2013staticgreedy}, IMRANK \cite{cheng2014imrank} and many more. Another kind of solution methodologies exist where the community structure of the network has been exploited. We call such methodologies as community-based solution approaches. There are many solution approaches which belongs to this category \cite{chen2014cim,bozorgi2016incim,shang2017cofim,bozorgi2017community}. There are another class of solution approaches that uses optimization tools from soft computing and nature inspired techniques. We call such approaches as soft computing-based approaches. Plenty of approaches have been proposed which belongs to this category \cite{gong2016influence,csimcsek2018using,bucur2016influence}.  
\par In past one decade, the problem of influence maximization has been studied in different variants. Nugyan et al. introduced the Budgeted Influence Maximization Problem where every node of the network is assigned with a selection cost and a fixed amount of budget has been allocated. This problem asks to choose a seed set such that the total selection cost of the seed set is less than or equal to the allocated budget. At the same time, the influence spread by the seed set is maximum among all the seed sets of the same budget. They proposed a $(1-\frac{1}{\sqrt{e}})$-factor approximation algorithm and a heuristic solution for this problem. Later, there are a few studies on this problem which includes a community-based solution approach proposed by Banerjee et al. \cite{banerjee2019combim}, a bunch of heuristic solution proposed by Han et al. \cite{han2014balanced} and many more. Another kind of influence diffusion problem that has been studied is the Target Set Selection Problem and its different variants. In this problem, every node of the network is assigned with a positive number which we call as threshold and this value signifies that this node will be influenced once that many nodes among its neighbors are already influenced. There are many studies on this problem \cite{raghavan2019branch,cicalese2014latency,chu2023fpt,hartmann2018target}.   
\subsection{Regret Analysis in Computational Advertisement}
To the best of our knowledge, there has been limited amount of literature available on the regret analysis in `computational advertisement'. The first study on this topic was done by Zhang et al. \cite{zhang2021minimizing} who formulated the regret minimization problem in the context of billboard advertisement. They posed this problem as a discrete optimization problem and showed that this problem is NP-hard and hard to approximate beyond any constant factor. They also proposed several heuristic solution and experimented with benchmark datasets. Later on there are several studies on this problem by Ali et al. \cite{DBLP:conf/sac/Ali0P24,DBLP:journals/corr/abs-2401-16464,DBLP:conf/aaai/AliB0P23}. In \cite{DBLP:conf/aaai/AliB0P23}, Ali et al. proposed a local search-based approach. In \cite{DBLP:conf/sac/Ali0P24}, Ali et al. studied the regret minimization problem in billboard advertisement under zonal influence constraint. They proposed a sampling-based approach to solve this problem.

\par However, to the best of our knowledge, the regret minimization problem has not been studied in the context of social media advertisement context. In this paper, we bridge this gap by studying the regret minimization problem in the context of social networks. Next, we proceed to describe the problem statement subsequently.

\section{Background and Problem Definition} \label{Sec:BPD}
In this section, we describe the background information and defines our problem formally. Initially, we start by describing the notion of social network. 
\subsection{Social Network and Its Diffusion Phenomena}  A \emph{social network} is defined as an interconnected structure among a group of users and often represented by a simple, weighted, undirected graph $\mathcal{G}(\mathcal{V}, \mathcal{E, \mathcal{P}})$ where the set of users $\{u_1, u_2, \ldots, u_m\}$ constitute the vertex set $\mathcal{V}$. There is an edge between the user $u_i$ and $u_j$, if $u_i$ and $u_j$ has social relationship. $\mathcal{P}$ is the edge weight function that maps each edge to its corresponding influence probability, i.e, $\mathcal{P}: \mathcal{E} \longrightarrow (0,1]$. For any edge $(u_iu_j)$, its influence probability is denoted by $\mathcal{P}(u_i,u_j)$ and defined as the probability by which the user $u_i$ will be able to influence $u_j$. For any user $u_i$, the set of neighbors is denoted as $N(u_i)$ and defined as the set of users with which $u_i$ is directly linked, i.e., $N(u_i)=\{u_j: (u_iu_j) \in \mathcal{E}\}$.
\par One of the important phenomenon of social network is the diffusion of information which says that the users of a social network tend to share the information and in this way information propagates through the network and become viral. This phenomenon has been adopted by the commercial houses for promoting their brands and creating influence among its prospective customers. Several models have been introduced in the literature to study the diffusion process in a social network. One of the widely studied is the Independent Cascade Model which has been stated in Definition \ref{Def:IC_Mdoel}.   


\begin{definition}[Independent Cascade Model]\cite{wang2012scalable}
\label{Def:IC_Mdoel}
In Independent Cascade Model, the rules for the diffusion are as follows:
\begin{description}
    \item[1.] The diffusion process starts from a set of initially active nodes (called seed nodes) information is diffused in discrete time stamps.
    \item[2.] A node can be in either of two states: \emph{active} (influenced) or \emph{inactive} (non-influenced).
    \item[3.] A node can change its state from inactive to active, but not vice versa.
    \item[4.] The diffusion process stops when no more node activation is possible.
\end{description}
\end{definition}

\subsection{Influence Maximization Problem}
In general, the users of a social network is formed by rational and self interested human beings, and hence, they need to incentivize if they are chosen as seed nodes. This has been formalized as the cost function $\mathcal{C}: \mathcal{V} \longrightarrow \mathbb{R}^{+}$. For any user $u \in \mathcal{V}$, the selection cost of $u$ is denoted by $\mathcal{C}(u)$. A fixed amount of budget $\mathcal{B}$ is available. Now, given this information, the social influence maximization problem (cost version) is stated in Definition \ref{Def:SIM_Problem}

\begin{definition}[Influence Maximization Problem (cost version)] \label{Def:SIM_Problem}
Given a social network $\mathcal{G}(\mathcal{V}, \mathcal{E}, \mathcal{P})$, a cost function $\mathcal{C}: \mathcal{V} \longrightarrow \mathbb{R}^{+}$, and a budget $\mathcal{B}$, this problem asks to choose a subset of the users such that the total selection cost is less than the budget and the influence in the network is maximized. Mathematically, this problem can be expressed as a discrete optimization problem and written in Equation \ref{Eq:1}.

\begin{equation} \label{Eq:1}
S^{OPT} \longleftarrow \underset{S \subseteq \mathcal{V} \text{ and }\mathcal{C}(S) \leq \mathcal{B} }{argmax} \ \sigma(S)
\end{equation}    

\end{definition}

Here, $\mathcal{C}(S)$ denotes the total selection cost of all the nodes in $S$, i.e., $\mathcal{C}(S)=\underset{u \in S}{\sum} \ \mathcal{C}(u)$. As mentioned in \cite{DBLP:conf/kdd/KempeKT03}, this has been reported that this is a computationally hard problem to solve. The main result is reported in Theorem \ref{Th:Hard_Influence}.

\begin{theorem} \label{Th:Hard_Influence}
Given a social network $\mathcal{G}(\mathcal{V}, \mathcal{E}, \mathcal{P})$, and two positive integers $\mathcal{B}$ and $\ell$, even if the selection cost of all the users is $1$,  still it is \textsf{NP-complete} to decide whether $\mathcal{G}$ has an influence at least $\ell$.

\end{theorem}

\subsection{Social Media Advertisement through External Agent}
In recent times, several influence providers have emerged who conducts social media advertisement on a payment basis and the process goes in the following way. A commercial house approaches to an influence provider to obtain certain number of views of an advertisement content (e.g., video, animation etc.) on a payment basis. The payment rule is as follows: If the required or more influence is provided then the full payment will be made else a partial payment. Now, from the influence provider's point of view whether it provides more or less influence it is loss for them. Because, if it provides less influence then it will have the partial payment. On the other hand, if it provides more influence then the influence provider does not get any extra incentive for it. This loss has been mathematically formalized as regret. Depending upon the mentioned two cases, two different kinds of regret is possible:

\begin{itemize}
\item \textbf{Revenue Regret}: This kind of regret happens when the influence provider provides less influence compared to the influence demand. For any seed set $S$, the revenue regret associated with the seed set $S$ is denoted by $\mathcal{R}_{r}$ and can be defined using Equation \ref{Eq:Rev_Regret}.

\begin{equation} \label{Eq:Rev_Regret}
\mathcal{R}_{r}(S_{i})= \mathcal{B}_{i} \cdot (1- \gamma \cdot \frac{I(S_i)}{\sigma_i}) + \delta \cdot |S_{i}| 
\end{equation}

\item  \textbf{Excessive Regret}: This kind of regret happens when the influence provider provides more influence than the required.

\begin{equation} \label{Eq:Ex_Regret}
\mathcal{R}_{e}(S_{i})= \mathcal{B}_{i} \cdot \frac{I(S_{i}) - I_{i}}{I_{i}} + \delta \cdot |S_{i}| 
\end{equation}
\end{itemize}

\paragraph{\textbf{Regret Model.}} Following the above, we formulate the regret model for the influence provider perspective to assign seed set $S_{i}$ to the advertisers $a_{i}$ and it is stated in Definition \ref{Def:Reg_Model}.

\begin{definition}[The Regret Model] \label{Def:Reg_Model}
    Let, for the advertiser $a_i \in \mathcal{A}$, the allocated seed set is $S_{i}$ and the regret associated with this allocation is denoted by $\mathcal{R}(S_{i})$. This quantity can be defined by the following conditional equation:
    
 \[
 \mathcal{R}(S_{i}) = 
\begin{cases}
    \mathcal{B}_{i} \cdot (1- \gamma \cdot \frac{I(S_i)}{I_i}) + \delta \cdot |S_{i}|,& \text{if }  a_{i} \cdot I_{i} > I(S_{i}) \\
   \mathcal{B}_{i} \cdot \frac{I(S_{i}) - I_{i}}{I_{i}} + \delta \cdot |S_{i}| ,              & \text{if }  a_{i} \cdot I_{i} < I(S_{i})\\
     0,              & \text{otherwise}
\end{cases}
\]
Here, the fraction $\frac{I(S_i)}{I_i})$ is part of the satisfied influence for the advertiser $a_i$ and $\gamma$ is a parameter which is called as the penalty ratio due to the unsatisfied demand. 
\end{definition}

Note that in the regret model definition, we discriminate between small and large seed sets $S_{1}$ and $S_{2}$ with the same influence value with $|S_{1}| << |S_{2}|$ or $|S_{1}|>> |S_{2}|$. In practice, achieving lower regret with fewer seed sets is desirable. We discourage using a larger seed set because it consumes a higher cost than the smaller one while both return the same amount of influence value. Hence, the parameter $\delta$ is there as cardinality penalty ratio.


\subsection{Problem Definition}
Now, we define our problem formally. Consider a set of $\ell$ advertisers $\mathcal{A}=\{a_1, a_2, \ldots, a_{\ell}\}$ are approaching to an influence provider $\mathcal{X}$ with their respective influence demand and corresponding budget. For the $i$-th advertiser $a_i$, we denote its corresponding influence demand and budget by $\sigma_i$ and $\mathcal{B}_{i}$, respectively. This has been diagrammatically represented in Figure \ref{fig:schematic_diagram}. The influence provider has the access of the social network $\mathcal{G}$ and it returns the seed sets and their corresponding influence value to the respective advertiser. Here, the constraint is that for any pair of advertisers $a_i,a_j \in \mathcal{A}$, $S_i \cap S_j = \emptyset$. Consider, $\Pi=\{S_1, S_2, \ldots, S_{\ell}\}$ is a selection of $\ell$ seed sets. Here, we define the notion of feasible selection in Definition 
\ref{Def:Feassible_Selection}.
\begin{definition}[Feasible Selection] \label{Def:Feassible_Selection}
 Given a set of $\ell$ advertisers with their respective influence demand and budget $(\sigma_i, \mathcal{B}_{i})^{\ell}_{i=1}$, an influence provider $\mathcal{X}$, a social network $\mathcal{G}(\mathcal{V}, \mathcal{E}, \mathcal{P})$, a selection $\Pi=\{S_1, S_2, \ldots, S_{\ell}\}$ is said to be a feasible selection when for all $i \in \{1,2, \ldots, \ell \}$, $\mathcal{C}(S_i) \leq \mathcal{B}_{i}$.

\end{definition}
For any problem instance, the set of all possible feasible selection is denoted by $\Sigma$. 
Now, for any selection $\Pi$, the regret associated with this is denoted by $\mathbb{Q}(\Pi)$ and stated in Definition \ref{Def:Regret}.

\begin{definition} [Regret Associated with a Selection] \label{Def:Regret}
Given a set of $\ell$ advertisers $\mathcal{A}=\{a_1, a_2, \ldots, a_{\ell}\}$ with  their respective influence demand and corresponding budget $(\sigma_i,\mathcal{B}_{i})_{i=1}^{\ell}$, a social network $\mathcal{G}$, and a feasible selection $\Pi=\{S_1, S_2, \ldots, S_{\ell}\} \in \Sigma$, the regret associated with the selection $\Pi$ is denoted by $\mathbb{Q}(\Pi)$ and defined as the sum of the regret for each of the seed sets which has been written in Equation \ref{Eq:Regret}.
\begin{equation} \label{Eq:Regret}
\mathbb{Q}(\Pi)= \underset{S \in \{S_1, S_2, \ldots, S_{\ell}\}}{\sum} \mathcal{R}(S)
\end{equation}
\end{definition}

Now, we formally define the Regret Minimization in Social Media Advertisement Problem in Definition \ref{Def:Problem}.
\begin{definition}[Regret Minimization in Social Media Advertisement Problem] \label{Def:Problem}
Given a set of $\ell$ advertisers $\mathcal{A}=\{a_1, a_2, \ldots, a_{\ell}\}$  with  their respective influence demand and corresponding budget $(\sigma_i,\mathcal{B}_{i})_{i=1}^{\ell}$, and a social network $\mathcal{G}$, the Regret Minimization in Social Media Advertisement Problem (RMSMA Problem) asks to find out a selection that leads to the least regret for the influence provider. Mathematically, this problem can be posed as a discrete optimization problem and expressed in Equation \ref{Eq:Problem}.

\begin{equation} \label{Eq:Problem}
\Pi^{*} \longleftarrow \underset{\Pi \in \Sigma}{argmin} \ \mathbb{R}(\Pi)
\end{equation}
\end{definition}
 Here, $\Pi^{*}$ denotes the optimal selection. From the computational point of view, this problem can be posed as follows:
 
\begin{tcolorbox}
\underline{\textbf{RMSMA Problem}}\\
\textbf{Input:} Advertisers $\mathcal{A}=\{a_1, a_2, \ldots, a_{\ell}\}$, The Social Network $\mathcal{G}(\mathcal{V}, \mathcal{E}, \mathcal{P})$, Influence Demand and Budget $(\sigma_i,\mathcal{B}_{i})_{i=1}^{\ell}$. \\
\textbf{Problem:} Find out the seed sets and the expected influence $(S_{i}, I(S_i))_{i=1}^{n}$ such that for any $a_i,a_j \in \mathcal{A}$, $S_i \cap S_j = \emptyset$ and the regret as defined in Equation \ref{Eq:Regret} is minimized.
\vspace{0.2 cm}\\
\textbf{Output:} For all $a_i \in \mathcal{A}$, the corresponding Seed Set $S_i$ and its expected influence $I(S_i)$. 
\end{tcolorbox} 
The symbols and notations used in this paper has been mentioned in Table \ref{Table1:Notations}. Next, we proceed to describe the hardness result of the problem.

\begin{table}[h]
    \caption{Symbols and Notations with Their Interpretations}
    \label{Table1:Notations}
    \centering
    \begin{tabular}{@{}p{0.20\linewidth}p{0.75\linewidth}@{}}
        \toprule
        \textbf{Notation} & \textbf{Description} \\
        \midrule
        $\mathcal{G}(\mathcal{V}, \mathcal{E}, \mathcal{P})$ & The social network \\
        \hline
        $\mathcal{V}, \mathcal{E}$ & Nodes and edges of $\mathcal{G}$ \\
        \hline
        $n, m$ & Number of nodes and edges of $\mathcal{G}$ \\
        \hline
        $\mathcal{P}$ & Edge weight function of $\mathcal{G}$ \\ 
        \hline
        $(u_i u_j)$ & An arbitrary edge of $\mathcal{G}$ \\
        \hline
        $Pr(u_i, u_j)$ & Influence probability between users $u_i$ and $u_j$ \\
        \hline
        $N(u_i)$ & Neighbor set of user $u_i$ \\
        \hline
        $\mathcal{A}$ & Set of advertisers \\
        \hline
        $\ell$ & Number of advertisers \\
        \hline
        $S_i$ & Seed set for advertiser $a_i$ \\
        \hline
        $I(S_i)$ & Influence of seed set $S_i$ \\ 
        \hline
        $\mathcal{R}(S_i)$ & Regret for advertiser $a_i$ with seed set $S_i$ \\
        \hline
        $\mathcal{R}(u|S_i)$ & Marginal regret decrease for user $u$ wrt $S_i$ \\
        \hline
        $\sigma_i$ & Influence demand of advertiser $a_i$ \\
        \hline
        $\mathcal{B}_i$ & Budget of advertiser $a_i$ \\
        \hline
        $\mathcal{C}(u_i)$ & Selection cost of user $u_i$ \\ 
        \hline
        $\gamma$ & Influence penalty ratio \\
        \hline
        $\delta$ & Cardinality penalty ratio \\
        \hline
        $\lambda$ & Demand-supply ratio \\
        \hline
        $\omega$ & Average individual demand ratio \\
        \hline
        $\Pi$ & Allocation of seed set to advertiser \\
        \hline
        $k$ & Regret tolerance parameter \\
        \hline
        $\mathbb{R}_0^+$ & Set of non-negative real numbers \\
        \bottomrule
    \end{tabular}
\end{table}

\section{Proposed Solution Approaches} \label{Sec:PS}
In this section we describe the proposed solution approaches. Initially, we start by defining the notion of marginal decrease in regret which has been stated in Definition \ref{Def:Dec_Regret}.

\begin{definition}[Marginal Decrease in Regret] \label{Def:Dec_Regret}
Given a Social Network $\mathcal{G}(\mathcal{V}, \mathcal{E, \mathcal{P}})$, for any advertiser $a_i \in \mathcal{A}$ the corresponding seed set is $S_i$. The marginal decrease in regret for the user $u$ with respect to the seed set $S_i$ is denoted by $\mathcal{R}(u|S_i)$ and defined as the difference between the regret values when the user $u$ has not been included and when not. Mathematically, this can be expressed in Equation 
\begin{equation}
\mathcal{R}(u|S_i)= \mathcal{R}(S_i)- \mathcal{R}(S_i \cup \{u\})
\end{equation}
\end{definition}
\subsection{\textbf{Budget Effective Greedy (BG) Approach}}
On the basis of Definition \ref{Def:Dec_Regret} first we describe budget effective greedy approach. The idea for this approach is to select the nodes which are within the budget and also gives us the minimum regret. Hence, it is named Budget Effective Greedy Approach. Here we explain how this approach actually works and it is as follows. \par Algorithm \ref{Algo:Greedy} is a heuristic greedy approach. In Line No. $1$, we order all the advertisers based on the budget effectiveness i.e., budget over influence demand. We also initialize empty set $\mathcal{Q}$ of seed sets for each advertiser (Line No. $2$). Next, in Line No. $3$ to $8$ a greedy heuristic is employed to select seed nodes that can reduce the regret (i.e., maximizing $(\mathcal{R}(\mathcal{Q}_{i}) - \mathcal{R}(\mathcal{Q}_{i} \cup \{u\})/\mathcal{I}(\{u\})))$ best to the advertisers. After all the advertisers are satisfied or influence provider runs out of seed nodes, Algorithm \ref{Algo:Greedy} will return seed nodes set $\mathcal{Q}$ (Line No. $9$).
\par Now, we analyze Algorithm \ref{Algo:Greedy} to understand its time and space requirement. For all the advertisers, computing their budget per unit influence will take $\mathcal{O}(n)$ time. Then sorting the advertisers based on this value will take $\mathcal{O}(n \log n)$ time. Hence, Line No. $1$ will take $\mathcal{O}(n \log n)$ time to execute. Line No. $2$ is an assignment statement and it will take $\mathcal{O}(1)$ time. The for loop at Line No. $3$,  will execute $\mathcal{O}(n)$ times. However, it is important to infer how many times the \texttt{while} loop in Line No. $4$ will execute. We have infrared this in the following way. Consider $\mathcal{C}_{min}$ denotes the minimum selection cost among all the users, i.e., $\mathcal{C}_{min}=\underset{u \in \mathcal{V}}{min} \ \mathcal{C}(u)$. So, for the $i$-th advertiser in the worst case, the \texttt{while} loop can execute at most $\lceil \frac{\mathcal{B}_{i}}{\mathcal{C}_{min}} \rceil$ times. Now, it can be observed that for any advertiser $a_j \in \mathcal{A}$, maximum number of times the \texttt{while} loop is executed is $\underset{a_j \in \mathcal{A}}{max} \ \lceil \frac{\mathcal{B}_{j}}{\mathcal{C}_{min}} \rceil$. We call this quantity as $x$. Now, considering the fact that the diffusion is happening based on the IC Model, computing the element which reduces the regret the most can be found in $\mathcal{O}(n^{2} \cdot (m+n))$ time. All the statements in Line No. $6$, $7$, and $8$ will take $\mathcal{O}(1)$ time to execute where $x=\underset{a_j \in \mathcal{A}}{max} \ \lceil \frac{\mathcal{B}_{j}}{\mathcal{C}_{min}} \rceil$. Hence, the total time requirement by Algorithm \ref{Algo:Greedy} will be of $\mathcal{O}(n \log n + x \cdot n^{3} \cdot (m+n))$ and this reduces to $\mathcal{O}(x \cdot n^{3} \cdot (m+n))$. The extra space consumed by Algorithm \ref{Algo:Greedy} to store the marginal gain of the users which will take $\mathcal{O}(n)$ in the worst case. Hence, Theorem \ref{Complexity-Algo1} holds.

\begin{theorem}\label{Complexity-Algo1}
The running time and space requirement of Algorithm \ref{Algo:Greedy} will be of $\mathcal{O}(x \cdot n^{3} \cdot (m+n))$ and $\mathcal{O}(n)$, respectively.
\end{theorem}

\begin{algorithm}
\caption{Budget Effective Greedy Approach for the RMSMA Problem}
\label{Algo:Greedy}
\SetKwInOut{Input}{\textit{Input}}
\SetKwInOut{Output}{\textit{Output}}
\Input{The Social Network $\mathcal{G}(\mathcal{V}, \mathcal{E, \mathcal{P}})$, The Set of  Advertisers $\mathcal{A}=\{a_1, a_2, \ldots, a_l\}$, Their Influence Demand and Corresponding Budget $(\sigma_i, \mathcal{B}_{i})_{i=1}^{l}$
}
\KwResult{An allocation $\Pi$}
Order each advertiser $a_{i} \in \mathcal{A}$ based on descending order of $\frac{\mathcal{B}_{i}}{\sigma_i}$\;
Initialize $\Pi = \{ S_{1}, S_{2}, \ldots S_{l} \}$\;
\For{$i=1 \text{ to }l$}{
\While{$\mathcal{I}(S_{i}) < \sigma_i~ \text{and}~ \mathcal{B}_{i} > 0 ~\text{and}~ |\mathcal{V}| \neq 0$}{

$u^{*} \longleftarrow \underset{u \in \mathcal{V}}{argmax} \ \frac{\mathcal{R}(S_{i}) - \mathcal{R}(S_{i} \cup \{u\})}{\mathcal{I}(\{u\})}$;
$S_{i} \leftarrow S_{i} \cup \{u^{*}\}$\;
$\mathcal{V} \leftarrow \mathcal{V} \setminus \{u^{*}\}$;
\(\mathcal{B}_{i} \leftarrow \mathcal{B}_{i} -  \mathcal{C}(u^{*})\)\;
}}
return $\Pi$\;
\end{algorithm}

\subsection{\textbf{Advertiser Elimination Approach (AEA)}}
Our second proposed approach is the Advertiser Elimination Approach for the RMSMA Problem which is a two step strategy. This approach is an improvised version of the Algorithm \ref{Algo:Greedy}. In first step, The idea of this approach is that it allocates resources in a social network to meet advertisers' influence demands while respecting their budget constraints. After that if multiple advertisers remain undersatisfied, the algorithm strategically eliminates the least promising candidate based on a secondary budget-to-demand sorting, ensuring computational efficiency and a more focused allocation. That's why we call it Advertiser Elimination Approach. Ultimately, this approach balances influence maximization with cost constraints, yielding an allocation that strives to satisfy maximum advertisers within the available network resources. Technically, in this approach, first the advertisers are sorted in descending order based on the budget per unit influence demand value. Next, we initialize the seed set for all the advertisers. Now, for each advertiser, we perform the following task. Until the influence demand is satisfied, budget and non seed node is available, we keep on finding the node that leads to the maximum decrease in regret within the available budget. The node is added to the seed set of the corresponding advertiser and the selection cost of the node is subtracted from the available budget. Next, we count the number of advertiser and the set of advertisers whose influence demand has not been satisfied. If this number of unsatisfied advertisers is more than $k$ (where $k$ is the input parameter called regret tolerance) then the unsatisfied advertisers are sorted in ascending order based on the budget per unit influence demand value. In this sorted list, the first advertiser is one who expects more influence by spending less budget, and of course, making such advertisers satisfied does not make much sense. Hence, we remove the first advertiser in the sorted list and the same process is repeated. Algorithm \ref{Algo:Greedy_Two} describes the process in the form of pseudocode.

\begin{algorithm} [!htb]
\caption{Advertiser Elimination Approach for the RMSMA Problem}
\label{Algo:Greedy_Two}
\SetKwInOut{Input}{\textit{Input}}
\SetKwInOut{Output}{\textit{Output}}
\Input{The Social Network $\mathcal{G}(\mathcal{V}, \mathcal{E, \mathcal{P}})$, The Set of  Advertisers $\mathcal{A}=\{a_1, a_2, \ldots, a_l\}$, Their Influence Demand and Corresponding Budget $(\sigma_i, \mathcal{B}_{i})_{i=1}^{l}$, Regret Tolerance parameter $k$}
\KwResult{An allocation $\Pi$}
Order each advertiser $a_{i} \in \mathcal{A}$ based on descending order of $\frac{\mathcal{B}_{i}}{\sigma_i}$\;
Initialize $\Pi = \{S_{1},S_{2}, \ldots S_{l}\}, k$\;
\While {$TRUE$}{
\For{$\text{each}~ a_{i}\in \mathcal{A}$}{
\While{$\mathcal{I}(S_{i}) < \sigma_i~ \text{and}~ \mathcal{B}_{i} > 0 ~\text{and}~ |\mathcal{V}| \neq 0$}{

$u^{*} \longleftarrow \underset{u \in \mathcal{V}}{argmax} \ \frac{\mathcal{R}(S_{i}) - \mathcal{R}(S_{i} \cup \{u\})}{\mathcal{I}(\{u\})}$;
$S_{i} \leftarrow S_{i} \cup \{u^{*}\}$\;
$\mathcal{V} \leftarrow \mathcal{V} \setminus \{u^{*}\}$;
\(\mathcal{B}_{i} \leftarrow \mathcal{B}_{i} -  \mathcal{C}(u^{*})\)\;
}}
$count = 0, \mathcal{A}_{u} =  [ ~]$\;
\For{$\text{each}~ a_{i}\in \mathcal{A}$}{
\If{$\mathcal{I}(S_{i}) < \sigma_i$}{
$count = count +1$; $\mathcal{A}_{u}[count] \leftarrow a_{i}$\;
}}
\If{$count \geq k$}{
$\mathcal{A}_{u} \leftarrow$ sort each advertiser $a_{i}\in \mathcal{A}_{u}$ in ascending order of $\frac{\mathcal{B}_{i}}{\sigma_i}$\;
$\mathcal{A} \leftarrow \mathcal{A} \setminus \mathcal{A}_{u}[1]$
}}
return $\Pi$\;
\end{algorithm}
\par We analyze Algorithm \ref{Algo:Greedy_Two} to understand its time and space requirement. As mentioned previously, execution of Line No. $1$ will take $\mathcal{O}(n \log n)$ time. It is easy to observe that in the worst case, the \texttt{while} loop at Line No. $3$ will execute $\mathcal{O}(n-k+1)$ times. In worst case scenario when \( k \ll n \) then it will run in $\mathcal{O}(n)$ time. As mentioned previously, similar to the Algorithm \ref{Algo:Greedy} Line No. $4$ to $8$ in the Algorithm \ref{Algo:Greedy_Two} will take $\mathcal{O}(x \cdot n^{3} \cdot (m+n))$ time. Next, in Line No. $9$ initializing $count, \mathcal{A}_{u}$ will take time $\mathcal{O}(n)$ each and in Line No. $11$ to $15$ will take $\mathcal{O}(n)$ time to execute. Line No. $16$ the \texttt{if} condition can be satisfies closer to $n$ times in the worst case and in Line No. $17$ sorting advertisers will take $\mathcal{O}(n\log n)$ time. So, Line No. $16$ to $19$ will take $\mathcal{O}(n \log n)$ time. Therefore, the total time taken by Algorithm \ref{Algo:Greedy_Two} will be of $\mathcal{O}(n((x \cdot n^{3}.(m+n)) + n + (n\log n)))$ and this will be reduced to $\mathcal{O}((x \cdot n^{4}(m+n))$. The additional space requirement for Algorithm \ref{Algo:Greedy_Two} will be of $\mathcal{O}(n)$. Hence, the following theorem statement holds.

\begin{theorem}
The running time and space requirement of Algorithm \ref{Algo:Greedy_Two} will be of $\mathcal{O}((x \cdot n^{4}(m+n))$ and $\mathcal{O}(n)$, respectively.\end{theorem}

\subsection{\textbf{Advertiser Driven Local Search (ADLS) Approach}}

Our third approach, the Advertiser Driven Local Search Approach for the RMSMA Problem is again a two step heuristic designed to efficiently allocate social network nodes to multiple advertisers while balancing their influence demands and budget constraints. In the initial step, advertisers are prioritized by their budget to influence demand ratios, and a greedy strategy is used to iteratively assign nodes each chosen to maximize the reduction in overall regret relative to the influence it provides until the advertiser’s influence requirement is met or budget exhausted. After this initial allocation, in second step the algorithm calculates a threshold based on the average regret and identifies advertisers whose regret exceeds this level. For these under satisfied advertisers, a localized reallocation is triggered, where the greedy process is reapplied using the remaining network nodes and updated budgets to further improve the allocation. This local search refinement helps enhance overall satisfaction, ensuring a more balanced and cost effective resource distribution across all advertisers. Hence, we call it an Advertiser Driven Local Search Approach. 

Here we understand our third approach line by line as written in Algorithm \ref{Algo:GreedyWithThreshold}. The first step will be to sort advertisers in decreasing order of their budget to demand ratio $\frac{\mathcal{B}_i}{\sigma_i}$ as done in our previous algorithms, Algorithm $\ref{Algo:Greedy}$ and Algorithm $\ref{Algo:Greedy_Two}$. This prioritizes advertisers with a higher ``value for money" ensuring more efficient budget usage. Next, we will initialize allocation set for each adveriser as empty set and initialize the regret for all advertiser with the initial regret value. After all the initialization we will do the further exploration by iterating over each advertiser $a_i$. For each advertiser, repeatedly allocate nodes until the total influence $\mathcal{I}(S_{i})$ of the nodes in $S_{i}$ meets or exceeds the influence demand $\sigma_i$ or the budget $\mathcal{B}_i$ of the advertiser is exhausted or there are no remaining nodes in $\mathcal{V}$. Identify the node $u^{*}$ from the remaining set $\mathcal{V}$ that provides the highest marginal gain per unit of influence when added to $S_{i}$. The marginal gain is measured by the decrease in regret $\mathcal{R}(S_{i})$ when $u^{}$ is added, divided by the influence $\mathcal{I}({u})$ provided by $u^{}$. Once all the advertisers are processed we will have their allocated seed set, influence and regret. Thereafter, we compute the threshold $\tau$ as the average of all advertisers' regret values. This threshold identifies advertisers with significantly higher regret. So, we identify advertisers whose regret exceeds the threshold $\tau$. These advertisers are considered underperforming and require further attention. If there are any high-regret advertisers then we rerun the greedy allocation (Lines 5–11) for these advertisers. For allocating the seed set for these advertisers we will allocate the seed nodes from the remaining nodes of the previous run and the allocated nodes of the advertisers having regret higher than the threshold. The outcome of this will be the $\Pi$ with the new allocations and hence, the final allocation $\Pi$ for all advertisers.

\begin{algorithm}[!htb]
\caption{Advertiser Driven Local Search Approach for the RMSMA Problem}
\label{Algo:GreedyWithThreshold}
\SetKwInOut{Input}{\textit{Input}}
\SetKwInOut{Output}{\textit{Output}} 
\Input{The Social Network $\mathcal{G}(\mathcal{V}, \mathcal{E}, \mathcal{P})$, 
       The Set of Advertisers $\mathcal{A}=\{a_1, a_2, \ldots, a_l\}$, 
       Their Influence Demand and Corresponding Budget $(\sigma_i, \mathcal{B}_{i})_{i=1}^{l}$}
\Output{An Allocation $\Pi$}
Order each advertiser $a_{i} \in \mathcal{A}$ based on descending order of $\frac{\mathcal{B}_{i}}{\sigma_i}$\;
Initialize $\Pi = \{S_{1}, S_{2}, \ldots, S_{l} \}$\;

\For{$i = 1$ \textbf{to} $l$}{
    \While{$\mathcal{I}(S_{i}) < \sigma_i$ \textbf{and} $\mathcal{B}_{i} > 0$ \textbf{and} $|\mathcal{V}| \neq 0$}{
        $u^{*} \longleftarrow \underset{u \in \mathcal{V}}{\arg\max} \ \frac{\mathcal{R}(S_{i}) - \mathcal{R}(S_{i} \cup \{u\})}{\mathcal{I}(\{u\})}$\;
        $S_{i} \leftarrow S_{i} \cup \{u^{*}\}$\;
        $\mathcal{V} \leftarrow \mathcal{V} \setminus \{u^{*}\}$\;
        $\mathcal{B}_{i} \leftarrow \mathcal{B}_{i} - \mathcal{C}(u^{*})$\;
    }
}

Calculate the threshold $\tau$ as a function of average regret: $\tau \longleftarrow  \text{mean}(\mathcal{R})$\;

Identify advertisers with high regret: $\mathcal{A}_{\text{high}} = \{a_i \in \mathcal{A} \mid r_i > \tau\}$\;

\If{$\mathcal{A}_{\text{high}} \neq \emptyset$}{
\For{$i = 1$ \textbf{to} $|\mathcal{A}_{\text{high}}|$}{
    Re-run the greedy algorithm (from Line $4$ - $9$) with the remaining nodes in $\mathcal{V}$ and their allocated budgets $\mathcal{B}$ respectively\;
    Update $\Pi$ accordingly\;
}
}
\Return $\Pi$\;
\end{algorithm}

We analyze Algorithm $\ref{Algo:GreedyWithThreshold}$ to understand its time and space complexity. In Line No. $1$, ordering all advertisers in decreasing manner on the basis of the ratio $\frac{\mathcal{B}_{i}}{\sigma_i}$ will take $\mathcal{O}(n \log n)$ time. The Line No. $2$ will take $\mathcal{O}(n)$ time to execute. The Line No. $3$ will execute $\mathcal{O}(n)$ times. The while loop in this algorithm will execute in $\mathcal{O}(x \cdot n^2 (m+n))$ times which is same as discussed in Algorithm $\ref{Algo:Greedy}$. In while loop $O(1)$ time will be taken to execute Lines $6$, $7$ and $8$ each. Hence, this part of the Algorithm $\ref{Algo:GreedyWithThreshold}$ from Line No. $3$ to $10$ will take $O(x \cdot n^3 (m+n))$  time to execute. The mean calculation at the Line No. $11$ will be in $O(n)$ time. The time taken to identify the advertisers with the high regret in Line No. $12$ will be $O(n)$. The Line No. $13$ will take $O(1)$ time for comparison. And afterwards, again we run the greedy method, so the same time $\mathcal{O}(x \cdot n^2 (m+n))$ will be taken from Line No. $15$. Overall, from Line No. $13$ to $18$, it takes $\mathcal{O}(x \cdot n^3 (m+n))$. Hence, the total time requirement is $O(n \log n + x \cdot n^3 (m+n) + x \cdot n^3 (m+n))$ which finally reduces to $O(x \cdot n^3 (m+n)$ time. The additional space requirement for Algorithm \ref{Algo:GreedyWithThreshold} will be of $\mathcal{O}(n)$. Hence, Theorem \ref{Th:Algowiththreshold} holds.

\begin{theorem} \label{Th:Algowiththreshold}
The running time and space requirement of Algorithm \ref{Algo:GreedyWithThreshold} will be of $O(x \cdot n^3 (m+n)$ and $\mathcal{O}(n)$, respectively.
\end{theorem}

\paragraph{\textbf{An Illustrative Example.}} Consider an influence provider with a social network $\mathcal{G}$, contains $12$ nodes $\mathcal{V} = \{ u_{1}, u_{2}, \ldots, u_{12}\}$ with corresponding individual influence value and six advertisers $\mathcal{A} = \{ a_{1}, a_{2}, \ldots, a_{6}\}$ with individual influence demands reported in Table \ref{ETable:1} and \ref{ETable:2}, respectively. So, the order of budget effective advertisers is $<a_{2},a_{1},a_{4}, a_{3},a_{5}, a_{6}>$. In the seed node allotment of Algorithm \ref{Algo:Greedy} advertiser $a_{1}, a_{2}, a_{3}, a_{4}$ are fully satisfied with no excessive regret (ER) or unsatisfied regret (UR) however advertiser $a_{5}$ and $a_{6}$ not satisfied due to the seed node runs out i.e., $|\mathcal{S}| = \emptyset$ as shown in Table \ref{ETable:3}. Next, when the social network and advertiser information shown in Table \ref{ETable:1} and \ref{ETable:2} used in Algorithm \ref{Algo:Greedy_Two} by setting the \texttt{regret tolerance} parameter $k = 1$, advertiser $a_{6}$ is removed from the advertiser set. The allocation of nodes is presented in Table \ref{ETable:4}. The ``ADLS" approach extends the ``AEA" approach to minimize the regret further. Table \ref{ETable:5} shows that the allocation from the ``AEA" approach advertiser $a_{3}, a_{4}, a_{5}$ and $a_{6}$ are reallocated because these advertiser generate regrets. After reallocation, advertiser $a_{6}$ only generates regret, i.e., unsatisfied regret as presented in Table \ref{ETable:5}. Hence, the ``ADLS" approach minimize the regret better.

\begin{table}[h]
    \centering
    \caption{Illustrative Example}
    \begin{subtable}{\textwidth}
        \centering
        \begin{tabular}{| c | c | c | c | c | c | c | c | c | c | c | c | c | c |}
            \hline
            $\mathcal{V}$ & $u_{1}$ & $u_{2}$ & $u_{3}$ & $u_{4}$ & $u_{5}$ & $u_{6}$ & $u_{7}$ & $u_{8}$ & $u_{9}$ & $u_{10}$ & $u_{11}$ & $u_{12}$ \\ \hline
            $\sigma(u_{i})$ & 4 & 6 & 5 & 4 & 5 & 2 & 3 & 2 & 3 & 2 &2 &5 \\ \hline
            $Cost(u_{i})$ & \$6 & \$9 & \$7.5 & \$6 & \$7.5 & \$3 & \$4.5 & \$3 & \$4.5 & \$3 & \$3 & \$7.5  \\ \hline
        \end{tabular}
        \subcaption{Nodes information.}
        \label{ETable:1}
    \end{subtable}
    
    \begin{subtable}{\textwidth}
        \centering
        \begin{tabular}{| c | c | c | c | c | c | c |}
            \hline
            $\mathcal{A}$ & $a_{1}$ & $a_{2}$ & $a_{3}$ & $a_{4}$ & $a_{5}$ & $a_{6}$\\ \hline
            $\sigma_{i}$ & $10$ & $8$ &$6$ & $10$ & $9$ & $5$\\ \hline
            $B_{i}$ & \$18 & \$17 & \$10 & \$17 & \$11 & \$5 \\ \hline
        \end{tabular}
        \subcaption{Advertisers information.}
        \label{ETable:2}
    \end{subtable}
    
    \begin{subtable}{\textwidth}
        \centering
        \begin{tabular}{ | c | c | c | c | c | c | c |}
            \hline
            $\mathcal{A}$ & $a_{1}$ & $a_{2}$ & $a_{3}$ & $a_{4}$ & $a_{5}$ & $a_{6}$\\ \hline
            $\mathcal{S}$ & $\{ u_{3}, u_{12} \}$ & $\{ u_{2}, u_{6}\}$ & $\{ u_{4}, u_{7}\}$ & $\{ u_{1}, u_{5}, u_{8} \}$ & $\{ u_{9}, u_{10}, u_{11}\}$ & $\{ \}$  \\ \hline
            Satisfied & Y & Y & Y & Y & N & N\\ \hline
            Regret & - & - & ER & ER & UR & UR\\ \hline
        \end{tabular}
        \subcaption{Allotment after the ``BG" approach.}
        \label{ETable:3}
    \end{subtable}
    
    \begin{subtable}{\textwidth}
        \centering
        \begin{tabular}{ | c | c | c | c | c | c |}
            \hline
            $\mathcal{A}$ & $a_{1}$ & $a_{2}$ & $a_{3}$ & $a_{4}$ & $a_{5}$ \\ \hline
            $\mathcal{S}$ & $\{ u_{3}, u_{12} \}$ & $\{ u_{2}, u_{6}\}$ & $\{ u_{4}, u_{7}\}$ & $\{ u_{1}, u_{5}, u_{8} \}$ & $\{ u_{9}, u_{10}, u_{11}\}$ \\ \hline
            Satisfied & Y & Y & Y & Y & N \\ \hline
            Regret & - & - & ER & ER & UR \\ \hline
        \end{tabular}
        \subcaption{Allotment after the ``AEA" approach.}
        \label{ETable:4}
    \end{subtable}
    
    \begin{subtable}{\textwidth}
        \centering
        \begin{tabular}{ | c | c | c | c | c | c | c |}
            \hline
            $\mathcal{A}$ & $a_{1}$ & $a_{2}$ & $a_{3}$ & $a_{4}$ & $a_{5}$ & $a_{6}$\\ \hline
            $\mathcal{S}$ & $\{ u_{3}, u_{12} \}$ & $\{ u_{2}, u_{6}\}$ & $\{ u_{1}, u_{8}\}$ & $\{ u_{5}, u_{9}, u_{10} \}$ & $\{ u_{4}, u_{7}, u_{11}\}$ & $\{ \}$\\ \hline
            Satisfied & Y & Y & Y & Y & Y & N\\ \hline
            Regret & - & - & - & - & - & UR\\ \hline
        \end{tabular}
        \subcaption{Allotment after the ``ADLS" approach.}
        \label{ETable:5}
    \end{subtable}
\end{table}

\section{Experimental Details} \label{Sec:Experimental_Details}
In this section, we describe the experimental evaluation of the proposed solution approach. Initially, we start by describing the datasets.
\subsection{Dataset Description}
Now, we describe the datasets that we have used in our experiments.

\begin{itemize}
\item \textbf{Congress-Twitter Dataset} \cite{fink2023twitter}: This network represents the Twitter interaction network for the 117th United States Congress, both House of Representatives and Senate. The base data was collected via the Twitter’s API, then the empirical transmission probabilities were quantified according to the fraction of times one member retweeted, quote tweeted, replied to, or mentioned another member’s tweet.

\item \textbf{Email-Eu-Core Dataset} \cite{10.1145/3097983.3098069} The network was generated using email data from a large European research institution. The information about all incoming and outgoing email between members of the research institution is anonymized by them. There is an edge $(u, v)$ in the network if person $u$ sent person $v$ at least one email. The e-mails only represent communication between institution members (the core), and the dataset does not contain incoming messages from or outgoing messages to the rest of the world.

\item \textbf{Facebook Dataset} \cite{leskovec2012learning}: This dataset consists of `circles' (or `friends lists') from Facebook. Facebook data was collected from survey participants using this Facebook app. The dataset includes node features (profiles), circles, and ego networks. However, in this study, we have not considered any node feature. 

\item \textbf{Wikivote Dataset} \cite{10.1145/1772690.1772756} The network contains all the Wikipedia voting data from the inception of Wikipedia till January 2008. Nodes in the network represent wikipedia users and a directed edge from node $i$ to node $j$ represents that user $i$ voted on user $j$.

\end{itemize}

The basic statistics of the datasets has been mentioned in Table \ref{table:graph_stats}.

\begin{table}[h!]
\centering
\small 
\begin{tabular}{|p{2.5cm}|p{1.5cm}|p{1.5cm}|p{1.5cm}|p{1.5cm}|p{1.5cm}|}
\hline
\textbf{Dataset} & \textbf{Type} & \textbf{Number of Nodes $|\mathcal{V}|$} & \textbf{Number of Edges $|\mathcal{E}|$} & \textbf{Average Degree $d_{avg}$} & \textbf{Maximum Degree $d_{max}$} \\ \hline
Congress-Twitter & Directed & 475 & 13289 & 55.95 & 284 \\ \hline
Email-Eu-Core & Directed & 1005 & 25571 & 50.88 & 546 \\ \hline
Facebook & Undirected & 4039 & 88234 & 43.69 & 1045 \\ \hline
Wikivote & Directed & 7115 & 103689 & 29.15 & 1167 \\ \hline
\end{tabular}
\caption{Description of Graph Datasets}
\label{table:graph_stats}
\end{table}

\subsection{Experimental Setup}
In our study, there are following parameters whose value need to be set up. 
\paragraph{\textbf{Influence Probability}} In our experiments, we have used the following two influence probability set up: (i) Uniform and (ii) Trivalency. In Uniform probability setting, all the edges of the network will have the same probability value, i.e., for all $(u_iu_j) \in \mathcal{E}$, $\mathcal{P}(u_i,u_j)=p_c$. In this study, we have considered value of $p_c = 0.1$. In the Trivalency setting, the influence probability of each edge is chosen uniformly at random from the set $\{0.1, 0.01, 0.001\}$.

\paragraph{\textbf{Selection Cost of the Users}} In this study, we have considered the \emph{degree proportional cost setting}. This setting has also been considered \cite{DBLP:journals/ton/NguyenTD17}. This means that more the degree of an user, more costly the user is to select as seed node. The selection cost of any user $u$ is defined as $\mathcal{C}(u)= h \cdot (|\mathcal{V}| /
\sum_{v \in \mathcal{V}} deg(v))
\cdot deg(u)$. Here, $h$ is a constant and in our study we have considered the value of $h$ is $1000$.

\paragraph{\textbf{The Number of Advertisers}} We have considered the following values as the  number of advertisers: $5$, $10$, $20$, $50$, and $100$.

\subsubsection{Key Parameters}
All the key parameters are summarized in Table \ref{Key-parameters}. This table includes the demand-supply ratio $(\lambda)$, average individual demand ratio $(\omega)$, influence penalty ratio $(\gamma)$, and cardinality penalty ratio $(\delta)$. In each experiment, we vary only one parameter and set the remaining parameter in the default setting (highlighted in bold).

\begin{table}[h!]
\caption{Key Parameters\label{Key-parameters}} 
\vspace{-0.15in} 
\centering 
\begin{tabular}{|p{2cm}|p{5.5cm}|}
\hline
Parameter & Values \\ \hline
$\lambda$ & $40\%$, $60\%$, $80\%$, \textbf{100\%}, $120\%$ \\ \hline
$\omega$ & $1\%$, $2\%$, \textbf{5\%}, $10\%$, $20\%$ \\ \hline
$\gamma$ & $0$, $0.25$, \textbf{0.5}, $0.75$, $1$ \\ \hline
$\delta$ & \textbf{0.01}, $0.02$, $0.03$, $0.04$, $0.05$ \\ \hline
\end{tabular}
\end{table}

\paragraph{\textbf{Demand Supply Ratio $\lambda$}} It refers to the ratio of global influence demand of the advertisers over the influence provider influence supply, i.e., $\lambda = \mathcal{I}^{d}/ \mathcal{I}^{s}$, where $\mathcal{I}^{d} = \sum_{i=1}^{n} \mathcal{I}_{i}$ represent the global influence demand and $\mathcal{I}^{s} = \sum_{s \in \mathcal{S}} \mathcal{I}(\{s\})$ is the influence provider supply. We have considered five values of $\lambda$, i.e., $40\%,$ $60\%,$ $80\%,$ $100\%,$ $120\%$, which denotes low, medium, high full, and excessive, respectively.

\paragraph{\textbf{Average Individual Demand Ratio $\omega$}} It is the percentage of average individual demand of the advertisers over the influence provider supply i.e., $\omega = \mathcal{I}^{avg}/ \mathcal{I}^{s}$, where $\mathcal{I}^{avg} = \mathcal{I}^{d}/|\mathcal{A}|$ is the average individual demand of the advertisers. This parameter is useful for simulating the individual demands of the advertiser.

\paragraph{\textbf{Advertisers Demand $\mathcal{I}$}} Once the $\mathcal{I}^{d}$ is fixed, $\mathcal{I}^{avg}$ can be easily computed i.e., $\mathcal{I}^{avg}= \omega \cdot \mathcal{I}^{s}$. Also, we can generate the influence demand of each advertiser using $\mathcal{I}^{a} = \lfloor \alpha \cdot \mathcal{I}^{s} \cdot \omega \rfloor$, where $\alpha$ is a factor randomly generated to simulate different demand setting.

\paragraph{\textbf{Advertiser Payment $\mathcal{B}$}} We set the payment of an advertiser as proportional to his influence demand, $\mathcal{B}^{a} = \lfloor \beta \cdot \mathcal{I}^{a} \rfloor$, where $\beta$ is a parameter to simulate different payment.

\paragraph{\textbf{Influence Penalty Ratio $\gamma$}} This parameter decides the penalty imposed on the influence provider due to the influence supply to the advertisers. We vary $\gamma$ from $0, 0.25, 0.5, 0.75, 1$, where two extreme cases when $\gamma = 0$, the influence provider cannot receive any payment if the advertisers required influence is satisfied and when $\gamma = 1$, the influence provider can receive the same fraction of payment the fraction of influence he provides to the advertisers.

\paragraph{\textbf{Cardinality Penalty Ratio $\delta$}} This parameter imposes a penalty if the influence provider provides less or more amount of influence to the advertiser based on the cardinality of the allocated seed set to the advertisers. We vary $\delta$ from $0.01,0.02,0.03,0.04,0.05$ to simulate different penalties imposed on the advertiser.

\subsubsection{Other Setups}

\paragraph{\textbf{Environment Setup}} All the codes are implemented in Python and executed on HPC cluster run under CentOS-7.8 (a specific version of Red Hat Enterprise Linux). The processor architecture of HPC cluster is for CPU node is Intel Xeon Gold 6248 with 32 GB memory.
\paragraph{\textbf{Performance Metrics}} The effectiveness metrics measure total regret, including unsatisfied and excessive regret. The efficiency metrics include the running time of the proposed approaches. 

\subsection{Baseline Methods}
We compared our proposed approaches with the following baseline methods.

\paragraph{\textbf{Random}}

In this baseline method we randomly select the nodes of the graph within the assigned payment $\mathcal{B}$.

\paragraph{\textbf{Top-$K$}}
The top influential users are selected within the assigned payment $\mathcal{B}$.

\subsection{Goals of the Experimentation}
In this study, we have fixed the following research questions (RQ) and through our experimental study we have found out the explanations of these questions.
\begin{itemize}
\item RQ1: What will happen if we increase the global influence demand $(\mathcal{I}^{d})$ of the advertisers to low, normal, high, full and excessive.
\item RQ2: Which kind of advertisers are more beneficial for the influence providers perspective. A small number of advertisers with high individual influence demand or large number of advertisers with small individual influence demand.
\item RQ3: How the computational time varies when we vary $\lambda, \omega$ and $\gamma$ for the proposed and baseline methods.
\end{itemize}

\subsection{Experimental Results and Discussion}\label{Sec:Experimental_Evaluations}
Now, we report the obtained results from our experiments and describe them in order to get the answers of the research questions.

\subsubsection{\textbf{Effectiveness Study}}

At first we evaluate how varying $\lambda$ and $\omega$ impacts the total regret. This total regret includes unsatisfied and excessive regret and we report each of them separately in our experiments. The experiments for each datasets are conducted for two different probability settings, Uniform ($0.1$) and Trivalency. Now, to ease our discussion we consider four different cases for each dataset. We examine four distinct cases that represent different combinations of global ($\lambda$) and individual ($\omega$) influence demands. In Case $1$, both the global and individual influence demands are low, meaning the influence provider serves a large number of advertisers with modest requirements. Case $2$ features low global but high individual influence demand, leading to fewer advertisers with significantly larger influence requirements and a unique interplay between excessive and unsatisfied regret. In Case $3$, high global influence demand coupled with low individual influence demand creates a scenario where most advertisers receive sufficient influence, though the overall allocation efficiency varies. Finally, Case $4$ considers the situation where both global and individual influence demands are high, resulting in a dynamic interplay: while increased individual influence demand can reduce excessive regret, it may also raise unsatisfied regret if advertiser requirements are not fully met. These cases provide a comprehensive framework for analyzing and comparing the performance of different allocation algorithms under diverse demand conditions.

\paragraph{\textbf{Case 1: low $\lambda$, low $\omega$}}

In Case $1$, we consider scenarios where $\lambda \leq 60\%$ and $\omega \leq 30\%$, indicating that both the global and individual influence demands from advertisers are low. This implies that the influence provider serves a large number of advertisers with relatively low influence requirements. Since $\lambda$ is small, the global influence demand remains significantly lower than the available supply. In both probability settings, we observe that as global influence demand increases, excessive regret also rises.

\begin{figure}[H]
\centering
\begin{tabular}{lclc}
    Excessive Regret & \includegraphics[width=0.11\linewidth]{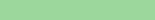} & 
    Unsatisfied Regret & \includegraphics[width=0.11\linewidth]{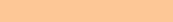} \\
\end{tabular}
\begin{tabular}{ccccc}
    \includegraphics[scale=0.10]{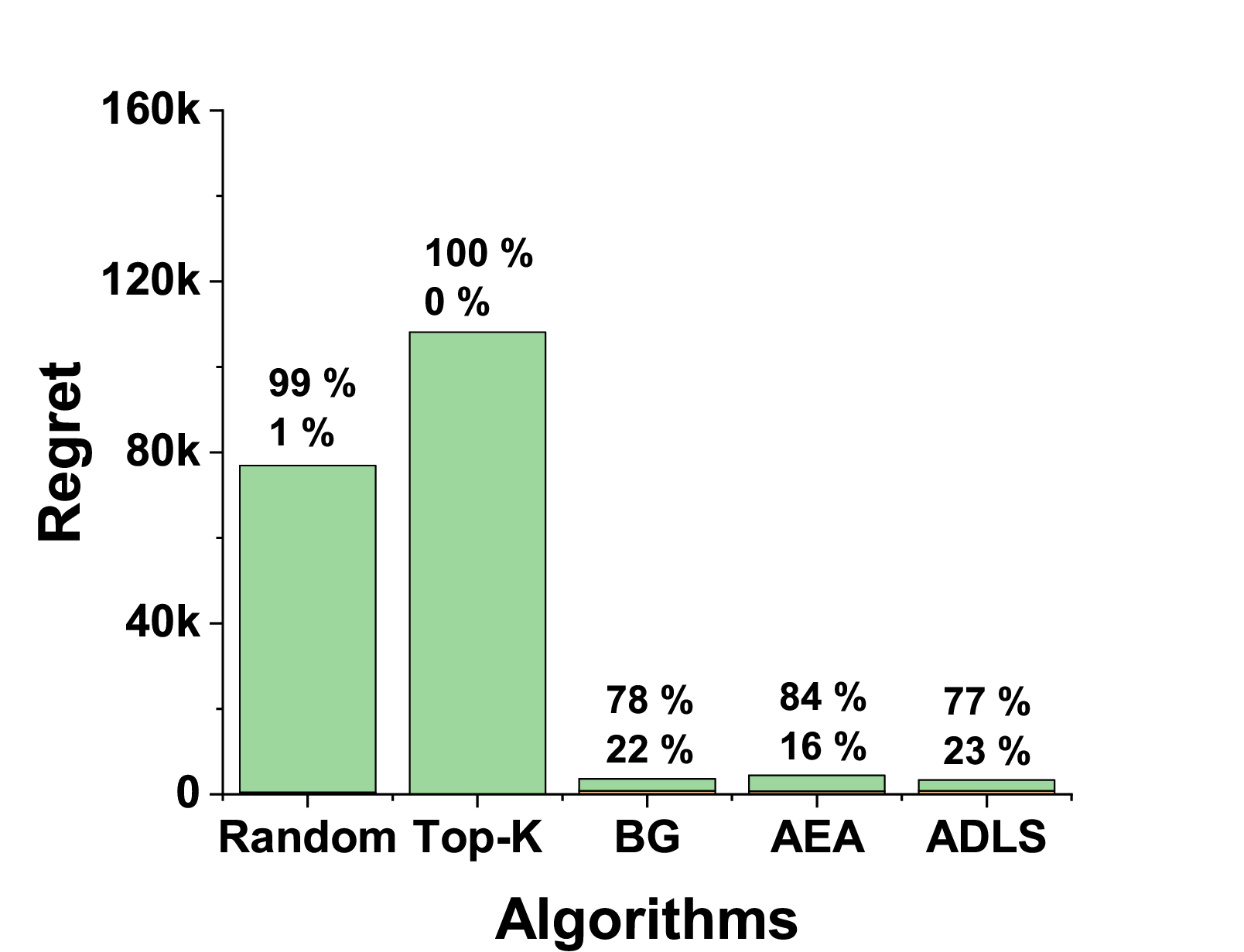} & 
    \includegraphics[scale=0.10]{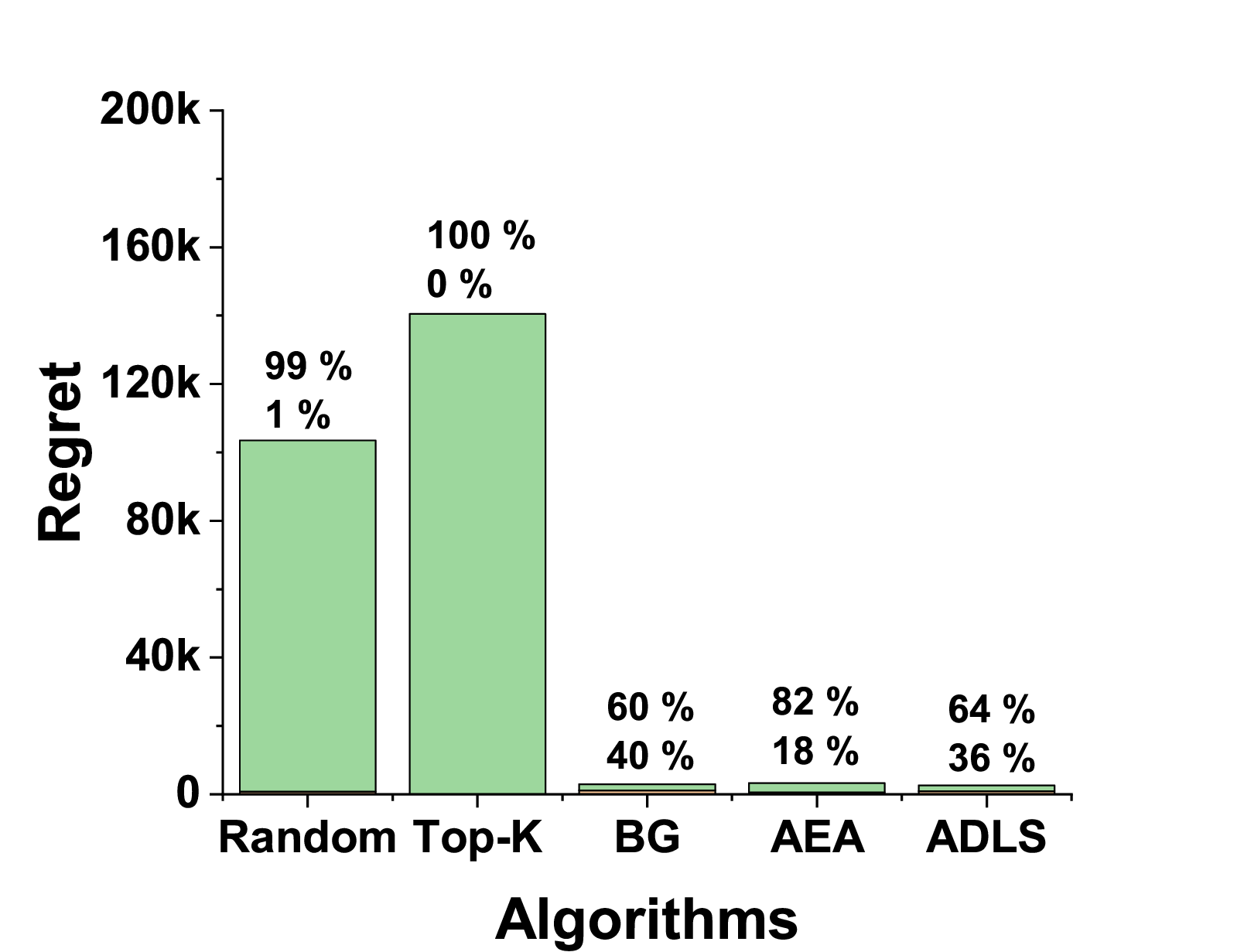} & 
    \includegraphics[scale=0.10]{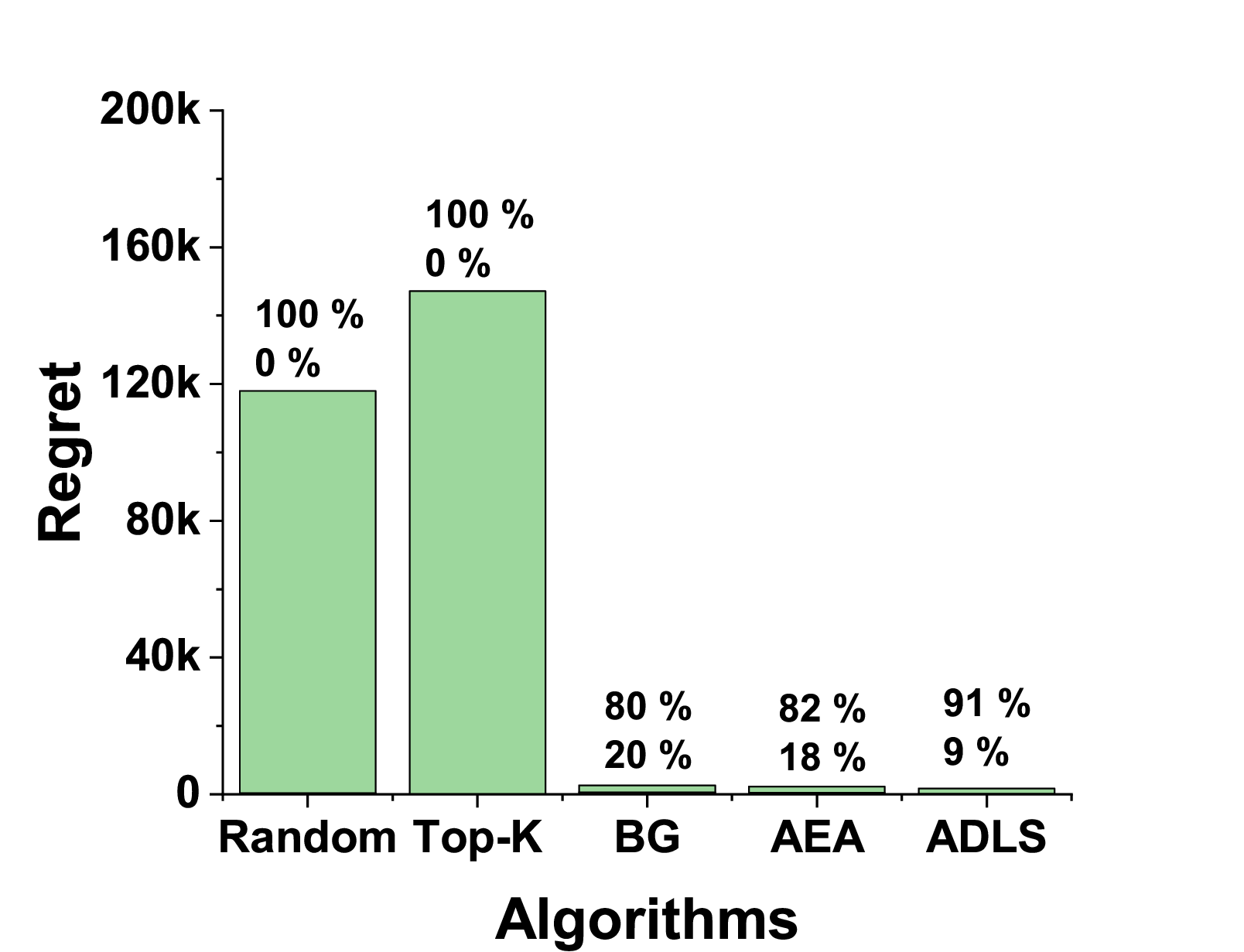} & 
    \includegraphics[scale=0.10]{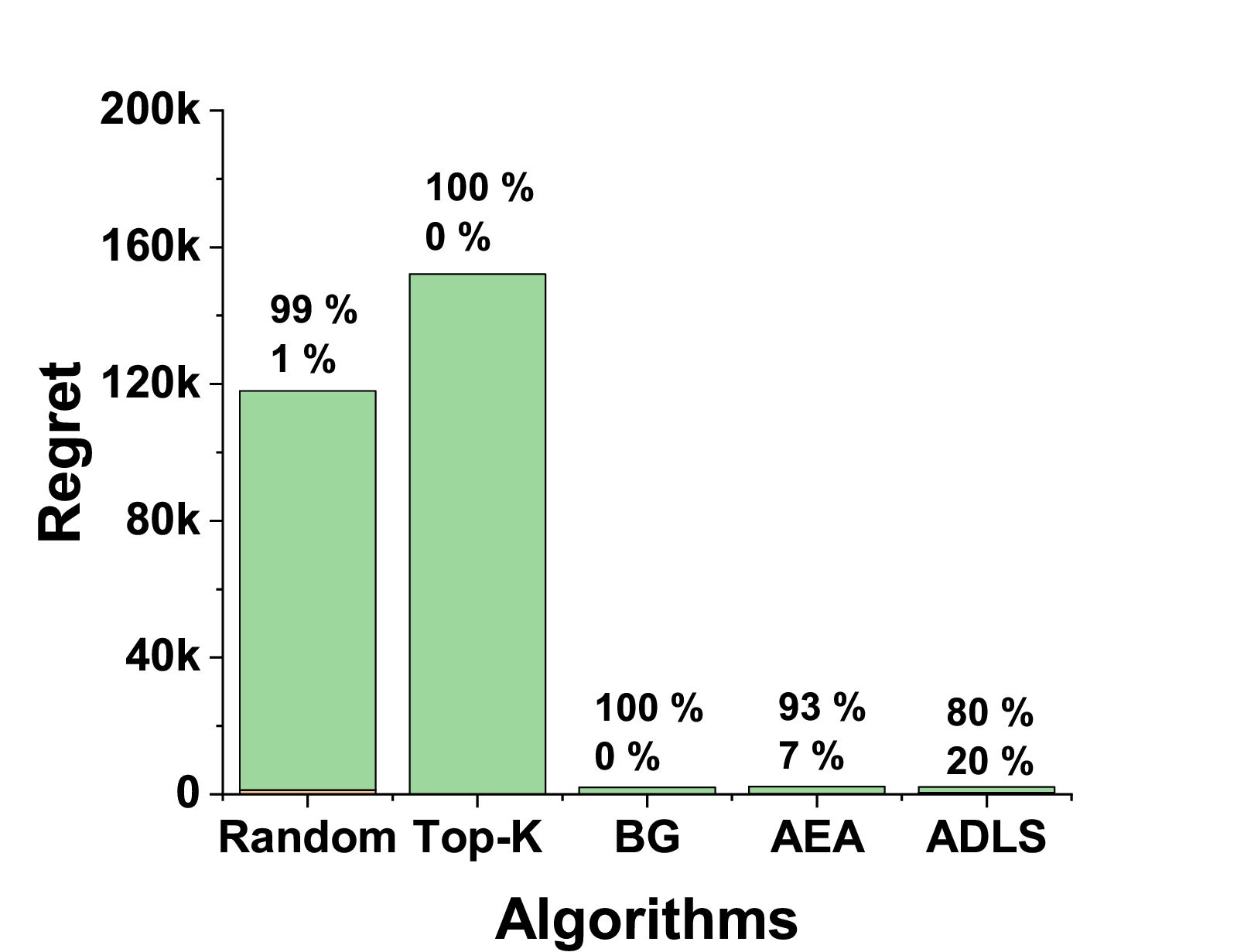} &
    \includegraphics[scale=0.10]{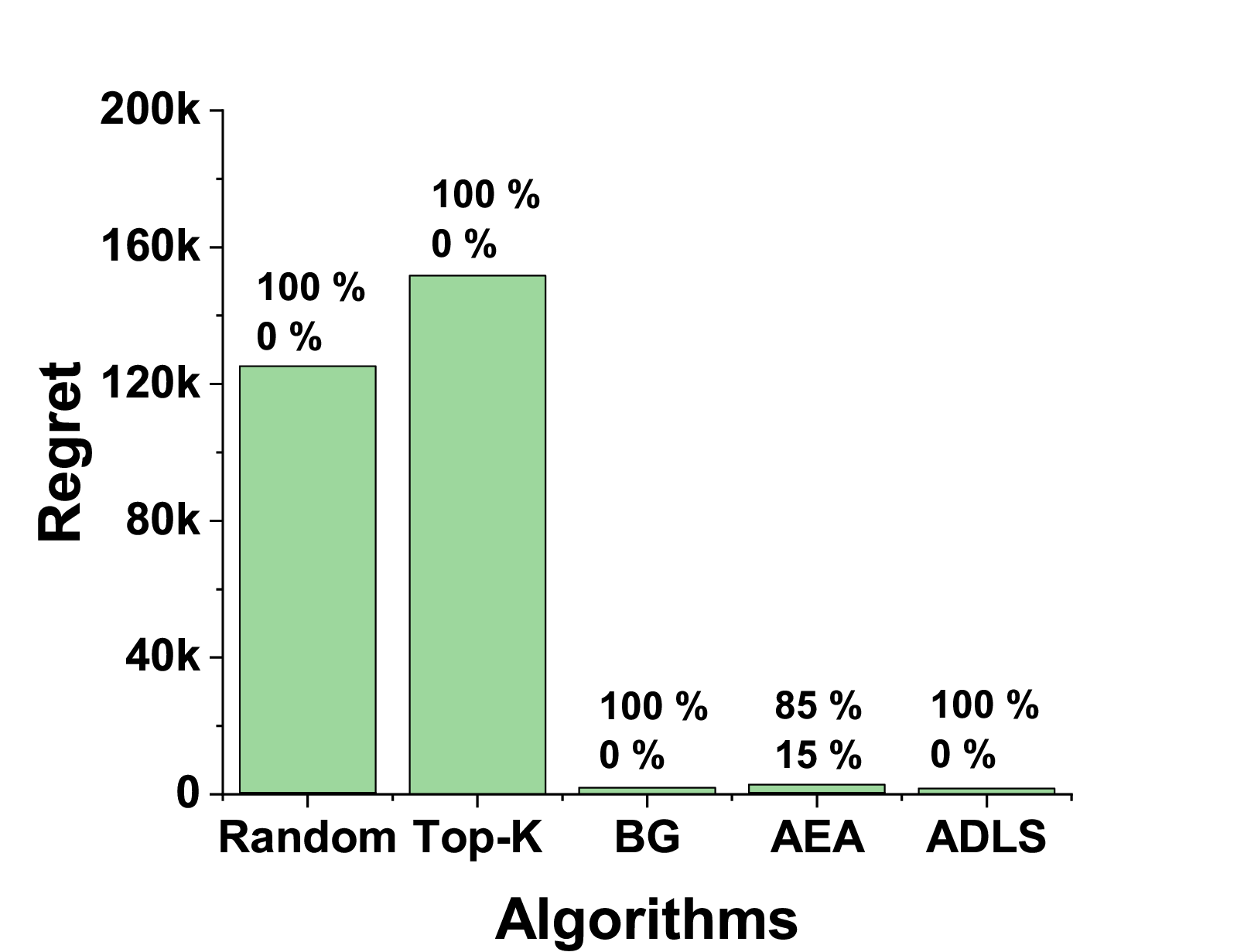} \\
    (a) $\lambda$ = $40\%$ & 
    (b) $\lambda$ = $60\%$ & 
    (c) $\lambda$ = $80\%$ & 
    (d) $\lambda$ = $100\%$ & 
    (e) $\lambda$ = $120\%$ \\
\end{tabular}
\caption{Regret of varying the demand-supply ratio $\lambda$ when $\omega = 10\%$ (Email-Eu-Core, $|\mathcal{A}| = 100$, $p_{c}: Uniform(0.1) $)}
\label{Fig:1_Influence_1}
\begin{tabular}{ccccc}
    \includegraphics[scale=0.10]{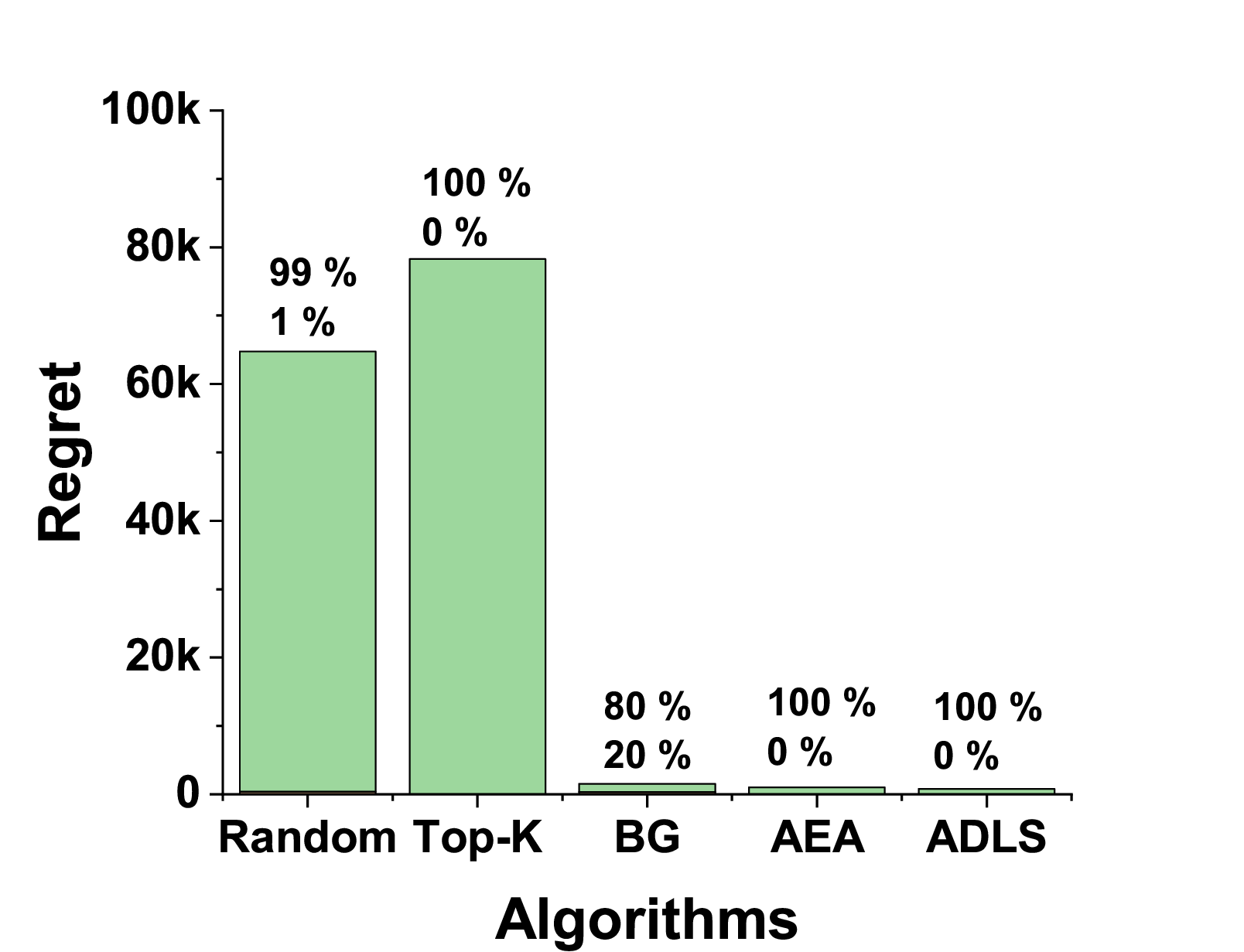} & 
    \includegraphics[scale=0.10]{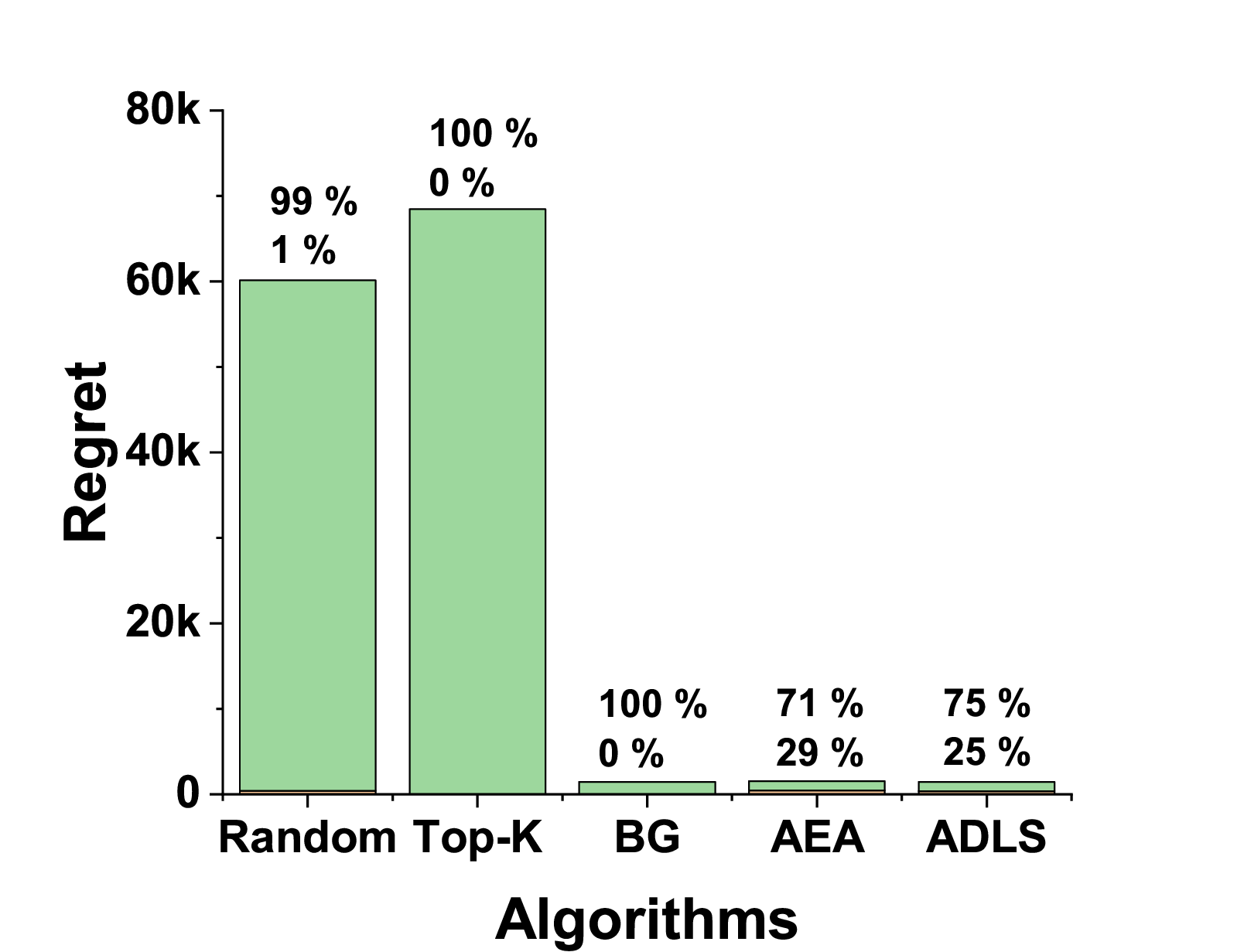} & 
    \includegraphics[scale=0.10]{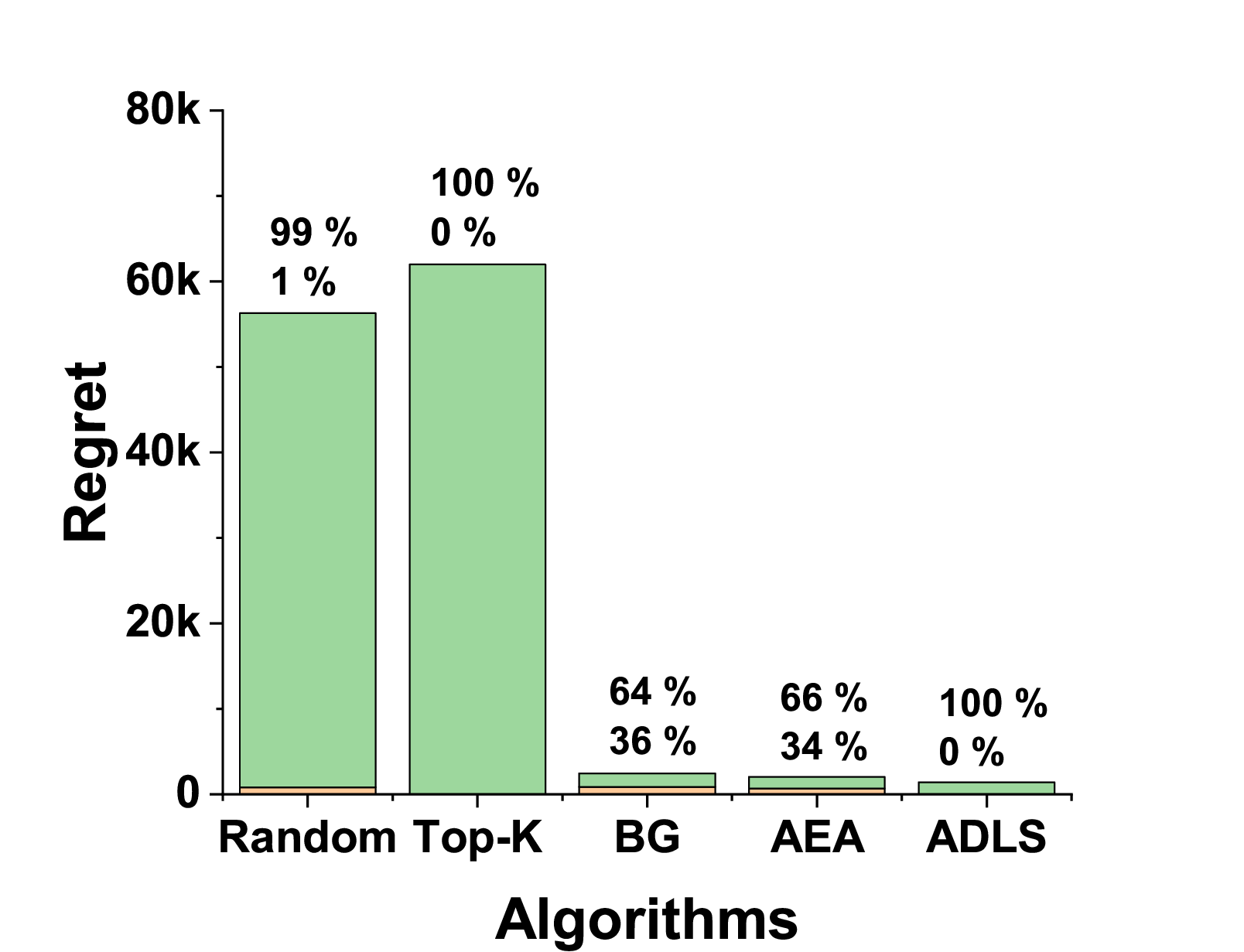} & 
    \includegraphics[scale=0.10]{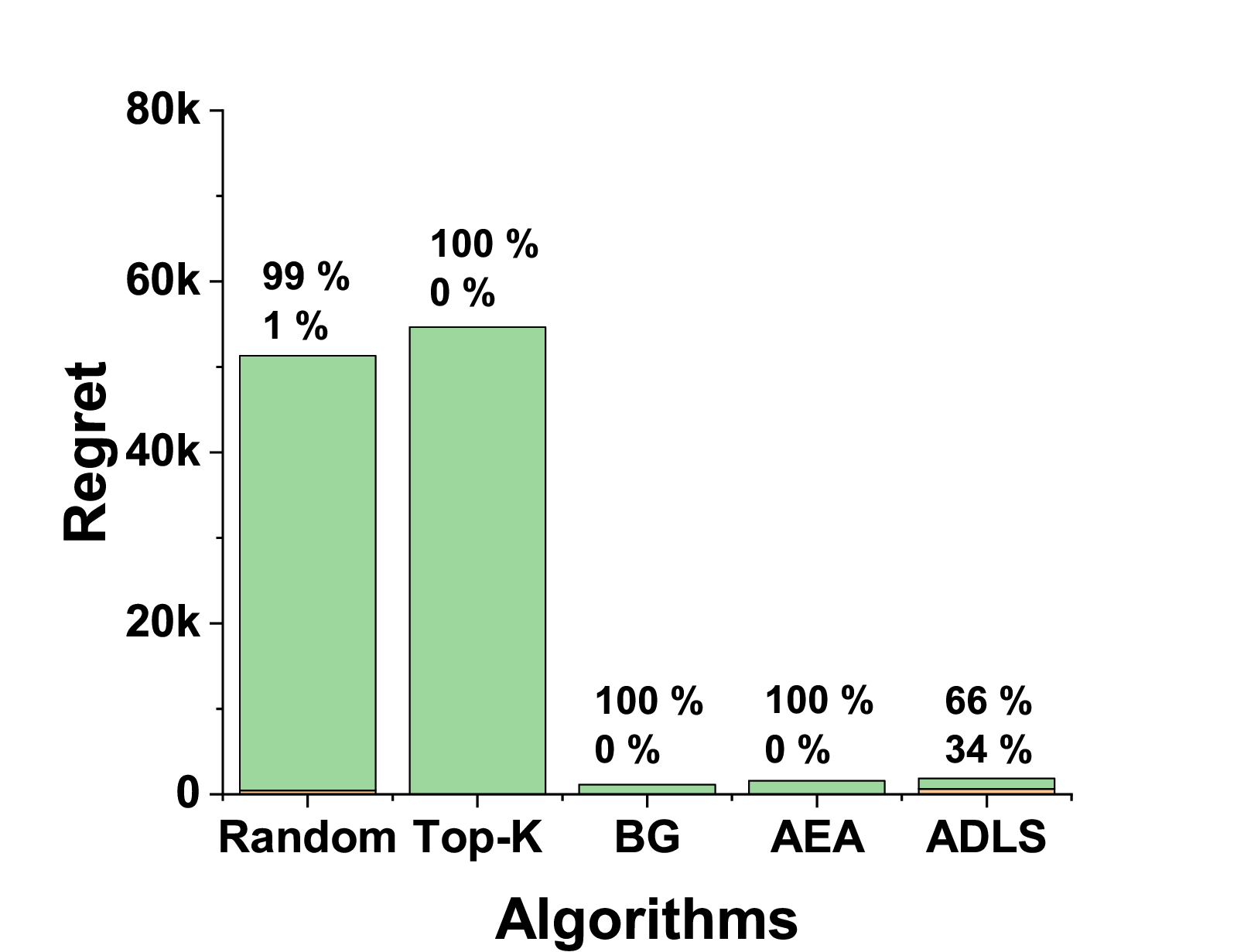} &
    \includegraphics[scale=0.10]{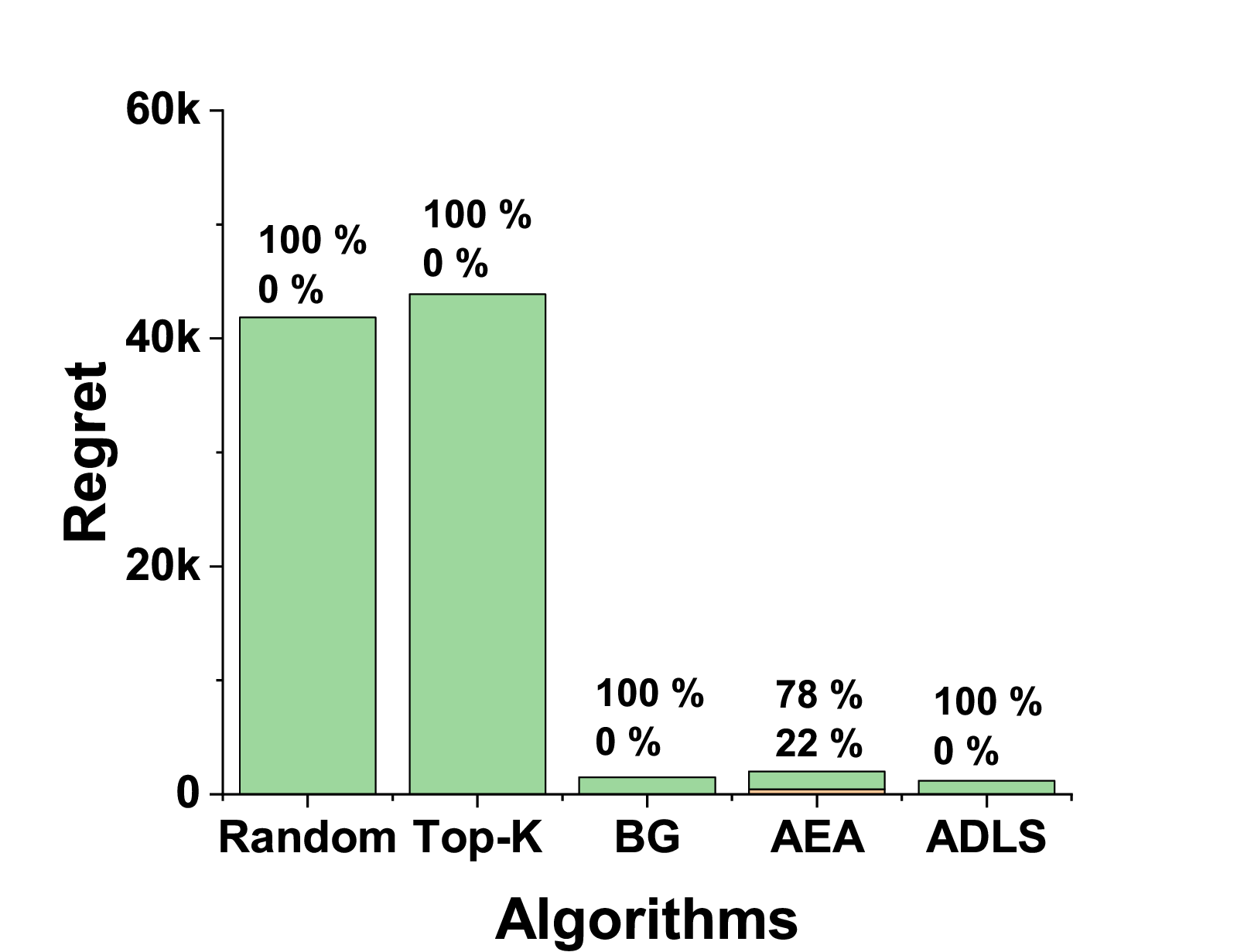} \\
    (a) $\lambda$ = $40\%$ & 
    (b) $\lambda$ = $60\%$ & 
    (c) $\lambda$ = $80\%$ & 
    (d) $\lambda$ = $100\%$ & 
    (e) $\lambda$ = $120\%$ \\
\end{tabular}
\caption{Regret of varying the demand-supply ratio $\lambda$ when $\omega = 30\%$ (Email-Eu-Core, $|\mathcal{A}| = 50$, $p_{c}: Uniform(0.1) $)}
\label{Fig:1_Influence_2}
\begin{tabular}{ccccc}
    \includegraphics[scale=0.10]{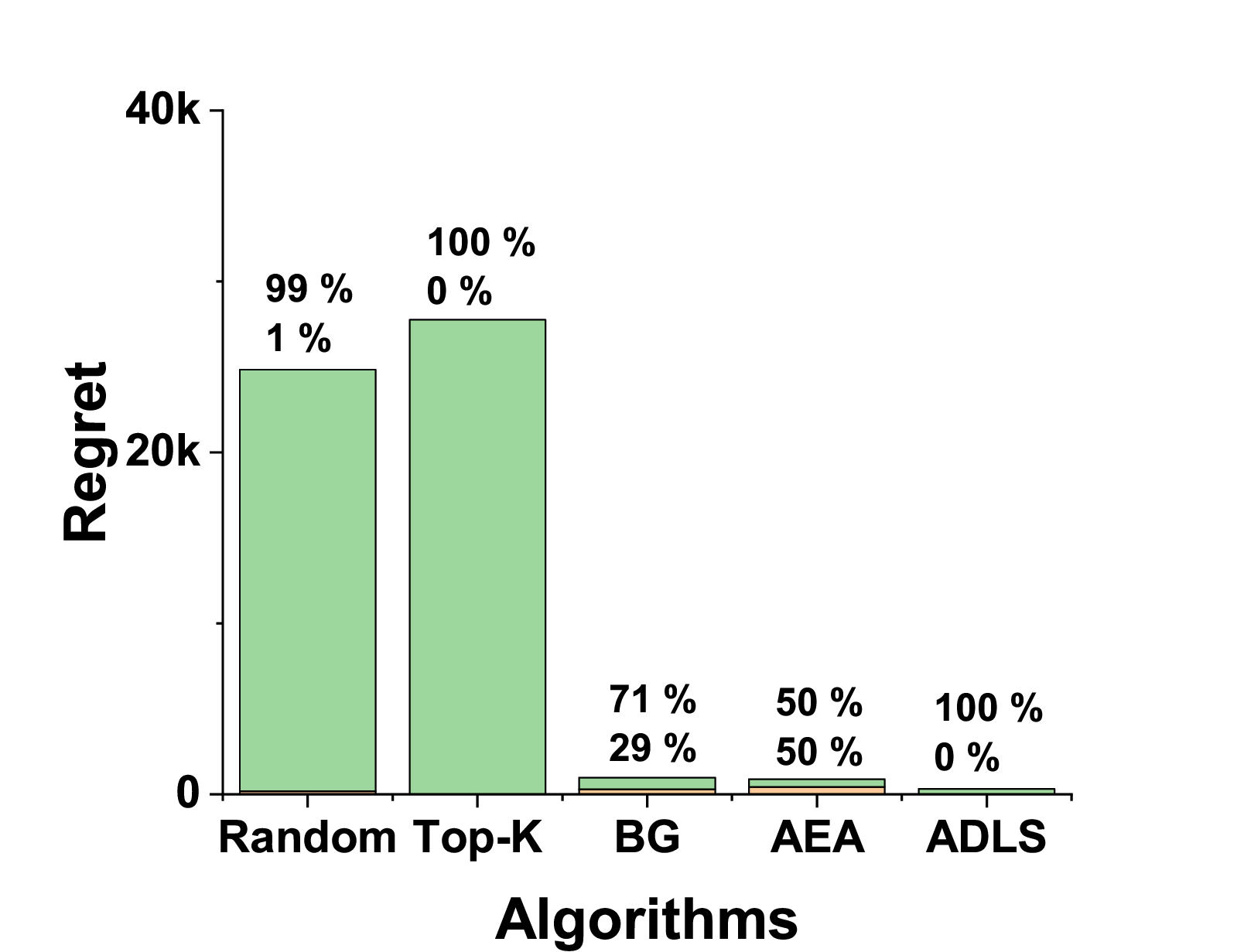} & 
    \includegraphics[scale=0.10]{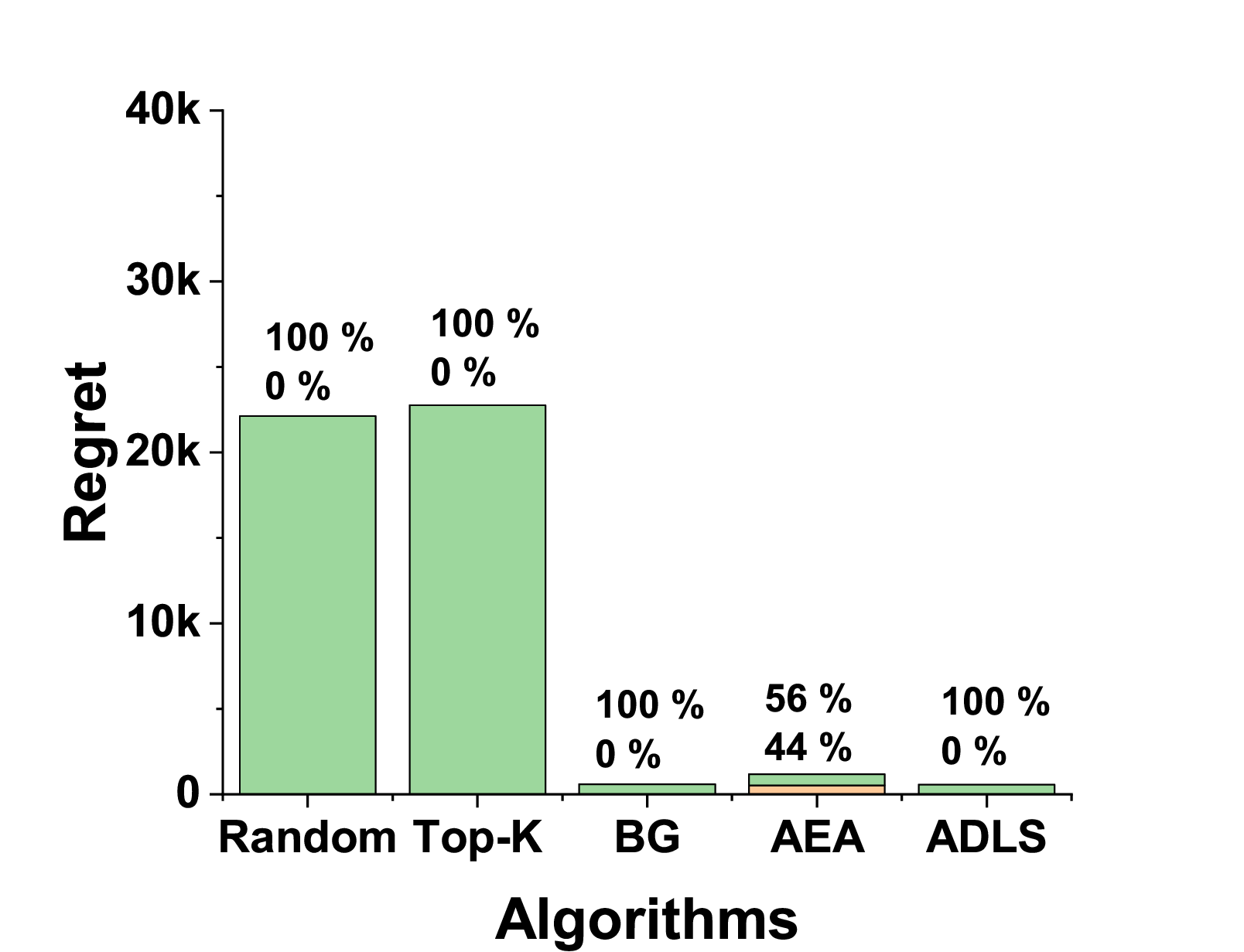} & 
    \includegraphics[scale=0.10]{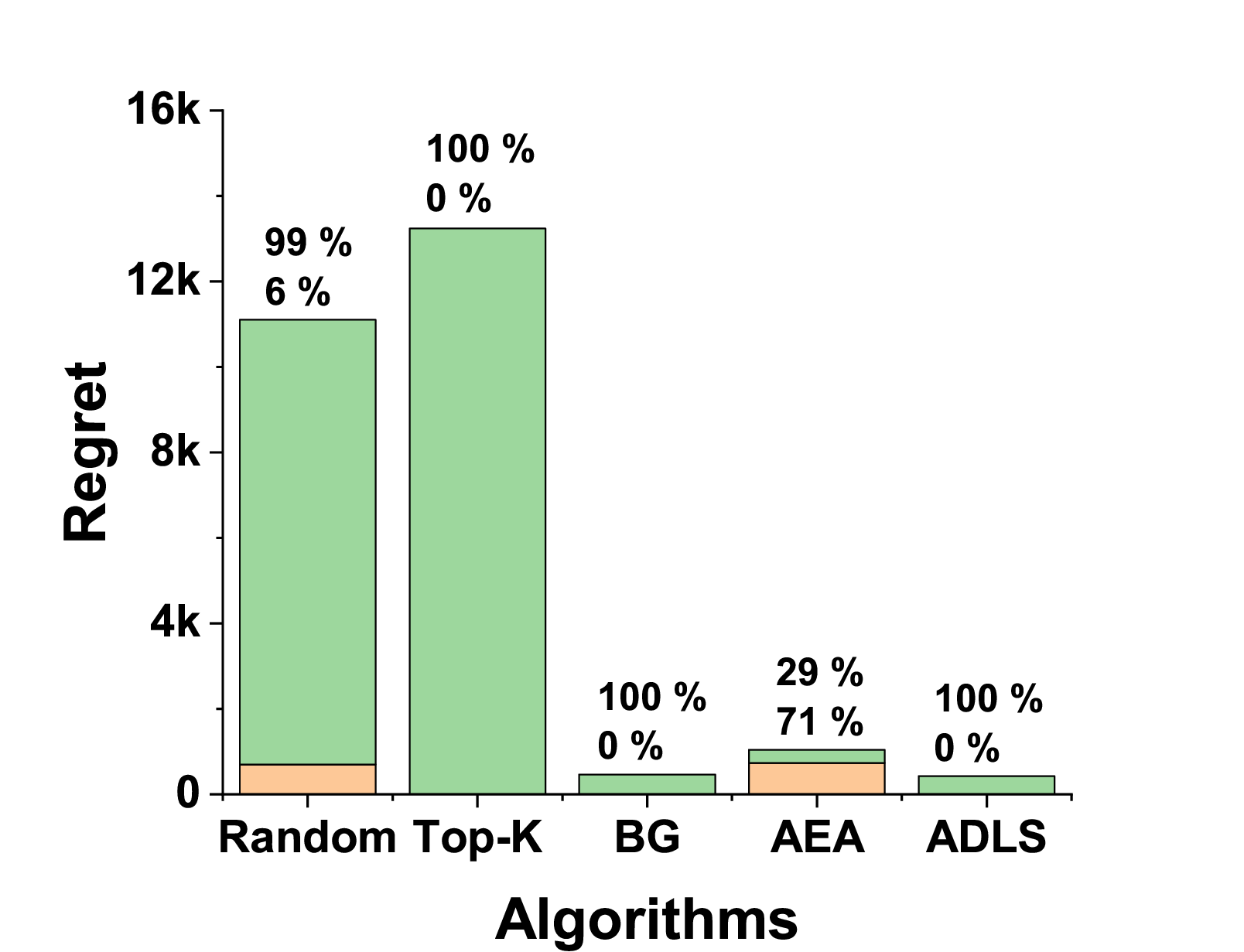} & 
    \includegraphics[scale=0.10]{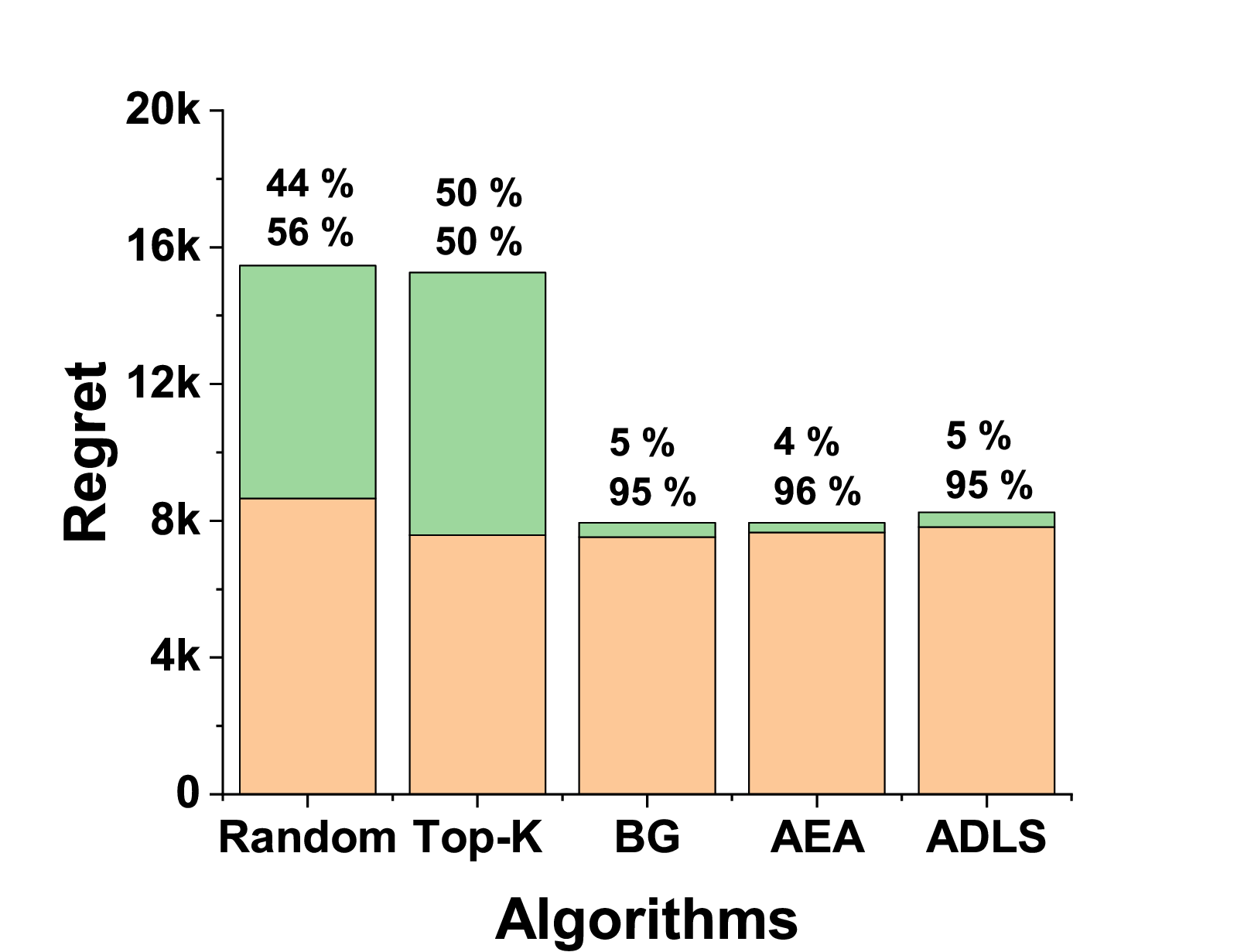} &
    \includegraphics[scale=0.10]{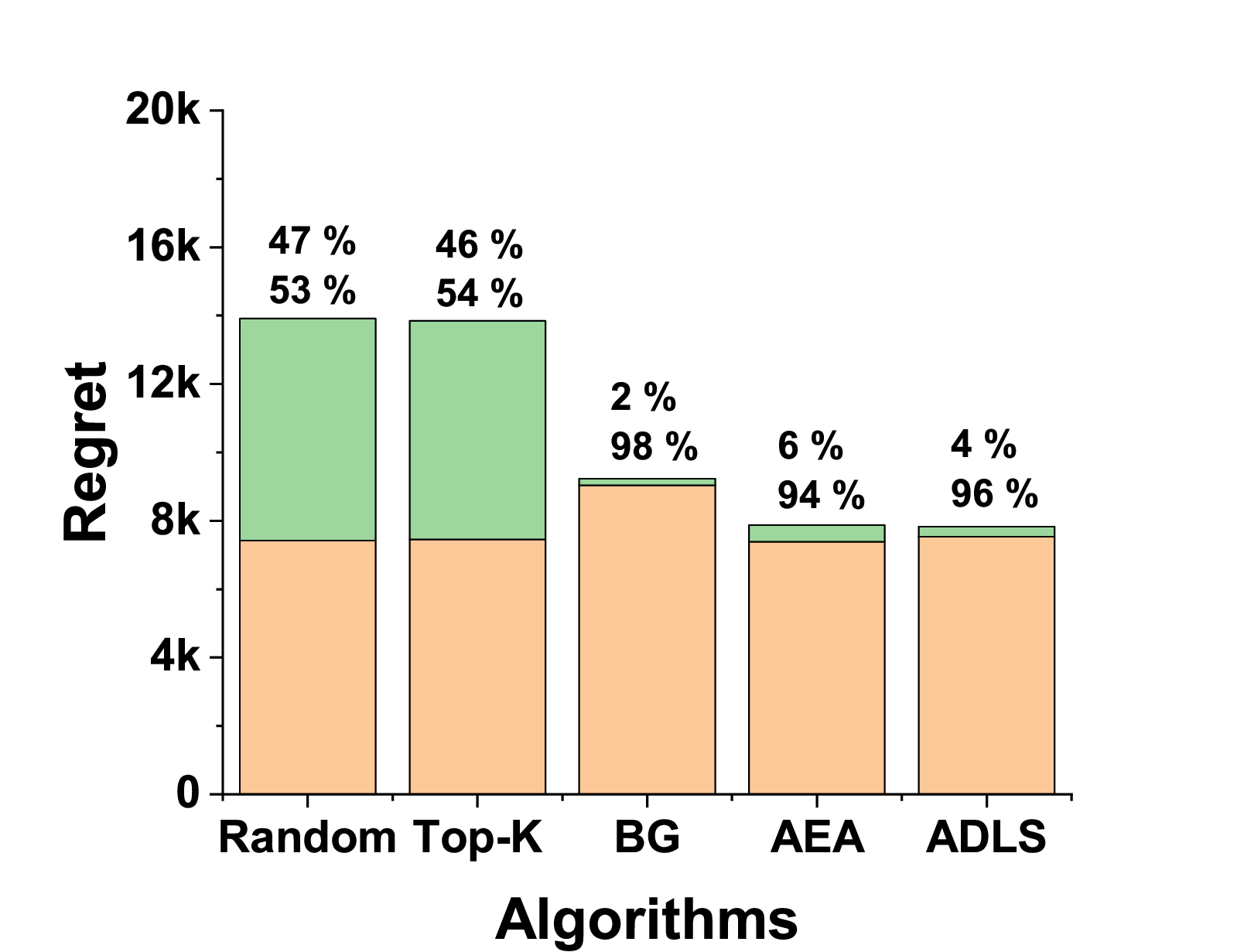} \\
    (a) $\lambda$ = $40\%$ & 
    (b) $\lambda$ = $60\%$ & 
    (c) $\lambda$ = $80\%$ & 
    (d) $\lambda$ = $100\%$ & 
    (e) $\lambda$ = $120\%$ \\
\end{tabular}
\caption{Regret of varying the demand-supply ratio $\lambda$ when $\omega = 50\%$ (Email-Eu-Core, $|\mathcal{A}| = 20$, $p_{c}: Uniform(0.1) $)}
\label{Fig:1_Influence_3}
\begin{tabular}{ccccc}
    \includegraphics[scale=0.10]{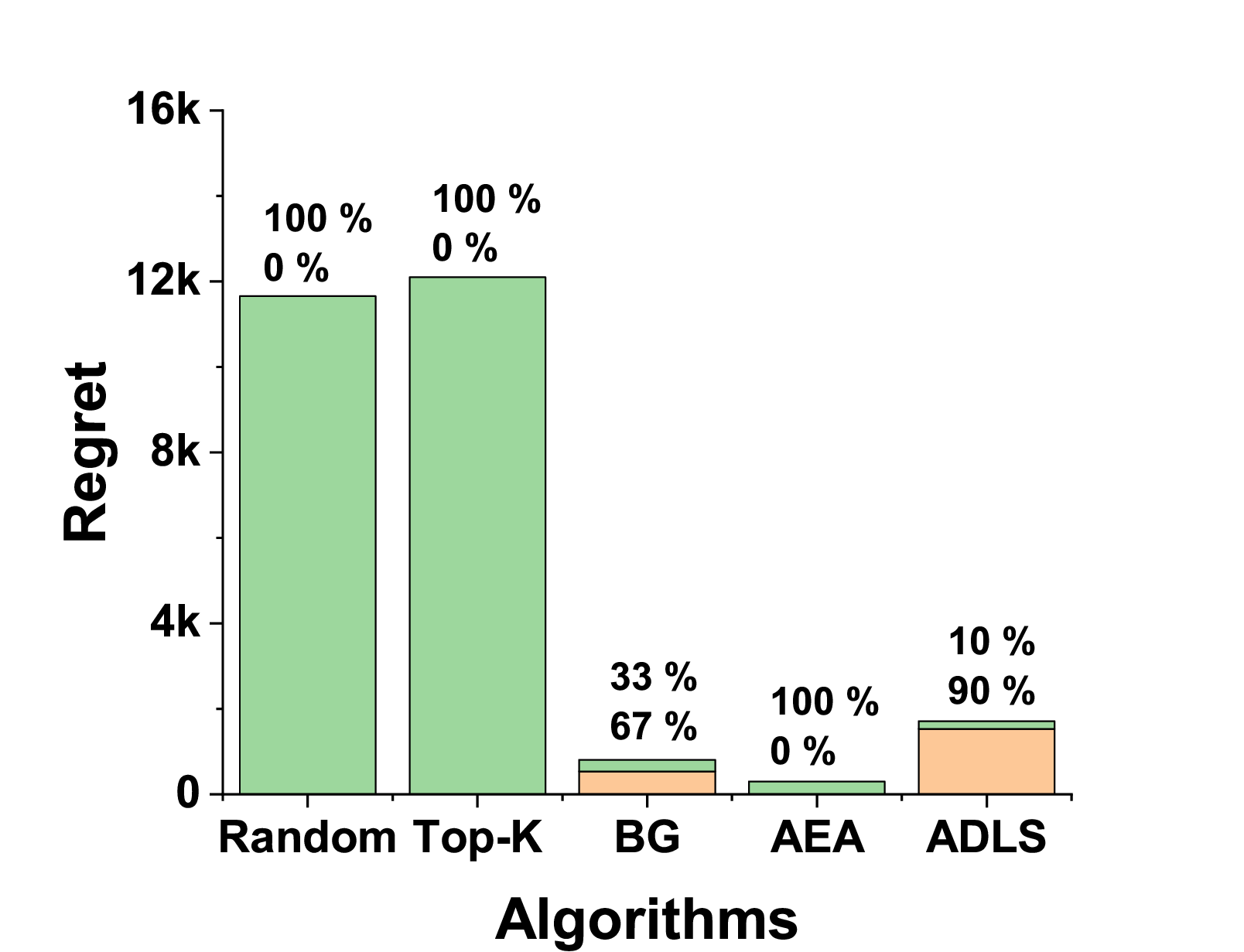} & 
    \includegraphics[scale=0.10]{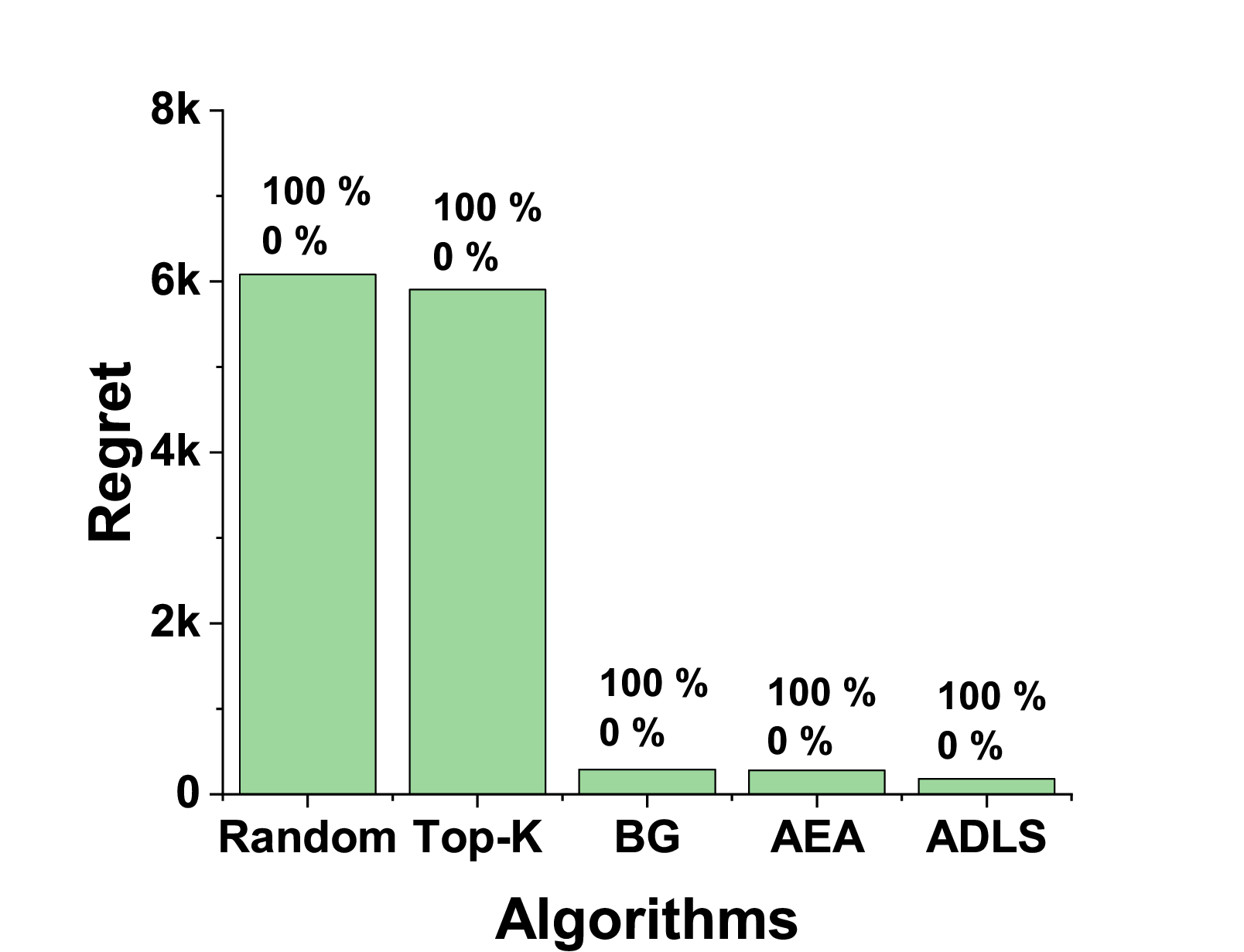} & 
    \includegraphics[scale=0.10]{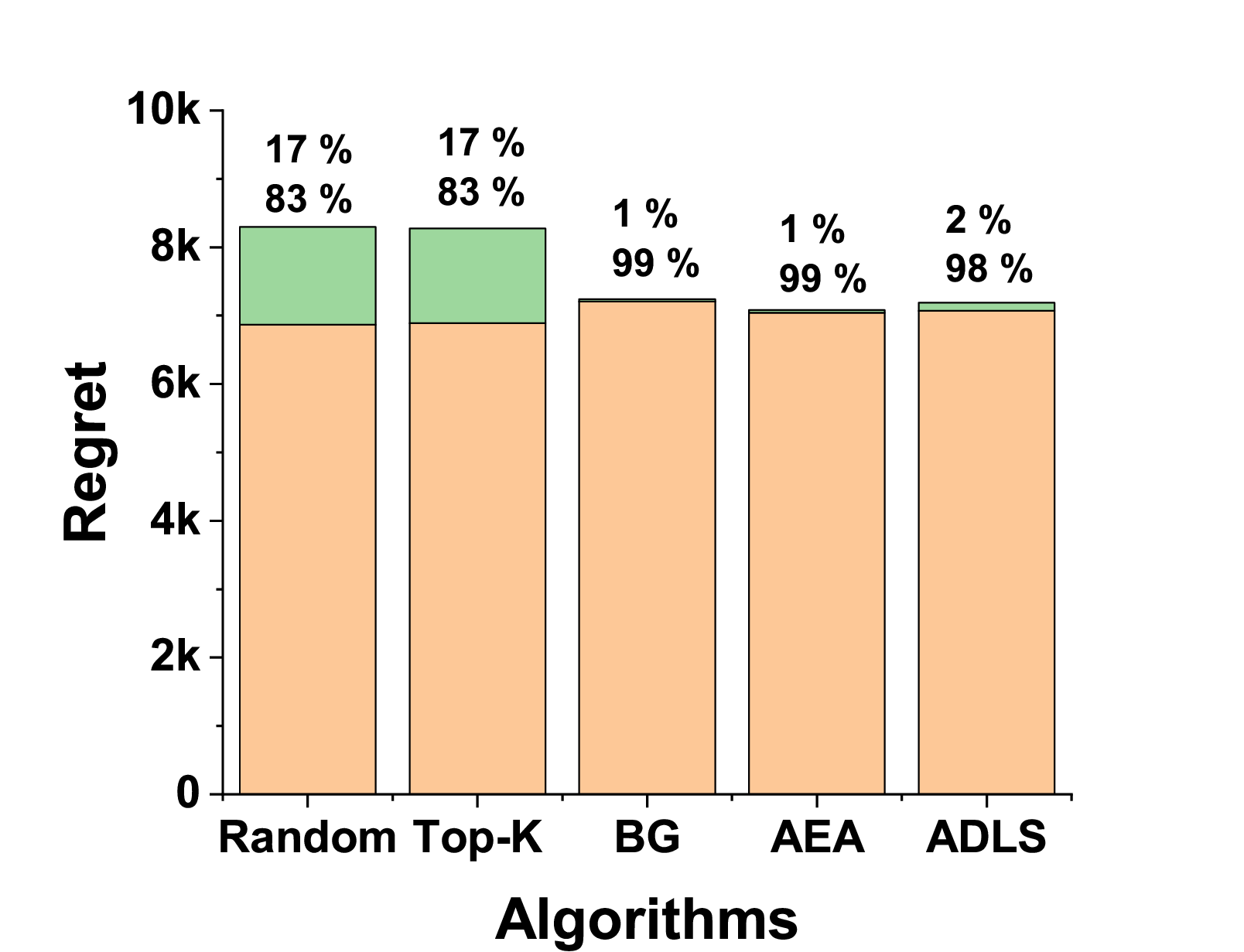} & 
    \includegraphics[scale=0.10]{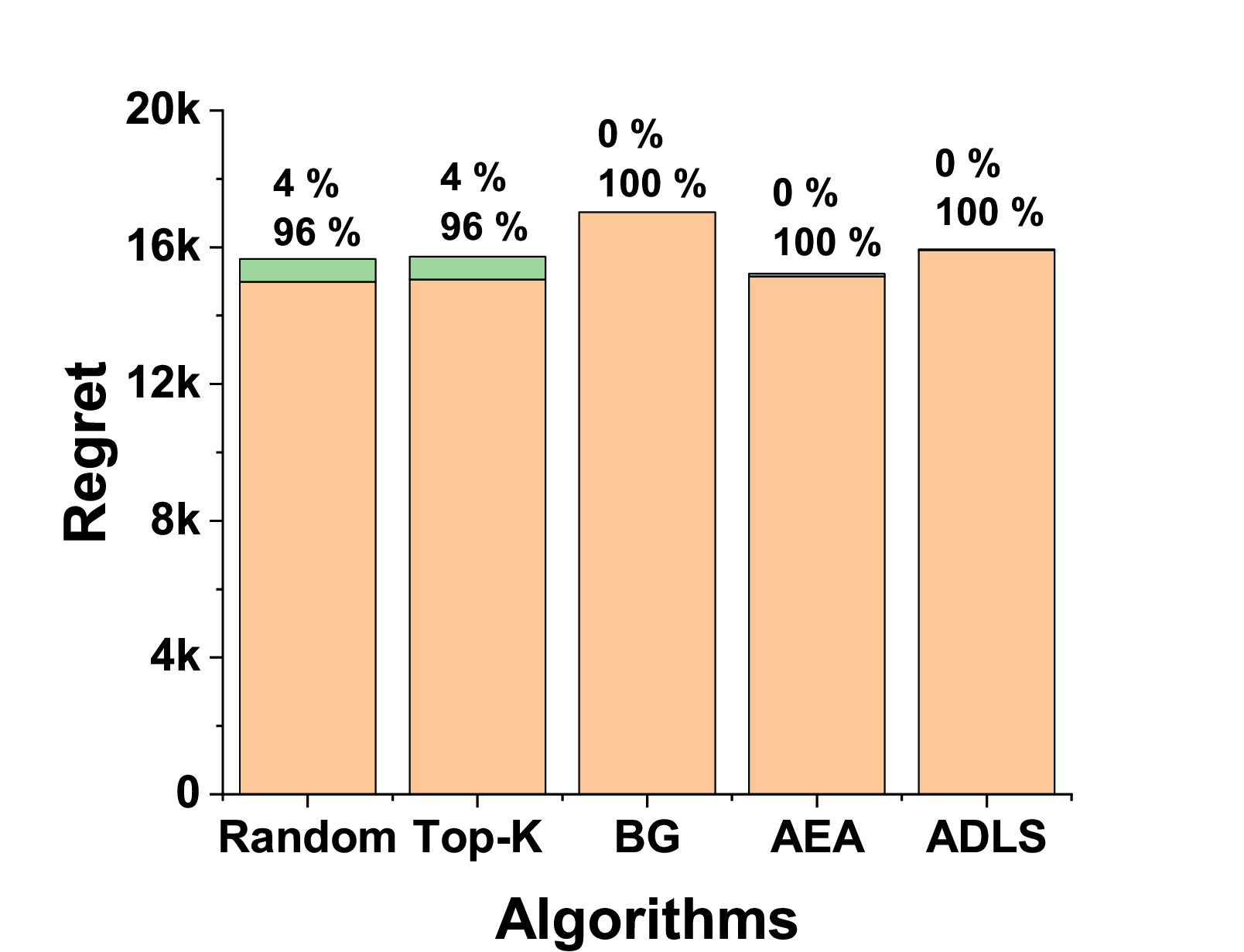} &
    \includegraphics[scale=0.10]{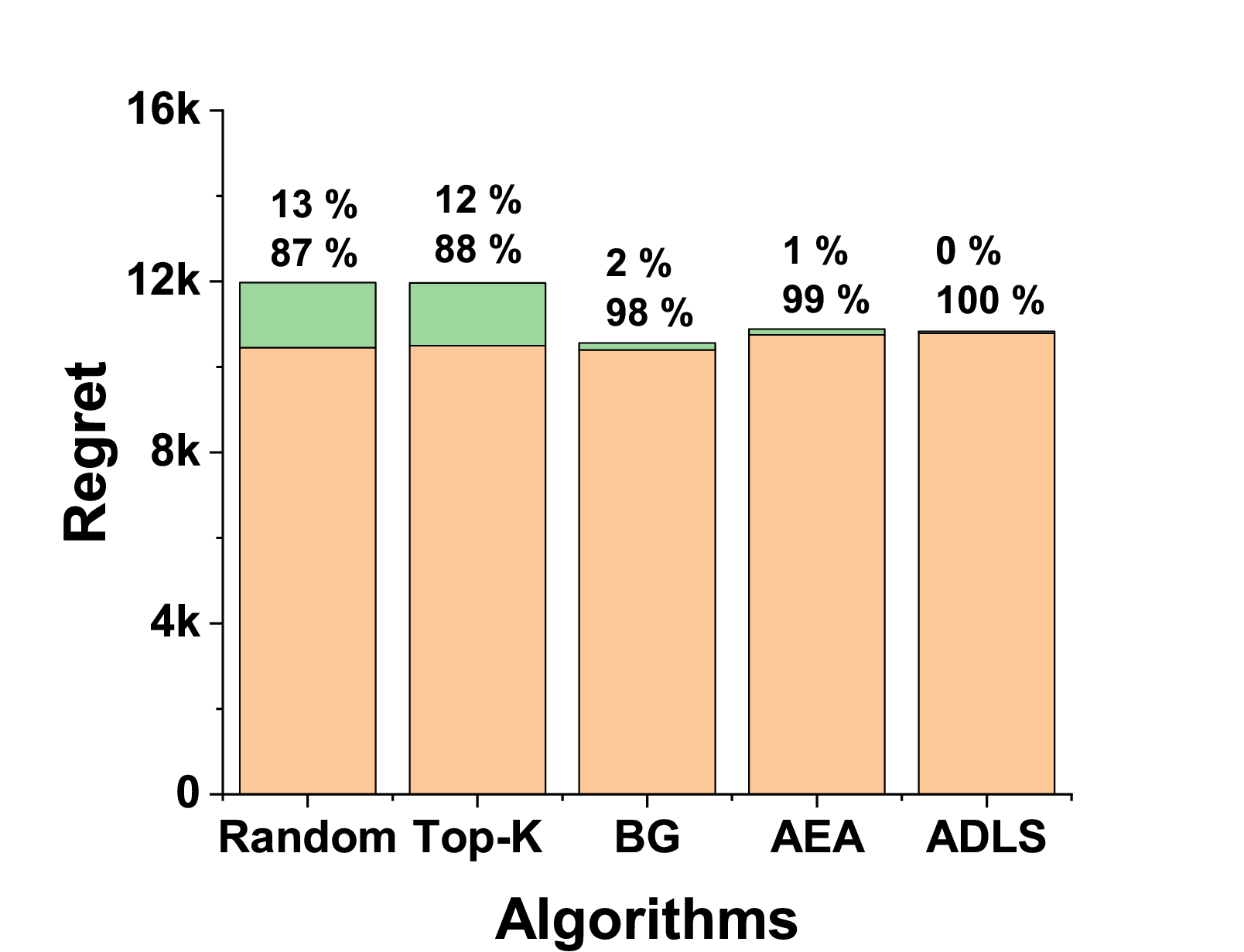} \\
    (a) $\lambda$ = $40\%$ & 
    (b) $\lambda$ = $60\%$ & 
    (c) $\lambda$ = $80\%$ & 
    (d) $\lambda$ = $100\%$ & 
    (e) $\lambda$ = $120\%$ \\
\end{tabular}
\caption{Regret of varying the demand-supply ratio $\lambda$ when $\omega = 70\%$ (Email-Eu-Core, $|\mathcal{A}| = 10$, $p_{c}: Uniform(0.1) $)}
\label{Fig:1_Influence_4}
\begin{tabular}{ccccc}
    \includegraphics[scale=0.10]{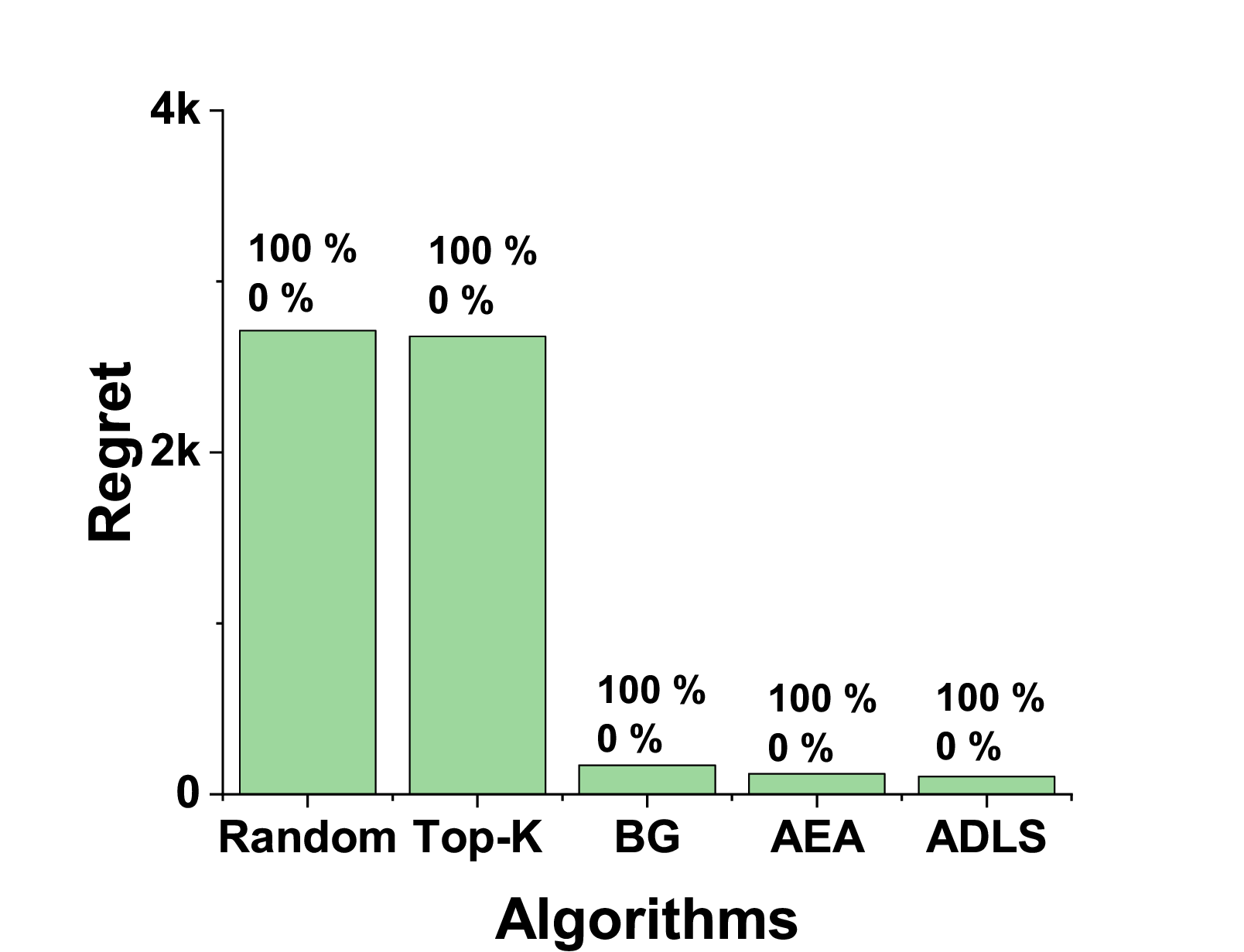} & 
    \includegraphics[scale=0.10]{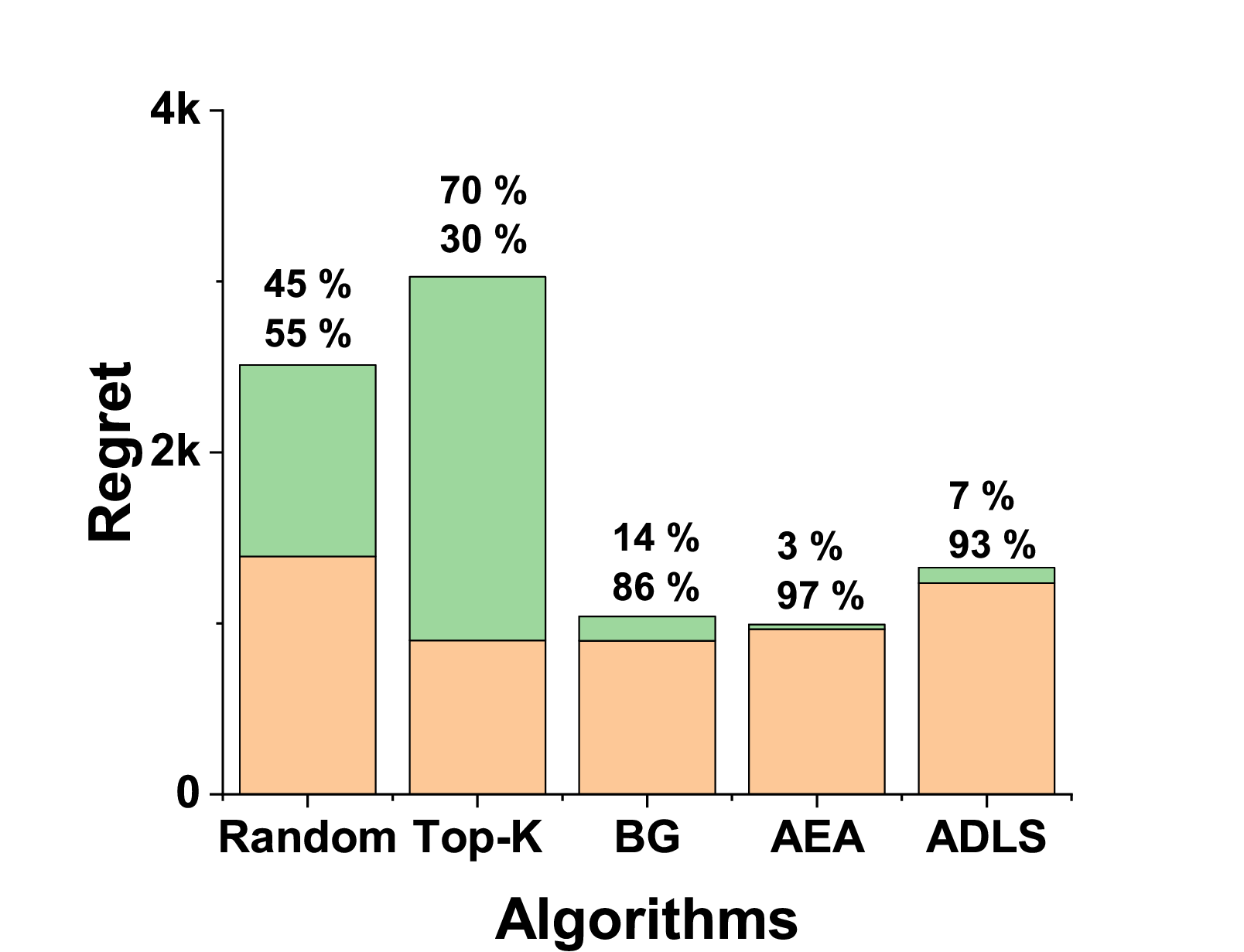} & 
    \includegraphics[scale=0.10]{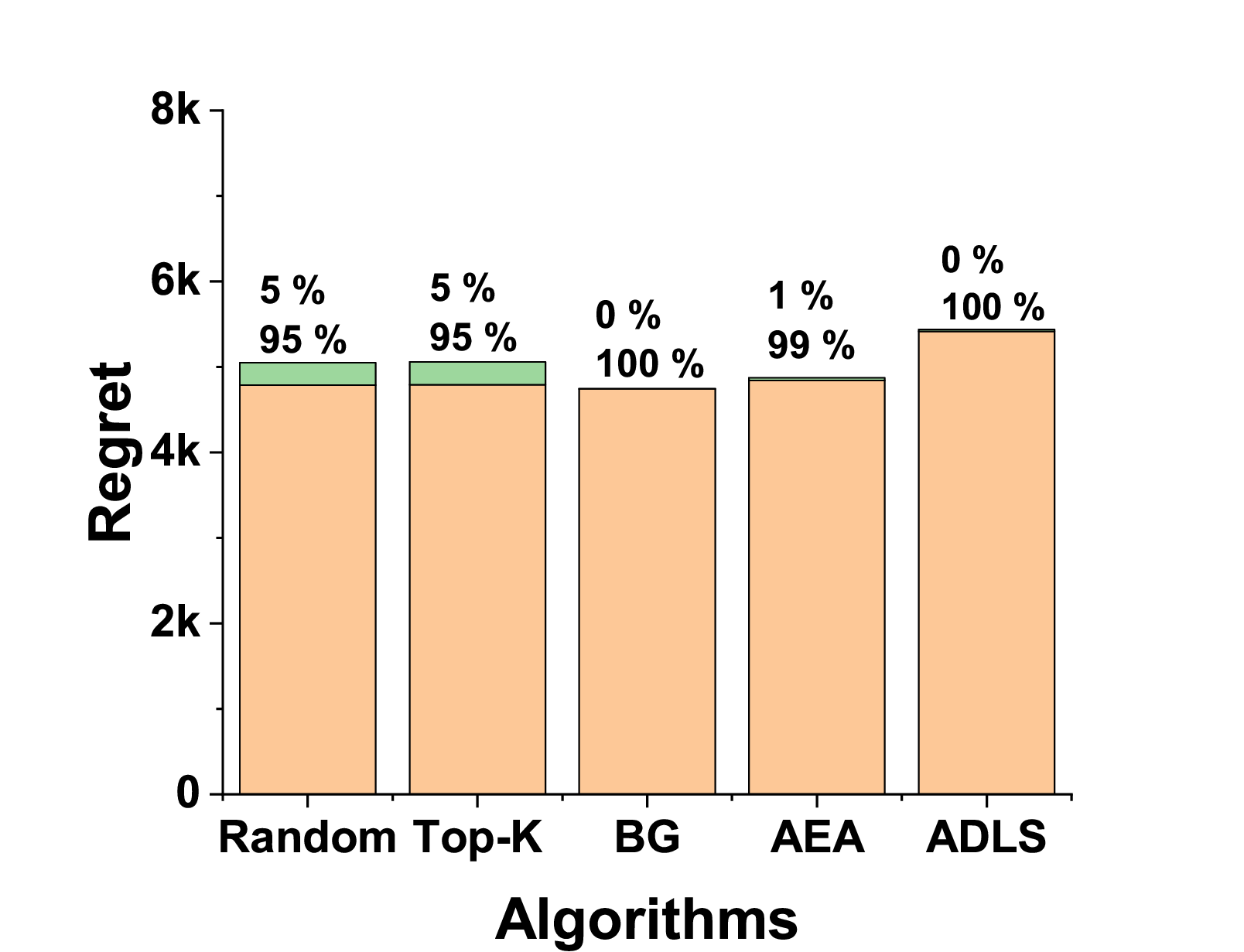} & 
    \includegraphics[scale=0.10]{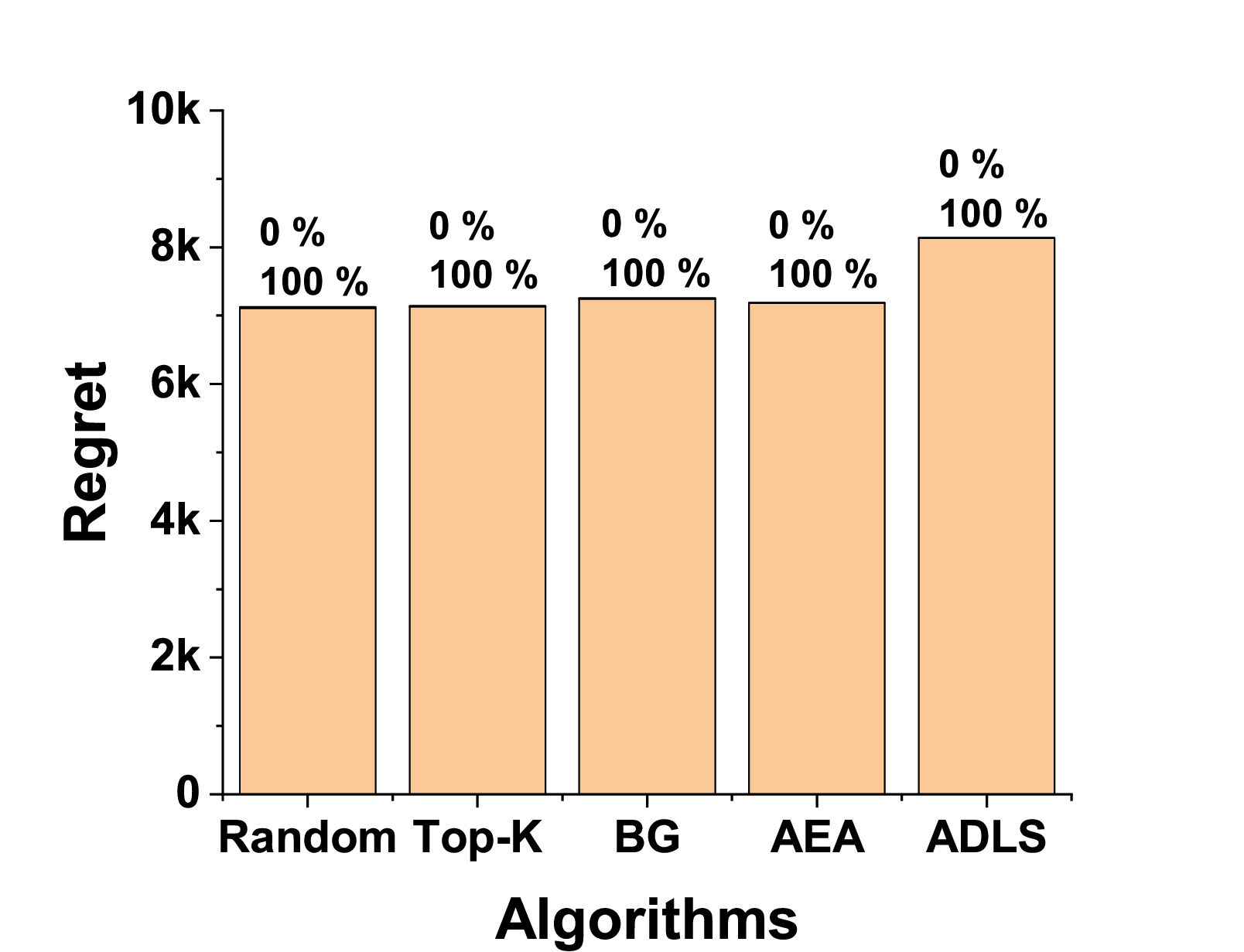} &
    \includegraphics[scale=0.10]{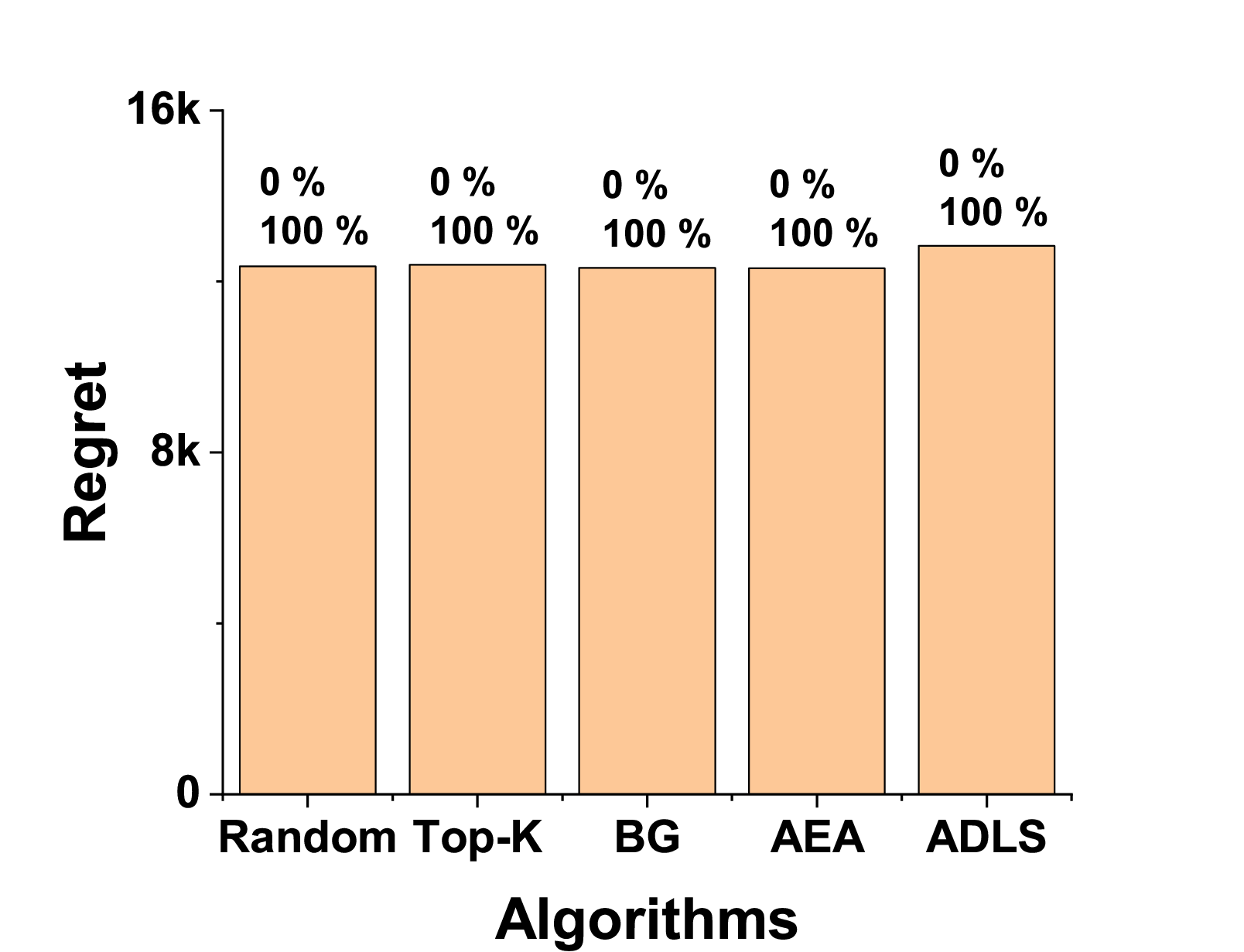} \\
    (a) $\lambda$ = $40\%$ & 
    (b) $\lambda$ = $60\%$ & 
    (c) $\lambda$ = $80\%$ & 
    (d) $\lambda$ = $100\%$ & 
    (e) $\lambda$ = $120\%$ \\
\end{tabular}
\caption{Regret of varying the demand-supply ratio $\lambda$ when $\omega = 90\%$ (Email-Eu-Core, $|\mathcal{A}| = 5$, $p_{c}: Uniform(0.1) $)}
\label{Fig:1_Influence_5}
\end{figure}

\begin{figure}[H]
\centering
\begin{tabular}{lclc}
    Excessive Regret & \includegraphics[width=0.11\linewidth]{Excessive.png} & 
    Unsatisfied Regret & \includegraphics[width=0.11\linewidth]{Unsatisfied.png} \\
\end{tabular}
\begin{tabular}{ccccc}
    \includegraphics[scale=0.10]{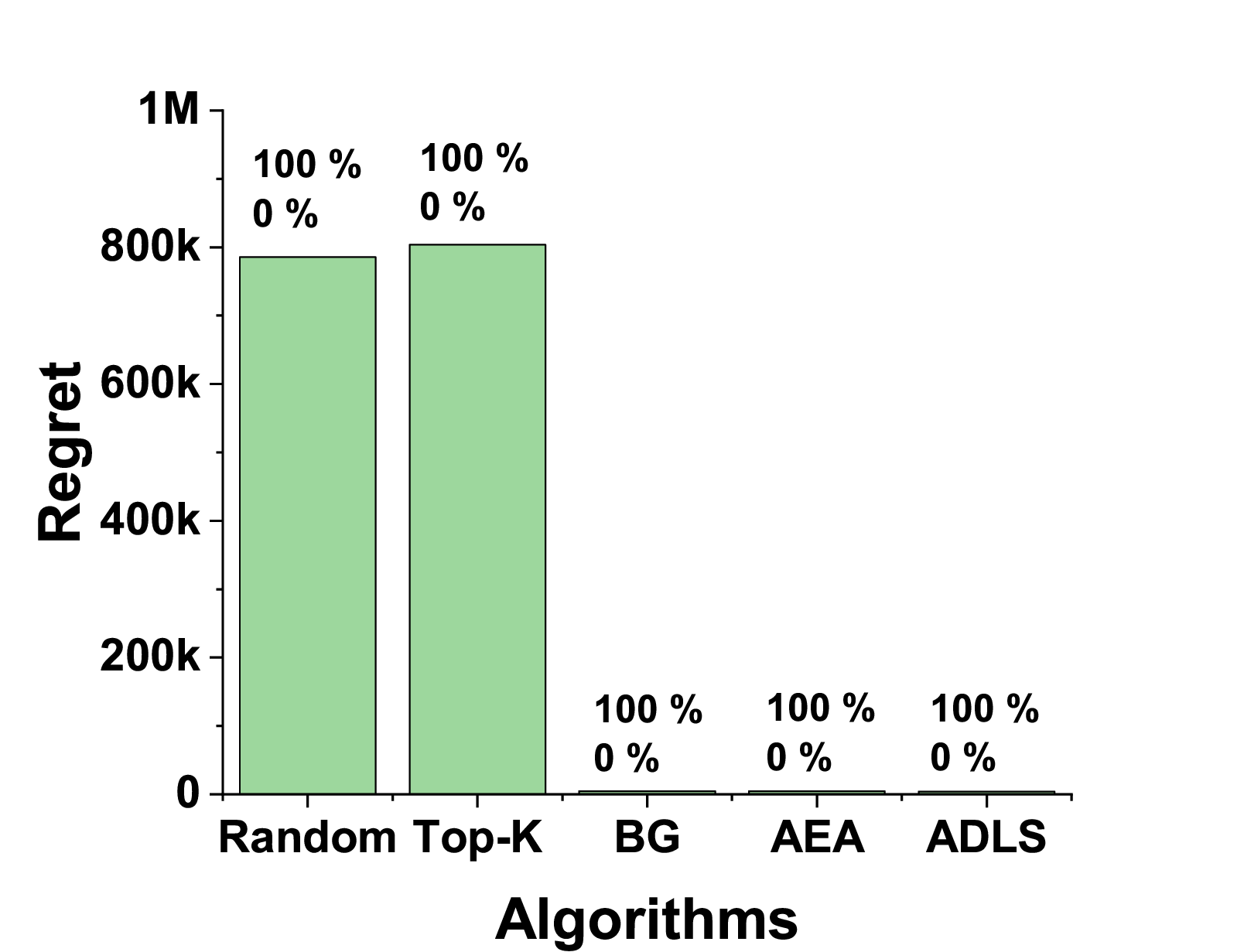} & 
    \includegraphics[scale=0.10]{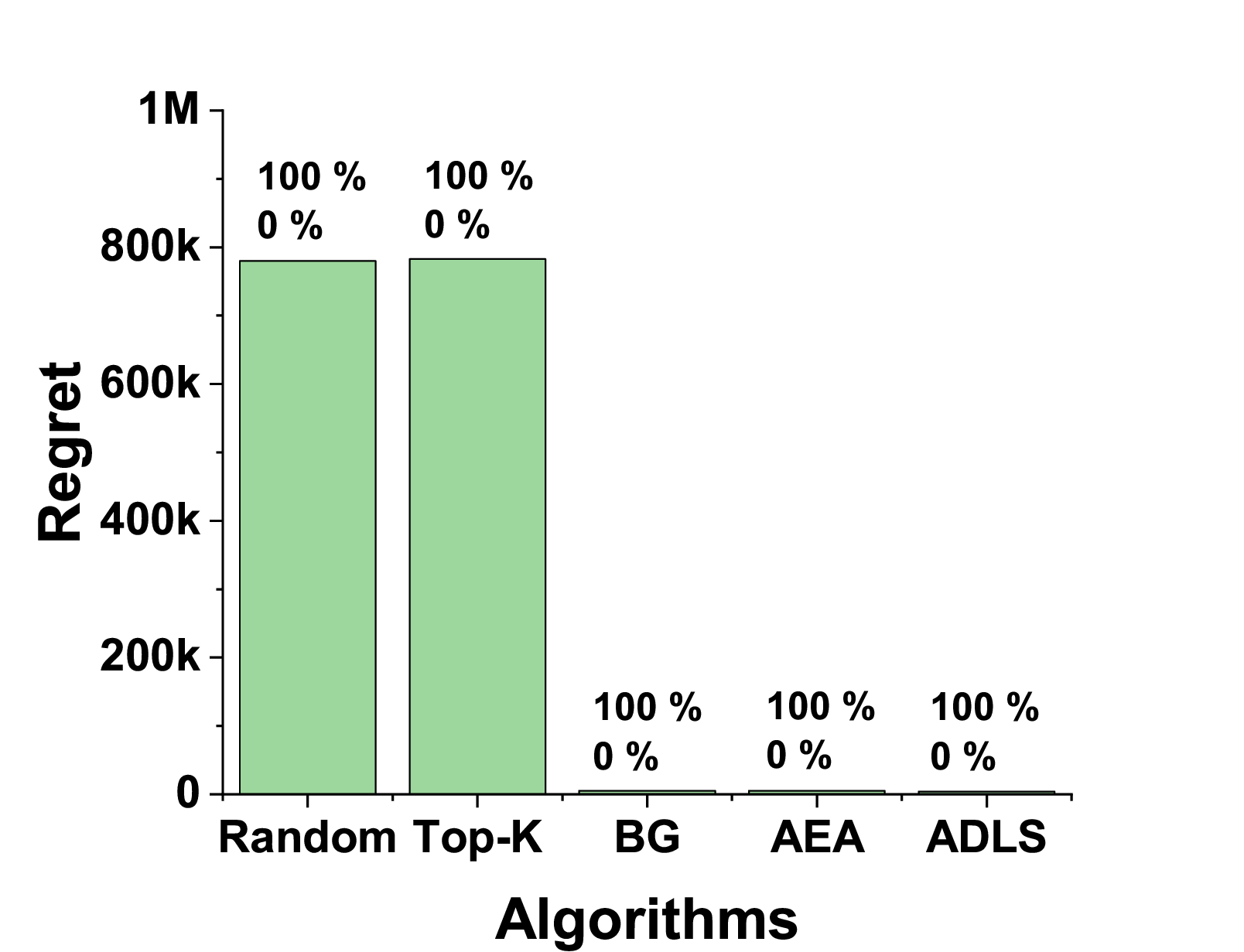} & 
    \includegraphics[scale=0.10]{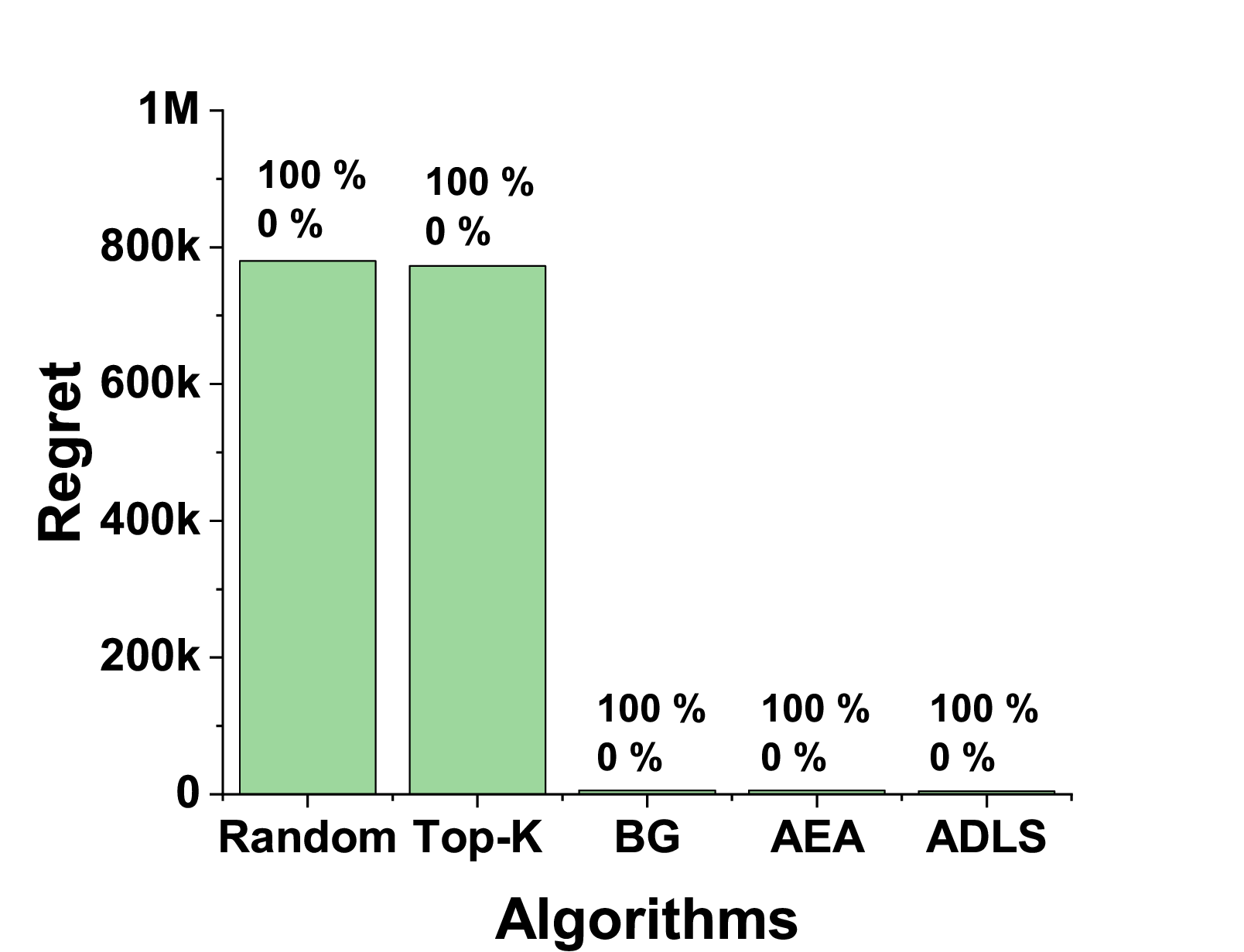} & 
    \includegraphics[scale=0.10]{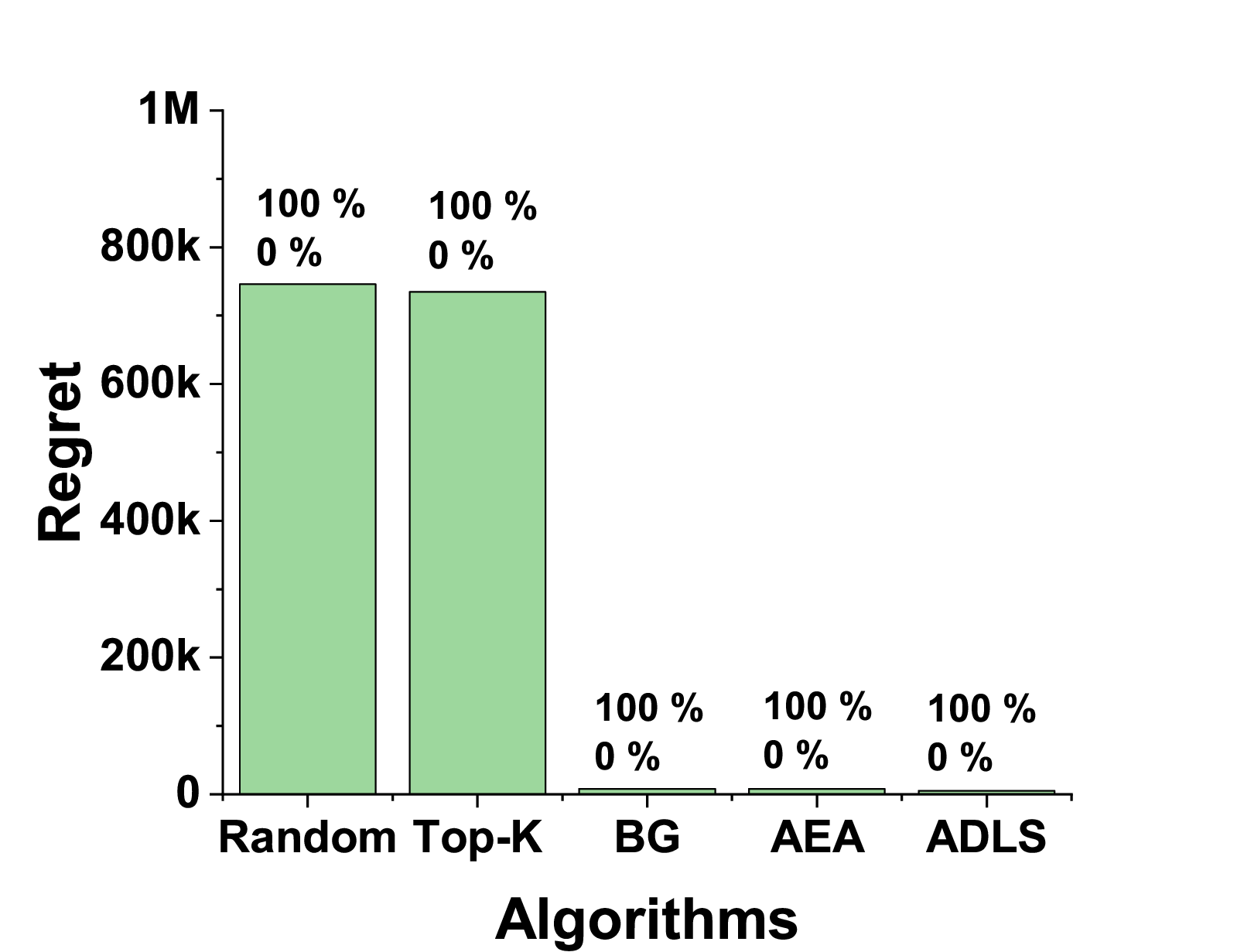} &
    \includegraphics[scale=0.10]{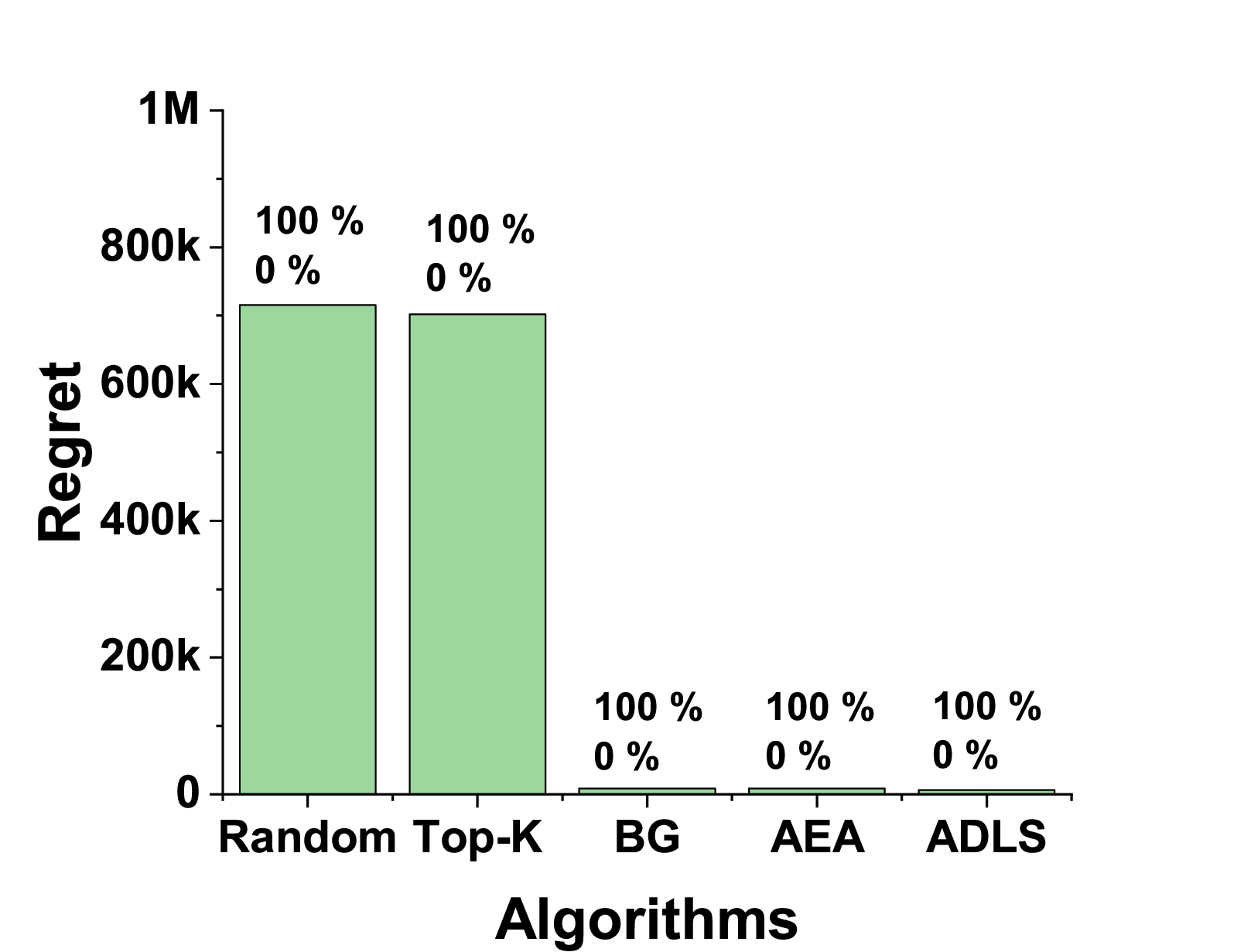} \\
    (a) $\lambda$ = $40\%$ & 
    (b) $\lambda$ = $60\%$ & 
    (c) $\lambda$ = $80\%$ & 
    (d) $\lambda$ = $100\%$ & 
    (e) $\lambda$ = $120\%$ \\
\end{tabular}
\caption{Regret of varying the demand-supply ratio $\lambda$ when $\omega = 10\%$ (Facebook, $|\mathcal{A}| = 100$, $p_{c}: Uniform(0.1) $)}
\label{Fig:3_Influence_1}
\begin{tabular}{ccccc}
    \includegraphics[scale=0.10]{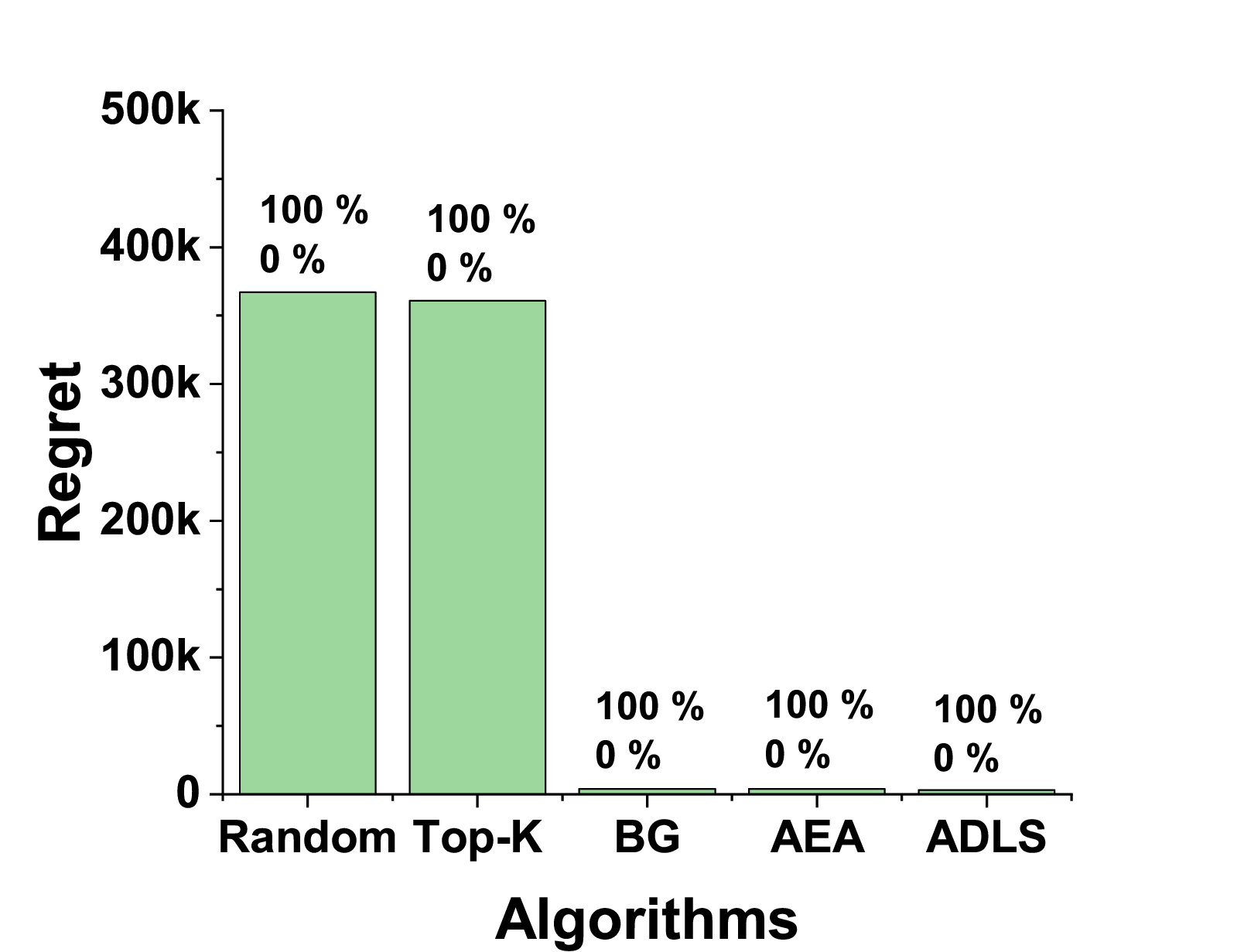} & 
    \includegraphics[scale=0.10]{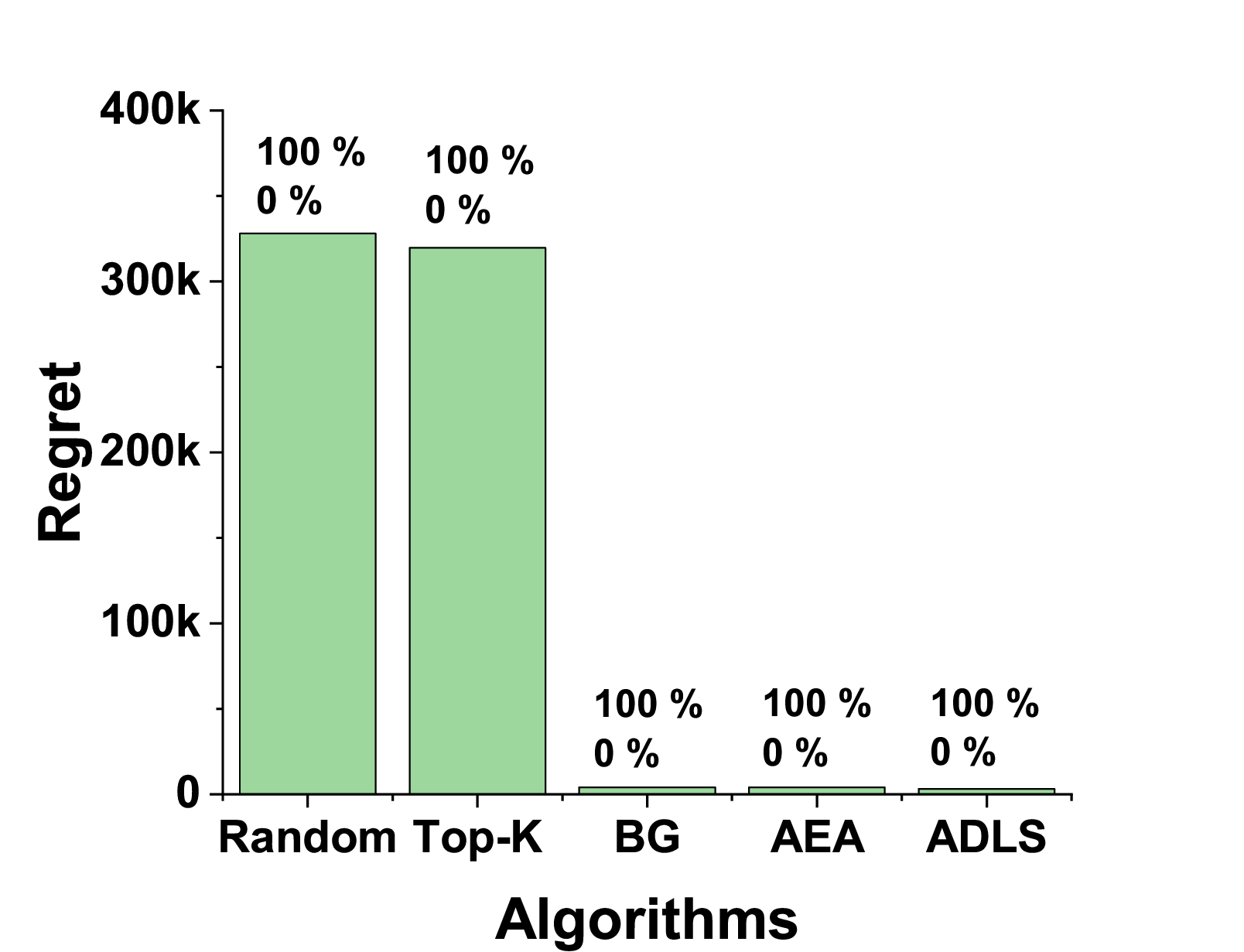} & 
    \includegraphics[scale=0.10]{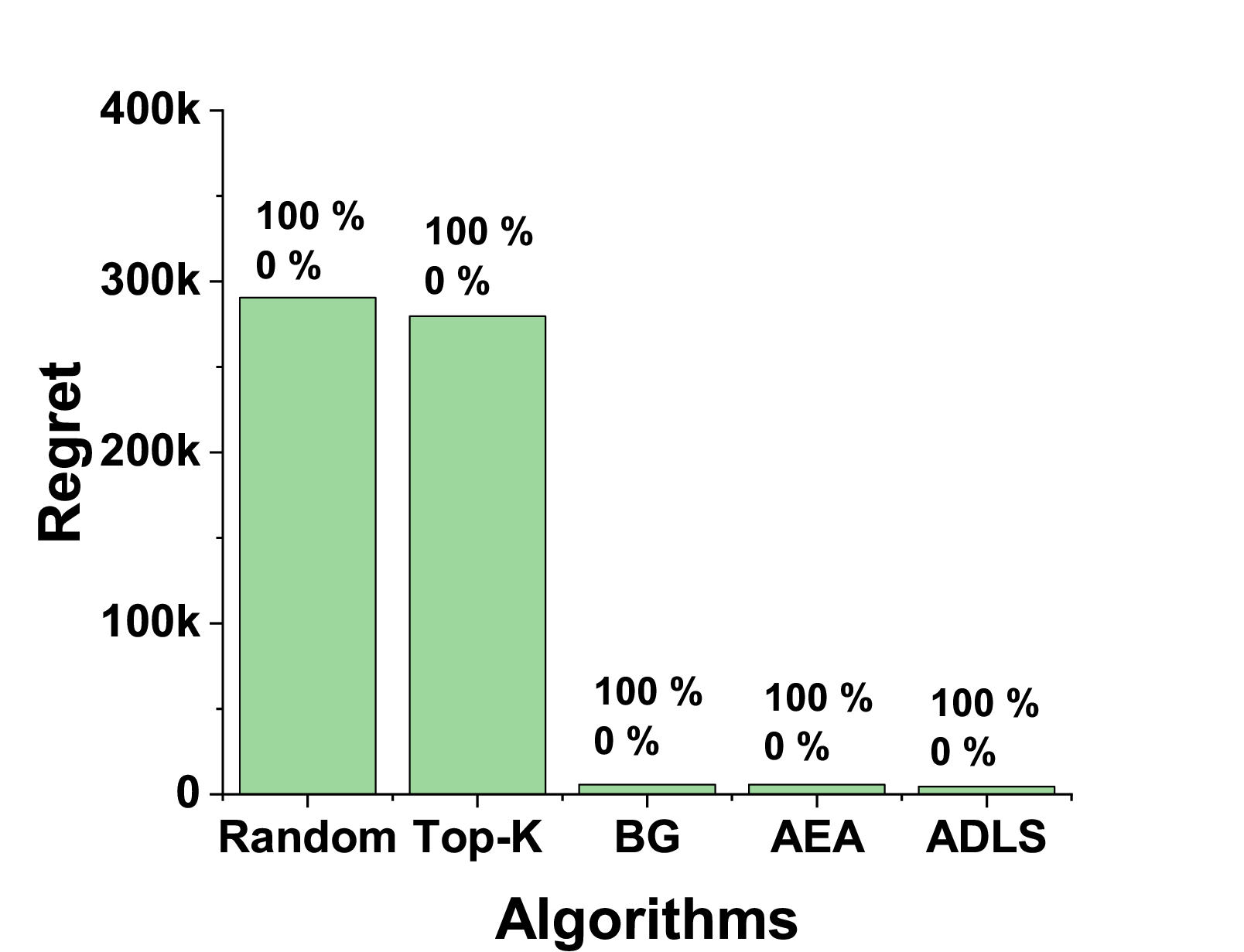} & 
    \includegraphics[scale=0.10]{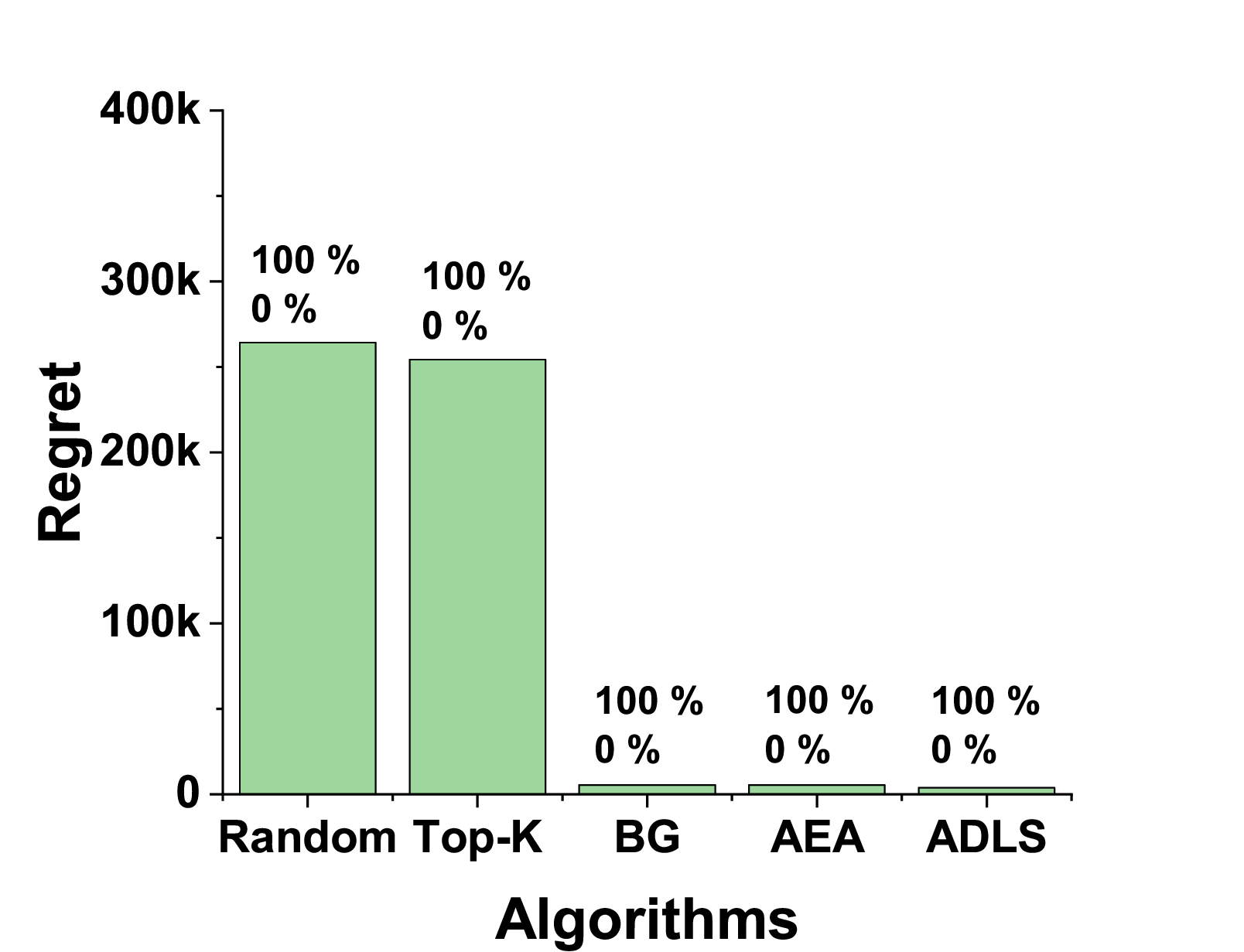} &
    \includegraphics[scale=0.10]{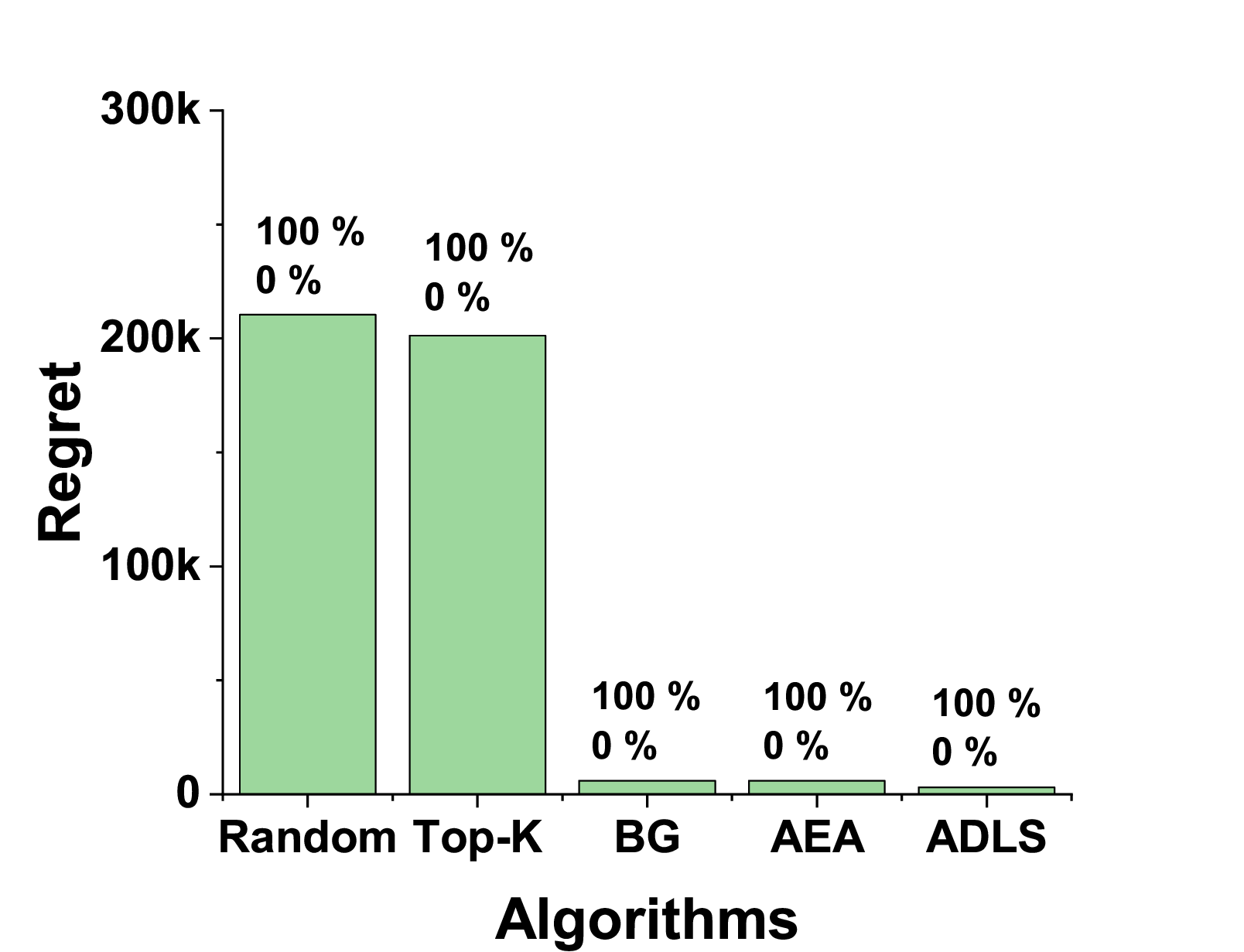} \\
    (a) $\lambda$ = $40\%$ & 
    (b) $\lambda$ = $60\%$ & 
    (c) $\lambda$ = $80\%$ & 
    (d) $\lambda$ = $100\%$ & 
    (e) $\lambda$ = $120\%$ \\
\end{tabular}
\caption{Regret of varying the demand-supply ratio $\lambda$ when $\omega = 30\%$ (Facebook, $|\mathcal{A}| = 50$, $p_{c}: Uniform(0.1) $)}
\label{Fig:3_Influence_2}
\begin{tabular}{ccccc}
    \includegraphics[scale=0.10]{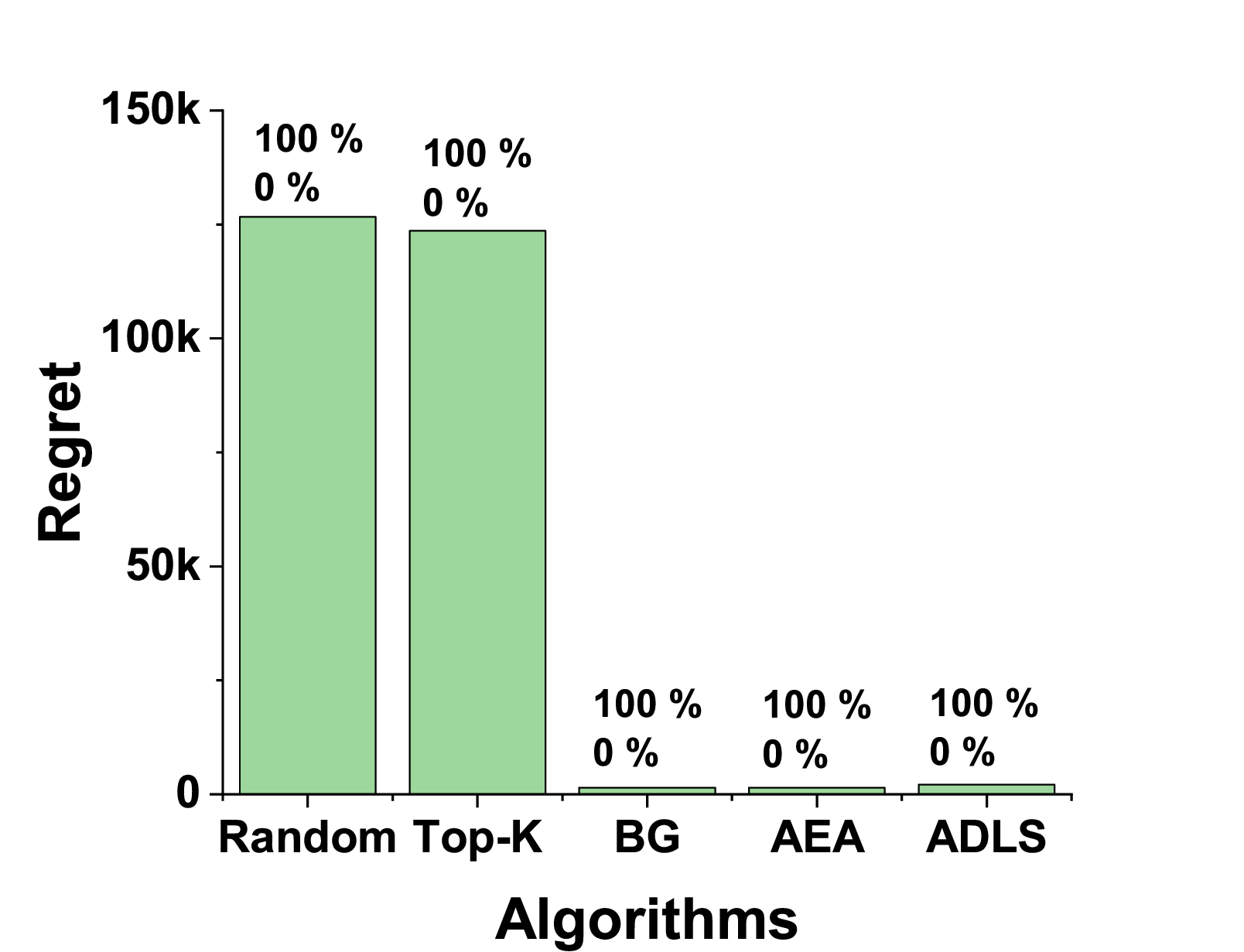} & 
    \includegraphics[scale=0.10]{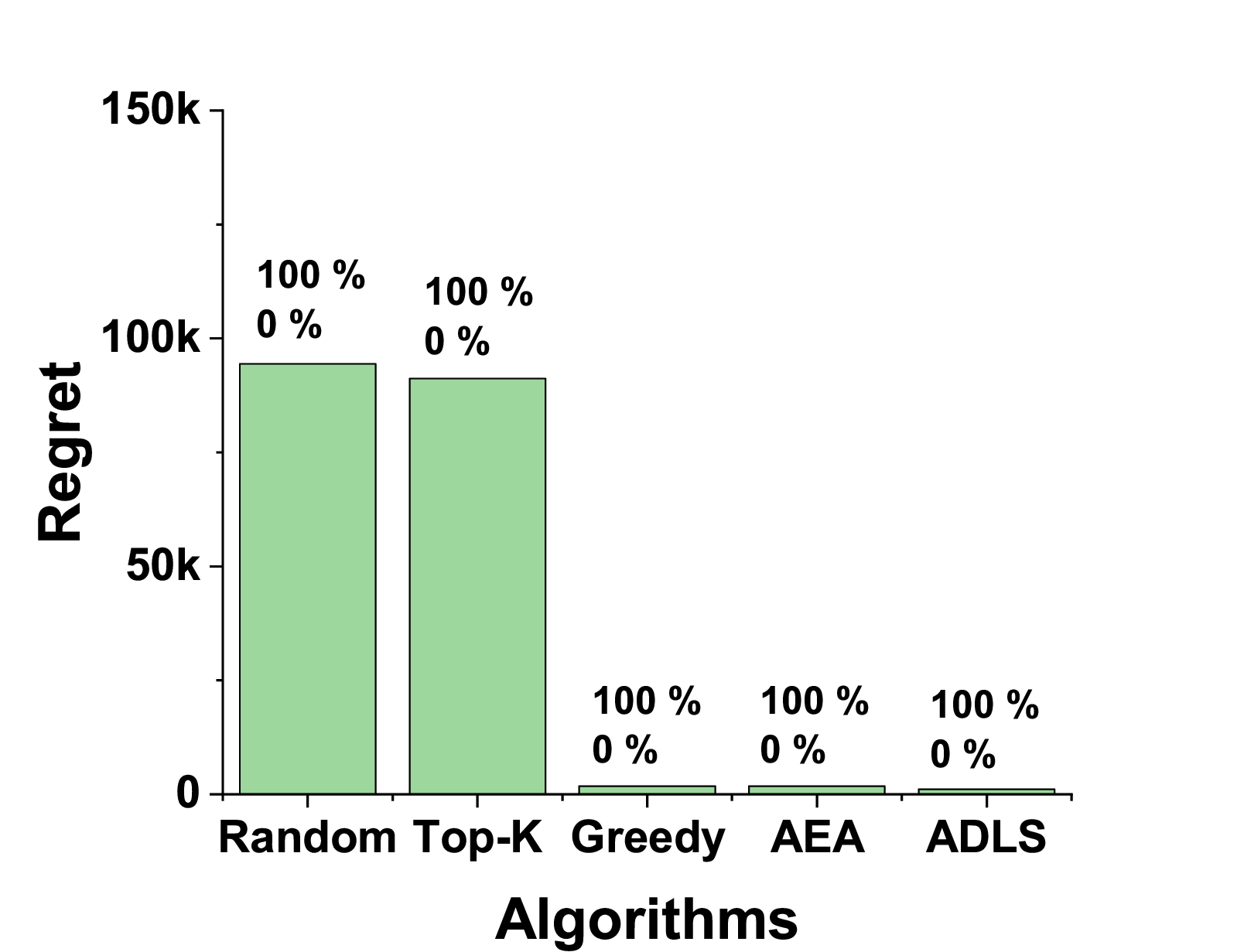} & 
    \includegraphics[scale=0.10]{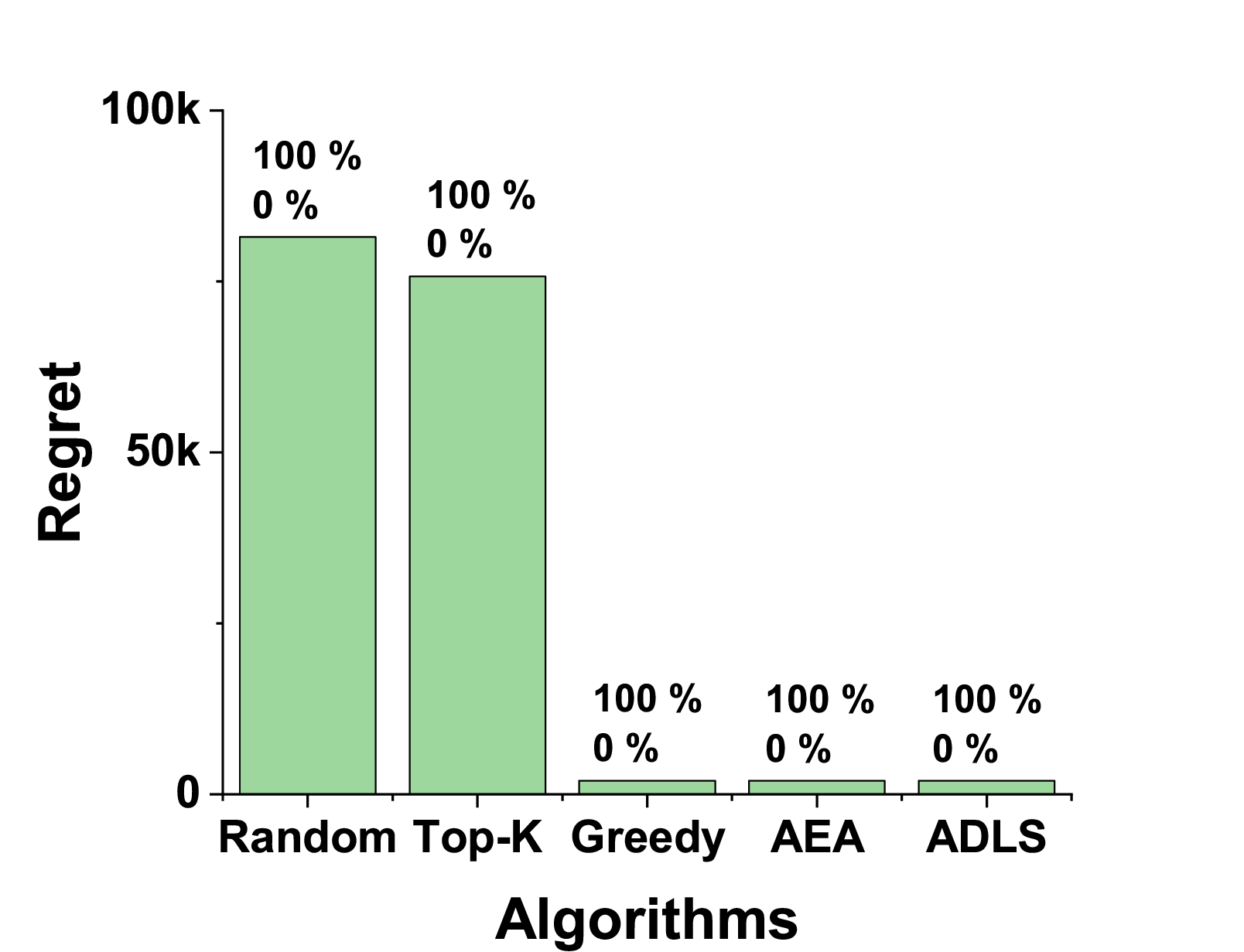} & 
    \includegraphics[scale=0.10]{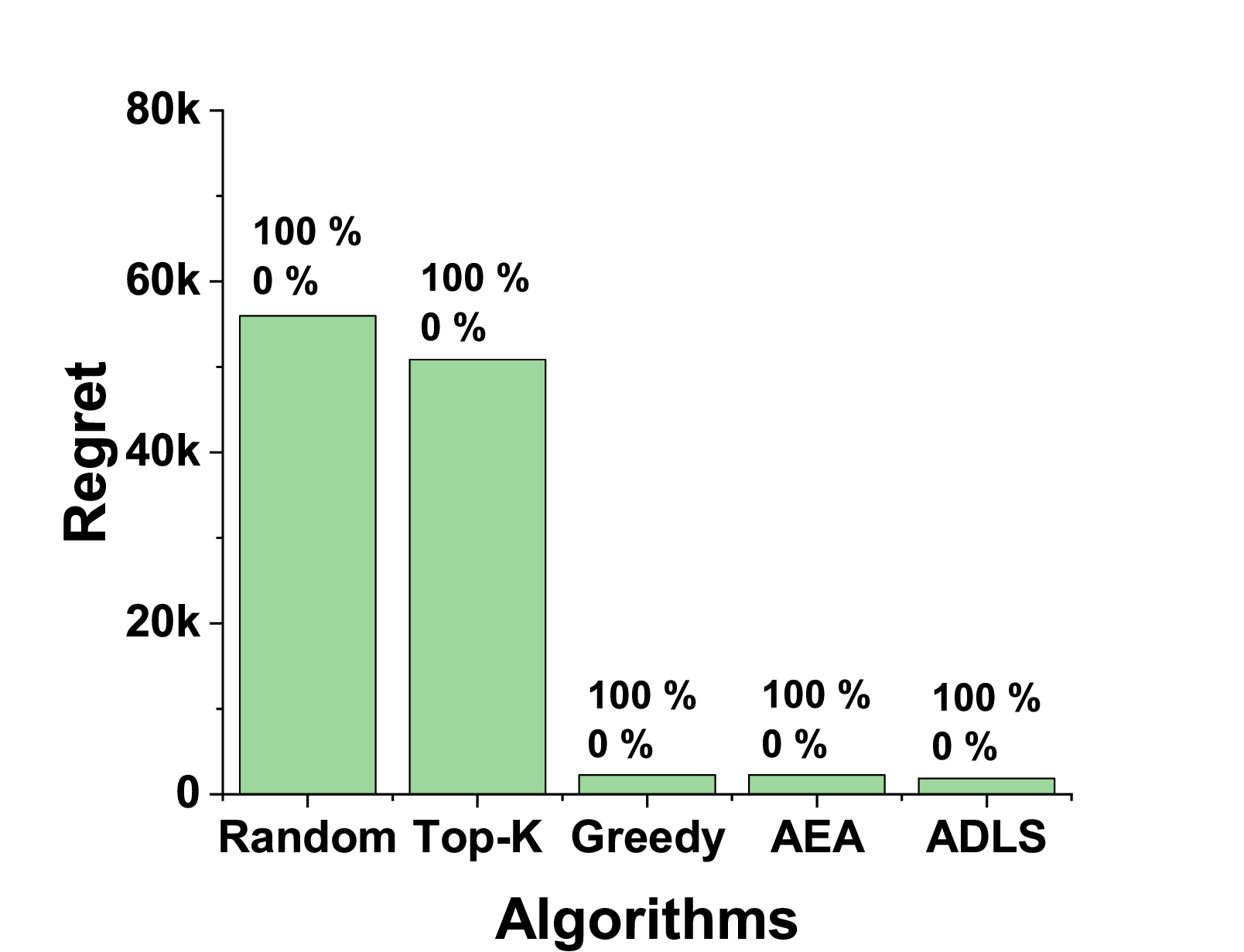} &
    \includegraphics[scale=0.10]{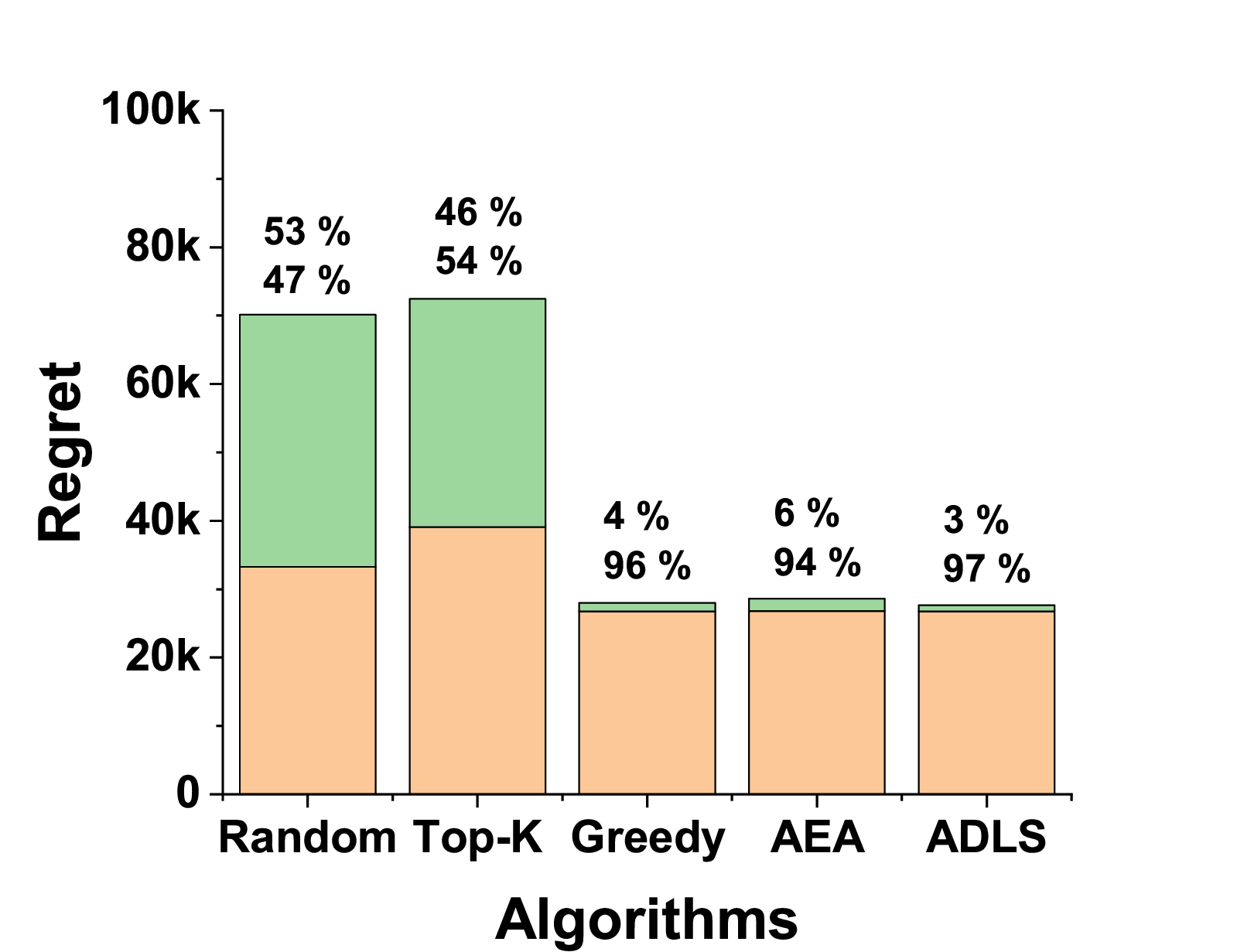} \\
    (a) $\lambda$ = $40\%$ & 
    (b) $\lambda$ = $60\%$ & 
    (c) $\lambda$ = $80\%$ & 
    (d) $\lambda$ = $100\%$ & 
    (e) $\lambda$ = $120\%$ \\
\end{tabular}
\caption{Regret of varying the demand-supply ratio $\lambda$ when $\omega = 50\%$ (Facebook, $|\mathcal{A}| = 20$, $p_{c}: Uniform(0.1) $)}
\label{Fig:3_Influence_3}
\begin{tabular}{ccccc}
    \includegraphics[scale=0.10]{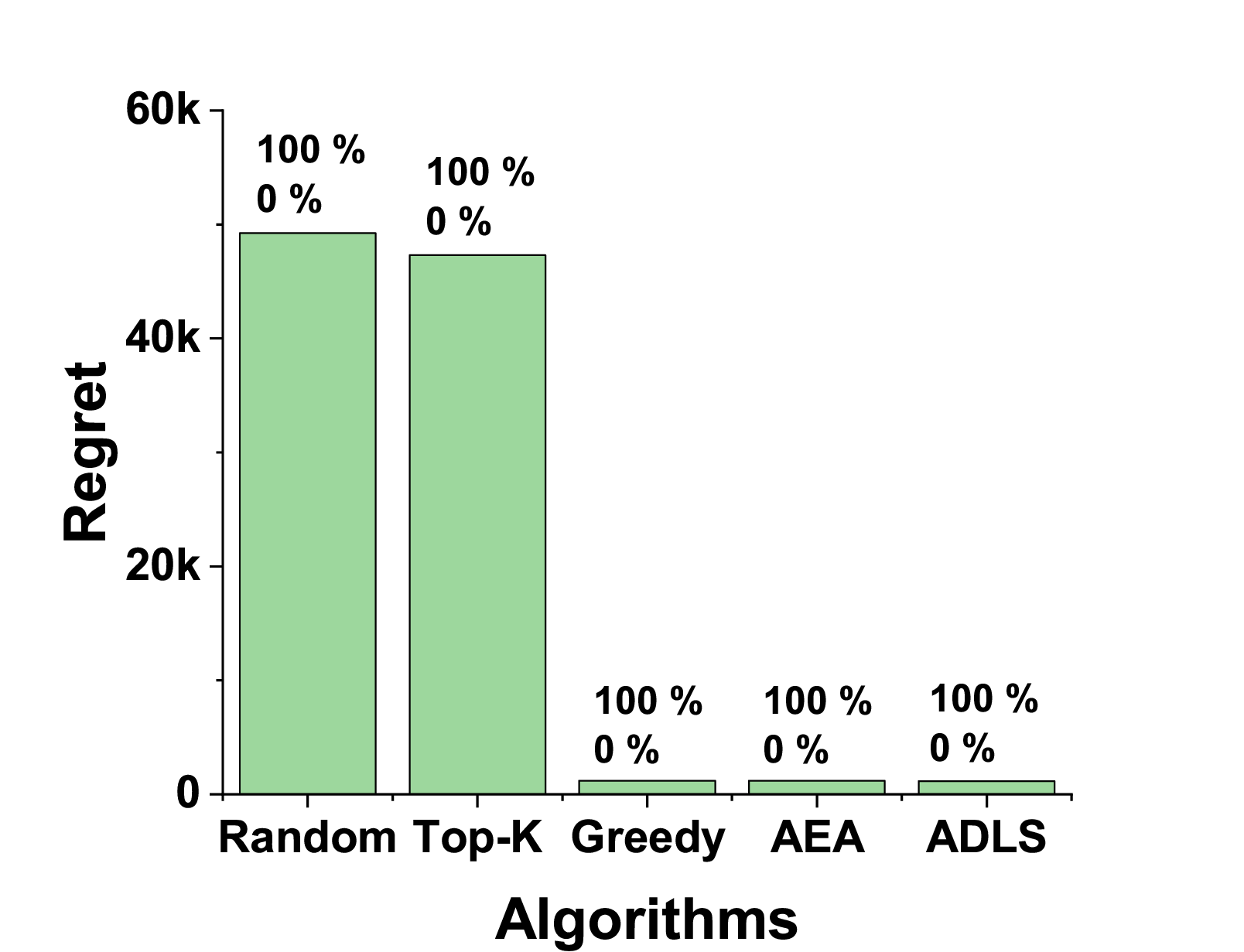} & 
    \includegraphics[scale=0.10]{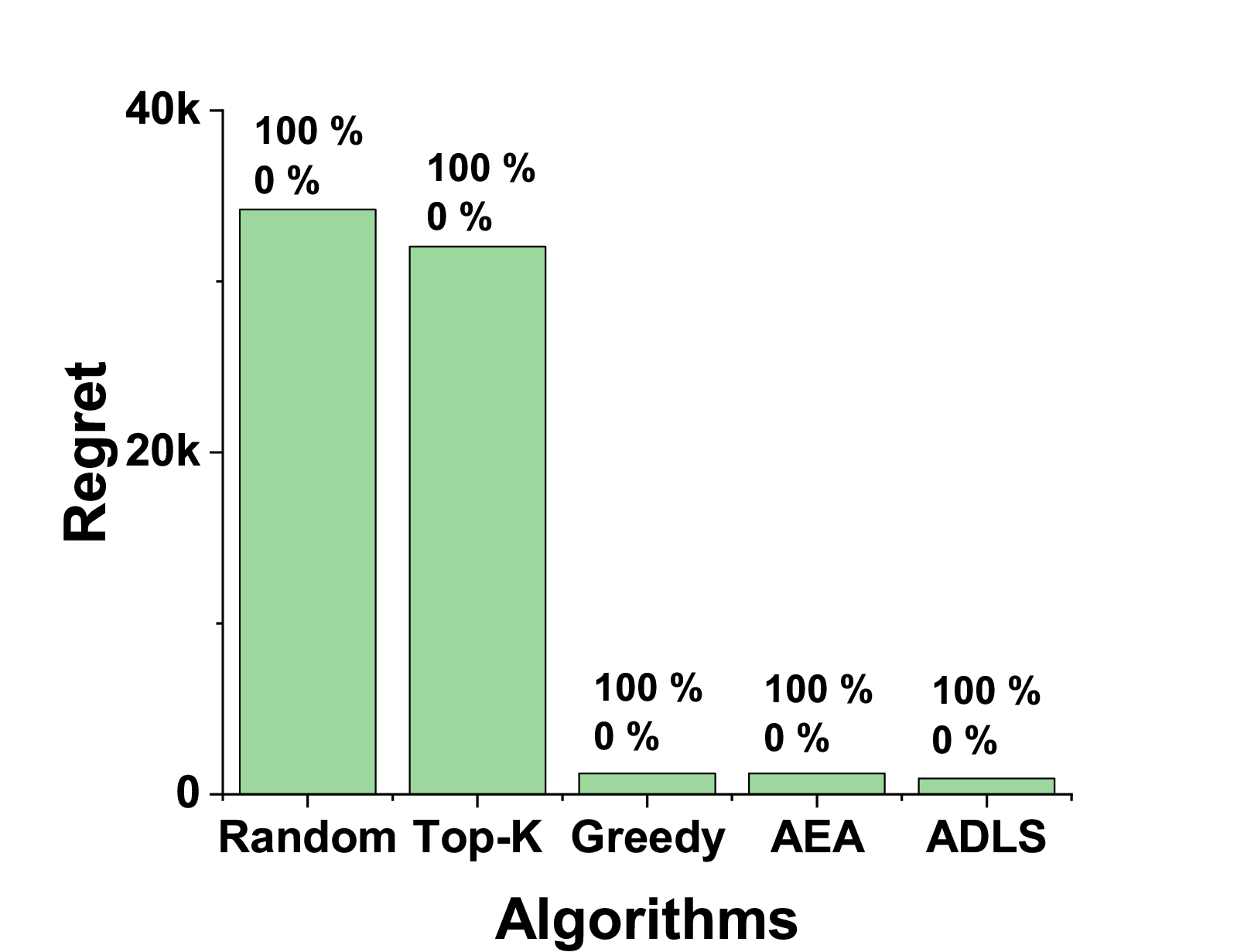} & 
    \includegraphics[scale=0.10]{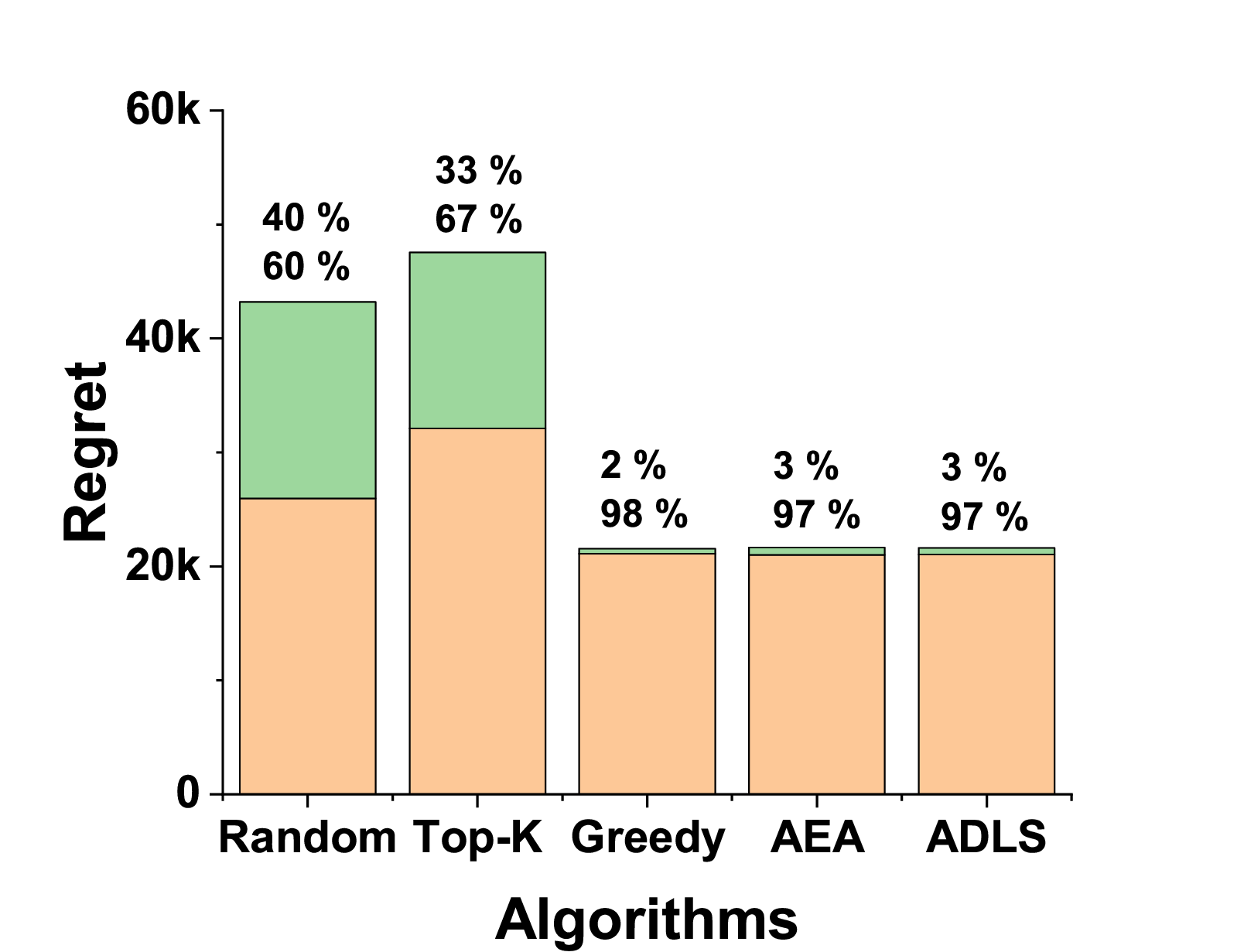} & 
    \includegraphics[scale=0.10]{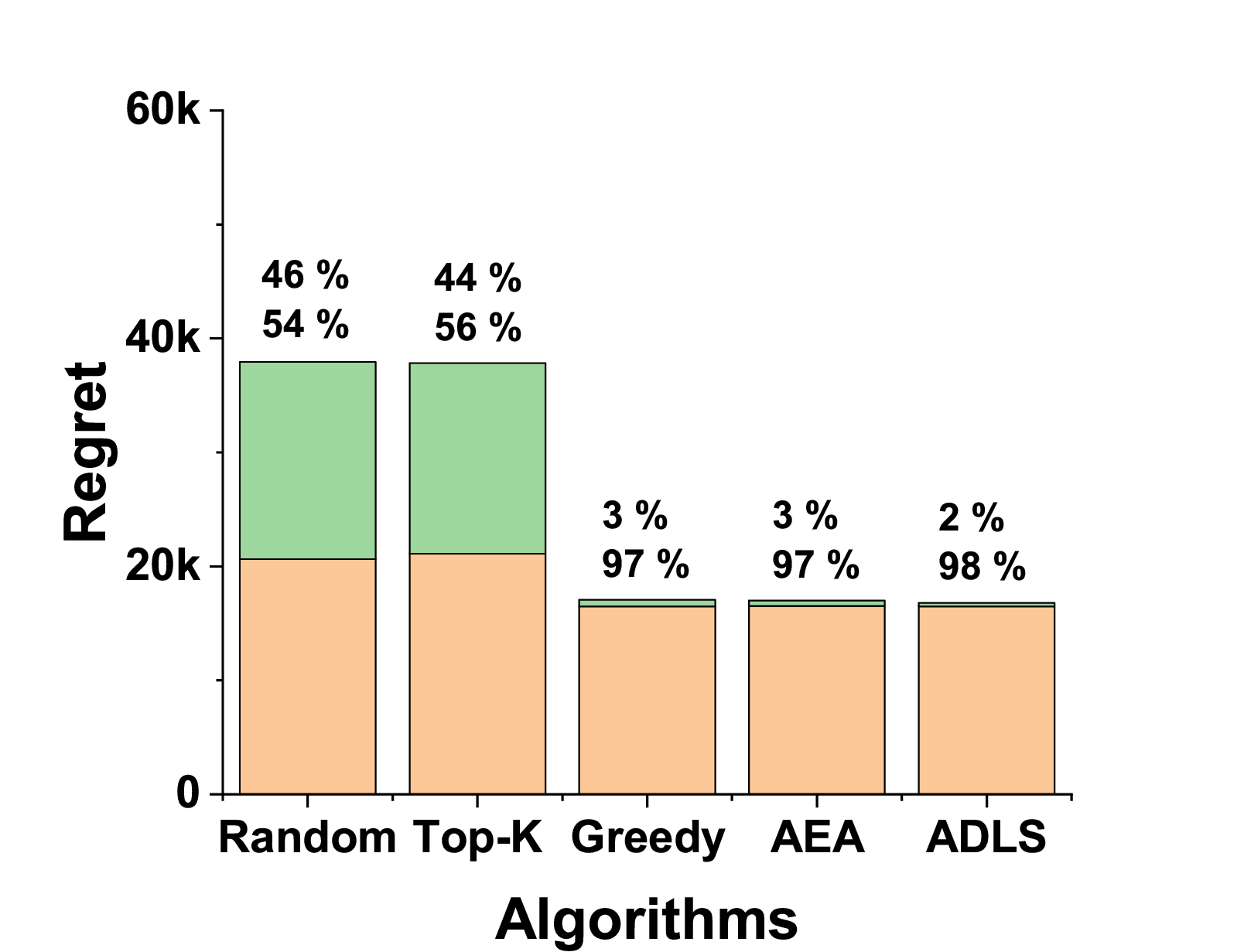} &
    \includegraphics[scale=0.10]{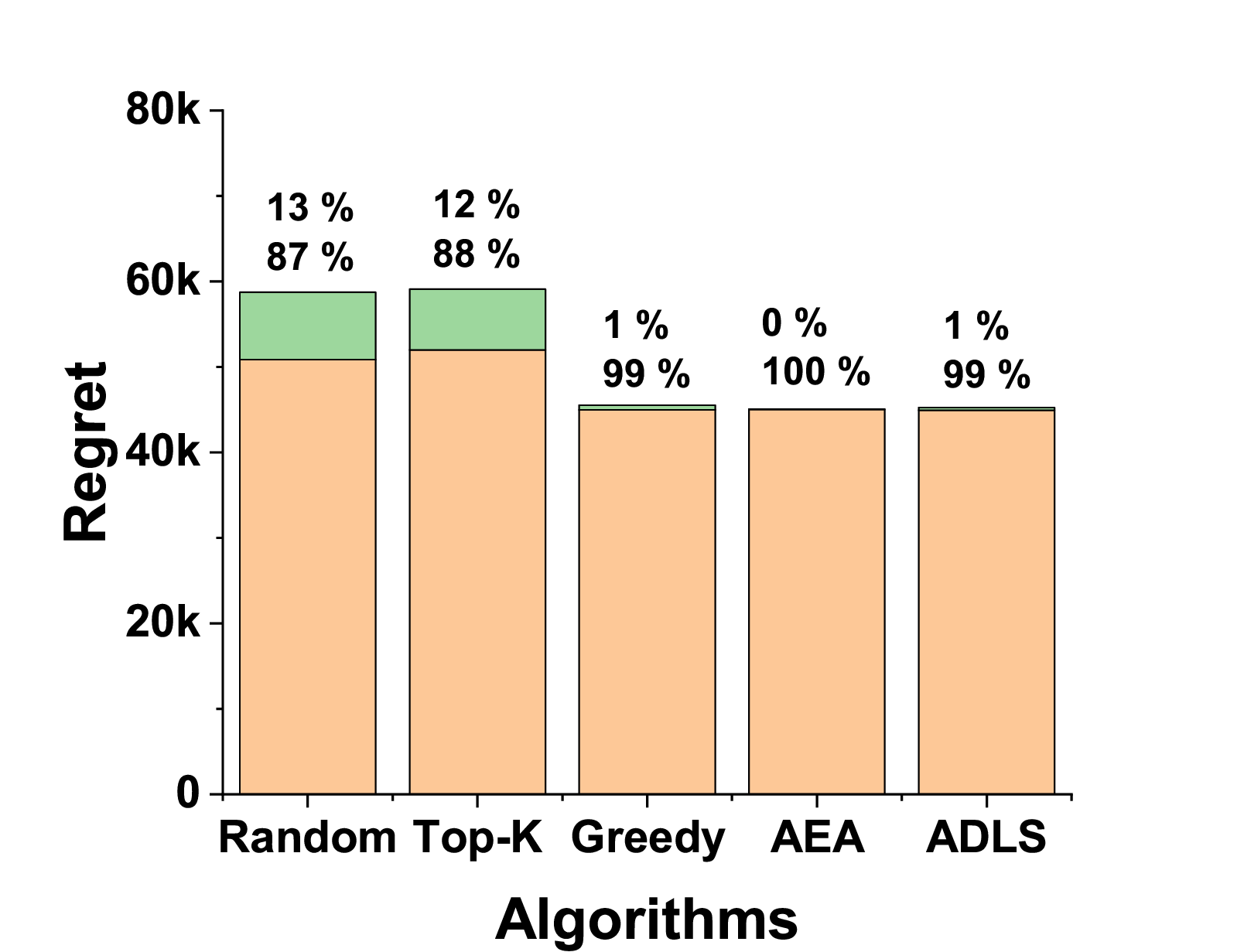} \\
    (a) $\lambda$ = $40\%$ & 
    (b) $\lambda$ = $60\%$ & 
    (c) $\lambda$ = $80\%$ & 
    (d) $\lambda$ = $100\%$ & 
    (e) $\lambda$ = $120\%$ \\
\end{tabular}
\caption{Regret of varying the demand-supply ratio $\lambda$ when $\omega = 70\%$ (Facebook, $|\mathcal{A}| = 10$, $p_{c}: Uniform(0.1) $)}
\label{Fig:3_Influence_4}
\begin{tabular}{ccccc}
    \includegraphics[scale=0.10]{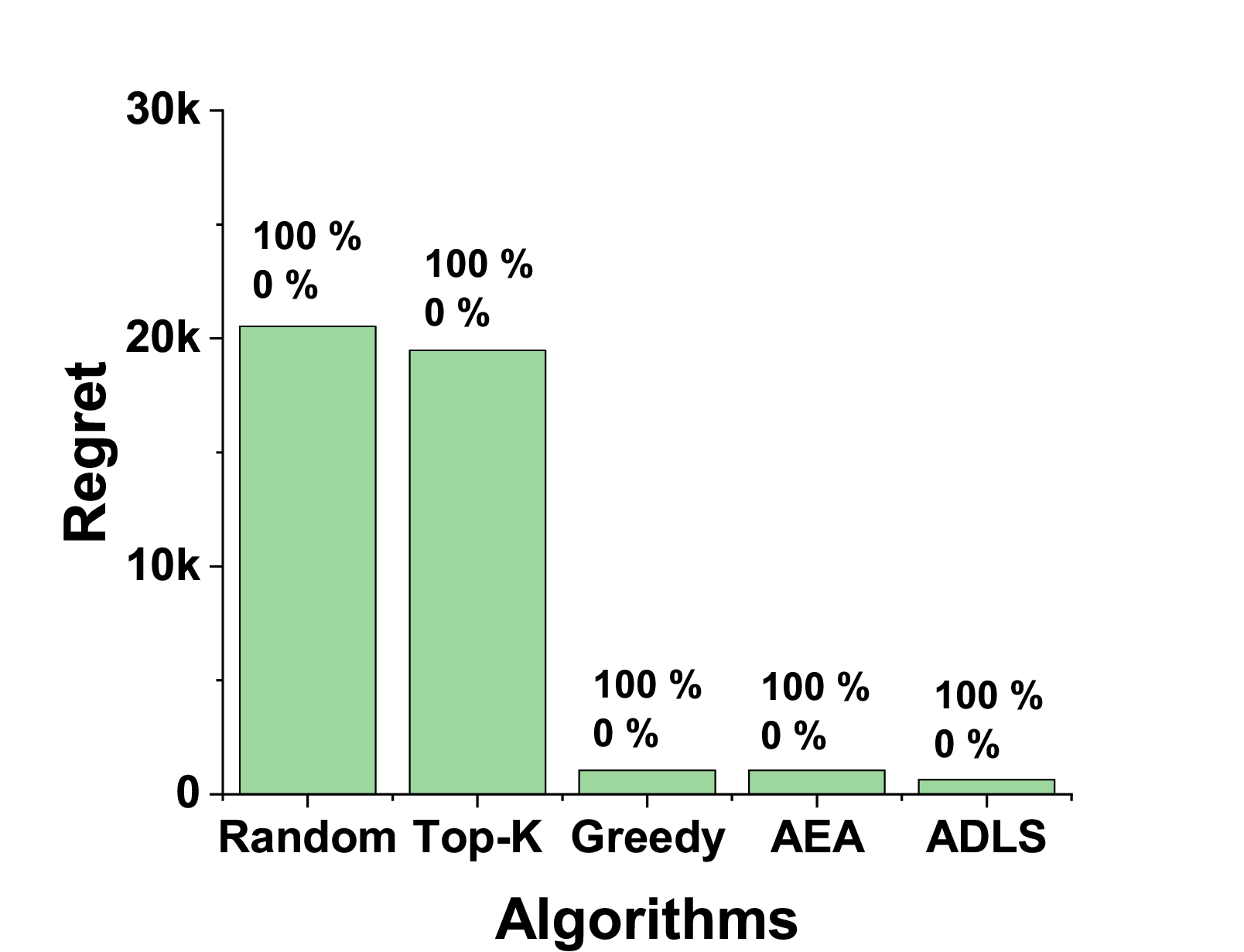} & 
    \includegraphics[scale=0.10]{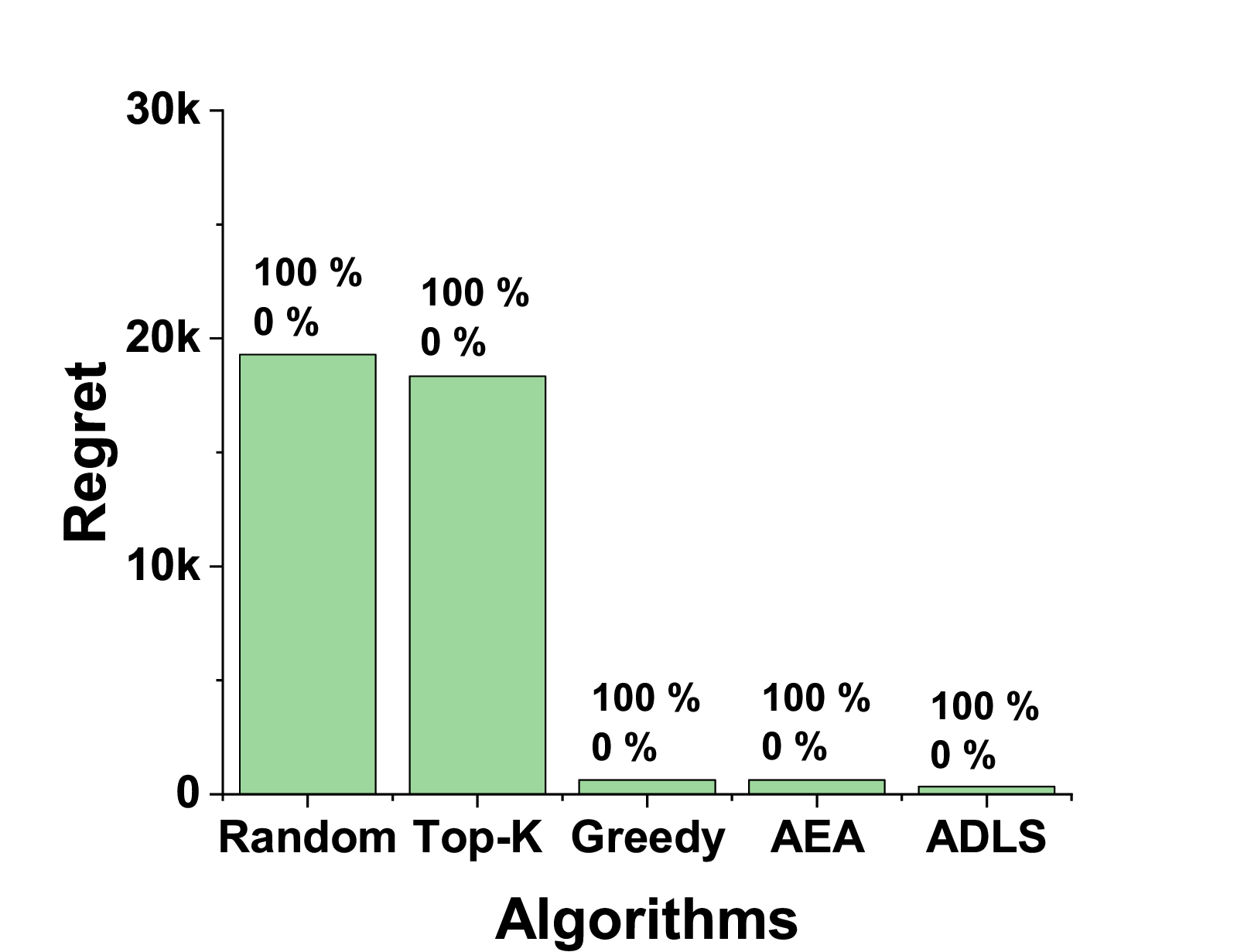} & 
    \includegraphics[scale=0.10]{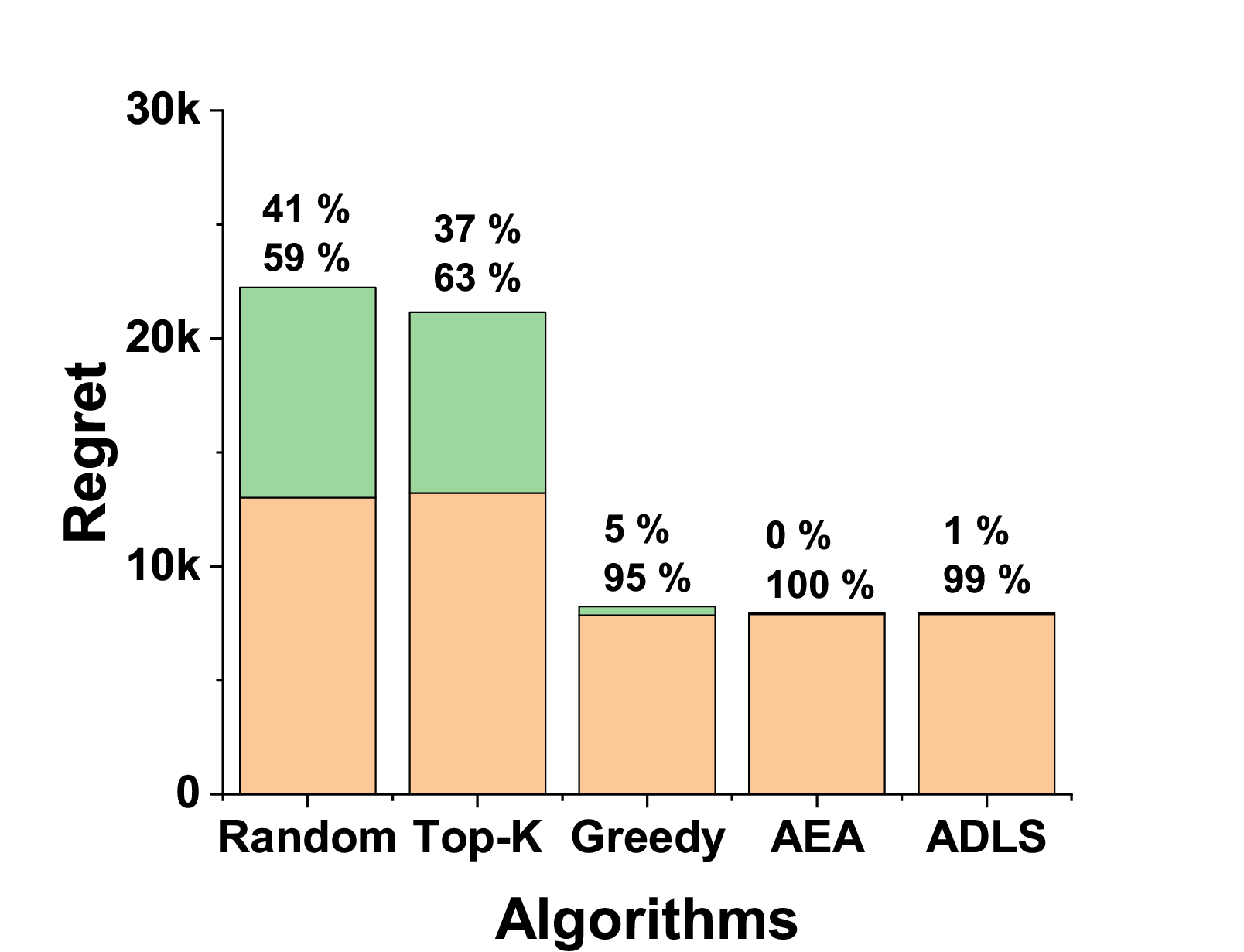} & 
    \includegraphics[scale=0.10]{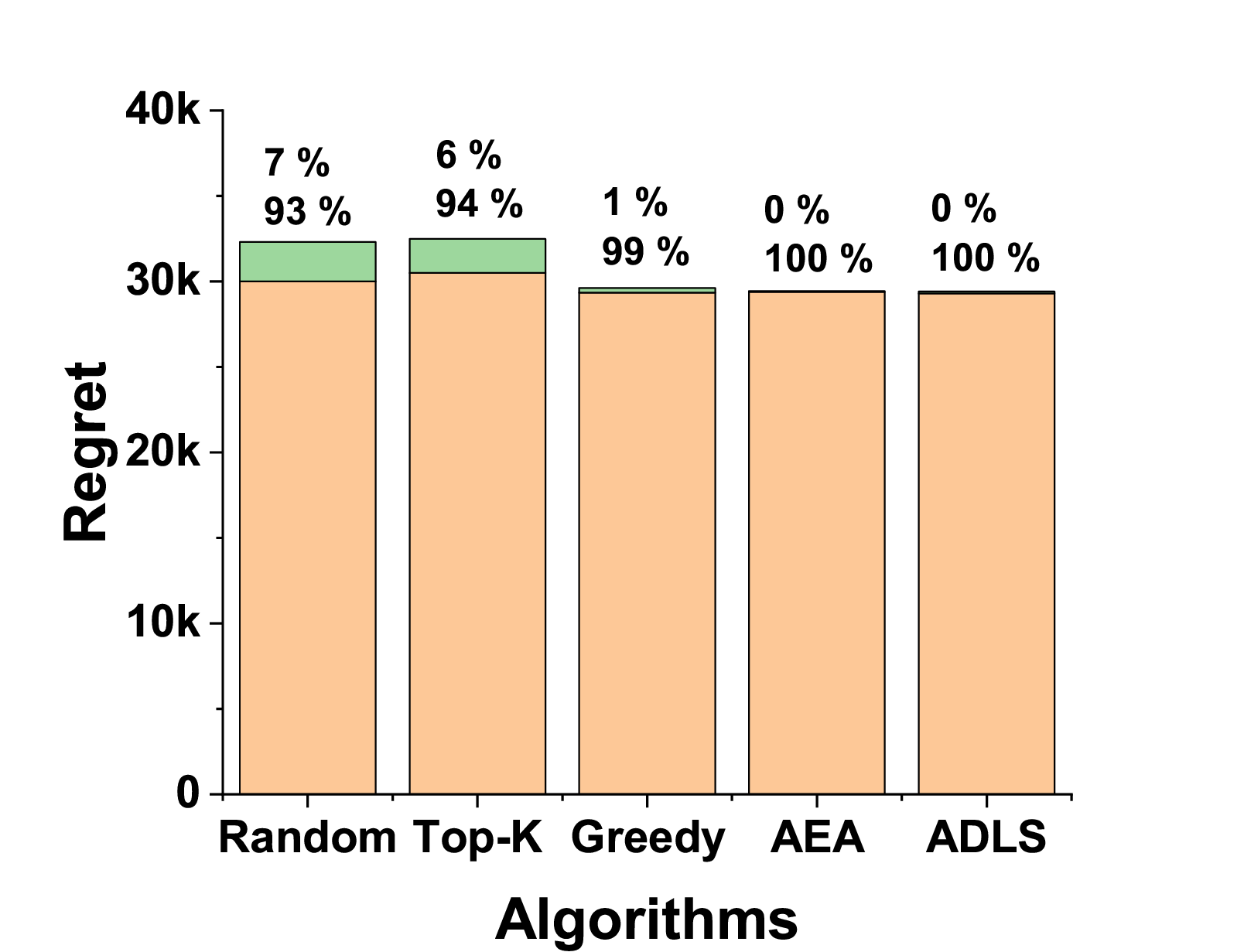} &
    \includegraphics[scale=0.10]{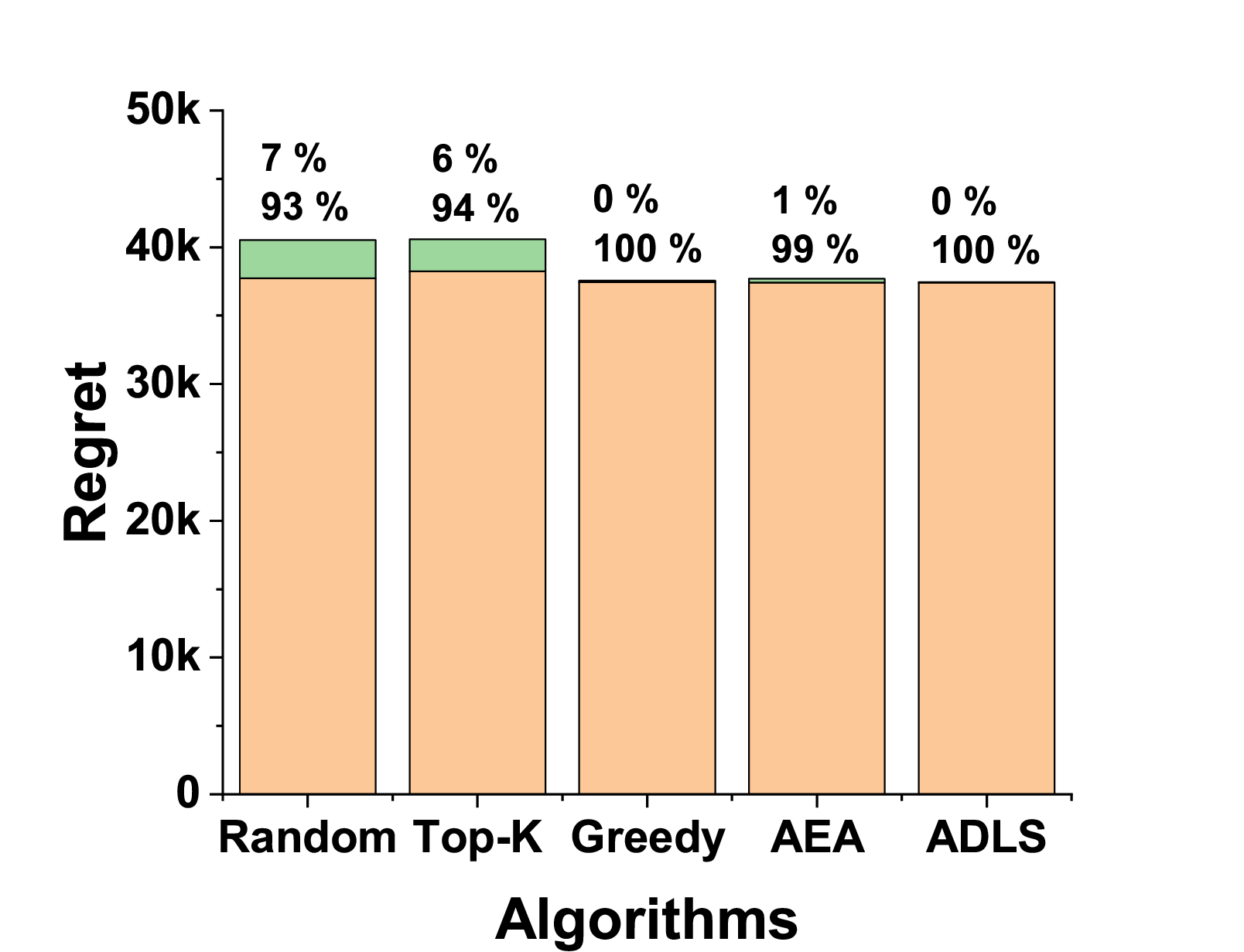} \\
    (a) $\lambda$ = $40\%$ & 
    (b) $\lambda$ = $60\%$ & 
    (c) $\lambda$ = $80\%$ & 
    (d) $\lambda$ = $100\%$ & 
    (e) $\lambda$ = $120\%$ \\
\end{tabular}
\caption{Regret of varying the demand-supply ratio $\lambda$ when $\omega = 90\%$ (Facebook, $|\mathcal{A}| = 5$, $p_{c}: Uniform(0.1) $)}
\label{Fig:3_Influence_5}
\end{figure}

\begin{figure}[H]
\centering
\begin{tabular}{lclc}
    Excessive Regret & \includegraphics[width=0.11\linewidth]{Excessive.png} & 
    Unsatisfied Regret & \includegraphics[width=0.11\linewidth]{Unsatisfied.png} \\
\end{tabular}
\begin{tabular}{ccccc}
    \includegraphics[scale=0.10]{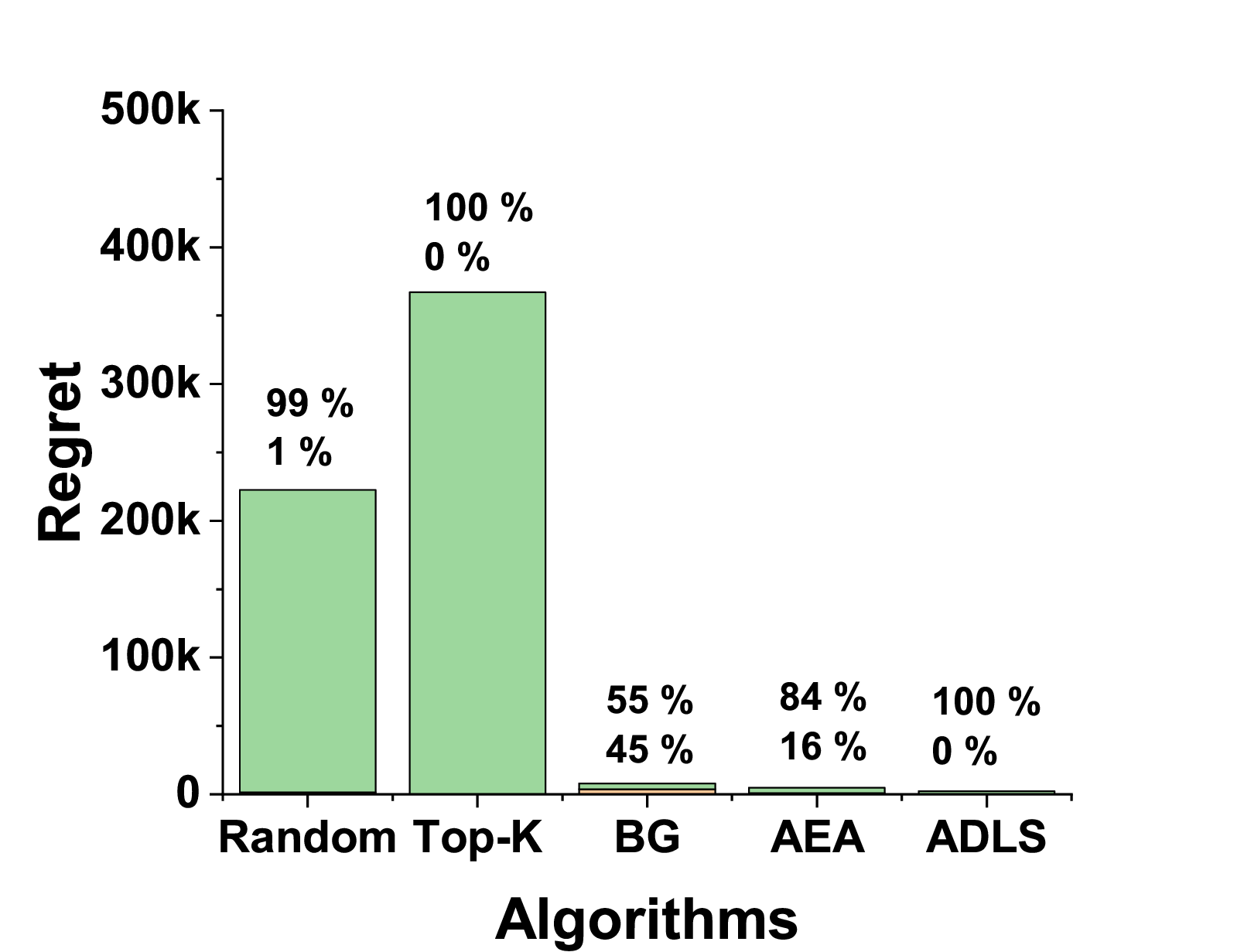} & 
    \includegraphics[scale=0.10]{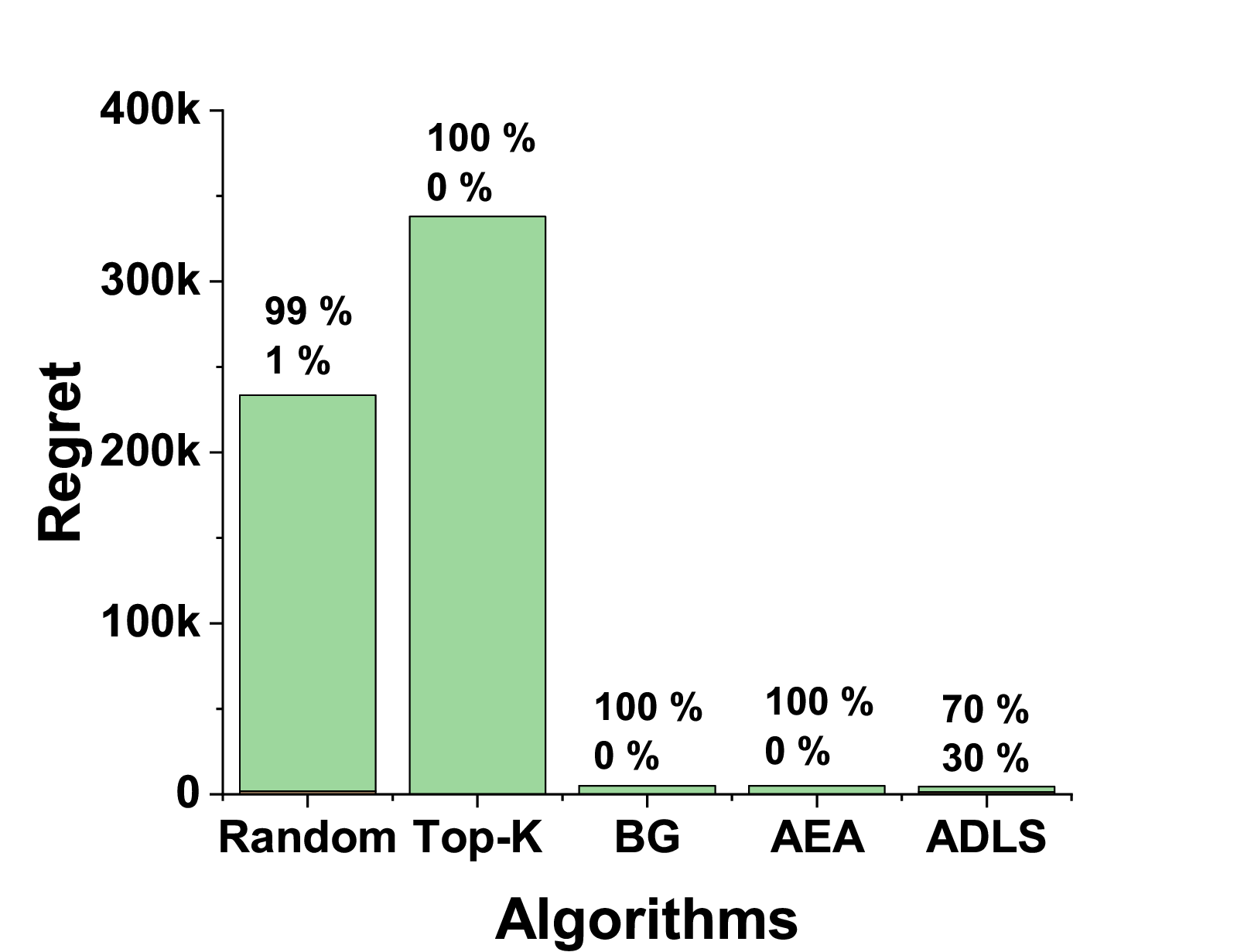} & 
    \includegraphics[scale=0.10]{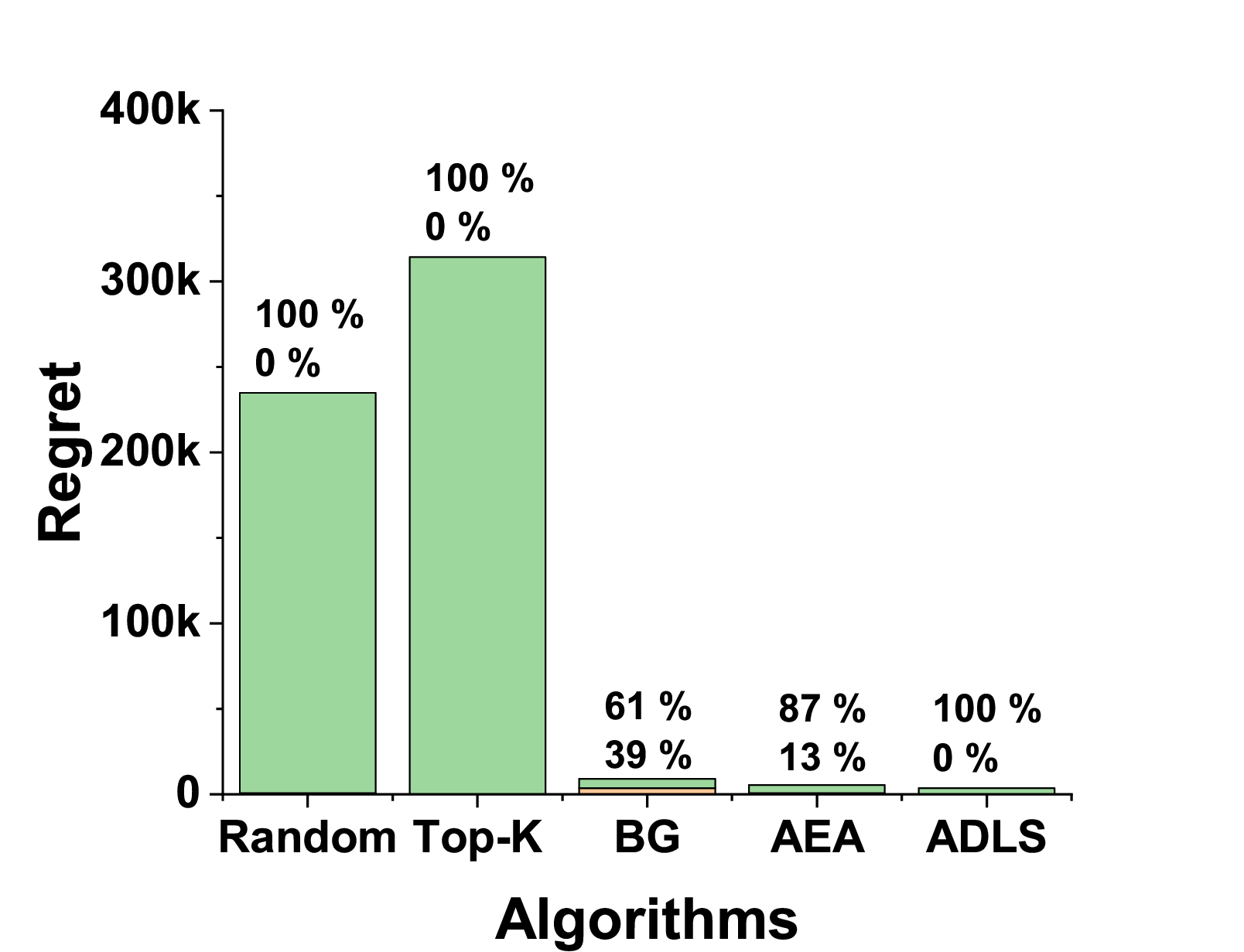} & 
    \includegraphics[scale=0.10]{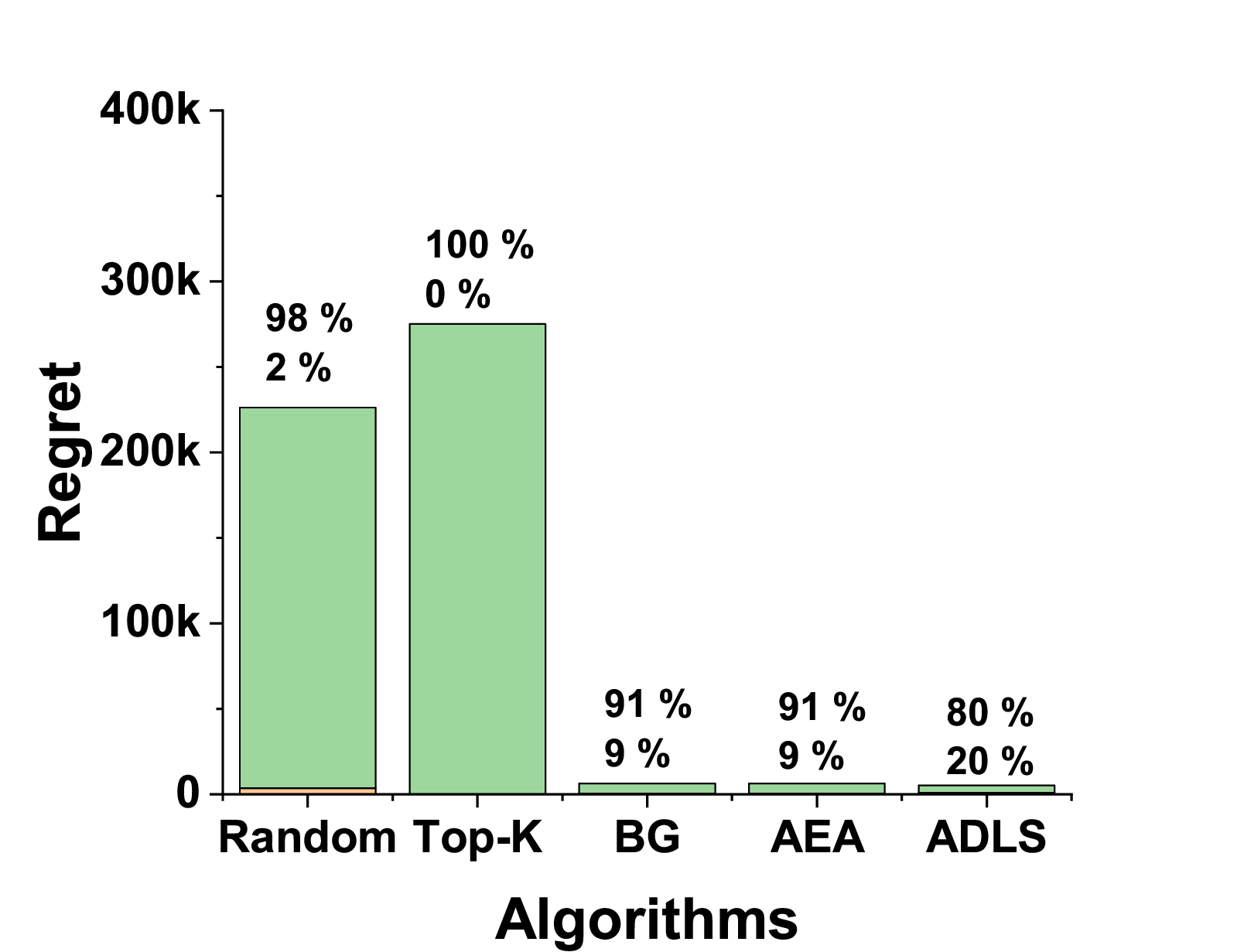} &
    \includegraphics[scale=0.10]{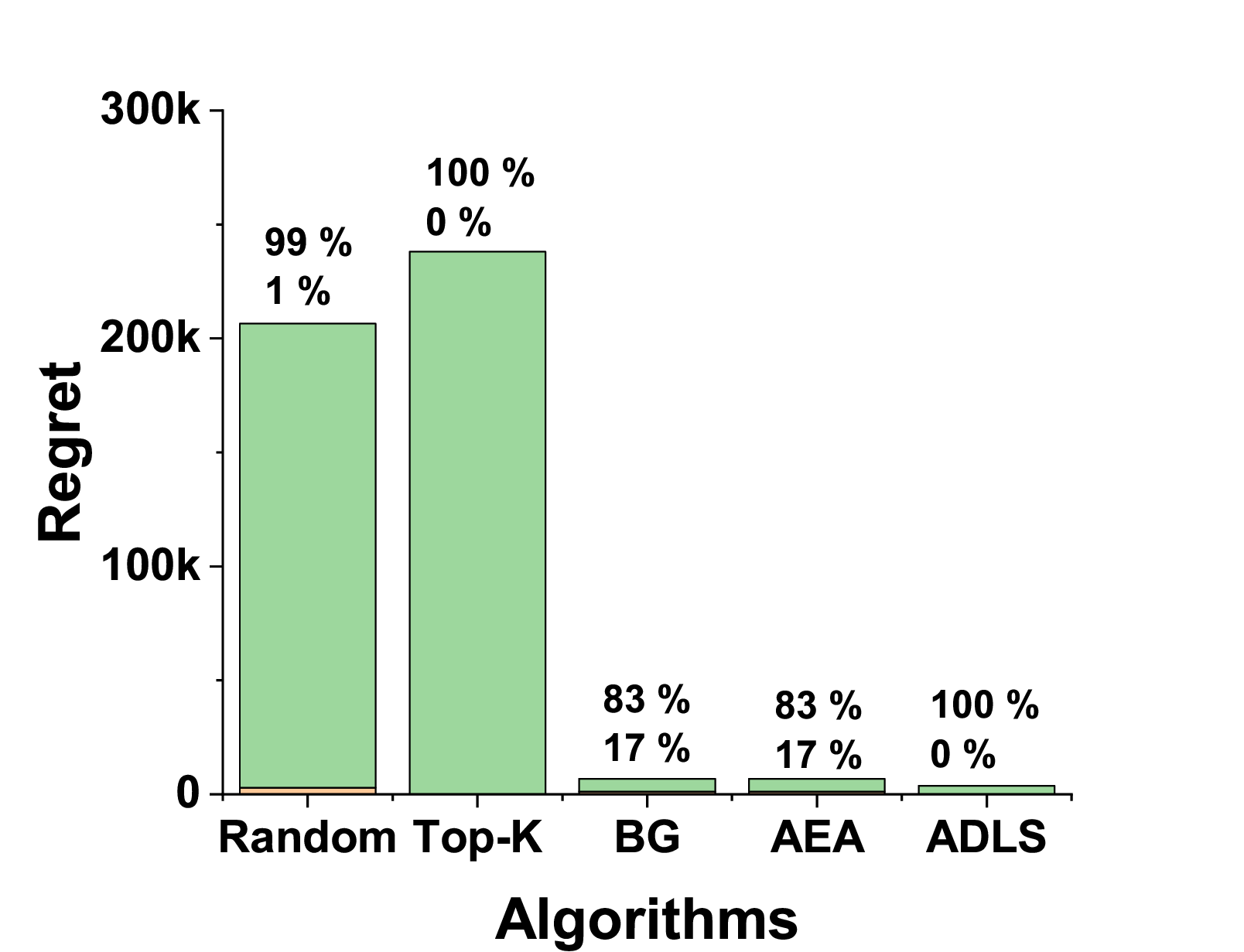} \\
    (a) $\lambda$ = $40\%$ & 
    (b) $\lambda$ = $60\%$ & 
    (c) $\lambda$ = $80\%$ & 
    (d) $\lambda$ = $100\%$ & 
    (e) $\lambda$ = $120\%$ \\
\end{tabular}
\caption{Regret of varying the demand-supply ratio $\lambda$ when $\omega = 10\%$ (Wikivote, $|\mathcal{A}| = 100$, $p_{c}: Uniform(0.1) $)}
\label{Fig:5_Influence_1}
\begin{tabular}{ccccc}
    \includegraphics[scale=0.10]{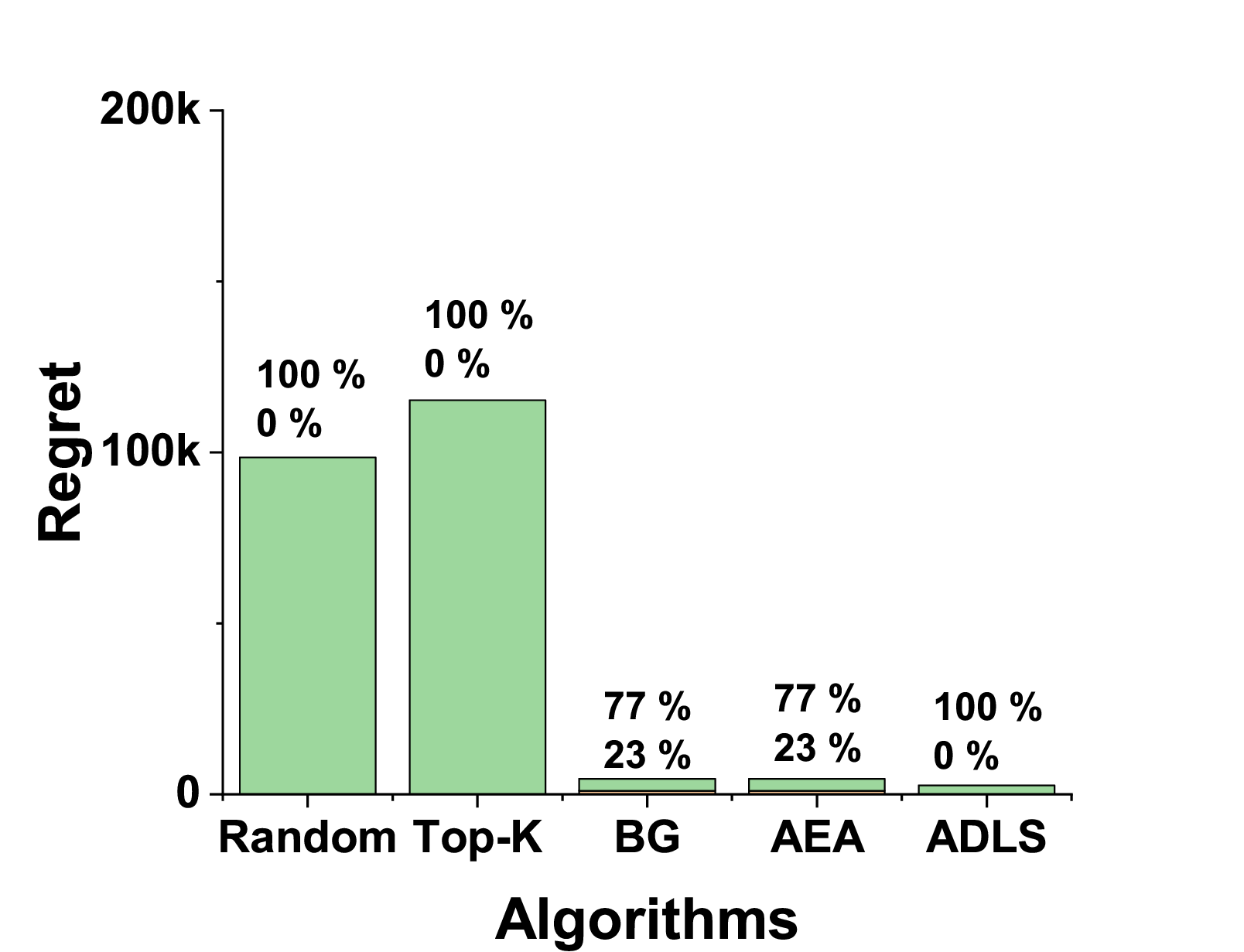} & 
    \includegraphics[scale=0.10]{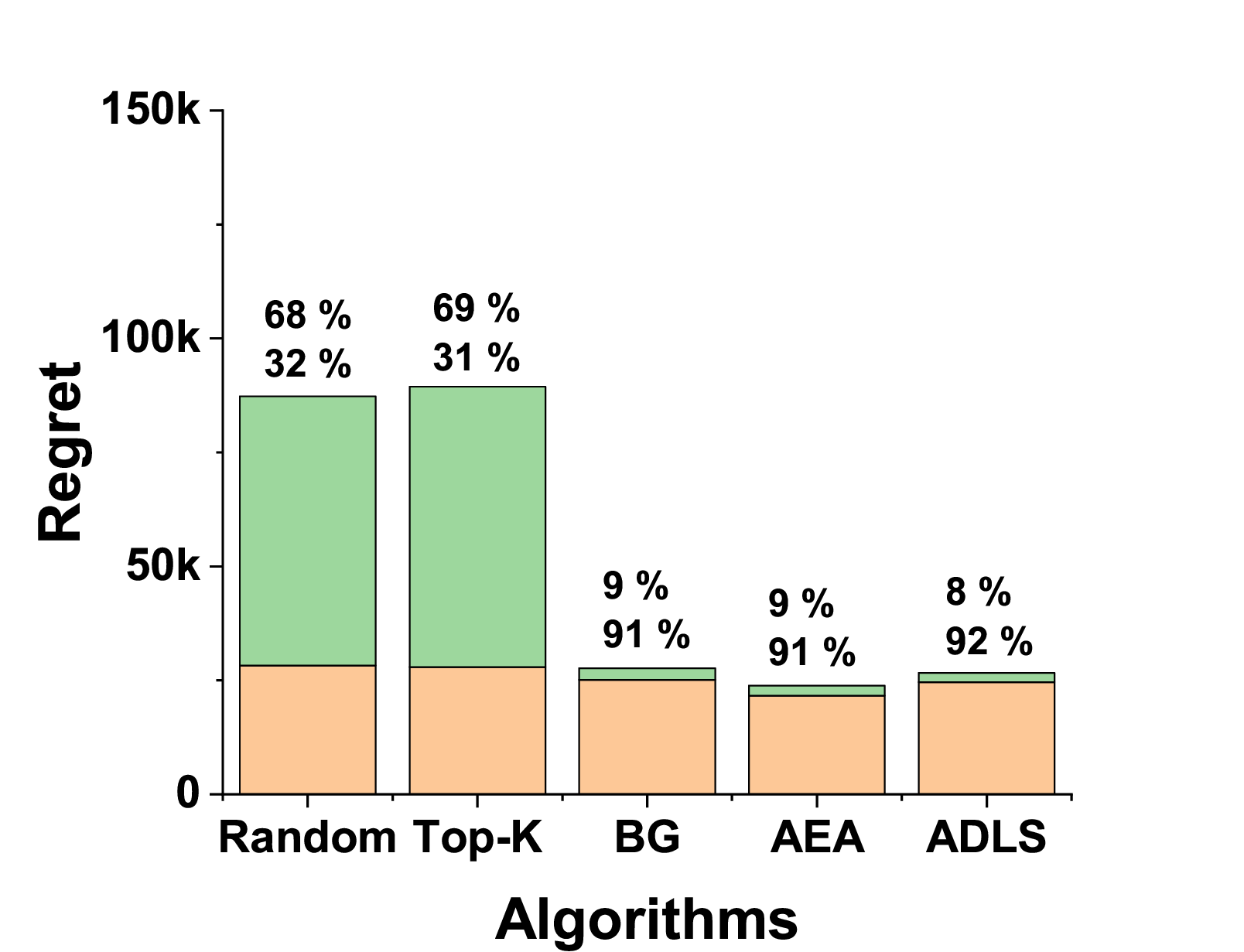} & 
    \includegraphics[scale=0.10]{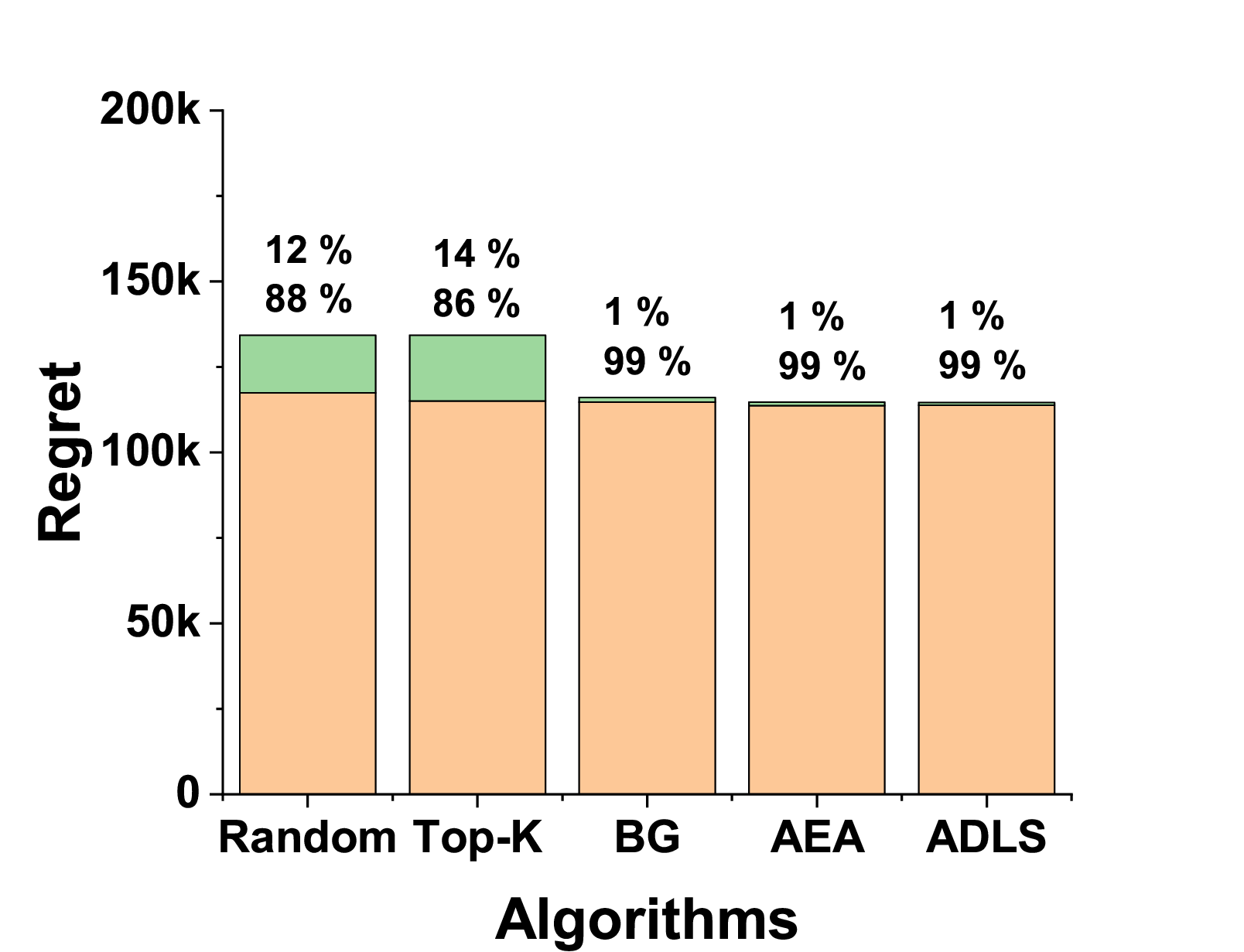} & 
    \includegraphics[scale=0.10]{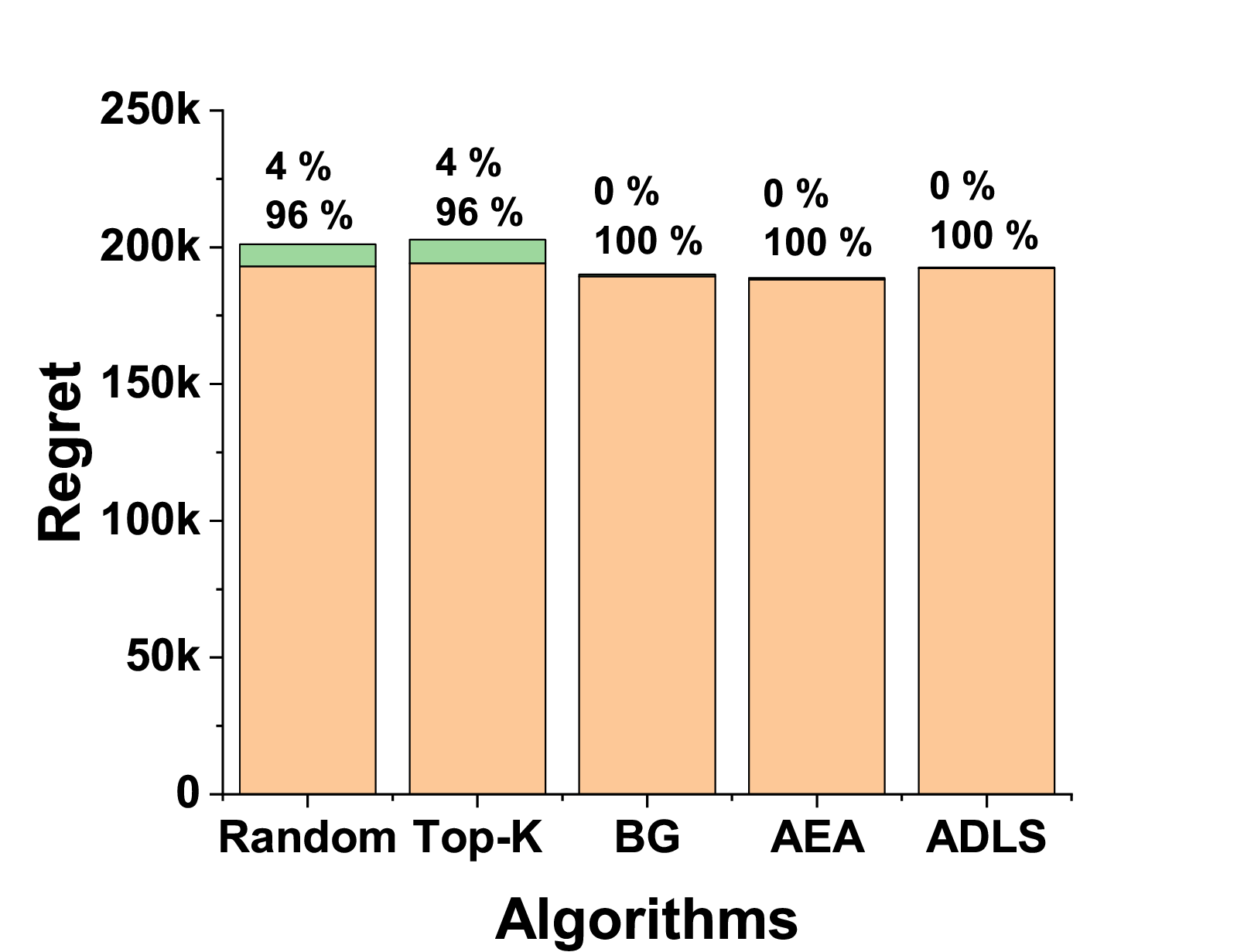} &
    \includegraphics[scale=0.10]{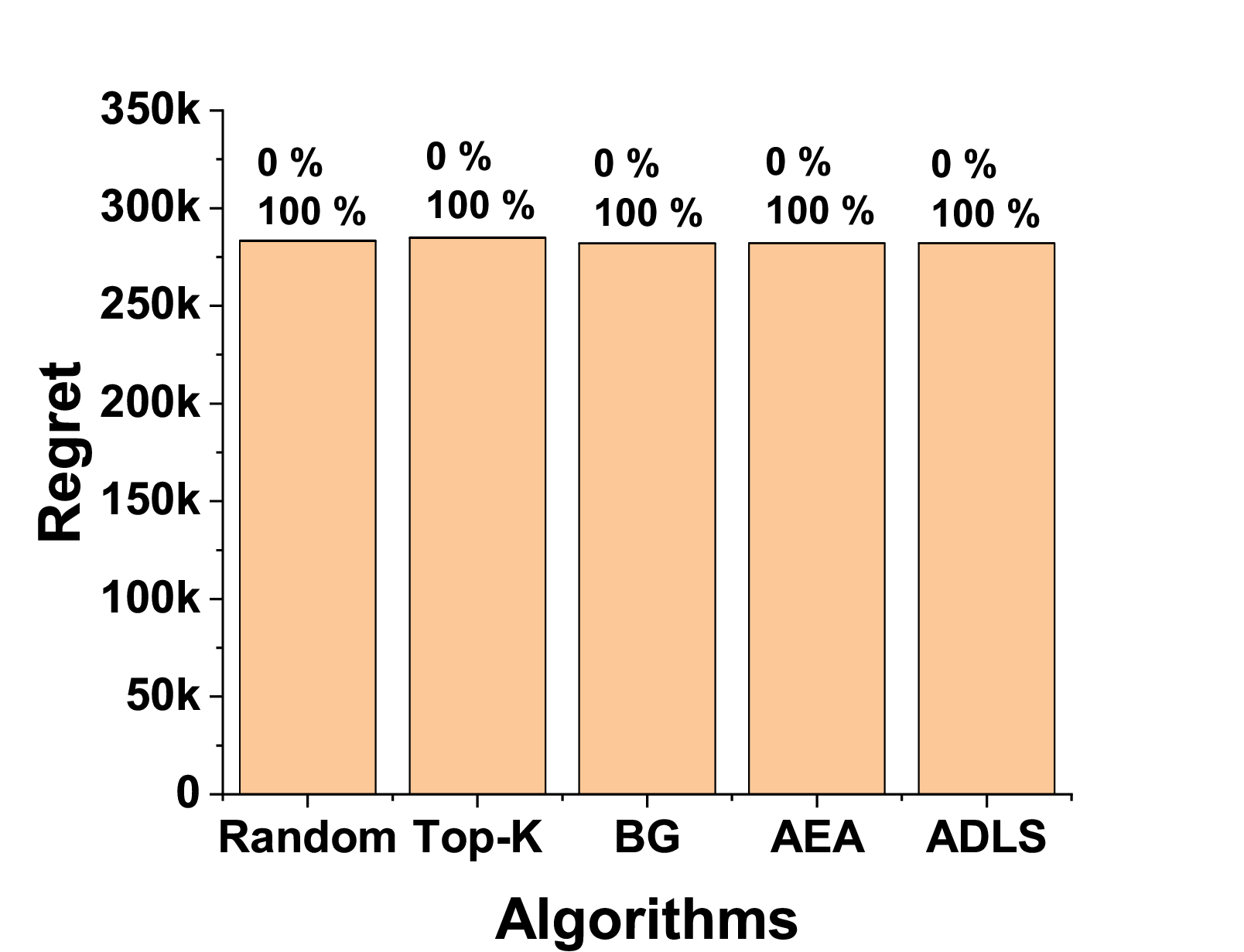} \\
    (a) $\lambda$ = $40\%$ & 
    (b) $\lambda$ = $60\%$ & 
    (c) $\lambda$ = $80\%$ & 
    (d) $\lambda$ = $100\%$ & 
    (e) $\lambda$ = $120\%$ \\
\end{tabular}
\caption{Regret of varying the demand-supply ratio $\lambda$ when $\omega = 30\%$ (Wikivote, $|\mathcal{A}| = 50$, $p_{c}: Uniform(0.1) $)}
\label{Fig:5_Influence_2}
\begin{tabular}{ccccc}
    \includegraphics[scale=0.10]{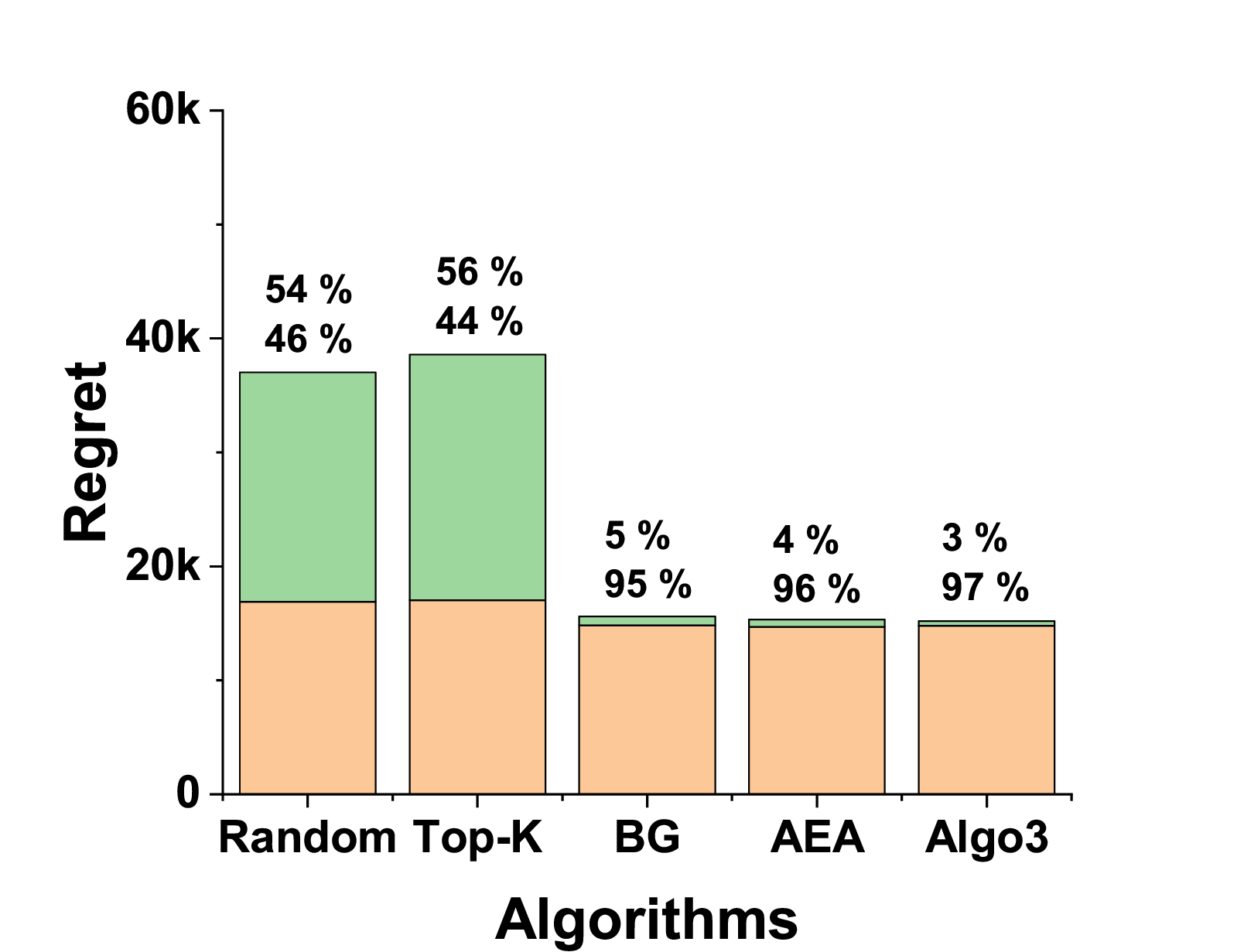} & 
    \includegraphics[scale=0.10]{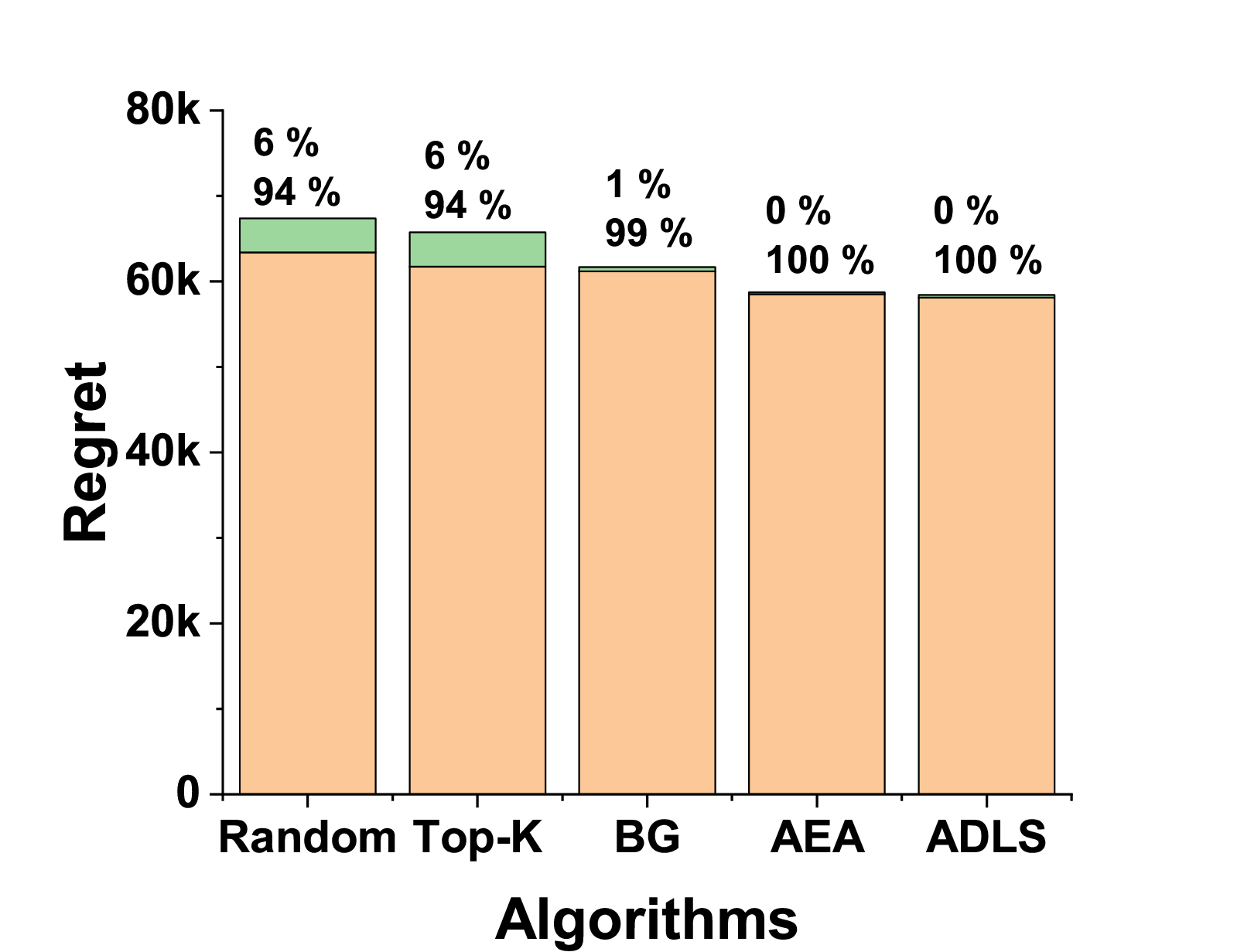} & 
    \includegraphics[scale=0.10]{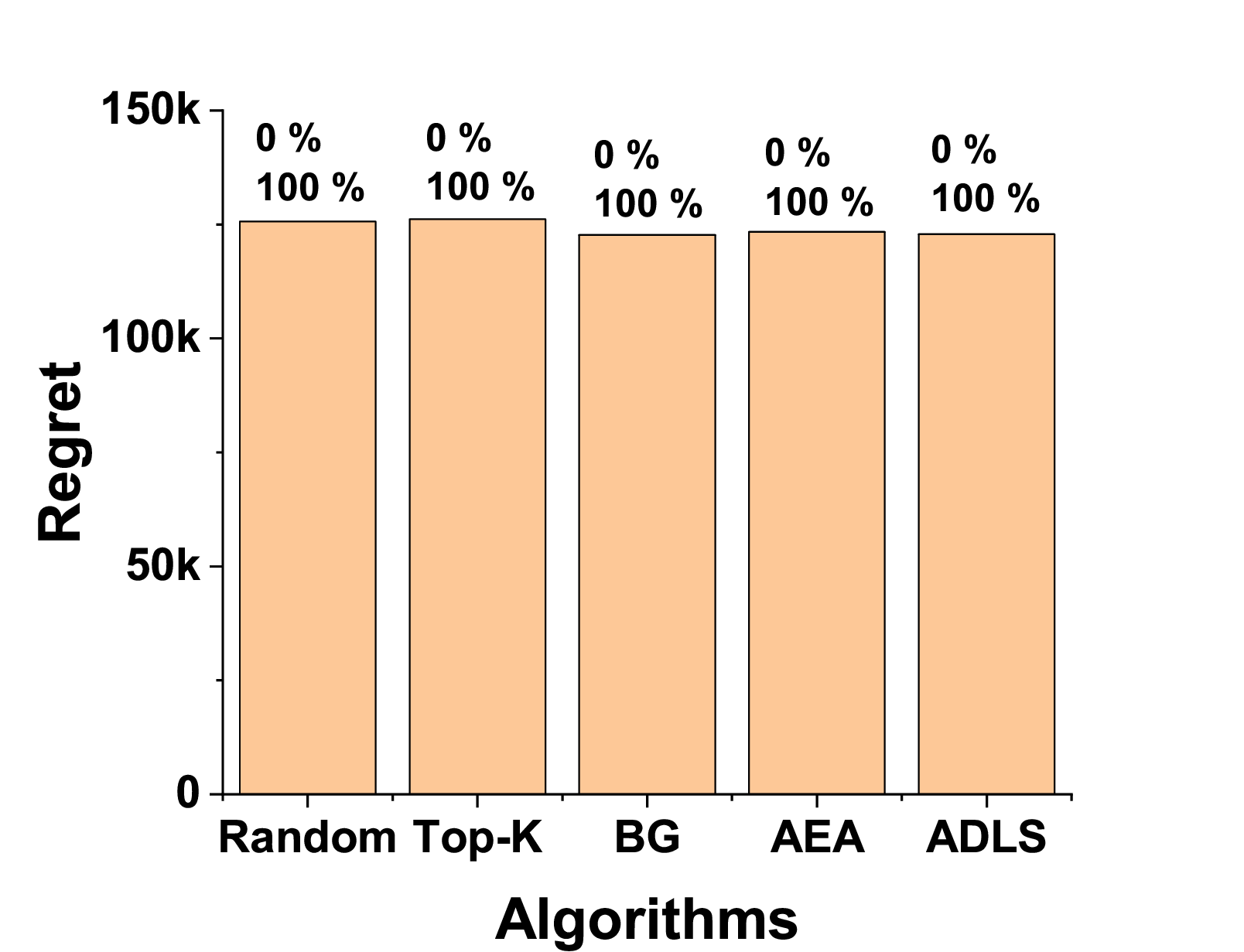} & 
    \includegraphics[scale=0.10]{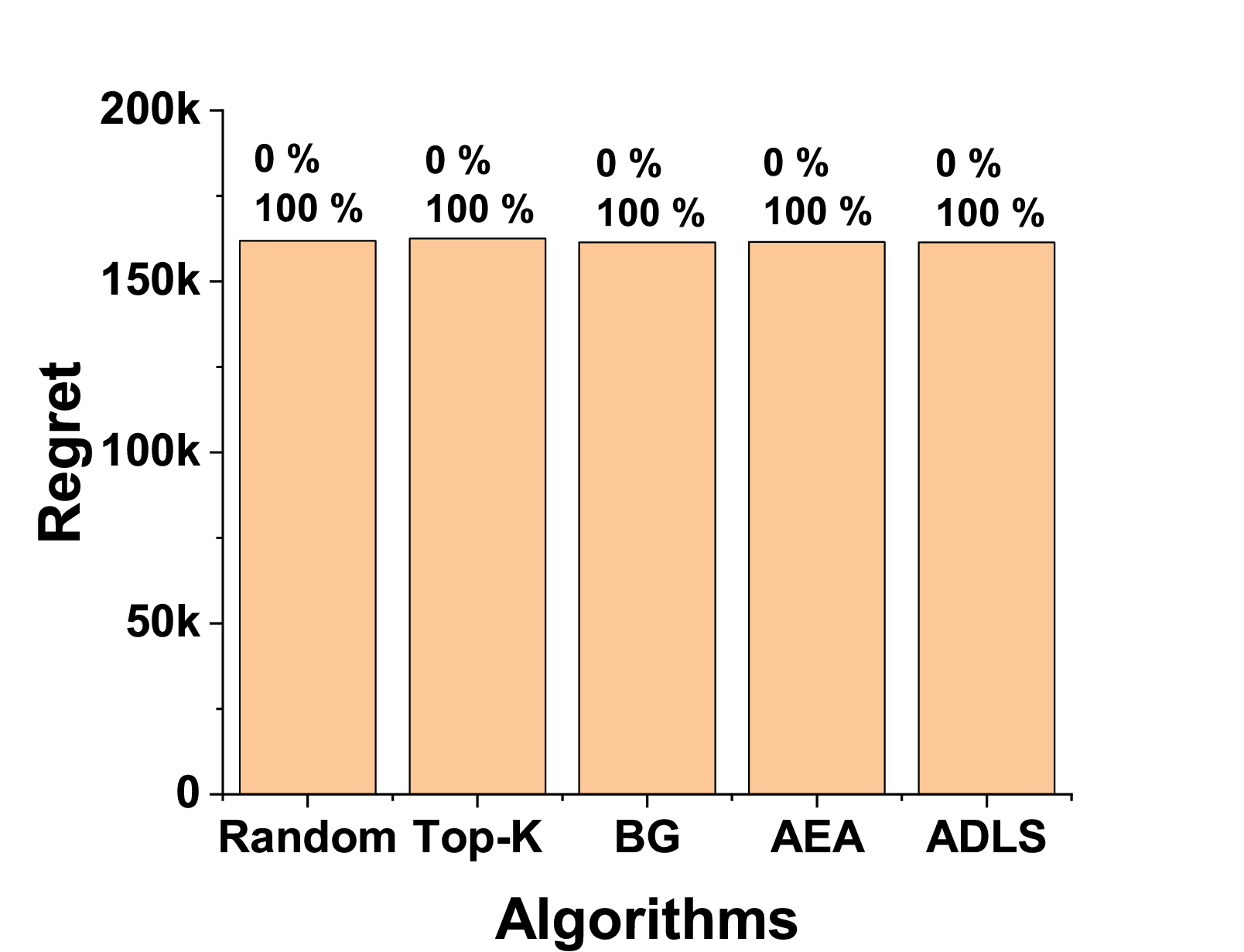} &
    \includegraphics[scale=0.10]{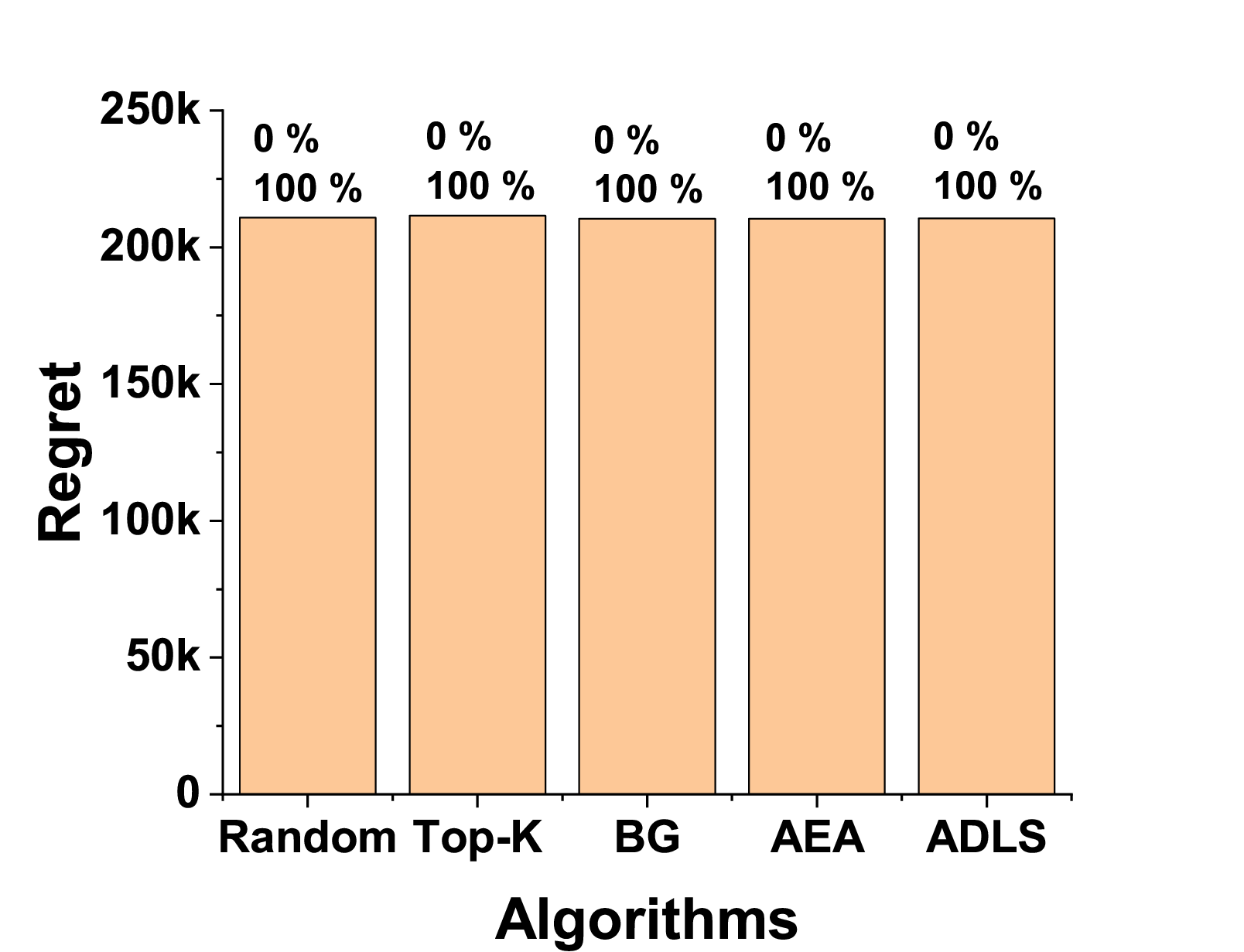} \\
    (a) $\lambda$ = $40\%$ & 
    (b) $\lambda$ = $60\%$ & 
    (c) $\lambda$ = $80\%$ & 
    (d) $\lambda$ = $100\%$ & 
    (e) $\lambda$ = $120\%$ \\
\end{tabular}
\caption{Regret of varying the demand-supply ratio $\lambda$ when $\omega = 50\%$ (Wikivote, $|\mathcal{A}| = 20$, $p_{c}: Uniform(0.1) $)}
\label{Fig:5_Influence_3}
\begin{tabular}{ccccc}
    \includegraphics[scale=0.10]{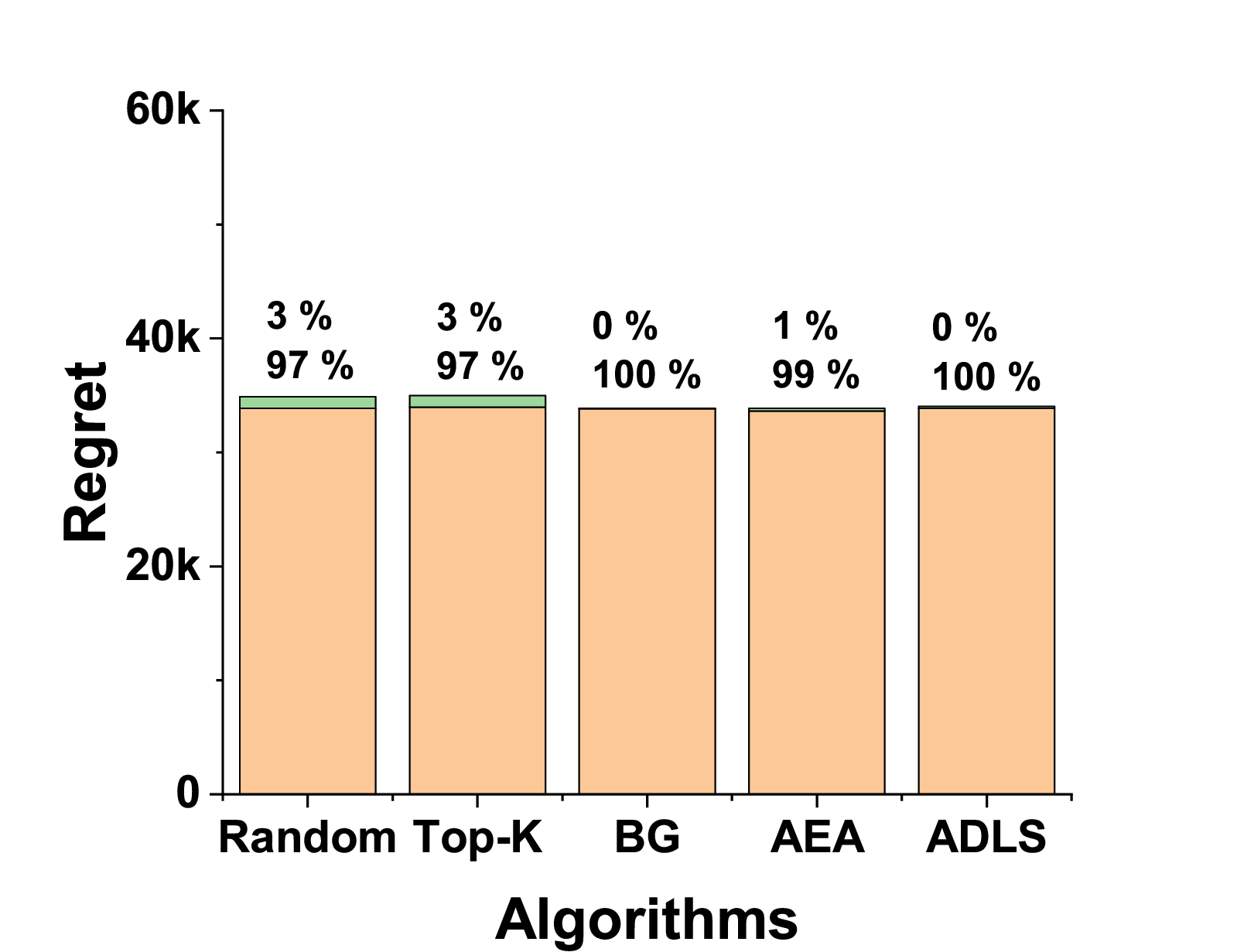} & 
    \includegraphics[scale=0.10]{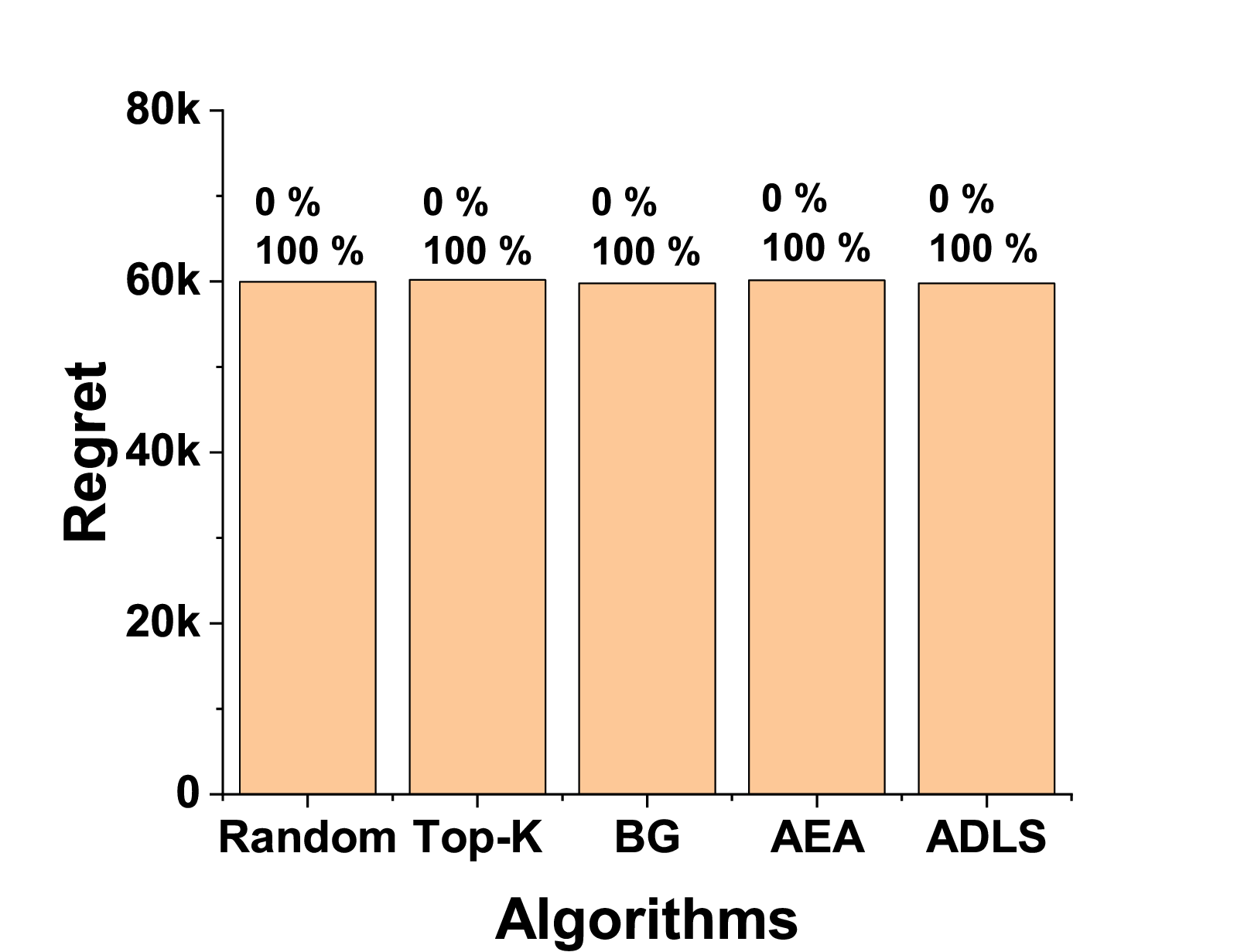} & 
    \includegraphics[scale=0.10]{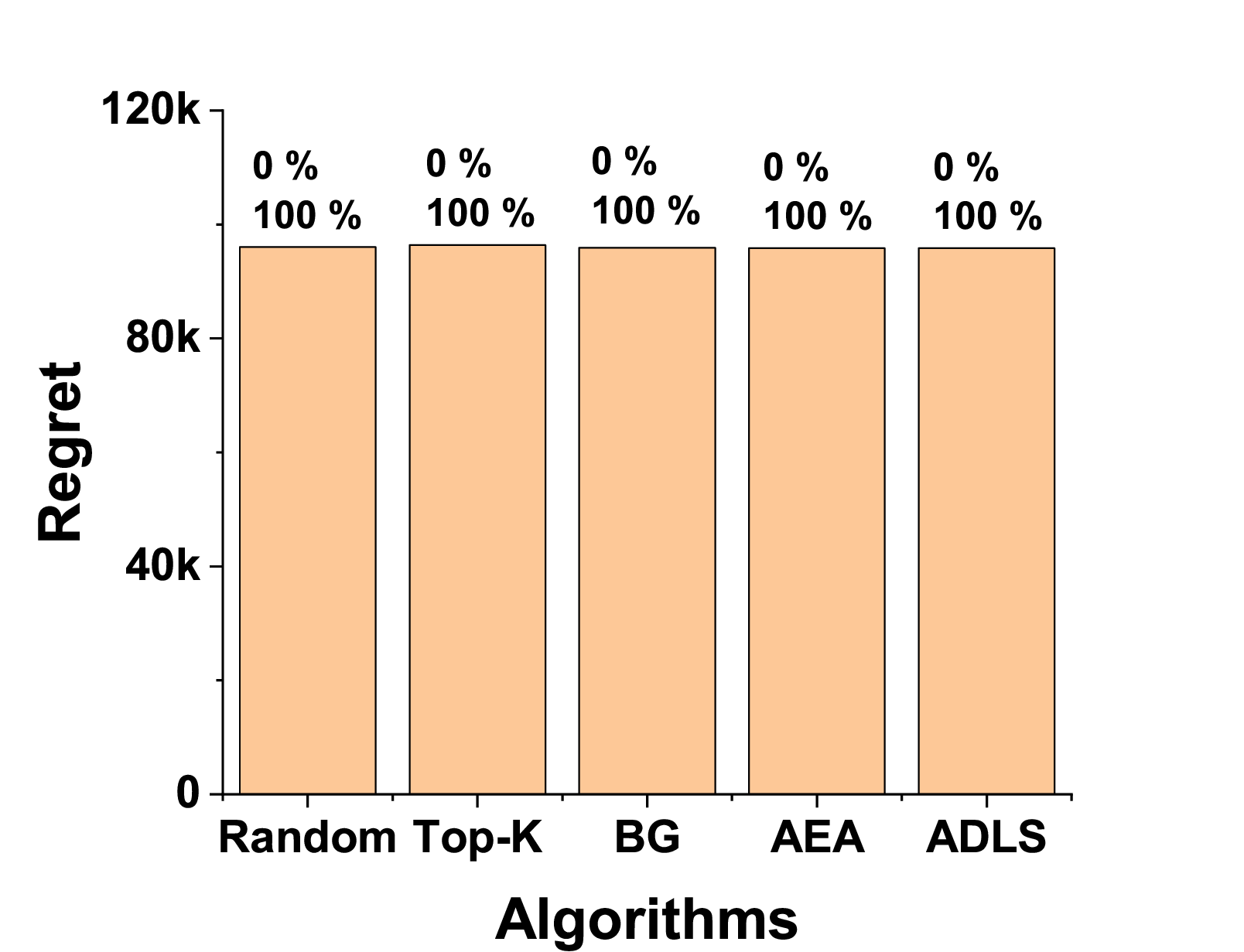} & 
    \includegraphics[scale=0.10]{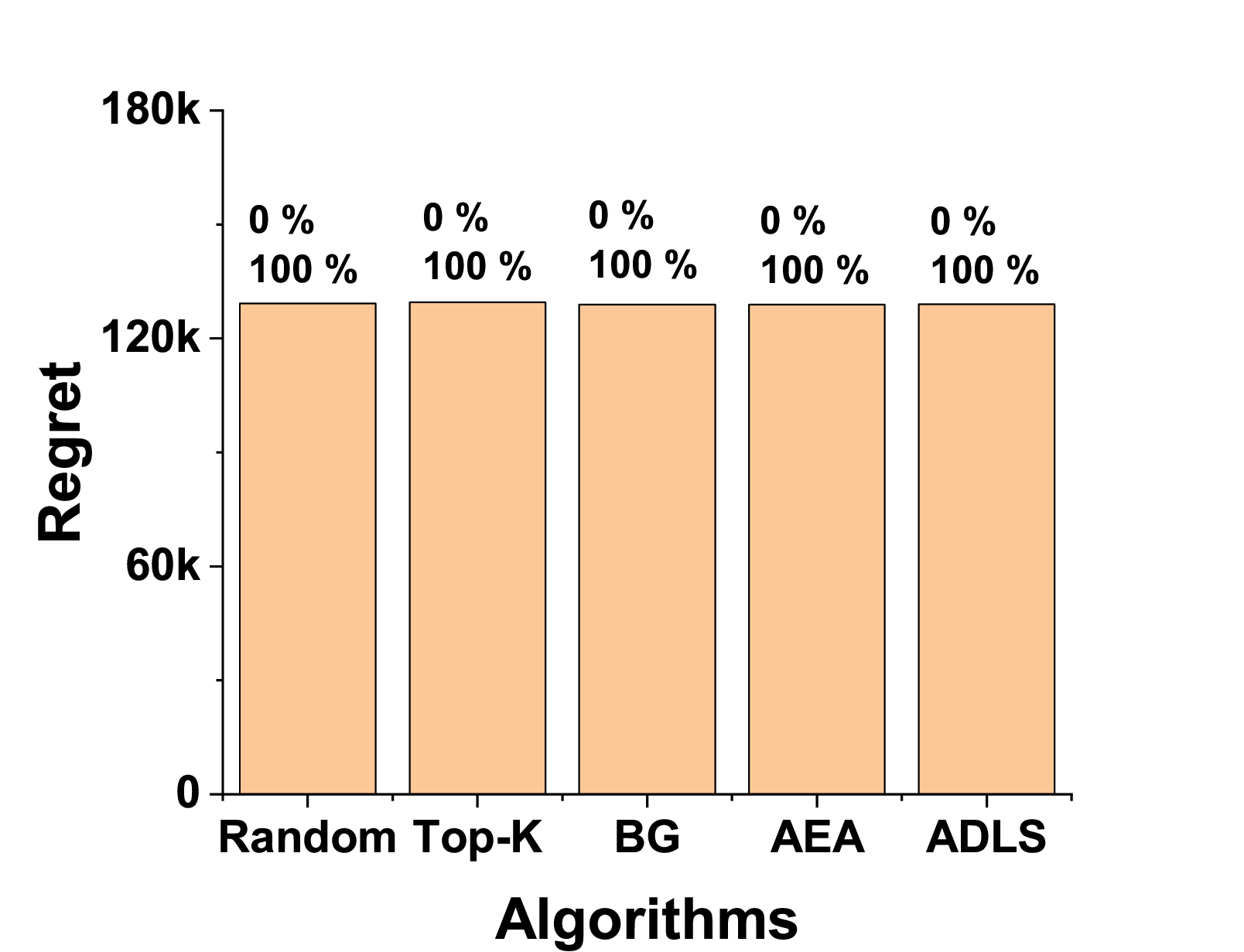} &
    \includegraphics[scale=0.10]{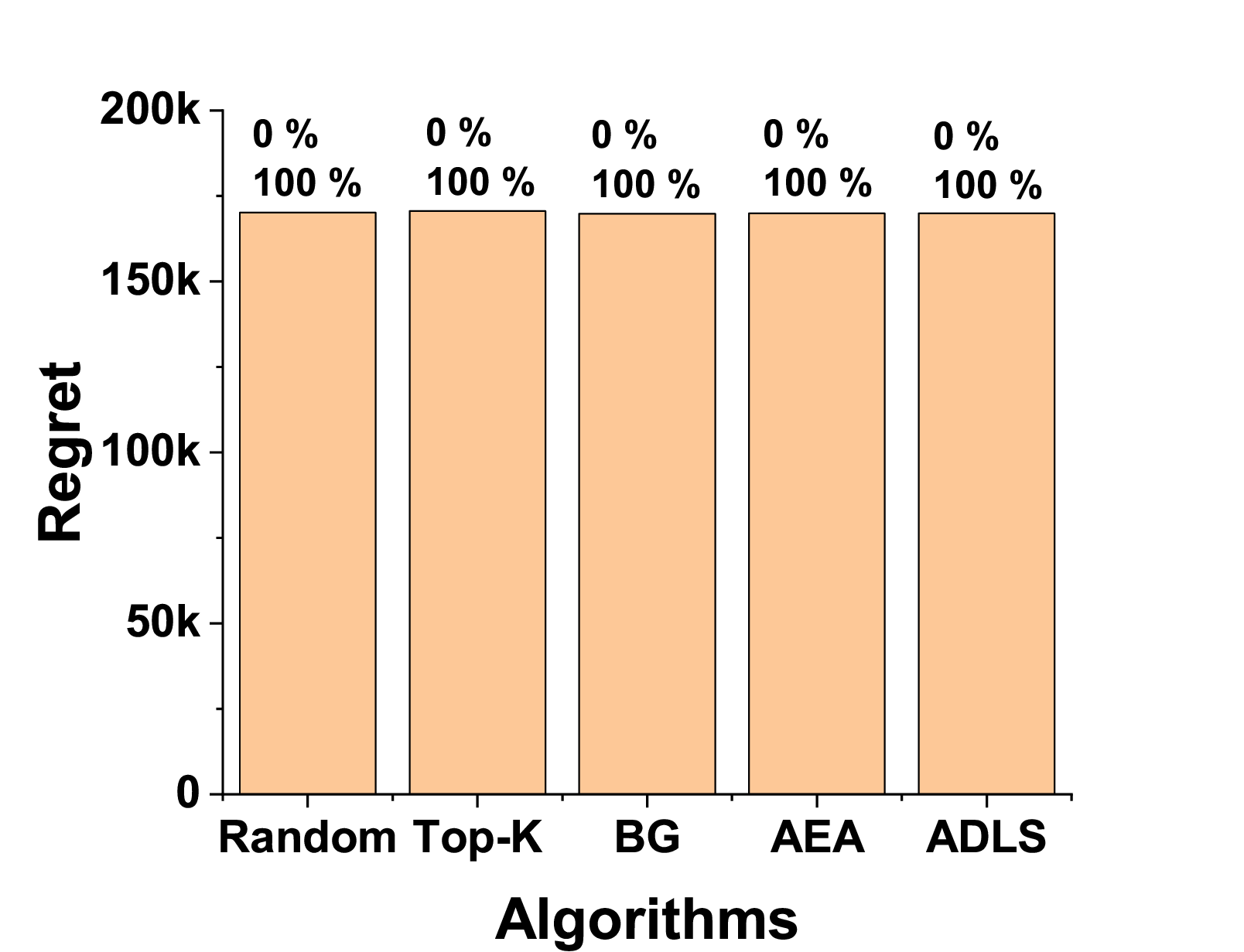} \\
    (a) $\lambda$ = $40\%$ & 
    (b) $\lambda$ = $60\%$ & 
    (c) $\lambda$ = $80\%$ & 
    (d) $\lambda$ = $100\%$ & 
    (e) $\lambda$ = $120\%$ \\
\end{tabular}
\caption{Regret of varying the demand-supply ratio $\lambda$ when $\omega = 70\%$ (Wikivote, $|\mathcal{A}| = 10$, $p_{c}: Uniform(0.1) $)}
\label{Fig:5_Influence_4}
\begin{tabular}{ccccc}
    \includegraphics[scale=0.10]{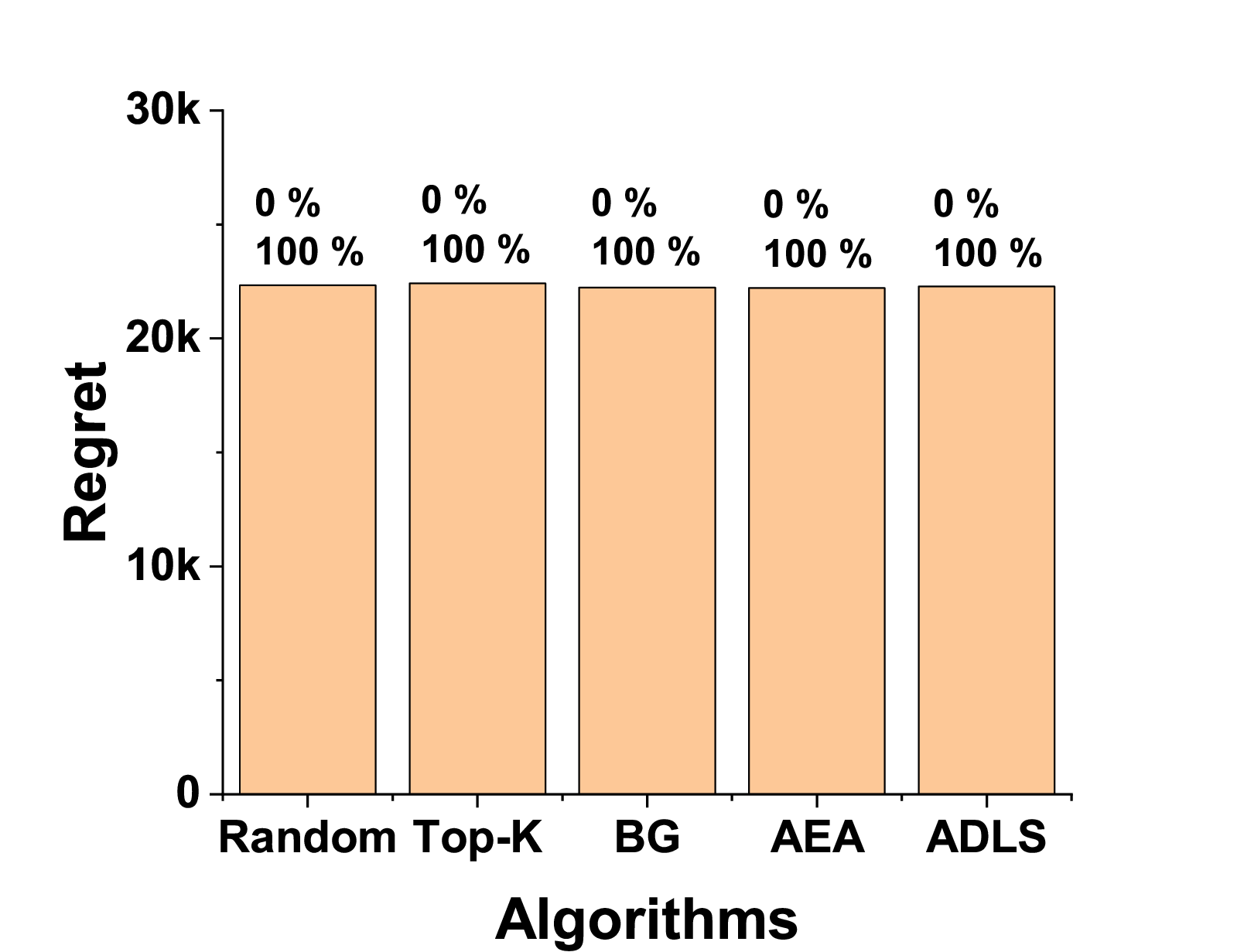} & 
    \includegraphics[scale=0.10]{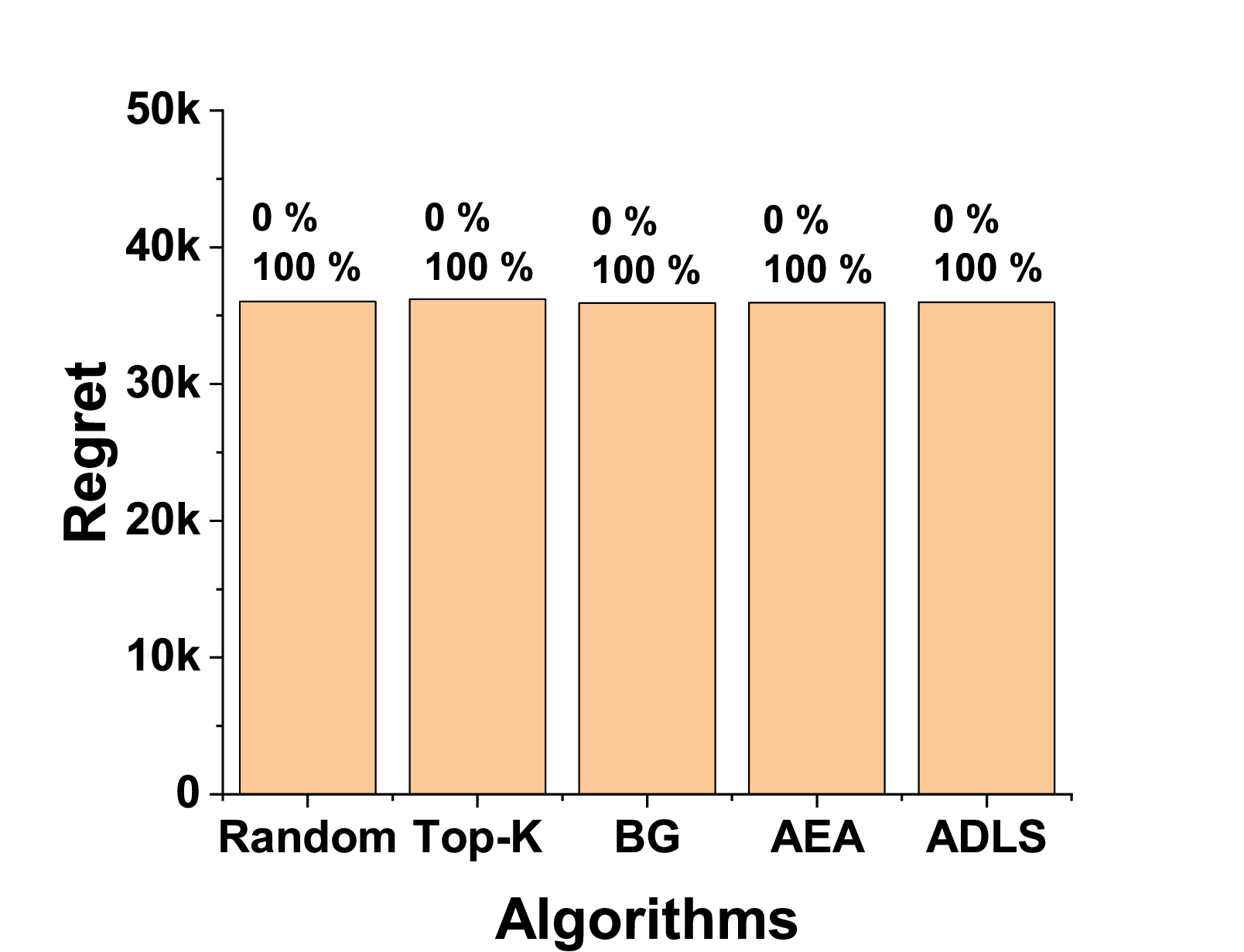} & 
    \includegraphics[scale=0.10]{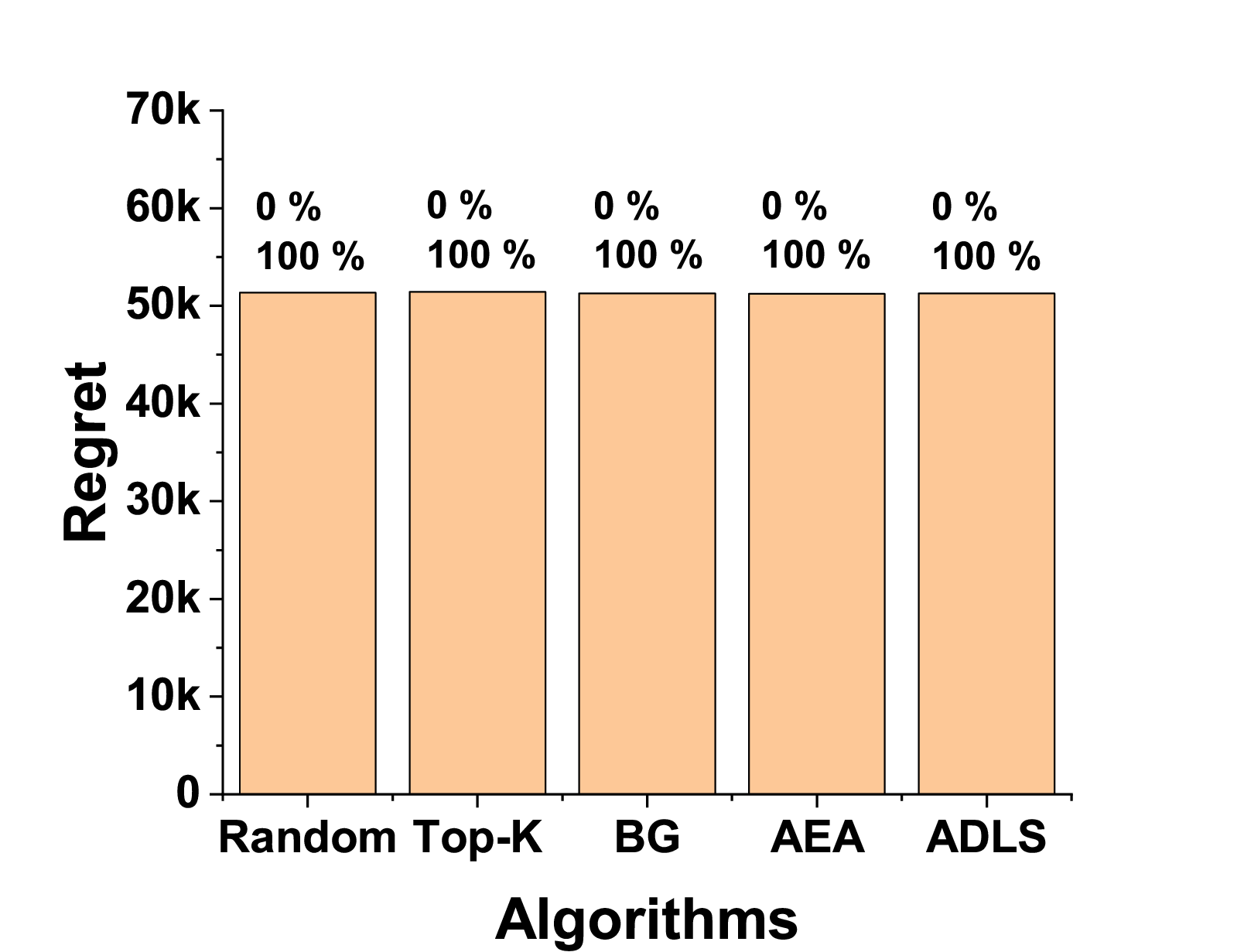} & 
    \includegraphics[scale=0.10]{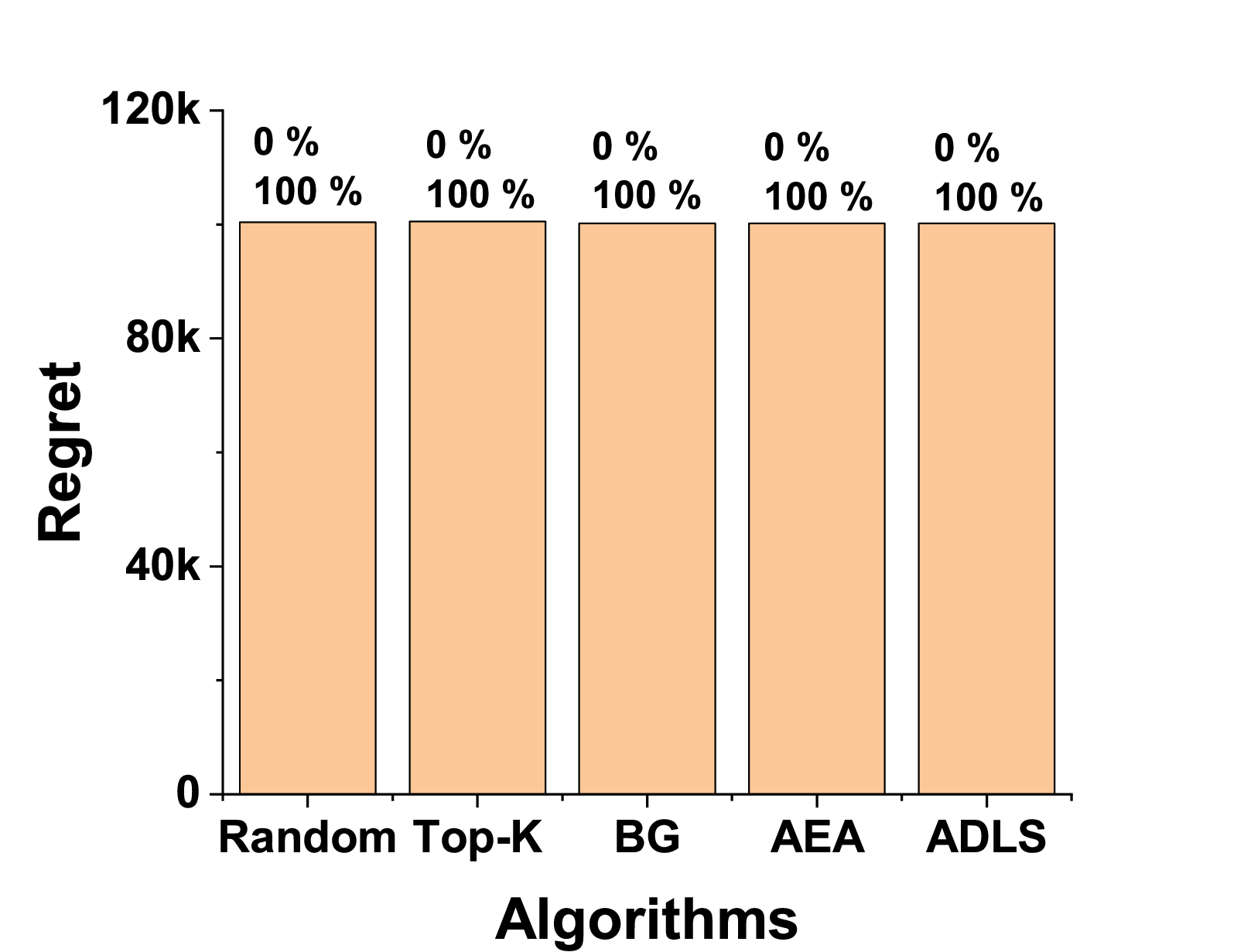} &
    \includegraphics[scale=0.10]{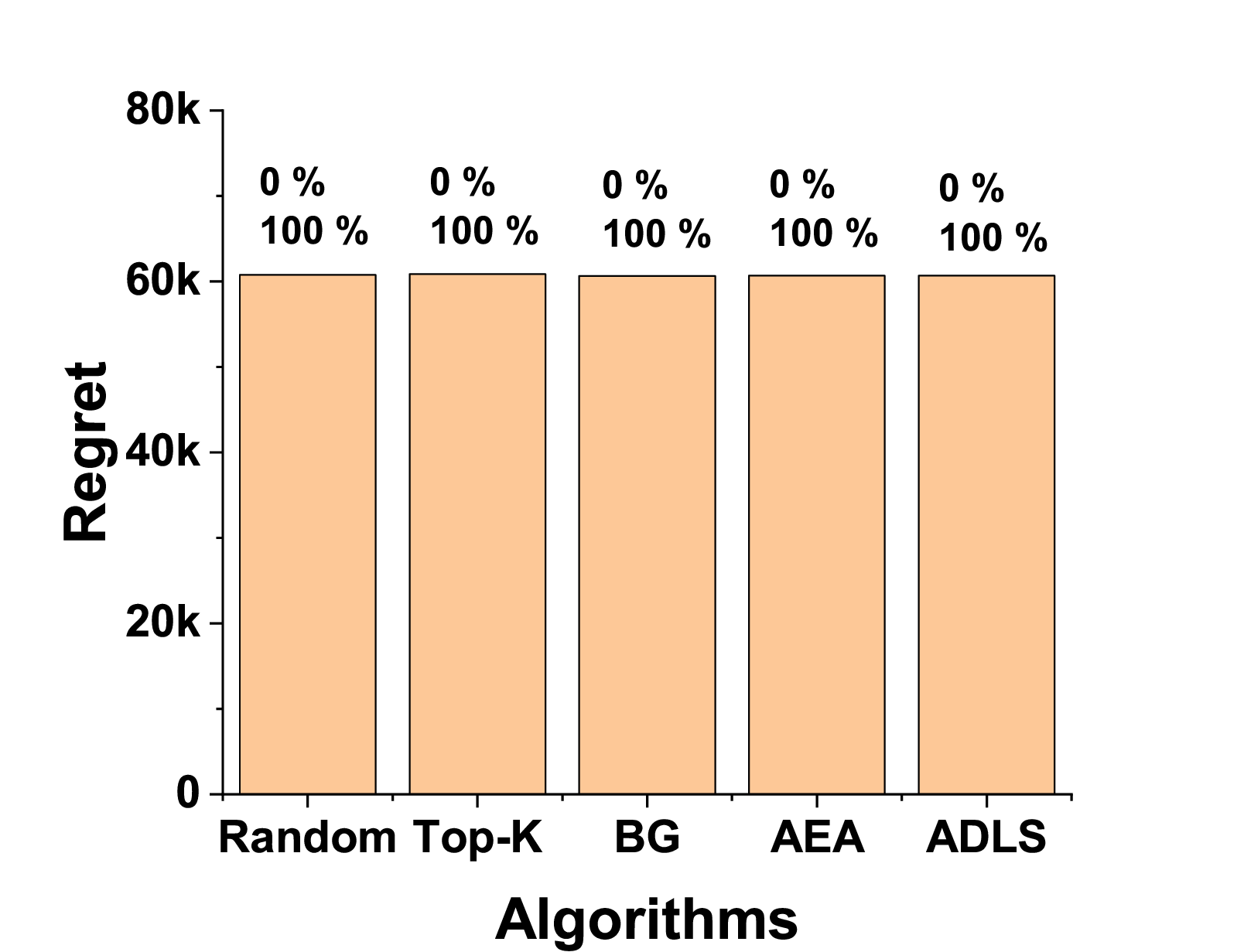} \\
    (a) $\lambda$ = $40\%$ & 
    (b) $\lambda$ = $60\%$ & 
    (c) $\lambda$ = $80\%$ & 
    (d) $\lambda$ = $100\%$ & 
    (e) $\lambda$ = $120\%$ \\
\end{tabular}
\caption{Regret of varying the demand-supply ratio $\lambda$ when $\omega = 90\%$ (Wikivote, $|\mathcal{A}| = 5$, $p_{c}: Uniform(0.1) $)}
\label{Fig:5_Influence_5}
\end{figure}

Now, firstly we analyse Uniform $(0.1)$ probability setting results. In Email-Eu-Core dataset (Figures $\ref{Fig:1_Influence_1}$ (a) \& (b) and $\ref{Fig:1_Influence_2}$ (a) \& (b)), the best performing algorithm is ``ADLS" with an average total regret of $2061.31$. The ``Top-$K$" algorithm performs the worst, with $4695.14\%$ higher regret than ``ADLS". ``BG" and ``AEA" perform relatively well but still show $16.01\%$ and $25.04\%$ higher regret than ``ADLS". In Facebook dataset (Figures $\ref{Fig:3_Influence_1}$ (a) \& (b) and $\ref{Fig:3_Influence_2}$ (a) \& (b)), the best performing algorithm is ``ADLS" with an average total regret of $3563.17$. The highest regret is from ``Top-$K$'', which is $15806.38\%$ worse than ``ADLS". ``BG" and ``AEA" perform moderately but still show $24.20\%$ and $24.20\%$ higher regret than ``ADLS". In Wikivote dataset (Figures $\ref{Fig:5_Influence_1}$ (a) \& (b) and $\ref{Fig:5_Influence_2}$ (a) \& (b)), ``ADLS" achieved the lowest total regret across multiple instances. For example, in one instance (Figure ($\ref{Fig:5_Influence_1}$ (a)), ``ADLS" had a total regret of $2294.16$, whereas ``Random" and ``Top-$K$" had significantly higher values ($222618.28$ and $367158.72$, respectively). This represents an improvement of over $98\%$ compared to ``Random", proving ``ADLS"'s effectiveness in minimizing regret. In Table $\ref{tab:congressregretuniform01}$, we observe Congress-Twitter dataset. In it ``Random" and ``Top-$K$" algorithms consistently yield the highest regret across all scenarios. For instance, at $\lambda = 60\%$, $\omega = 10\%$, the regret for ``Top-$K$" is $29737.52$, whereas ``BG" achieves a significantly lower regret of $19828.20$. The ``BG", ``AEA", and ``ADLS" algorithms consistently outperform ``Random" and ``Top-$K$", reducing regret by $20-75\%$ in most cases.

\begin{figure}[H]
\centering
\begin{tabular}{lclc}
    Excessive Regret & \includegraphics[width=0.11\linewidth]{Excessive.png} &
    Unsatisfied Regret & \includegraphics[width=0.11\linewidth]{Unsatisfied.png}\\
\end{tabular}
\begin{tabular}{ccccc}
    \includegraphics[scale=0.10]{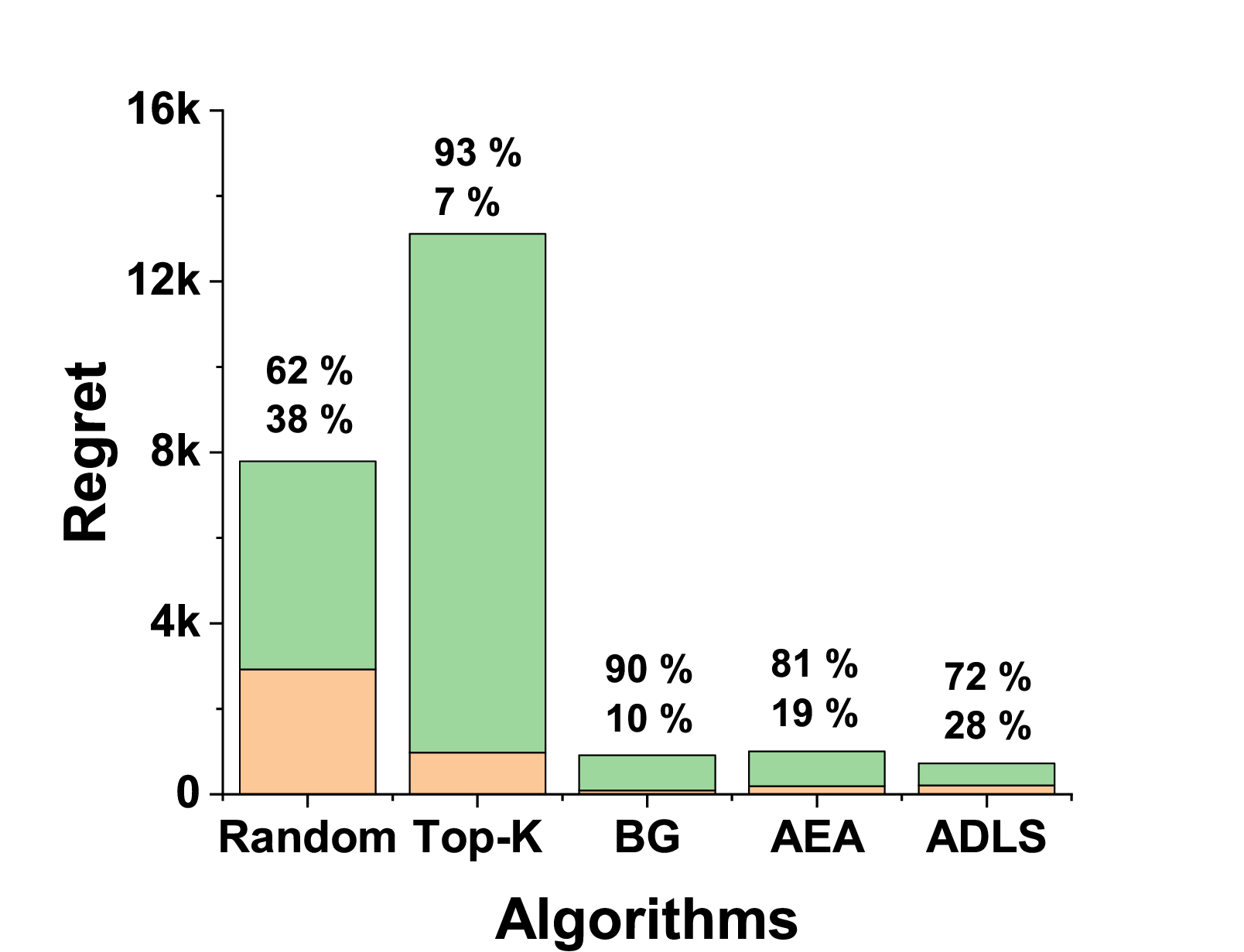} & 
    \includegraphics[scale=0.10]{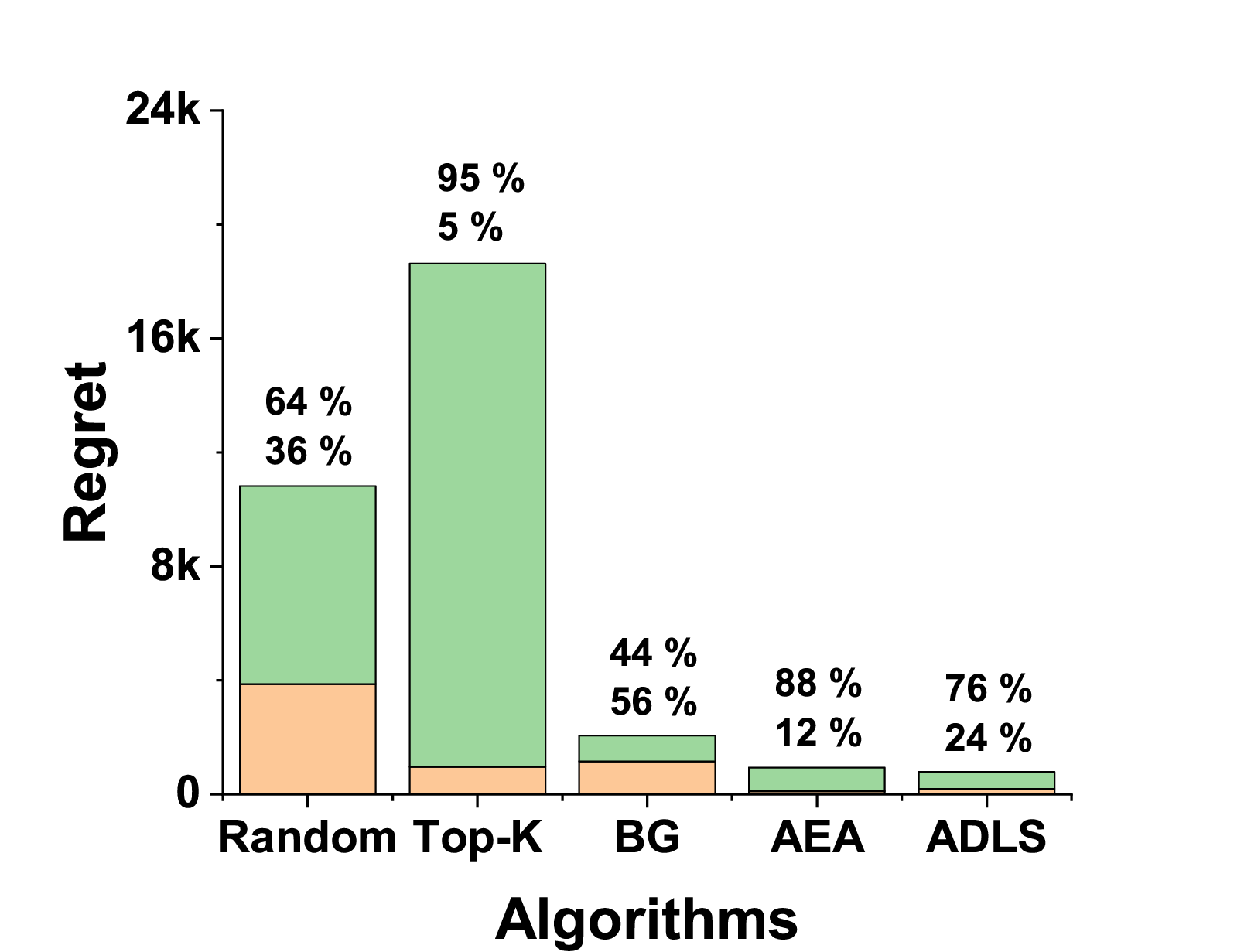} & 
    \includegraphics[scale=0.10]{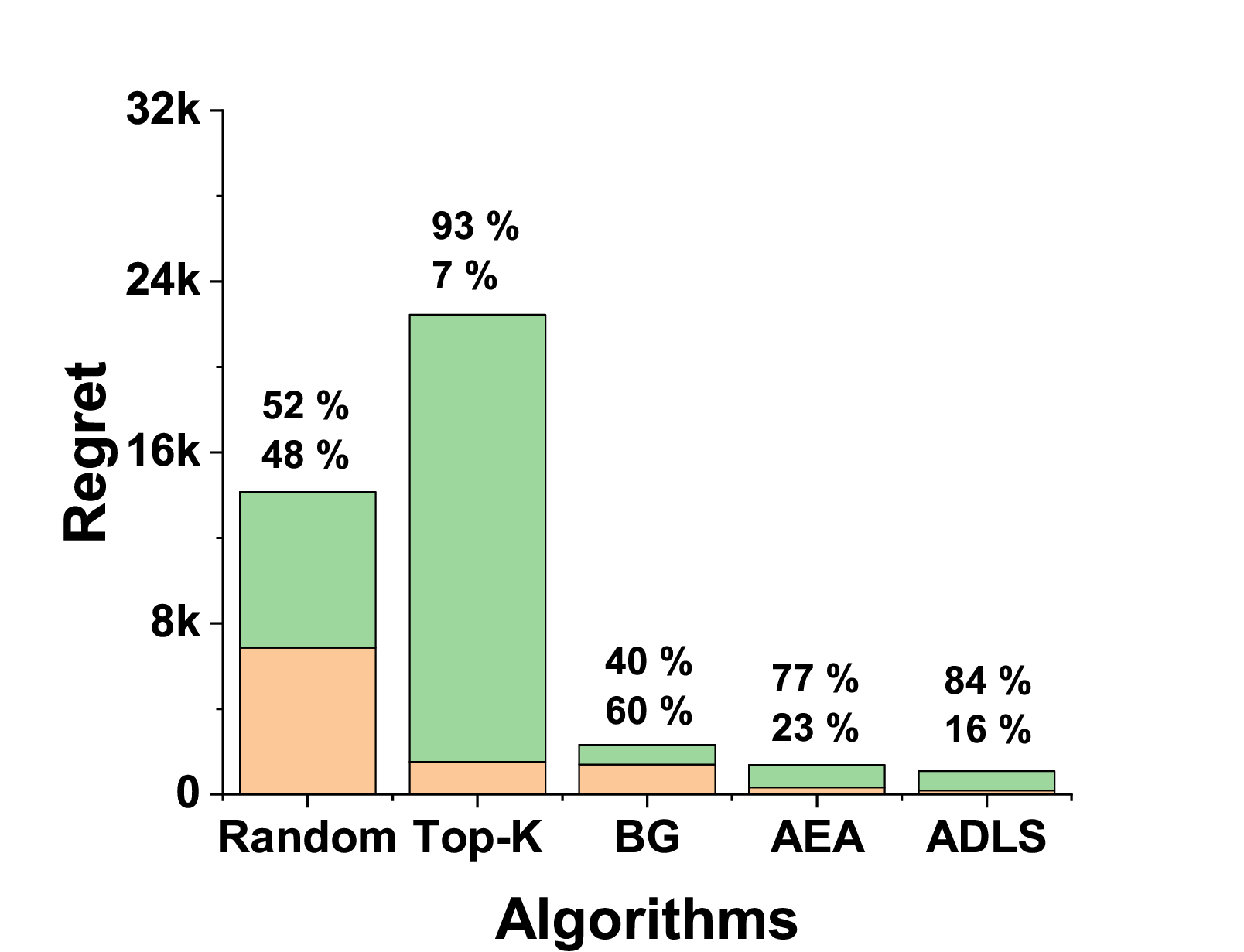} & 
    \includegraphics[scale=0.10]{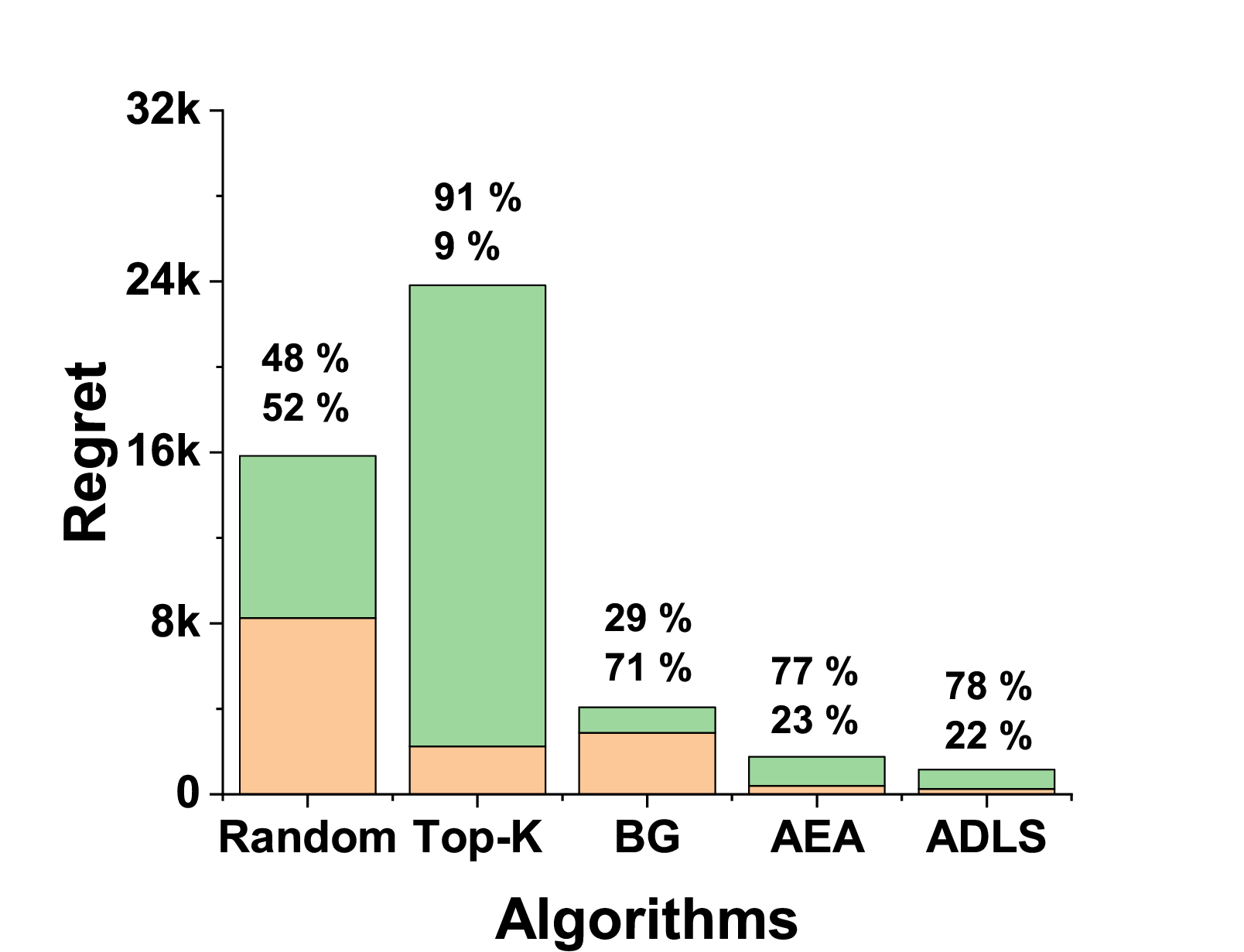} &
    \includegraphics[scale=0.10]{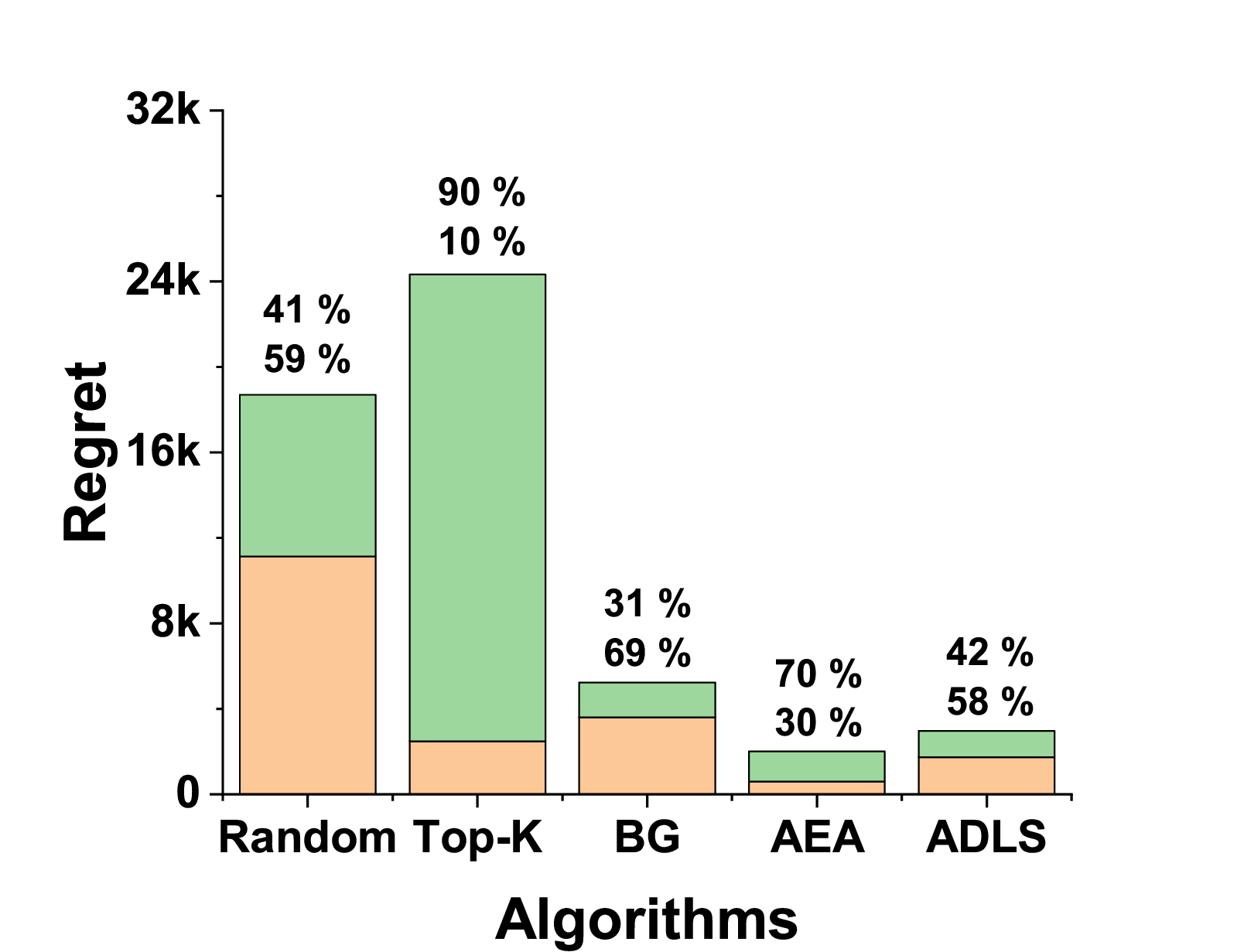} \\
    (a) $\lambda$ = $40\%$ & 
    (b) $\lambda$ = $60\%$ & 
    (c) $\lambda$ = $80\%$ & 
    (d) $\lambda$ = $100\%$ & 
    (e) $\lambda$ = $120\%$ \\
\end{tabular}
\caption{Regret of varying the demand-supply ratio $\lambda$ when $\omega = 10\%$ (Email-Eu-Core, $|\mathcal{A}| = 100$, $p_{c}: Trivalency$)}
\label{Fig:2_Influence_1}
\begin{tabular}{ccccc}
    \includegraphics[scale=0.10]{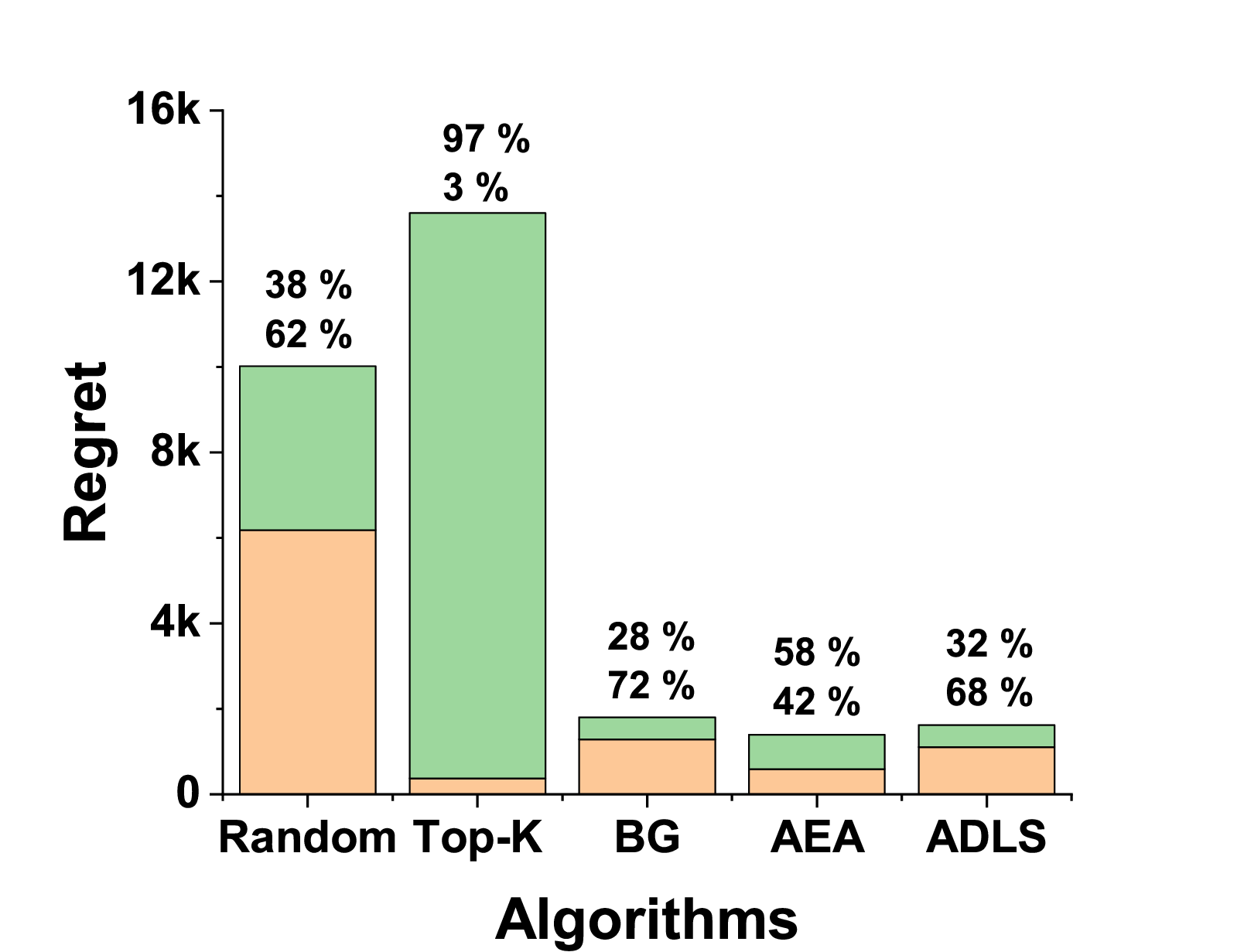} & 
    \includegraphics[scale=0.10]{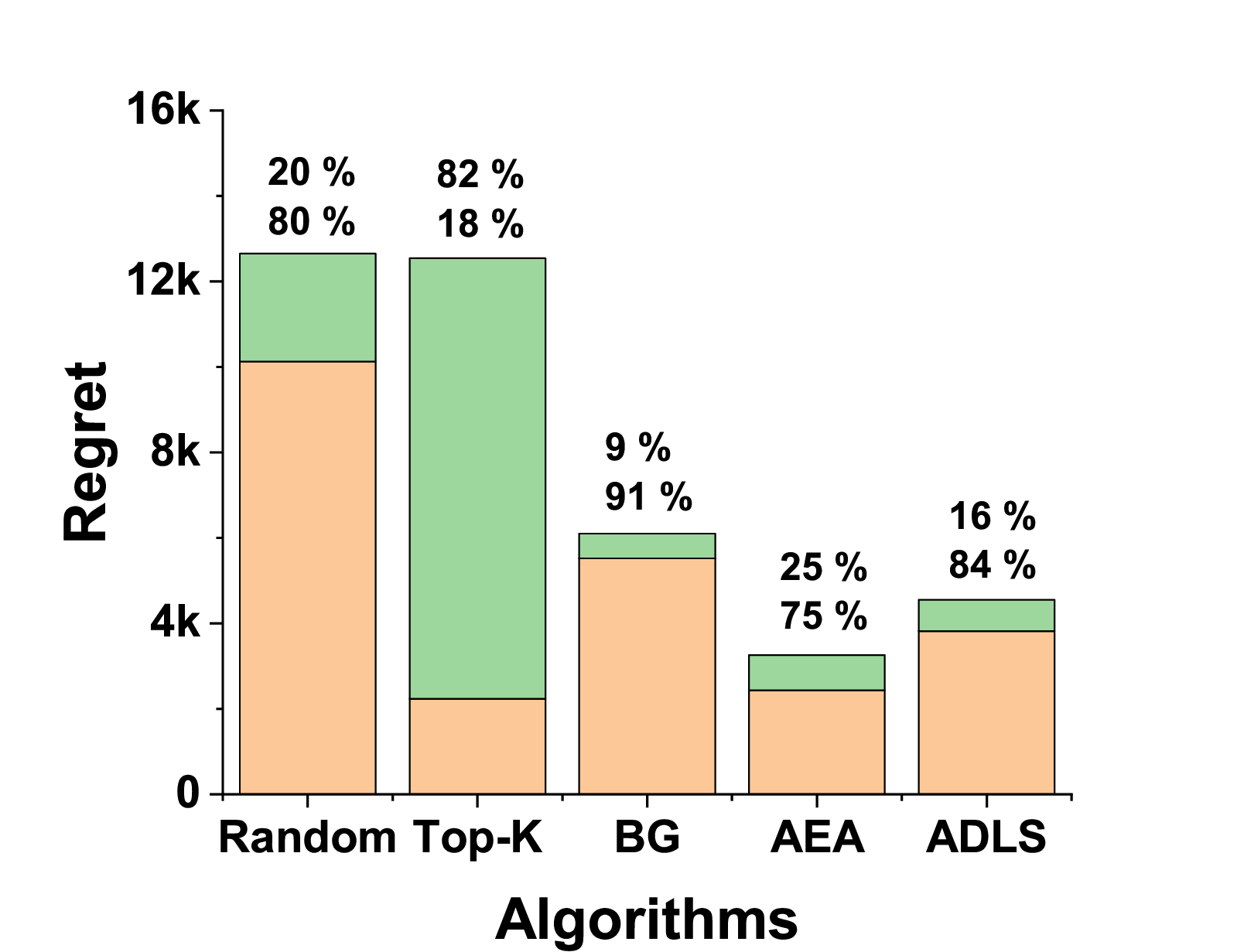} & 
    \includegraphics[scale=0.10]{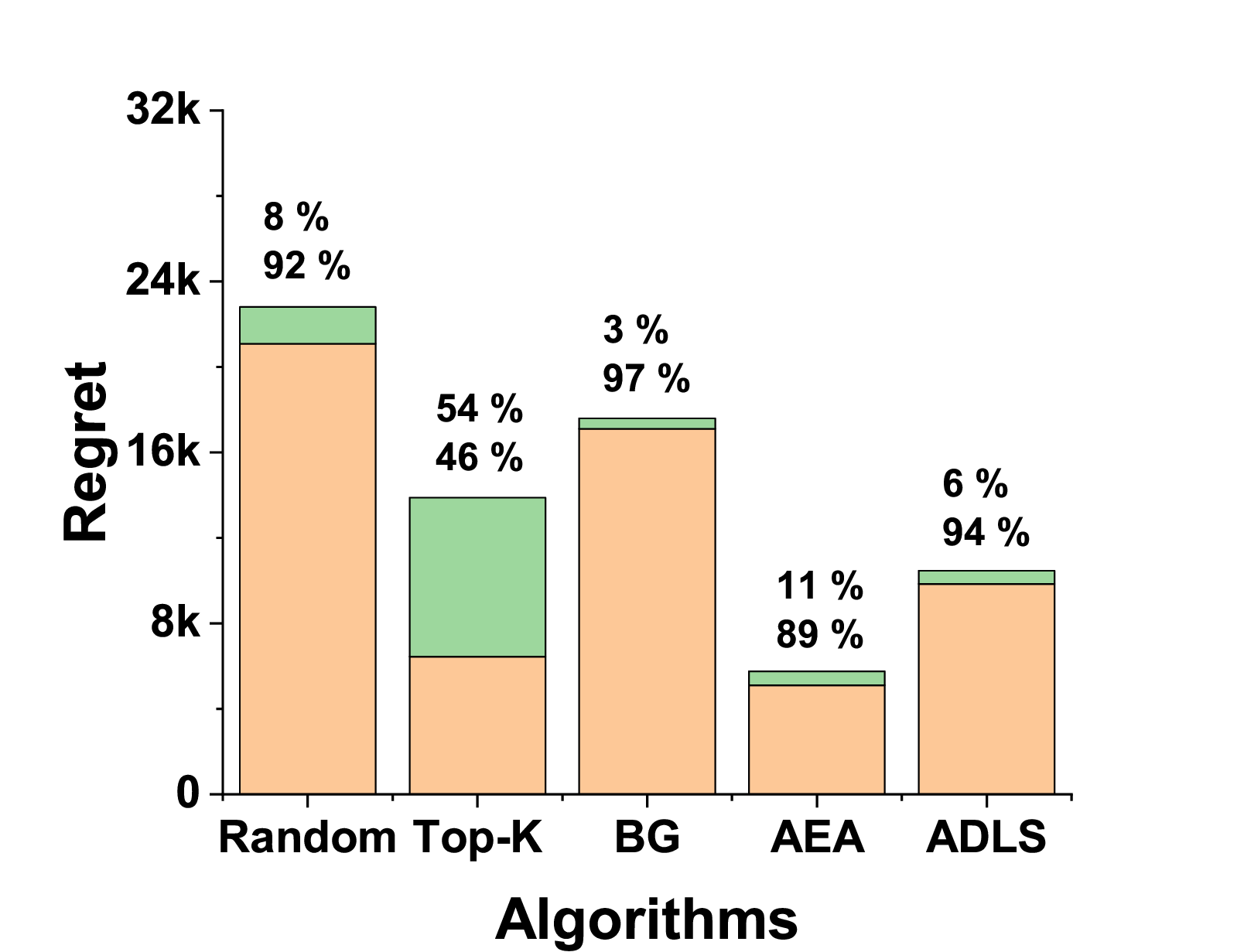} & 
    \includegraphics[scale=0.10]{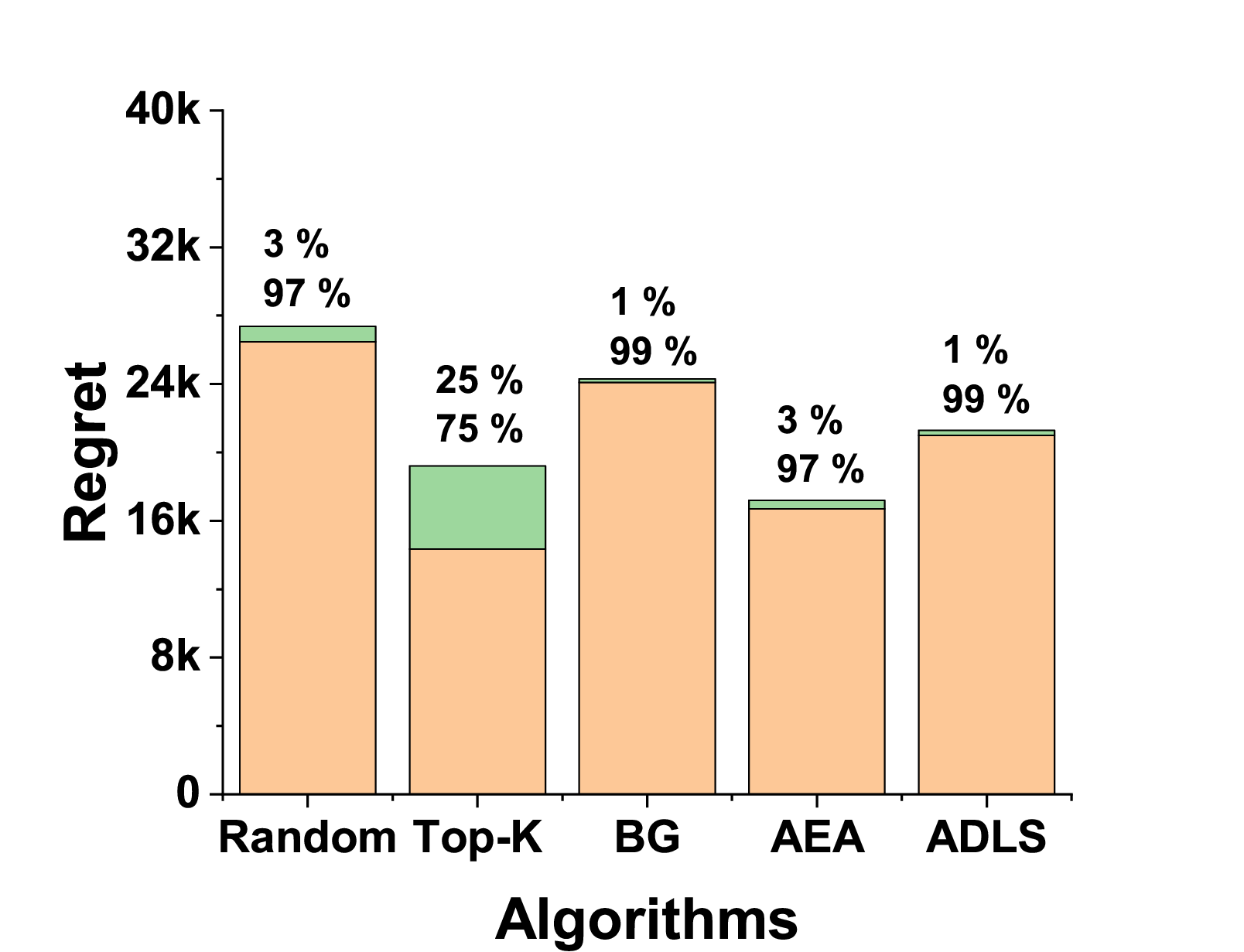} &
    \includegraphics[scale=0.10]{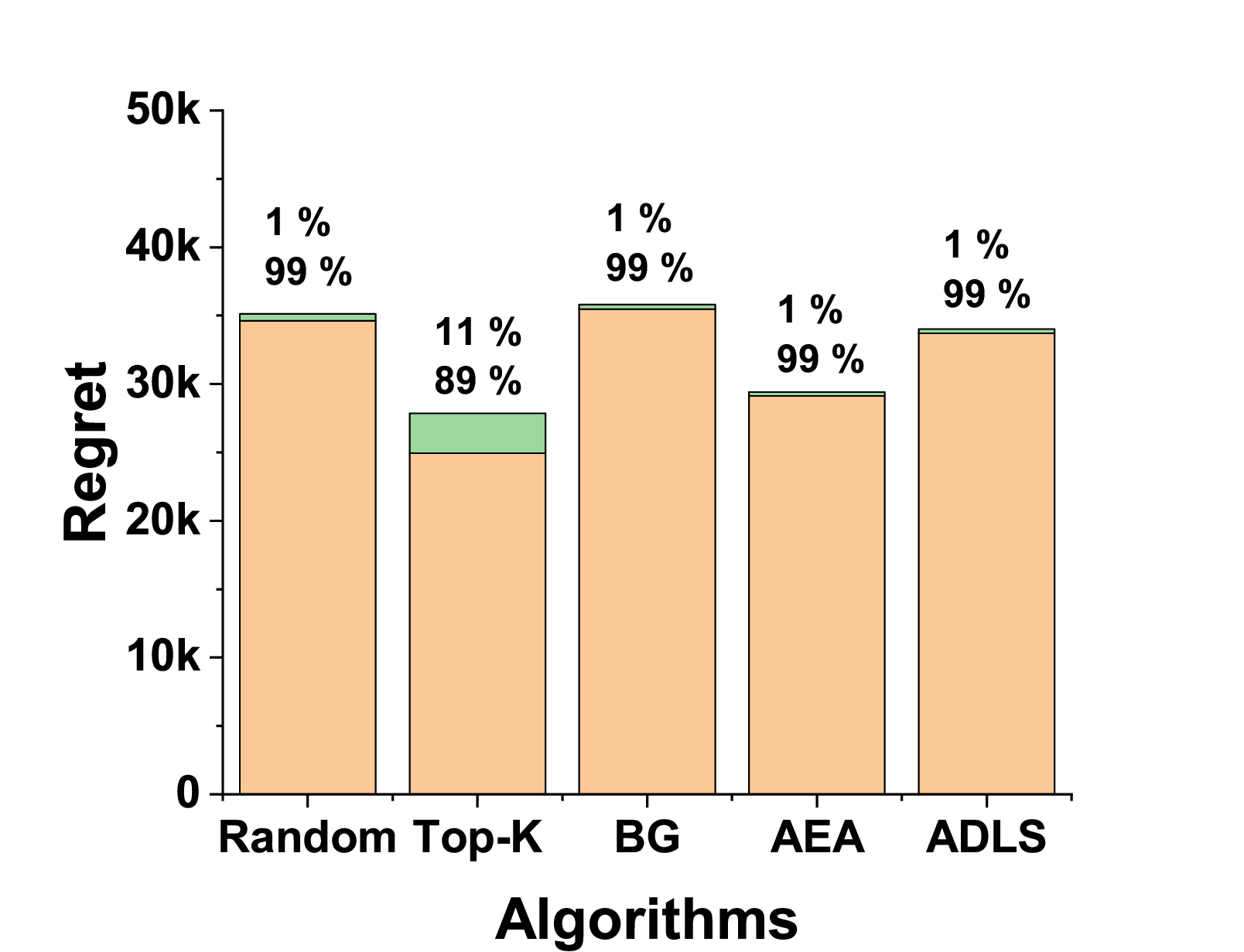} \\
    (a) $\lambda$ = $40\%$ & 
    (b) $\lambda$ = $60\%$ & 
    (c) $\lambda$ = $80\%$ &
    (d) $\lambda$ = $100\%$ & 
    (e) $\lambda$ = $120\%$ \\
\end{tabular}
\caption{Regret of varying the demand-supply ratio $\lambda$ when $\omega = 30\%$ (Email-Eu-Core, $|\mathcal{A}| = 50$, $p_{c}: Trivalency$)}
\label{Fig:2_Influence_2}
\begin{tabular}{ccccc}
    \includegraphics[scale=0.10]{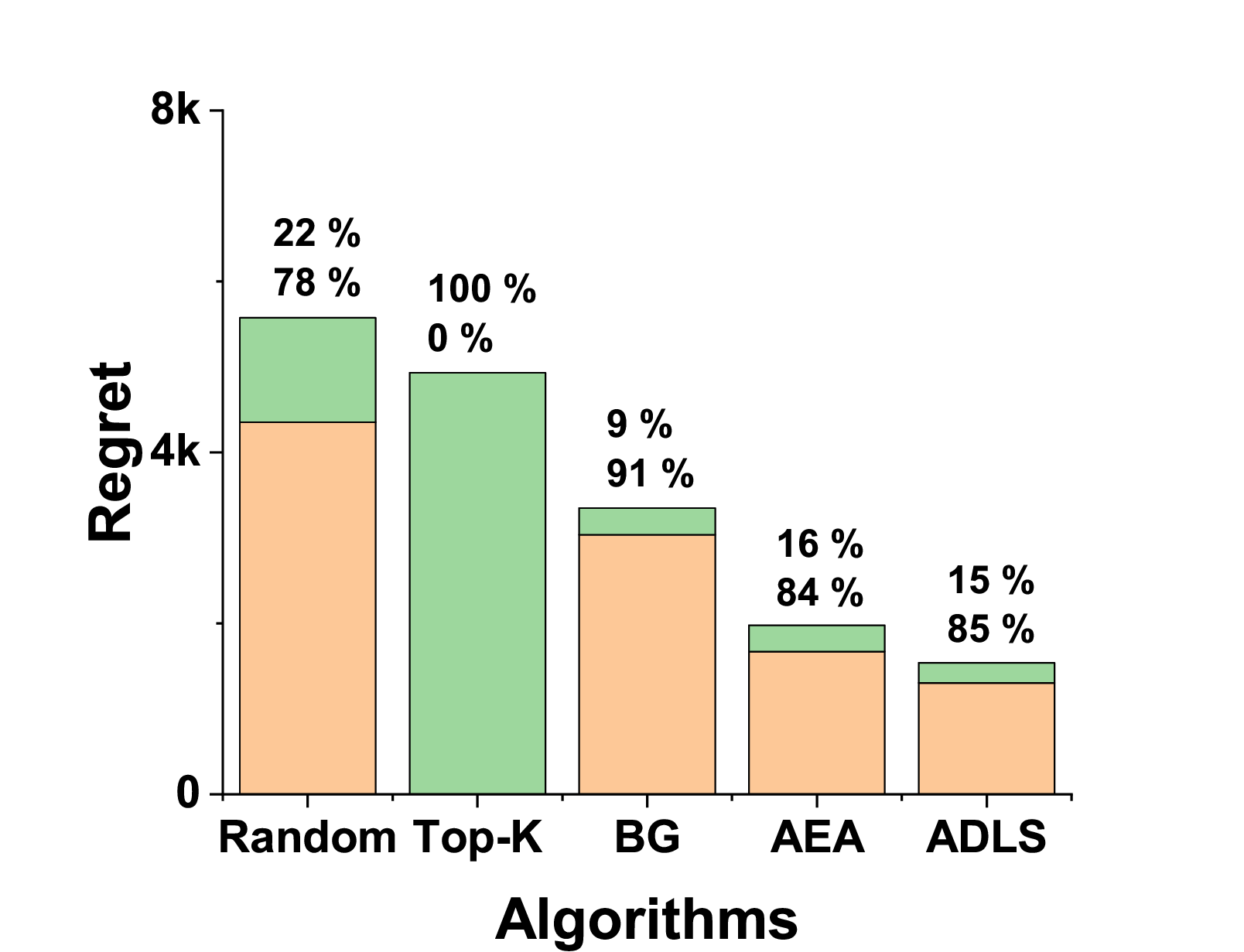} & 
    \includegraphics[scale=0.10]{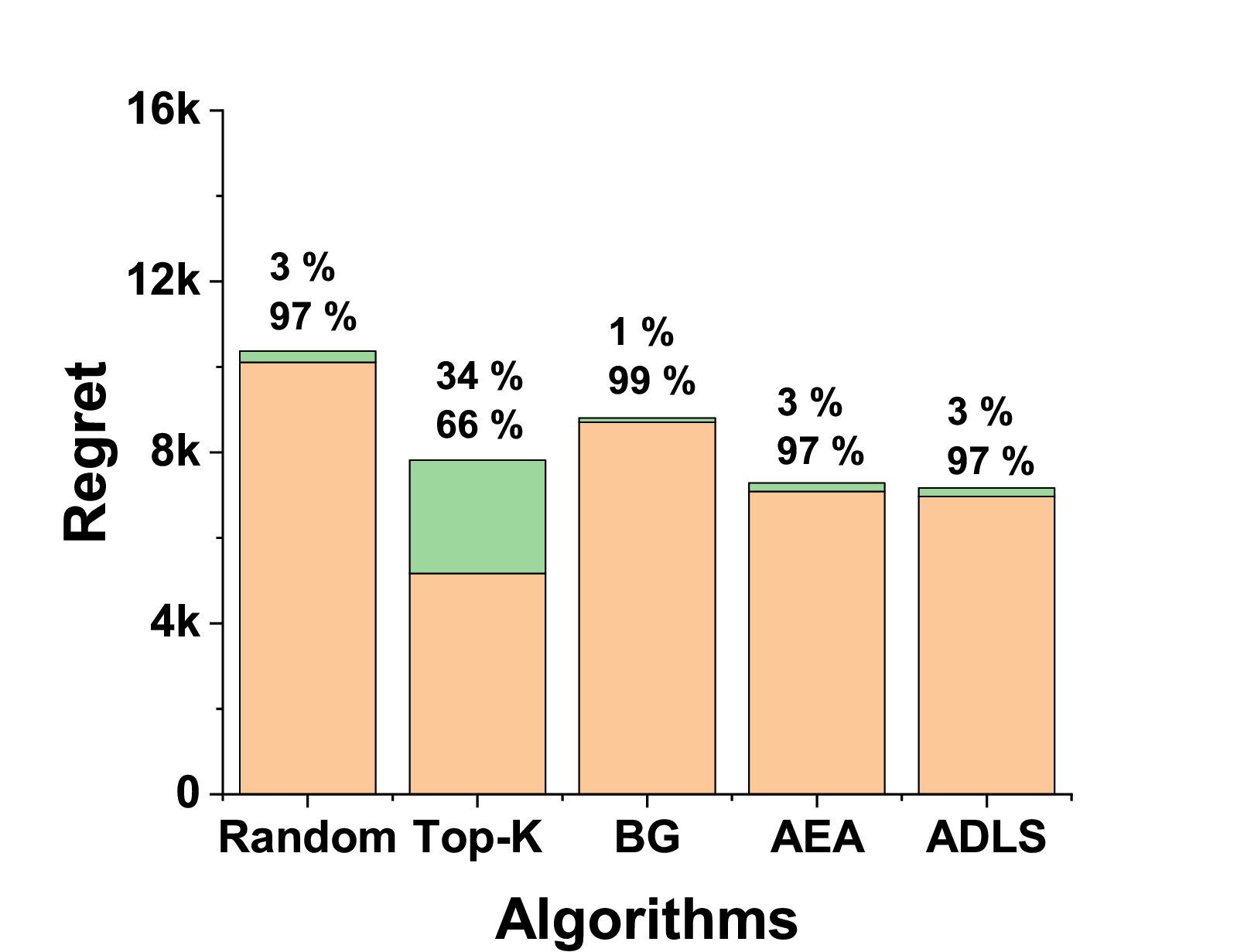} & 
    \includegraphics[scale=0.10]{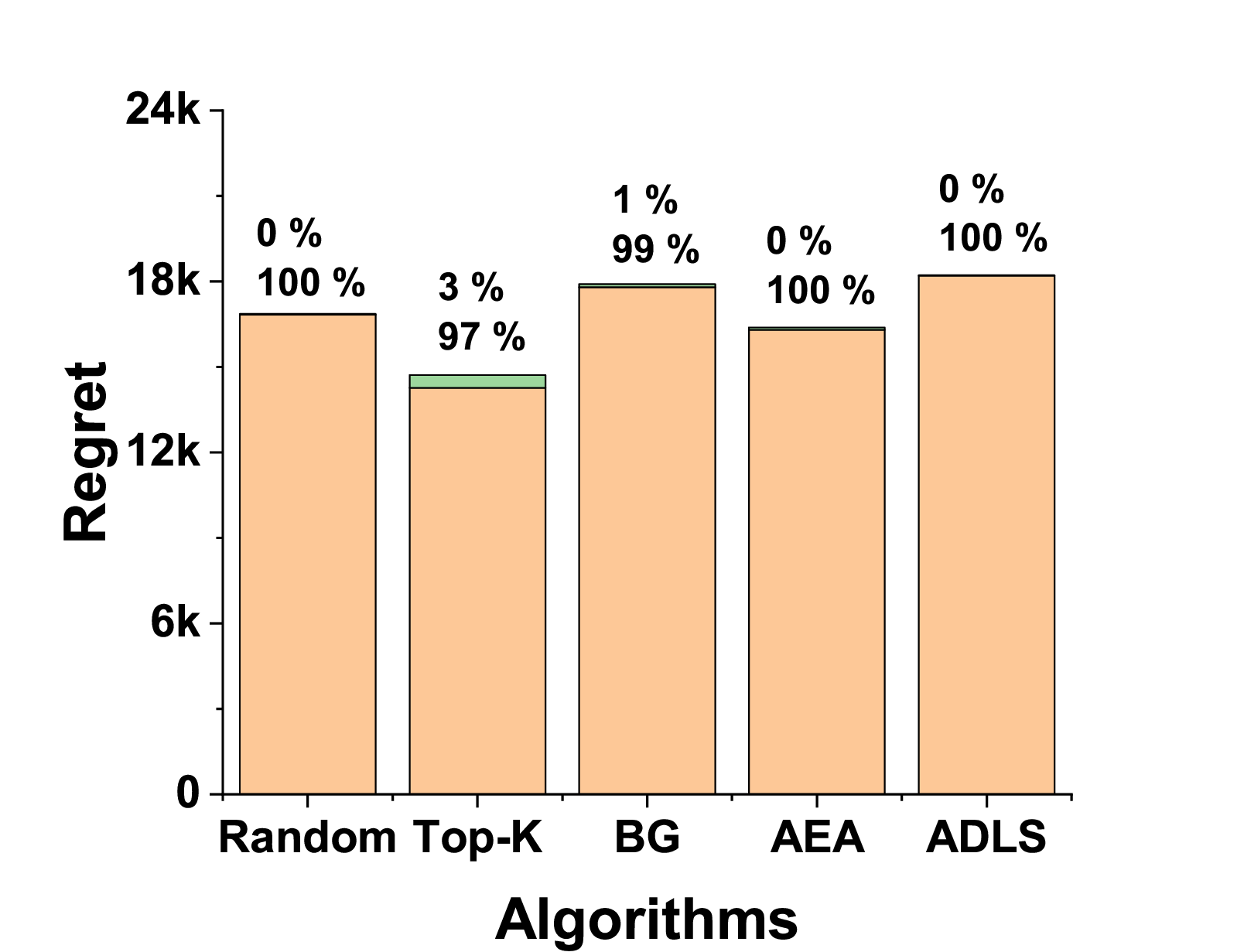} & 
    \includegraphics[scale=0.10]{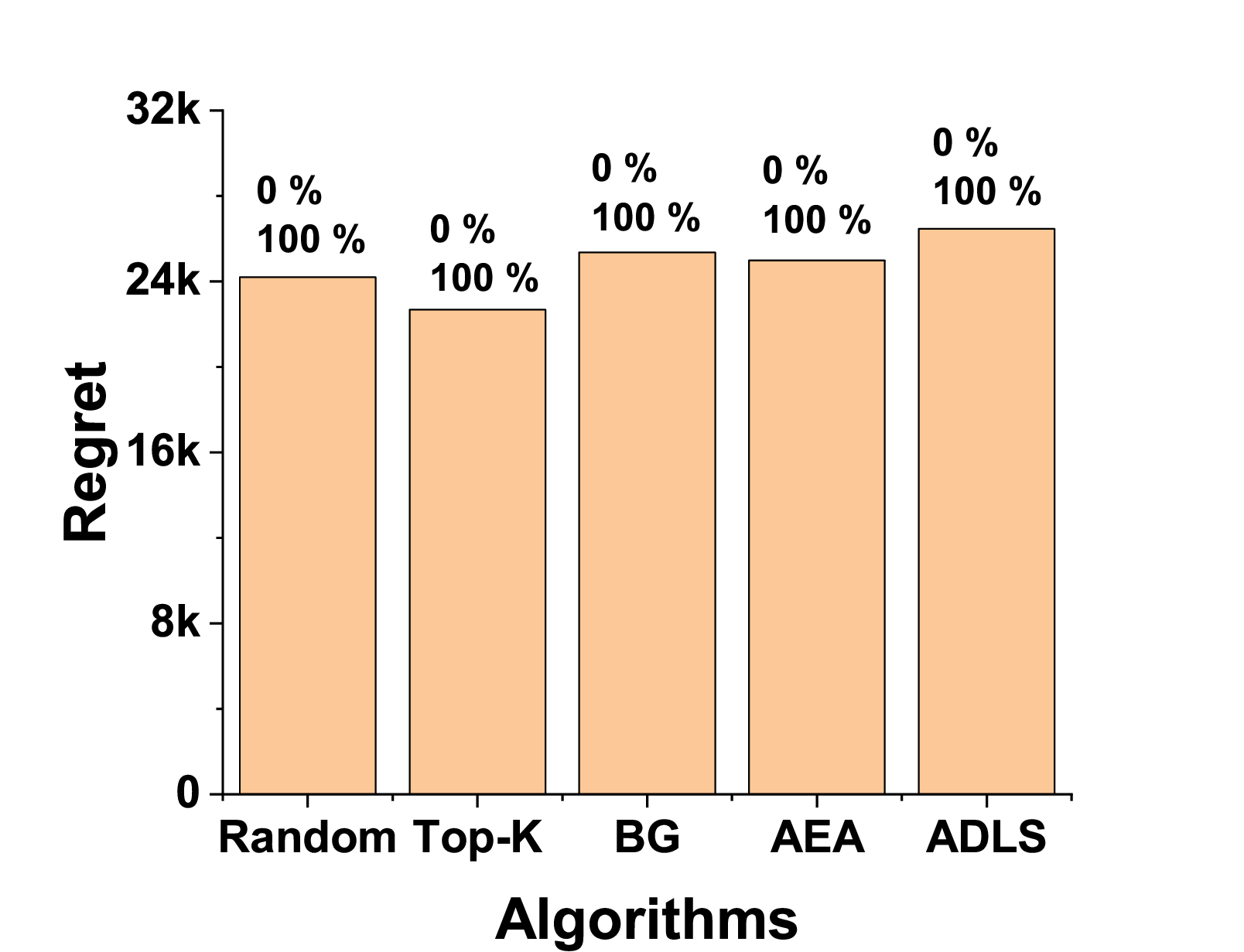} &
    \includegraphics[scale=0.10]{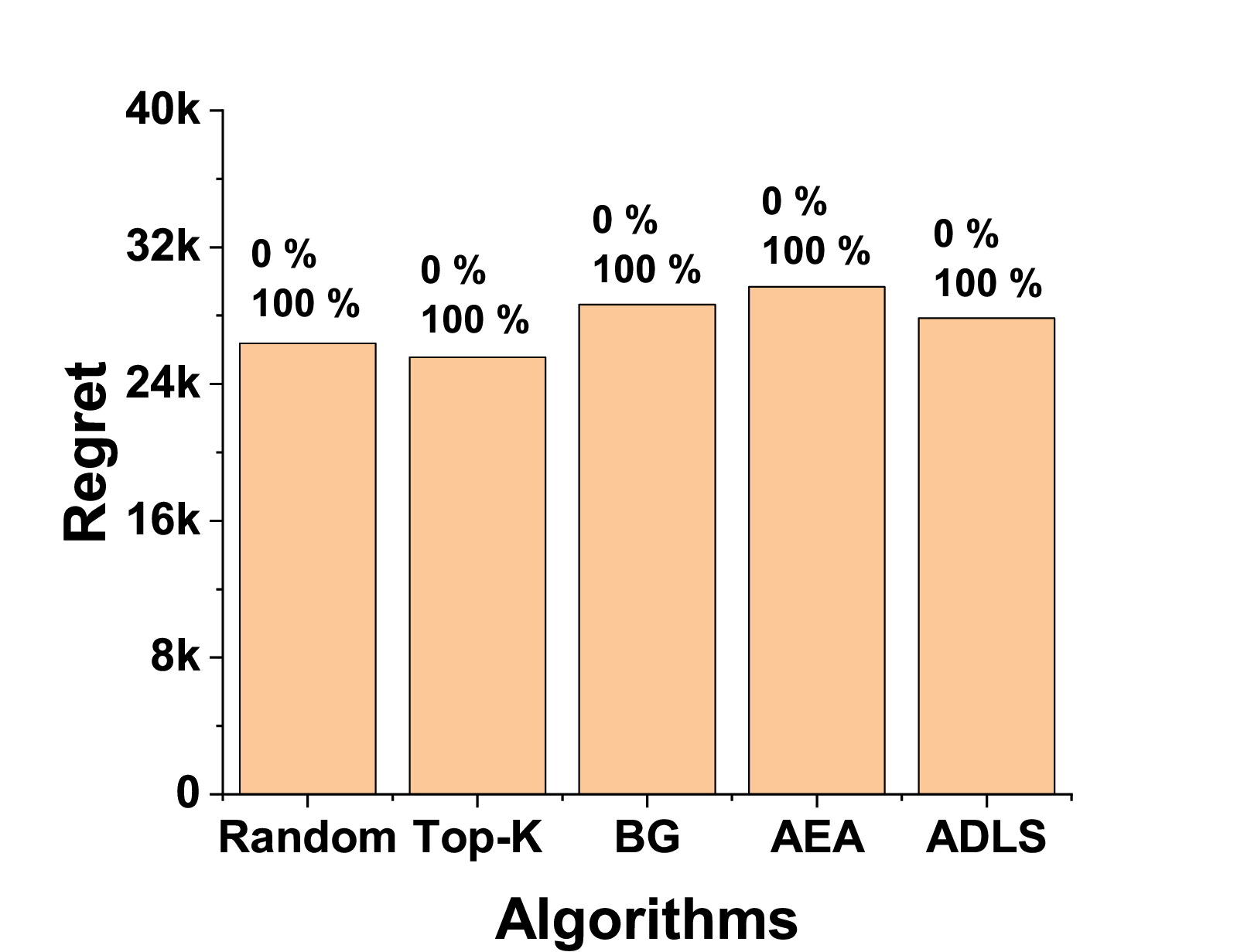} \\
    (a) $\lambda$ = $40\%$ & 
    (b) $\lambda$ = $60\%$ & 
    (c) $\lambda$ = $80\%$ & 
    (d) $\lambda$ = $100\%$ & 
    (e) $\lambda$ = $120\%$ \\
\end{tabular}
\caption{Regret of varying the demand-supply ratio $\lambda$ when $\omega = 50\%$ (Email-Eu-Core, $|\mathcal{A}| = 20$, $p_{c}: Trivalency$)}
\label{Fig:2_Influence_3}
\begin{tabular}{ccccc}
    \includegraphics[scale=0.10]{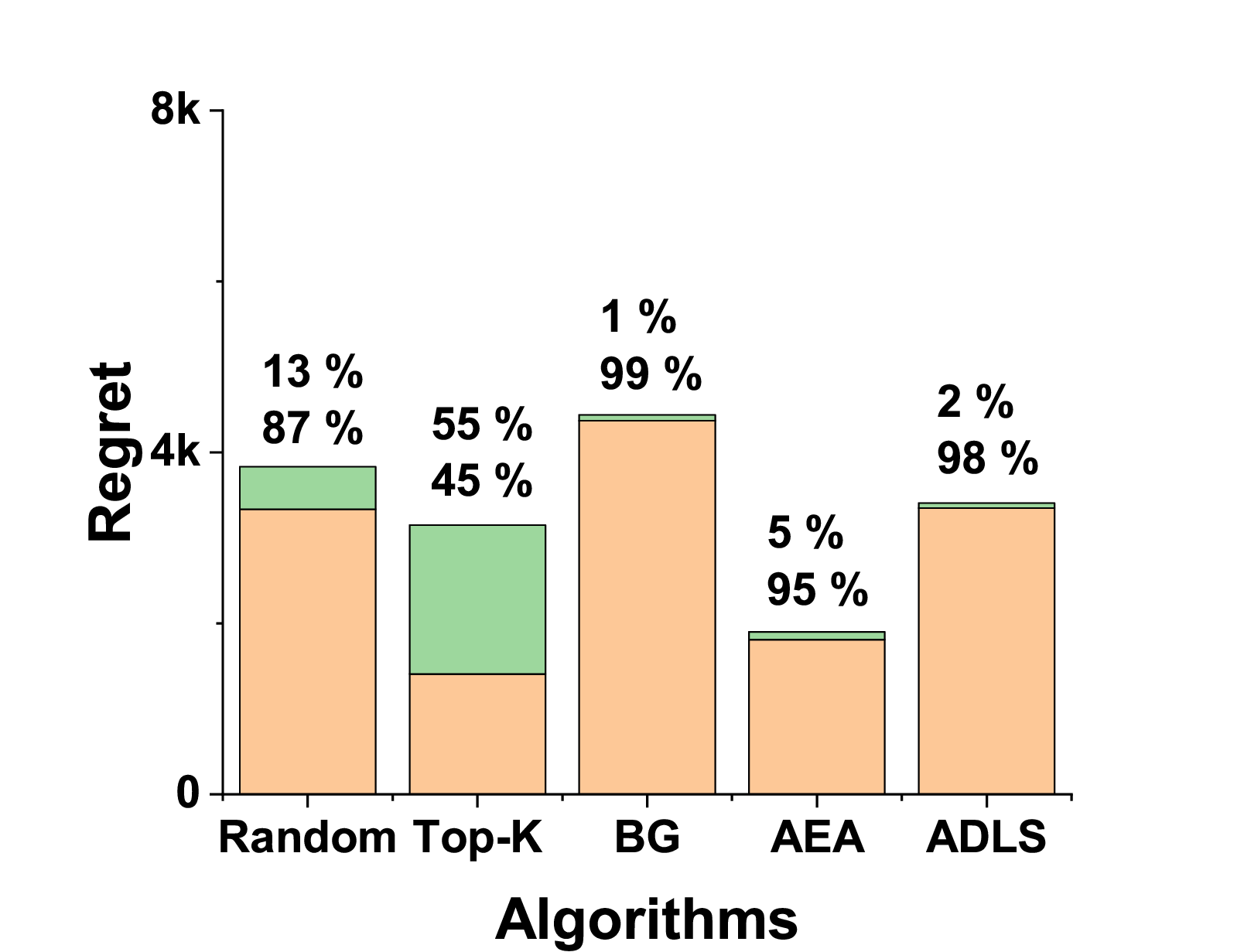} & 
    \includegraphics[scale=0.10]{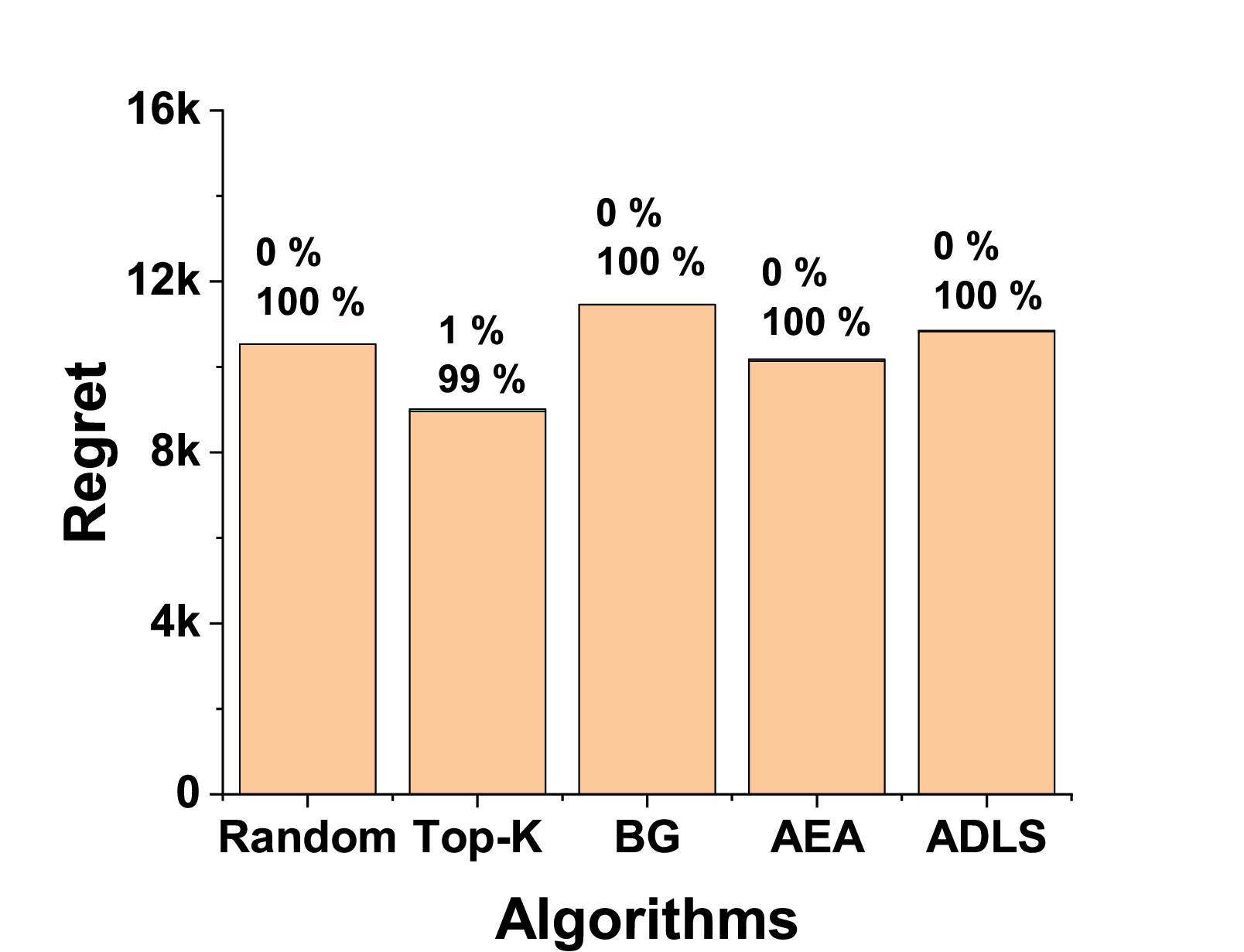} & 
    \includegraphics[scale=0.10]{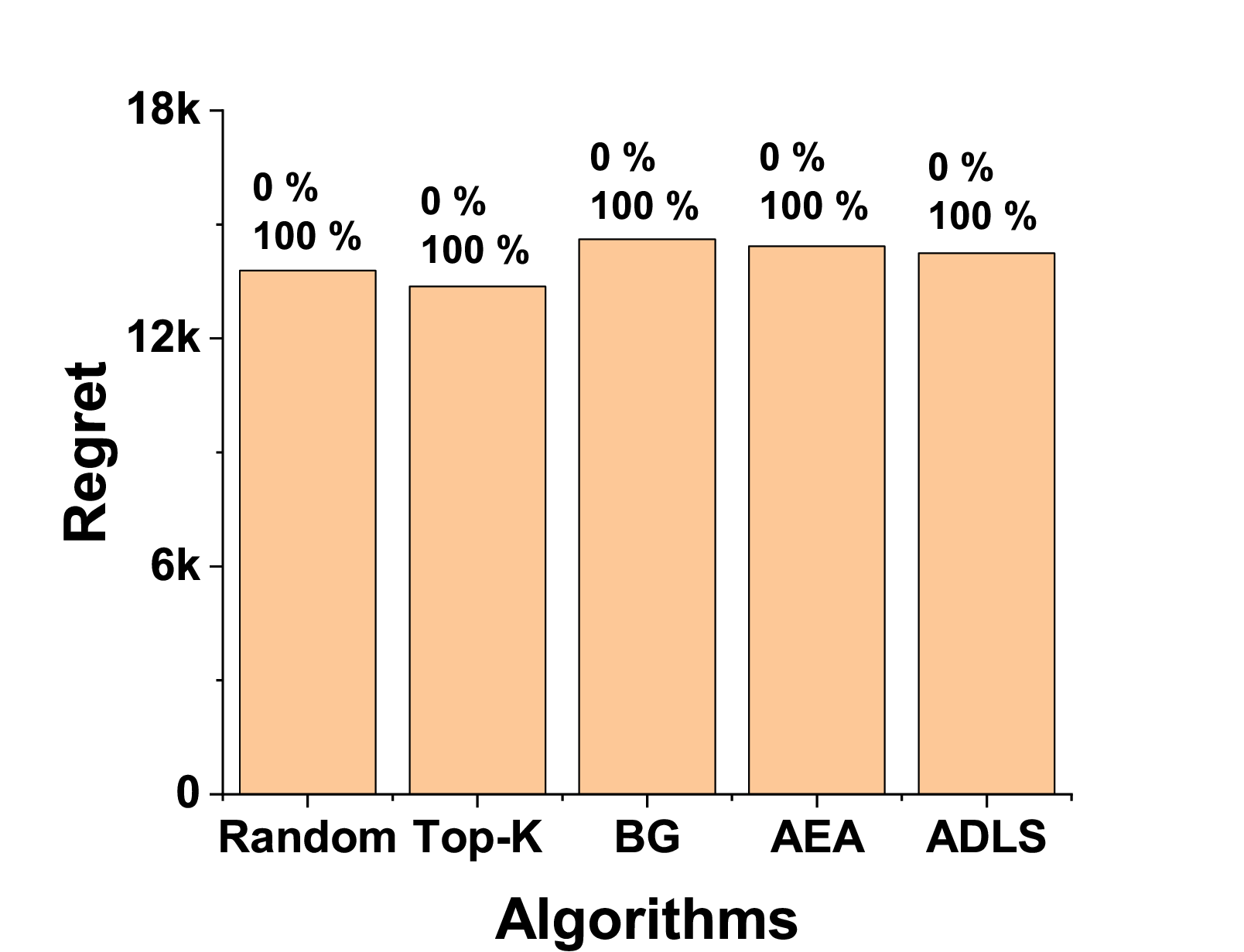} & 
    \includegraphics[scale=0.10]{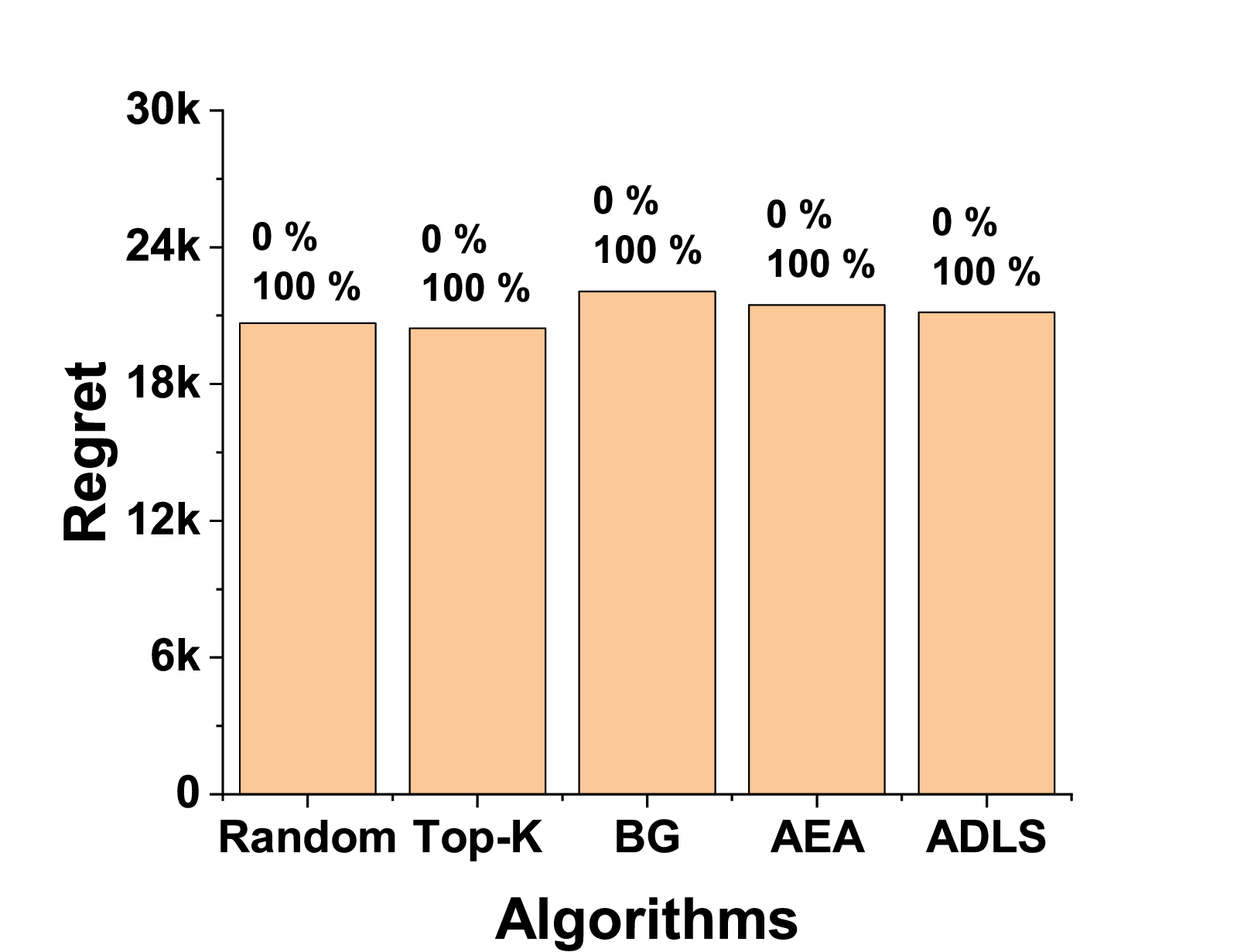} &
    \includegraphics[scale=0.10]{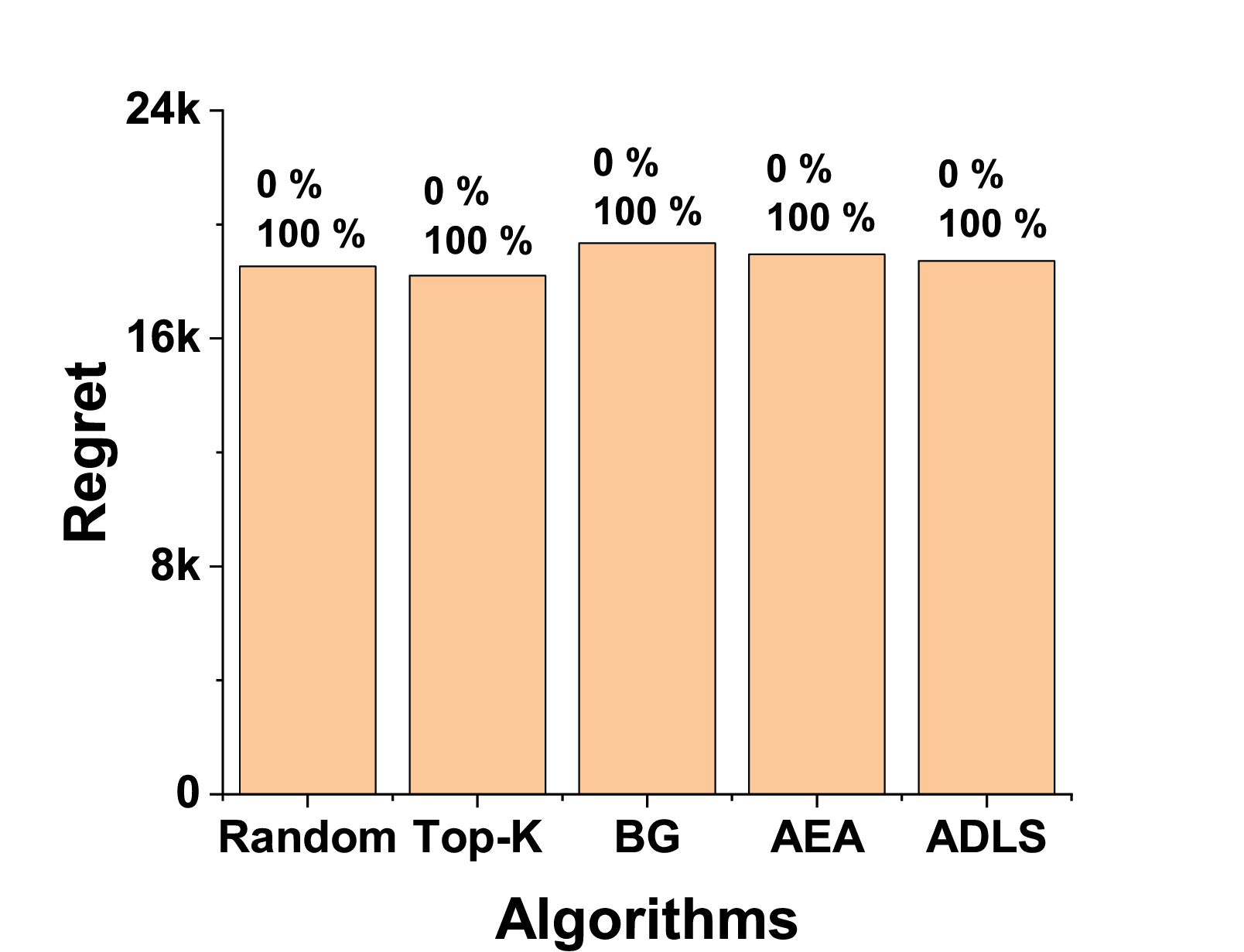} \\
    (a) $\lambda$ = $40\%$ & 
    (b) $\lambda$ = $60\%$ & 
    (c) $\lambda$ = $80\%$ & 
    (d) $\lambda$ = $100\%$ & 
    (e) $\lambda$ = $120\%$ \\
\end{tabular}
\caption{Regret of varying the demand-supply ratio $\lambda$ when $\omega = 70\%$ (Email-Eu-Core, $|\mathcal{A}| = 10$, $p_{c}: Trivalency$)}
\label{Fig:2_Influence_4}
\begin{tabular}{ccccc}
    \includegraphics[scale=0.10]{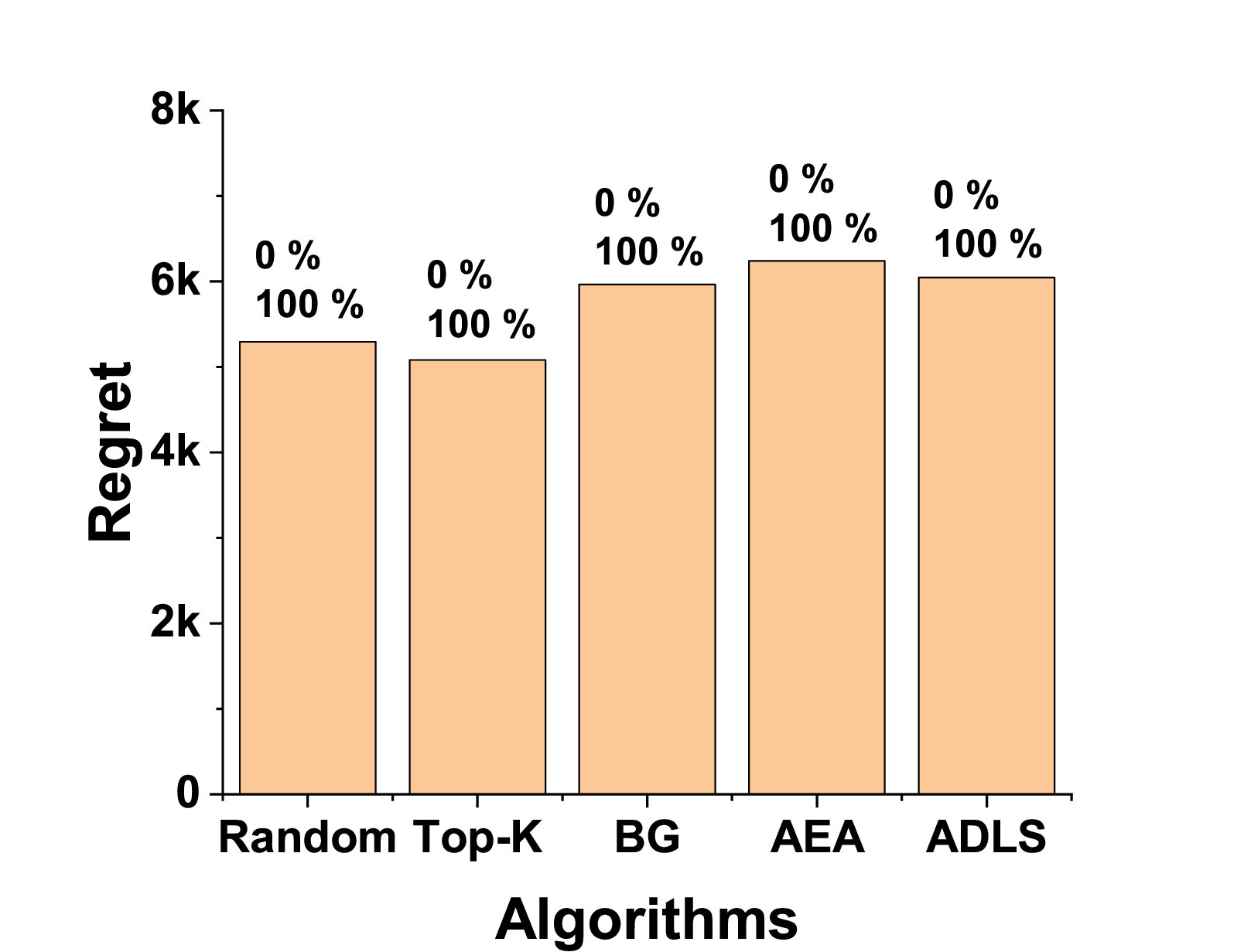} & 
    \includegraphics[scale=0.10]{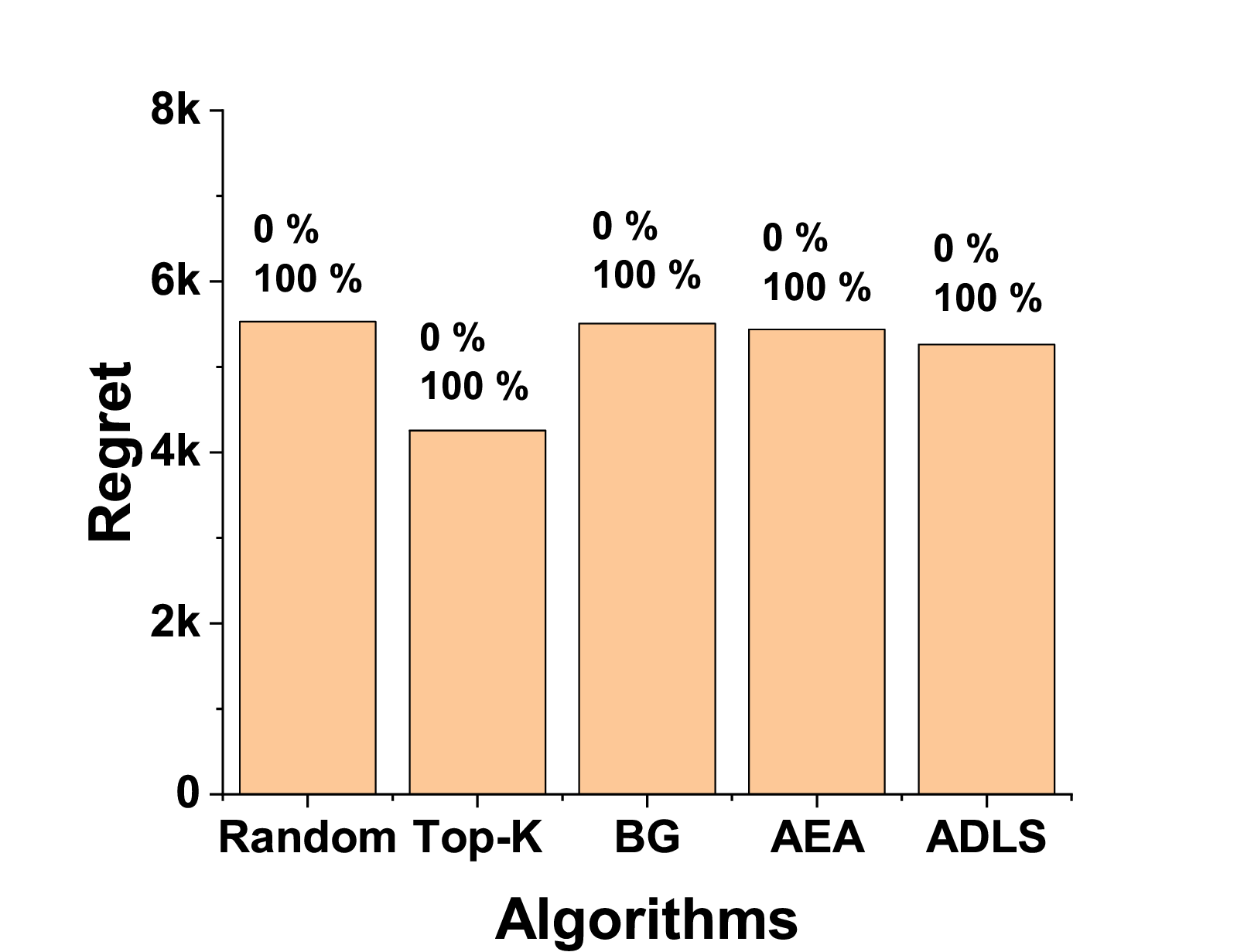} & 
    \includegraphics[scale=0.10]{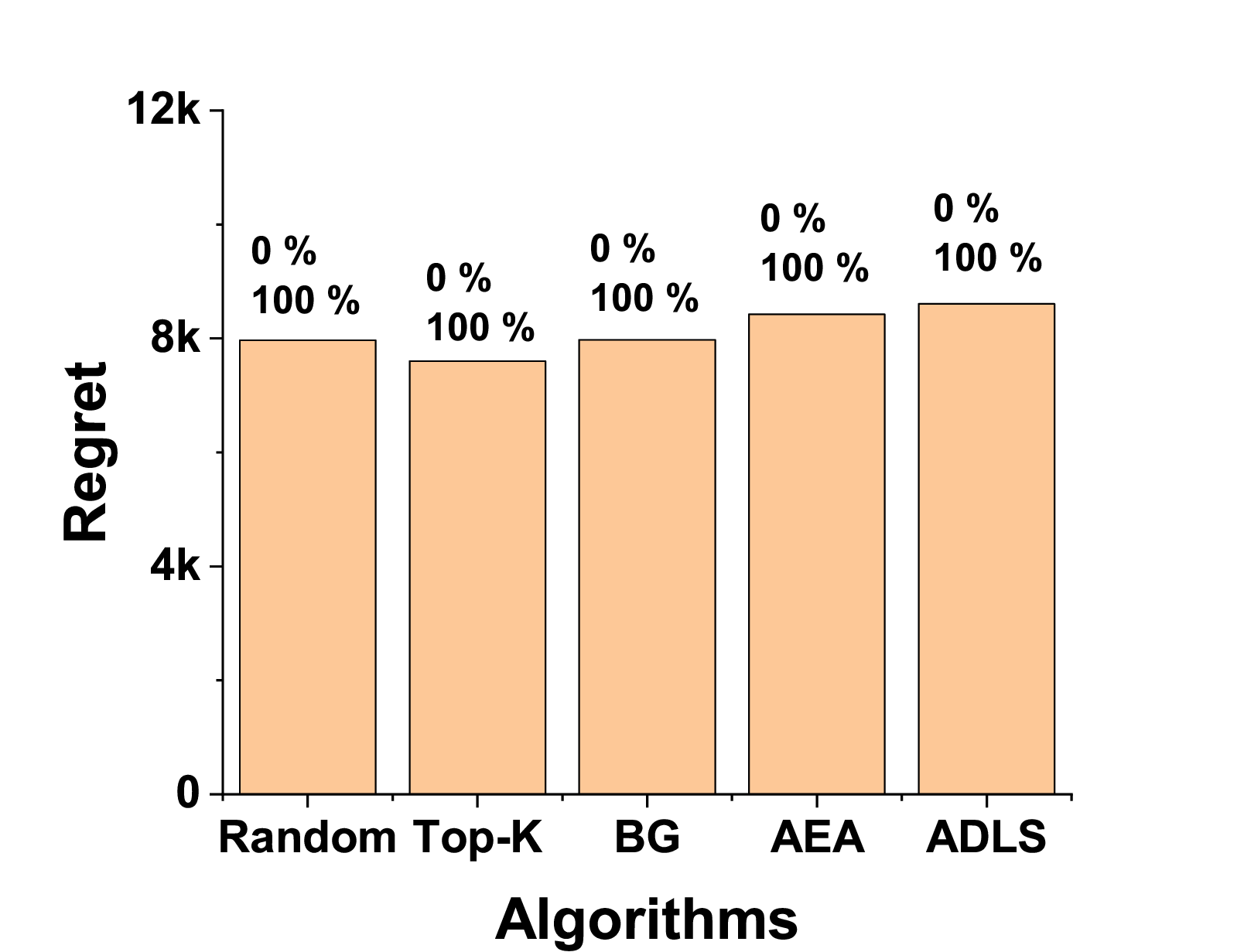} & 
    \includegraphics[scale=0.10]{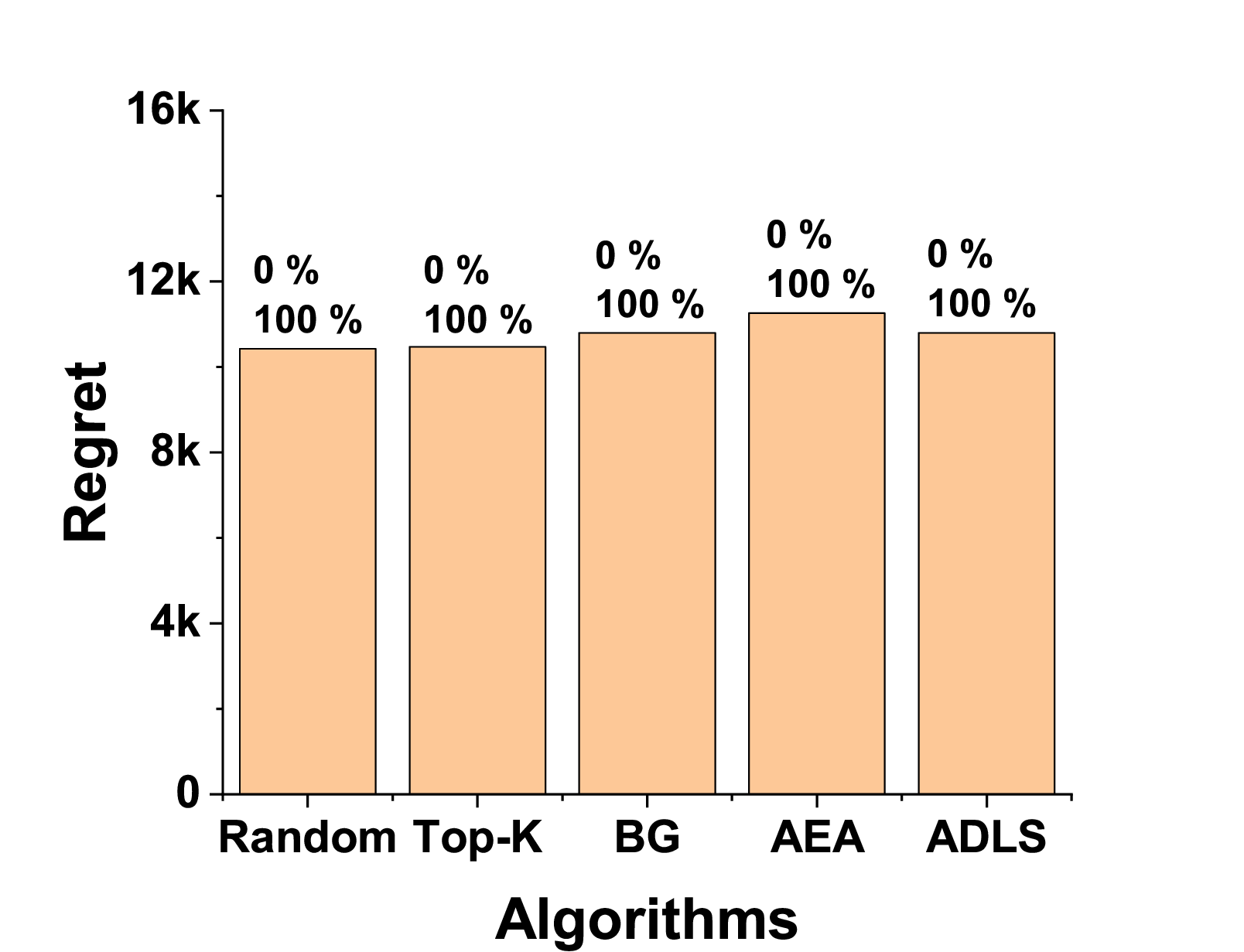} &
    \includegraphics[scale=0.10]{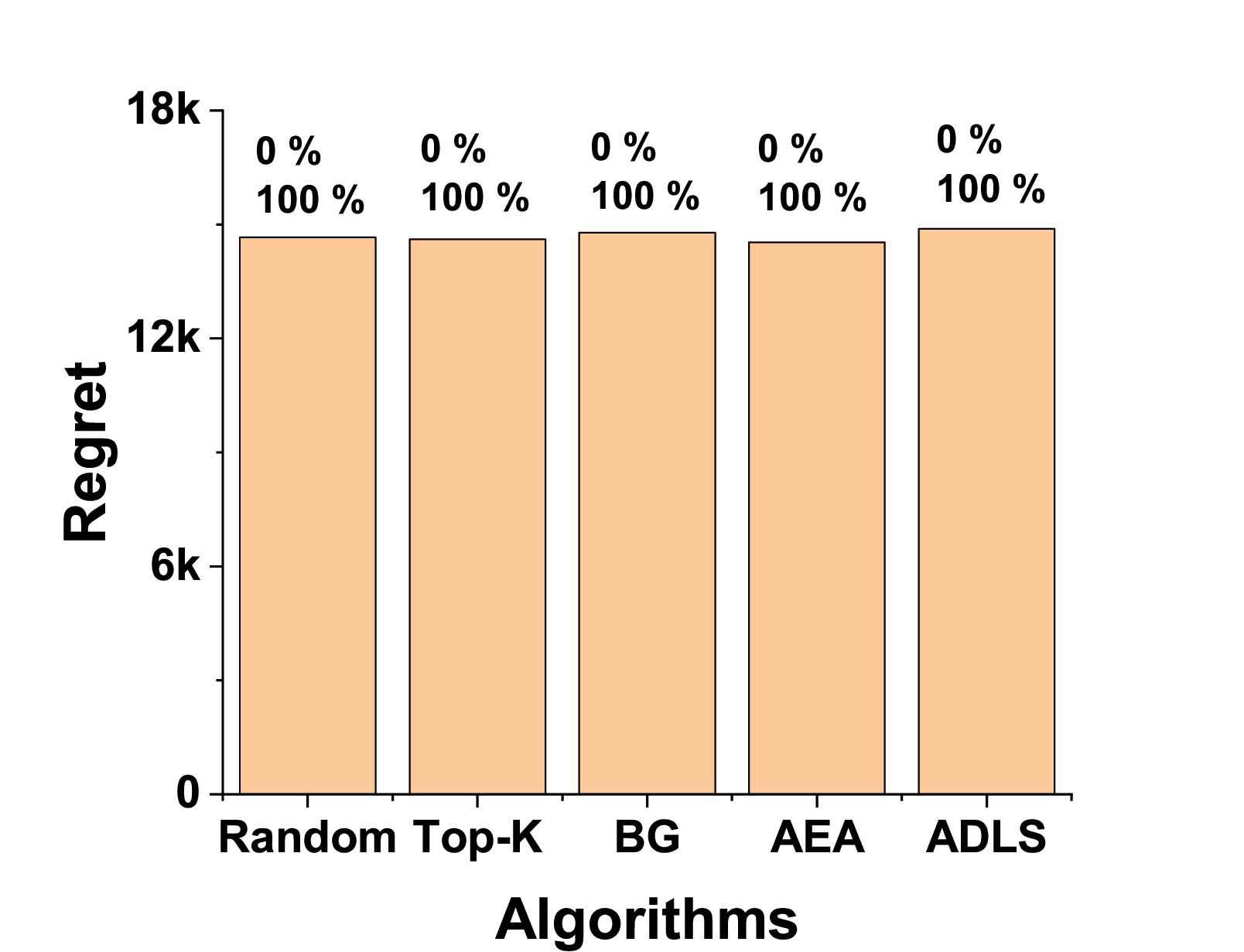} \\
    (a) $\lambda$ = $40\%$ & 
    (b) $\lambda$ = $60\%$ & 
    (c) $\lambda$ = $80\%$ & 
    (d) $\lambda$ = $100\%$ & 
    (e) $\lambda$ = $120\%$ \\
\end{tabular}
\caption{Regret of varying the demand-supply ratio $\lambda$ when $\omega = 90\%$ (Email-Eu-Core, $|\mathcal{A}| = 5$, $p_{c}: Trivalency$)}
\label{Fig:2_Influence_5}
\end{figure}

\begin{figure}[H]
\centering
\begin{tabular}{lclc}
    Excessive Regret & \includegraphics[width=0.11\linewidth]{Excessive.png} &
    Unsatisfied Regret & \includegraphics[width=0.11\linewidth]{Unsatisfied.png}\\
\end{tabular}
\begin{tabular}{ccccc}
    \includegraphics[scale=0.10]{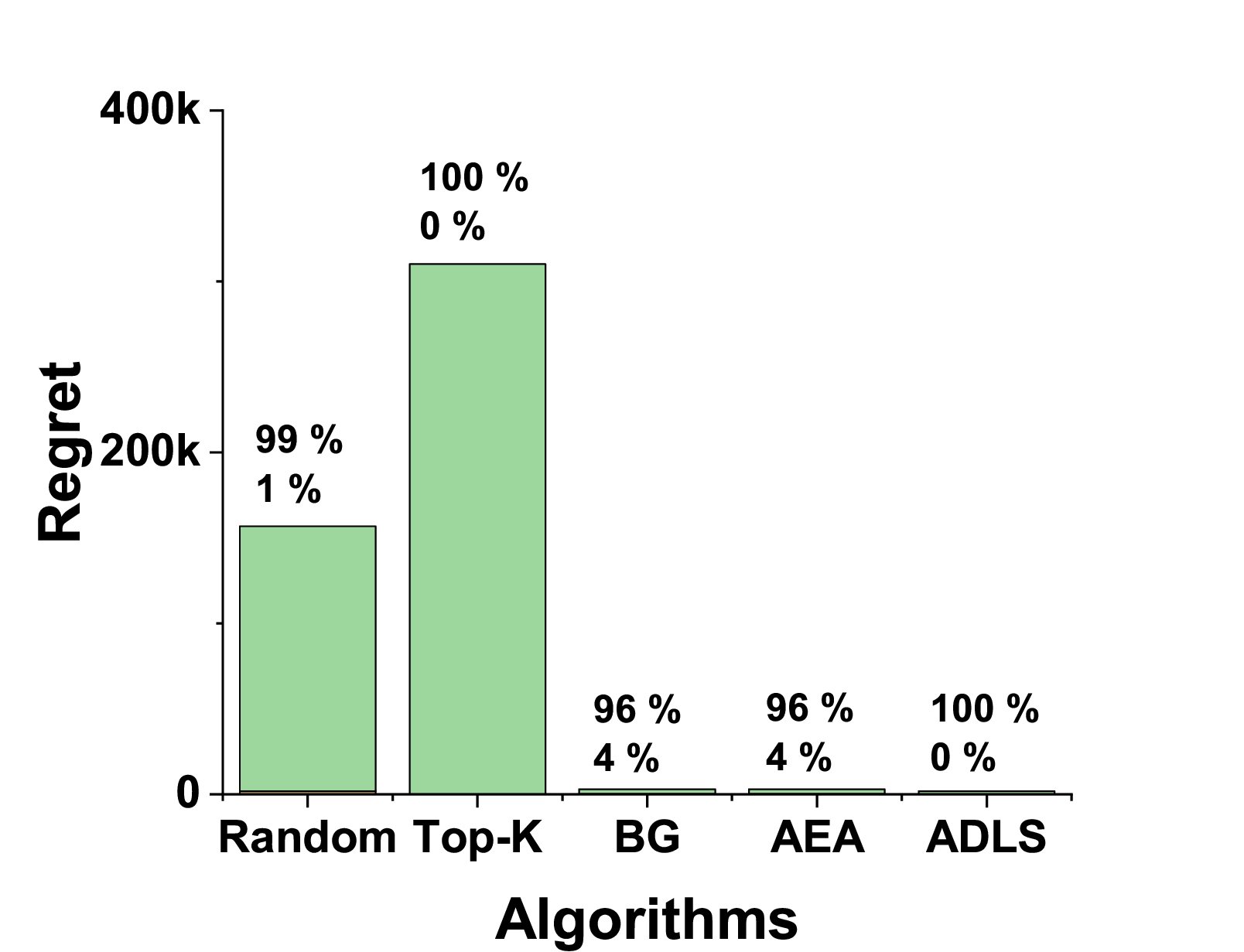} & 
    \includegraphics[scale=0.10]{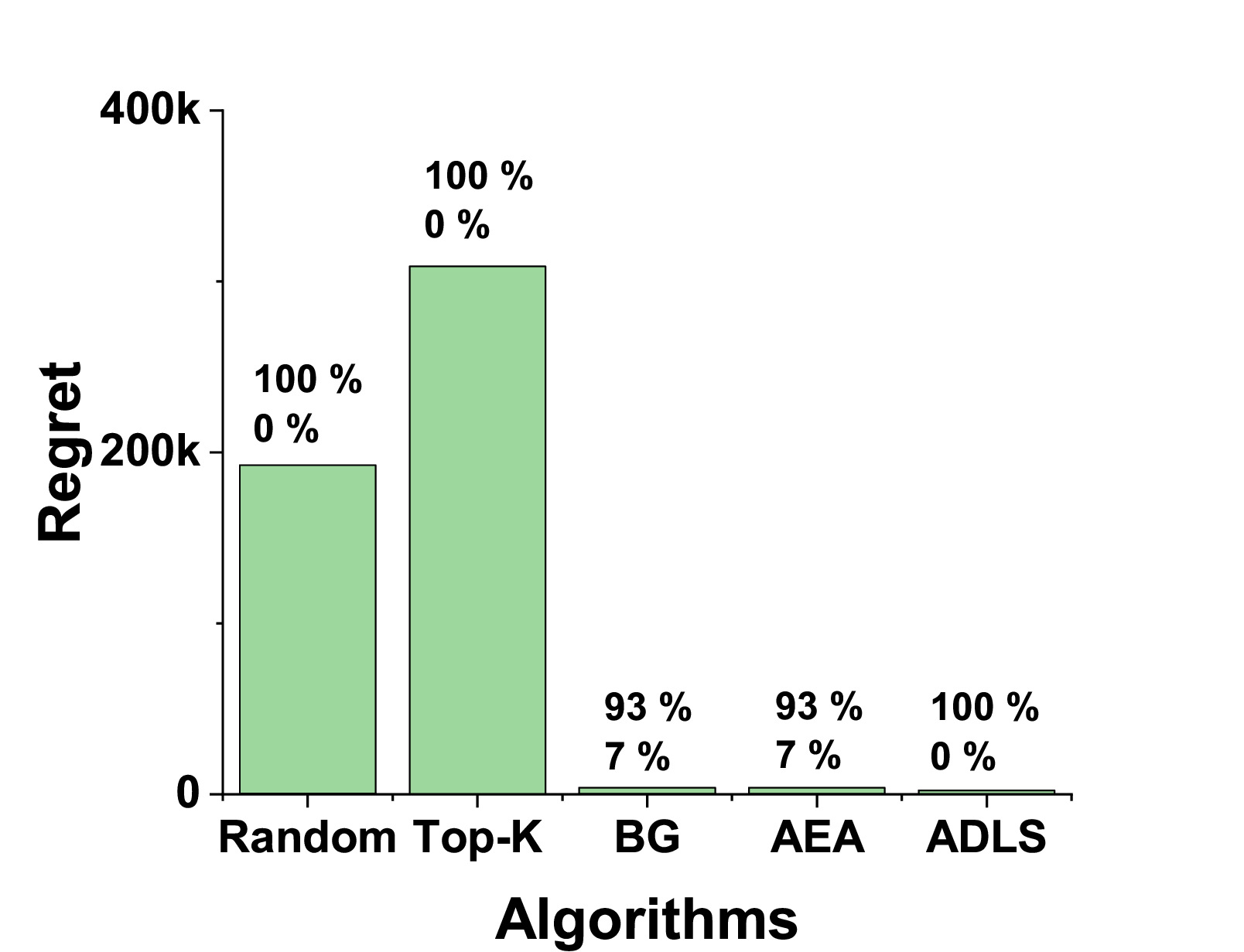} & 
    \includegraphics[scale=0.10]{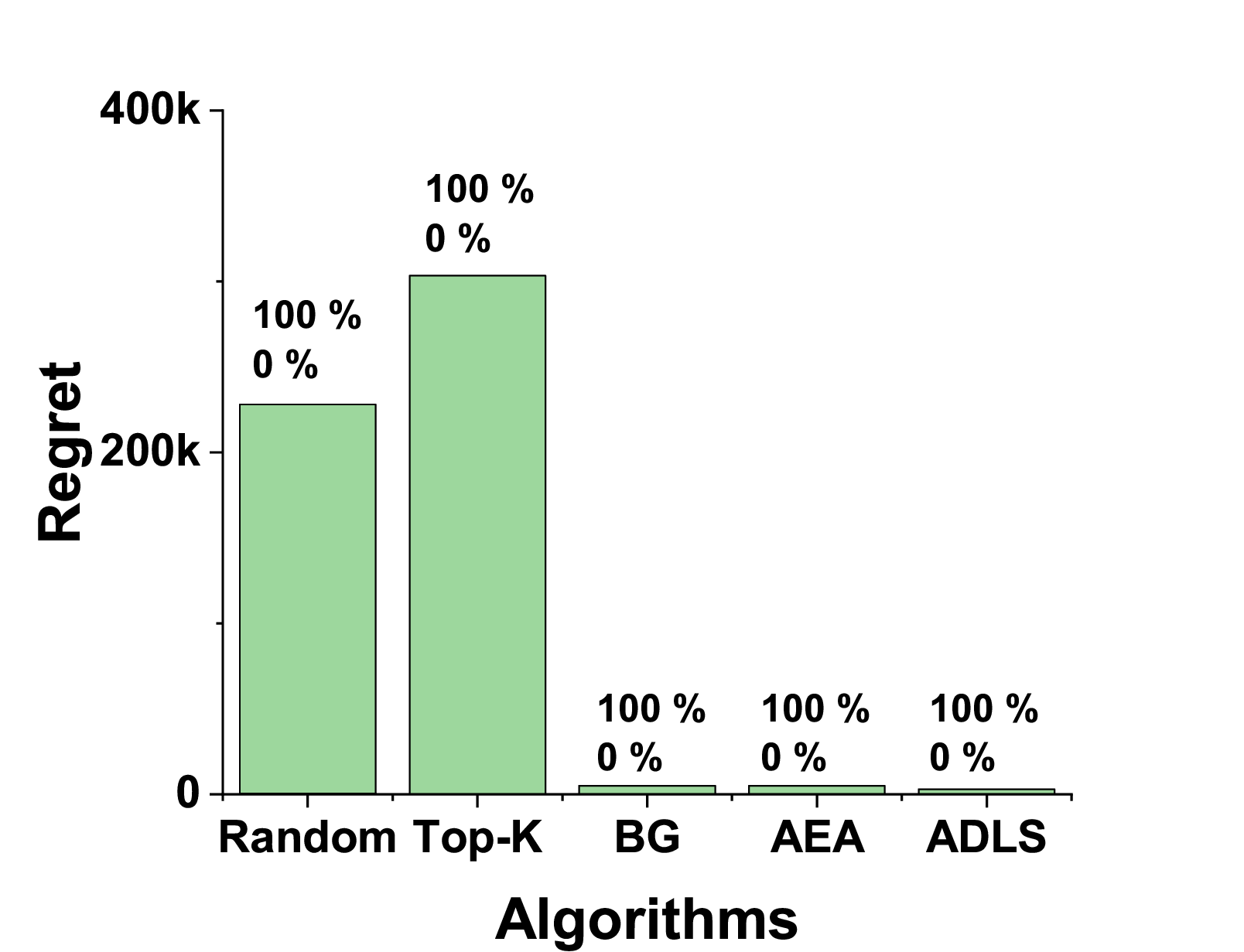} & 
    \includegraphics[scale=0.10]{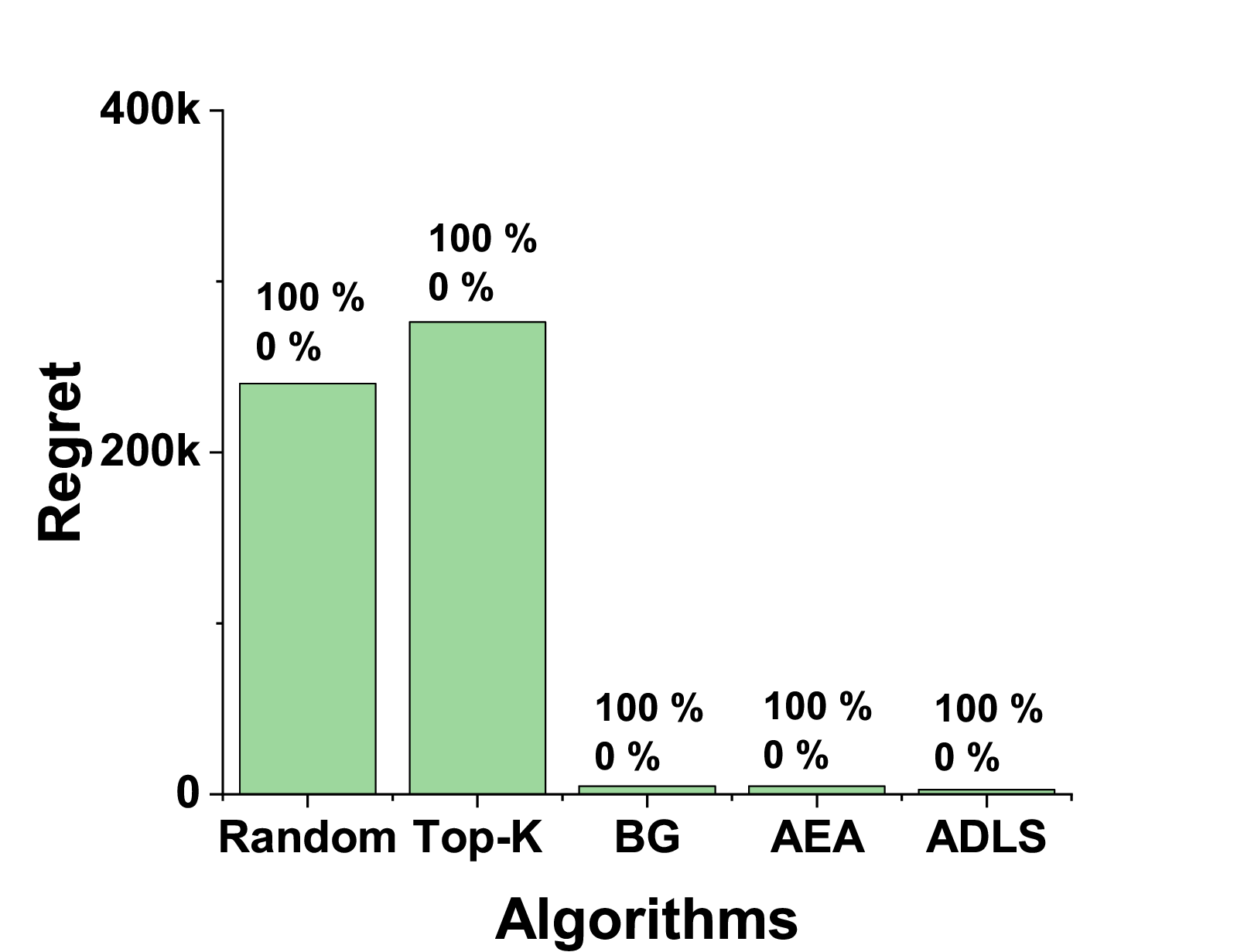} &
    \includegraphics[scale=0.10]{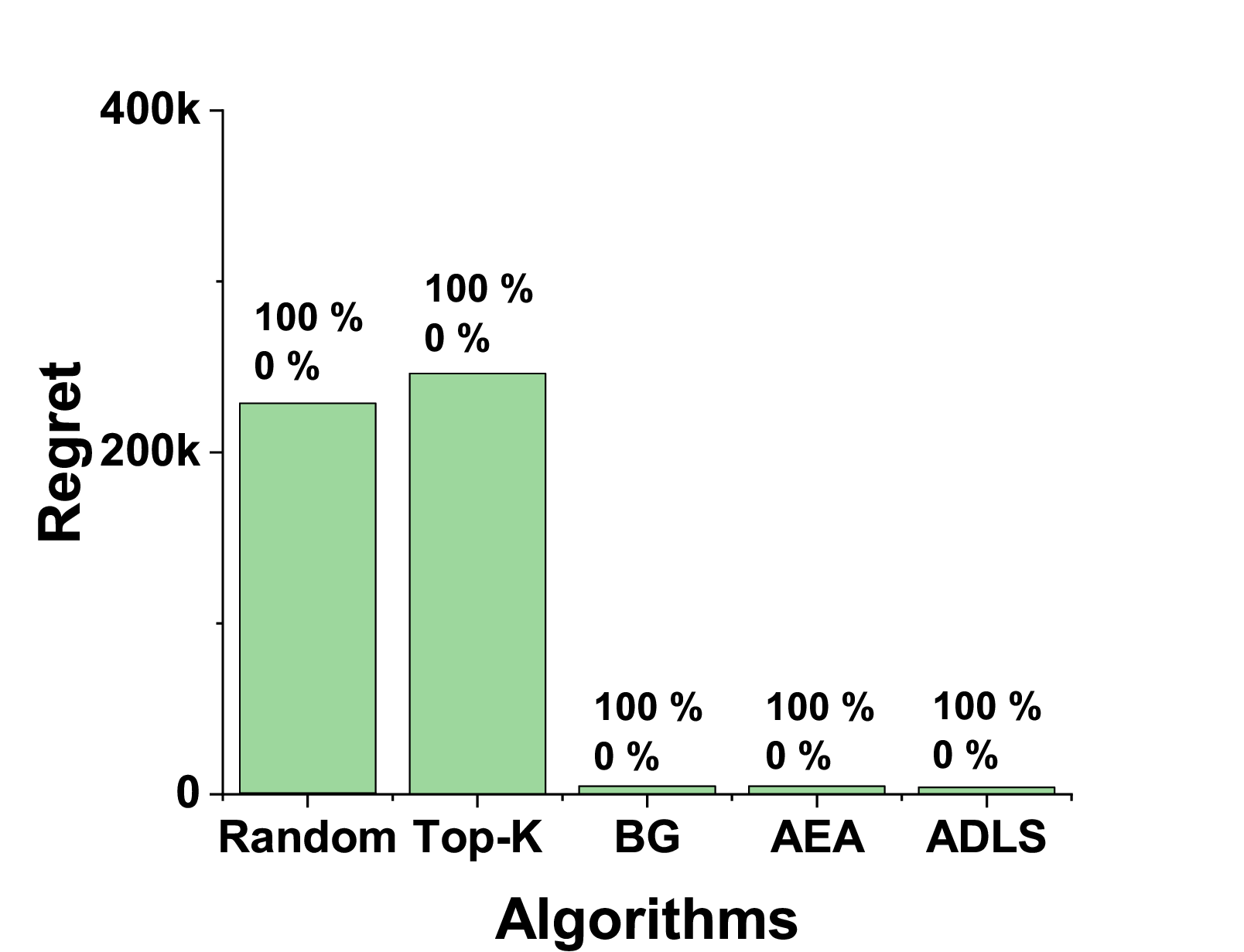} \\
    (a) $\lambda$ = $40\%$ & 
    (b) $\lambda$ = $60\%$ & 
    (c) $\lambda$ = $80\%$ & 
    (d) $\lambda$ = $100\%$ & 
    (e) $\lambda$ = $120\%$ \\
\end{tabular}
\caption{Regret of varying the demand-supply ratio $\lambda$ when $\omega = 10\%$ (Facebook, $|\mathcal{A}| = 100$, $p_{c}: Trivalency$)}
\label{Fig:4_Influence_1}
\begin{tabular}{ccccc}
    \includegraphics[scale=0.10]{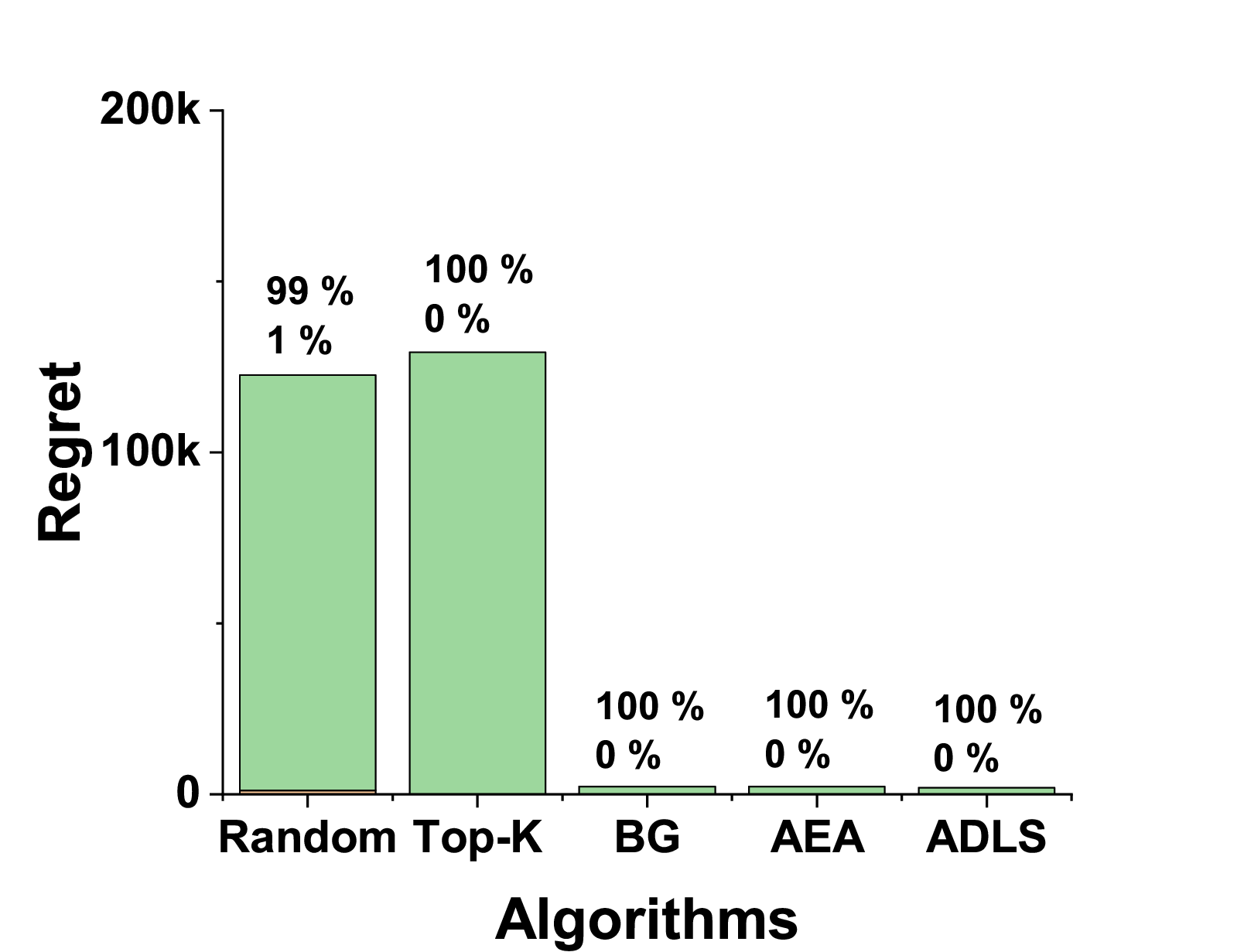} & 
    \includegraphics[scale=0.10]{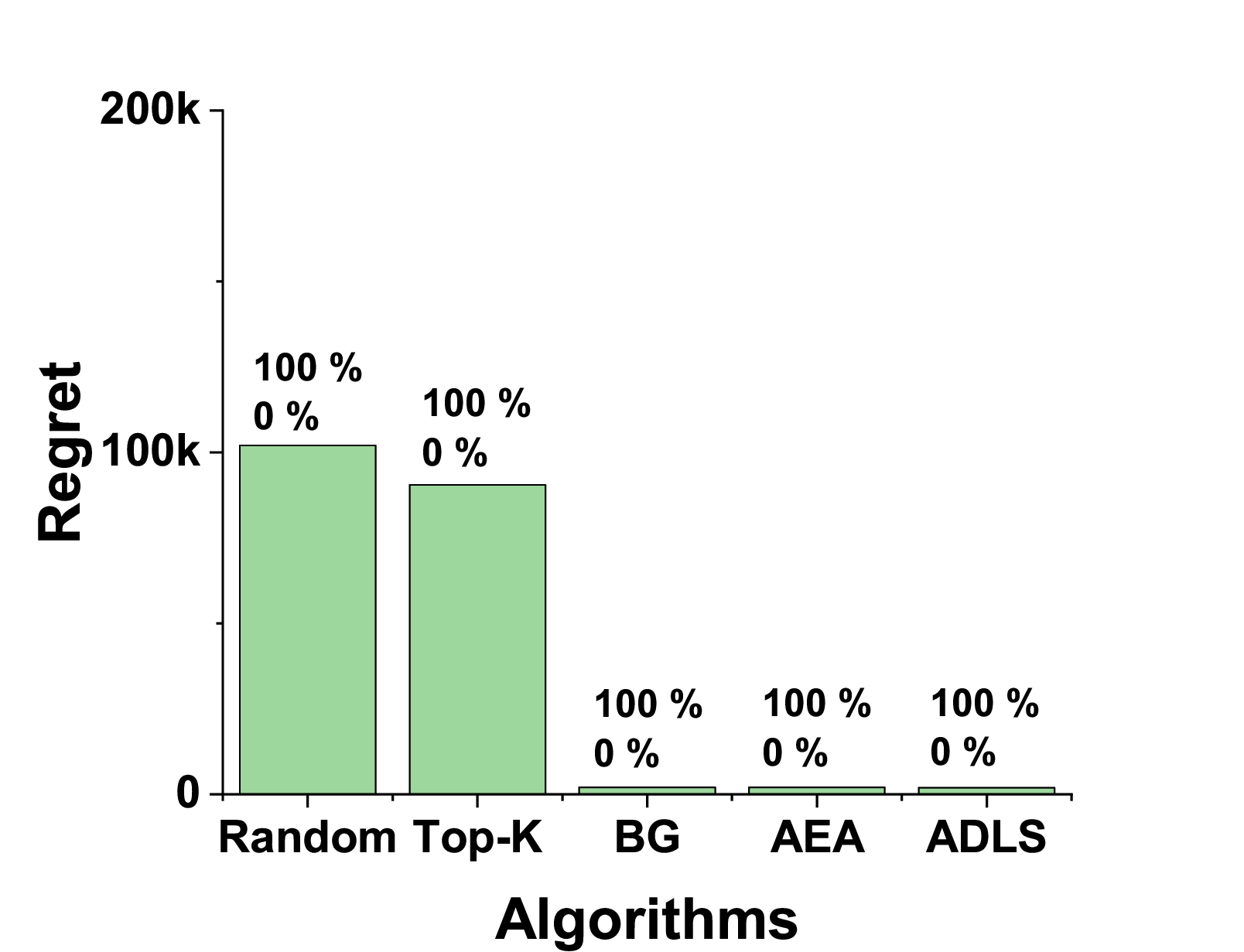} & 
    \includegraphics[scale=0.10]{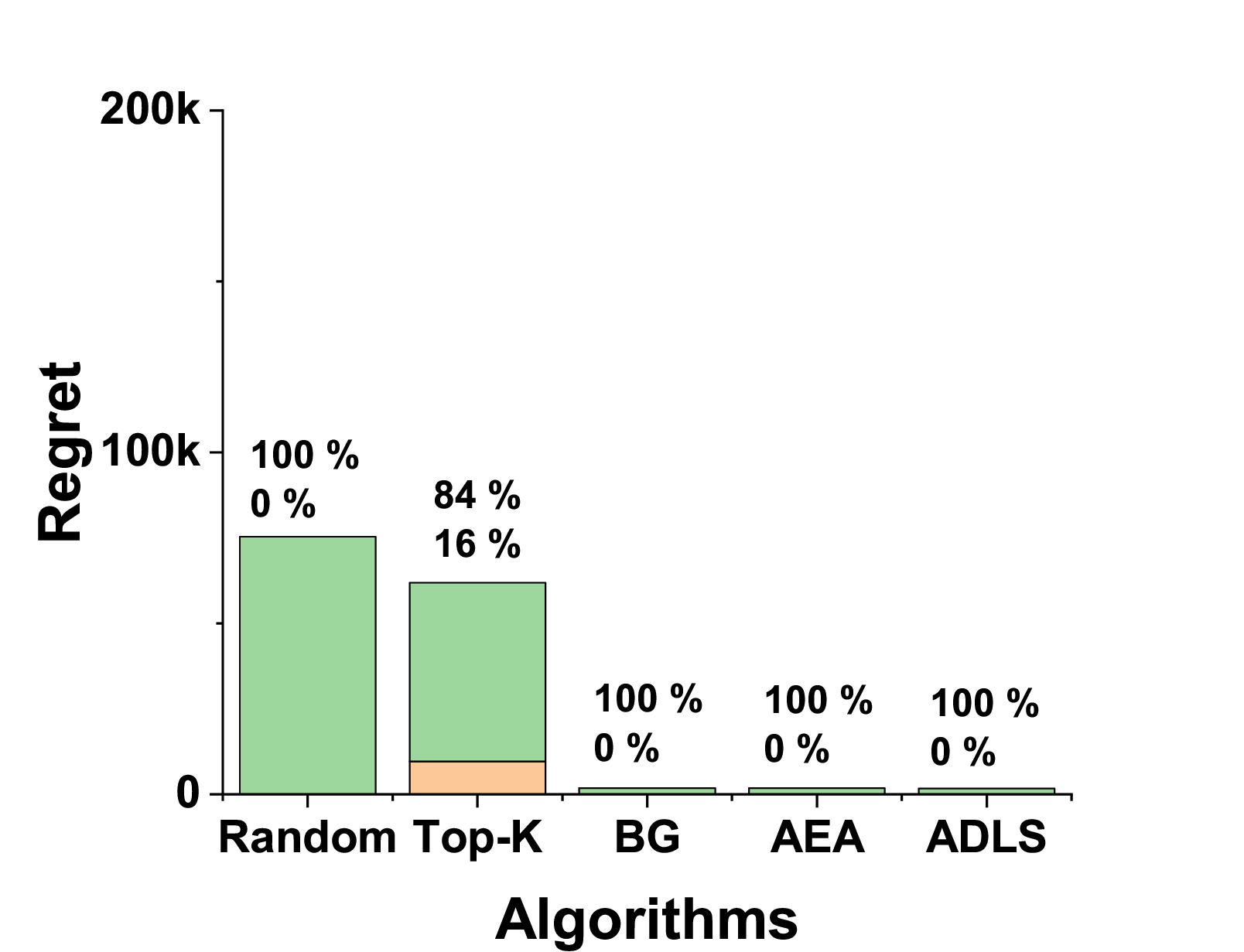} & 
    \includegraphics[scale=0.10]{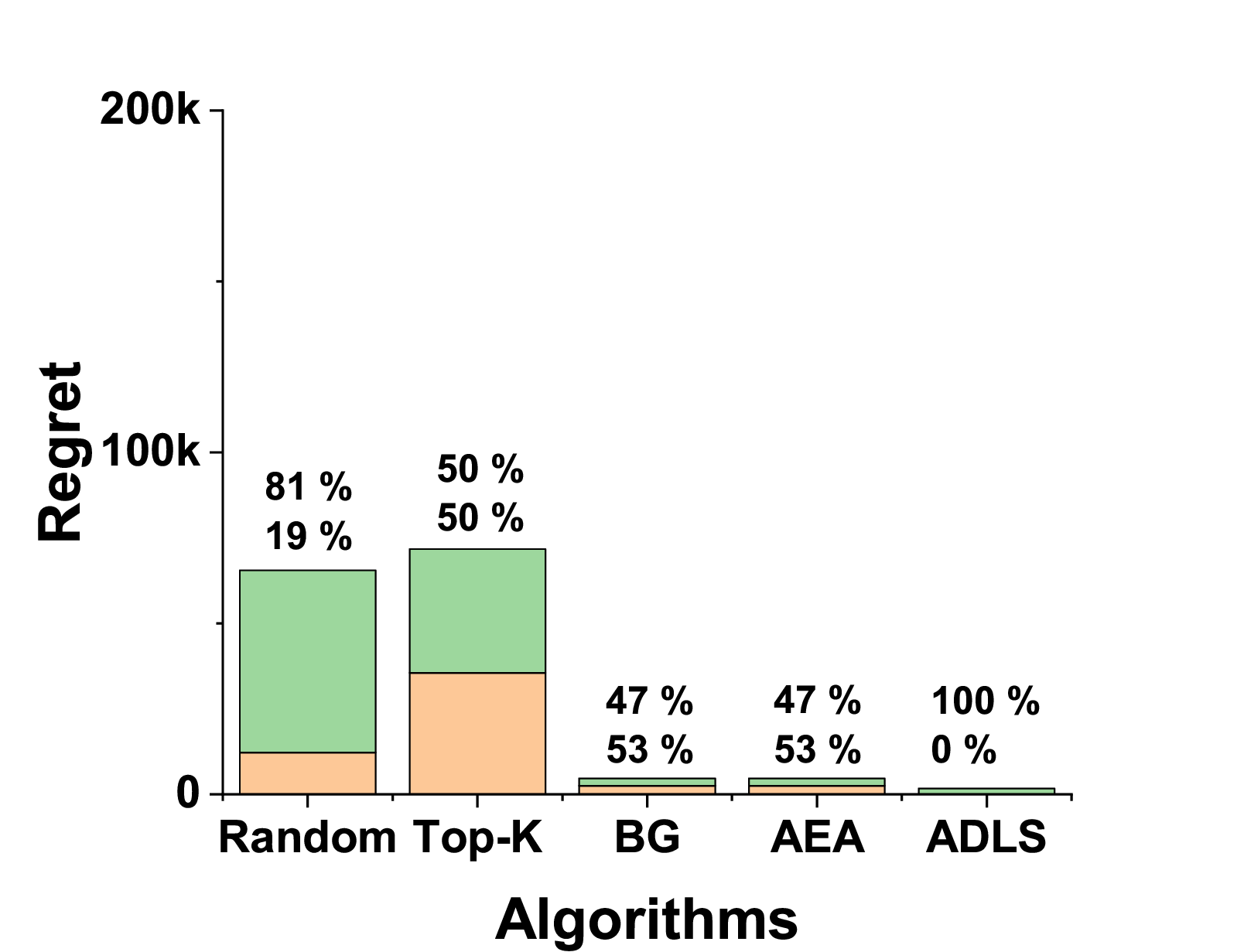} &
    \includegraphics[scale=0.10]{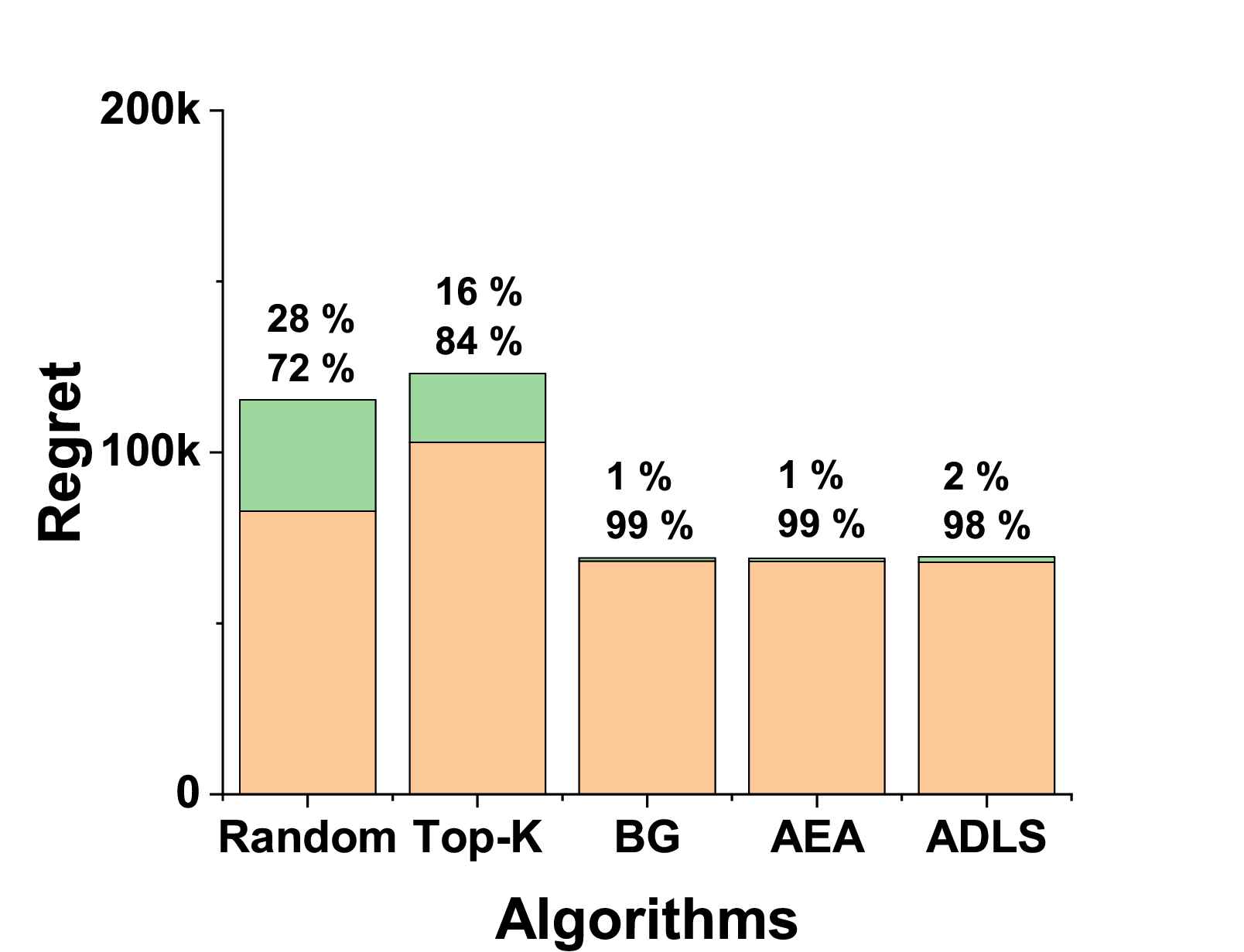} \\
    (a) $\lambda$ = $40\%$ & 
    (b) $\lambda$ = $60\%$ & 
    (c) $\lambda$ = $80\%$ &
    (d) $\lambda$ = $100\%$ & 
    (e) $\lambda$ = $120\%$ \\
\end{tabular}
\caption{Regret of varying the demand-supply ratio $\lambda$ when $\omega = 30\%$ (Facebook, $|\mathcal{A}| = 50$, $p_{c}: Trivalency$)}
\label{Fig:4_Influence_2}
\begin{tabular}{ccccc}
    \includegraphics[scale=0.10]{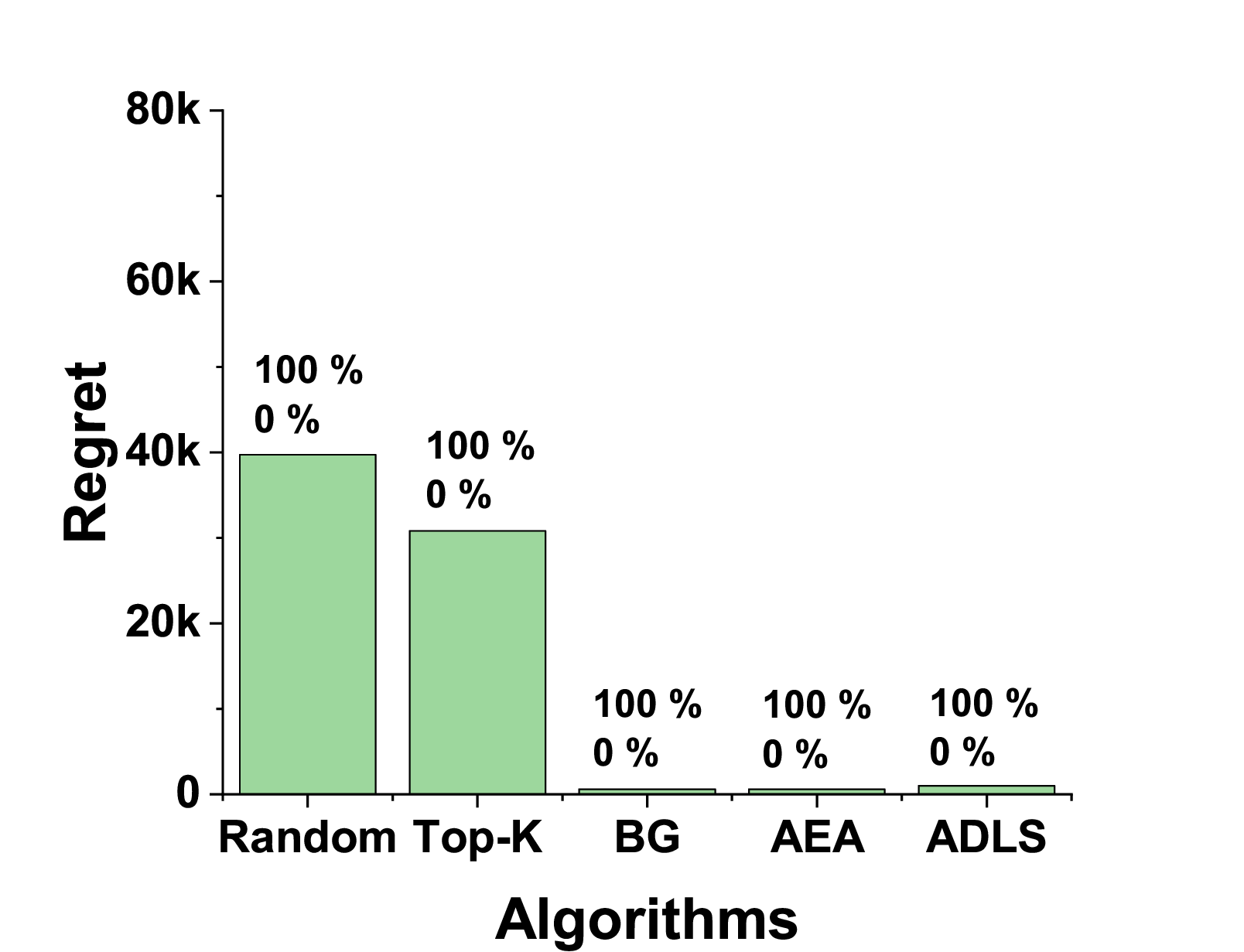} & 
    \includegraphics[scale=0.10]{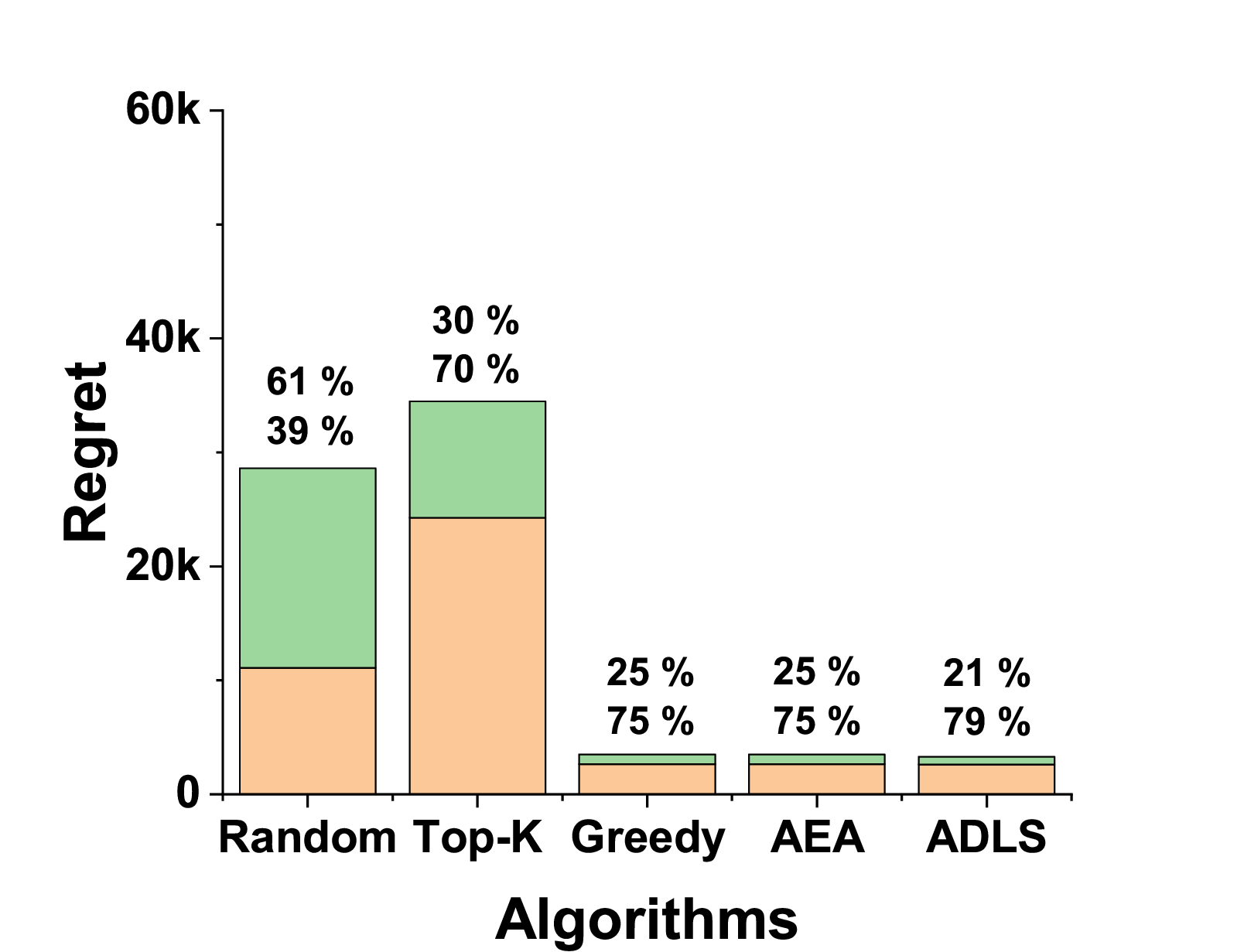} & 
    \includegraphics[scale=0.10]{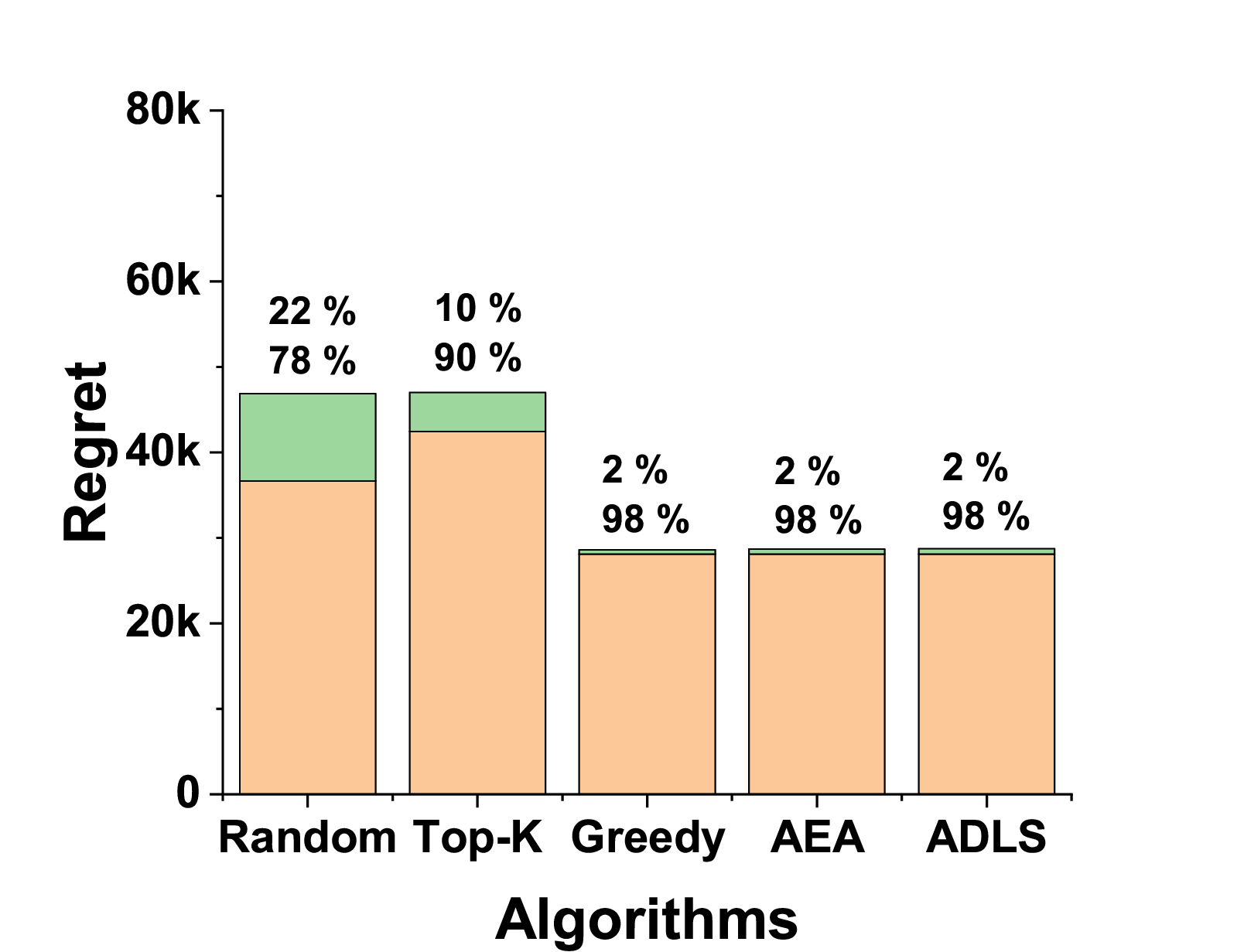} & 
    \includegraphics[scale=0.10]{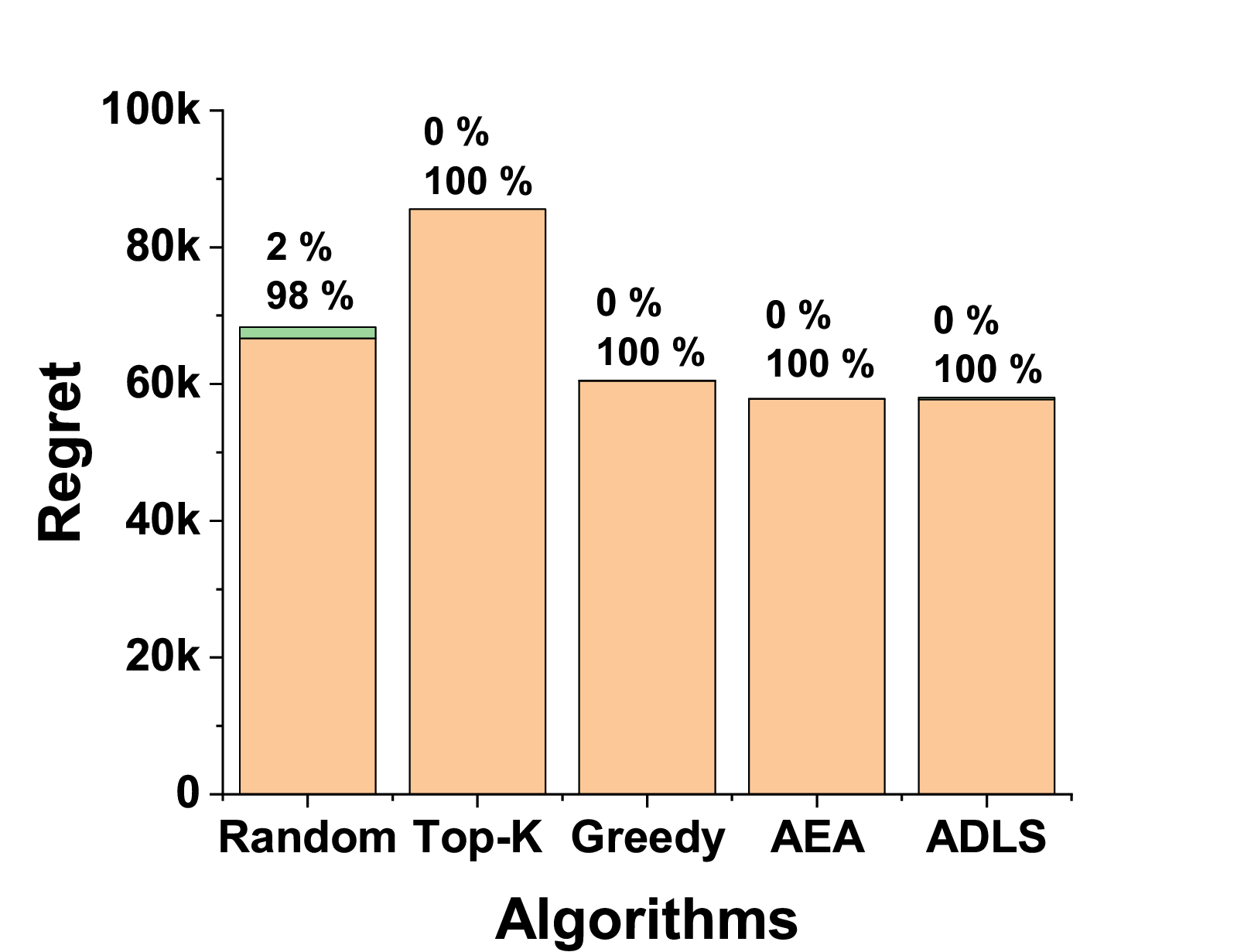} &
    \includegraphics[scale=0.10]{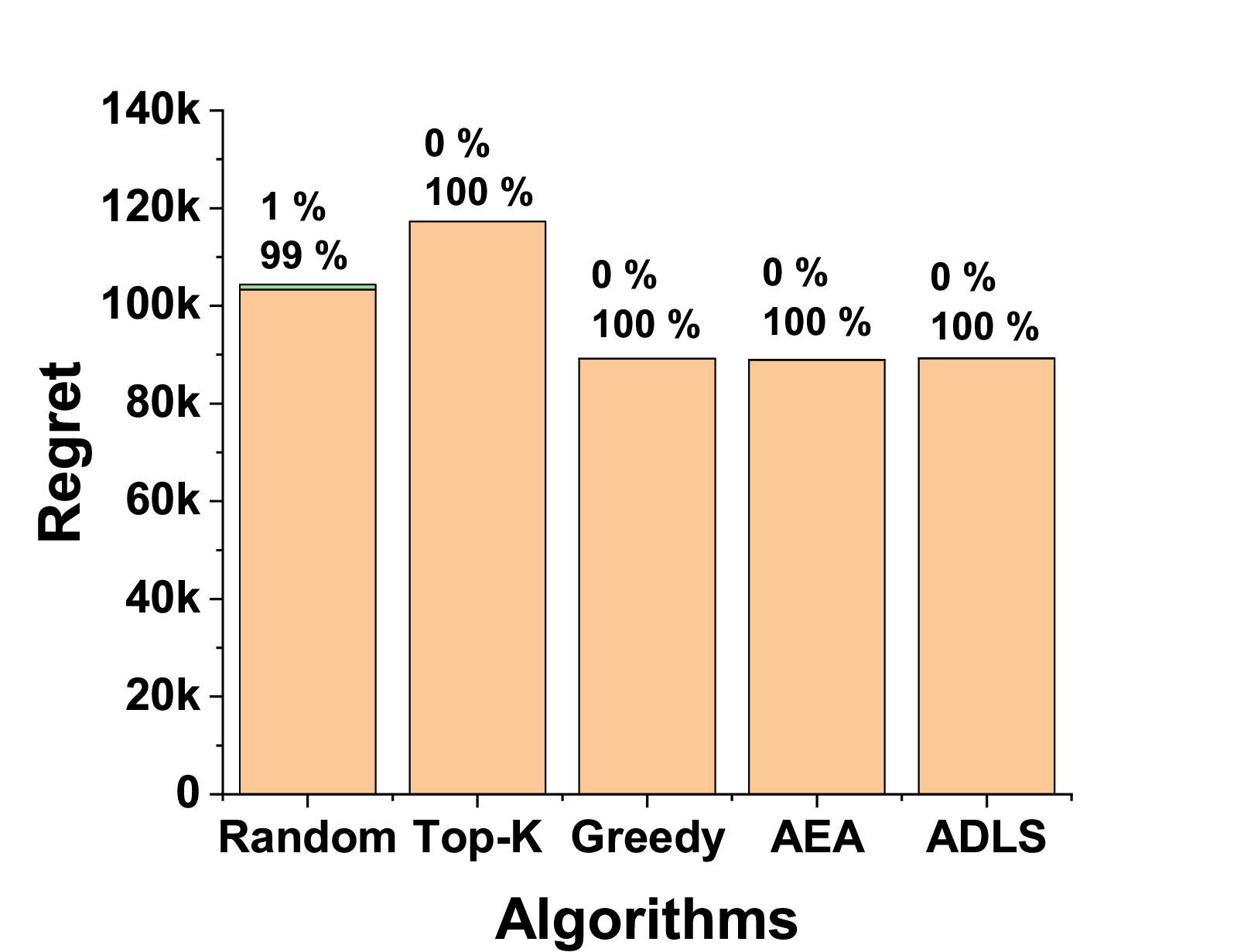} \\
    (a) $\lambda$ = $40\%$ & 
    (b) $\lambda$ = $60\%$ & 
    (c) $\lambda$ = $80\%$ & 
    (d) $\lambda$ = $100\%$ & 
    (e) $\lambda$ = $120\%$ \\
\end{tabular}
\caption{Regret of varying the demand-supply ratio $\lambda$ when $\omega = 50\%$ (Facebook, $|\mathcal{A}| = 20$, $p_{c}: Trivalency$)}
\label{Fig:4_Influence_3}
\begin{tabular}{ccccc}
    \includegraphics[scale=0.10]{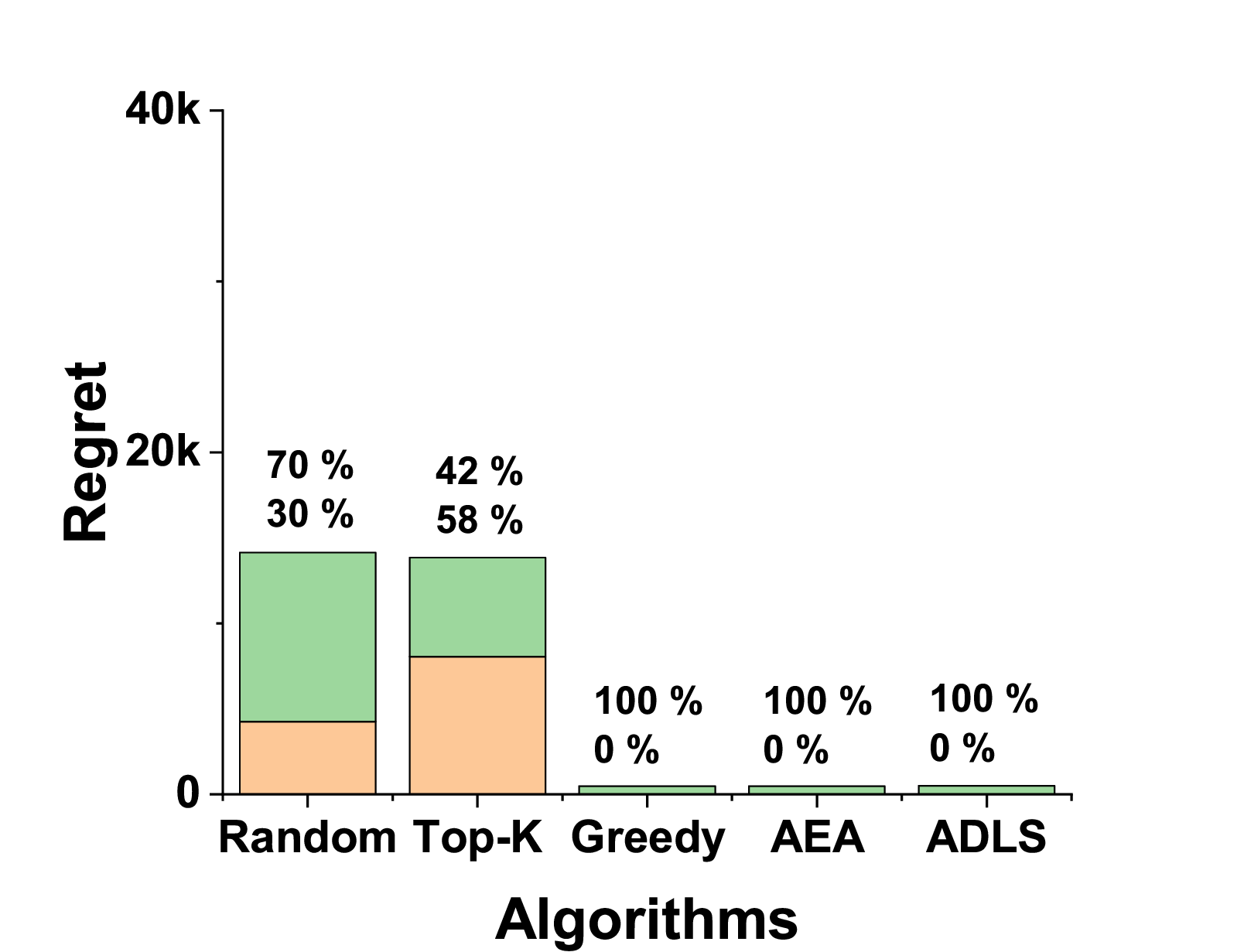} & 
    \includegraphics[scale=0.10]{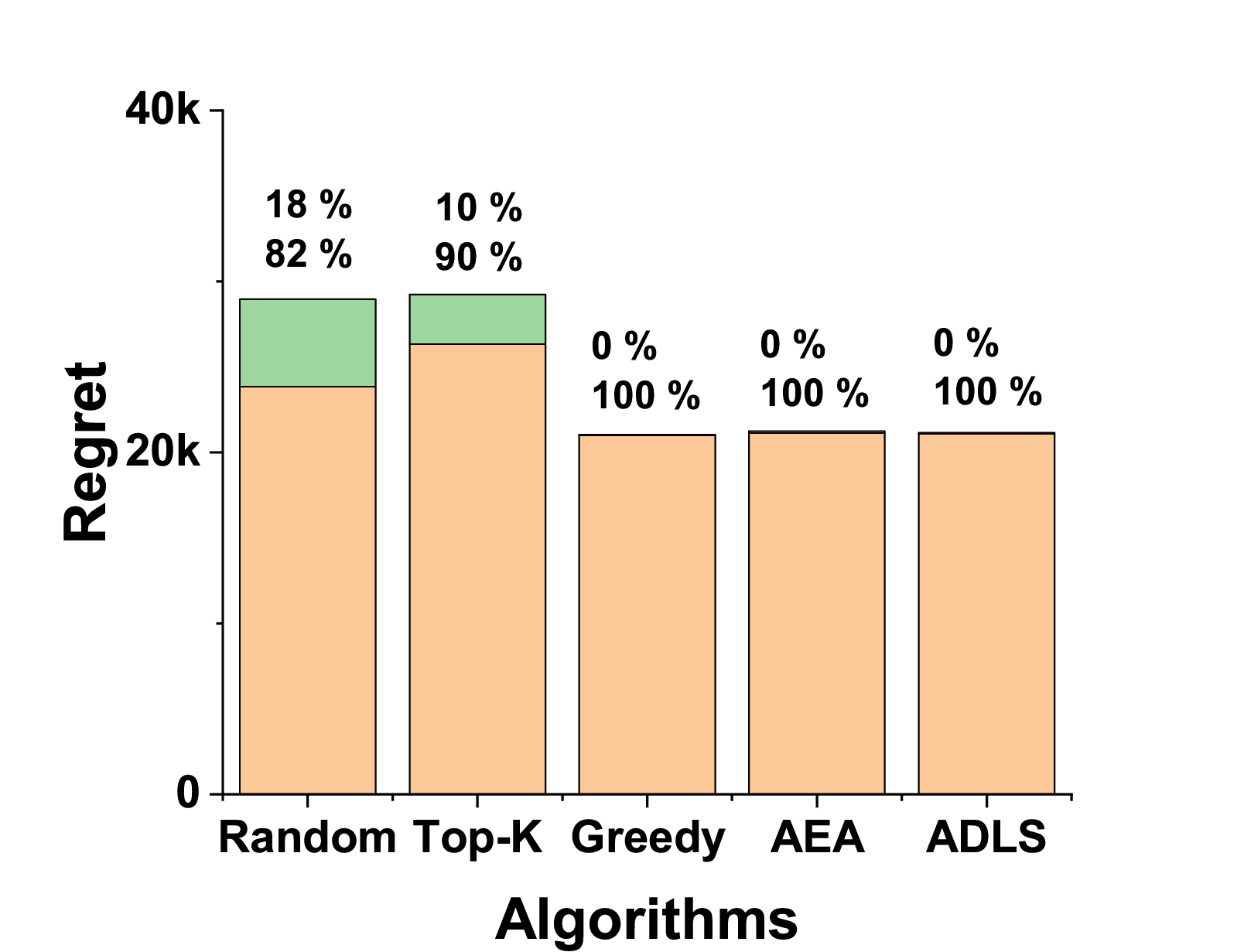} & 
    \includegraphics[scale=0.10]{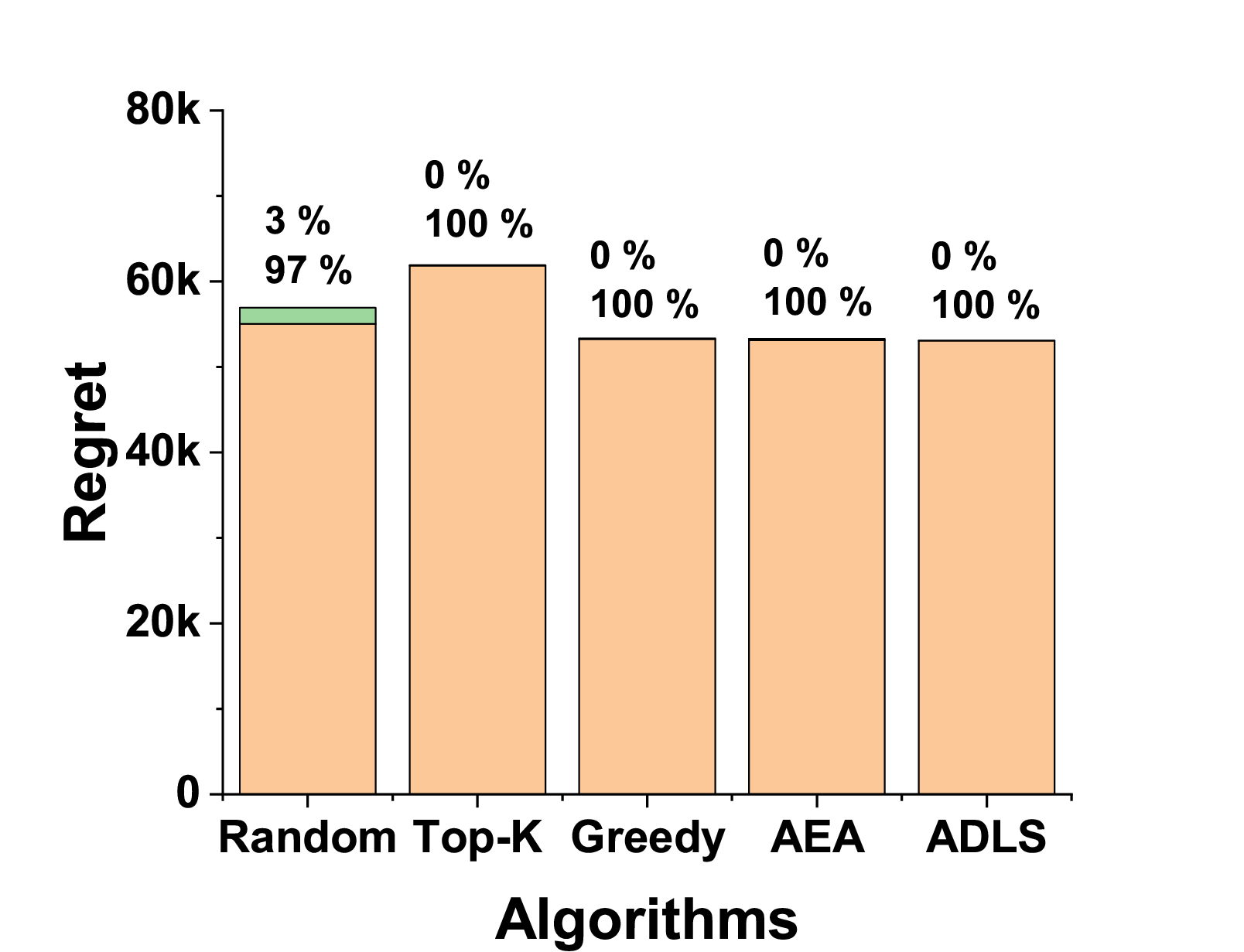} & 
    \includegraphics[scale=0.10]{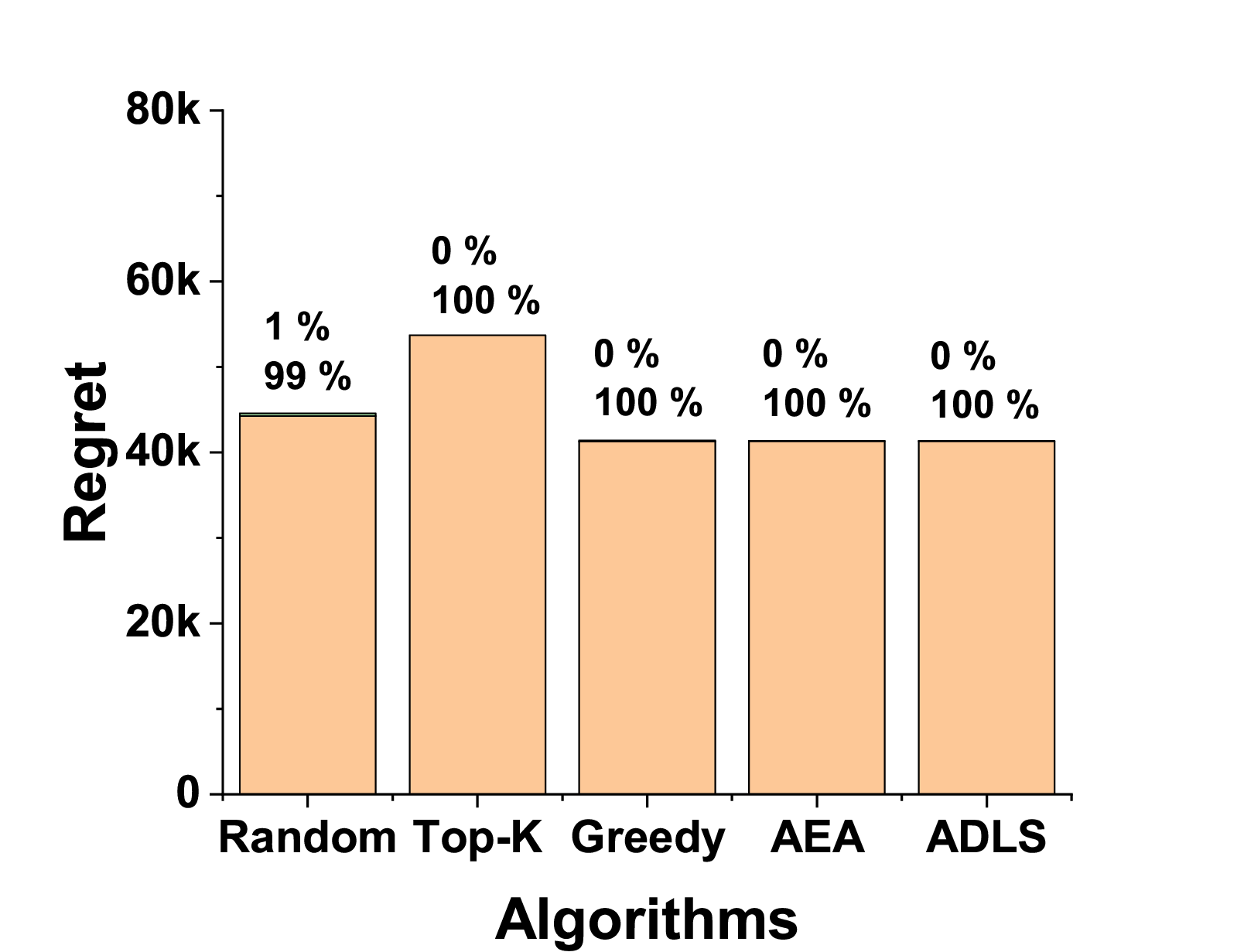} &
    \includegraphics[scale=0.10]{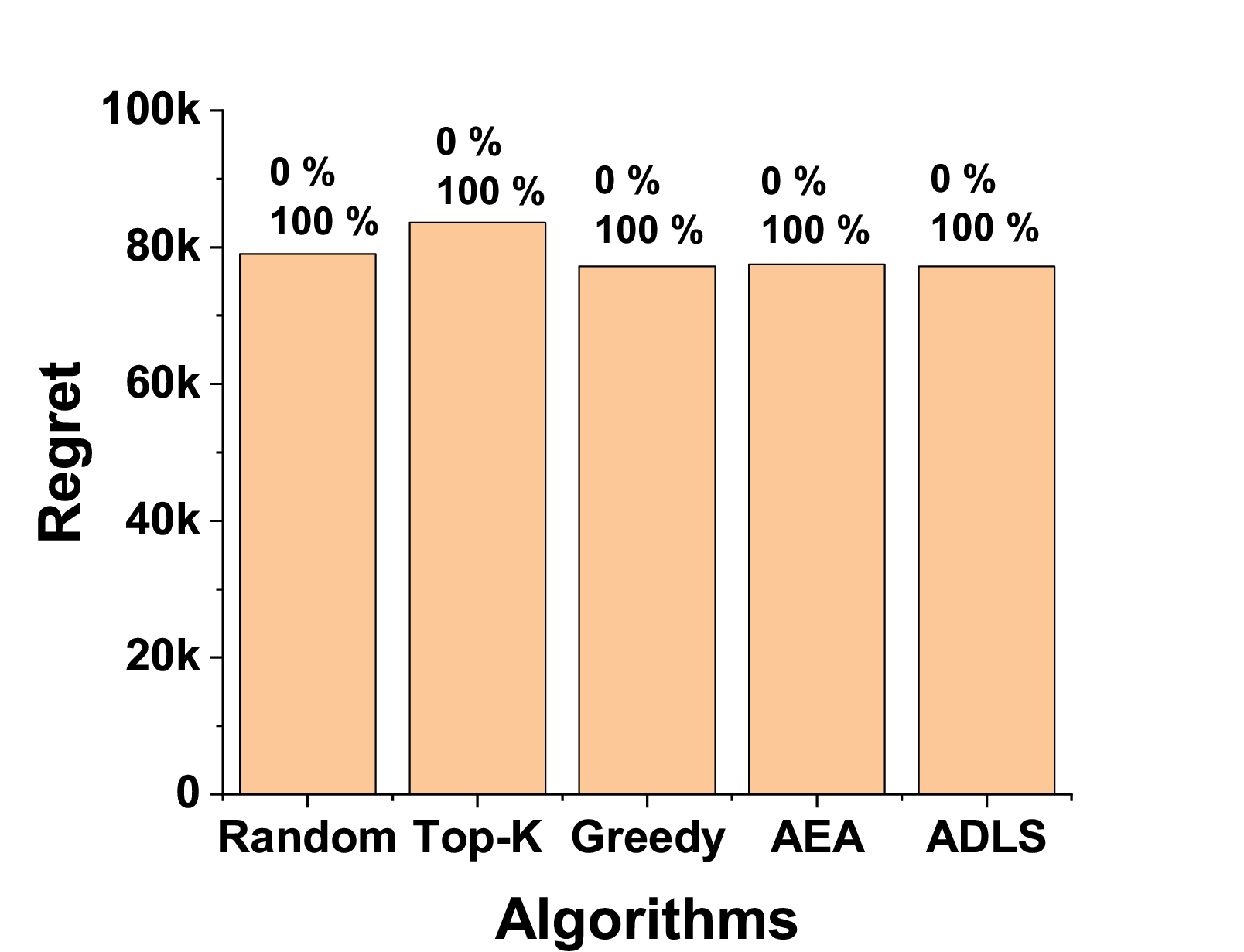} \\
    (a) $\lambda$ = $40\%$ & 
    (b) $\lambda$ = $60\%$ & 
    (c) $\lambda$ = $80\%$ & 
    (d) $\lambda$ = $100\%$ & 
    (e) $\lambda$ = $120\%$ \\
\end{tabular}
\caption{Regret of varying the demand-supply ratio $\lambda$ when $\omega = 70\%$ (Facebook, $|\mathcal{A}| = 10$, $p_{c}: Trivalency$)}
\label{Fig:4_Influence_4}
\begin{tabular}{ccccc}
    \includegraphics[scale=0.10]{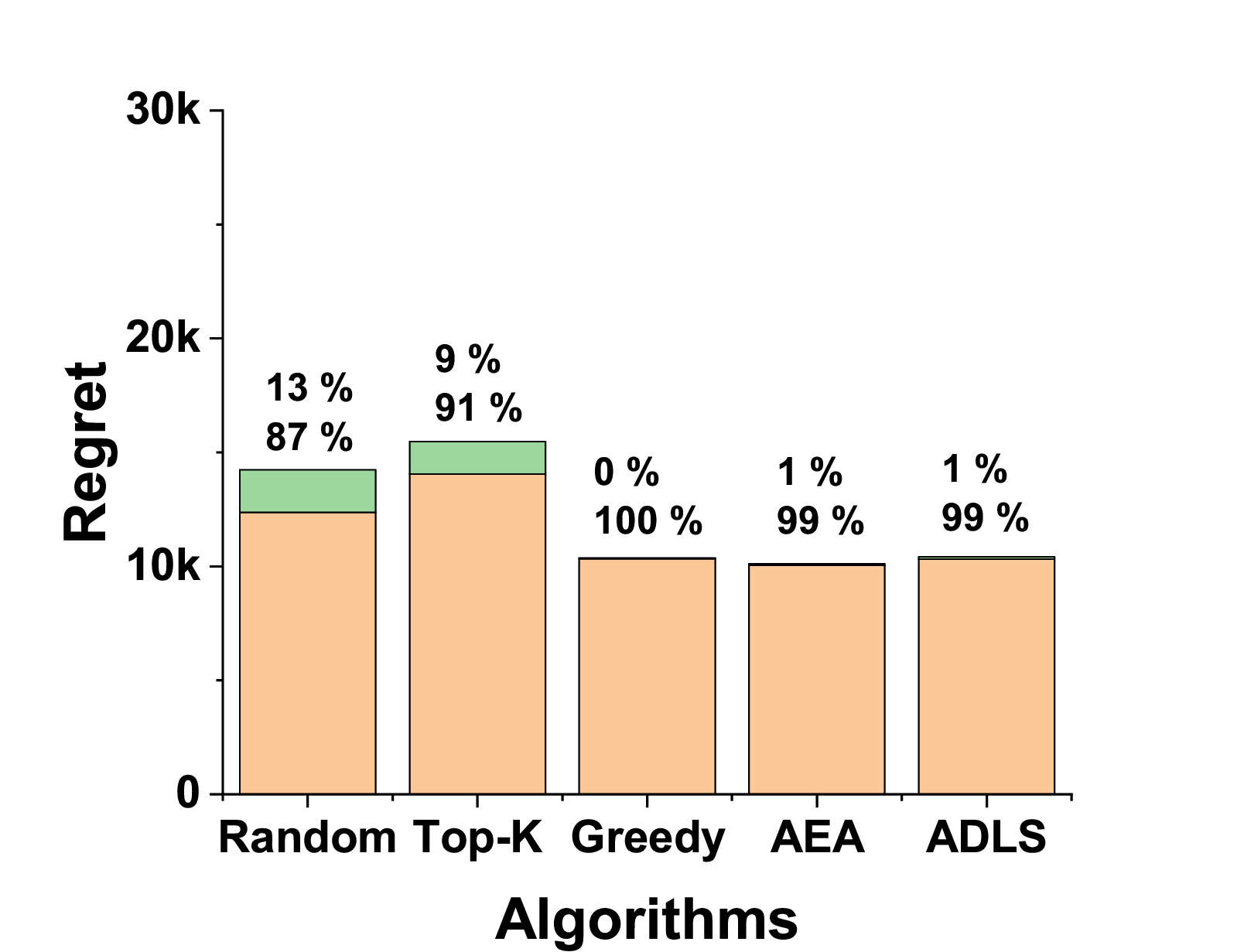} & 
    \includegraphics[scale=0.10]{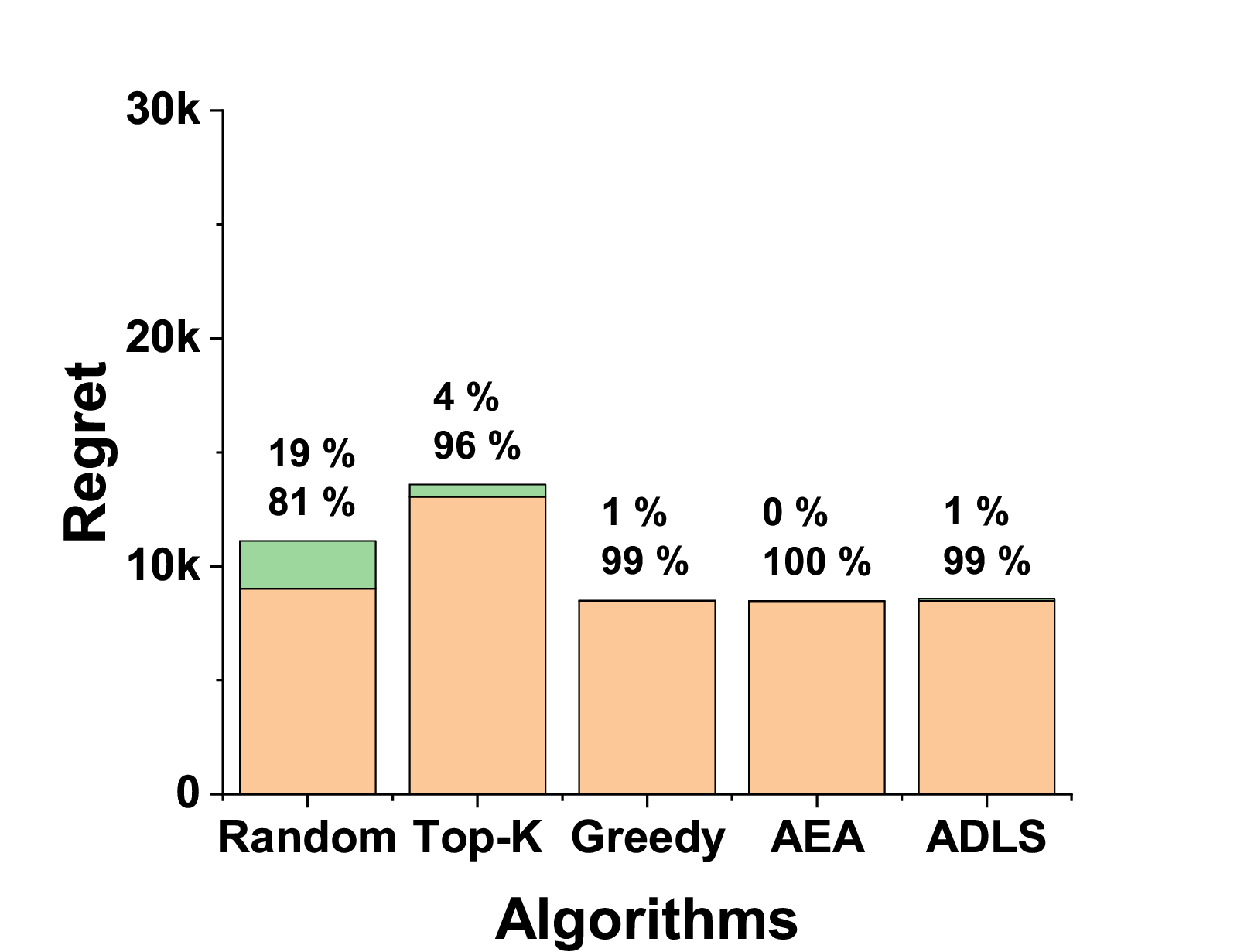} & 
    \includegraphics[scale=0.10]{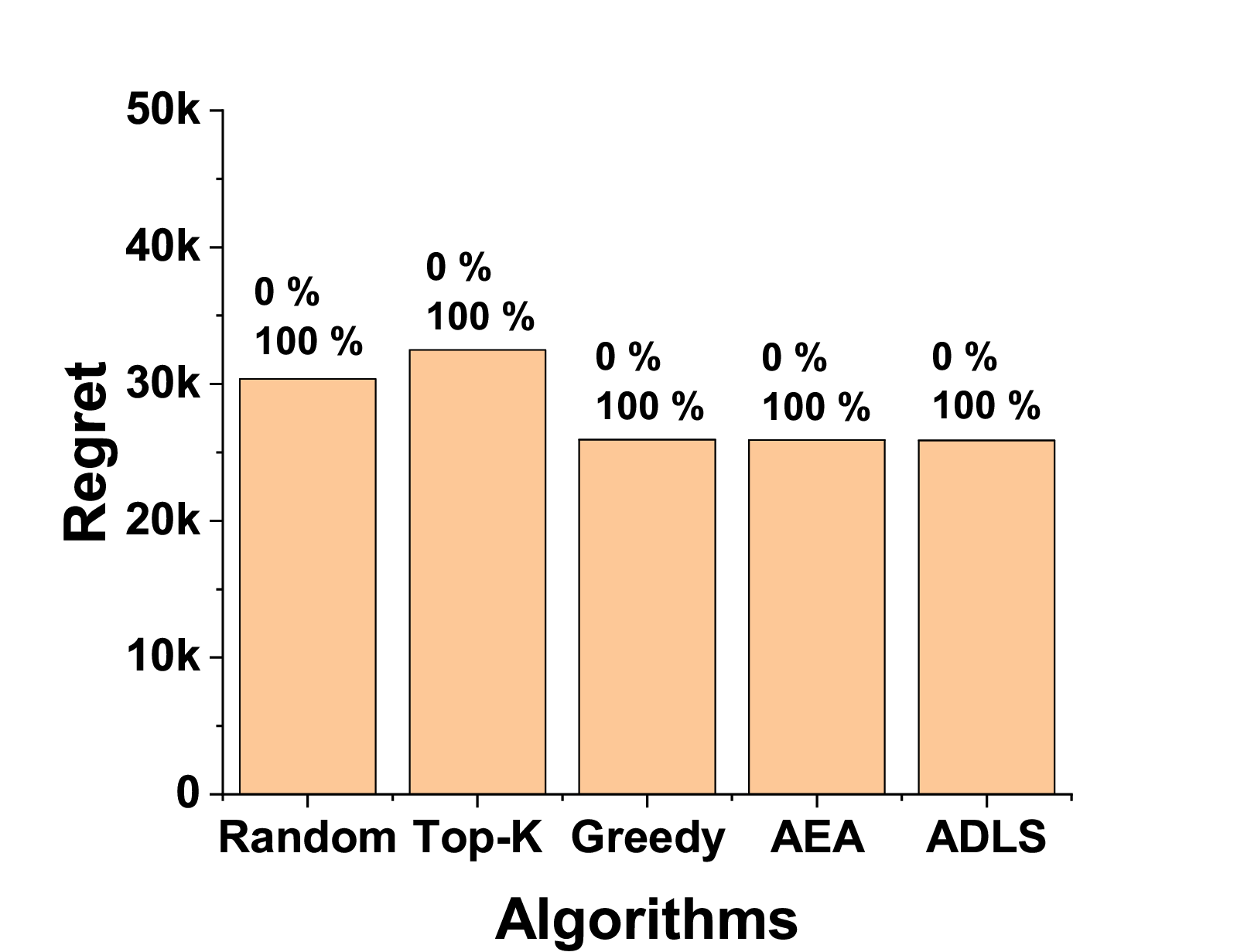} & 
    \includegraphics[scale=0.10]{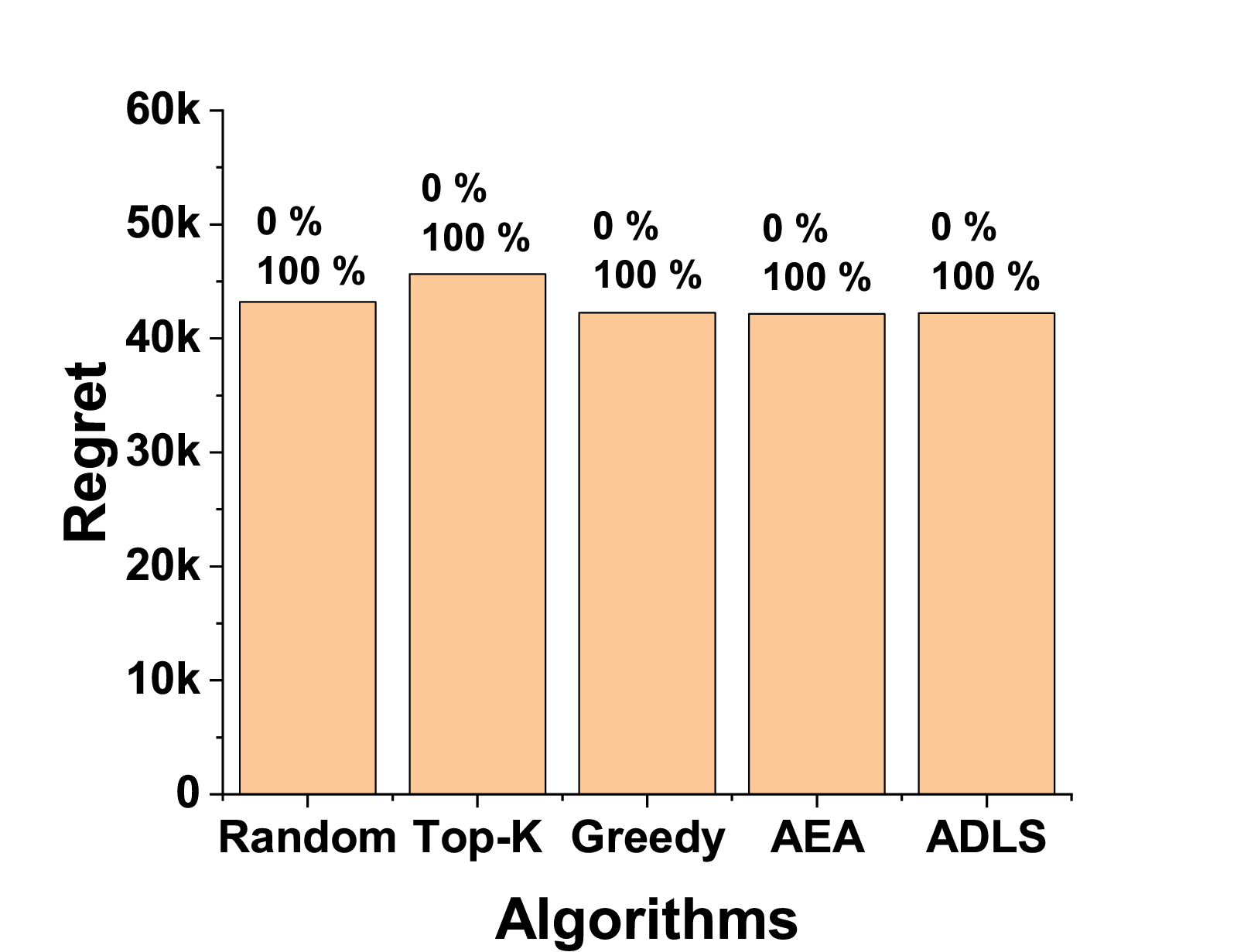} &
    \includegraphics[scale=0.10]{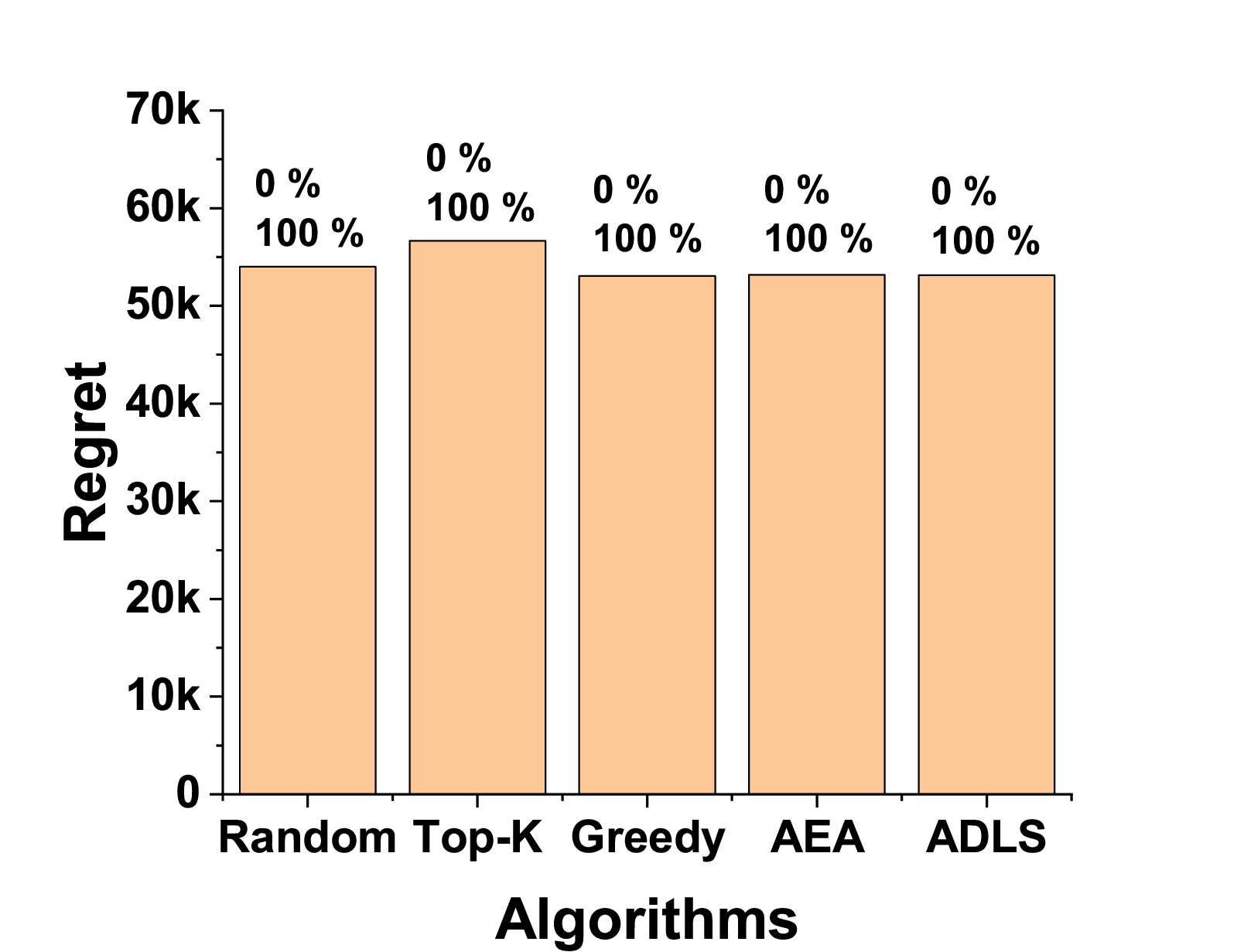} \\
    (a) $\lambda$ = $40\%$ & 
    (b) $\lambda$ = $60\%$ & 
    (c) $\lambda$ = $80\%$ & 
    (d) $\lambda$ = $100\%$ & 
    (e) $\lambda$ = $120\%$ \\
\end{tabular}
\caption{Regret of varying the demand-supply ratio $\lambda$ when $\omega = 90\%$ (Facebook, $|\mathcal{A}| = 5$, $p_{c}: Trivalency$)}
\label{Fig:4_Influence_5}
\end{figure}

\begin{figure}[H]
\centering
\begin{tabular}{lclc}
    Excessive Regret & \includegraphics[width=0.11\linewidth]{Excessive.png} &
    Unsatisfied Regret & \includegraphics[width=0.11\linewidth]{Unsatisfied.png}\\
\end{tabular}
\begin{tabular}{ccccc}
    \includegraphics[scale=0.10]{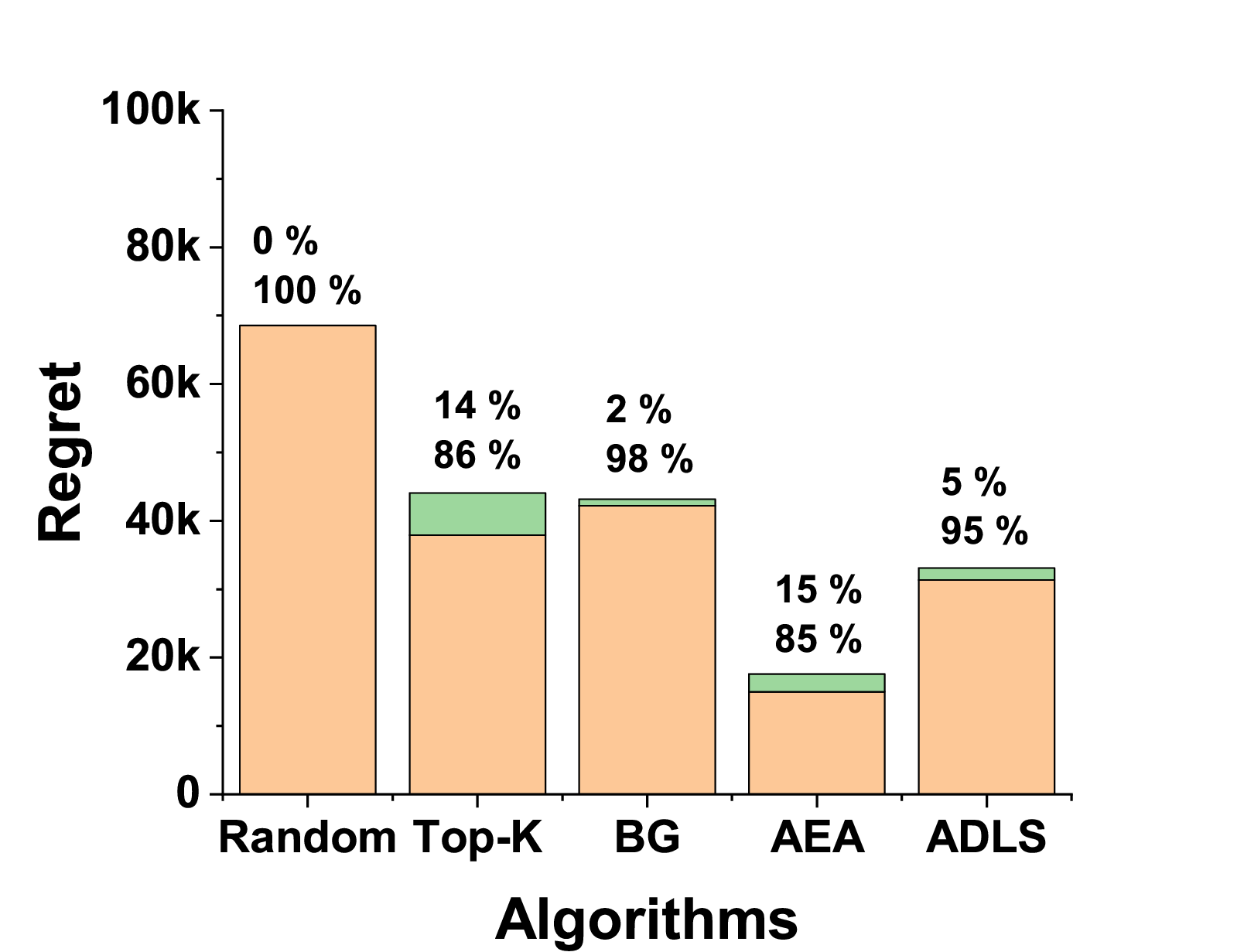} & 
    \includegraphics[scale=0.10]{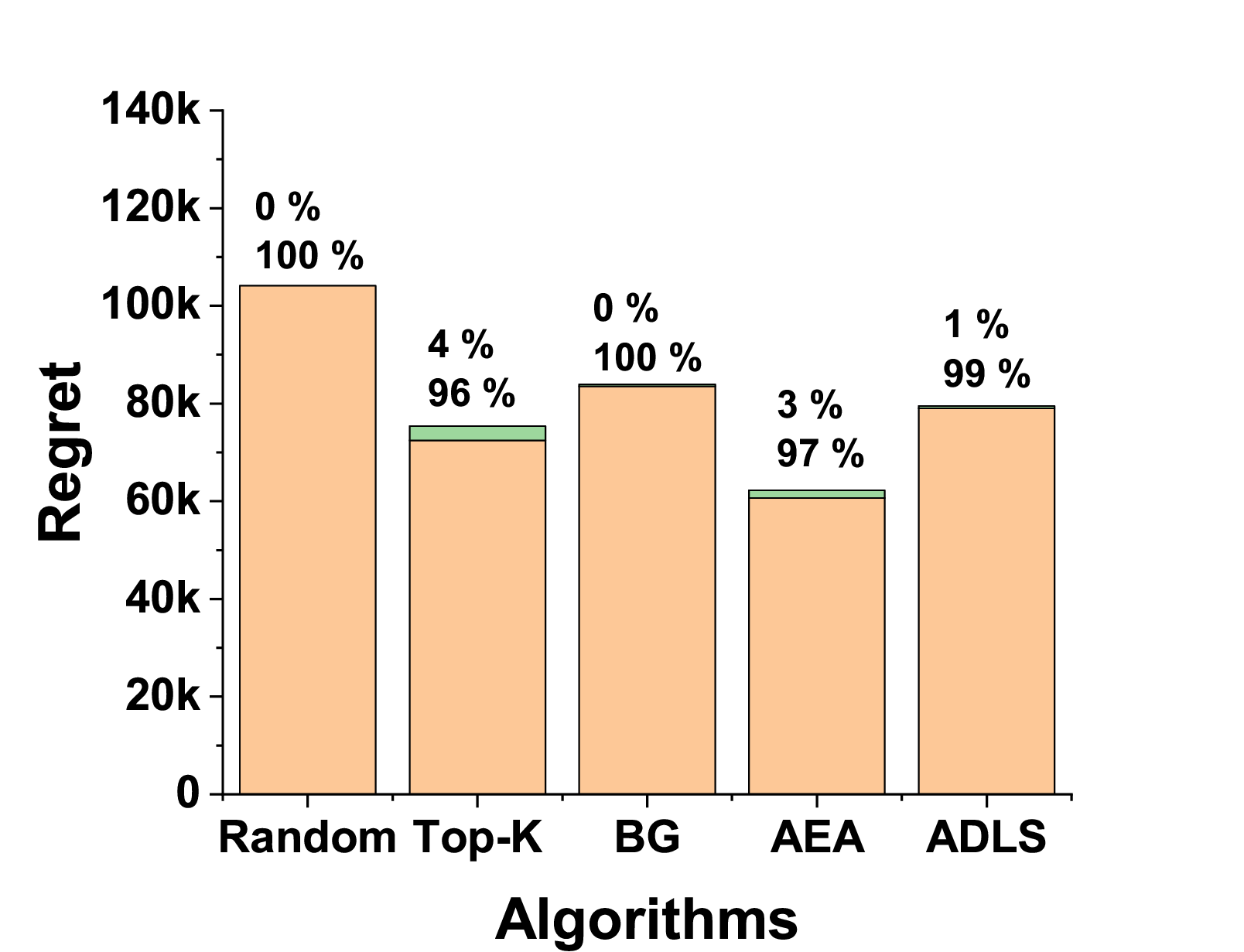} & 
    \includegraphics[scale=0.10]{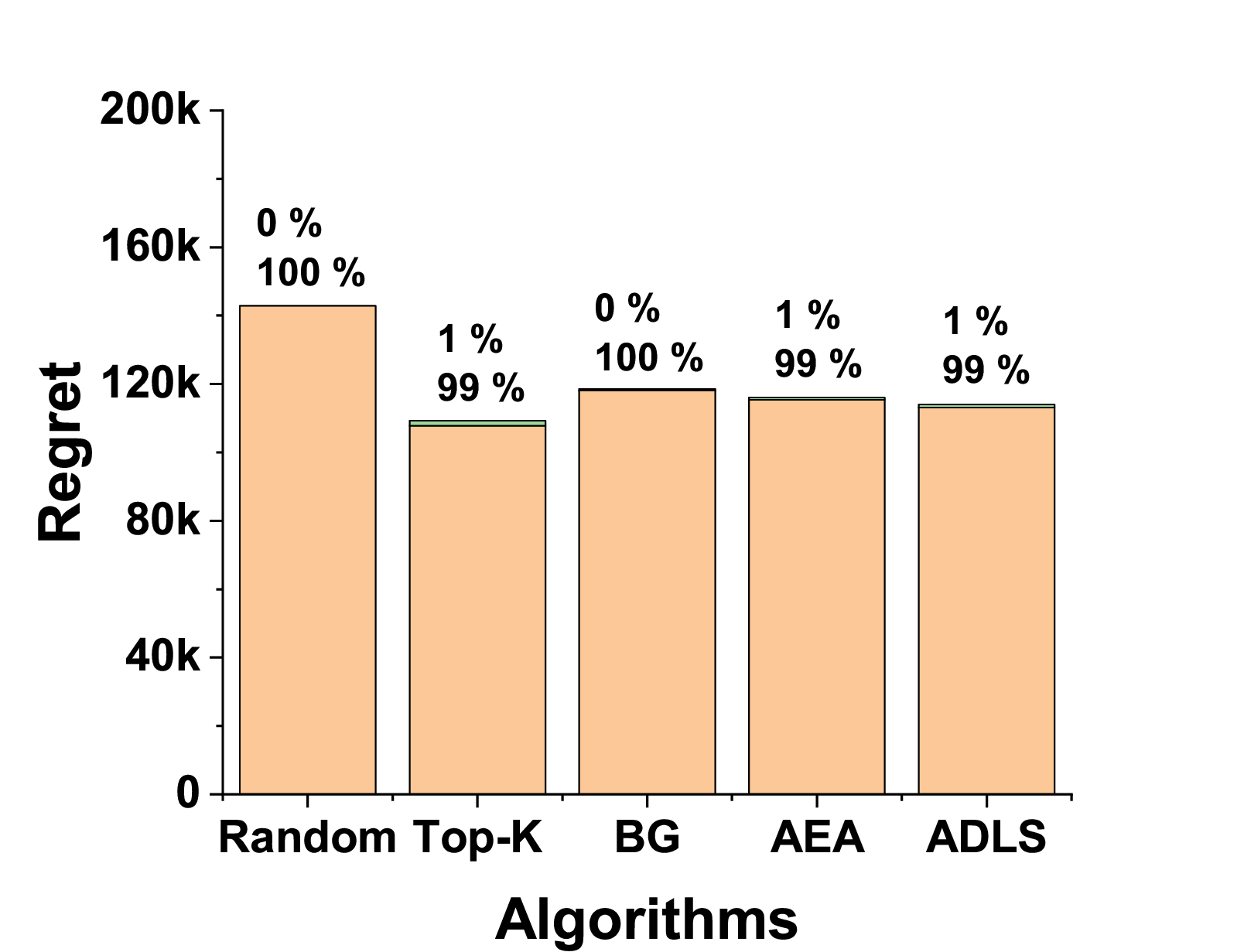} & 
    \includegraphics[scale=0.10]{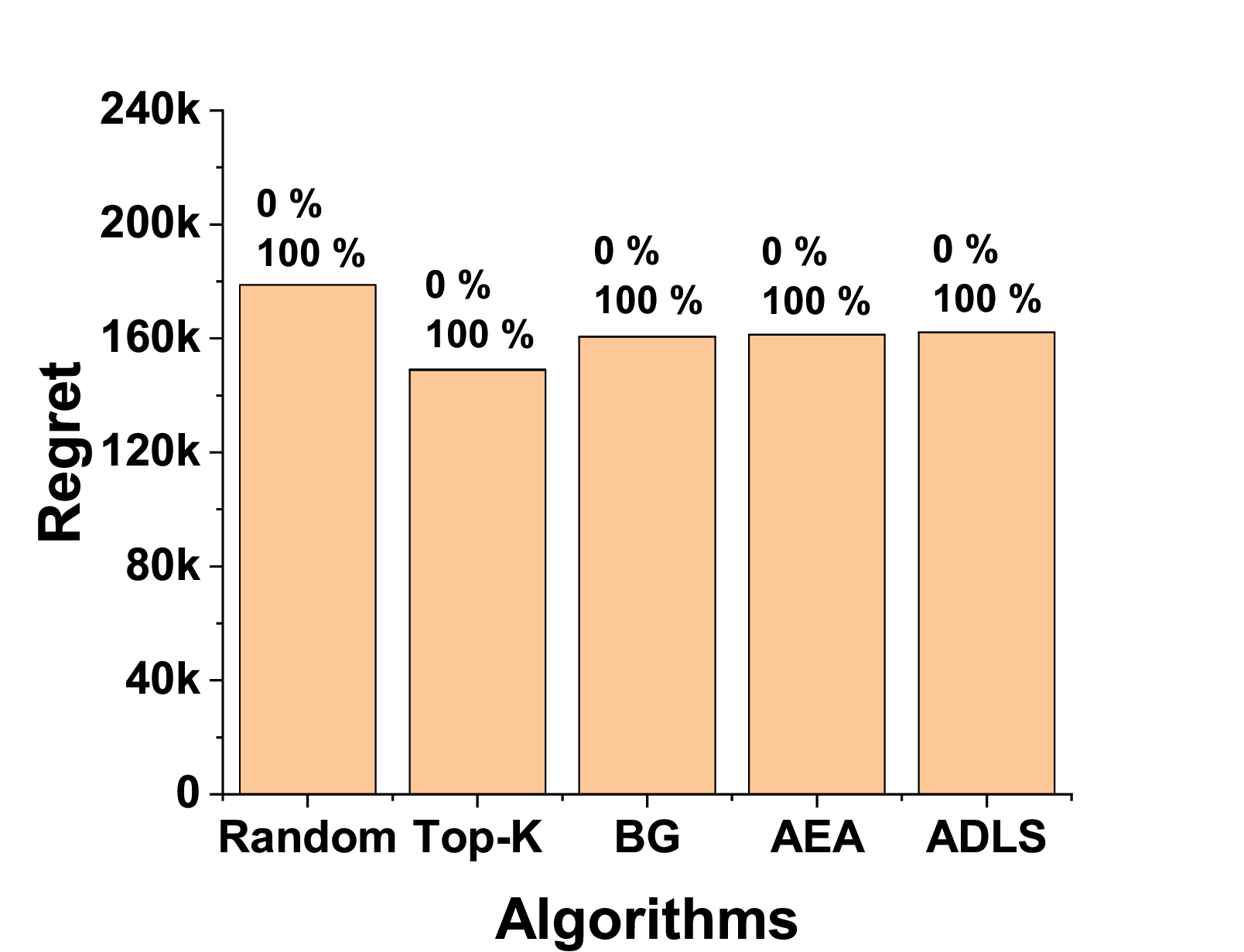} &
    \includegraphics[scale=0.10]{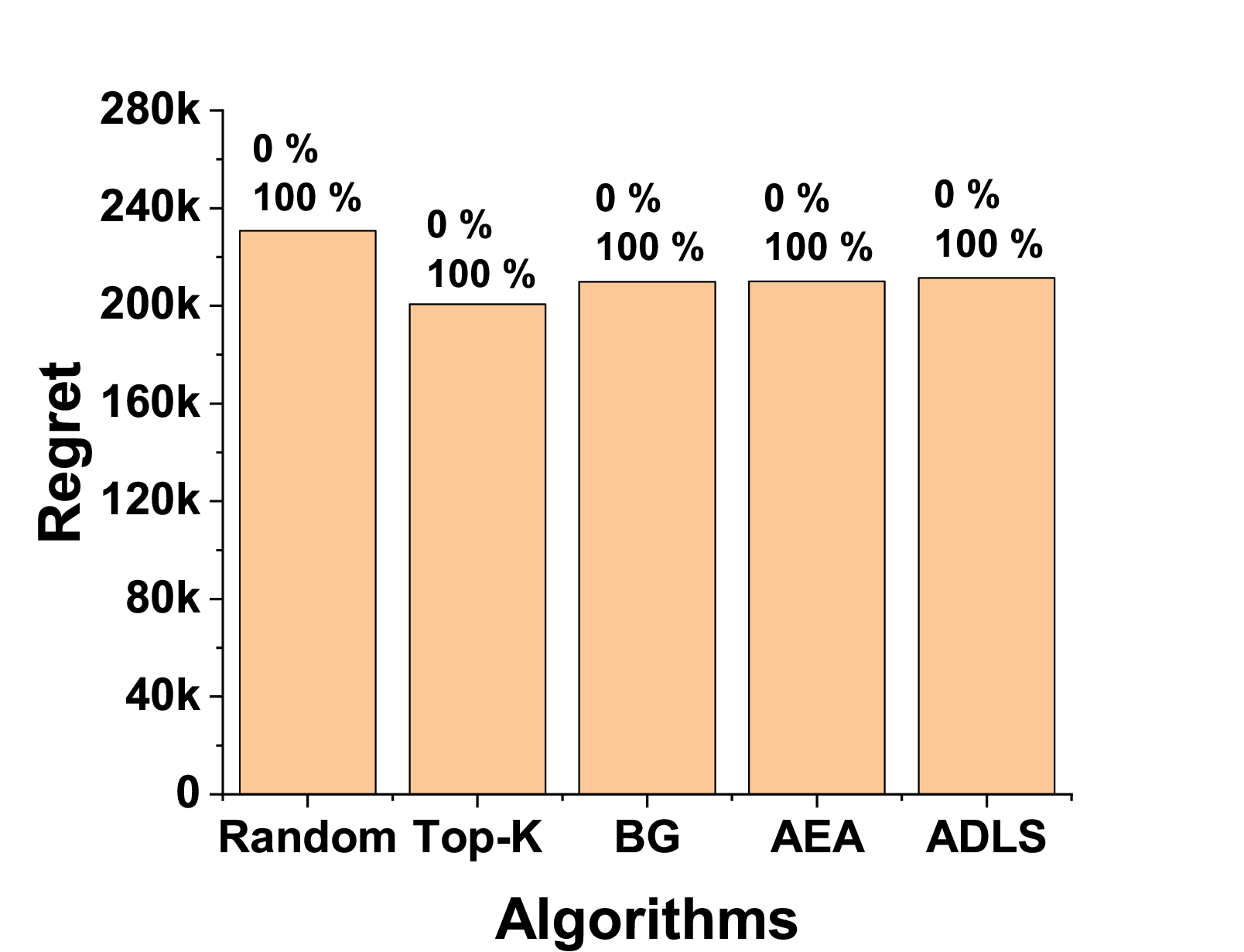} \\
    (a) $\lambda$ = $40\%$ & 
    (b) $\lambda$ = $60\%$ & 
    (c) $\lambda$ = $80\%$ & 
    (d) $\lambda$ = $100\%$ & 
    (e) $\lambda$ = $120\%$ \\
\end{tabular}
\caption{Regret of varying the demand-supply ratio $\lambda$ when $\omega = 10\%$ (Wikivote, $|\mathcal{A}| = 100$, $p_{c}: Trivalency$)}
\label{Fig:6_Influence_1}
\begin{tabular}{ccccc}
    \includegraphics[scale=0.10]{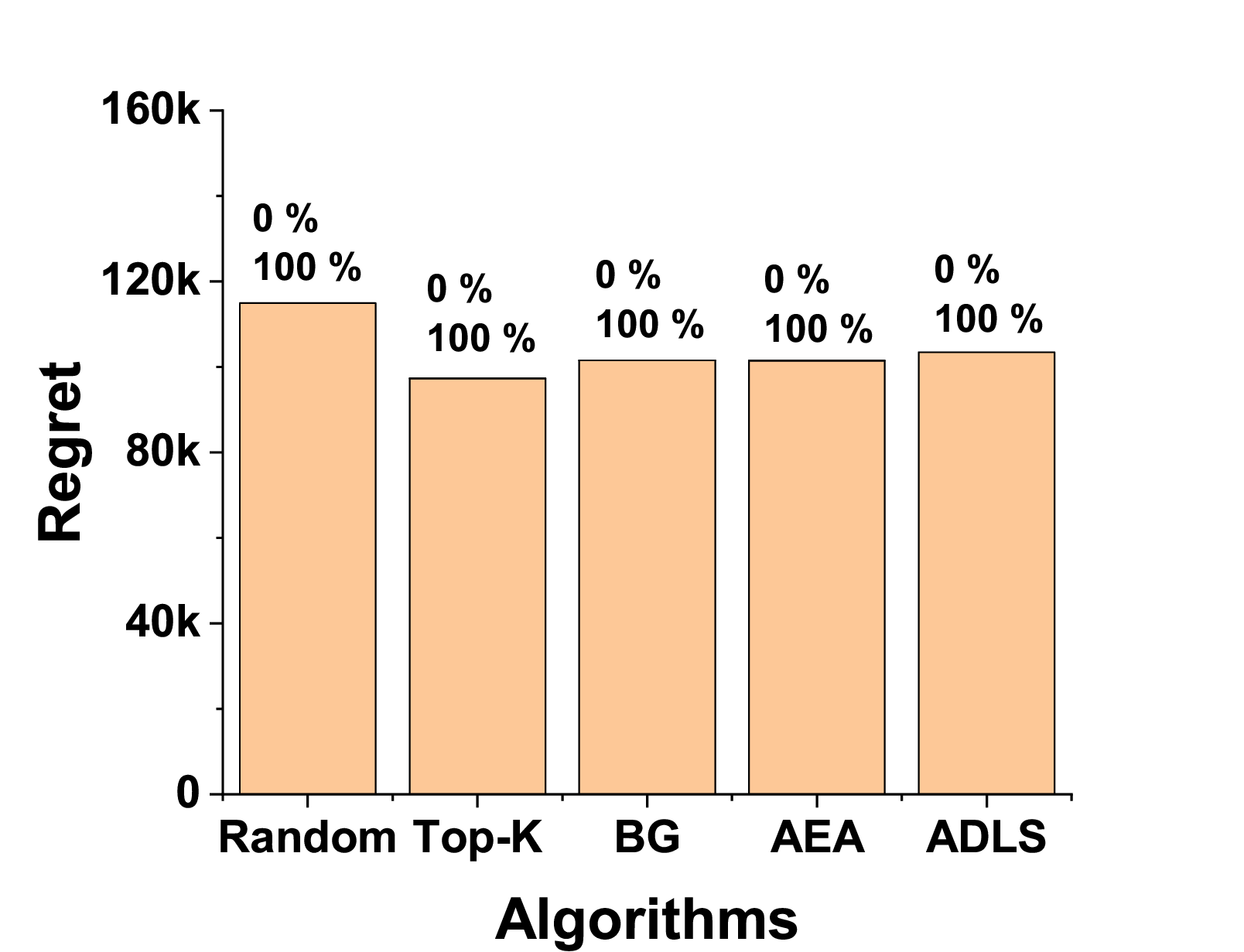} & 
    \includegraphics[scale=0.10]{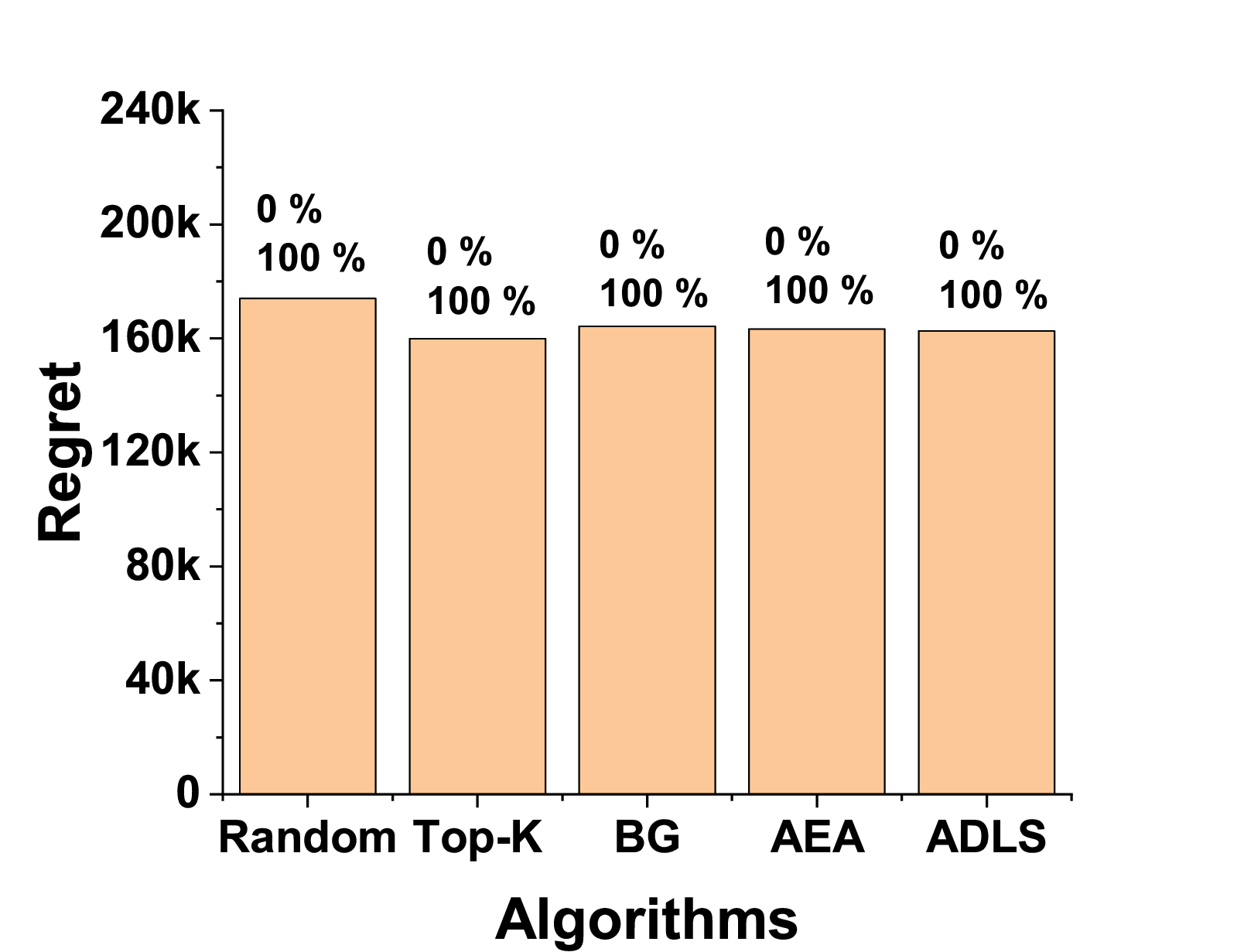} & 
    \includegraphics[scale=0.10]{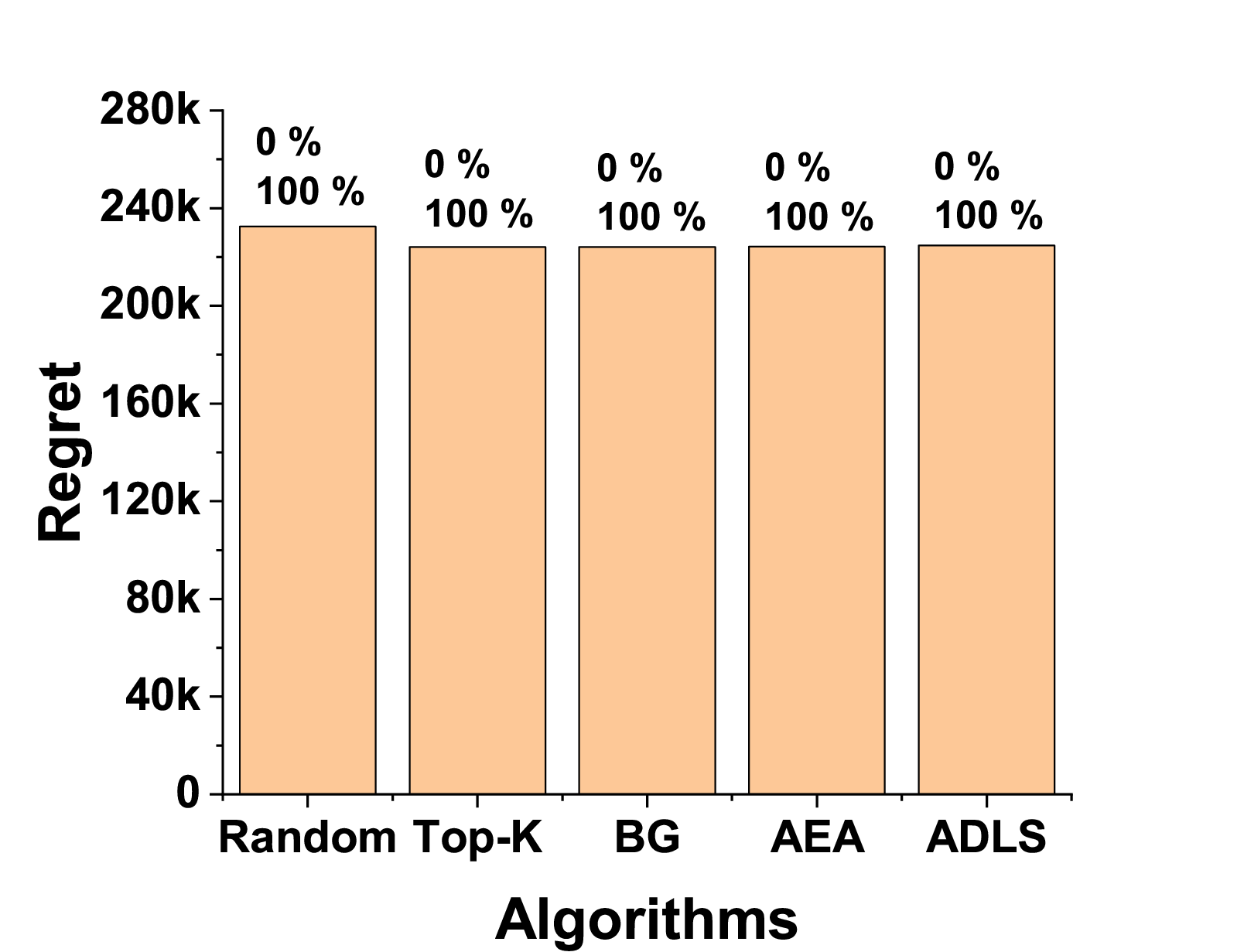} & 
    \includegraphics[scale=0.10]{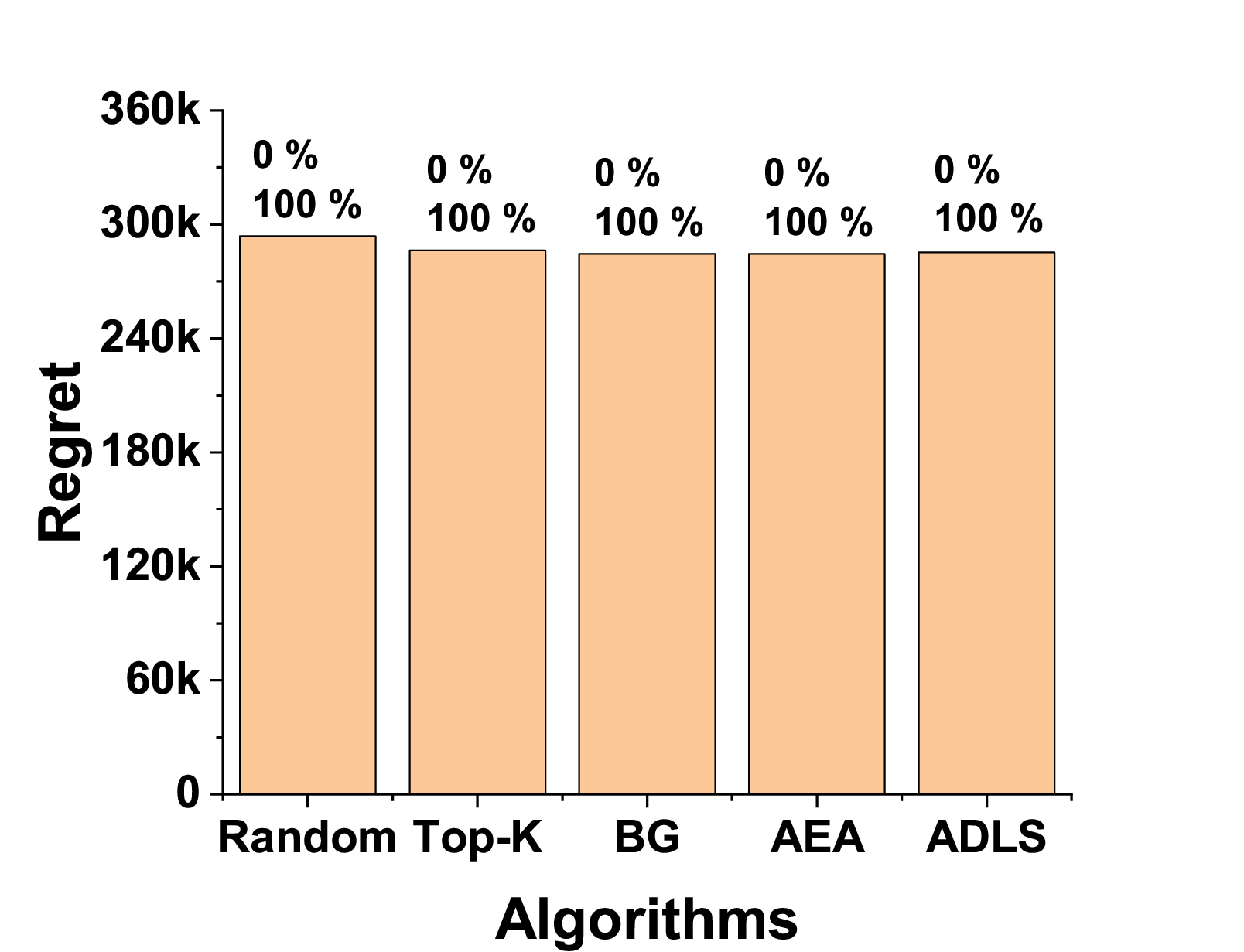} &
    \includegraphics[scale=0.10]{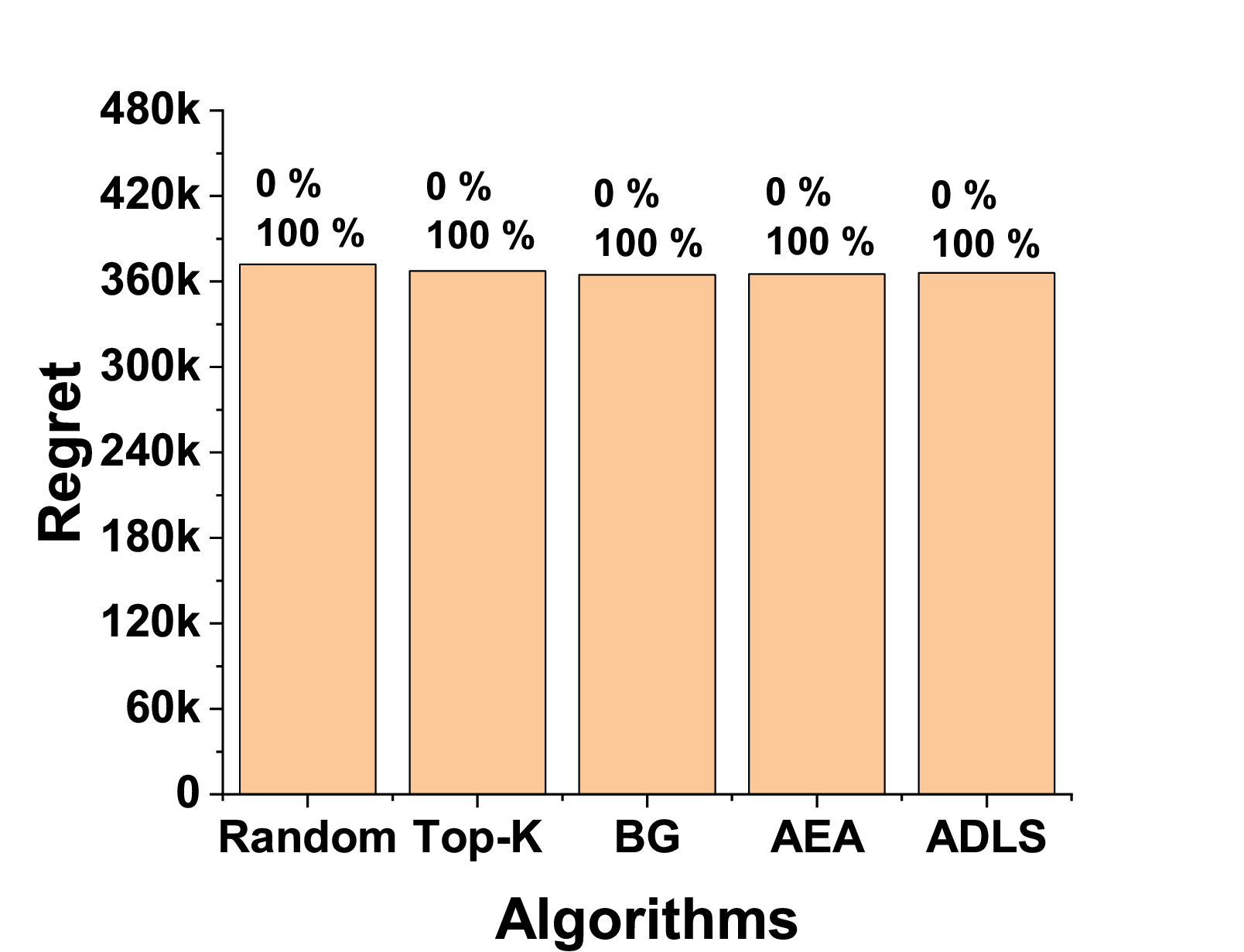} \\
    (a) $\lambda$ = $40\%$ & 
    (b) $\lambda$ = $60\%$ & 
    (c) $\lambda$ = $80\%$ &
    (d) $\lambda$ = $100\%$ & 
    (e) $\lambda$ = $120\%$ \\
\end{tabular}
\caption{Regret of varying the demand-supply ratio $\lambda$ when $\omega = 30\%$ (Wikivote, $|\mathcal{A}| = 50$, $p_{c}: Trivalency$)}
\label{Fig:6_Influence_2}
\begin{tabular}{ccccc}
    \includegraphics[scale=0.10]{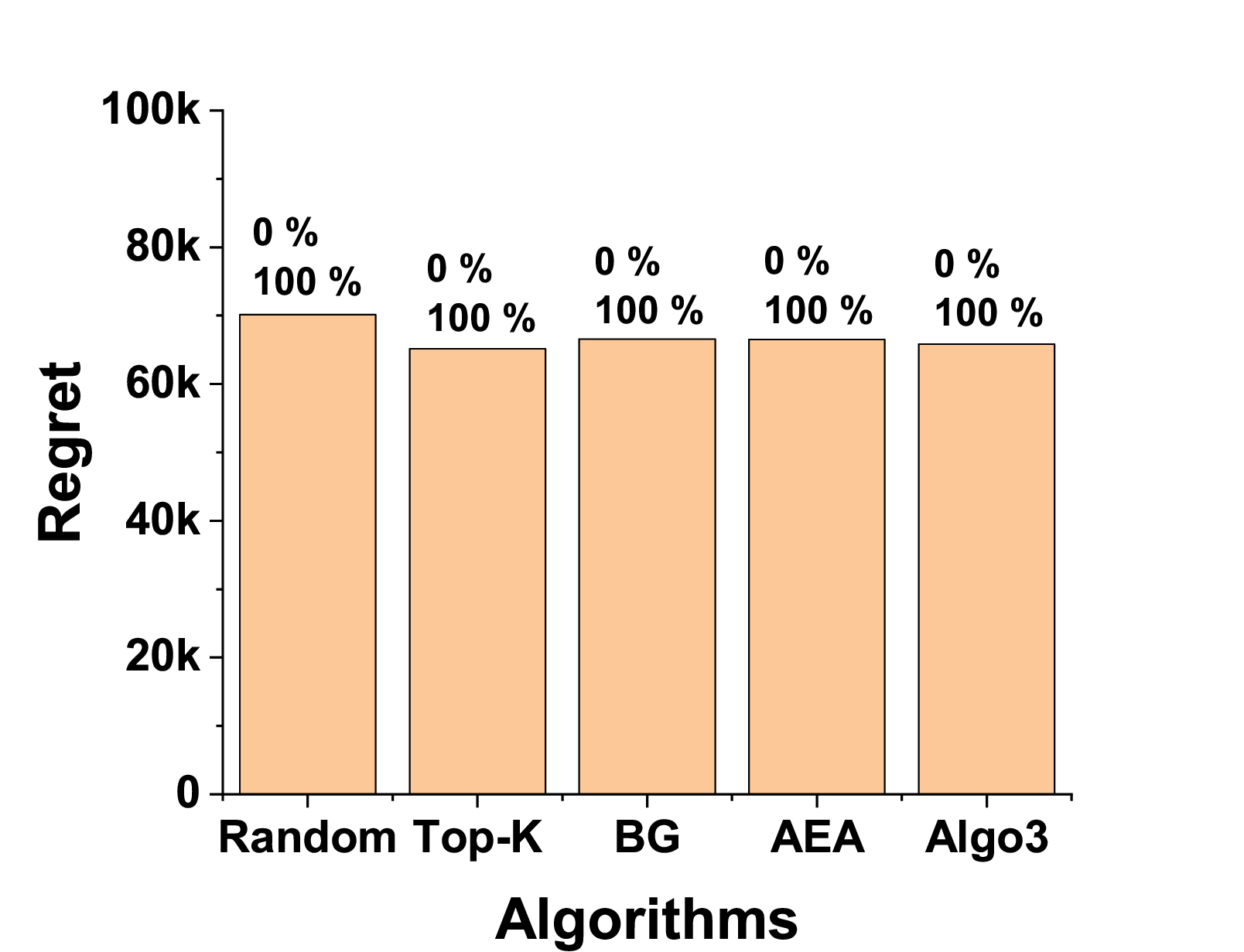} & 
    \includegraphics[scale=0.10]{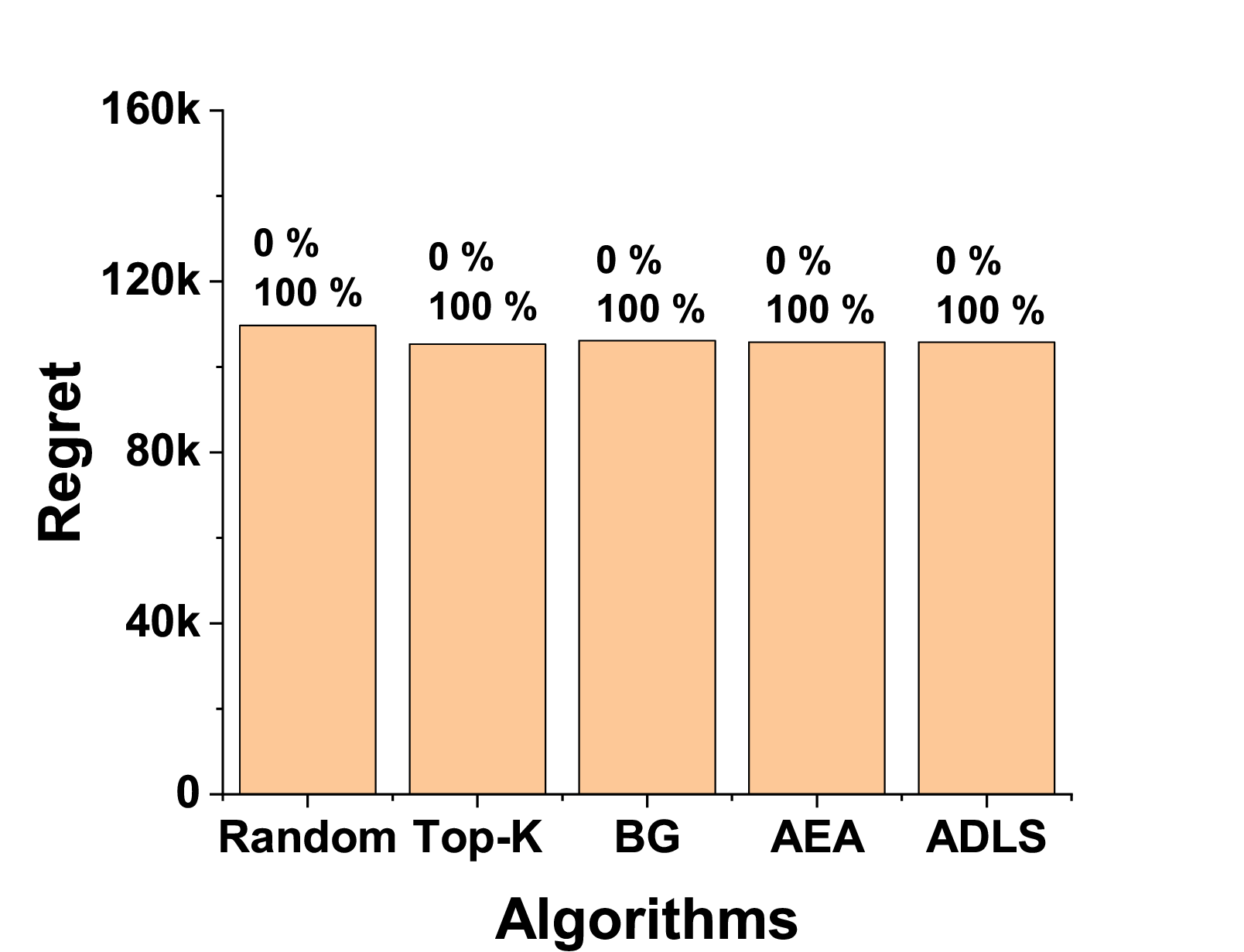} & 
    \includegraphics[scale=0.10]{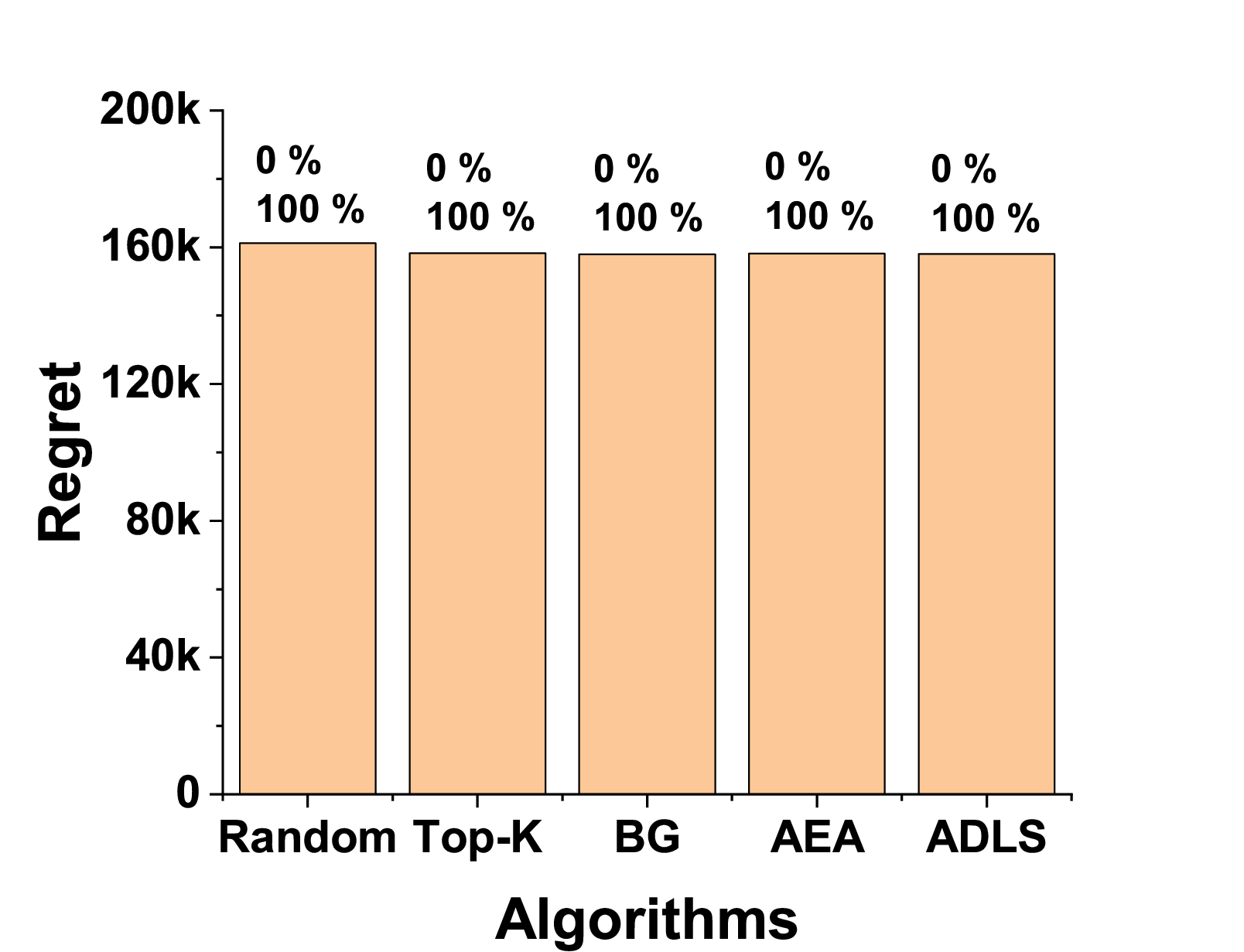} & 
    \includegraphics[scale=0.10]{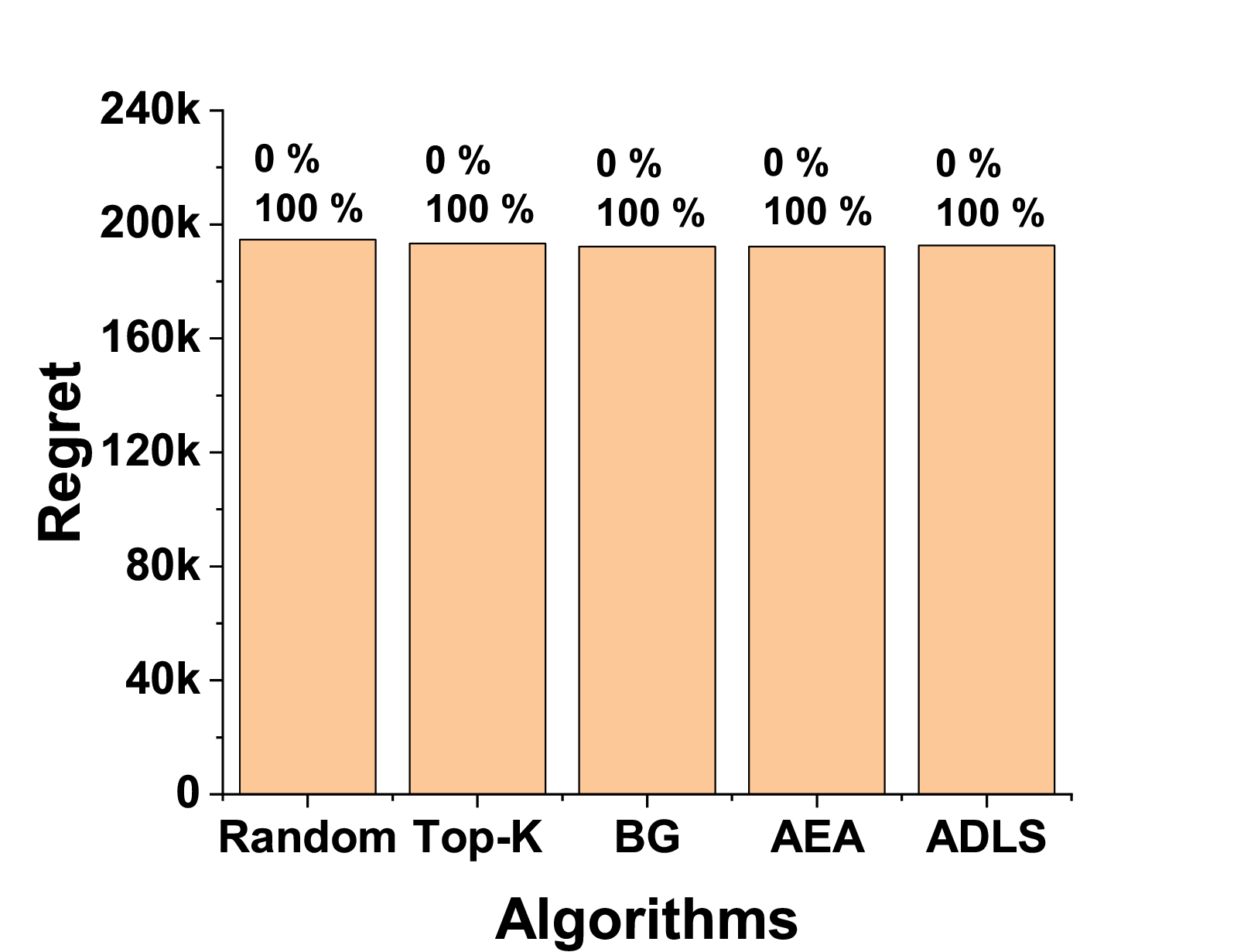} &
    \includegraphics[scale=0.10]{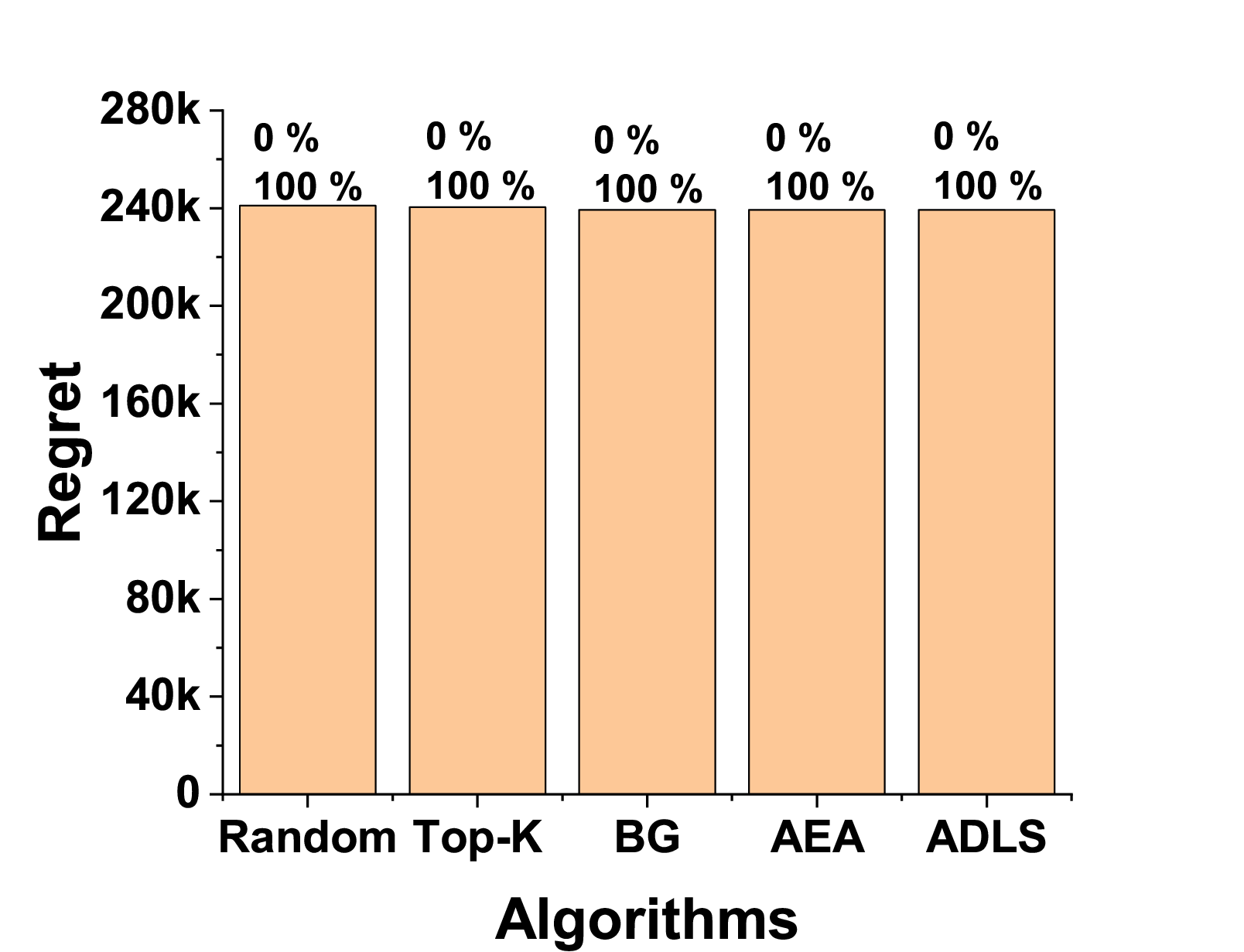} \\
    (a) $\lambda$ = $40\%$ & 
    (b) $\lambda$ = $60\%$ & 
    (c) $\lambda$ = $80\%$ & 
    (d) $\lambda$ = $100\%$ & 
    (e) $\lambda$ = $120\%$ \\
\end{tabular}
\caption{Regret of varying the demand-supply ratio $\lambda$ when $\omega = 50\%$ (Wikivote, $|\mathcal{A}| = 20$, $p_{c}: Trivalency$)}
\label{Fig:6_Influence_3}
\begin{tabular}{ccccc}
    \includegraphics[scale=0.10]{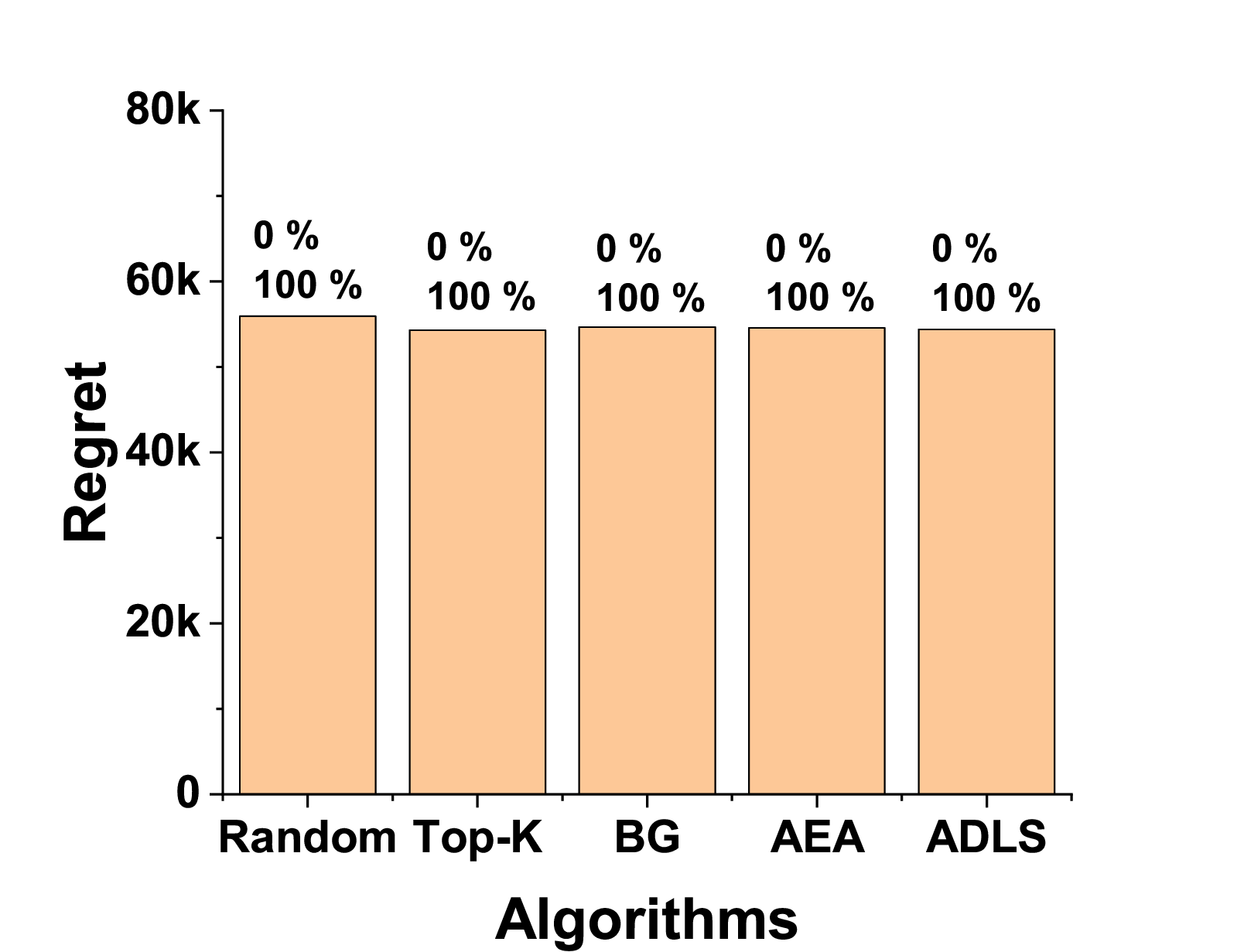} & 
    \includegraphics[scale=0.10]{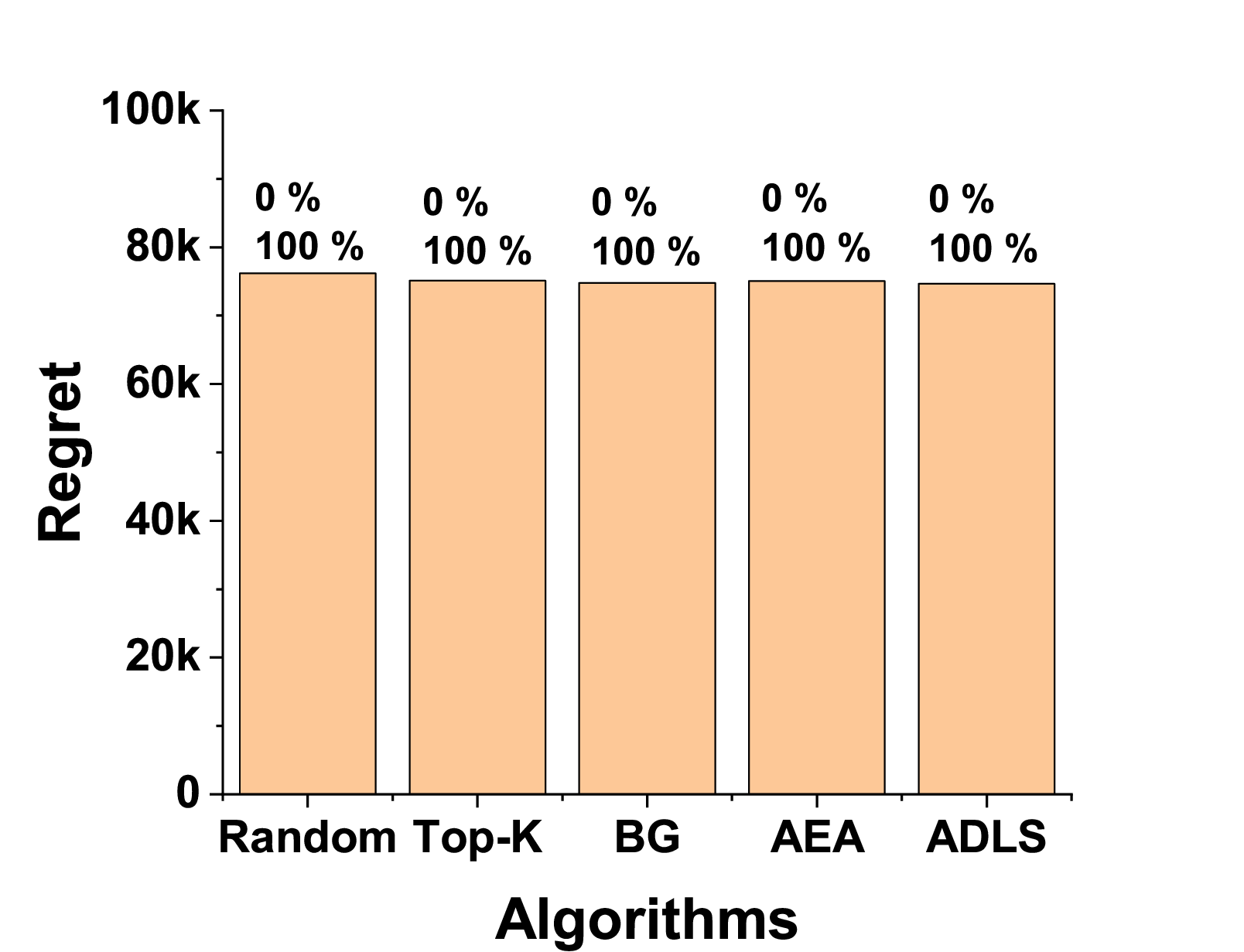} & 
    \includegraphics[scale=0.10]{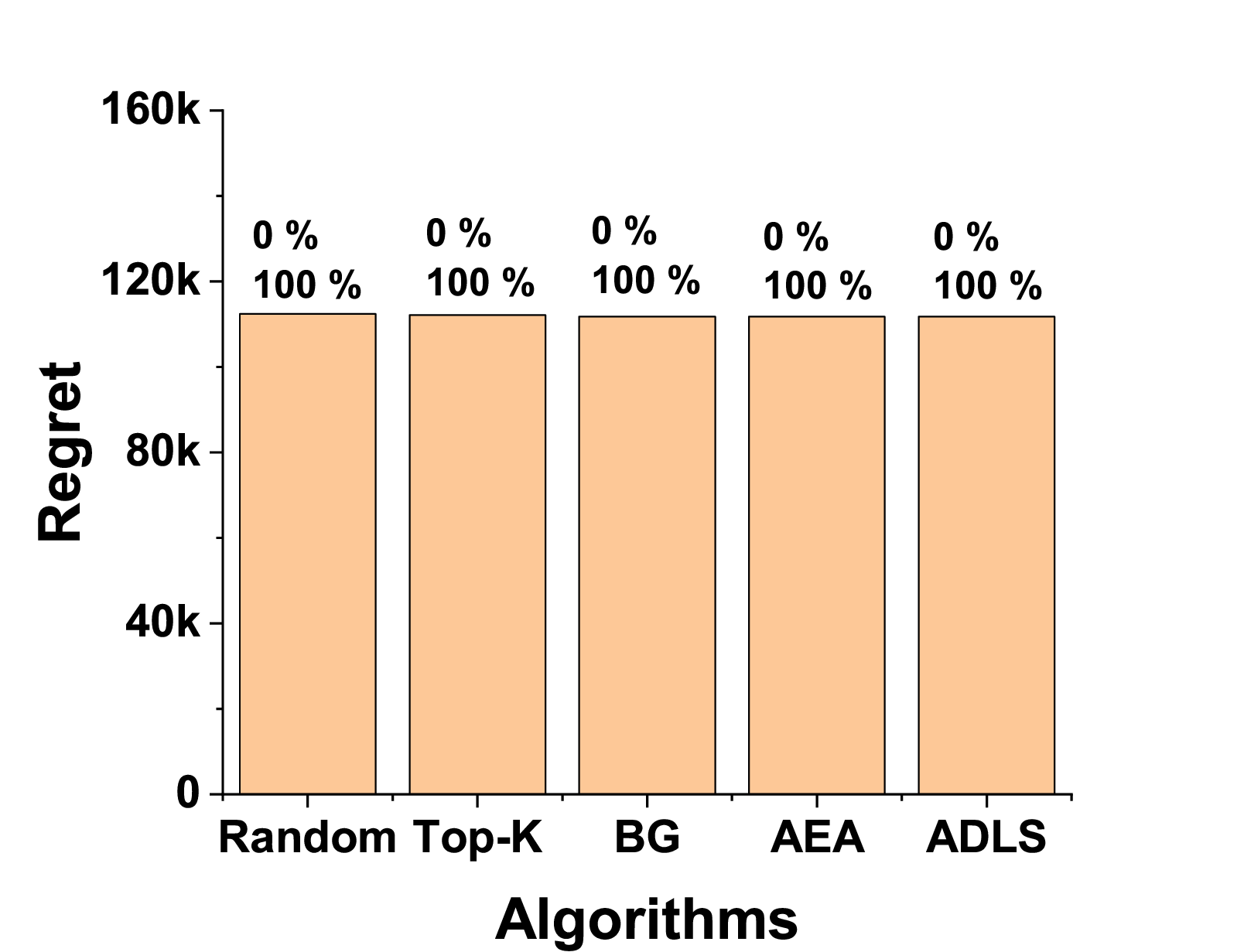} & 
    \includegraphics[scale=0.10]{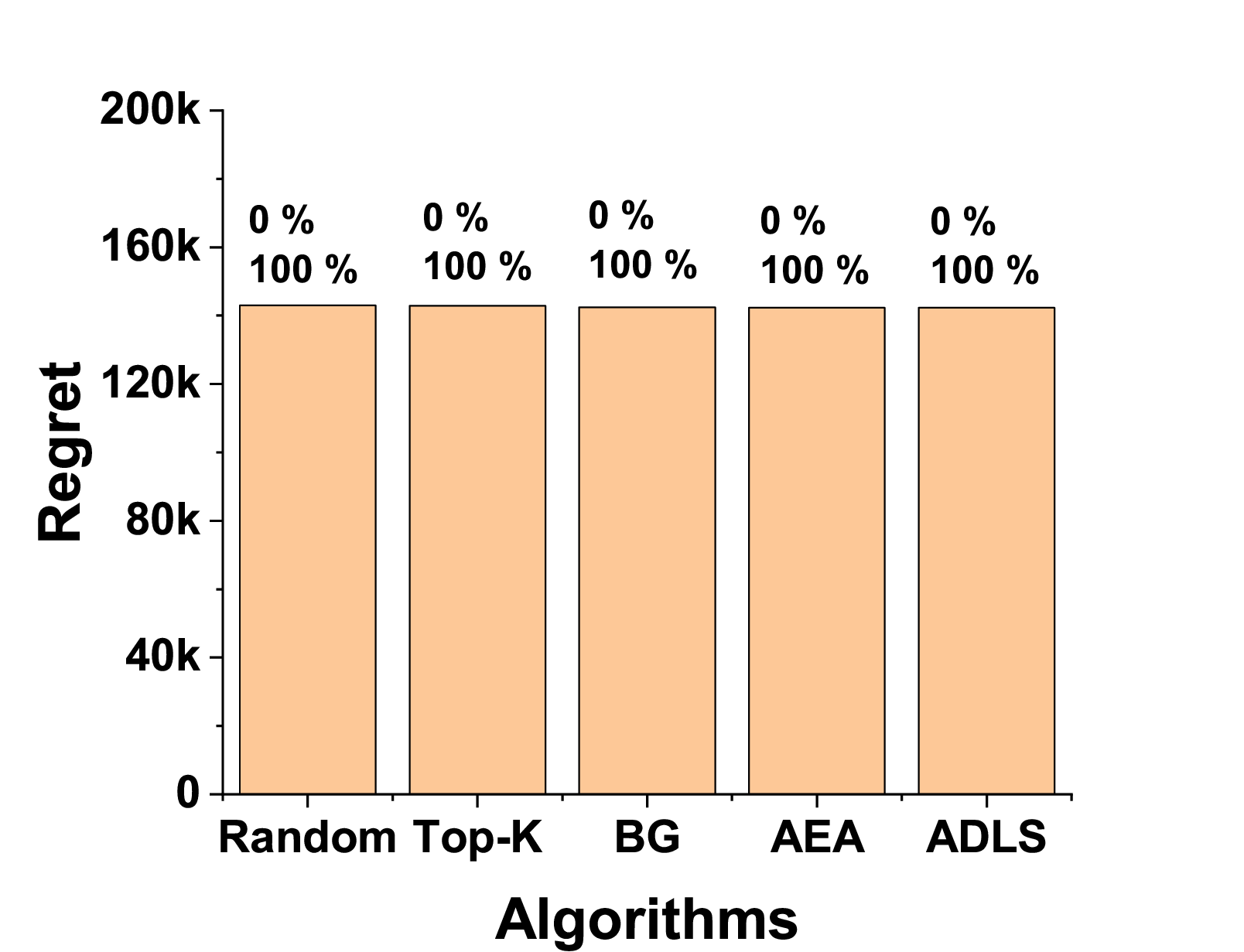} &
    \includegraphics[scale=0.10]{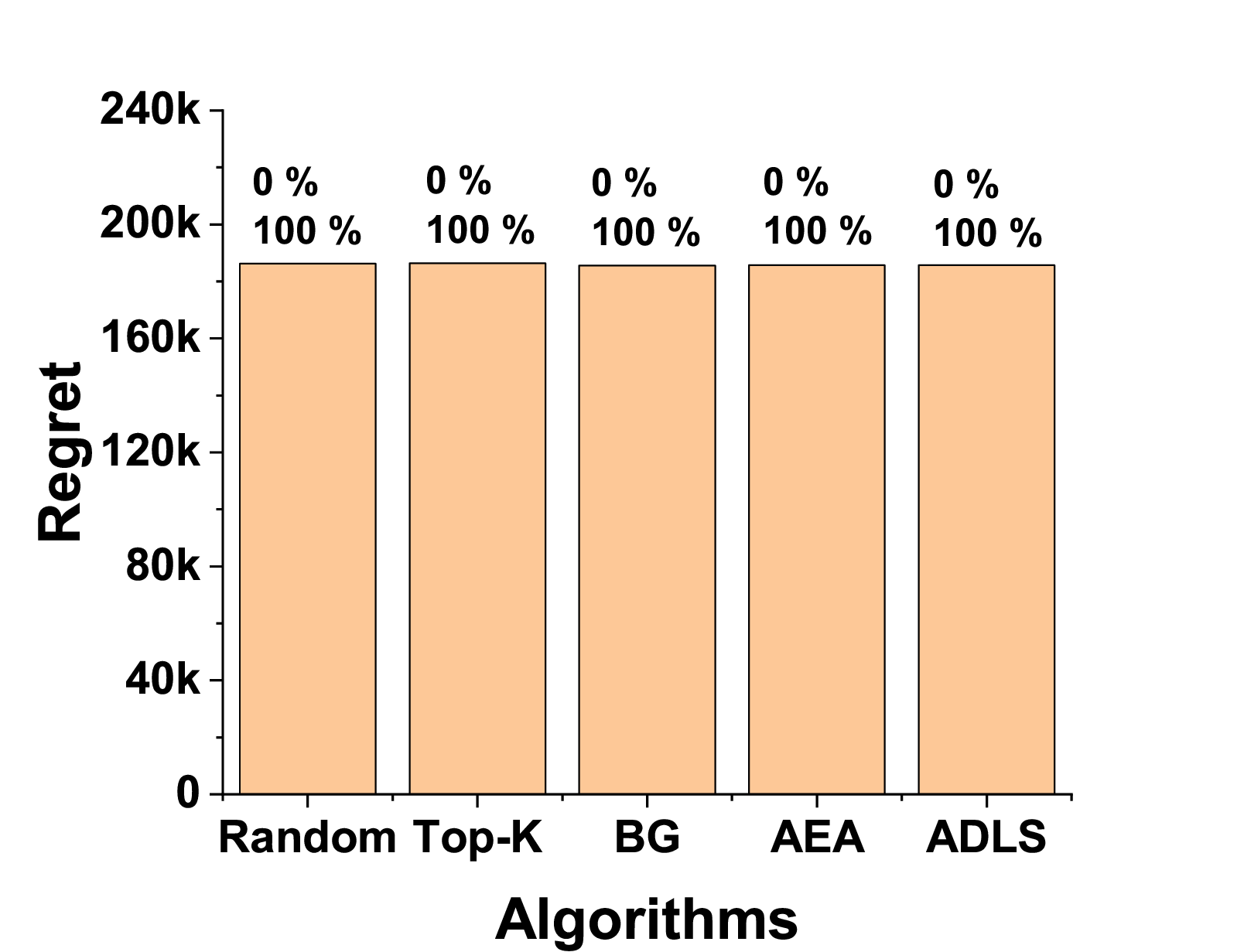} \\
    (a) $\lambda$ = $40\%$ & 
    (b) $\lambda$ = $60\%$ & 
    (c) $\lambda$ = $80\%$ & 
    (d) $\lambda$ = $100\%$ & 
    (e) $\lambda$ = $120\%$ \\
\end{tabular}
\caption{Regret of varying the demand-supply ratio $\lambda$ when $\omega = 70\%$ (Wikivote, $|\mathcal{A}| = 10$, $p_{c}: Trivalency$)}
\label{Fig:6_Influence_4}
\begin{tabular}{ccccc}
    \includegraphics[scale=0.10]{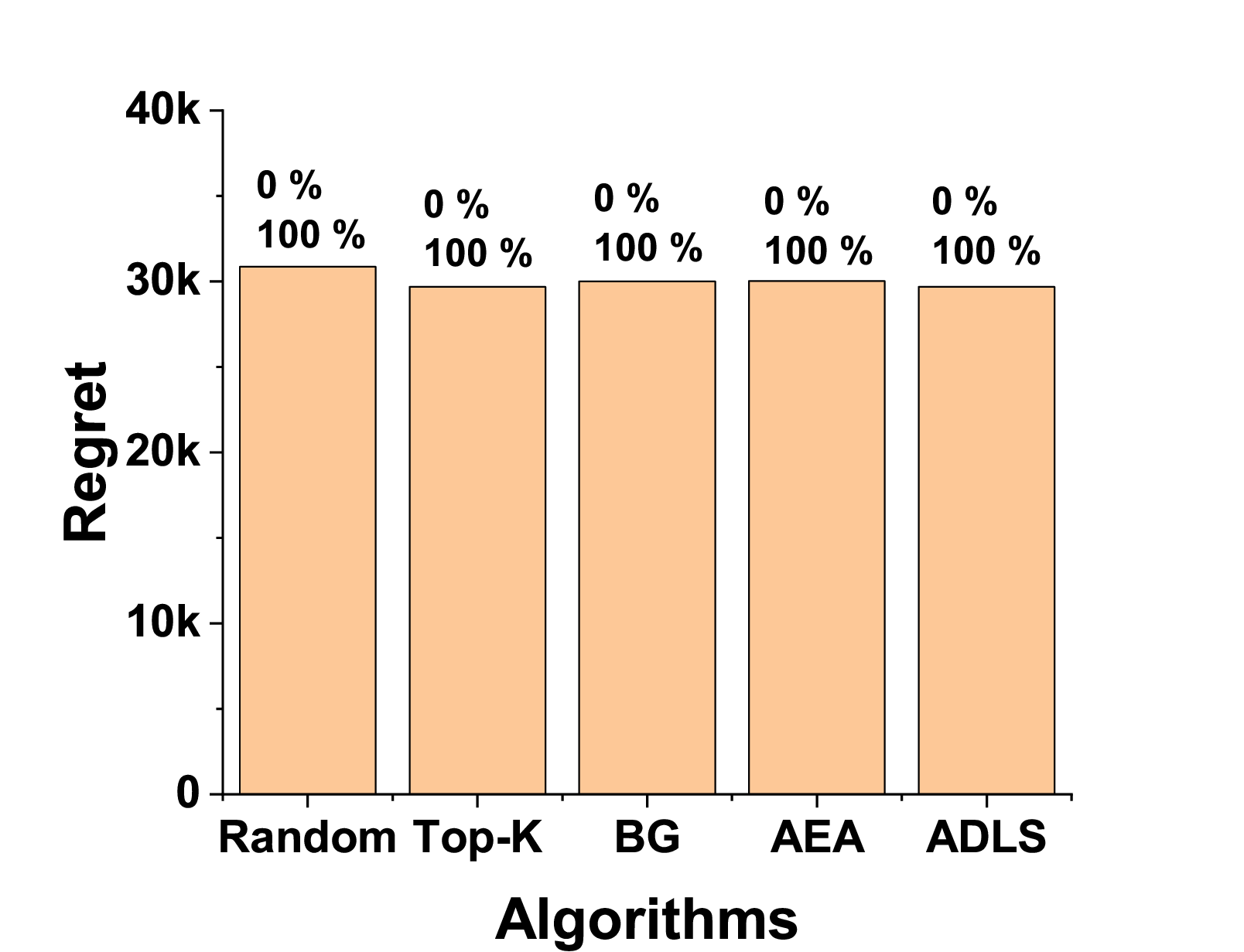} & 
    \includegraphics[scale=0.10]{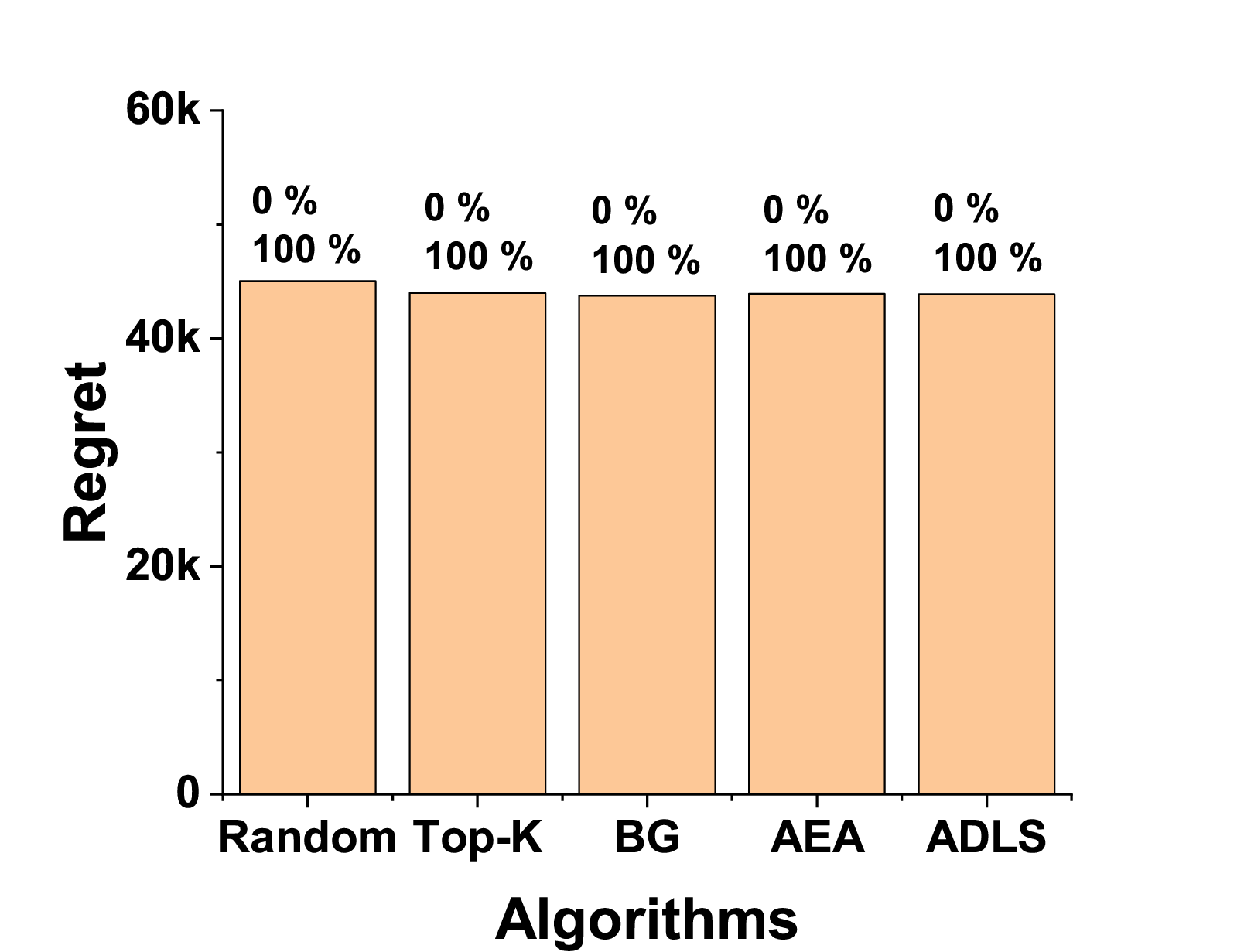} & 
    \includegraphics[scale=0.10]{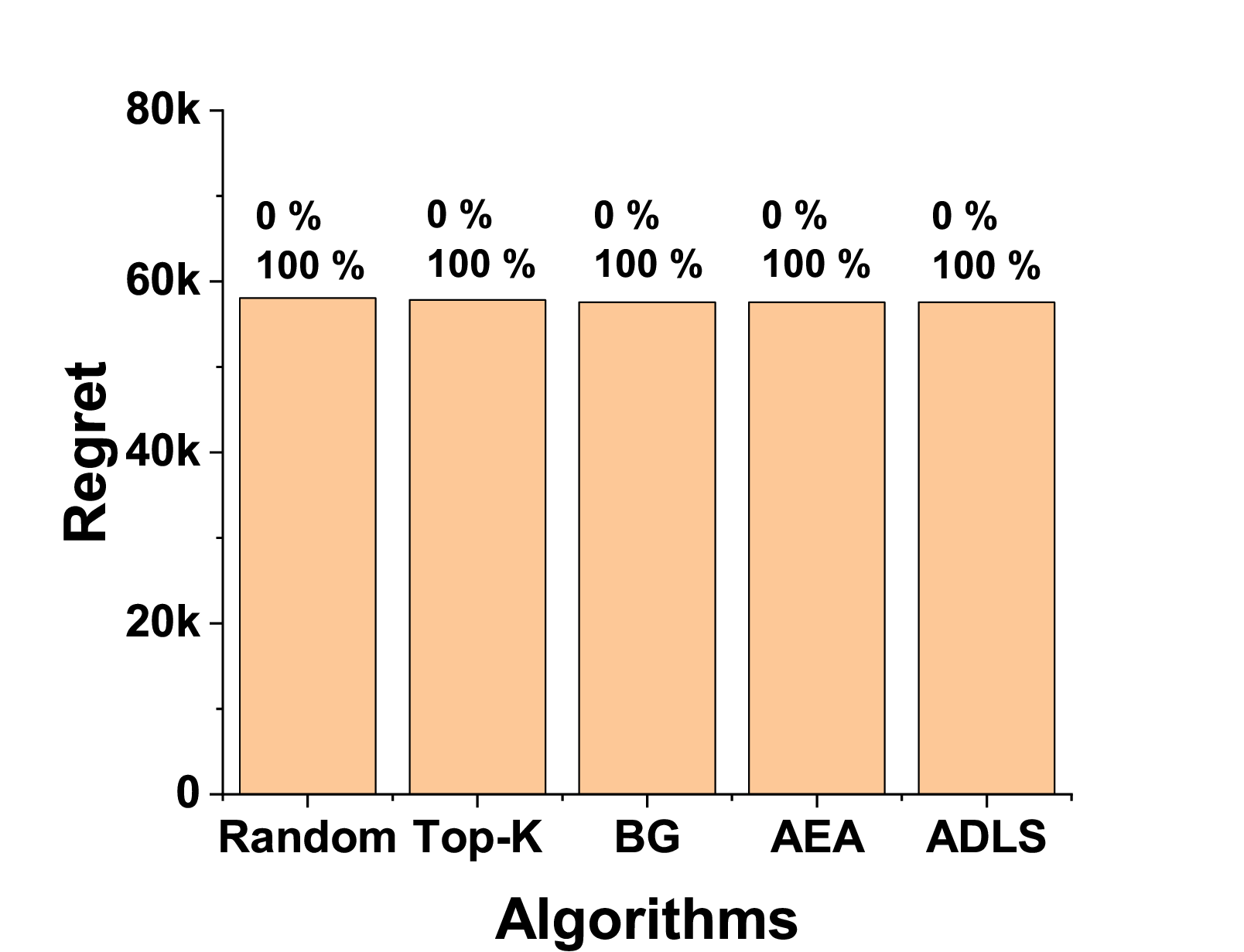} & 
    \includegraphics[scale=0.10]{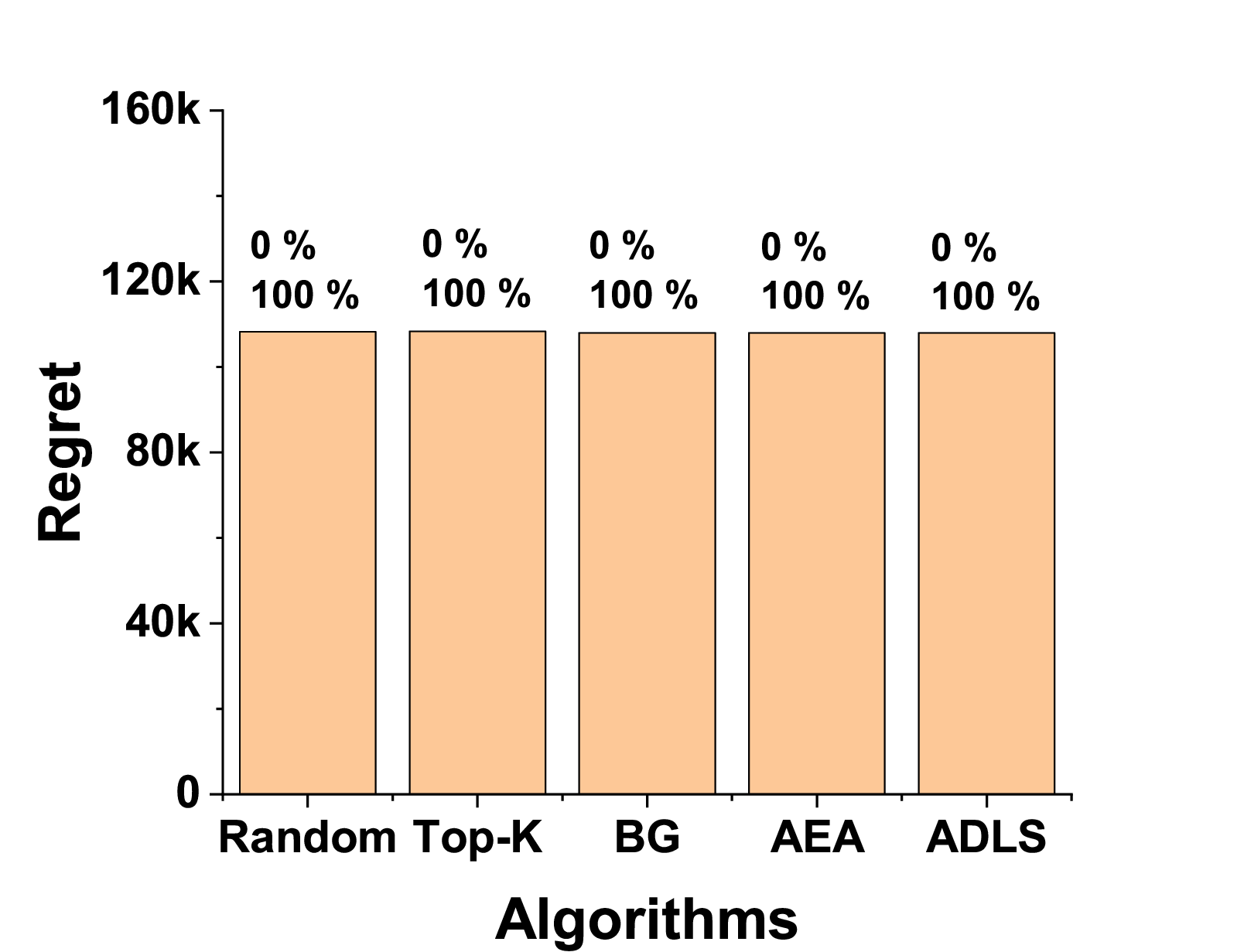} &
    \includegraphics[scale=0.10]{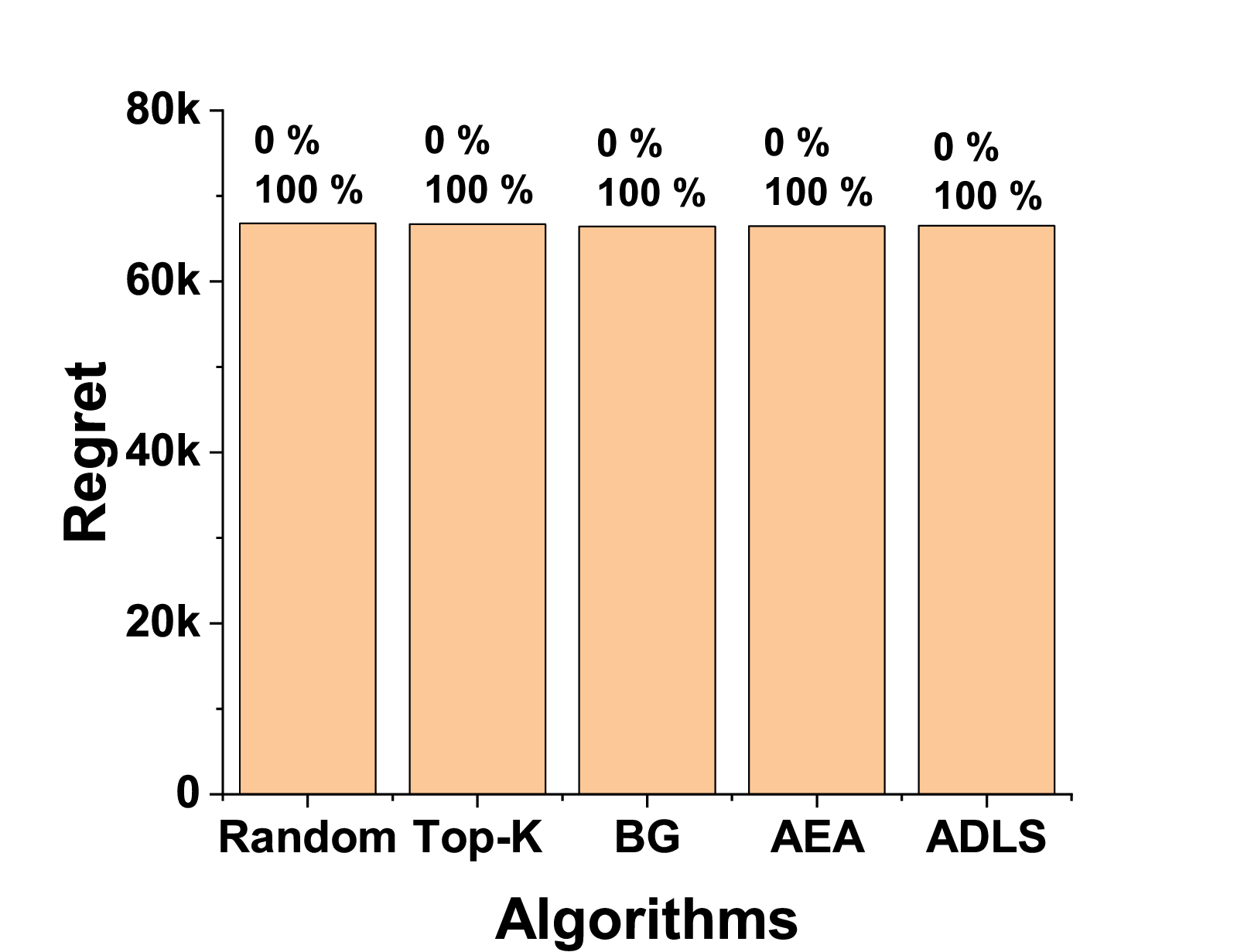} \\
    (a) $\lambda$ = $40\%$ & 
    (b) $\lambda$ = $60\%$ & 
    (c) $\lambda$ = $80\%$ & 
    (d) $\lambda$ = $100\%$ & 
    (e) $\lambda$ = $120\%$ \\
\end{tabular}
\caption{Regret of varying the demand-supply ratio $\lambda$ when $\omega = 90\%$ (Wikivote, $|\mathcal{A}| = 5$, $p_{c}: Trivalency$)}
\label{Fig:6_Influence_5}
\end{figure}

Now we analyse Trivalency probability setting results of all datasets for this case. In Email-Eu-Core dataset (Figures $\ref{Fig:2_Influence_1}$ (a) \& (b) and $\ref{Fig:2_Influence_2}$ (a) \& (b)), the best performing algorithm is ``ADLS" with an average total regret of $1920.50$. The highest regret is from ``Top-$K$", which is $779.16\%$ worse than ``ADLS". ``BG" and ``AEA" perform moderately but still show approximately $65.16\%$ higher regret than ``ADLS". In Facebook dataset (Figures $\ref{Fig:4_Influence_1}$ (a) \& (b) and $\ref{Fig:4_Influence_2}$ (a) \& (b)), the best performing algorithm is ``ADLS" with an average total regret of $1998.71$. The highest regret is from ``Top-$K$", which is $10391.13\%$ worse than ``ADLS". ``BG" and ``AEA" perform moderately but still show $38.53\%$ and $38.53\%$ higher regret than ``ADLS". In Wikivote dataset, in this case, the total regret is analyzed across different algorithms and shown in Figures $\ref{Fig:6_Influence_1}$ (a) \& (b) and $\ref{Fig:6_Influence_2}$ (a) \& (b). From the dataset results, the ``ADLS" algorithm consistently achieves the lowest total regret compared to others. For example, in the Wikivote dataset, in percentage terms, ``ADLS" achieves an improvement of approximately $10-25\%$ compared to ``BG" and ``AEA", making it the best performing algorithm in this case. In Table $\ref{tab:congressregretTrivalency}$ for Congress-Twitter dataset, ``BG" algorithm consistently achieves the lowest regret, followed closely by ``AEA" and ``ADLS". In contrast, ``Random" and ``Top-$K$" exhibit significantly higher regret, making them less effective choices. As $\lambda$ increases, regret increases across all methods, but ``BG" maintains its advantage, demonstrating superior optimization.

\paragraph{\textbf{Case 2: Low $\lambda$, High $\omega$}}  
In Case $2$, we consider scenarios where $\lambda \leq 60\%$ and $\omega \geq 50\%$. Here, the global demand remains low, but the individual influence demand is significantly higher. This implies that the influence provider serves a smaller number of advertisers with larger influence requirements. In both probability settings, we observe that as individual influence demand increases, the overall regret decreases. While global demand remains constant, the increasing individual demand leads to a reduction in excessive regret, thereby lowering total regret. However, we also notice the emergence of unsatisfied regret, indicating that some advertiser demands are not being fully met. Our algorithms—``BG", ``AEA", and ``ADLS"—consistently outperform the baseline methods. Our observations under Uniform probability setting is as follows. In Email-Eu-Core dataset (Figures $\ref{Fig:1_Influence_3}$ (a) \& (b), $\ref{Fig:1_Influence_4}$ (a) \& (b) and $\ref{Fig:1_Influence_5}$ (a) \& (b)), ``ADLS" remains the best with an average regret of $691.83$. ``BG" performs better than ``Random" and ``Top-$K$" but is $0.06\%$ worse than ``ADLS". The ``Random" algorithm has the highest regret, performing $2347.09\%$ worse than ``ADLS". In Facebook dataset (Figures $\ref{Fig:3_Influence_3}$ (a) \& (b), $\ref{Fig:3_Influence_4}$ (a) \& (b) and $\ref{Fig:3_Influence_5}$ (a) \& (b)), ``ADLS" is the best with an average regret of $1340.51$. ``BG" performs better than ``Random" and ``Top-$K$" but is $5.26\%$ worse than ``ADLS". The ``Random" algorithm has significantly high regret, performing ``5580.58\%" worse than ``ADLS". In Wikivote dataset (Figures $\ref{Fig:5_Influence_3}$ (a) \& (b), $\ref{Fig:5_Influence_4}$ (a) \& (b) and $\ref{Fig:5_Influence_5}$ (a) \& (b)), for higher individual advertiser demand, ``ADLS" still performed the best, with minimal regret values. For example, in one instance in Figure $\ref{Fig:5_Influence_3}$ (a), ``ADLS" had a total regret of $15212.98$, whereas ``Random" had $37012.21$, showing a decrease of approximately $58.9\%$. This consistent reduction highlights ``ADLS"'s ability to manage higher advertiser demands effectively. In Congress dataset, from Table $\ref{tab:congressregretuniform01}$, we observe that ``Random" and ``Top-$K$" continue to yield significantly higher regret compared to ``BG", ``AEA", and ``ADLS" in most instances. Among all methods, ``BG" consistently achieves the lowest regret, making it the best-performing algorithm in this setting. At $\omega = 90\%$, regret is negligible across all methods, indicating efficient allocation at high influence demand.

These performance trends are consistent across all datasets under Trivalency probability settings. In Email-Eu-Core (Figures $\ref{Fig:2_Influence_3}$ (a) \& (b), $\ref{Fig:2_Influence_4}$ (a) \& (b) and $\ref{Fig:2_Influence_5}$ (a) \& (b)), ``ADLS" is the best with an average regret of $5741.57$. ``BG" performs better than ``Random" and ``Top-$K$" but is $31.42\%$ worse than ``ADLS". The ``Random" algorithm has significantly high regret, performing $41.99\%$ worse than ``ADLS". In Facebook dataset (Figures $\ref{Fig:4_Influence_3}$ (a) \& (b), $\ref{Fig:4_Influence_4}$ (a) \& (b) and $\ref{Fig:4_Influence_5}$ (a) \& (b)), ``ADLS" is the best with an average regret of $6485.85$. ``BG" performs better than ``Random" and ``Top-$K$". The ``Random" algorithm has significantly high regret, performing $334.91\%$ worse than ``ADLS". In Wikivote dataset (Figures $\ref{Fig:6_Influence_3}$ (a) \& (b), $\ref{Fig:6_Influence_4}$ (a) \& (b) and $\ref{Fig:6_Influence_5}$ (a) \& (b)), for advertisers with low global demand but high individual demand, the ``ADLS" algorithm again reports the lowest regret. For Congress-Twitter dataset in Table $\ref{tab:congressregretTrivalency}$, algorithms ``BG", ``AEA", and ``ADLS" perform better than ``Random" and ``Top-$K$" in this setting. At $\omega = 90\%$, regret is minimal across all methods, indicating that higher individual influence demand leads to efficient allocation. Despite this, ``Top-$K$" and ``Random" remain the worst-performing methods.

\paragraph{\textbf{Case 3: High $\lambda$, Low $\omega$}}  
In Case $3$, we consider scenarios where $\lambda \geq 80\%$ and $\omega \leq 30\%$. Here, the global demand is high, but the individual influence demand from advertisers remains low. As $\lambda$ increases, the influence provider has access to the maximum or entire supply available to distribute among advertisers. Since $\omega$ is still low, it is expected that most or all advertisers can be provided with sufficient influence. Our observations indicate that as individual influence demand increases, regret decreases when the global influence demand remains constant. Conversely, when individual demand is held constant, increasing the global influence demand does not significantly impact regret, either positively or negatively. We will firstly understand the results from the Uniform probaility setting. In Email-Eu-Core dataset (Figures $\ref{Fig:1_Influence_1}$ (c)-(e) and $\ref{Fig:1_Influence_2}$ (c)-(e)), ``ADLS" continues to outperform with the lowest regret of $1803.08$. ``BG" and ``AEA" show moderate results but still have $13.33\%$ and $12.83\%$ higher regret than ``ADLS". ``Top-$K$" has the worst performance, with regret being $5667.82\%$ higher than ``ADLS". In Facebook dataset (Figures $\ref{Fig:3_Influence_1}$ (c)-(e) and $\ref{Fig:3_Influence_2}$ (c)-(e)), ``ADLS" continues to be the best with the lowest regret of $4537.30$. ``BG" and ``AEA" show higher regret by $34.12\%$ and $34.12\%$ compared to ``ADLS". ``Top-$K$" has the worst performance, with regret being $11147.15\%$ higher than ``ADLS". In Wikivote dataset (Figures $\ref{Fig:5_Influence_1}$ (c)-(e) and $\ref{Fig:5_Influence_2}$ (c)-(e)), when global influence demand is high but individual demand is low, ``ADLS" remained the best performing algorithm. Anotable instance in Figure $\ref{Fig:5_Influence_1}$ (c) showed ``ADLS" achieving a total regret of $3716.52$, whereas ``Random" and ``Top-$K$" had much higher values of $234796.95$ and $314244.96$, respectively. This represents a reduction of over $98\%$ compared to ``Random", emphasizing ``ADLS"’s efficiency. Both ``Random" and ``Top-$K$" exhibit extremely high regret across all scenarios, making them unsuitable choices in Congress-Twiiter dataset as listed in Table $\ref{tab:congressregretuniform01}$. For example, at $\lambda = 80\%$, $\omega = 10\%$, with $|A|=100$, the regret for ``Top-$K$" is $49987.67$, while ``AEA" achieves a significantly lower regret of $29574.21$. In most cases, ``AEA" and ``ADLS" outperform ``BG" by $10-30\%$, demonstrating superior regret minimization.

Under the Trivalency probability setting, our proposed algorithms continue to perform better across all datasets, further validating their robustness in minimizing regret. To support our statement we see in Figures $\ref{Fig:2_Influence_1}$ (c)-(e) and $\ref{Fig:2_Influence_2}$ (c)-(e) of Email-Eu-Core dataset that ``ADLS" continues to be the best with the lowest regret. ``BG" and ``AEA" show higher regret by maximum $85.22\%$ compared to ``ADLS". ``Top-$K$" has the worst performance, with regret being $204.37\%$ higher than ``ADLS". In Facebook dataset (Figures $\ref{Fig:4_Influence_1}$ (c)-(e) and $\ref{Fig:4_Influence_2}$ (c)-(e)), ADLS continues to be the best with the lowest average regret of $2237.80$. ``BG" and ``AEA" show higher regret by $79.02\%$ and $79.02\%$ compared to ``ADLS". ``Top-$K$" has the worst performance, with regret being $7868.41\%$ higher than ``ADLS". In Wikivote dataset (Figures $\ref{Fig:6_Influence_1}$ (c)-(e) and $\ref{Fig:6_Influence_2}$ (c)-(e)) for advertisers with high global demand but low individual demand, ``ADLS" remains the best performer. ``Regret" values are the highest across all methods in Table $\ref{tab:congressregretTrivalency}$ for Congress-Twitter dataset, suggesting that low individual influence demand ($\omega$) results in inefficient allocation. The ``BG" algorithm performs better than ``Top-$K$" and ``Random", but at $\lambda = 120\%$, regret reaches its peak across all methods, confirming that high global influence demand with low individual demand is problematic.

\paragraph{\textbf{Case 4: High $\lambda$, High $\omega$}}  
In Case $4$, we consider scenarios where $\lambda \geq 80\%$ and $\omega \geq 50\%$. Here, both global demand and individual influence demand are high. As the global demand $\lambda$ increases, excessive regret decreases, but unsatisfied regret increases. Consequently, each unsatisfied advertiser contributes to a higher total regret. When individual demand increases while keeping the global influence demand constant, regret decreases. However, when global influence demand increases, regret also increases. This is because a higher global influence demand results in a proportional increase in individual influence demand, thereby contributing to higher regret. The results of Uniform probbility setting are as follows. In Email-Eu-Core dataset (Figures $\ref{Fig:1_Influence_3}$ (c)-(e), $\ref{Fig:1_Influence_4}$ (c)-(e) and $\ref{Fig:1_Influence_5}$ (c)-(e)), ``ADLS" achieves the lowest average regret at $7949.21$. ``BG" and ``AEA" are significantly worse, with regrets maximum to $4.43\%$ higher than ``ADLS". The ``Random" algorithm performs the worst again, with a regret $61.48\%$ higher than ``ADLS". In Facebook dataset (Figures $\ref{Fig:3_Influence_3}$ (c)-(e), $\ref{Fig:3_Influence_4}$ (c)-(e) and $\ref{Fig:3_Influence_5}$ (c)-(e)), ``ADLS" achieves the lowest average regret at $10556.87$. ``BG" and ``AEA" perform significantly worse, with regrets $1.46\%$ and $1.60\%$ higher than ``ADLS". The ``Random" algorithm is again the worst, with a regret that is $417.71\%$ higher than ``ADLS". In Wikivote dataset (Figures $\ref{Fig:5_Influence_3}$ (c)-(e), $\ref{Fig:5_Influence_4}$ (c)-(e) and $\ref{Fig:5_Influence_5}$ (c)-(e)), in the scenario where both global and individual influence demands are high, ``ADLS" continued to outperform the other algorithms. In one case (Figure $\ref{Fig:5_Influence_5}$ (e)), it achieved a total regret of $210489.62$, whereas ``Random" had $210875.27$, indicating an overall improvement. While the percentage difference is lower here, ``ADLS" still maintains its superiority in regret minimization. In Congress-Twitter dataset, from the Table $\ref{tab:congressregretuniform01}$ we see that as $\omega$ increases from $50\%$ to $70\%$, regret values drop significantly across all methods, indicating that higher individual influence demand leads to better optimization. For example, at $\omega = 50\%$, $\lambda = 80\%$, the regret values are as follows.``BG" has lowest regret of $612.18$, ``AEA" has $544.19$, and ``ADLS" has $183.18$. While ``ADLS" is not always the best, it achieves the lowest regret in certain instances, particularly when $\lambda = 80\%$, $\omega = 70\%$. ``Top-$K$" should be avoided, as it consistently exhibits the highest regret across all scenarios.

Our observation for Trivalency probability setting is as follows. For Email-Eu-Core dataset (Figures $\ref{Fig:2_Influence_3}$ (c)-(e), $\ref{Fig:2_Influence_4}$ (c)-(e) and $\ref{Fig:2_Influence_5}$ (c)-(e)), ``ADLS" achieves the lowest average regret at $20015.18$. ``BG" and ``AEA" perform significantly worse, with regrets $12.25\%$ and $8.49\%$ higher than ``ADLS". The ``Random" algorithm is again the worst, with a regret that is $6.05\%$ higher than ``ADLS". In Facebook dataset (Figures $\ref{Fig:4_Influence_3}$ (c)-(e), $\ref{Fig:4_Influence_4}$ (c)-(e) and $\ref{Fig:4_Influence_5}$ (c)-(e)), ``ADLS" achieves the lowest average regret at $45297.33$. ``BG" and ``AEA" perform significantly worse, with regrets $1.51\%$ and $0.02\%$ higher than ``ADLS". The ``Random" algorithm is again the worst, with a regret that is $19.59\%$ higher than ``ADLS". In Wikivote dataset (Figures $\ref{Fig:6_Influence_3}$ (c)-(e), $\ref{Fig:6_Influence_4}$ (c)-(e) and $\ref{Fig:6_Influence_5}$ (c)-(e)), when both global and individual demand are high, the results indicate that ``ADLS" continues to be the optimal algorithm. For instance, in the Wikivote dataset, ``ADLS" records a total regret of around $1294$, while ``BG" and ``AEA" exceed $1500$. The percentage difference highlights a $20-25\%$ improvement, reinforcing the robustness of ``ADLS" across different demand scenarios. In Trivalency probability setting, for Congress-Twitter dataset, regret values drop significantly as observed in Table $\ref{tab:congressregretTrivalency}$, reinforcing the idea that higher $\omega$ leads to better optimization. The algorithms ``BG", ``AEA", and ``ADLS" perform similarly, with minimal differences in regret values. At $\omega = 70\%$, all methods converge to similar regret values, indicating diminishing returns of increasing individual demand. Despite the improvement in regret, ``Top-K" and ``Random" remain the least effective choices.

\paragraph{\textbf{Comprehensive Observation}} 
In conclusion, our analysis across all four cases reveals that the performance of the evaluated algorithms is highly sensitive to the variations in global ($\lambda$) and individual ($\omega$) influence demands. In scenarios with low $\lambda$ and $\omega$ (Case $1$), where the overall demand is modest, the ``ADLS" algorithm consistently minimizes regret significantly better than alternatives such as ``Top-$K$" and ``Random", while ``BG" and ``AEA" perform moderately well. When individual influence demand is high despite low global demand (Case $2$), the overall regret decreases, yet unsatisfied advertiser demands emerge, with ``ADLS" maintaining strong performance and ``BG" often leading in specific instances. In cases where global demand is high but individual demand remains low (Case $3$), the abundance of available influence results in lower regret levels, again favoring ``ADLS", while ``BG" and ``AEA" show competitive performance. Finally, when both global and individual demands are high (Case $4$), although excessive regret decreases and unsatisfied regret becomes a concern, ``ADLS" consistently achieves the lowest regret, and the performance gap between ``ADLS" and the other methods narrows, with ``BG" and ``AEA" trailing only slightly in some settings. Overall, these findings underscore the robustness of the ``ADLS" algorithm in efficiently allocating influence across diverse demand scenarios, while simpler approaches such as ``Top-$K$" and ``Random" prove to be considerably less effective.

\begin{figure}[H]
    \centering
    \begin{tabular}{ccc}
        \includegraphics[scale=0.15]{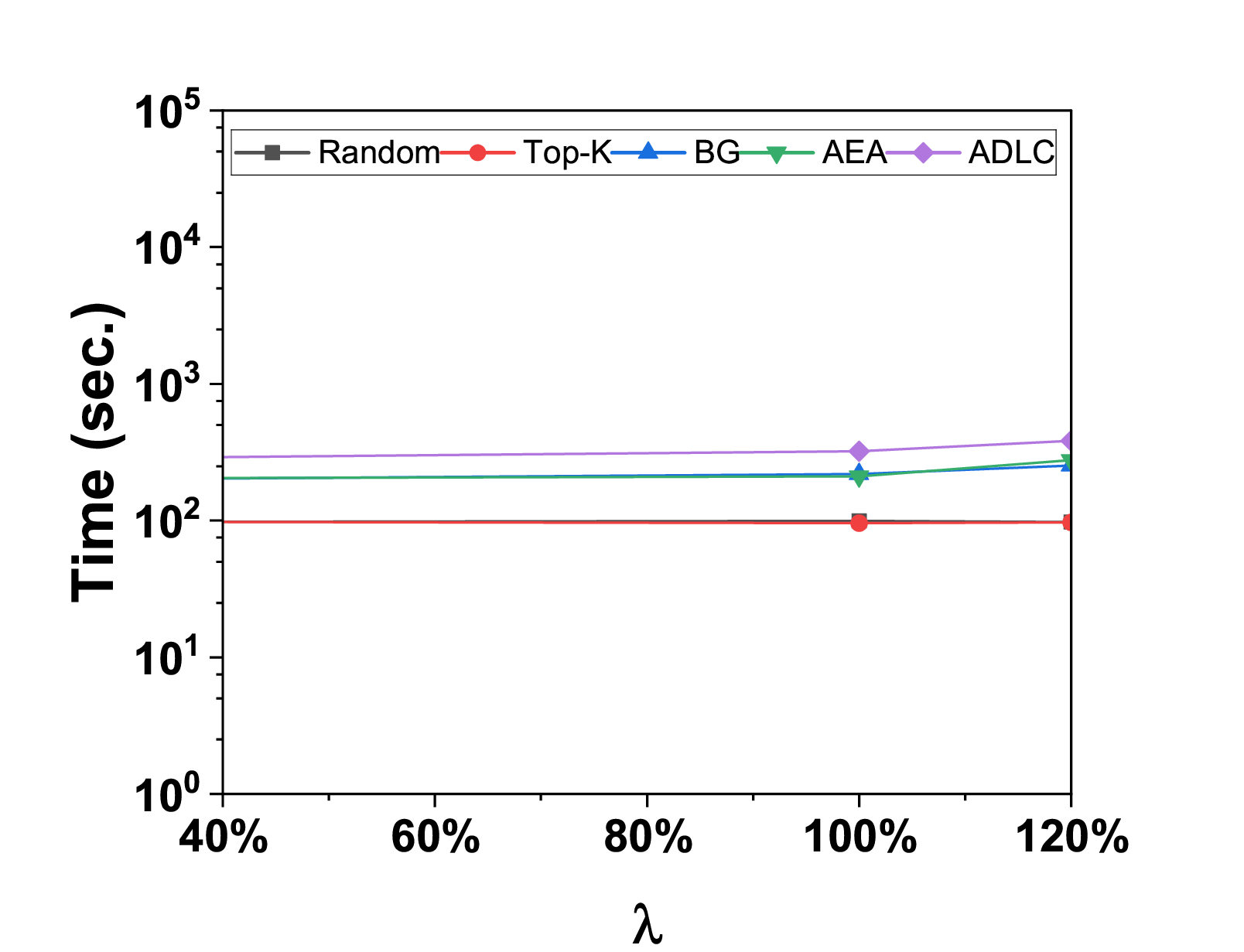} &
        \includegraphics[scale=0.15]{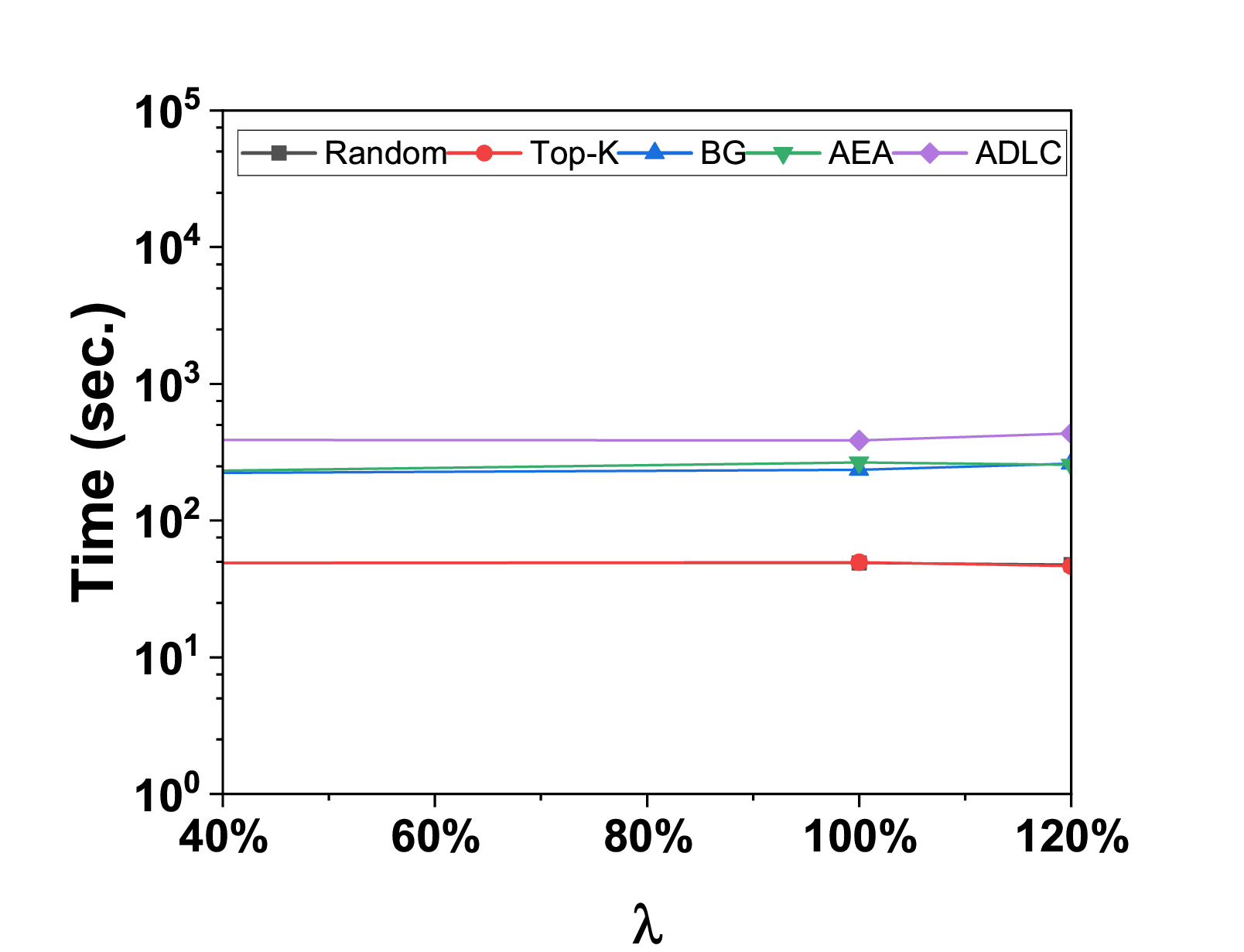} &
        \includegraphics[scale=0.15]{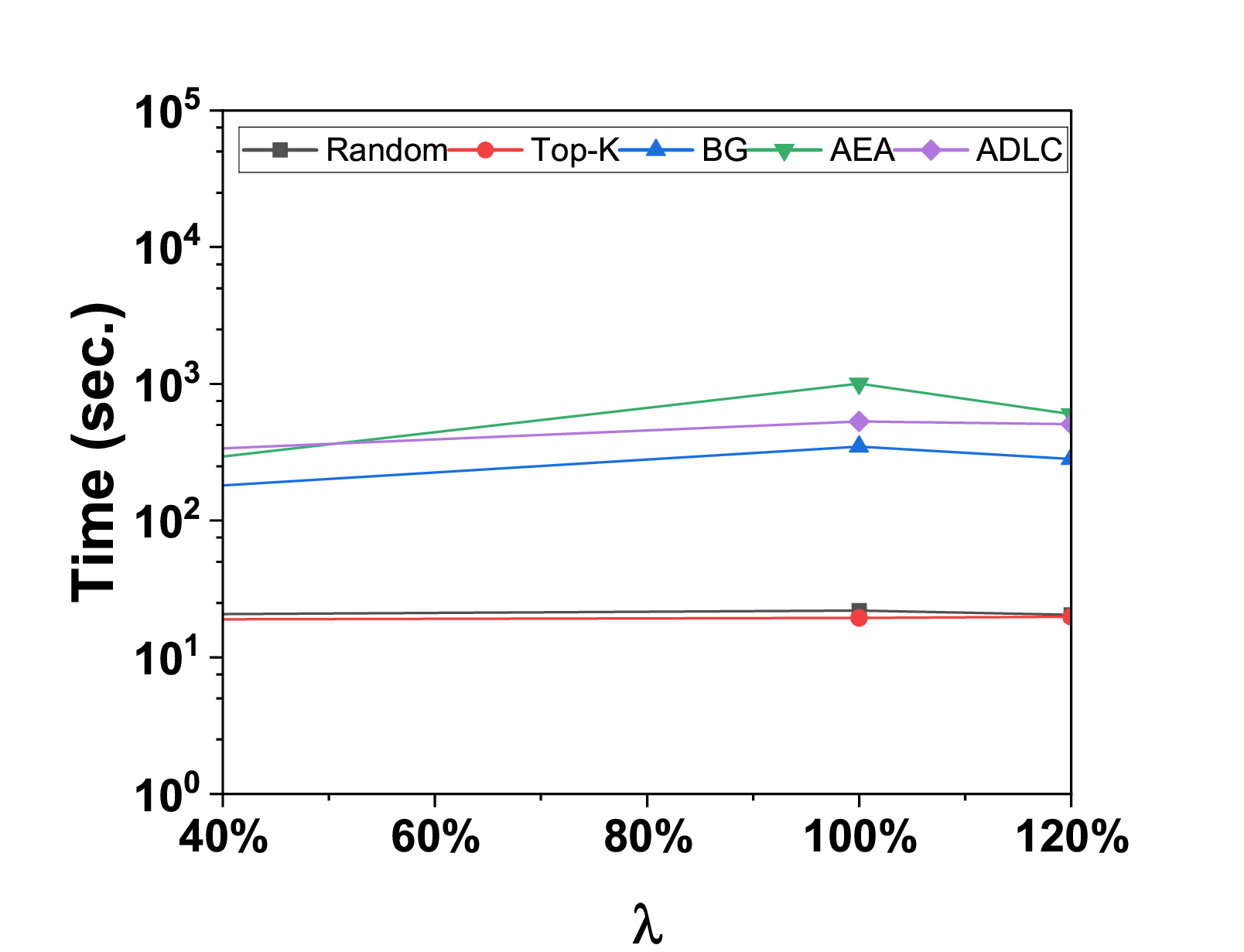} \\
        {\tiny (a) $\omega = 10\% (|\mathcal{A}| = 100)$} & 
        {\tiny (b) $\omega = 30\% (|\mathcal{A}| = 50)$} &
        {\tiny (c) $\omega = 50\% (|\mathcal{A}| = 20)$} \\
    \end{tabular}
    \vspace{0.5cm} 
    \begin{tabular}{cc}
        \includegraphics[scale=0.15]{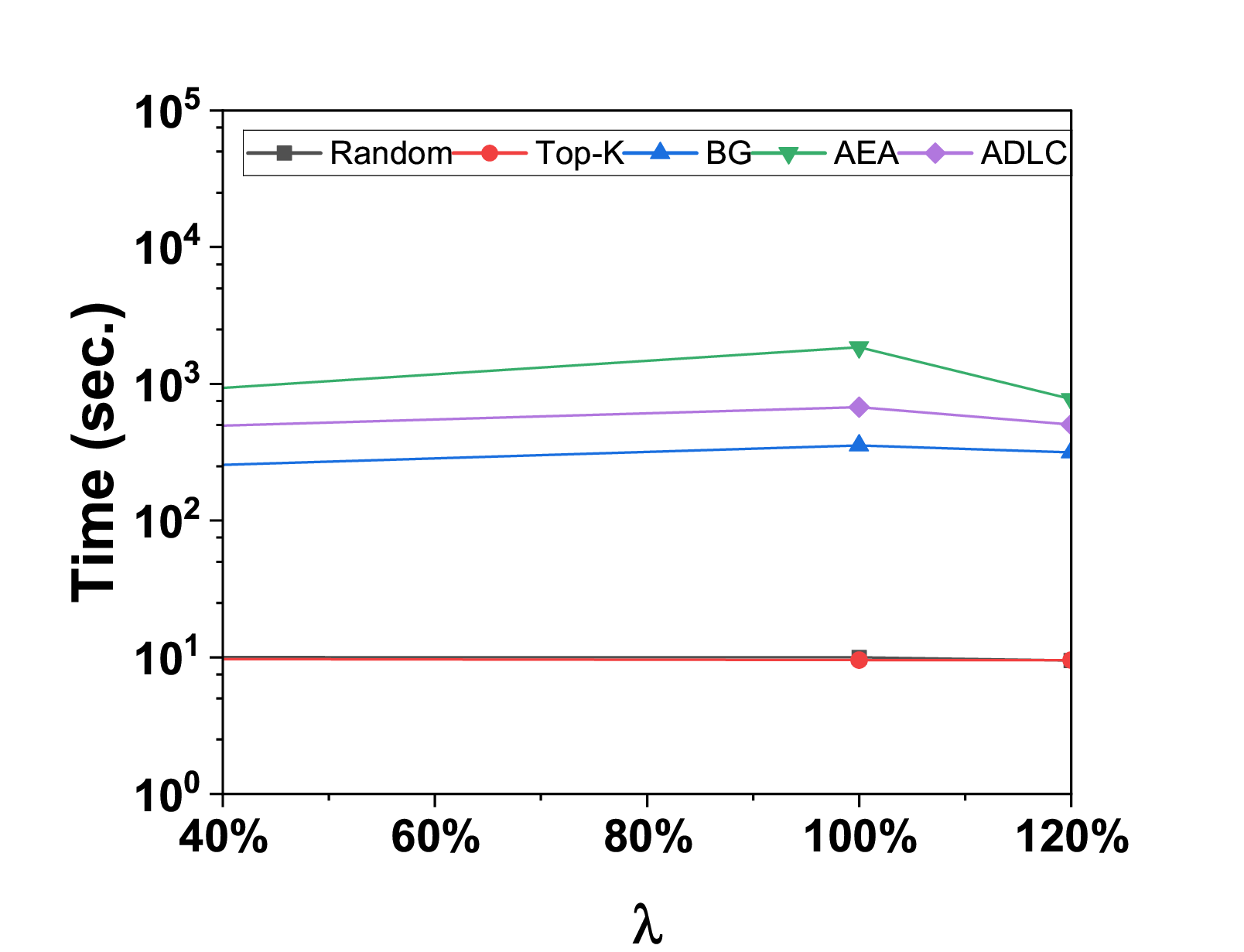} &
        \includegraphics[scale=0.15]{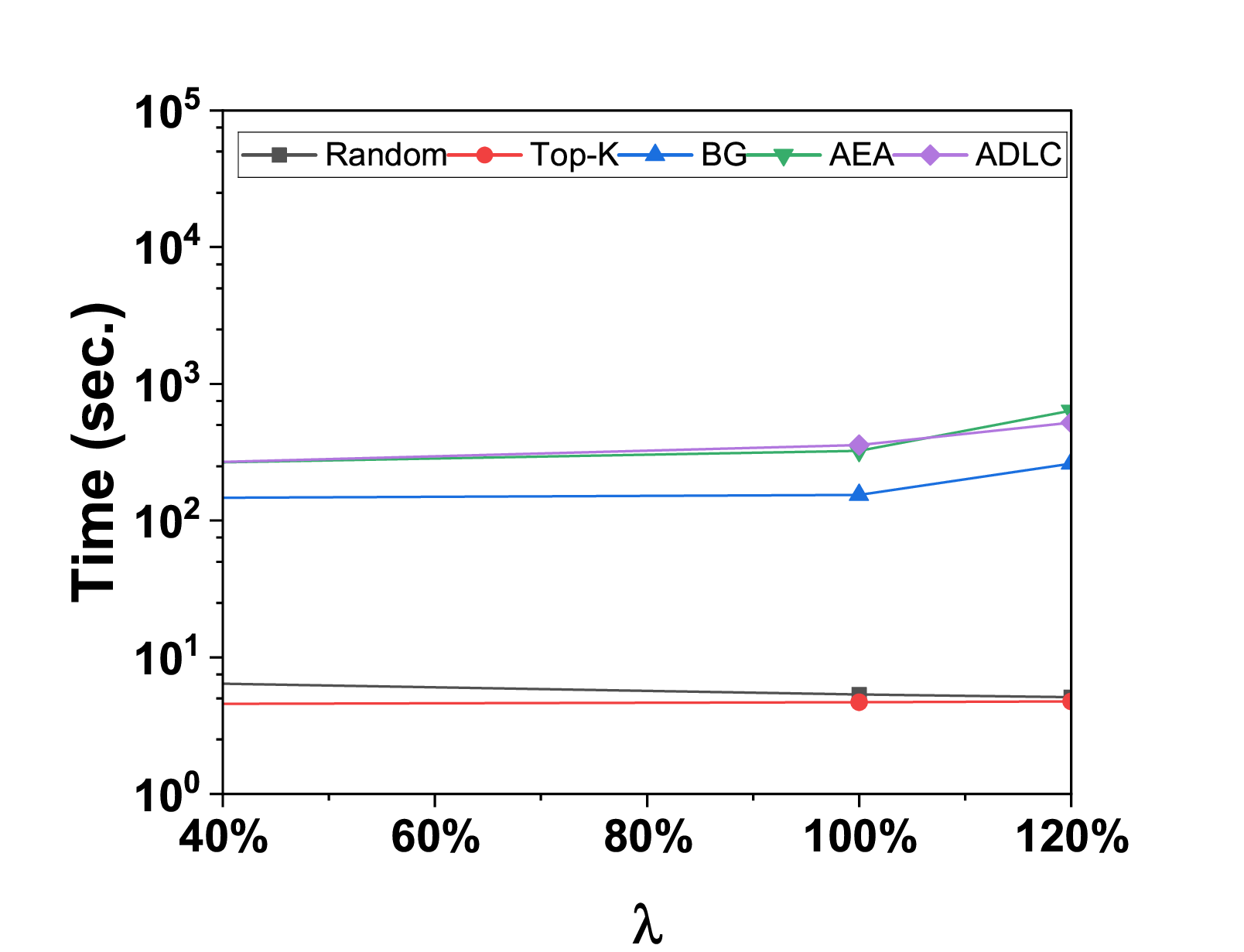} \\
        {\tiny (d) $\omega = 70\% (|\mathcal{A}| = 10)$} &
        {\tiny (e) $\omega = 90\% (|\mathcal{A}| = 5)$} \\
    \end{tabular}

    \caption{Efficiency Study Uniform (0.1) Probability Setting (Email-Eu-Core)}
    \label{Fig:7_Time0.1}
\end{figure}

\begin{figure}[H]
 \centering
\begin{tabular}{ccc}
   \includegraphics[scale=0.15]{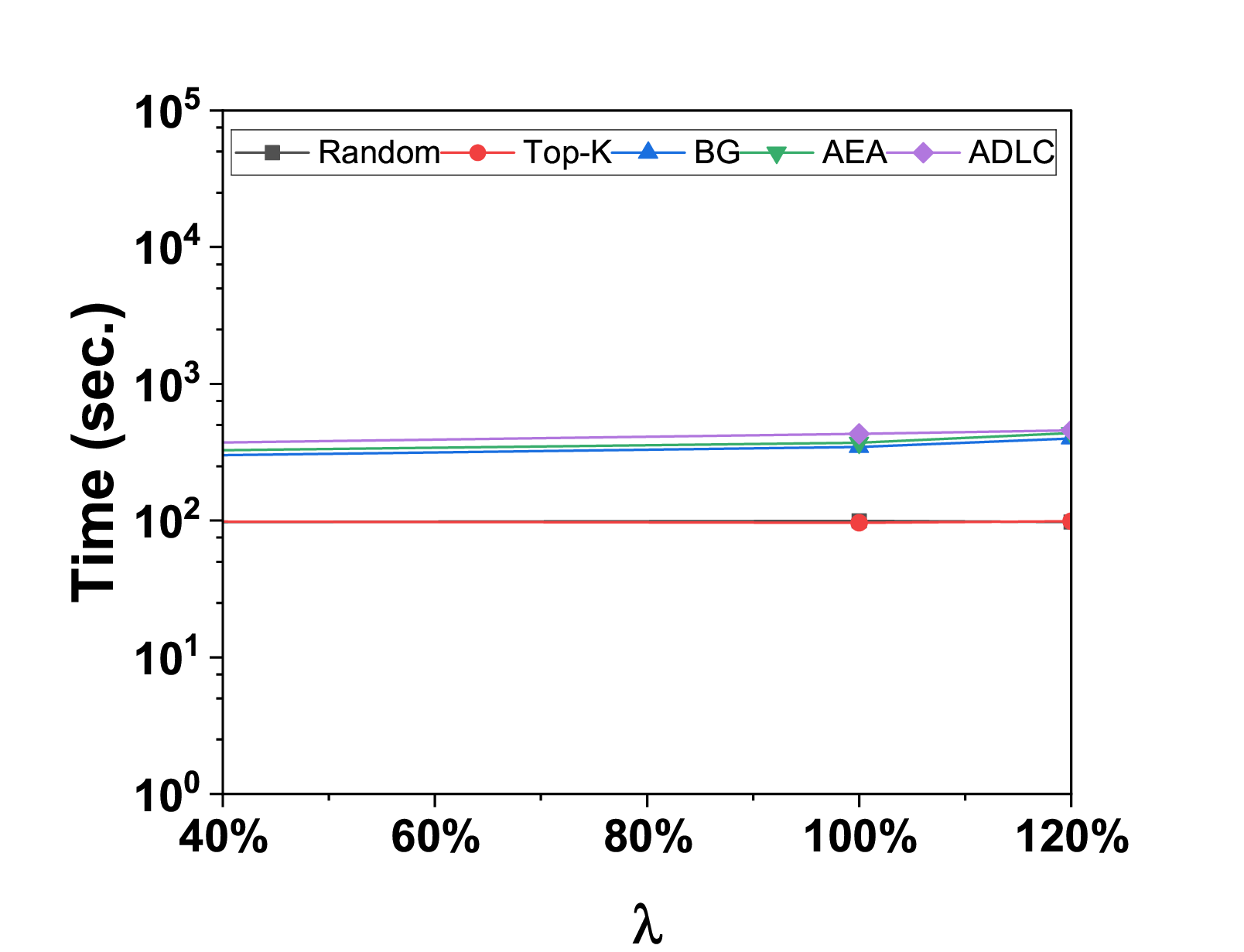} & 
   \includegraphics[scale=0.15]{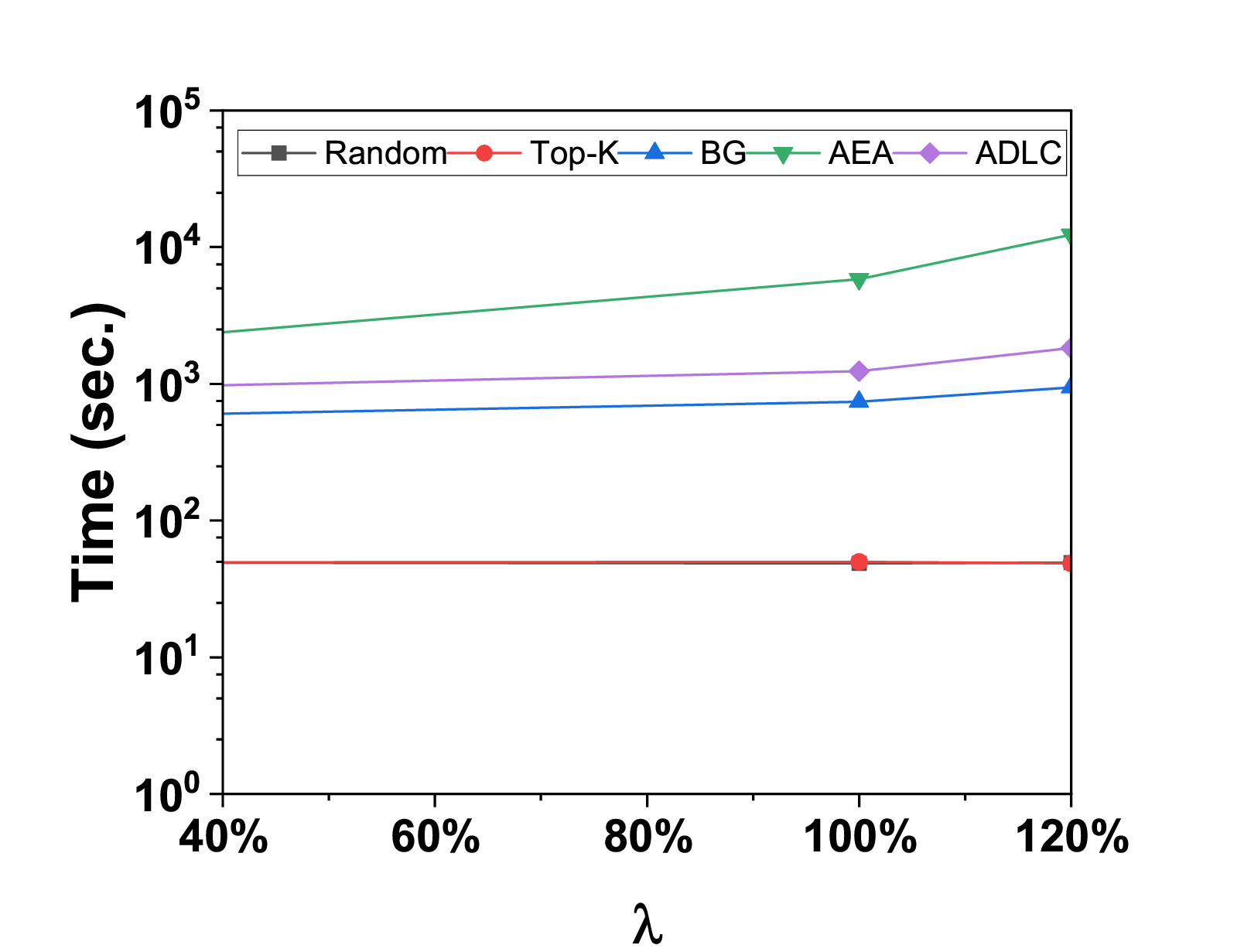} & 
   \includegraphics[scale=0.15]{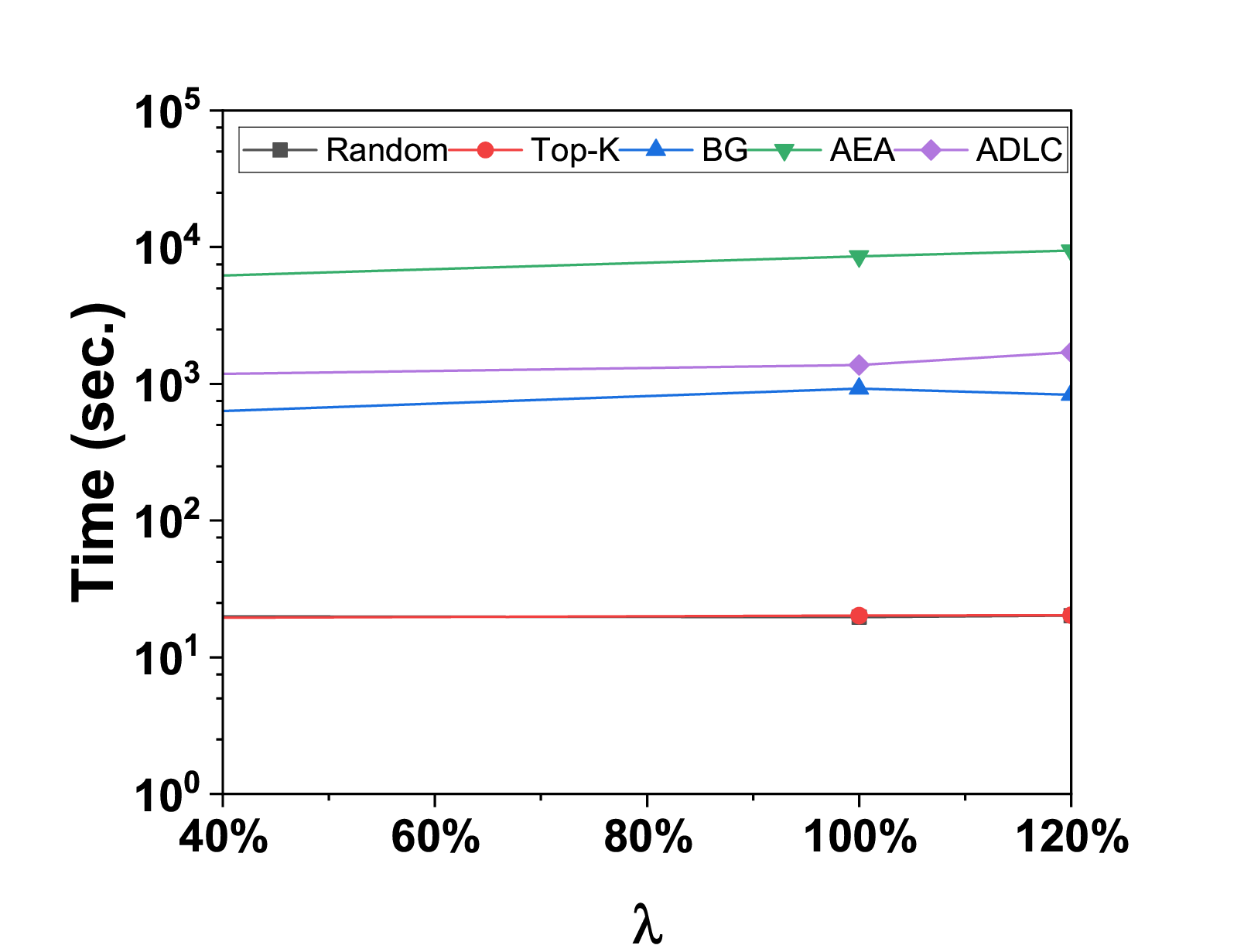} \\
         {\tiny (a) $\omega = 10\% (|\mathcal{A}| = 100)$} & 
        {\tiny (b) $\omega = 30\% (|\mathcal{A}| = 50)$} &
        {\tiny (c) $\omega = 50\% (|\mathcal{A}| = 20)$} \\
    \end{tabular} 
   
	\vspace{0.5cm} 
   
   \begin{tabular}{cc}
   \includegraphics[scale=0.15]{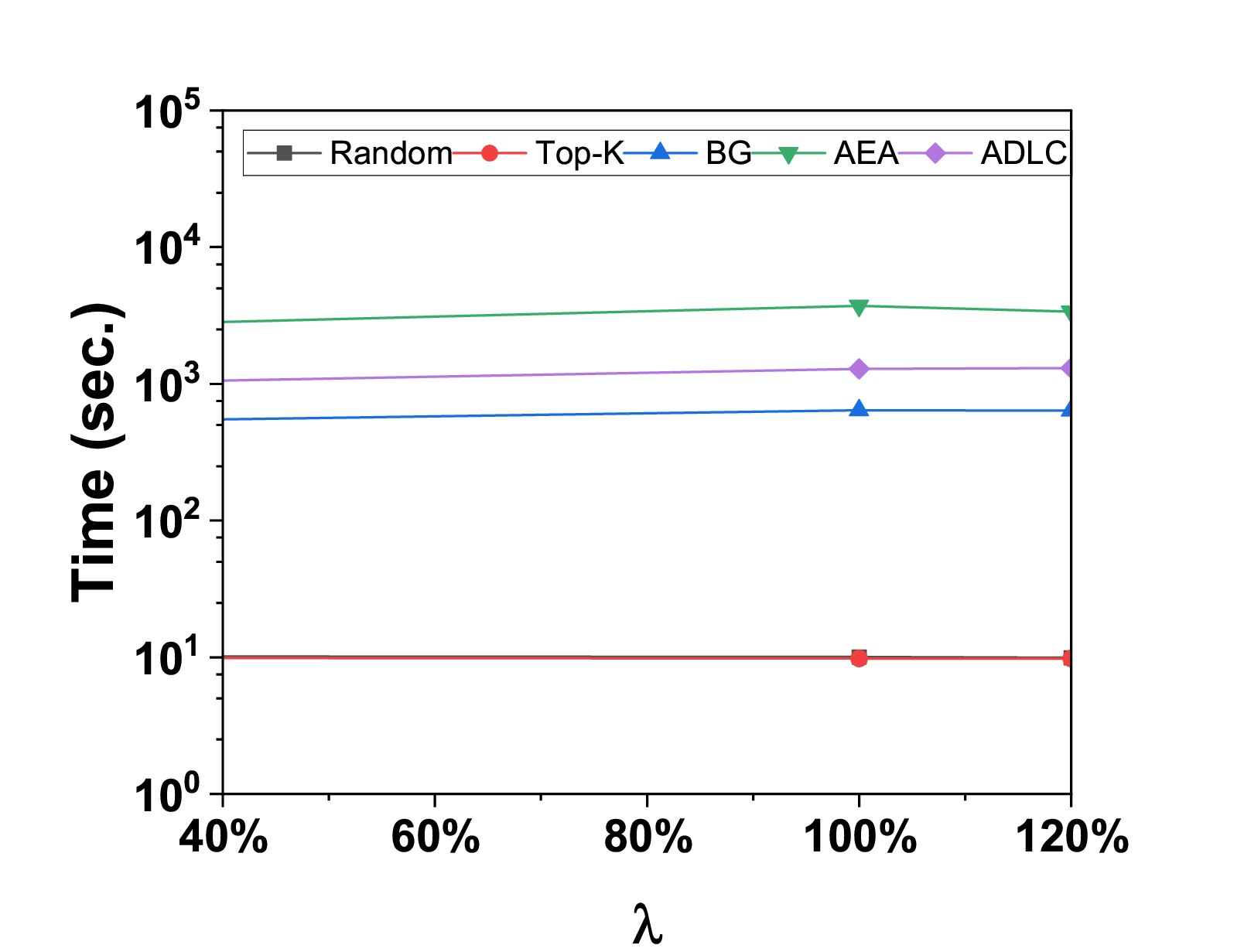} &
   \includegraphics[scale=0.15]{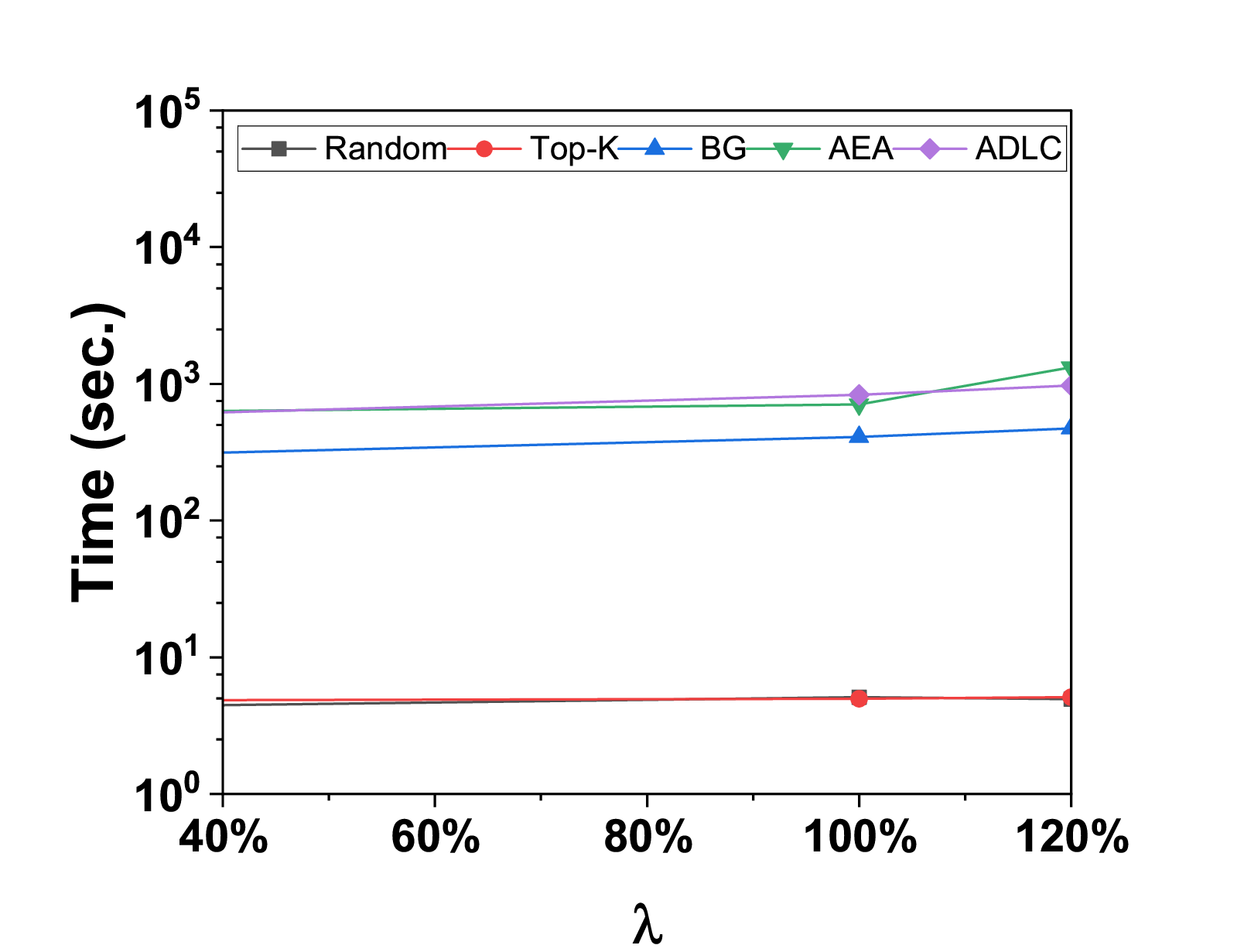} \\
    {\tiny (d) $\omega = 70\% (|\mathcal{A}| = 10)$} & 
    {\tiny (e) $\omega = 90\% (|\mathcal{A}| = 5)$} \\
\end{tabular}
\caption{Efficiency Study for Trivalency Probability Setting (Email-Eu-Core)}
\label{Fig:8_TimeTV}
\end{figure}

\begin{figure}[H]
    \centering
    \begin{tabular}{ccc}
        \includegraphics[scale=0.15]{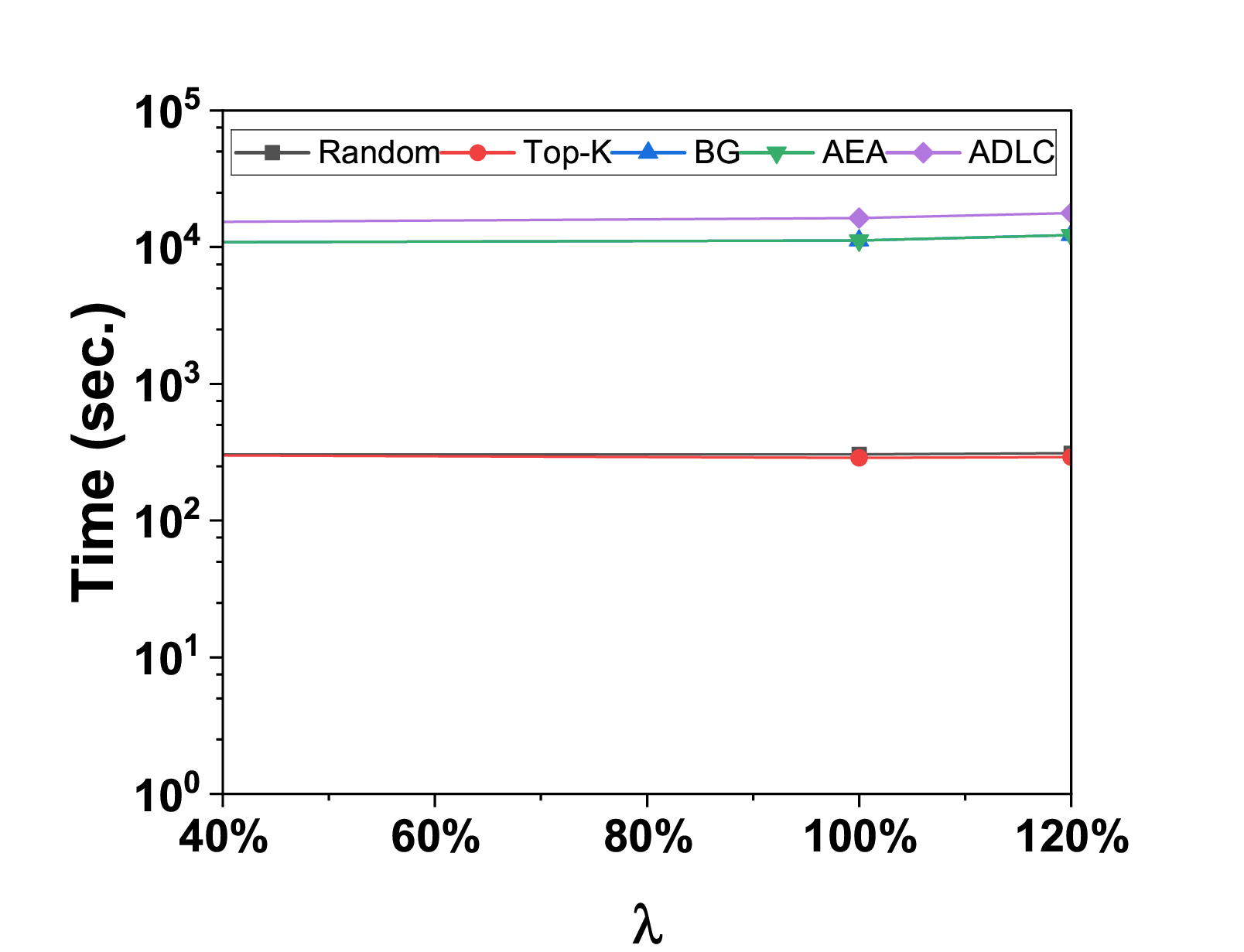} &
        \includegraphics[scale=0.15]{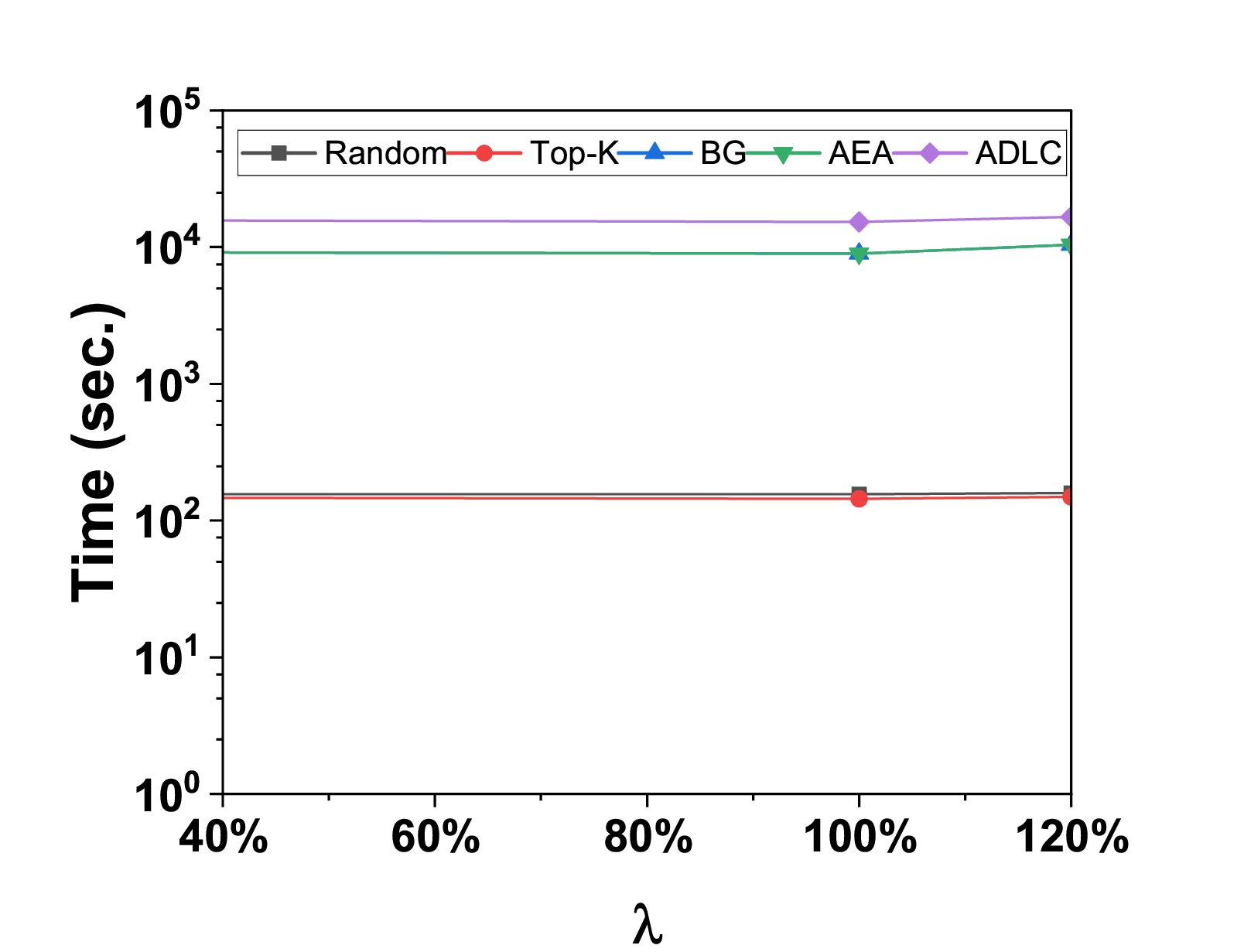} &
        \includegraphics[scale=0.15]{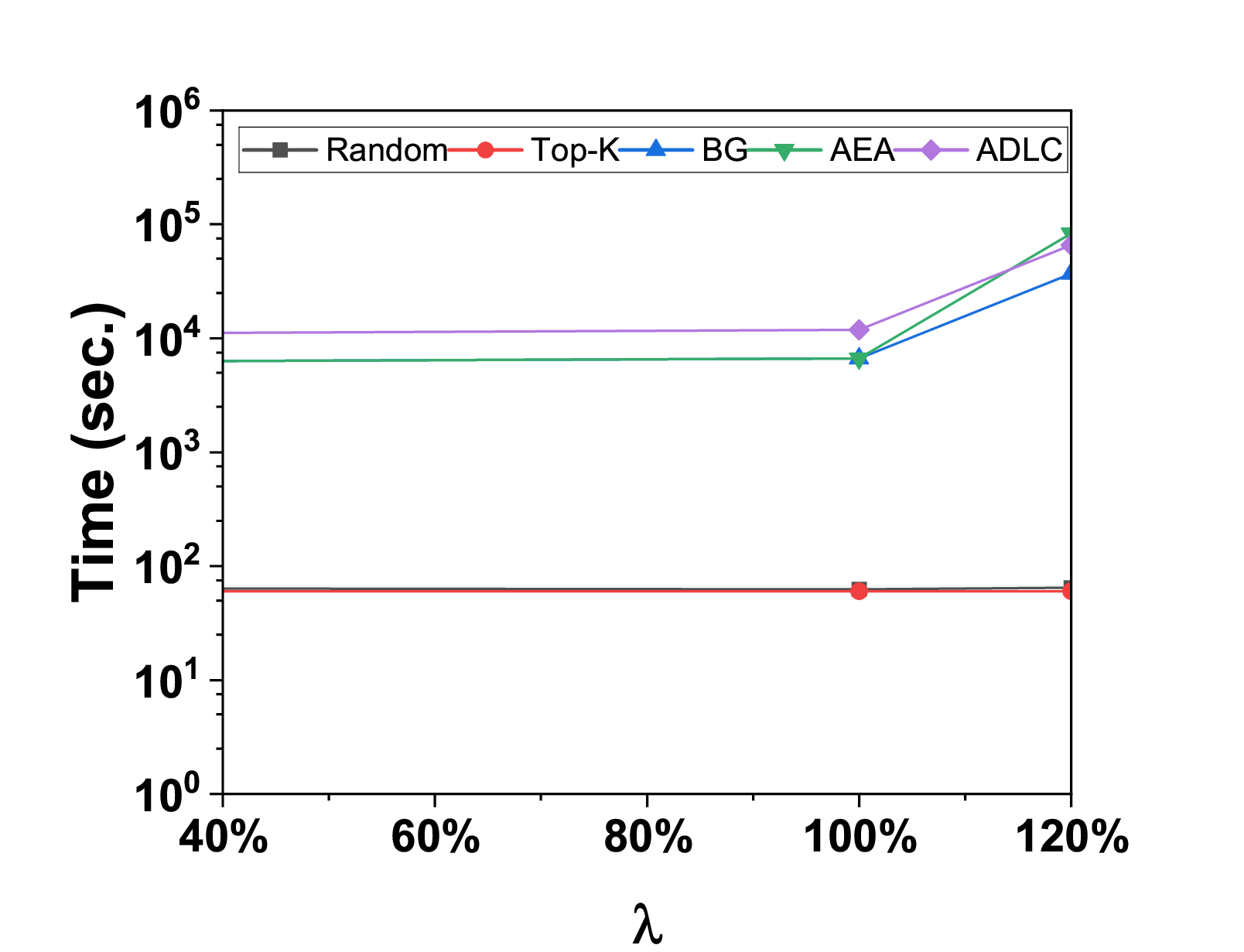} \\
        {\tiny (a) $\omega = 10\% (|\mathcal{A}| = 100)$} & 
        {\tiny (b) $\omega = 30\% (|\mathcal{A}| = 50)$} &
        {\tiny (c) $\omega = 50\% (|\mathcal{A}| = 20)$} \\
    \end{tabular}

    \vspace{0.5cm} 

    \begin{tabular}{cc}
        \includegraphics[scale=0.15]{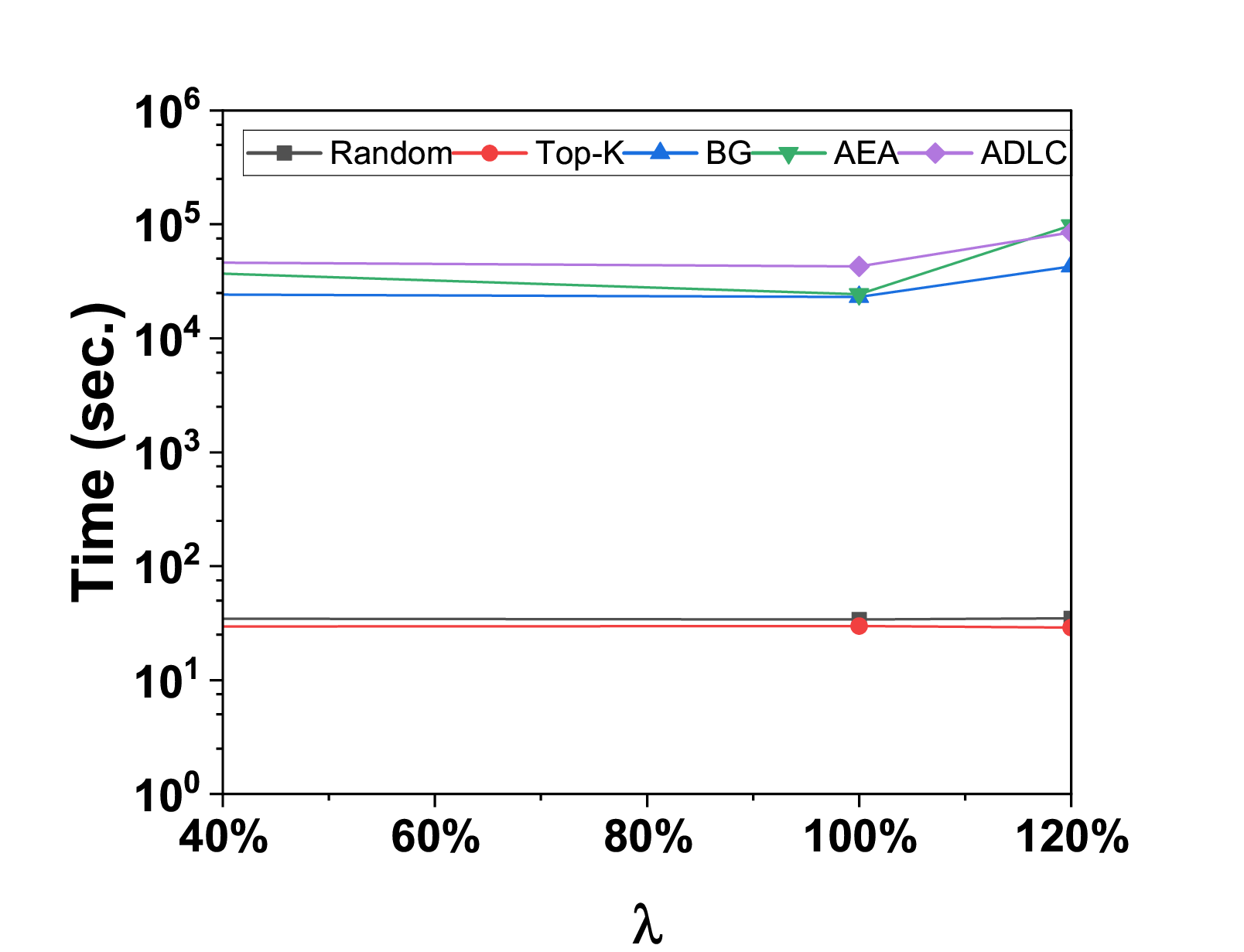} &
        \includegraphics[scale=0.15]{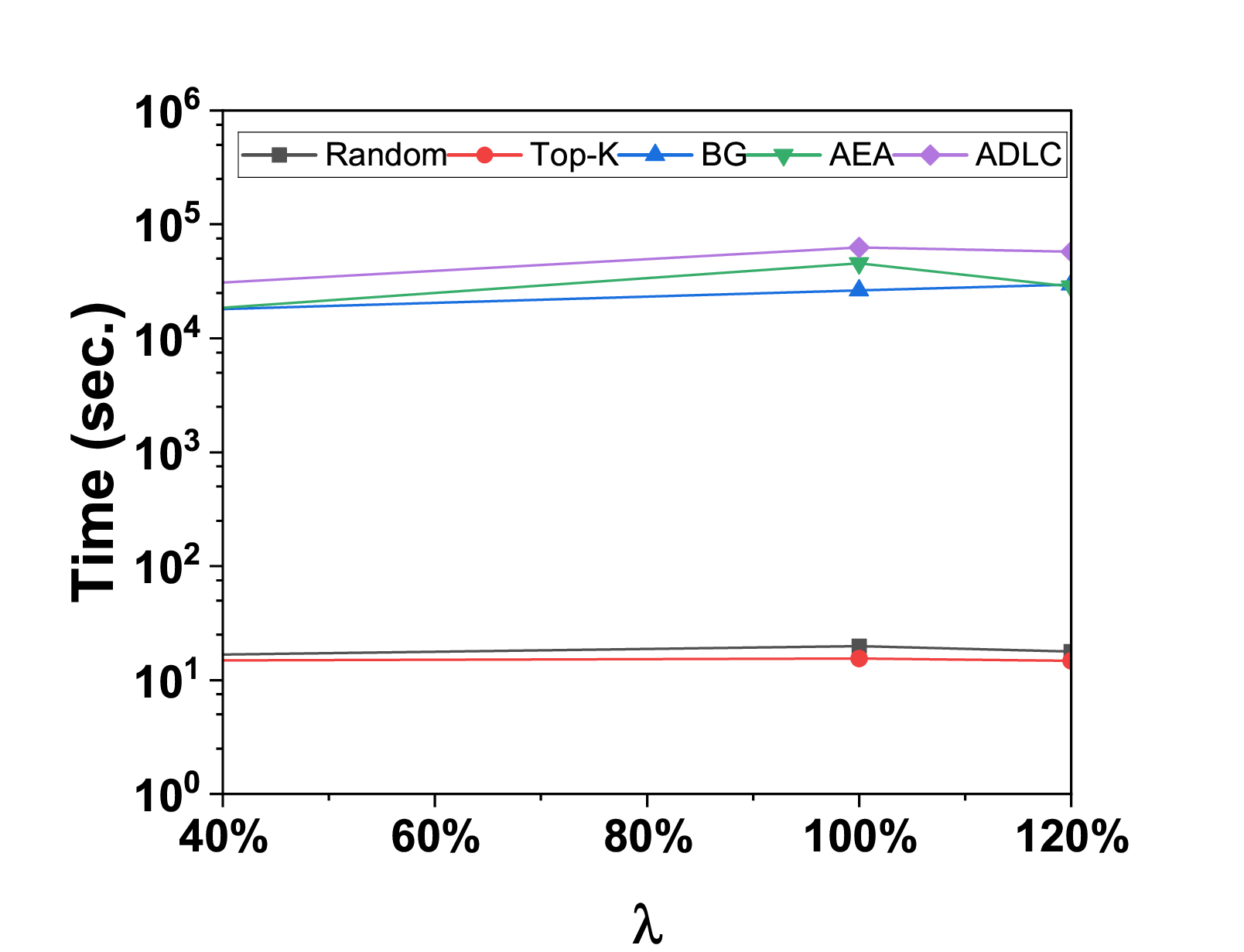} \\
        {\tiny (d) $\omega = 70\% (|\mathcal{A}| = 10)$} &
        {\tiny (e) $\omega = 90\% (|\mathcal{A}| = 5)$} \\
    \end{tabular}

    \caption{Efficiency Study Uniform (0.1) Probability Setting (Facebook)}
    \label{Fig:9_Time0.1}
\end{figure}

\begin{figure}[H]
 \centering
\begin{tabular}{ccc}
   \includegraphics[scale=0.15]{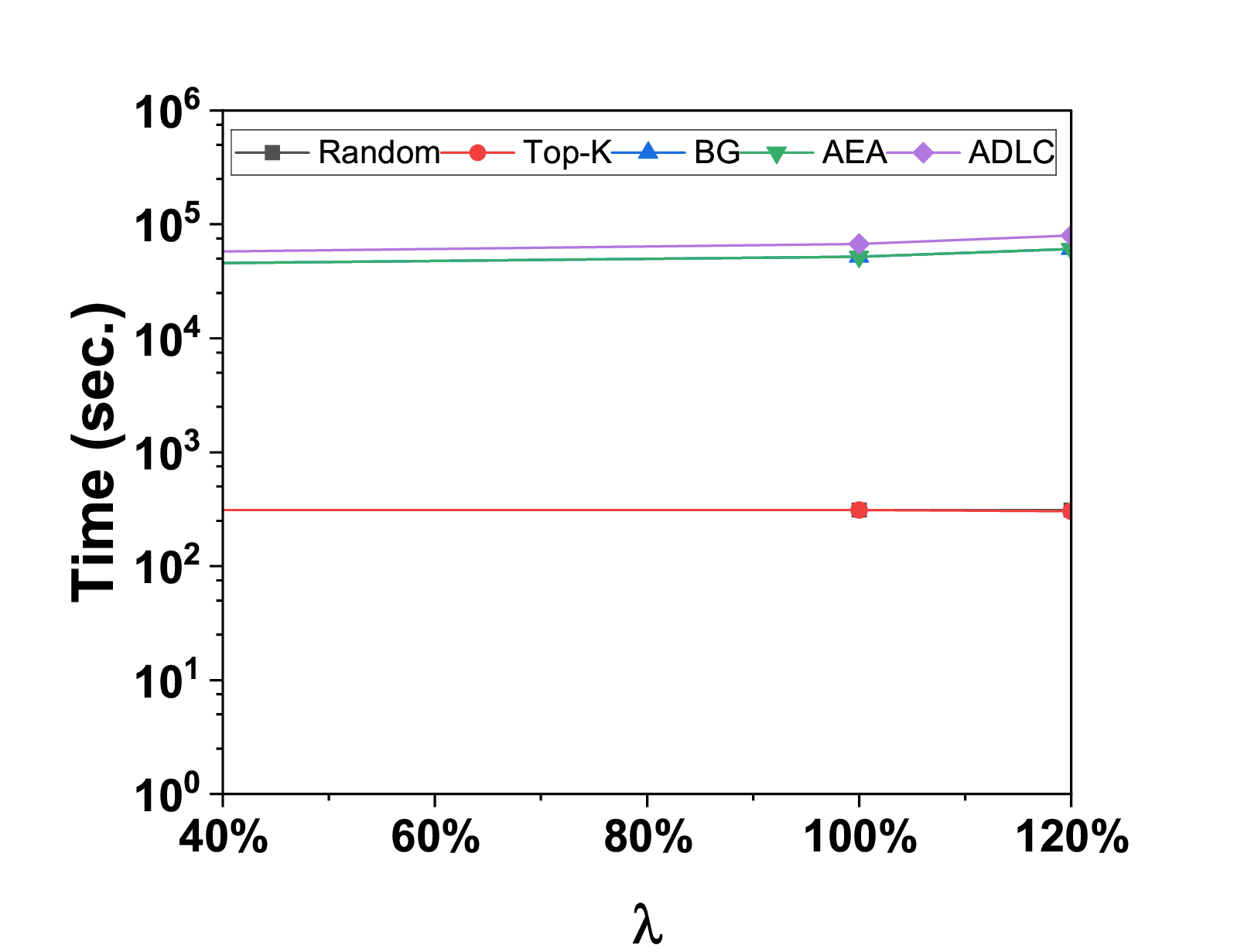} & 
   \includegraphics[scale=0.15]{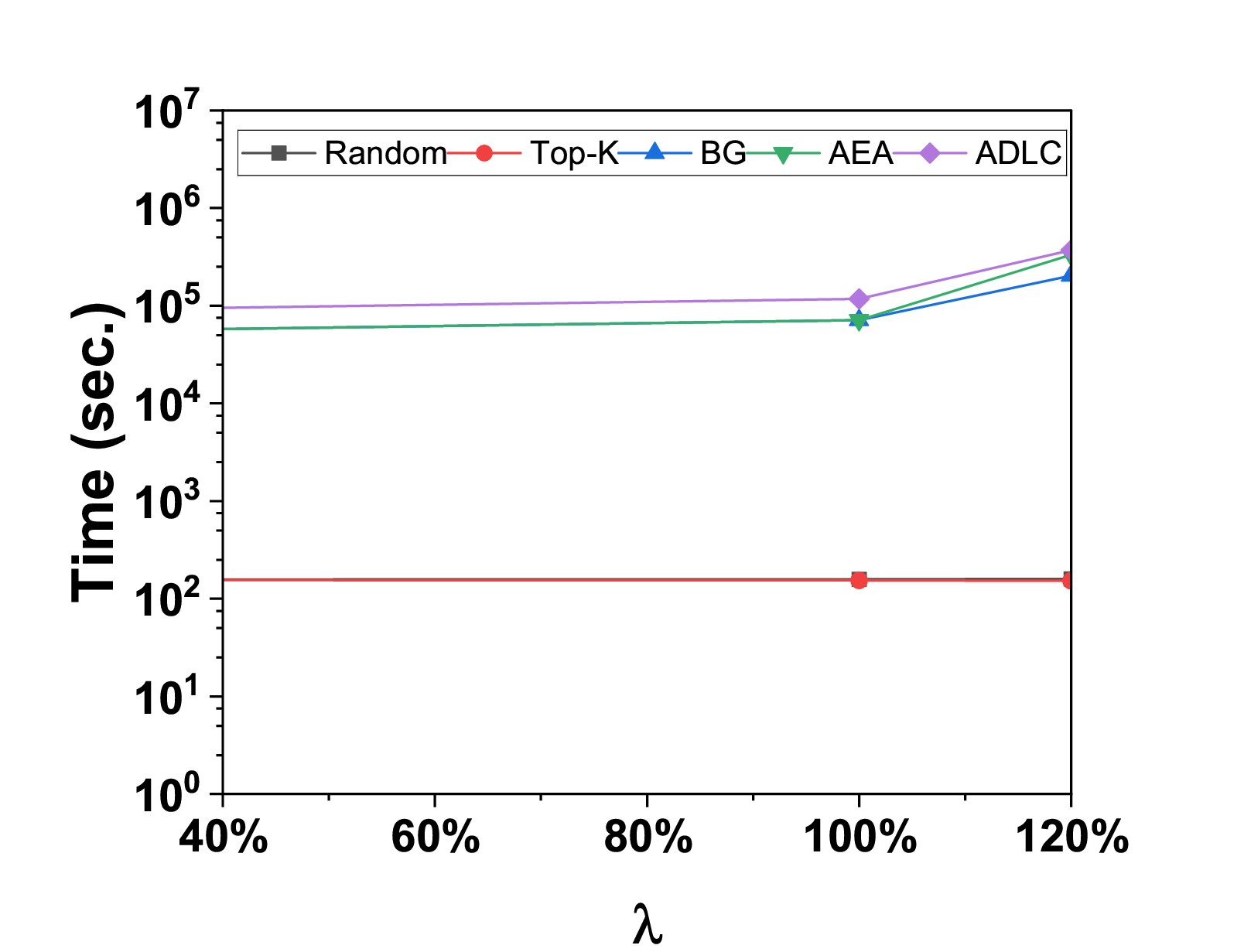} & 
   \includegraphics[scale=0.15]{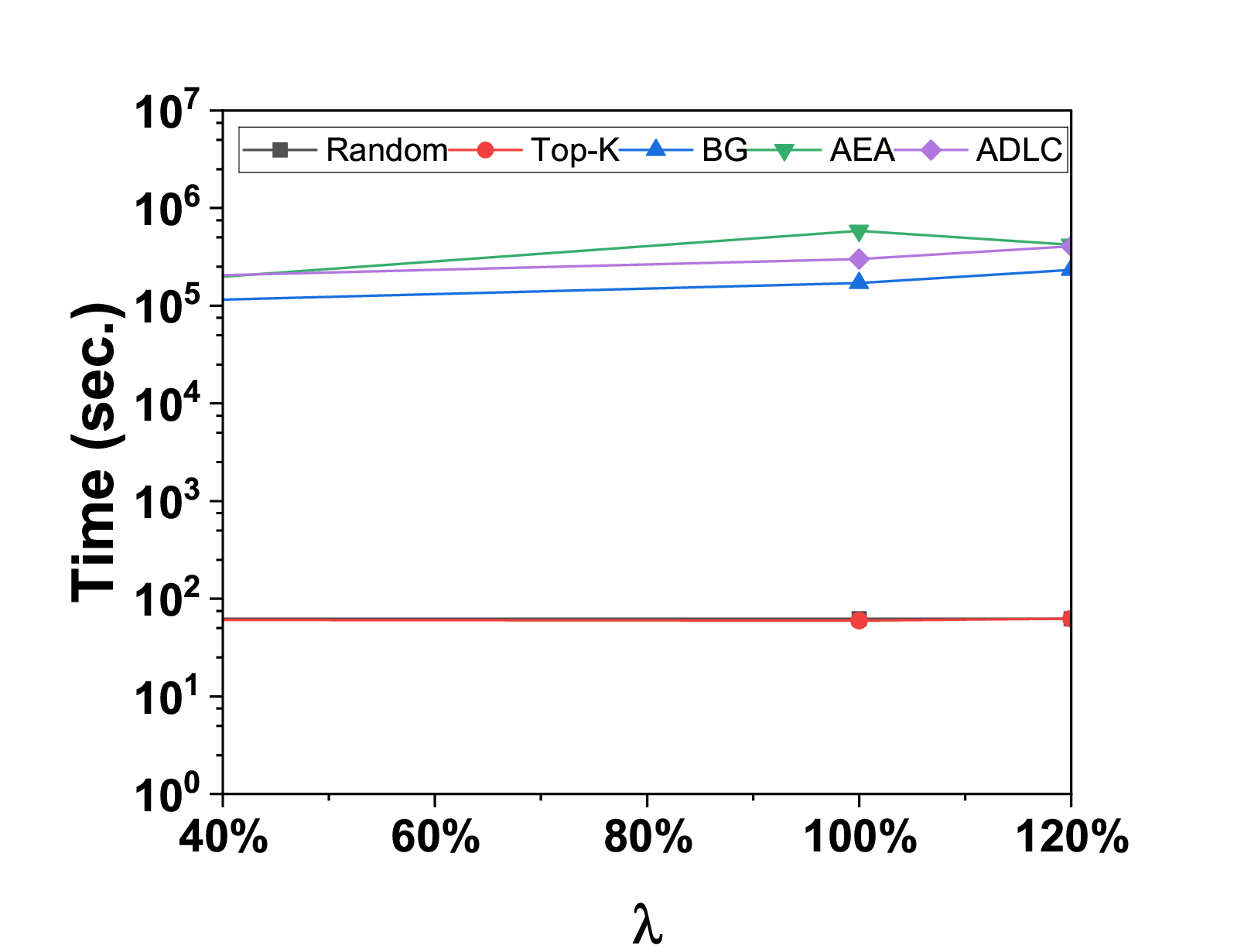} \\
         {\tiny (a) $\omega = 10\% (|\mathcal{A}| = 100)$} & 
        {\tiny (b) $\omega = 30\% (|\mathcal{A}| = 50)$} &
        {\tiny (c) $\omega = 50\% (|\mathcal{A}| = 20)$} \\
    \end{tabular} 
   
	\vspace{0.5cm} 
   
   \begin{tabular}{cc}
   \includegraphics[scale=0.15]{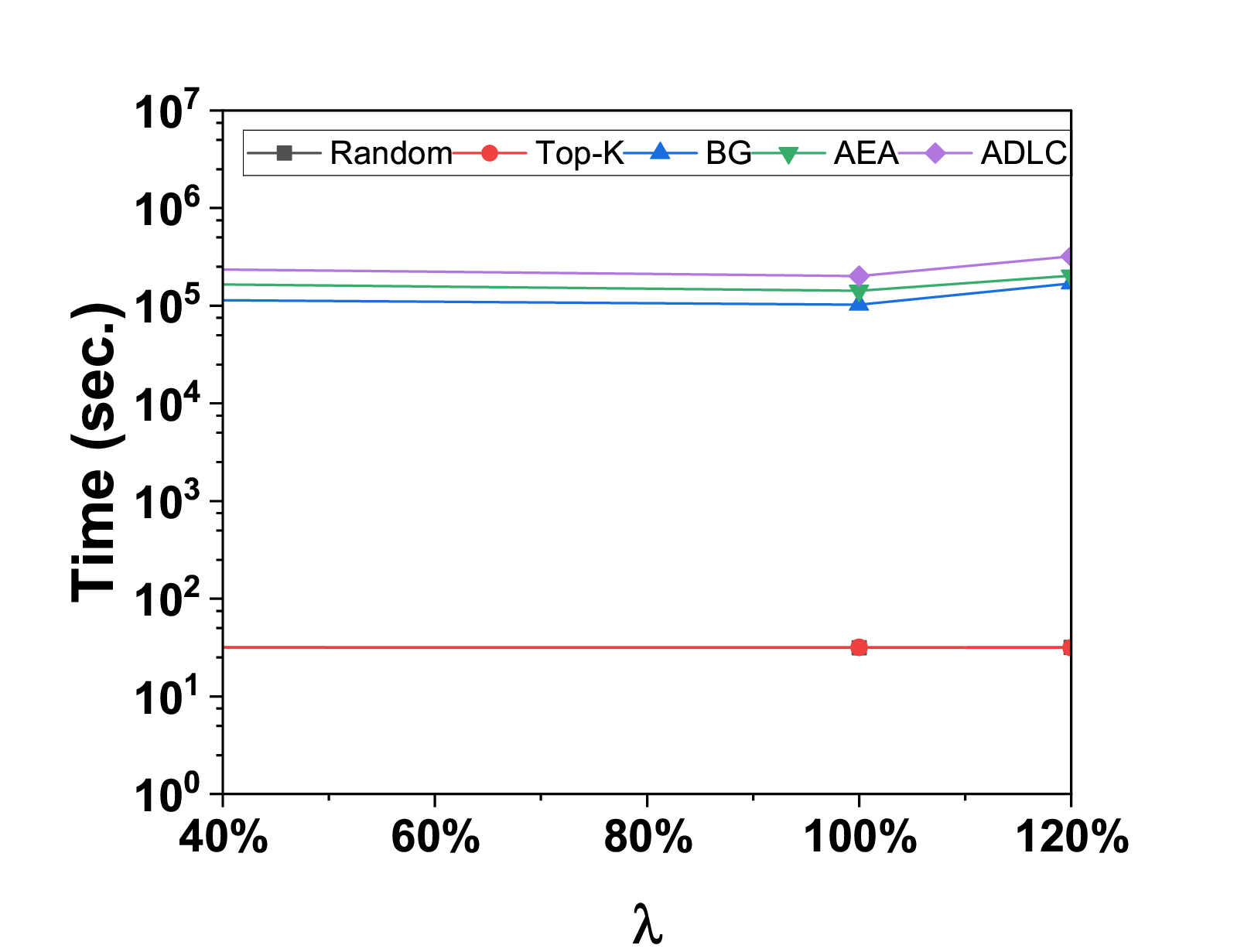} &
   \includegraphics[scale=0.15]{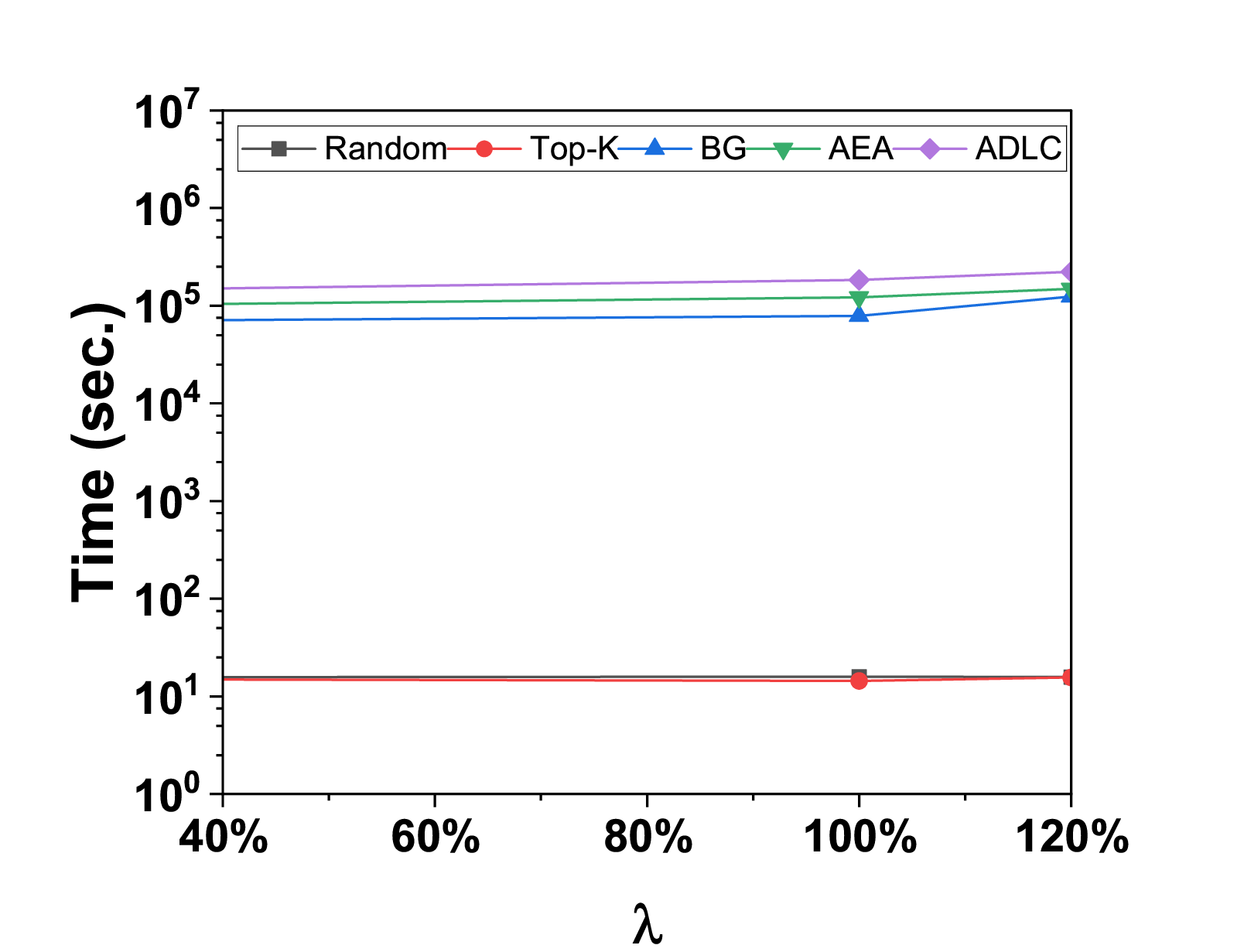} \\
    {\tiny (d) $\omega = 70\% (|\mathcal{A}| = 10)$} & 
    {\tiny (e) $\omega = 90\% (|\mathcal{A}| = 5)$} \\
\end{tabular}
\caption{Efficiency Study for Trivalency Probability Setting (Facebook)}
\label{Fig:10_TimeTV}
\end{figure}

\begin{figure}[H]
    \centering
    \begin{tabular}{ccc}
        \includegraphics[scale=0.15]{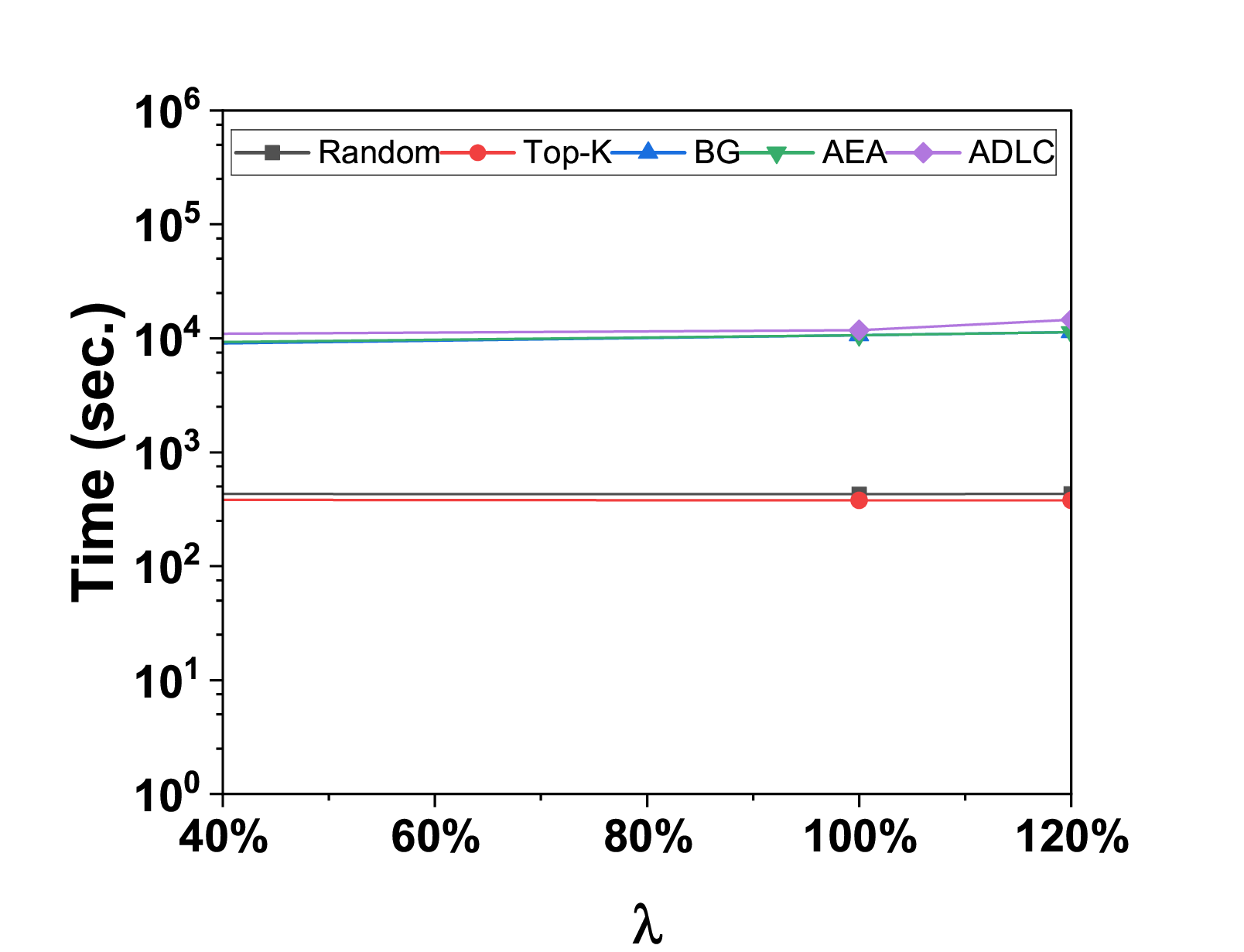} &
        \includegraphics[scale=0.15]{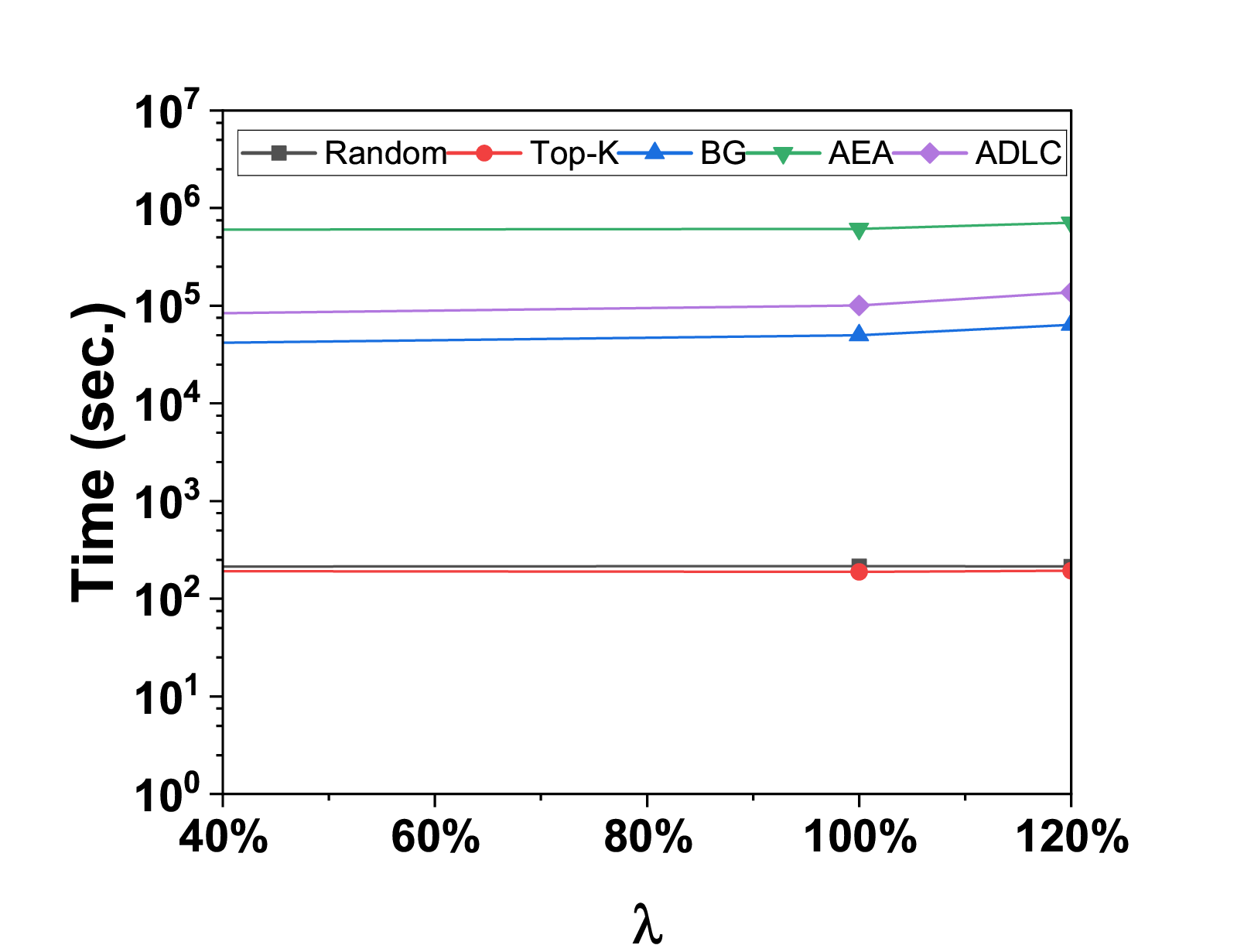} &
        \includegraphics[scale=0.15]{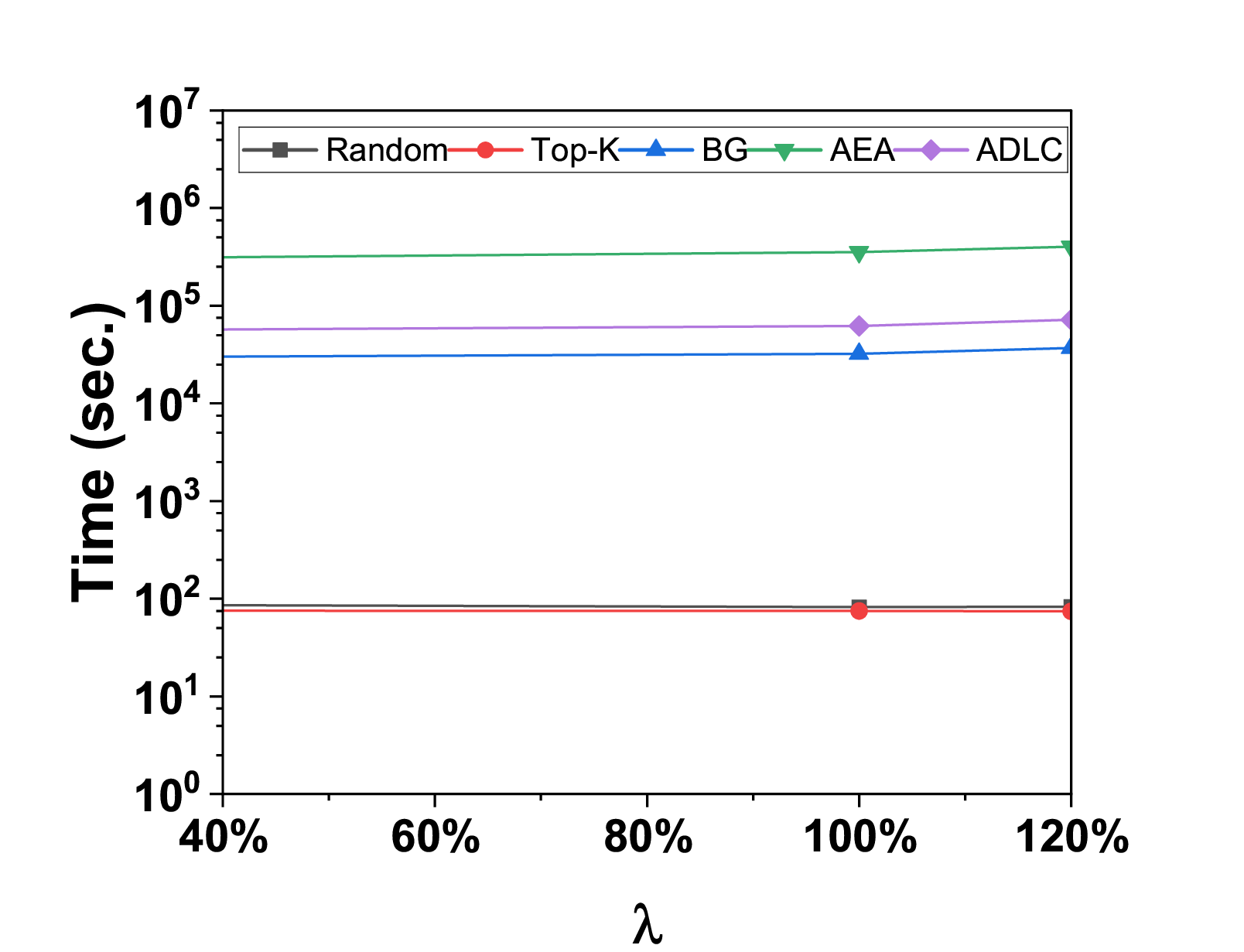} \\
        {\tiny (a) $\omega = 10\% (|\mathcal{A}| = 100)$} & 
        {\tiny (b) $\omega = 30\% (|\mathcal{A}| = 50)$} &
        {\tiny (c) $\omega = 50\% (|\mathcal{A}| = 20)$} \\
    \end{tabular}

    \vspace{0.5cm} 

    \begin{tabular}{cc}
        \includegraphics[scale=0.15]{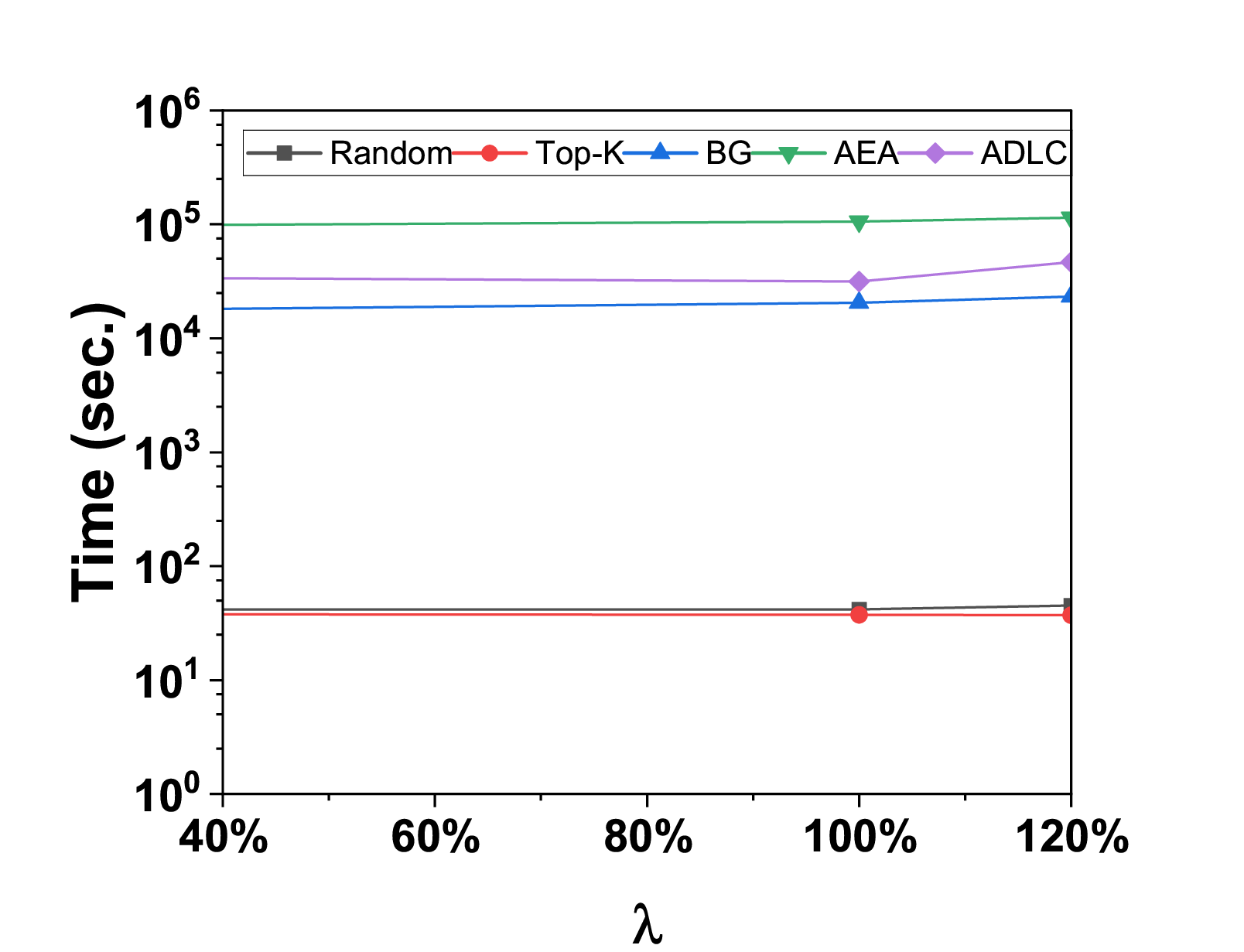} &
        \includegraphics[scale=0.15]{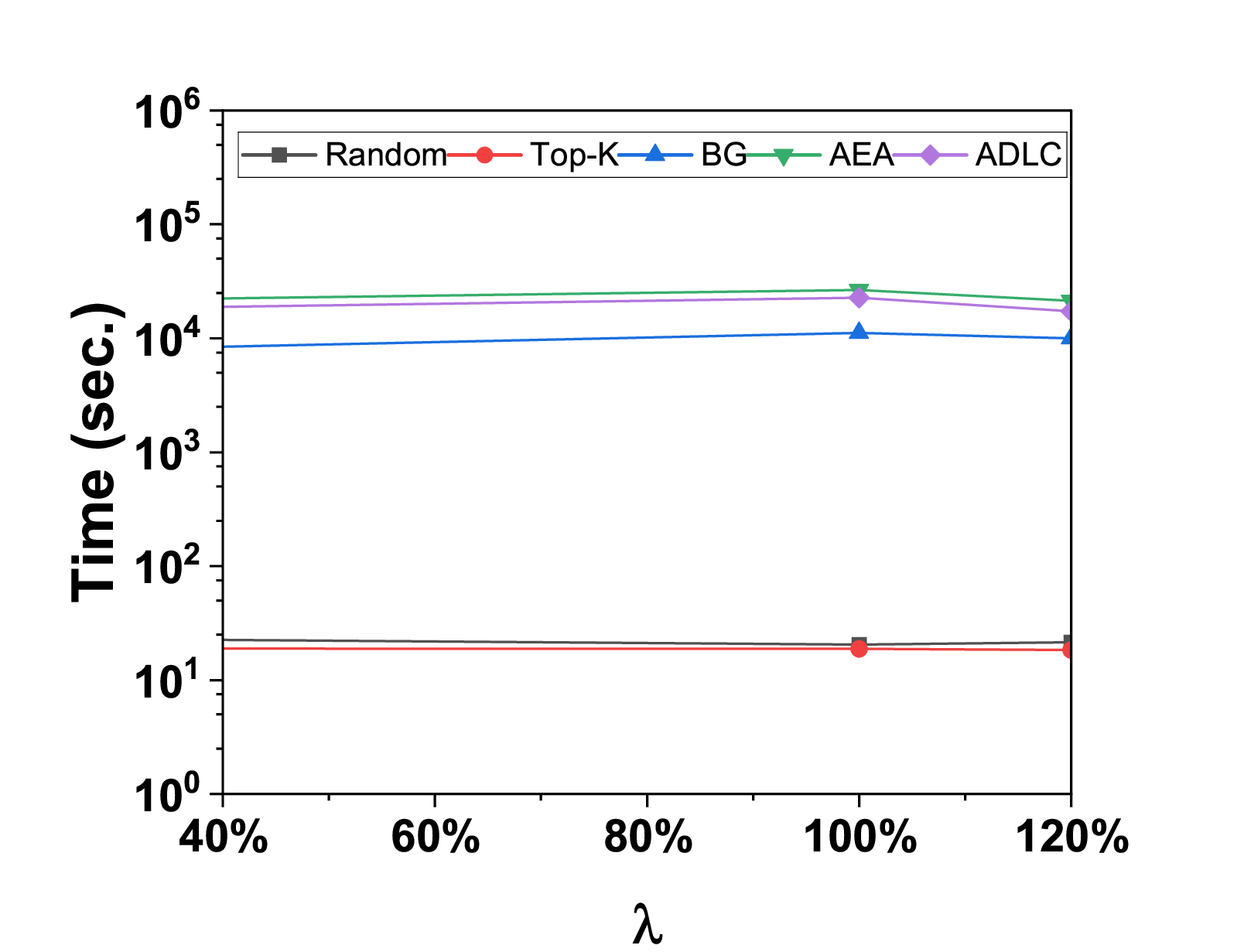} \\
        {\tiny (d) $\omega = 70\% (|\mathcal{A}| = 10)$} &
        {\tiny (e) $\omega = 90\% (|\mathcal{A}| = 5)$} \\
    \end{tabular}

    \caption{Efficiency Study Uniform (0.1) Probability Setting (Wikivote)}
    \label{Fig:11_Time0.1}
\end{figure}

\begin{figure}[H]
 \centering
\begin{tabular}{ccc}
   \includegraphics[scale=0.15]{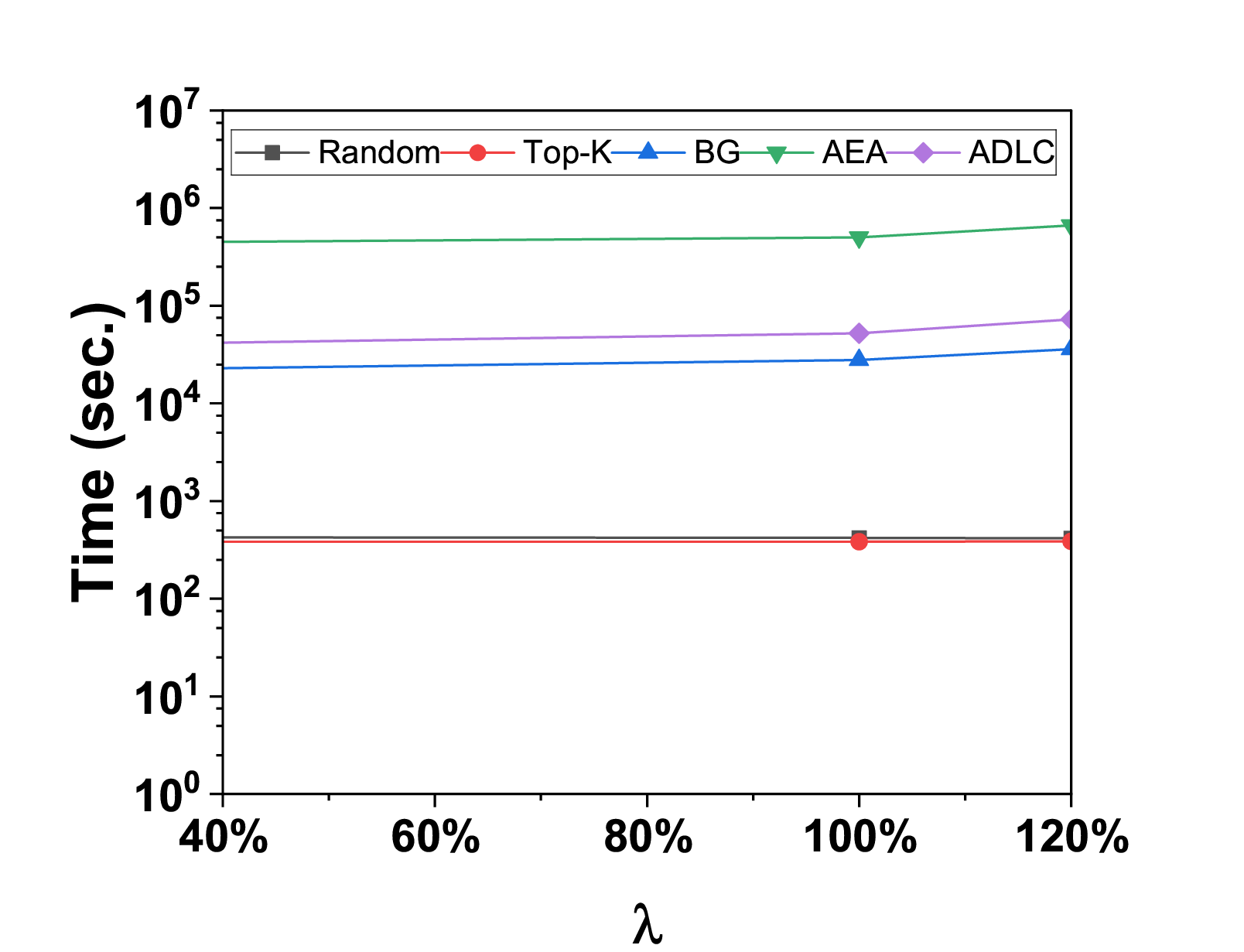} & 
   \includegraphics[scale=0.15]{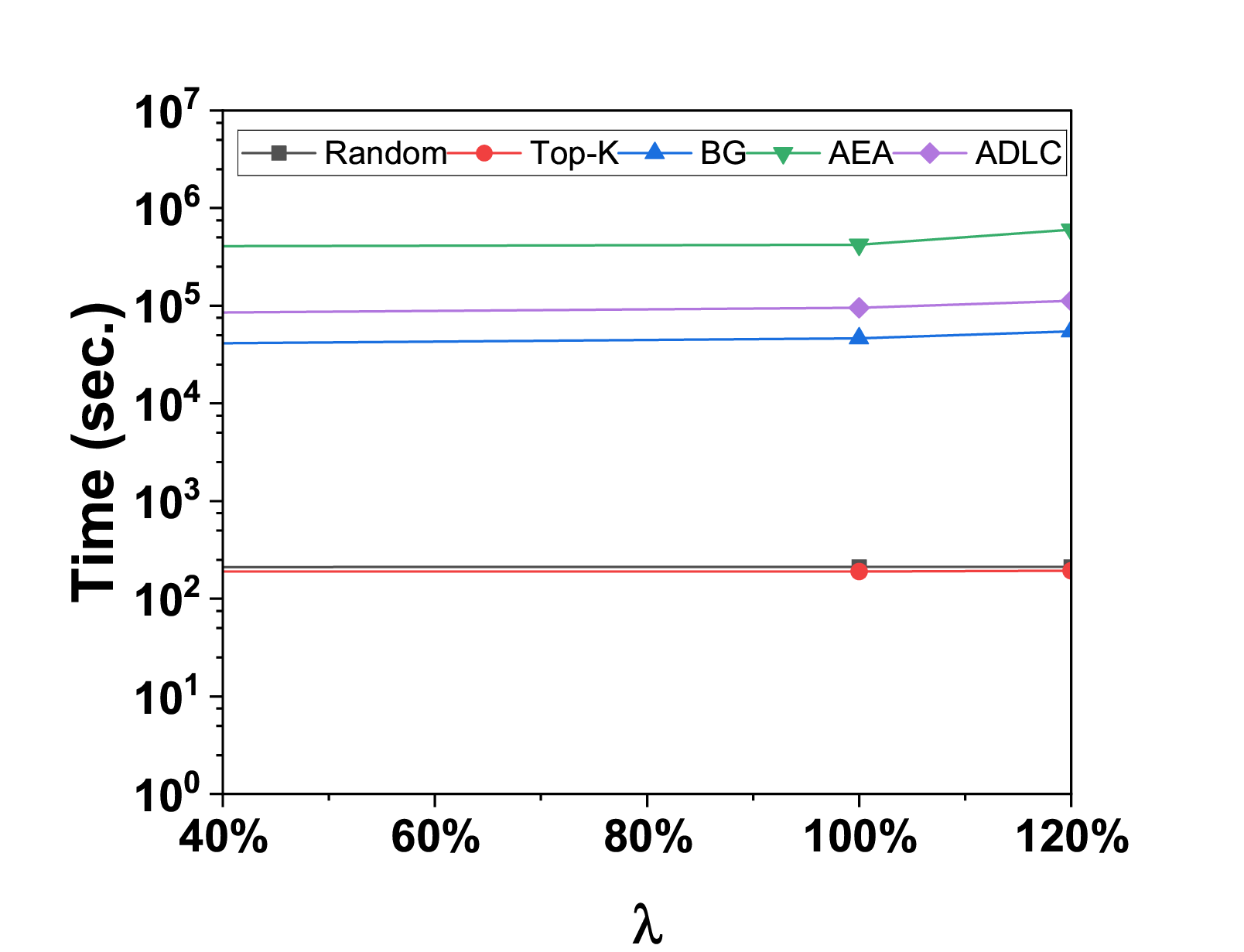} & 
   \includegraphics[scale=0.15]{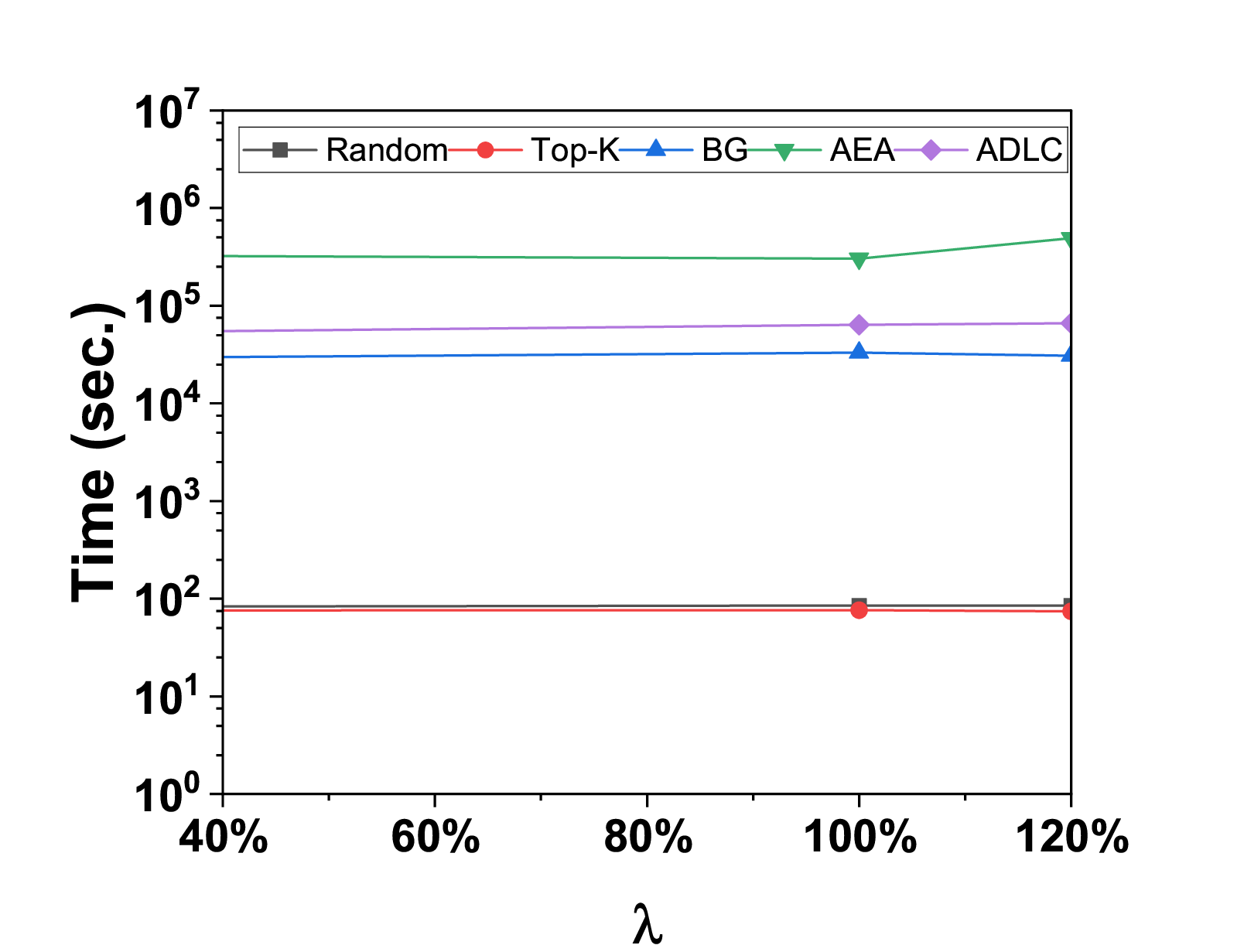} \\
         {\tiny (a) $\omega = 10\% (|\mathcal{A}| = 100)$} & 
        {\tiny (b) $\omega = 30\% (|\mathcal{A}| = 50)$} &
        {\tiny (c) $\omega = 50\% (|\mathcal{A}| = 20)$} \\
    \end{tabular} 
   
	\vspace{0.5cm} 
   
   \begin{tabular}{cc}
   \includegraphics[scale=0.15]{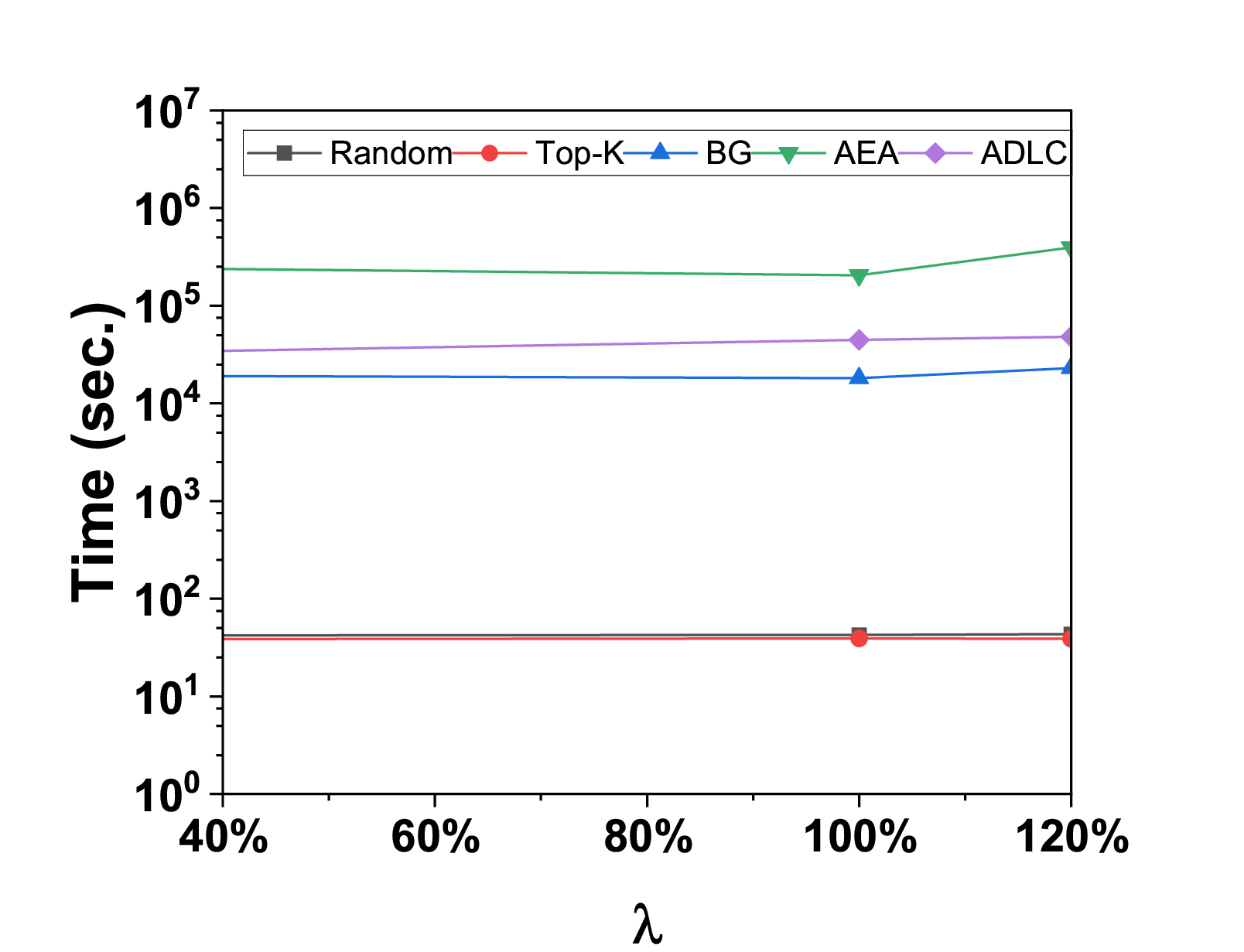} &
   \includegraphics[scale=0.15]{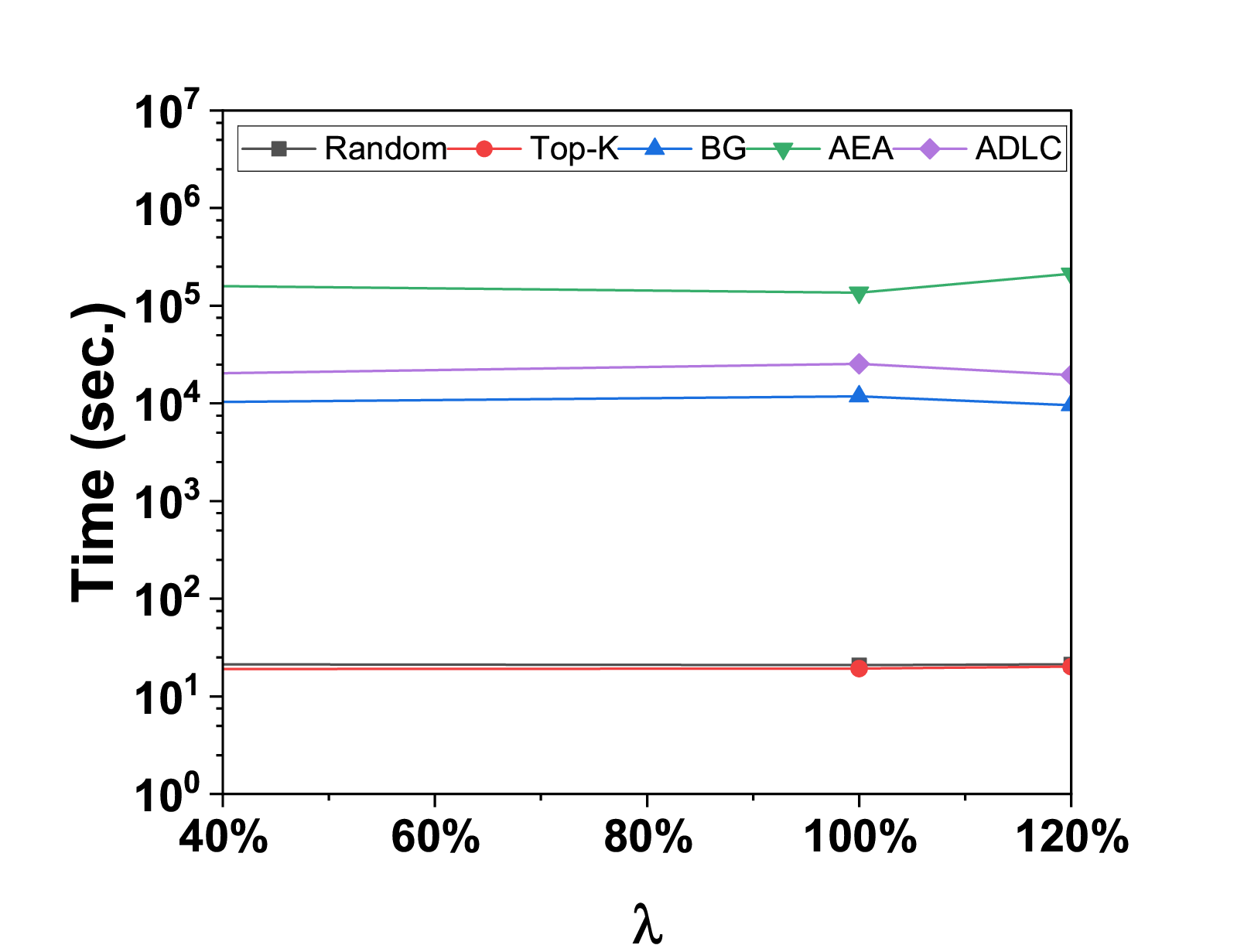} \\
    {\tiny (d) $\omega = 70\% (|\mathcal{A}| = 10)$} & 
    {\tiny (e) $\omega = 90\% (|\mathcal{A}| = 5)$} \\
\end{tabular}
\caption{Efficiency Study for Trivalency Probability Setting (Wikivote)}
\label{Fig:12_TimeTV}
\end{figure}

\begin{figure}[H]
 \centering
\begin{tabular}{ccccc}
   \includegraphics[scale=0.1]{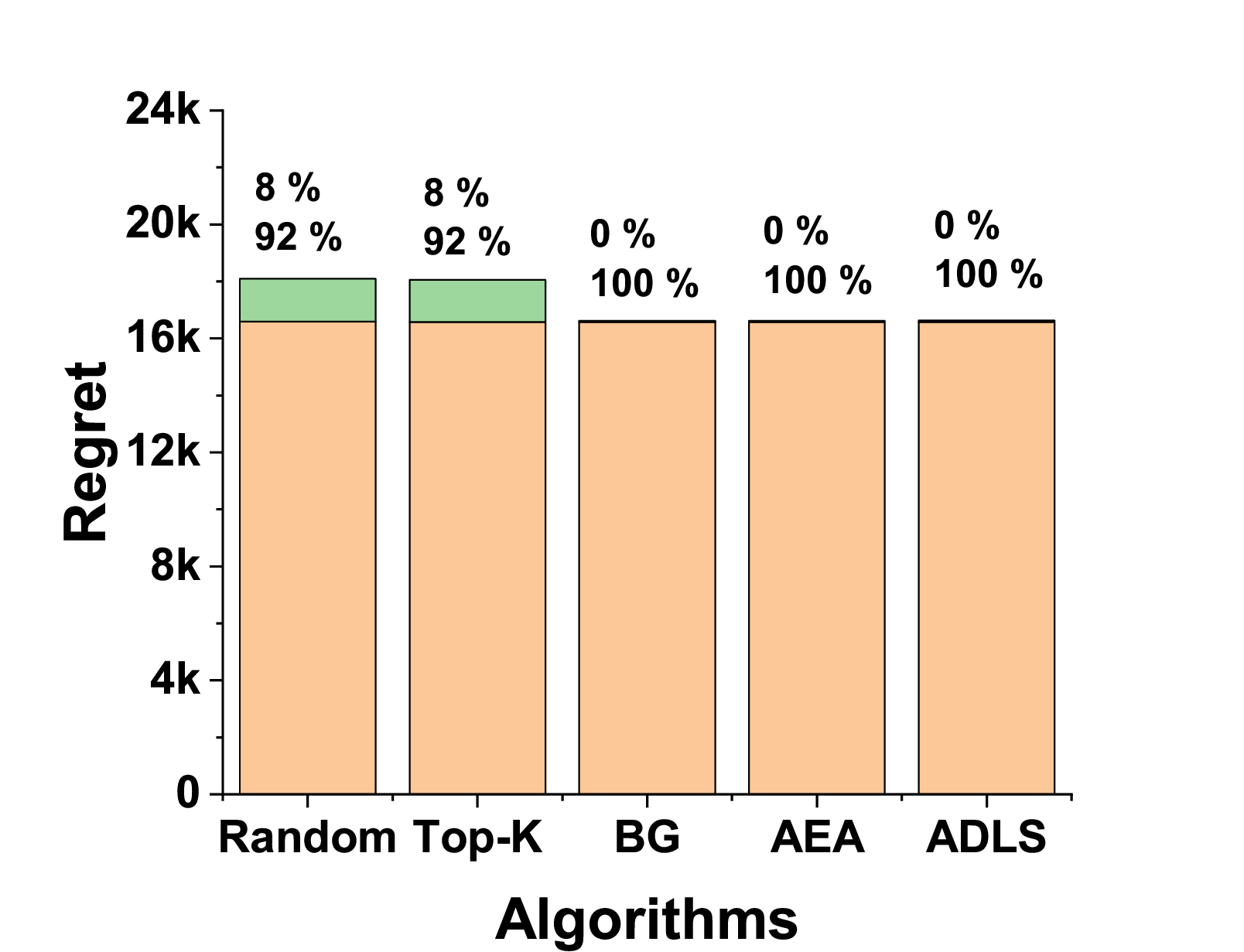} & 
   \includegraphics[scale=0.1]{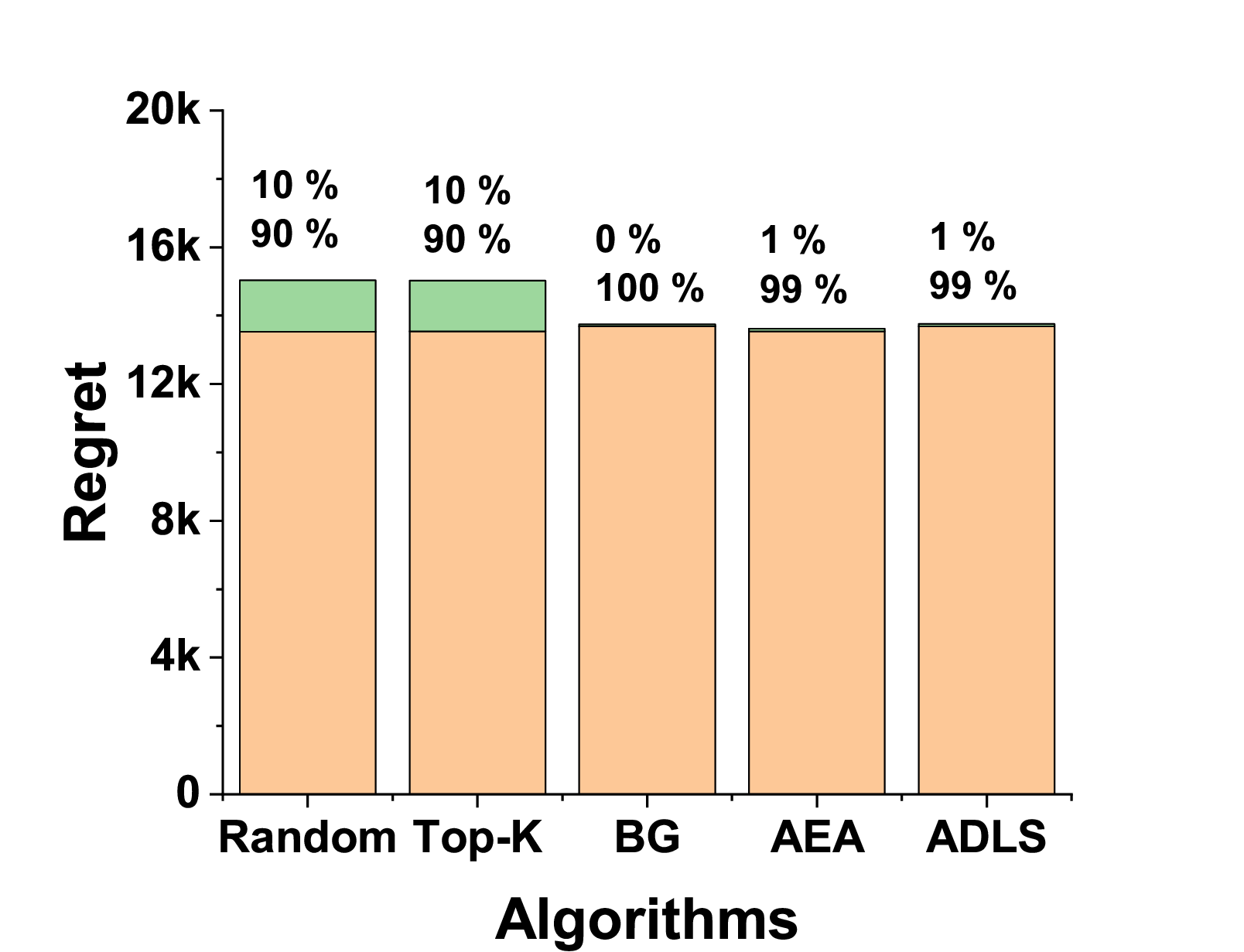} & 
   \includegraphics[scale=0.1]{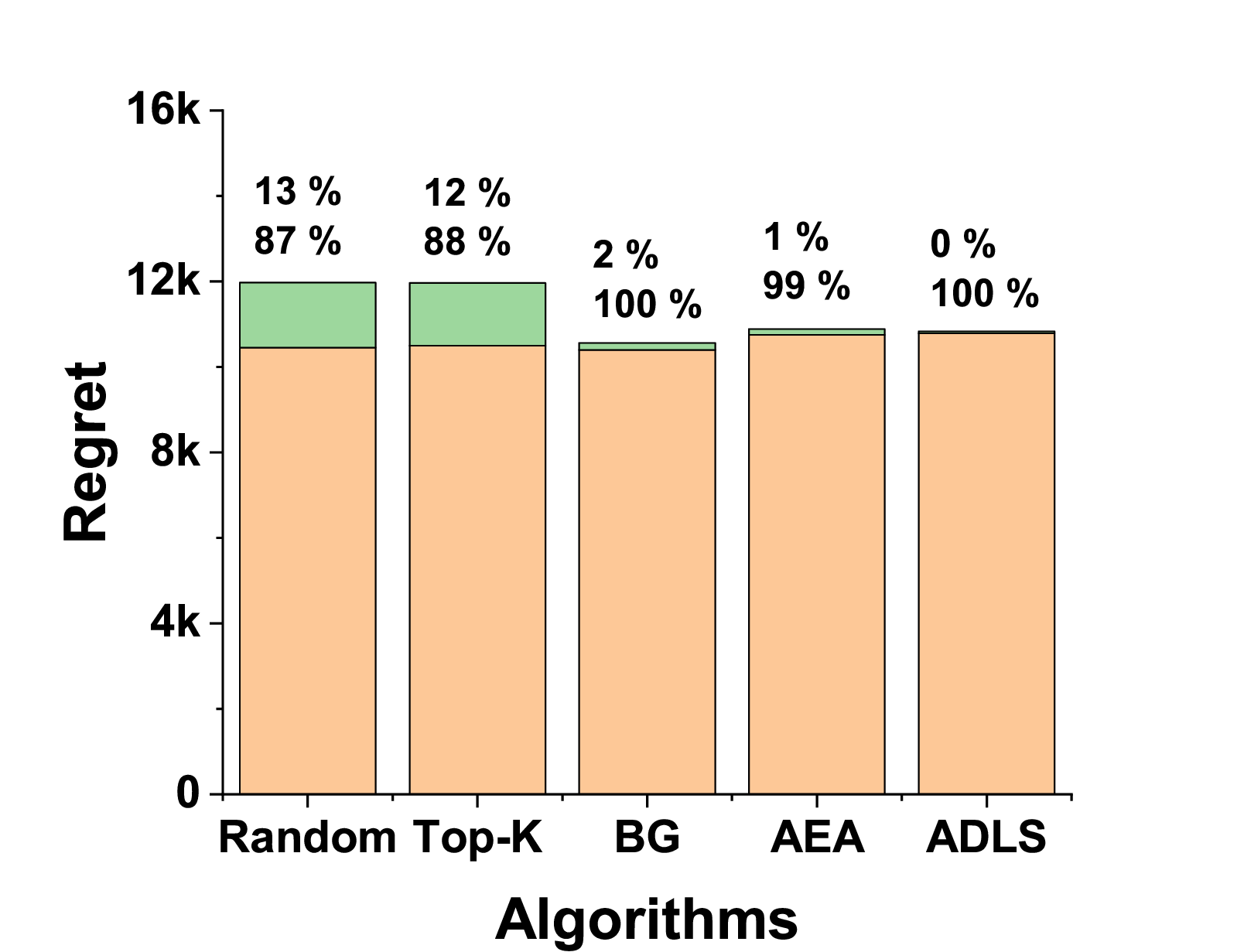} &
   \includegraphics[scale=0.1]{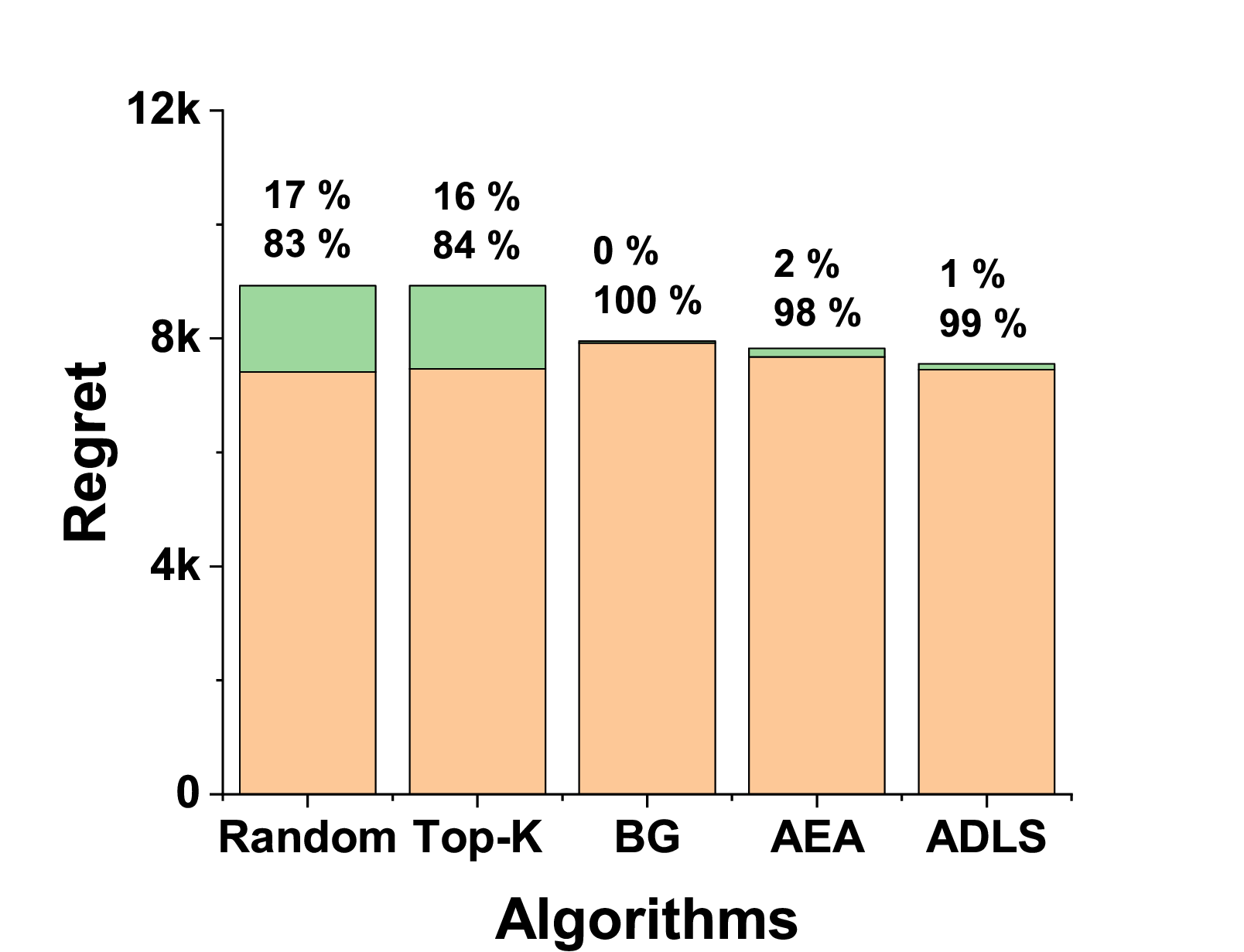} &
   \includegraphics[scale=0.1]{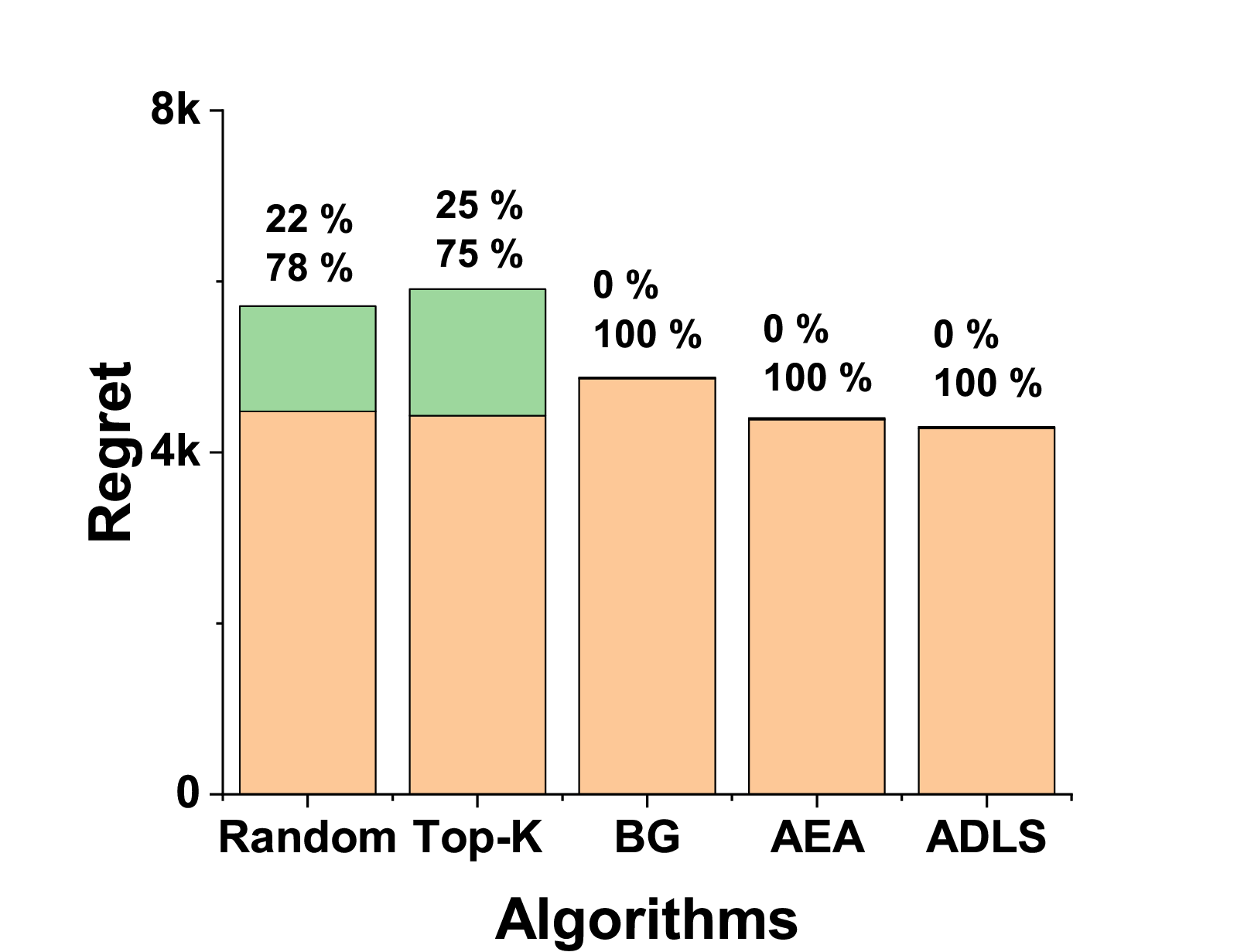} \\
    {\tiny (a) $\gamma = 0$ } & 
    {\tiny (b) $\gamma = 0.25$} &
    {\tiny (c) $\gamma = 0.5$} &
    {\tiny (d) $\gamma = 0.75$} & 
    {\tiny (e) $\gamma = 1 $} \\
\end{tabular}
\caption{Parameter Study for $\gamma$ $(|\mathcal{A}| = 20)$ Uniform(0.1) Probability Setting (Email-Eu-Core)}
\label{Fig:13_Gamma0.1}
\end{figure}

\begin{figure}[H]
 \centering
\begin{tabular}{ccccc}
   \includegraphics[scale=0.1]{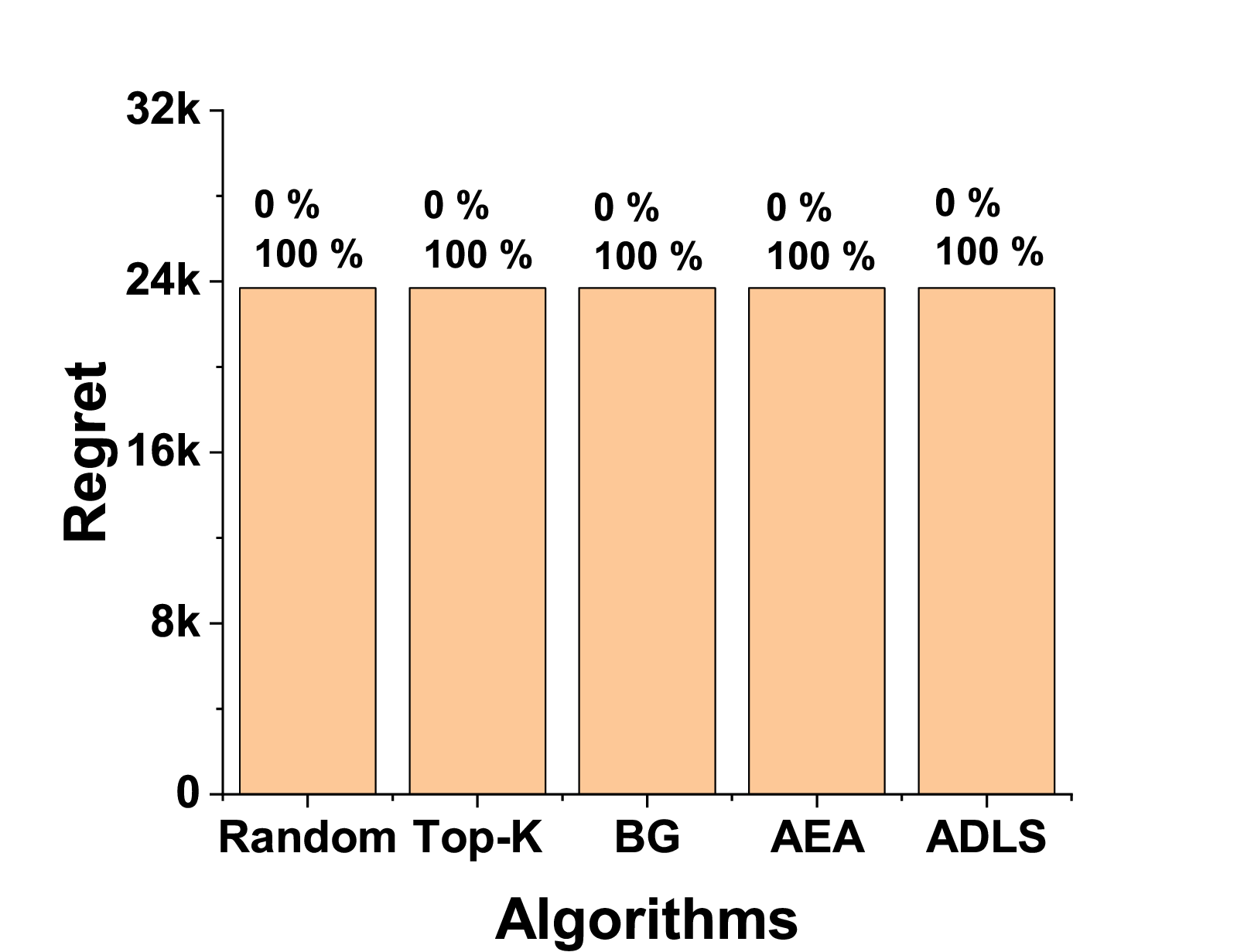} & 
   \includegraphics[scale=0.1]{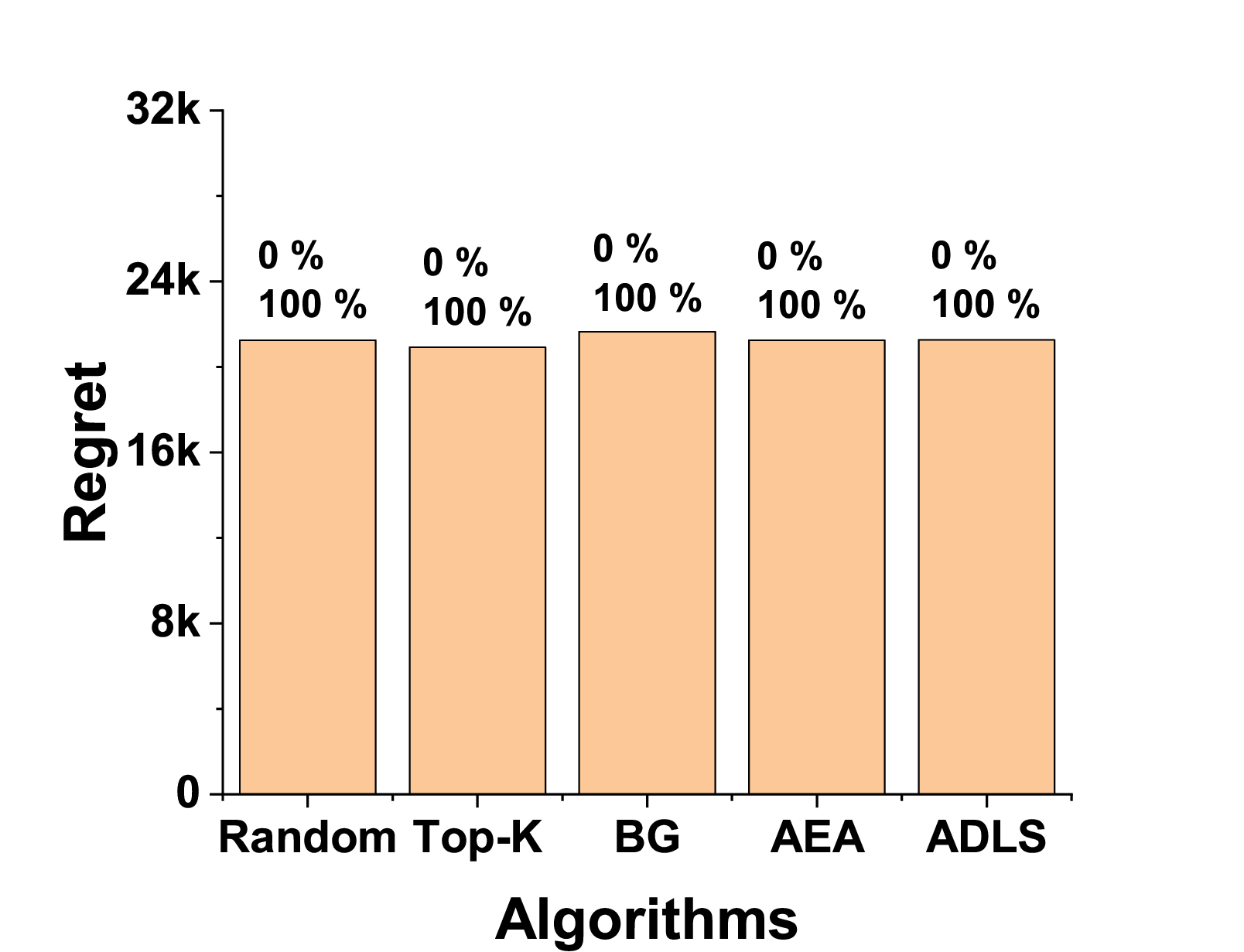} & 
   \includegraphics[scale=0.1]{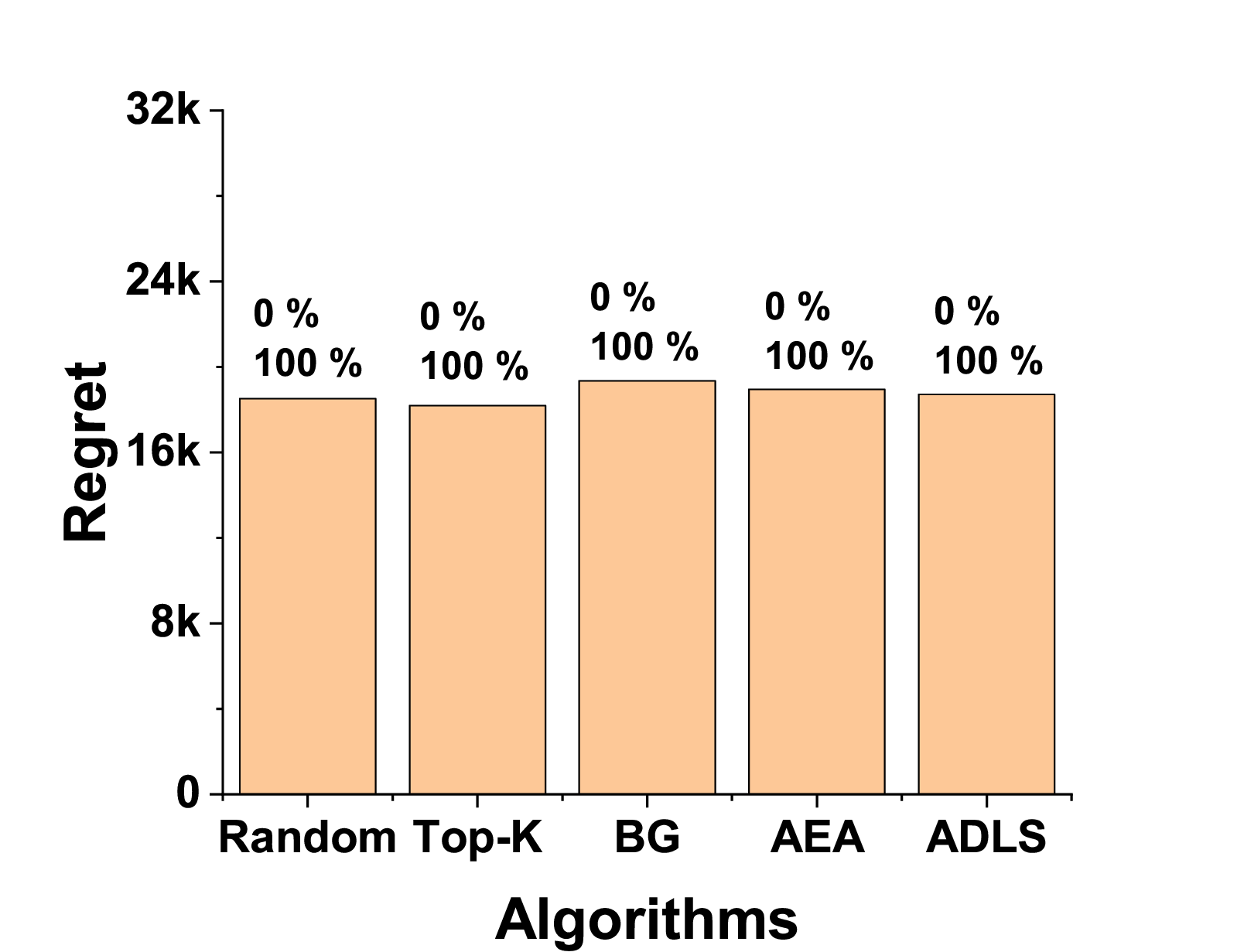} &
   \includegraphics[scale=0.1]{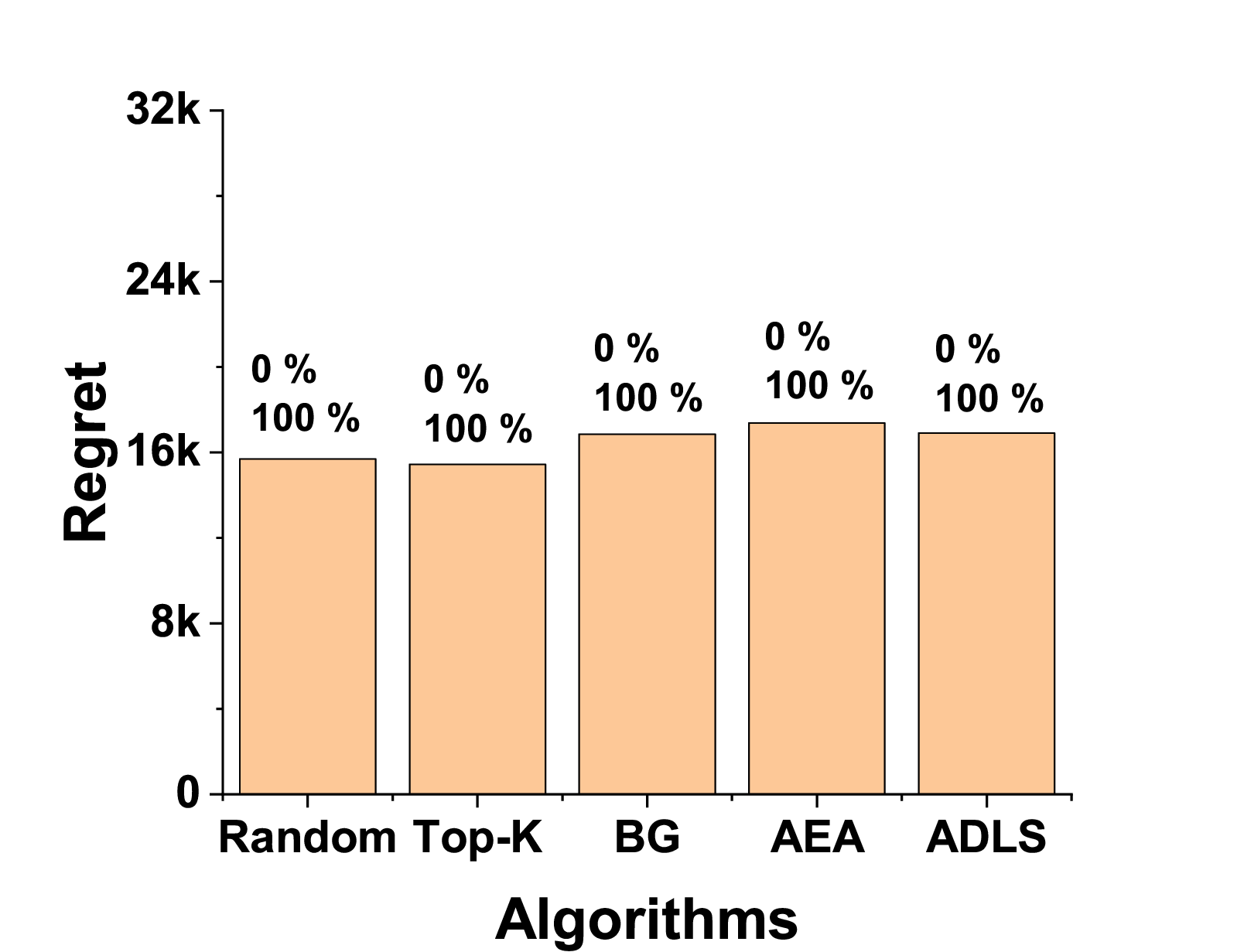} &
   \includegraphics[scale=0.1]{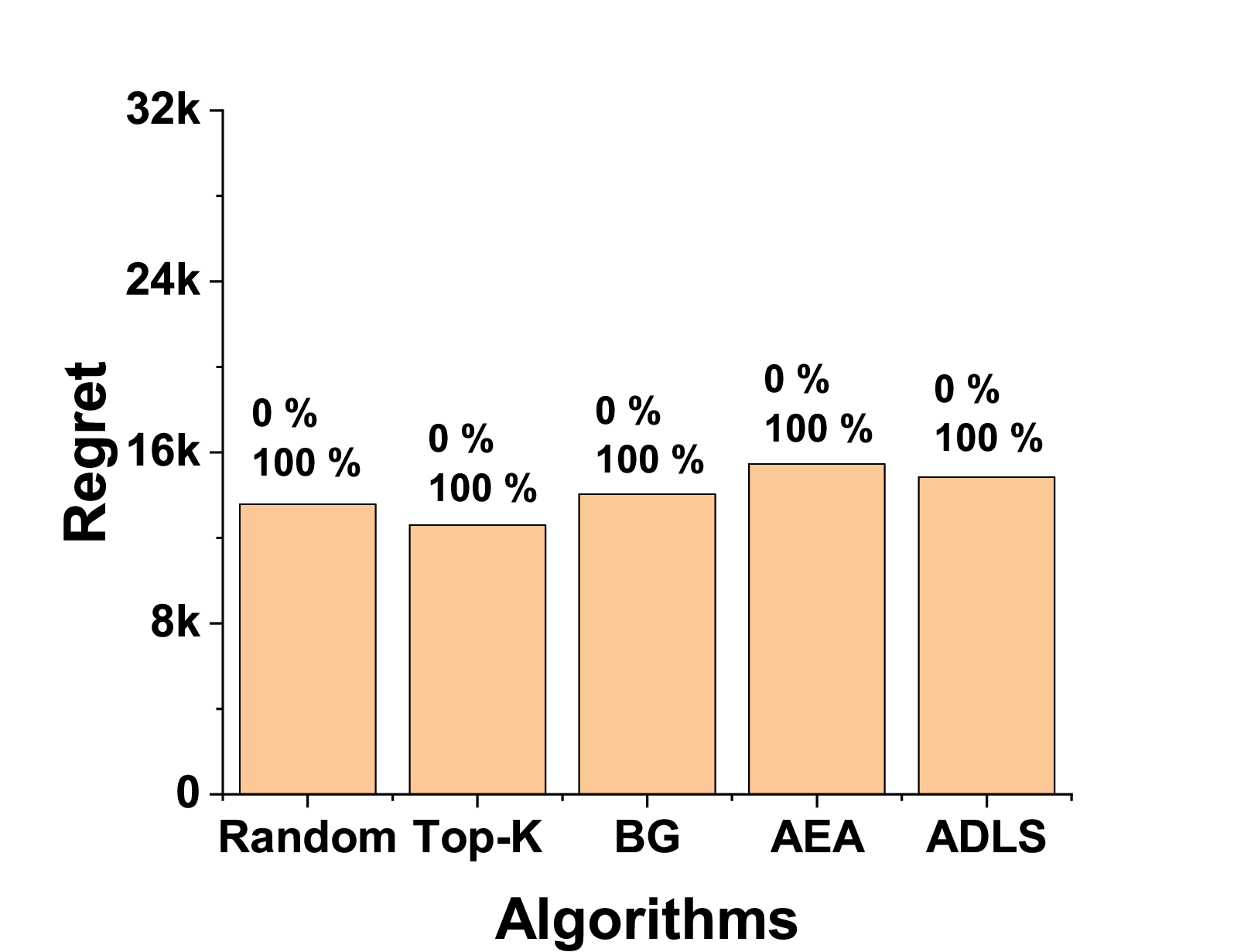} \\
    {\tiny (a) $\gamma = 0$ } & 
    {\tiny (b) $\gamma = 0.25$} &
    {\tiny (c) $\gamma = 0.5$} &
    {\tiny (d) $\gamma = 0.75$} & 
    {\tiny (e) $\gamma = 1 $} \\
\end{tabular}
\caption{Parameter Study for $\gamma$ $(|\mathcal{A}| = 20)$ Trivalency Probability Setting (Email-Eu-Core)}
\label{Fig:14_GammaTV}
\end{figure}

\begin{figure}[H]
 \centering
\begin{tabular}{ccccc}
   \includegraphics[scale=0.1]{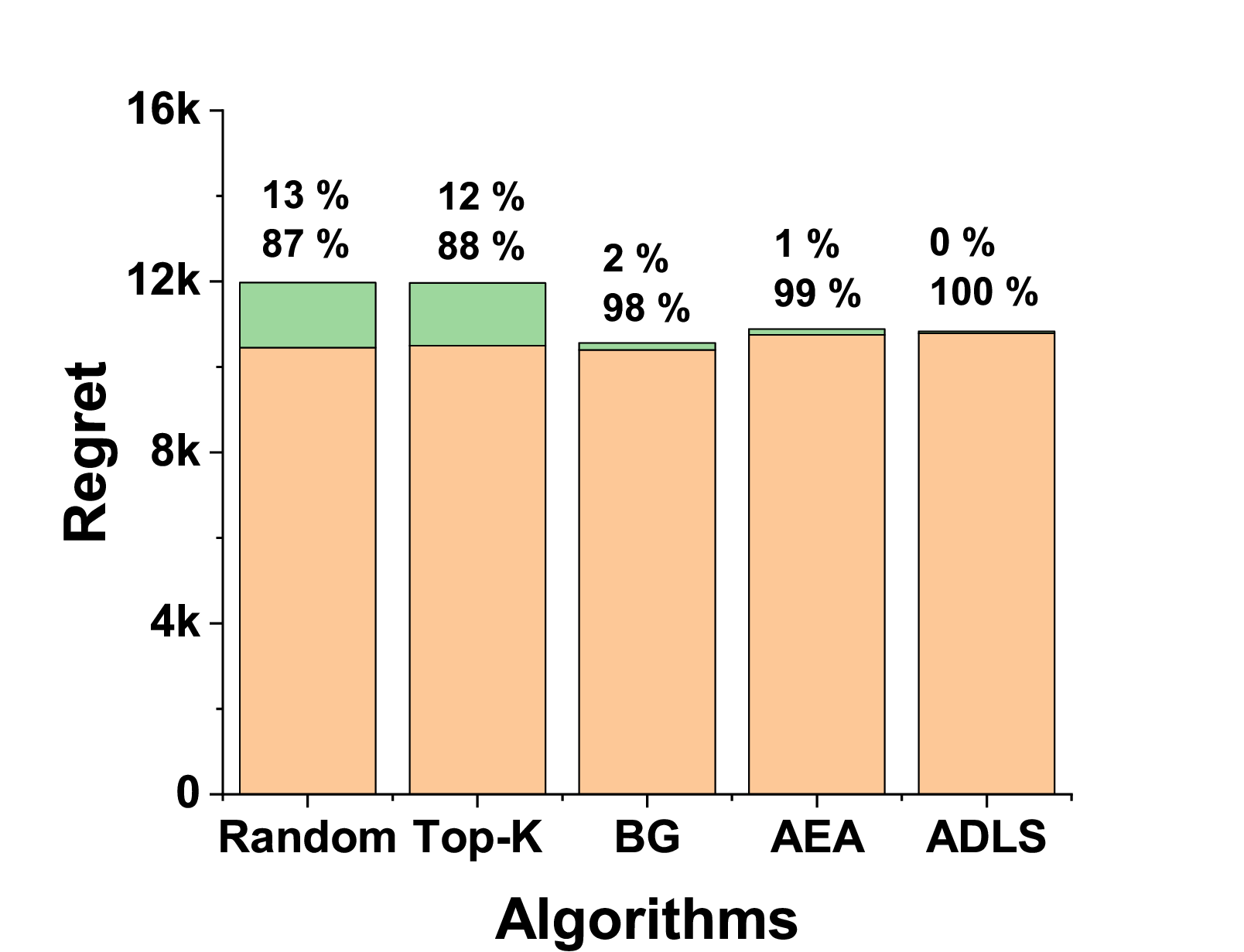} & 
   \includegraphics[scale=0.1]{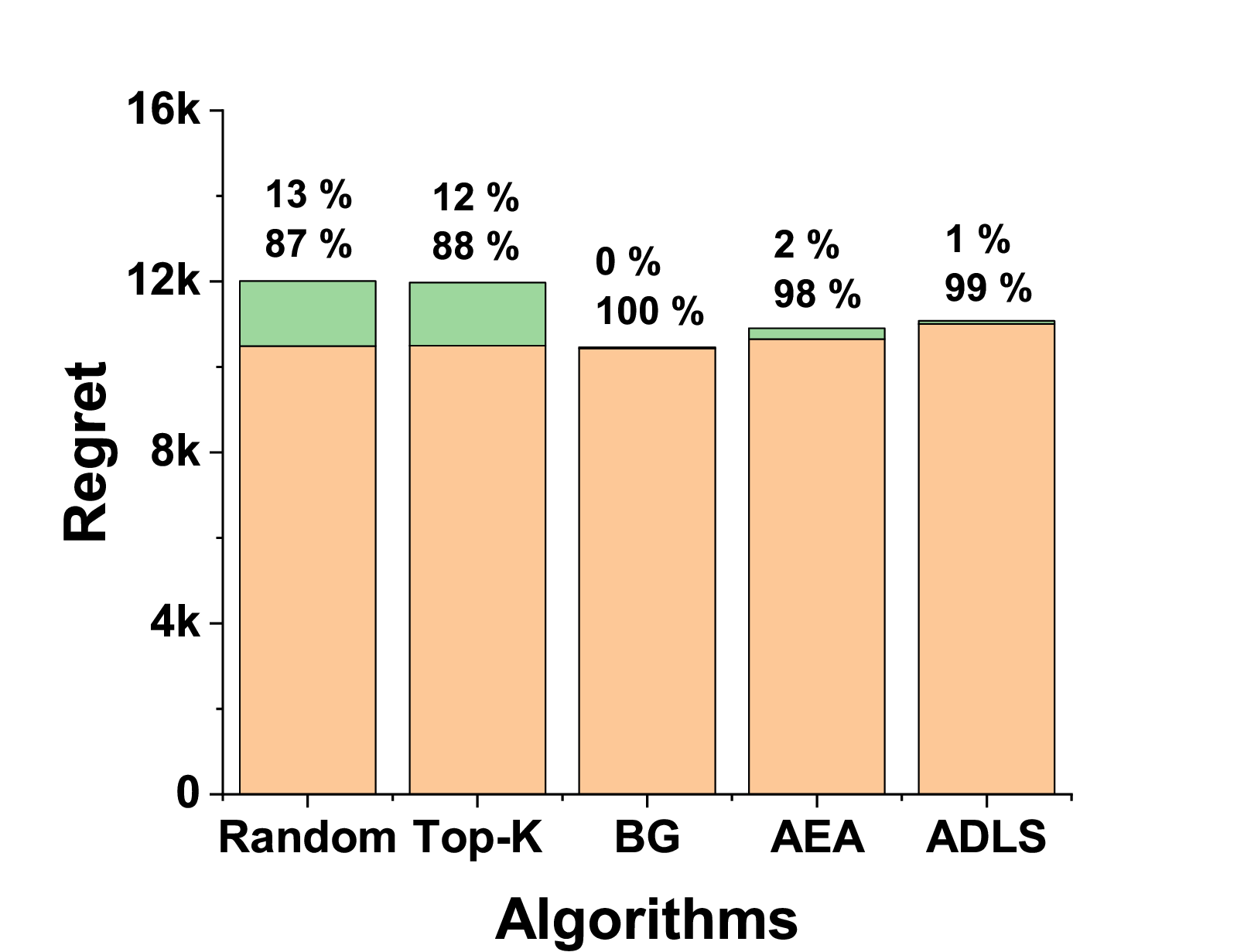} & 
   \includegraphics[scale=0.1]{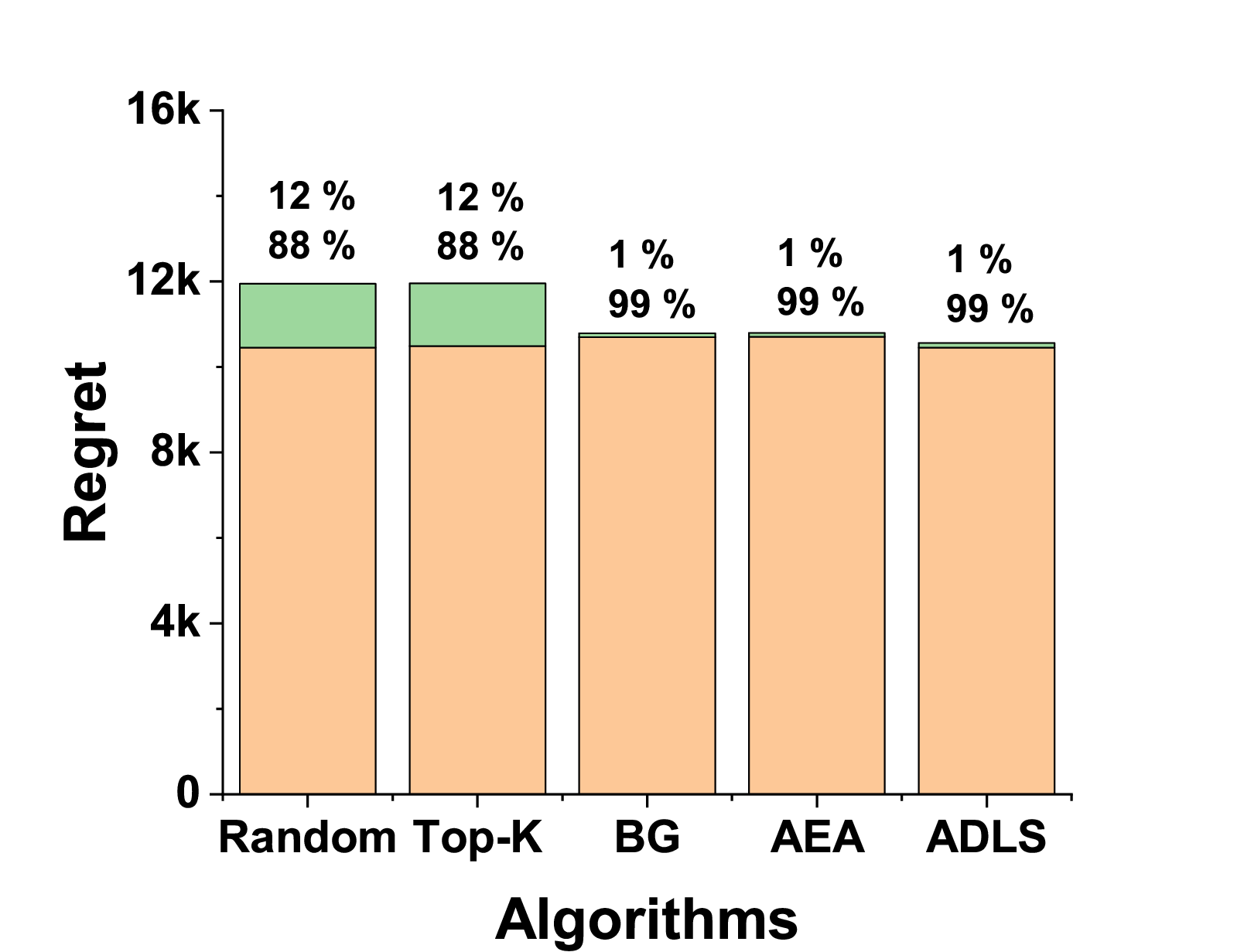} &
   \includegraphics[scale=0.1]{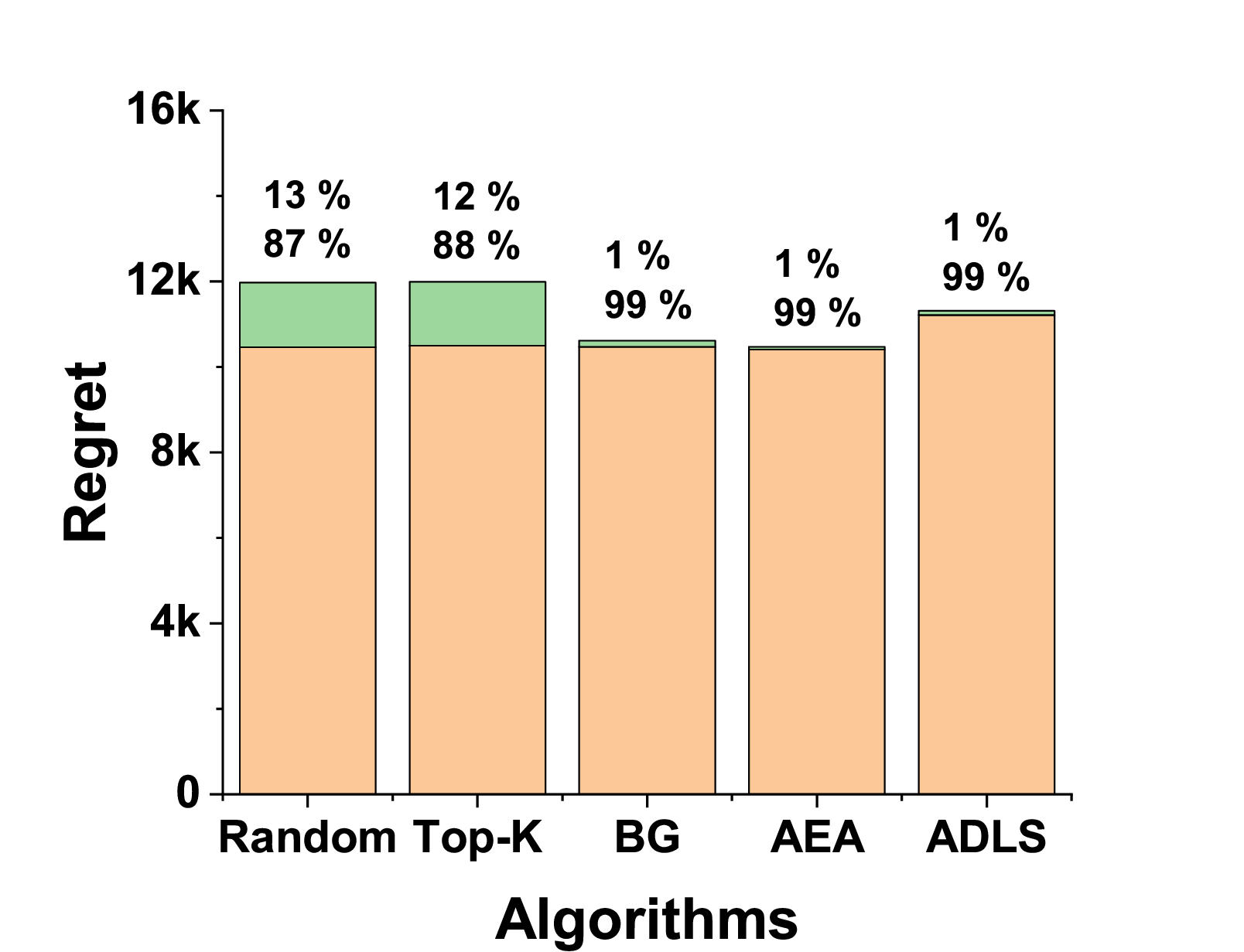} &
   \includegraphics[scale=0.1]{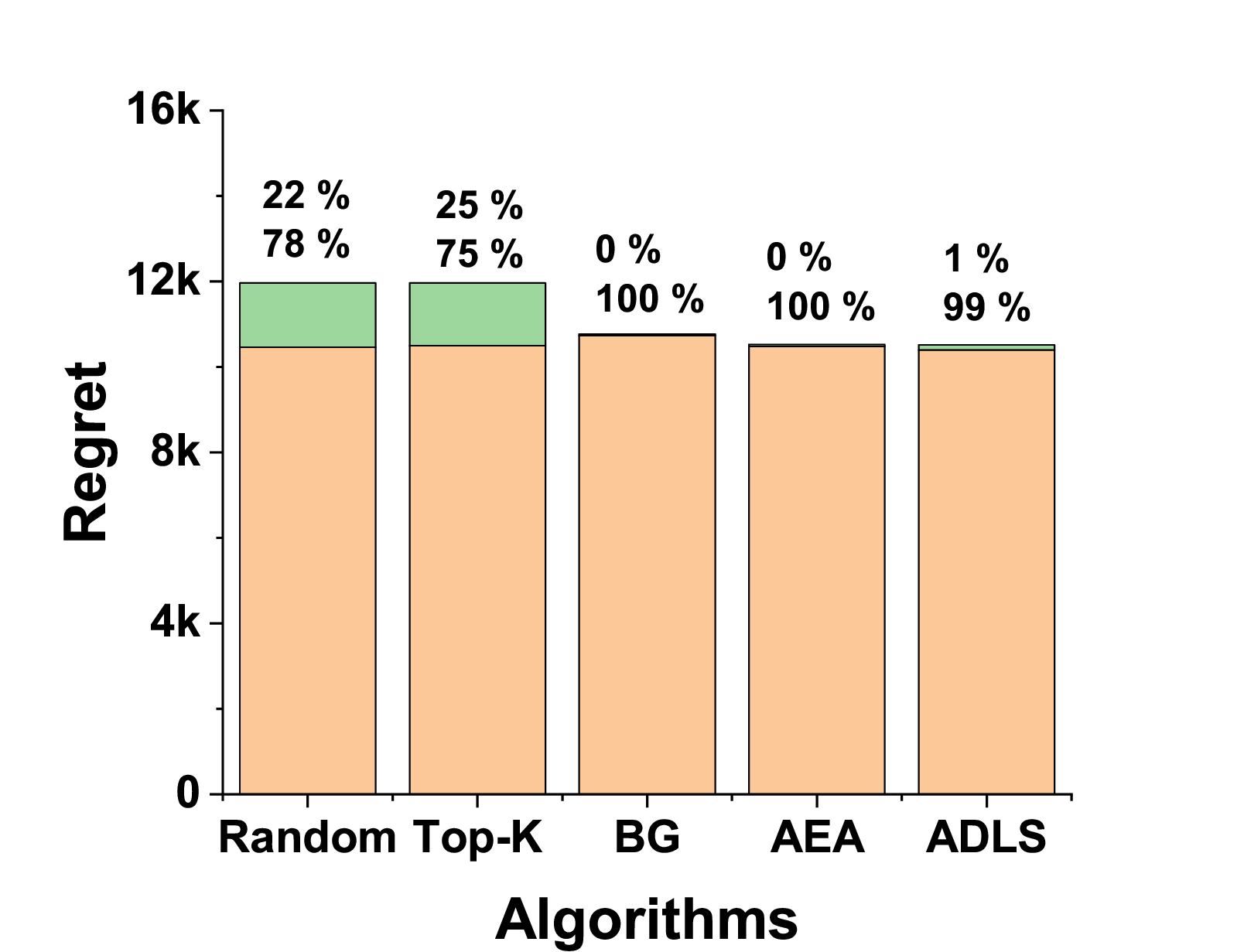} \\
    {\tiny (a) $\delta = 0.01$ } & 
    {\tiny (b) $\delta = 0.02$} &
    {\tiny (c) $\delta = 0.03$} &
    {\tiny (d) $\delta = 0.04$} & 
    {\tiny (e) $\delta = 0.05$} \\
\end{tabular}
\caption{Parameter Study for $\delta$ $(|\mathcal{A}| = 20)$ Uniform(0.1) Probability Setting (Email-Eu-Core)}
\label{Fig:15_Delta0.1}
\end{figure}

\begin{figure}[H]
 \centering
\begin{tabular}{ccccc}
   \includegraphics[scale=0.1]{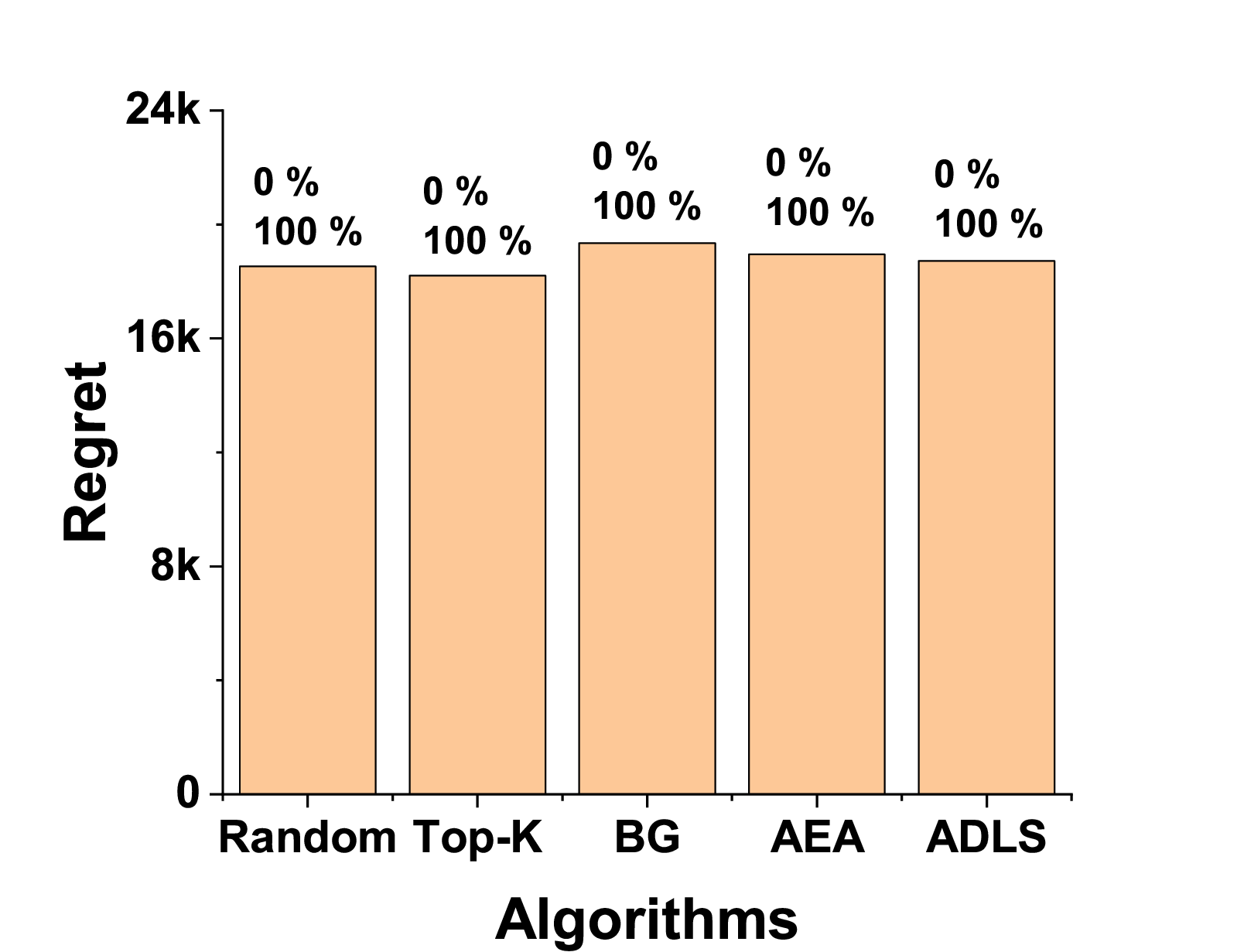} & 
   \includegraphics[scale=0.1]{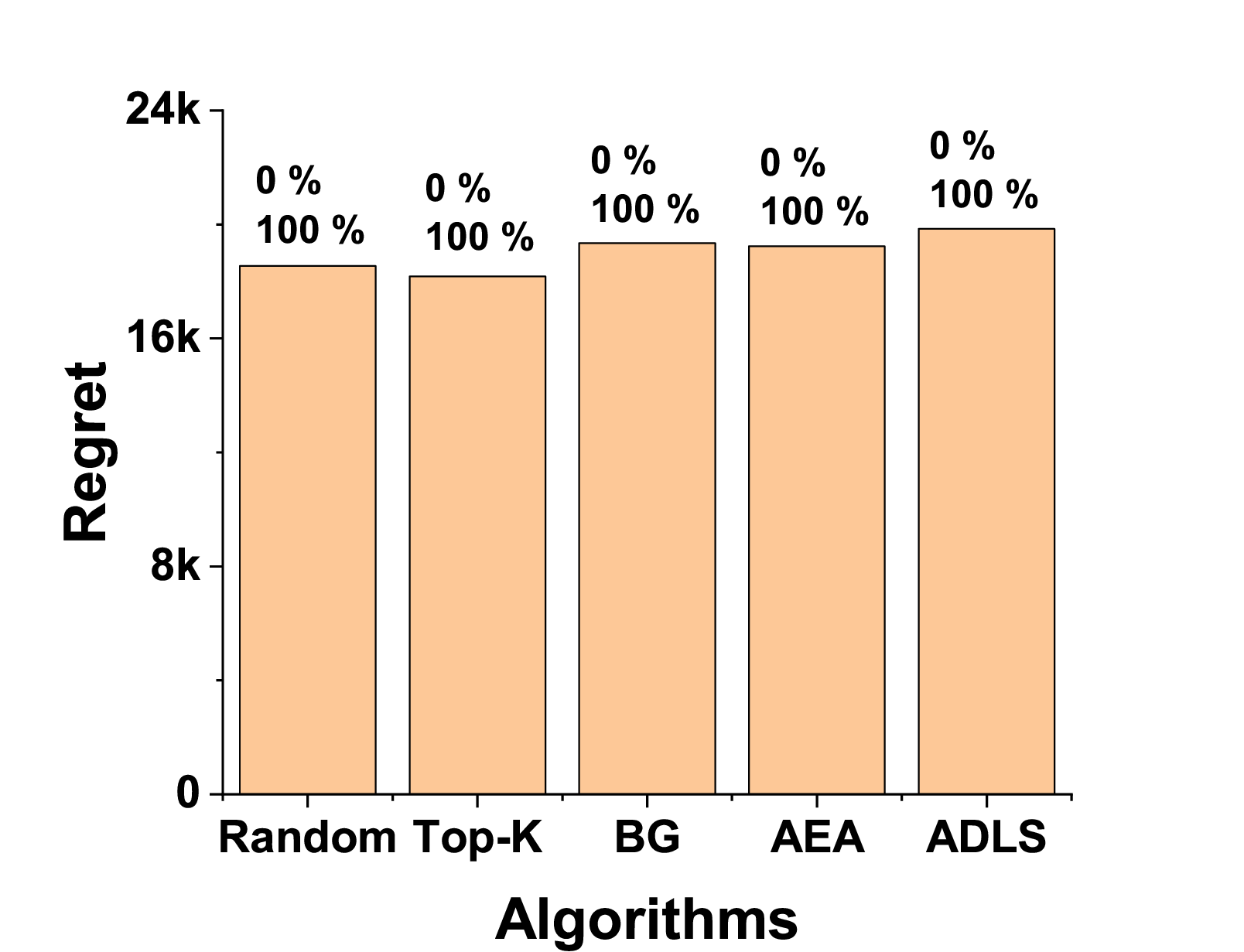} & 
   \includegraphics[scale=0.1]{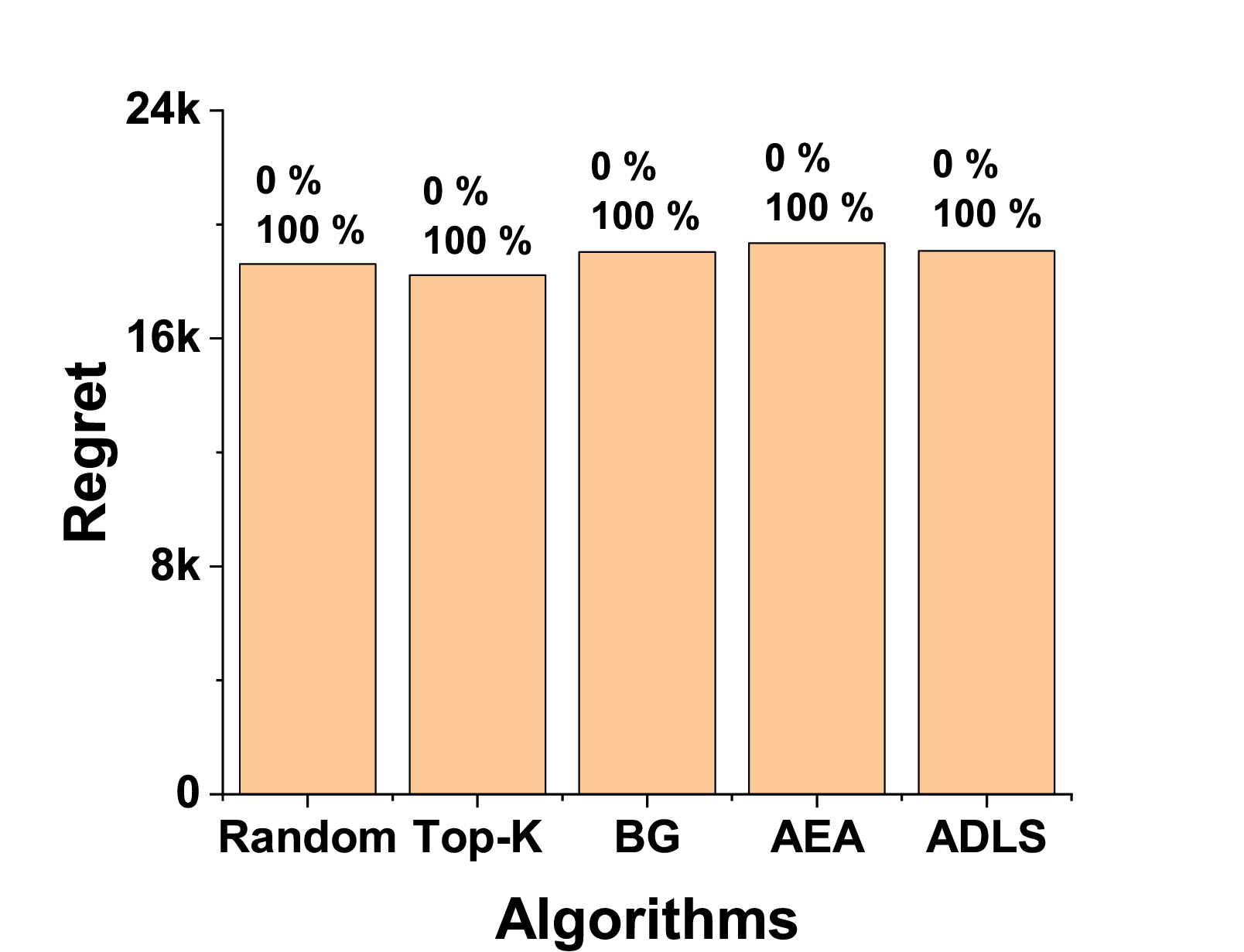} &
   \includegraphics[scale=0.1]{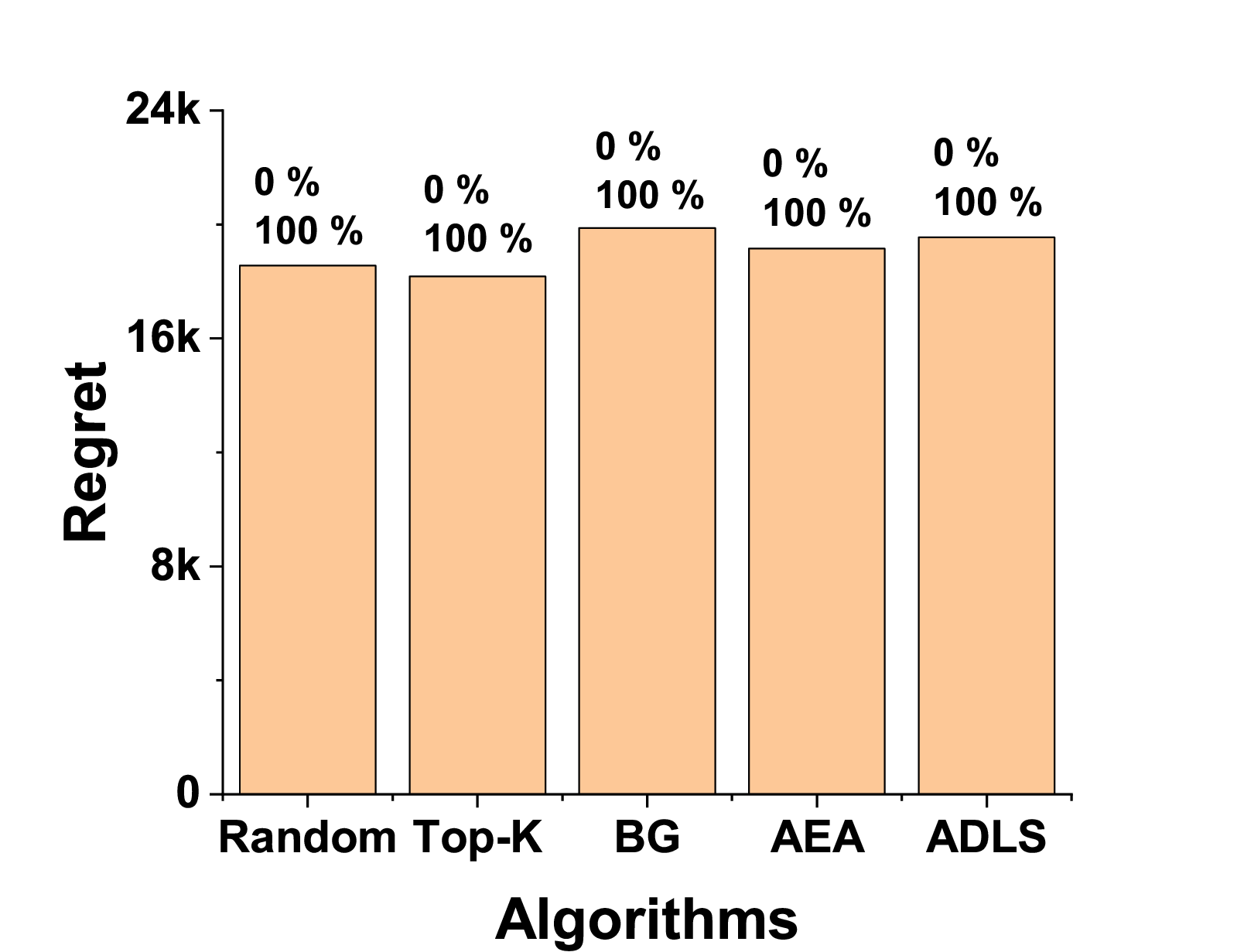} &
   \includegraphics[scale=0.1]{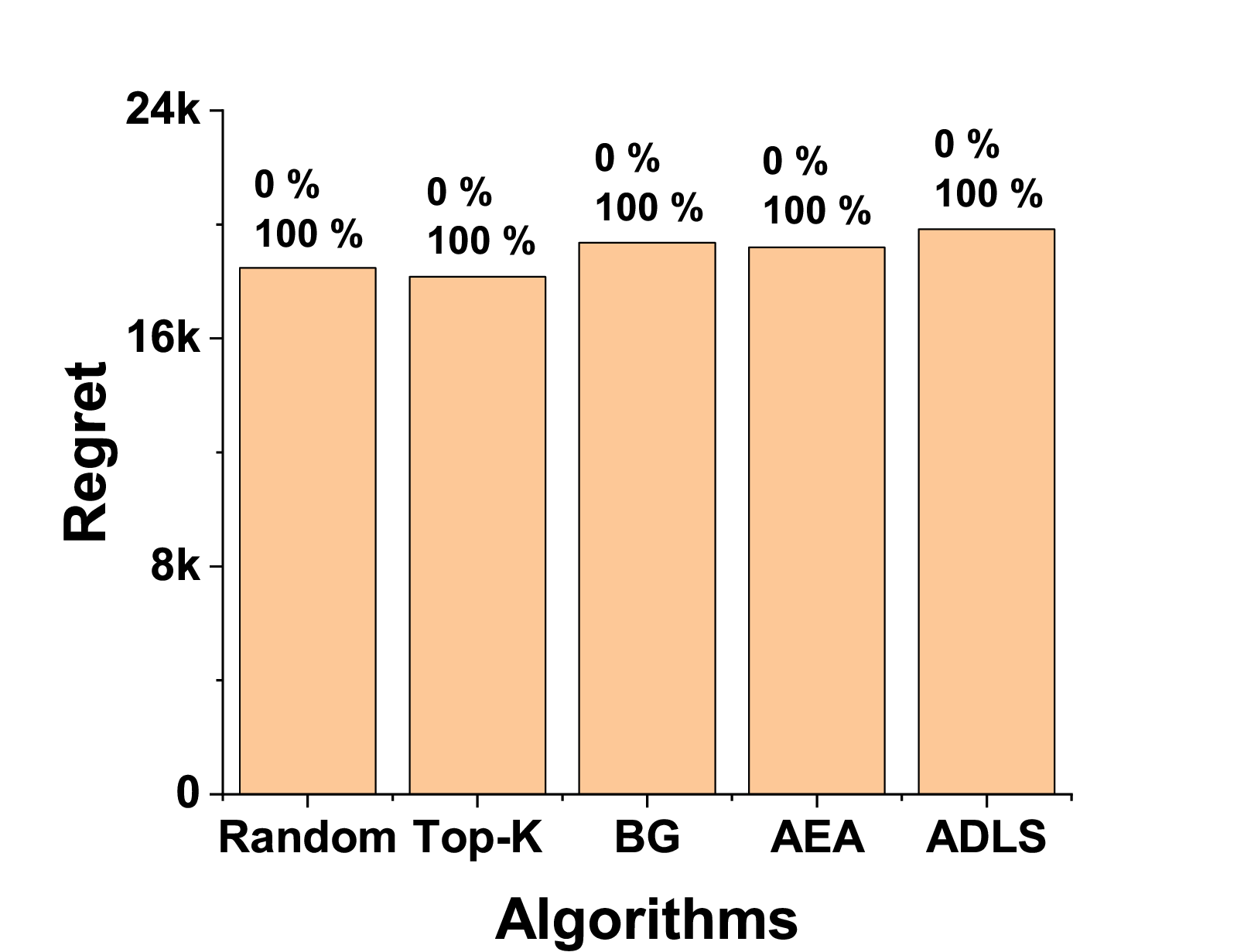} \\
    {\tiny (a) $\delta = 0.01$ } & 
    {\tiny (b) $\delta = 0.02$} &
    {\tiny (c) $\delta = 0.03$} &
    {\tiny (d) $\delta = 0.04$} & 
    {\tiny (e) $\delta = 0.05$} \\
\end{tabular}
\caption{Parameter Study for $\delta$ $(|\mathcal{A}| = 20)$ Trivalency Probability Setting (Email-Eu-Core)}
\label{Fig:16_DeltaTV}
\end{figure}

\begin{longtable}{|ccc|ccccc|}
\caption{Congress-Twitter dataset Regret Values for Uniform (0.1) probability setting}
\label{tab:congressregretuniform01}\\
\hline
\multicolumn{3}{|c|}{\textbf{Parameters}} &
  \multicolumn{5}{c|}{\textbf{Total Regret \([ \text{Uniform}(0.1) ]\)}} \\ \hline
\multicolumn{1}{|c|}{\(\boldsymbol{\lambda} \, (\%)\)} &
  \multicolumn{1}{c|}{\(\boldsymbol{\omega} \, (\%)\)} &
  \(\boldsymbol{|A|}\) &
  \multicolumn{1}{c|}{\textbf{Random}} &
  \multicolumn{1}{c||}{\textbf{Top-K}} &
  \multicolumn{1}{c|}{\textbf{BG}} &
  \multicolumn{1}{c|}{\textbf{AEA}} &
  \textbf{ADLC} \\ \hline
\endfirsthead

\hline
\multicolumn{3}{|c|}{\textbf{Parameters}} &
  \multicolumn{5}{c|}{\textbf{Total Regret \([ \text{Uniform}(0.1) ]\)}} \\ \hline
\multicolumn{1}{|c|}{\(\boldsymbol{\lambda} \, (\%)\)} &
  \multicolumn{1}{c|}{\(\boldsymbol{\omega} \, (\%)\)} &
  \(\boldsymbol{|A|}\) &
  \multicolumn{1}{c|}{\textbf{Random}} &
  \multicolumn{1}{c||}{\textbf{Top-K}} &
  \multicolumn{1}{c|}{\textbf{BG}} &
  \multicolumn{1}{c|}{\textbf{AEA}} &
  \textbf{ADLC} \\ \hline
\endhead
\multicolumn{1}{|c|}{40} &
  \multicolumn{1}{c|}{10} &
  100 &
  \multicolumn{1}{c|}{13376.81} &
  \multicolumn{1}{c||}{13291.20} &
  \multicolumn{1}{c|}{12585.92} &
  \multicolumn{1}{c|}{\textbf{12233.51}} &
  12385.94 \\ \hline
\multicolumn{1}{|c|}{60} &
  \multicolumn{1}{c|}{10} &
  100 &
  \multicolumn{1}{c|}{25391.24} &							  
  \multicolumn{1}{c||}{29737.52} &
  \multicolumn{1}{c|}{\textbf{19828.20}} &
  \multicolumn{1}{c|}{20616.84} &
  20321.09 \\ \hline
\multicolumn{1}{|c|}{80} &
  \multicolumn{1}{c|}{10} &
  100 &
  \multicolumn{1}{c|}{40645.32} &
  \multicolumn{1}{c||}{49987.67} &
  \multicolumn{1}{c|}{29950.00} &
  \multicolumn{1}{c|}{\textbf{29574.21}} &
  30191.59 \\ \hline
\multicolumn{1}{|c|}{100} &
  \multicolumn{1}{c|}{10} &
  100 &
  \multicolumn{1}{c|}{48480.66} &
  \multicolumn{1}{c||}{62314.09} &
  \multicolumn{1}{c|}{34377.04} &
  \multicolumn{1}{c|}{32877.58} &
  \textbf{32745.78} \\ \hline
\multicolumn{1}{|c|}{120} &
  \multicolumn{1}{c|}{10} &
  100 &
  \multicolumn{1}{c|}{48100.90} &
  \multicolumn{1}{c||}{62781.68} &
  \multicolumn{1}{c|}{\textbf{32706.01}} &
  \multicolumn{1}{c|}{33093.89} &
  32975.50 \\ \hline
\multicolumn{1}{|c|}{40} &
  \multicolumn{1}{c|}{30} &
  50 &
  \multicolumn{1}{c|}{29930.27} &
  \multicolumn{1}{c||}{35879.33} &
  \multicolumn{1}{c|}{\textbf{14809.39}} &
  \multicolumn{1}{c|}{15269.76} &
  15060.02 \\ \hline
\multicolumn{1}{|c|}{60} &
  \multicolumn{1}{c|}{30} &
  50 &
  \multicolumn{1}{c|}{30837.08} &
  \multicolumn{1}{c||}{36511.97} &
  \multicolumn{1}{c|}{11417.71} &
  \multicolumn{1}{c|}{11343.79} &
  \textbf{11059.94} \\ \hline
\multicolumn{1}{|c|}{80} &
  \multicolumn{1}{c|}{30} &
  50 &
  \multicolumn{1}{c|}{31338.21} &
  \multicolumn{1}{c||}{34550.39} &
  \multicolumn{1}{c|}{7838.85} &
  \multicolumn{1}{c|}{\textbf{7446.80}} &
  8054.58 \\
\hline
\end{longtable}

\begin{longtable}{|ccc|ccccc|}
\caption{Congress-Twitter dataset Regret Values for Trivalency probability setting}
\label{tab:congressregretTrivalency}\\
\hline
\multicolumn{3}{|c|}{\textbf{Parameters}} &
  \multicolumn{5}{c|}{\textbf{Total Regret \([ \text{Trivalency} ]\)}} \\ \hline
\multicolumn{1}{|c|}{\(\boldsymbol{\lambda} \, (\%)\)} &
  \multicolumn{1}{c|}{\(\boldsymbol{\omega} \, (\%)\)} &
  \(\boldsymbol{|A|}\) &
  \multicolumn{1}{c|}{\textbf{Random}} &
  \multicolumn{1}{c||}{\textbf{Top-K}} &
  \multicolumn{1}{c|}{\textbf{BG}} &
  \multicolumn{1}{c|}{\textbf{AEA}} &
  \textbf{ADLC} \\ \hline
\endfirsthead

\hline
\multicolumn{3}{|c|}{\textbf{Parameters}} &
  \multicolumn{5}{c|}{\textbf{Total Regret \([ \text{Trivalency} ]\)}} \\ \hline
\multicolumn{1}{|c|}{\(\boldsymbol{\lambda} \, (\%)\)} &
  \multicolumn{1}{c|}{\(\boldsymbol{\omega} \, (\%)\)} &
  \(\boldsymbol{|A|}\) &
  \multicolumn{1}{c|}{\textbf{Random}} &
  \multicolumn{1}{c||}{\textbf{Top-K}} &
  \multicolumn{1}{c|}{\textbf{BG}} &
  \multicolumn{1}{c|}{\textbf{AEA}} &
  \textbf{ADLC} \\ \hline
\endhead
\multicolumn{1}{|c|}{40} &
  \multicolumn{1}{c|}{10} &
  100 &
  \multicolumn{1}{c|}{5408.10} &
  \multicolumn{1}{c||}{5392.73} &
  \multicolumn{1}{c|}{5341.50} &
  \multicolumn{1}{c|}{\textbf{5212.88}} &
  5333.35 \\ \hline
\multicolumn{1}{|c|}{60} &
  \multicolumn{1}{c|}{10} &
  100 &
  \multicolumn{1}{c|}{7625.98} &
  \multicolumn{1}{c||}{7505.07} &
  \multicolumn{1}{c|}{7529.86} &
  \multicolumn{1}{c|}{\textbf{7348.92}} &
  7533.65 \\ \hline
\multicolumn{1}{|c|}{80} &
  \multicolumn{1}{c|}{10} &
  100 &
  \multicolumn{1}{c|}{10822.32} &
  \multicolumn{1}{c||}{10296.99} &
  \multicolumn{1}{c|}{10307.54} &
  \multicolumn{1}{c|}{\textbf{10187.87}} &
  10360.18 \\ \hline
\multicolumn{1}{|c|}{100} &
  \multicolumn{1}{c|}{10} &
  100 &
  \multicolumn{1}{c|}{13672.16} &
  \multicolumn{1}{c||}{13205.79} &
  \multicolumn{1}{c|}{13174.69} &
  \multicolumn{1}{c|}{13180.11} &
  \textbf{13144.07} \\ \hline
\multicolumn{1}{|c|}{120} &
  \multicolumn{1}{c|}{10} &
  100 &
  \multicolumn{1}{c|}{14720.62} &
  \multicolumn{1}{c||}{\textbf{13965.16}} &
  \multicolumn{1}{c|}{14134.05} &
  \multicolumn{1}{c|}{14074.29} &
  14085.10 \\ \hline
\multicolumn{1}{|c|}{40} &
  \multicolumn{1}{c|}{30} &
  50 &
  \multicolumn{1}{c|}{7627.89} &
  \multicolumn{1}{c||}{\textbf{7202.01}} &
  \multicolumn{1}{c|}{7366.88} &
  \multicolumn{1}{c|}{7272.52} &
  7244.92 \\ \hline
\multicolumn{1}{|c|}{60} &
  \multicolumn{1}{c|}{30} &
  50 &
  \multicolumn{1}{c|}{11601.32} &
  \multicolumn{1}{c||}{\textbf{11078.86}} &
  \multicolumn{1}{c|}{11139.29} &
  \multicolumn{1}{c|}{11223.31} &
  11166.68 \\ \hline
\multicolumn{1}{|c|}{80} &
  \multicolumn{1}{c|}{30} &
  50 &
  \multicolumn{1}{c|}{16490.07} &
  \multicolumn{1}{c||}{16026.52} &
  \multicolumn{1}{c|}{15949.91} &
  \multicolumn{1}{c|}{16035.33} &
  \textbf{15907.54} \\
\hline
\end{longtable}

\begin{longtable}{|ccc|ccccc|}
\caption{Congress-Twitter dataset Execution Time for Uniform (0.1) probability setting}
\label{tab:congressexecutiontimeuniform}\\
\hline
\multicolumn{3}{|c|}{\textbf{Parameters}} &
  \multicolumn{5}{c|}{\textbf{Execution Time (in seconds) \([ \text{Uniform}(0.1) ]\)}} \\ \hline
\multicolumn{1}{|c|}{\(\boldsymbol{\lambda}\) $(\%)$} &
  \multicolumn{1}{c|}{\(\boldsymbol{\omega}\) $(\%)$ } &
  \(\boldsymbol{|A|}\) &
  \multicolumn{1}{c|}{\textbf{Random}} &
  \multicolumn{1}{c||}{\textbf{Top-K}} &
  \multicolumn{1}{c|}{\textbf{BG}} &
  \multicolumn{1}{c|}{\textbf{AEA}} &
  \textbf{ADLC} \\ \hline
\endfirsthead

\hline
\multicolumn{3}{|c|}{\textbf{Parameters}} &
  \multicolumn{5}{c|}{\textbf{Execution Time (in seconds) \([ \text{Uniform}(0.1) ]\)}} \\ \hline
\multicolumn{1}{|c|}{\(\boldsymbol{\lambda}\)$(\%)$} &
  \multicolumn{1}{c|}{\(\boldsymbol{\omega}\)$(\%)$} &
  \(\boldsymbol{|A|}\) &
  \multicolumn{1}{c|}{\textbf{Random}} &
  \multicolumn{1}{c||}{\textbf{Top-K}} &
  \multicolumn{1}{c|}{\textbf{BG}} &
  \multicolumn{1}{c|}{\textbf{AEA}} &
  \textbf{ADLC} \\ \hline
\endhead\multicolumn{1}{|c|}{40} &
  \multicolumn{1}{c|}{10} &
  100 &
  \multicolumn{1}{c|}{56.700} &
  \multicolumn{1}{c||}{58.294} &
  \multicolumn{1}{c|}{21.560} &
  \multicolumn{1}{c|}{617.561} &
  42.581 \\ \hline
\multicolumn{1}{|c|}{60} &
  \multicolumn{1}{c|}{10} &
  100 &
  \multicolumn{1}{c|}{59.154} &
  \multicolumn{1}{c||}{57.799} &
  \multicolumn{1}{c|}{27.865} &
  \multicolumn{1}{c|}{410.538} &
  50.779 \\ \hline
\multicolumn{1}{|c|}{80} &
  \multicolumn{1}{c|}{10} &
  100 &
  \multicolumn{1}{c|}{59.630} &
  \multicolumn{1}{c||}{59.489} &
  \multicolumn{1}{c|}{36.364} &
  \multicolumn{1}{c|}{217.782} &
  65.914 \\ \hline
\multicolumn{1}{|c|}{100} &
  \multicolumn{1}{c|}{10} &
  100 &
  \multicolumn{1}{c|}{60.598} &
  \multicolumn{1}{c||}{55.892} &
  \multicolumn{1}{c|}{41.825} &
  \multicolumn{1}{c|}{120.250} &
  79.034 \\ \hline
\multicolumn{1}{|c|}{120} &
  \multicolumn{1}{c|}{10} &
  100 &
  \multicolumn{1}{c|}{59.907} &
  \multicolumn{1}{c||}{60.782} &
  \multicolumn{1}{c|}{43.828} &
  \multicolumn{1}{c|}{96.583} &
  73.505 \\ \hline
\multicolumn{1}{|c|}{40} &
  \multicolumn{1}{c|}{30} &
  50 &
  \multicolumn{1}{c|}{28.110} &
  \multicolumn{1}{c||}{28.026} &
  \multicolumn{1}{c|}{26.351} &
  \multicolumn{1}{c|}{26.746} &
  48.043 \\ \hline
\multicolumn{1}{|c|}{60} &
  \multicolumn{1}{c|}{30} &
  50 &
  \multicolumn{1}{c|}{29.416} &
  \multicolumn{1}{c||}{28.035} &
  \multicolumn{1}{c|}{36.856} &
  \multicolumn{1}{c|}{37.348} &
  67.160 \\ \hline
\multicolumn{1}{|c|}{80} &
  \multicolumn{1}{c|}{30} &
  50 &
  \multicolumn{1}{c|}{30.369} &
  \multicolumn{1}{c||}{28.308} &
  \multicolumn{1}{c|}{46.156} &
  \multicolumn{1}{c|}{46.348} &
  73.318 \\
\hline
\end{longtable}

\begin{longtable}{|ccc|ccccc|}
\caption{Congress-Twitter dataset Execution Time for Trivalency probability setting}
\label{tab:congressexecutiontimetrivalency}\\
\hline
\multicolumn{3}{|c|}{\textbf{Parameters}} &
  \multicolumn{5}{c|}{\textbf{Execution Time (in seconds) \([ \text{Trivalency}]\)}} \\ \hline
\multicolumn{1}{|c|}{\(\boldsymbol{\lambda}\)$(\%)$} &
  \multicolumn{1}{c|}{\(\boldsymbol{\omega}\)$(\%)$} &
  \(\boldsymbol{|A|}\) &
  \multicolumn{1}{c|}{\textbf{Random}} &
  \multicolumn{1}{c||}{\textbf{Top-K}} &
  \multicolumn{1}{c|}{\textbf{BG}} &
  \multicolumn{1}{c|}{\textbf{AEA}} &
  \textbf{ADLC} \\ \hline
\endfirsthead

\hline
\multicolumn{3}{|c|}{\textbf{Parameters}} &
  \multicolumn{5}{c|}{\textbf{Execution Time (in seconds) \([ \text{Trivalency}]\)}} \\ \hline
\multicolumn{1}{|c|}{\(\boldsymbol{\lambda}\)$(\%)$} &
  \multicolumn{1}{c|}{\(\boldsymbol{\omega}\)$(\%)$} &
  \(\boldsymbol{|A|}\) &
  \multicolumn{1}{c|}{\textbf{Random}} &
  \multicolumn{1}{c||}{\textbf{Top-K}} &
  \multicolumn{1}{c|}{\textbf{BG}} &
  \multicolumn{1}{c|}{\textbf{AEA}} &
  \textbf{ADLC} \\ \hline
\endhead
\multicolumn{1}{|c|}{40} &
  \multicolumn{1}{c|}{10} &
  100 &
  \multicolumn{1}{c|}{53.777} &
  \multicolumn{1}{c||}{54.330} &
  \multicolumn{1}{c|}{20.077} &
  \multicolumn{1}{c|}{1138.233} &
  42.125 \\ \hline
\multicolumn{1}{|c|}{60} &
  \multicolumn{1}{c|}{10} &
  100 &
  \multicolumn{1}{c|}{55.575} &
  \multicolumn{1}{c||}{53.374} &
  \multicolumn{1}{c|}{24.601} &
  \multicolumn{1}{c|}{1589.903} &
  43.460 \\ \hline
\multicolumn{1}{|c|}{80} &
  \multicolumn{1}{c|}{10} &
  100 &
  \multicolumn{1}{c|}{56.355} &
  \multicolumn{1}{c||}{54.696} &
  \multicolumn{1}{c|}{28.813} &
  \multicolumn{1}{c|}{2028.303} &
  56.383 \\ \hline
\multicolumn{1}{|c|}{100} &
  \multicolumn{1}{c|}{10} &
  100 &
  \multicolumn{1}{c|}{56.469} &
  \multicolumn{1}{c||}{52.144} &
  \multicolumn{1}{c|}{36.997} &
  \multicolumn{1}{c|}{2276.420} &
  75.952 \\ \hline
\multicolumn{1}{|c|}{120} &
  \multicolumn{1}{c|}{10} &
  100 &
  \multicolumn{1}{c|}{55.759} &
  \multicolumn{1}{c||}{55.878} &
  \multicolumn{1}{c|}{38.417} &
  \multicolumn{1}{c|}{2375.799} &
  77.629 \\ \hline
\multicolumn{1}{|c|}{40} &
  \multicolumn{1}{c|}{30} &
  50 &
  \multicolumn{1}{c|}{26.555} &
  \multicolumn{1}{c||}{26.205} &
  \multicolumn{1}{c|}{23.594} &
  \multicolumn{1}{c|}{737.068} &
  46.575 \\ \hline
\multicolumn{1}{|c|}{60} &
  \multicolumn{1}{c|}{30} &
  50 &
  \multicolumn{1}{c|}{27.850} &
  \multicolumn{1}{c||}{26.171} &
  \multicolumn{1}{c|}{37.201} &
  \multicolumn{1}{c|}{1115.152} &
  74.822 \\ \hline
\multicolumn{1}{|c|}{80} &
  \multicolumn{1}{c|}{30} &
  50 &
  \multicolumn{1}{c|}{28.025} &
  \multicolumn{1}{c||}{26.693} &
  \multicolumn{1}{c|}{53.038} &
  \multicolumn{1}{c|}{1558.651} &
  100.934 \\
\hline
\end{longtable}

\subsubsection{\textbf{Efficiency Study}}
This section analyzes the execution time performance of various algorithms, ``Top-$K$", ``Random", ``BG", ``AEA", and ``ADLS"— across multiple datasets under both Uniform and Trivalency influence probability settings. The evaluation considers the effects of parameters $\omega$ (individual influence demand) and $\lambda$ (global influence demand) on computational efficiency. The execution time under Uniform (0.1) probability setting is observed as follows. In Figures $\ref{Fig:7_Time0.1}$, for the Email-Eu-Core dataset, as $\omega$ increases, execution times decrease for all methods. The ``Top-$K$" algorithm is the fastest across all settings, followed closely by ``Random". The algorithms ``BG", ``AEA", and ``ADLS" exhibit significantly higher execution times, particularly at low $\omega$ values. Among these, ``ADLS" is generally the slowest algorithm, with execution times up to $292\%$ longer than ``Random" at $\omega = 10\%$. The ``BG" algorithm is also slower than ``Random", with execution times ranging from $50\%$ to $150\%$ higher. Both ``AEA" and ``ADLS" show extreme delays, particularly under high $\lambda$ and low $\omega$ conditions. For instance, at $\omega = 50\%$ and $\lambda = 100\%$, the execution time of ``AEA" reaches $1002$ seconds, exceeding that of ``Random" by more than $1000\%$. Similarly, ``ADLS" is over $500\%$ slower than ``Random" in several cases. In Figures $\ref{Fig:9_Time0.1}$, for the Facebook dataset, across all methods, execution times decrease as $\omega$ increases, indicating that as individual influence demand grows, computations become less intensive. The ``Top-$K$" algorithm consistently achieves the lowest execution times, followed closely by ``Random". Conversely, ``ADLS" consistently exhibits the highest execution times, particularly at low $\omega$ values, where it is up to $292\%$ slower than ``Random". The ``BG" algorithm is generally $50\%$ to $150\%$ slower than ``Random". Both ``AEA" and ``ADLS" experience extreme delays, especially when $\lambda$ is high and $\omega$ is low. The ``BG", ``AEA", and ``ADLS" algorithms incur significantly higher computation times, with ``ADLS" reaching $17753$ seconds at $\lambda = 120\%$, which is over $57$ times slower than ``Random" ($310.67$ seconds). For the Wikivote dataset, in Figure $\ref{Fig:11_Time0.1}$, the ``Top-$K$" and ``Random" algorithms are the fastest across all settings. In contrast, ``BG", ``AEA", and ``ADLS" exhibit significant increases in execution time, particularly at low $\omega$ values. Among all methods, ``AEA" is the slowest overall, especially when $\lambda$ is high. Execution times decrease as $\omega$ increases, indicating that higher $\omega$ leads to more efficient computations. At low values of $\omega = 10\%$ and $\omega = 30\%$, execution times remain significantly high across all algorithms. For instance, at $\omega = 10\%$ and $\lambda = 120\%$, the execution times are ``BG": $\sim 11288$ seconds, ``AEA": $\sim 11288$ seconds, and ``ADLS": $\sim 14600$ seconds (extremely slow). At $\omega = 30\%$ and $\lambda = 120\%$, ``AEA" reaches an enormous execution time of $712164$ seconds. Conversely, at high values of $\omega = 70\%$ and $\omega = 90\%$, execution times drop significantly. At $\omega = 90\%$ and $\lambda = 120\%$, the execution times are ``Random": $\sim 21$ seconds, ``Top-$K$": $\sim 18$ seconds, ``BG": $\sim 9993$ seconds, ``AEA": $\sim 21387$ seconds, and ``ADLS": $\sim 17351$ seconds. These results suggest that higher values of $\omega$ significantly reduce computational overhead. From Table $\ref{tab:congressexecutiontimeuniform}$, for Congress-Twitter dataset we analyze that ``ADLS" performs better than ``AEA" but worse than ``BG", ``Top-$K$", and ``Random". The ``Top-$K$" algorithm is slightly faster than ``Random" at high $\omega$ values.

We also observe the performance of all algorithms for all datasets under the Trivalency probability setting. From Figures $\ref{Fig:8_TimeTV}$, our observations for the Email-Eu-Core dataset are as follows. The ``Random" and ``Top-$K$" algorithms perform best. In contrast, ``BG", ``AEA", and ``ADLS" scale poorly as $\lambda$ increases. ``AEA" exhibits extremely high execution times for large $\lambda$ values, making it highly inefficient. From Figures $\ref{Fig:10_TimeTV}$, for the Facebook dataset, we observe that the algorithms ``BG", ``AEA", and ``ADLS" experience significant increases in execution time for high $\lambda$. Among them, ``AEA" has the worst performance, particularly at low $\omega$. Execution times are higher across all algorithms compared to other datasets in the Wikivote dataset, as shown in Figure $\ref{Fig:12_TimeTV}$. ``AEA" incurs extreme computation times, rendering it highly inefficient. ``ADLS" also suffers from large execution times, although it performs slightly better than ``AEA". From Table $\ref{tab:congressexecutiontimetrivalency}$, for the Congress-Twitter dataset, our observations indicate that ``ADLS" is slower than ``BG", ``Random", and ``Top-$K$", but much faster than ``AEA". ``Random" and ``Top-$K$" have similar performance, with ``Random" being slightly faster at high $\omega$.

In general, ``Top-$K$" and ``Random" are the fastest methods, while ``AEA" and ``ADLS" exhibit significantly higher execution times, especially under low $\omega$ and high $\lambda$ conditions. The results highlight the computational challenges faced by more sophisticated algorithms like ``AEA" and ``ADLS", particularly in large-scale or high-demand scenarios.

\subsubsection{\textbf{Other Parameter Study}}
The additional parameters we have used in our experiments are $\gamma$ and $\delta$. First we discuss the effect of $\gamma$ in our experiment. We have taken five different values of this parameter. From the Figures $\ref{Fig:13_Gamma0.1}$ and $\ref{Fig:14_GammaTV}$, we see that as $\gamma$ increases, the total regret for all algorithms decreases. In the Figure $\ref{Fig:13_Gamma0.1}$ (a), when $\gamma = 0$, that means there is no penalty on the provided influence. In this situation the regret is highest as the full budget contribute to the regret. On the other hand, when $\gamma = 1$, the total regret is lowest, which means that the influence provider receives a payment proportional to the ratio of the influence it provided. The other parameter we have considered is $\delta$, for its five different values. In the Figures $\ref{Fig:15_Delta0.1}$ and $\ref{Fig:16_DeltaTV}$, the effect of increasing the value of $\delta$ is nearly same. There is no significant increase or decrease with these values of $\delta$. Another parameter which is already discussed is the influence probability setting. We have used Uniform ($0.1$) and Trivalency influence probability settings for our experiments on all datasets. Our approach ``ADLS" dominates across all cases in both probability settings giving lower regrets. The only difference is between the two probability settings is that under Uniform, the regret values are generally higher than that of Trivalency.

\section{Concluding Remarks and Future Research Directions} \label{Sec:CFD}
In this paper we have studied the Regret Minimization in Social Media Advertisement Problem and posed this problem as a discrete optimization problem. We have shown that this problem is computationally hard and also hard to approximate within a constant factor. We have proposed three efficient heuristic solution approaches with a detailed complexity analysis. The proposed methodologies have been elaborated on with an example. A number of experiments have been conducted to show the effectiveness and efficiency of all the algorithms.  From the reported results, it has been observed that the proposed solution approaches lead to much less regret compared to the baseline methods. ``ADLS" consistently performs best in minimizing regret, making it the optimal choice for advertiser allocation under different probability settings. ``BG" and ``AEA" are competitive but fail in high demand scenarios where ``ADLS" excels. ``Random" and ``Top-$K$" methods are ineffective, leading to significantly higher total regret. Now, this study can be extended in the following directions. More efficient solution methodologies are of thrust. Also, a new regret model may be proposed, which will capture the loss in a much better way. The scalable algorithms for regret minimization in large-scale online social networks are to be designed.
\section*{Acknowledgement}
The authors would like to acknowledge Indian Institute of Technology (IIT) Jammu, India, with grant number SG100047, for providing the funds required for the project.
\bibliography{bibliography}
\end{document}